\documentclass{emulateapj}

\slugcomment{AAS}

\shorttitle{ALMA multiple-transition observations of high density
molecular tracers in ULIRGs}
\shortauthors{Imanishi et al.}

\begin{document}


\title{ALMA Multiple-Transition Observations of High Density Molecular
Tracers in Ultraluminous Infrared Galaxies}    


\author{Masatoshi Imanishi\altaffilmark{1}}
\affil{National Astronomical Observatory of Japan, National Institutes 
of Natural Sciences (NINS), 2-21-1 Osawa, Mitaka, Tokyo 181-8588, Japan}
\email{masa.imanishi@nao.ac.jp}

\author{Kouichiro Nakanishi\altaffilmark{1}}
\affil{National Astronomical Observatory of Japan, National Institutes 
of Natural Sciences (NINS), 2-21-1 Osawa, Mitaka, Tokyo 181-8588, Japan}

\and

\author{Takuma Izumi}
\affil{National Astronomical Observatory of Japan, National Institutes 
of Natural Sciences (NINS), 2-21-1 Osawa, Mitaka, Tokyo 181-8588, Japan}

\altaffiltext{1}{Department of Astronomy, School of Science, Graduate
University for Advanced Studies (SOKENDAI), Mitaka, Tokyo 181-8588}

\begin{abstract} 
We present the results of our ALMA observations of eleven
(ultra)luminous infrared galaxies ((U)LIRGs) at J=4--3 of HCN,
HCO$^{+}$, HNC and J=3--2 of HNC.   
This is an extension of our previously published HCN and HCO$^{+}$
J=3--2 observations to multiple rotational J-transitions of
multiple molecules, to investigate how molecular emission line
flux ratios vary at different J-transitions.
We confirm that ULIRGs that contain or may contain luminous obscured AGNs
tend to show higher HCN-to-HCO$^{+}$ flux ratios than
starburst galaxies, both at J=4--3 and J=3--2.
For selected HCN-flux-enhanced AGN-important ULIRGs, our isotopologue
H$^{13}$CN, H$^{13}$CO$^{+}$, and HN$^{13}$C J=3--2 line observations
suggest a higher abundance of HCN than HCO$^{+}$ and HNC, which is
interpreted to be primarily responsible for the elevated HCN flux in
AGN-important galaxies. 
For such sources, the intrinsic HCN-to-HCO$^{+}$ flux ratios after line
opacity correction will be higher than the observed ratios, 
making the separation between AGNs and starbursts even larger.
The signature of the vibrationally excited (v$_{2}$=1f) HCN J=4--3
emission line is seen in one ULIRG, IRAS 12112$-$0305 NE. 
P Cygni profiles are detected in the HCO$^{+}$ J=4--3 and
J=3--2 lines toward IRAS 15250$+$3609, with an estimated molecular outflow  
rate of $\sim$250--750 M$_{\odot}$ yr$^{-1}$.
The SiO J=6--5 line also exhibits a P Cygni profile in IRAS
12112$+$0305 NE, suggesting the presence of shocked outflow activity.   
Shock tracers are detected in many sources, suggesting
ubiquitous shock activity in the nearby ULIRG population.
\end{abstract} 
 
\keywords{galaxies: active --- galaxies: nuclei --- quasars: general ---
galaxies: Seyfert --- galaxies: starburst --- submillimeter: galaxies}

\section{Introduction} 

Ultraluminous infrared galaxies (ULIRGs) and luminous infrared galaxies
(LIRGs) radiate very large infrared luminosity with L$_{\rm IR}$ 
$>$ 10$^{12}$L$_{\odot}$ and $>$ 10$^{11}$L$_{\odot}$, respectively.
They are mostly gas-rich galaxy mergers \citep{sam96}.
The large infrared luminosity means that powerful energy sources, either
starbursts and/or active galactic nuclei (AGNs), are obscured and/or
surrounded by dust.  
Clarifying how starbursts and AGNs energetically contribute to (U)LIRG's
bolometric luminosities is fundamental to understanding the physical
nature of the (U)LIRG population. 
However, because AGNs are spatially compact compared with starburst
activity, the putative AGNs in merging (U)LIRGs can easily be hidden deep
inside nuclear gas and dust. 
It is highly desirable to establish a solid method to disentangle 
between the two energy sources, based on observations at wavelengths
where dust extinction effects are small.

In a starburst, radiative energy is released by a nuclear fusion, while
a mass-accreting supermassive black hole (SMBH) is the main
energy-generating mechanism in the case of an AGN.   
The surrounding molecular gas is expected to receive different
radiative and/or mechanical effects from the different energy sources,
so that chemical and physical conditions can be different. 
These chemically and physically different molecular gases are expected
to show different emission line fluxes among different molecular
species.  
As the effects of dust extinction are usually negligible in the
(sub)millimeter wavelength range where rotational J-transition lines of
many molecules exist, (sub)millimeter molecular line observations can
be used to scrutinize deeply obscured energy sources in the (U)LIRG's
nuclei. To minimize possible ambiguities about the interpretation of
molecular emission line flux ratios, using molecules with comparable
critical density is recommended, because the differences in their
spatial distributions are expected to be much smaller than that among
molecules with large differences in their critical densities.
A comparison of the flux ratios of HCN, HCO$^{+}$, and HNC emission
lines is a good choice, because 
(1) they have similar dipole moments ($\mu$ = 3.0, 3.9, and 3.1 debye for
HCN, HCO$^{+}$, and HNC, respectively); thus, the critical densities of
HCN, HCO$^{+}$, and HNC do not differ a lot at the same J-transitions,  
(2) these molecular lines probe dense (n$_{\rm H2}$ $>$10$^{4}$
cm$^{-3}$) molecular gas, which usually dominates (U)LIRG's nuclear
regions \citep{gao04a,gao04b}, and 
(3) these emission lines are modestly bright, so that their detection
is feasible in a large number of (U)LIRGs, including distant ones.

Based on millimeter interferometric observations of the nuclear regions
of nearby bright starburst and Seyfert galaxies (= modestly luminous
AGNs) before the ALMA era, a trend of higher HCN-to-HCO$^{+}$ J=1--0
(rotational transition) flux ratios in AGNs than in starbursts was found 
and argued to be a good AGN indicator \citep{koh05,kri08}.
This method was applied to dusty (U)LIRG nuclei, and it was found that 
sources with infrared-detectable luminous buried AGN signatures tend to
show higher HCN-to-HCO$^{+}$ J=1--0 flux ratios than those without such
signatures \citep{ima04,ima06b,in06,ima07a,ima09,pri15}. 
With the advent of ALMA, similarly enhanced HCN-to-HCO$^{+}$ flux
ratios in sources with infrared-identified AGNs have been confirmed also
at higher-J transitions, J=3--2 and J=4--3  
\citep{ima13a,ima13b,ion13,ima14b,gar14,izu15,izu16,ima16a,ima16b,ima16c,ima17,ima18}.
Although a small fraction of the observed sources display some
exceptions and need further detailed investigations
\citep{ima14b,pri15}, the flux comparison of HCN and HCO$^{+}$ could
potentially be an effective means to detect elusive buried AGNs.     
Further pursuit of this method is highly desirable.

A few possible origins of the higher HCN-to-HCO$^{+}$ J-transition line 
flux ratios in AGNs, compared with starbursts, are proposed. 
First, an enhanced HCN abundance in an AGN could be a reason for the
higher HCN flux \citep{yam07,izu16}, because increased abundance usually
results in increased flux, unless the flux saturates by line opacity
effects. 
In fact, such an HCN abundance enhancement in molecular gas illuminated
by a luminous AGN is predicted by some chemical calculations in some 
parameter ranges \citep{mei05,har10}.
Second, higher HCN excitation in AGNs could produce higher
HCN-to-HCO$^{+}$ J-transition line flux ratios, even without introducing
an HCN abundance enhancement. 
The critical density of HCN is a factor of $\sim$5 higher than
that of HCO$^{+}$ at the same J-transition line under the same line
opacity \citep{mei07,gre09}.
Because the radiative energy generation efficiency of a mass-accreting
SMBH in an AGN is much higher than that of a nuclear fusion 
inside stars in a starburst, molecular gas temperature in close vicinity
to an AGN can be higher than a starburst, which could result in higher
HCN excitation.  
Consequently, the observed HCN-to-HCO$^{+}$ flux ratios could be higher
in an AGN than in a starburst, particularly at higher-J transition lines.
Third, an infrared radiative pumping mechanism \citep{aal95,ran11} 
could enhance the fluxes of the HCN J-transition lines,
as compared with collisional excitation alone. Namely, HCN can be
vibrationally excited to the v$_{2}$=1 level by absorbing infrared 14
$\mu$m photons; and through the decay back to the vibrational ground
level (v=0), HCN J-transition line fluxes at v=0 can be higher. 
As an AGN can produce a larger amount of mid-infrared 14-$\mu$m
continuum-emitting hot dust than a starburst with the same bolometric
luminosity, this infrared radiative pumping mechanism is expected to
work more efficiently in an AGN than in a starburst.  
This could increase HCN flux more in an AGN than in a starburst; 
however, the infrared radiative pumping mechanism also works for
HCO$^{+}$, by absorbing infrared 12-$\mu$m photons. 
For the ULIRG IRAS 20551$-$4250, based on the observed infrared spectral
energy distribution at 10--20 $\mu$m, the infrared radiative pumping
rates are estimated to be comparable for HCN and HCO$^{+}$
under the same abundance \citep{ima16b}. 
We need further investigation of whether this infrared radiative pumping 
mechanism strongly increases the observed HCN-to-HCO$^{+}$ flux ratios
in luminous-AGN-containing (U)LIRGs in general. 

If the second excitation effects are largely responsible for the higher
HCN-to-HCO$^{+}$ flux ratios in AGNs than in starbursts, the flux ratios
can be different at different J-transitions, in such a way that the
ratios will decrease with increasing J-transitions in both AGNs 
and starbursts, but will decrease more drastically in starbursts than in
AGNs.
\citet{ima16c} presented the results of HCN-to-HCO$^{+}$ flux ratios at
J=3--2 for a large number of (U)LIRGs.
However, the excitation effects can be hardly disentangled from 
abundance effects, given only one J-transition.  
Extension to additional J-transition lines is needed. 
In this paper, we present the results of the HCN-to-HCO$^{+}$ flux
ratios both at J=4--3 and J=3--2, to see how the ratios vary at
different J-transitions. 
HNC J=3--2 and J=4--3 line data were also taken to investigate their
fluxes relative to HCN and HCO$^{+}$, and to constrain the third
infrared radiative pumping effects, because the infrared radiative
pumping rate for HNC, by absorbing infrared 21.5-$\mu$m photons, is
expected to be much higher than for HCN and HCO$^{+}$ in (U)LIRGs 
\citep{aal07a,ima16b}.    
Interesting features of detected molecular lines will also be discussed.
Throughout this paper, we adopt H$_{0}$ $=$ 71 km s$^{-1}$ Mpc$^{-1}$, 
$\Omega_{\rm M}$ = 0.27, and $\Omega_{\rm \Lambda}$ = 0.73, to be
consistent with our previous studies on related scientific
topics. 
We use molecular parameters derived from the Cologne Database of Molecular
Spectroscopy (CDMS) \citep{mul05} via Splatalogue
(http://www.splatalogue.net). 
HCN, HCO$^{+}$, and HNC represent H$^{12}$C$^{14}$N,
H$^{12}$C$^{16}$O$^{+}$, and H$^{14}$N$^{12}$C, respectively. 
When we say ``molecular line flux ratio'', we are referring to the
``rotational J-transition line flux ratio at v = 0'', unless otherwise
stated. 

\section{Targets} 

We observed eleven (U)LIRGs for which HCN J=3--2 and HCO$^{+}$ J=3--2
data had been taken with ALMA \citep{ima16c}. 
Their basic information is summarized in Table 1, and detailed
observational properties are described by \citet{ima16c}.
All galaxies are categorized as ULIRGs (L$_{\rm IR}$ $>$
10$^{12}$L$_{\odot}$), except NGC 1614 (L$_{\rm IR}$ $\sim$ 5 $\times$ 
10$^{11}$L$_{\odot}$; Table 1), which is classified as a LIRG
\citep{sam96}.
In brief, these galaxies have various types of infrared-diagnosed primary
energy sources, from starburst-dominated to AGN-dominated
\citep{ima16c}, based on the strengths of polycyclic aromatic
hydrocarbon (PAH) emission and dust absorption features in infrared
spectra \citep{ima07b,vei09,ima10,nar10}.  
The LIRG NGC 1614 is classified as starburst-dominated.
All ULIRGs, except IRAS 12112$+$0305, IRAS 22491$-$1808, IRAS
13509$+$0442, and IRAS 20414$-$1651, show infrared-detectable luminous
AGN signatures \citep{ima16c}.  
We aim to investigate the variation of molecular line flux ratios as a
function of infrared-estimated primary energy sources at J=3--2 and
J=4--3, and to obtain information on how the molecular line flux ratios
differ at different J-transitions due to excitation effects. 

\section{Observations and Data Analysis} 

Our observations of eleven (U)LIRGs in the HCN, HCO$^{+}$, and HNC
J=4--3 lines in band 7 (275--373 GHz) and HNC J=3--2 line in band
6 (211--275 GHz) were conducted within our ALMA Cycle 2 program
2013.1.00032.S (PI = M. Imanishi) and Cycle 3 program 2015.1.00027.S (PI
= M. Imanishi). 
Table 2 summarizes our observation details. 
The widest 1.875-GHz width mode was adopted.
In our approved ALMA proposals, we requested to take HCN J=4--3,
HCO$^{+}$ J=4--3, HNC J=4--3, and HNC J=3--2 line data for all galaxies 
in Table 1. 
However, some targeted line data are not available for a fraction of
(U)LIRGs due to the scheduling constraints of ALMA observations.

In band 7, HCN J=4--3 (rest-frame frequency $\nu_{\rm rest}$ = 
354.505 GHz) and HCO$^{+}$ J=4--3 ($\nu_{\rm rest}$ = 356.734 GHz) lines
were simultaneously covered, as well as the vibrationally excited HCN
v$_{2}$=1, l=1f (hereafter v$_{2}$=1f) J=4--3 line ($\nu_{\rm rest}$ =
356.256 GHz).  
HCN J=4--3 and HCO$^{+}$ J=4--3 lines were covered in USB or LSB of
ALMA, depending on Earth's atmospheric transmission at the redshifted
frequencies of targeted lines. 
When they were observed in USB, the CS J=7--6 ($\nu_{\rm rest}$ =
342.883 GHz) line was included in LSB. 

To obtain HNC J=4--3 ($\nu_{\rm rest}$ = 362.630 GHz) line data in band
7, we needed independent observations.
The vibrationally excited HNC v$_{2}$=1f J=4--3 ($\nu_{\rm rest}$ =
365.147 GHz) line data were taken simultaneously.
In band 6, data of the HNC J=3--2 ($\nu_{\rm rest}$ = 271.981 GHz) line, 
together with the vibrationally excited HNC v$_{2}$=1f J=3--2 
($\nu_{\rm rest}$ = 273.870 GHz) line, were obtained.

In addition to these lines, the isotopologue H$^{13}$CN J=3--2
($\nu_{\rm rest}$ = 259.012 GHz), 
H$^{13}$CO$^{+}$ J=3--2 ($\nu_{\rm rest}$ = 260.255 GHz), and 
HN$^{13}$C J=3--2 ($\nu_{\rm rest}$ = 261.263 GHz) line data were taken
independently for three ULIRGs, which show HCN flux excesses and modestly 
bright molecular emission lines (i.e., IRAS 08572$+$3915, 
IRAS 12112$+$0305, and IRAS 22491$-$1808) in our ALMA Cycle 4 program
2016.1.00051.S (PI = M. Imanishi).  
The details of these isotopologue observations are tabulated also in
Table 2.
The primary aim was to estimate possible flux attenuation by line
opacity (not dust extinction) for HCN, HCO$^{+}$, and HNC, based on a
comparison of HCN-to-H$^{13}$CN, HCO$^{+}$-to-H$^{13}$CO$^{+}$, 
and HNC-to-HN$^{13}$C flux ratios at J=3--2 \citep{jim17,ima17}, 
to test the high HCN abundance scenario in HCN-flux-enhanced
AGN-containing ULIRGs, and to obtain the intrinsic flux ratios among HCN,
HCO$^{+}$, and HNC in this small sample. 
The H$^{13}$CN J=3--2 line was included in our HNC J=3--2 observations
in Cycles 2 and 3. 
However, for the three ULIRGs, the integration time in Cycle 2 was much
shorter than the new isotopologue observations in Cycle 4.
We will use H$^{13}$CN J=3--2 line data taken in Cycle 4 for these three
ULIRGs.

We used CASA (https://casa.nrao.edu) for our data reduction, starting
from calibrated data provided by ALMA. 
All data delivered to us passed the quality check by ALMA, except for
the HNC J=4--3 data of IRAS 15250$+$3609, for which the achieved rms
noise level was about twice as large as that requested in our proposal. 
However, the detected HNC J=4--3 emission line in IRAS 15250$+$3609 was
brighter than our original conservative estimate; thus, we decided to
include the data for our scientific discussion. 
We first checked the visibility plots and chose channels free from
strong emission lines to extract continuum information.
After subtracting the continuum emission component using the CASA task
``uvcontsub'', we applied the ``clean'' procedure to create
continuum-subtracted molecular line data.  
We employed 40-channel spectral binning, or 20-channel binning if data
were acquired and delivered with 2-channel binning (Spec Avg. = 2 in
the ALMA Observing Tool).
The final velocity resolutions were $\sim$20 km s$^{-1}$. 
Band 7 spectra of a few sources were very spiky and did not have 
sufficient signal-to-noise ratios; in this 
case, we applied additional two spectral element binning to 
make molecular emission lines and continuum profiles smoother.
The employed pixel scale for the ``clean'' procedure ranged from 
0$\farcs$03 pixel$^{-1}$ to 0$\farcs$1 pixel$^{-1}$, depending on the 
actually achieved synthesized beam sizes. 
We set the pixel scale smaller than one-third of the beam sizes. 
The ``clean'' procedure was also applied to the continuum data, 
with the 0$\farcs$03--0$\farcs$1 pixel$^{-1}$ scale.

As the nuclear AGN-affected dense molecular gas of ULIRGs is usually
concentrated to $\sim$kpc or sub kpc, a $<$1 arcsec angular resolution
($<$1.8 kpc at $z \sim$ 0.1) was desired to investigate 
the physical properties, with minimum contamination from
spatially extended molecular gas predominantly affected by starburst
activity. 
At the same time, to recover the entire nuclear molecular gas
emission with the spatial extent of a maximum of a few kpc, we avoided
lengthy baseline array configurations.   
In band 6, the requested array configurations resided 
between C34--2 (0$\farcs$61 angular resolution [AR] and 7$\farcs$8
maximum recoverable scale [MRS]) and C34--7 (AR $\sim$ 0$\farcs$18 and
MRS $\sim$ 4$\farcs$0) in Cycle 2, 
between C36--2 (AR $\sim$ 0$\farcs$8 and  MRS $\sim$ 11$\farcs$0) and
C36--5 (AR $\sim$ 0$\farcs$2 and MRS $\sim$ 3$\farcs$4) in Cycle 3, and 
between C40--2 (AR $\sim $1$\farcs$0 and MRS $\sim $ 9$\farcs$6) and 
C40--6 (AR $\sim$ 0$\farcs$15 and MRS $\sim$ 1$\farcs$3) in Cycle 4. 
In band 7, the requested array configurations were 
between C34--2 (AR $\sim$ 0$\farcs$59 and MRS $\sim$ 7$\farcs$6) and 
C34--7 (AR $\sim$ 0$\farcs$12 and MRS $\sim$ 2$\farcs$6) in Cycle 2, 
and between C36--1 (AR $\sim$ 1$\farcs$0 and MRS $\sim$ 7$\farcs$3) 
and
C36--4 (AR $\sim$ 0$\farcs$2 and MRS $\sim$ 2$\farcs$8) in Cycle 3.
The maximum recoverable scale was smaller with the higher angular
resolution configuration.  
HCN, HCO$^{+}$, and HNC emission lines at J=3--2 and J=4--3
data were taken in Cycles 2 and 3. As such, we should be able to recover
the total molecular emission within a spatial scale of a few arcsec. 
For NGC 1614, molecular emission is spatially more extended than other
ULIRGs \citep{ima13a}. 
ALMA band 6 observation of NGC 1614 was conducted in Cycle 2; thus,
emission with a spatial extent of $\sim$4'' was recovered.  
According to the ALMA Cycles 2 and 3 Proposer's Guides, the
absolute flux calibration uncertainty of bands 6 and 7 data is expected
to be $<$10\%.  
The position reference frames were FK5 for objects observed in ALMA
Cycle 2 and ICRS for those observed in ALMA Cycles 3 and 4.

\section{Results} 
 
Figure 1 presents newly obtained continuum emission images of IRAS
12112$+$0305, NGC 1614, and IRAS 13509$+$0442, all of which show some
morphological structures, in addition to nuclear compact emission. 
The continuum emission of the remaining ULIRGs is dominated by a single
nuclear compact source, as seen in the continuum images at different
frequencies \citep{ima16c}, and is also shown in the Appendix 
(Figure 29).   
The continuum emission properties of all sources are summarized in 
Table 3. 
Except for NGC 1614, the estimated continuum flux levels in band 6 
taken during HNC J=3--2 observations (32b) (Table 3) agree within
$\sim$20\% with those taken during HCN J=3--2 and HCO$^{+}$ J=3--2
observations (32a) \citep{ima16c}.
The small difference in the flux can be explained by the slight
frequency difference and maximum $\sim$10\% absolute calibration
uncertainty in individual ALMA observations.
For NGC 1614, the continuum flux difference is larger than other
ULIRGs, because it shows spatially extended structures and the continuum 
flux measurements were made at slightly different positions.  
For IRAS 08572$+$3915, IRAS 12112$+$0305, and IRAS 22491$-$1808, the 
continuum flux levels in band 6 taken during the isotopologue
H$^{13}$CN, H$^{13}$CO$^{+}$, and HN$^{13}$C J=3--2 observations (iso32)
were systematically smaller than the other band 6 data (Table 3).
This can be explained by the lower frequency (=longer wavelength) of the
former data, because dust thermal radiation is usually dominant in this
frequency range at $\nu_{\rm rest}$ $>$ 250 GHz. 
Continuum flux measurements in band 7 were also consistent within
$<$20$\%$ between data taken during HCN J=4--3 and HCO$^{+}$ J=4--3
observations (43a) and HNC J=4--3 observations (43b), except IRAS
15250$+$3609, which shows $\sim$30\% discrepancy. 
For the IRAS 15250$+$3609 data taken with HCN J=4--3 and HCO$^{+}$
J=4--3, LSB data as a whole were flagged by the pipeline due
to poor Earth atmospheric transmission, which may cause some
systematic differences. 

Figures 2--12 show newly acquired ALMA spectra, within the beam size, at
the continuum emission peak positions of individual galaxy nuclei. 
For IRAS 12112$+$0305, molecular emission line signatures were present
for both the north-eastern (NE) and south-western (SW) nuclei,
whose spectra are shown separately in Figure 4.  
The expected observed frequencies of J=3--2 of ``HNC, H$^{13}$CN,
H$^{13}$CO$^{+}$, HN$^{13}$C, HNC v$_{2}$=1f'' and 
J=4--3 of ``HCN, HCO$^{+}$, HNC, HCN v$_{2}$=1f, HNC v$_{2}$=1f'' as
well as other bright CS J=5--4 ($\nu_{\rm rest}$ = 244.936 GHz), 
CS J=7--6 ($\nu_{\rm rest}$ = 342.883 GHz), 
SiO J=6--5 ($\nu_{\rm rest}$ = 260.518 GHz), 
HC$_{3}$N J=27--26 ($\nu_{\rm rest}$ = 245.606 GHz), and 
HC$_{3}$N J=30--29 ($\nu_{\rm rest}$ = 272.885 GHz) lines are indicated
with downward arrows. 
Other serendipitously detected emission lines are also tentatively
identified. 

In the continuum images of IRAS 13509$+$0442 in Figure 1, a bright
source was detected at $\sim$1$''$ east and 8--9$''$ north of the IRAS
13509$+$0442 nucleus.
We denote this north-eastern source to as IRAS 13509$+$0442 NE.
It was detected in the continuum emission map at $\sim$235 GHz 
with a flux of $\sim$1.9 mJy \citep{ima16c}.
The continuum fluxes at $\sim$310 GHz in band 7 (3.6--4.1 mJy; Table 3)
are a factor of $\sim$2 higher than those at $\sim$235 GHz in band 6
(1.9--2.1 mJy; Table 3 and \citet{ima16c}), which suggests that
emission in this frequency range comes from dust thermal radiation,
rather than synchrotron or thermal free-free radiation. 
No clear emission lines were detected in the ALMA spectra of
IRAS 13509$+$0442 NE (see Figure 11 and \citet{ima16c}).
At $\sim$5$''$ from IRAS 13509$+$0442 NE in the north-east
direction, a galaxy (SDSS J135331.89$+$042816.4; z=0.186) was 
detected in the optical range (0.3--1.0 $\mu$m) with the Sloan Digital
Sky Survey (SDSS DR7) \citep{sch10}, as well as in the near-infrared
(1.0--2.4 $\mu$m) with the UKIRT Infrared Deep Sky Survey (UKIDSS)
\citep{law07}. However, we regard that this SDSS optical and UKIDSS
near-infrared source is not directly related to IRAS 13509$+$0442 NE,
because there is a $\sim$5$''$ positional offset, while the peak
positions of the SDSS optical, UKIDSS near-infrared, and ALMA
(sub)millimeter data agree well for ULIRG IRAS 13509$+$0442 itself.
Thus, IRAS 13509$+$0442 NE is a (sub)millimeter bright, but optically
and near-infrared faint object.
No source is catalogued at the position of IRAS 13509$+$0442 NE in the
Faint Images of the Radio Sky at Twenty-cm (FIRST) Survey data at radio
1.4 GHz (21 cm) \citep{whi97}. 
This kind of spectral energy distribution is often seen in distant dusty
(U)LIRGs at z $>$ 1 \citep{tam14}.
Thus, IRAS 13509$+$0442 NE could be a distant (U)LIRG. 

In Figure 13, we display integrated intensity (moment 0) maps 
of the detected HNC J=3--2, HCN J=4--3, HCO$^{+}$ J=4--3, and HNC J=4--3
emission lines for the observed ULIRGs, by summing spectral elements
with significant detection. 
For the HNC J=3--2 and HCO$^{+}$ J=4--3 lines of IRAS 12112$+$0305, the
moment 0 maps at NE and SW nuclei are shown separately, because spectral
elements used for the creation of the moment 0 maps differ slightly
between NE and SW nuclei, due to a velocity shift.
HCN J=4--3 and HNC J=4--3 emission lines are not
clearly detected in the SW nucleus. 
The HNC J=3--2 emission line of NGC 1614 shows spatially extended 
morphology and its moment 0 map, together with spectra at individual
positions, are shown in Figure 14 separately. 
Tables 4, 5, 6, and 7 summarize the properties of moment 0 maps of HNC
J=3--2, HCN J=4--3, HCO$^{+}$ J=4--3, and HNC J=4--3, respectively.

The integrated intensity (moment 0) maps of the isotopologue H$^{13}$CN
J=3--2, H$^{13}$CO$^{+}$ J=3--2, and HN$^{13}$C J=3--2 emission lines 
based on Cycle 4 data for IRAS 08572$+$3915, IRAS 12112$+$0305
NE, and IRAS 22491$-$1808 are shown in Figure 15.
Their properties are summarized in Table 8.
No clear isotopologue emission line was detected at IRAS 12112$+$0305 SW.

To estimate the intrinsic continuum emission sizes, after deconvolution,
of the (U)LIRG's nuclei, we use the CASA task ``imfit''.
With the exception of NGC 1614, for which the signatures of spatially
extended structures are evident in Figure 1 and the task ``imfit''
indeed provides intrinsic sizes substantially larger than the beam
sizes, the continuum emission of the remaining ULIRGs is dominated by
nuclear compact components.
Table 9 lists the deconvolved continuum emission sizes in apparent
scale in (mas) and physical scale in (pc).
Here, we note that the deconvolved sizes are in practice dependent
on the achieved synthesized beam sizes, because it is usually difficult
to constrain sizes much smaller than the beam sizes. 
Thus, the synthesized beam size information is shown in Table 9 for
reference. 
In the case of spatially unresolved continuum emission, more stringent
constraints are placed based on data with smaller synthesized beam sizes.
From Table 9, we can see that the intrinsic continuum emission sizes of
ULIRGs in band 7 ($\sim$850 $\mu$m) and band 6 ($\sim$1.2 mm) are
constrained to $\lesssim$500 pc, except IRAS 13509$+$0442, supporting
the previous suggestion based on high spatial-resolution mid-infrared
$\sim$10 $\mu$m imaging observations \citep{soi00} that the luminosities
of nearby ULIRGs are generally dominated by compact nuclear energy
sources with $\lesssim$500 pc.   

For IRAS 13509$+$0442, the sizes of the continuum-emitting regions 
at $\sim$850 $\mu$m in band 7 and $\sim$1.2 mm in band 6 are estimated to
be $\sim$1 kpc. 
Although the synthesized beam sizes for the observations of this ULIRG
are not very small, smaller apparent scales for the continuum emission
are derived with comparable beam sizes for other ULIRGs (e.g., IRAS
15250$+$3609 and PKS 1345$+$12).
The continuum emission of IRAS 13509$+$0442 is estimated to be 
spatially more extended than the majority of other ULIRGs.
IRAS 13509$+$0442 does not exhibit obvious buried AGN signatures in the
infrared \citep{ima07b,ima10,nar10}.
IRAS 12112$+$0305 and IRAS 22491$-$1808 similarly show starburst-like
spectra without any discernible AGN signature in the infrared, and yet
show signatures of vibrationally excited HCN v$_{2}$=1f emission lines  
at J=3--2, most likely originating in the infrared radiative pumping by
absorbing infrared photons coming from AGN-heated hot dust emission
\citep{ima16c}.  
IRAS 20414$-$1651 also has a similar starburst-dominated infrared
spectral shape and shows some possible signature of the HCN v$_{2}$=1f
J=3--2 emission line, although the detection significance is
$<$3$\sigma$ \citep{ima16c}.  
IRAS 13509$+$0442 does not display even the signature of the HCN
v$_{2}$=1f emission line at J=3--2 \citep{ima16c}.
No putative AGN in IRAS 13509$+$0442 is discernible, and this ULIRG
may be dominated by $\sim$1 kpc-scale starburst activity.  

Table 10 summarizes the estimated intrinsic sizes of molecular emission
lines, after deconvolution, using the CASA task ``imfit'' in the
integrated intensity (moment 0) maps. 
Except NGC 1614, the molecular emission sizes whenever constrained are
smaller than the synthesized beam sizes (Tables 4, 5, 6, 7).
To estimate the line fluxes, we fitted Gaussian curves to the 
continuum-subtracted spectra, within the beam size, at the individual nuclear
positions.
Gaussian fitting results for HNC J=3--2 in band 6 for NGC 1614 and
ULIRGs are shown in Figures 14 and 16, respectively.
Those for HCN J=4--3, HCO$^{+}$ J=4--3, and HNC J=4--3 emission
lines in band 7 are displayed in Figure 17.
In Figure 18, Gaussian fits to the isotopologue 
H$^{13}$CN J=3--2, H$^{13}$CO$^{+}$ J=3--2, and HN$^{13}$C J=3--2
emission lines taken in Cycle 4 are shown.
The following emission lines are clearly double-peaked: 
HNC J=3--2 of IRAS 12127$-$1412; 
HCN J=4--3, HCO$^{+}$ J=4--3, and H$^{13}$CN J=3--2 of IRAS 12112$+$0305 NE; 
HCN J=4--3 and HCO$^{+}$ J=4--3 of IRAS 15250$+$3609; and 
HCN J=4--3, HCO$^{+}$ J=4--3, and HNC J=4--3 of IRAS 20414$-$1651.
We employ two Gaussians to fit these double-peaked emission lines.
The Gaussian fit fluxes of HNC J=3--2, HCN J=4--3, HCO$^{+}$ J=4--3, HNC
J=4--3, and isotopologue emission lines are summarized also in Tables 4,
5, 6, 7, and 8, respectively.  
Except NGC 1614, thanks to the spatially compact molecular emission, the
Gaussian fits of the detected molecular emission lines in the spectra
within the beam sizes should recover the bulk of the fluxes. 
These flux estimates will be used for our quantitative discussion of the
molecular line flux ratios, except NGC 1614, for which we will use
Gaussian fit fluxes in the spatially integrated spectra.   

In addition to the targeted main emission lines, signatures of other
molecular emission lines are discernible in some objects. 
By using the Splatalogue database (http://www.splatalogue.net), 
we attempted to identify these faint molecular emission
lines; however, this process is complicated by the presence of multiple
candidates.
Our tentative identifications of these lines, their integrated intensity
(moment 0) maps, Gaussian fits in the spectra within the beam sizes, and
emission line properties are summarized in the Appendix (Figures 30 and
31, and Table 16).  

For IRAS 08572$+$3915, IRAS 12112$+$0305, and IRAS 22491$-$1808, 
H$^{13}$CN J=3--2 lines were included in our ALMA Cycle 2 data, in
addition to the Cycle 4 deeper data. 
For IRAS 08572$+$3915, the flux upper limit in Cycle 2 data (Table 16)
is higher than the detected flux in deep Cycle 4 data (Table 8).
For IRAS 12112$+$0305 NE and IRAS 22491$-$1808, the H$^{13}$CN J=3--2
fluxes based on Gaussian fits are consistent between Cycle 2 and 4 data
(Tables 8 and 16), if we allow $\sim$10\% uncertainty in the absolute
flux calibration in individual ALMA observations.  
For these sources, we adopt deeper Cycle 4 data, because 
(1) the detection significance is higher, and 
(2) not only H$^{13}$CN J=3--2 but also H$^{13}$CO J=3--2 and
HN$^{13}$C J=3--2 data were taken simultaneously with the same
synthesized beam pattern, allowing a reliable comparison of 
the flux attenuation by line opacity among HCN, HCO$^{+}$, and HNC,
based on the HCN-to-H$^{13}$CN, HCO$^{+}$-to-H$^{13}$CO$^{+}$, and 
HNC-to-HN$^{13}$C flux 
ratios at J=3--2 \citep{jim17,ima17}.
For other galaxies, H$^{13}$CN J=3--2 emission line fluxes or upper limits
obtained from shallow Cycle 2 or 3 data will be used for discussion.

Table 11 tabulates the ratios of J=4--3 to J=3--2 flux in (Jy km
s$^{-1}$) for HCN, HCO$^{+}$, and HNC.
The ratios of HCN-to-HCO$^{+}$ and HCN-to-HNC flux in (Jy km s$^{-1}$)
at J=4--3 and J=3--2 are summarized in Table 12.

In the spectrum of IRAS 12112$+$0305 NE in Figure 4c, we recognize 
an emission tail at the lower frequency side of the bright HCO$^{+}$
J=4--3 emission line. 
A similar profile was seen for the HCO$^{+}$ J=3--2 emission line 
in IRAS 12112$+$0305 NE \citep{ima16c}, as well as in other (U)LIRGs 
\citep{aal15a,aal15b,ima16c}.
This is usually attributed to the contribution from the
vibrationally excited (v$_{2}$=1f) HCN J=4--3 or J=3--2 emission line 
\citep{aal15a,aal15b,ima16c}.
We make an integrated intensity (moment 0) map, by integrating the
continuum-subtracted emission within the frequency range expected for
the HCN v$_{2}$=1f J=4--3 emission line 
(331.5--332.2 GHz; shown as a horizontal solid straight line in
Figure 4c), which is displayed in Figure 19. 
The HCN v$_{2}$=1f J=4--3 emission line may be detected at 
$\sim$3$\sigma$ level (Table 16 in Appendix) at the continuum peak
position of IRAS 12112$+$0305 NE. 

IRAS 15250$+$3609 also displays signatures of the vibrationally excited
HNC v$_{2}$=1f J=4--3 emission line in Figure 8d. 
Its integrated intensity (moment 0) map is shown in Figure 19, and
emission properties are summarized in Table 16 in the Appendix. 
We obtain $\sim$5$\sigma$ detection, assuming that the excess
emission tail at the expected observed frequency of HNC v$_{2}$=1f
J=4--3 as originating in this line.
Some signatures of the HNC v$_{2}$=1f J=4--3 emission line may be seen
in Superantennae (Figure 3e) and IRAS 12112$+$0305 NE (Figure 4e);
however, their flux estimates are difficult, due to contamination by the  
nearby strong emission line identified as 
CH$_{3}$OH 16(2,14)--16($-$1,16) ($\nu_{\rm rest}$ = 364.859 GHz).

During the observations of HNC J=3--2, the vibrationally excited
(v$_{2}$=1f) HNC J=3--2 line was also covered. 
However, the signatures of the HNC v$_{2}$=1f J=3--2 emission are not
clear in any of the spectra in Figures 2--12.
We created integrated intensity (moment 0) maps of the HNC v$_{2}$=1f
J=3--2 emission line, by summing several spectral elements around 
its expected frequency, depending on the actual signal profile. 
However, detection was $<$3$\sigma$ for all sources, and the 
HNC v$_{2}$=1f J=3--2 emission line with $>$0.3 
(Jy beam$^{-1}$ km s$^{-1}$) was not observed in any source. 

Figures 20 and 21 display intensity-weighted mean velocity (moment 1)
and intensity-weighted velocity dispersion (moment 2) maps of the
strongly detected ($>$10$\sigma$) HNC J=3--2, HCN J=4--3, HCO$^{+}$
J=4--3, and HNC J=4--3 emission lines in selected galaxies. 
In addition to these targeted lines, the moment 1 and 2 maps of the
serendipitously detected bright CS J=5--4 emission lines in our deep
Cycle 4 data of IRAS 12112$+$0305 NE and IRAS 22491$-$1808 are
shown in the Appendix (Figures 32 and 33).

\section{Discussion} 

\subsection{P Cygni Profiles and Molecular Outflow}

\subsubsection{IRAS 15250$+$3609}

IRAS 15250$+$3609 shows a broad negative absorption feature at the
0.4--0.5 GHz higher frequency (= blueshifted) side of the HCO$^{+}$
J=4--3 emission line in Figure 8c. 
A similar broad negative absorption feature was detected at the
higher frequency (blueshifted) side of the HCO$^{+}$ J=3--2 emission
line \citep{ima16c}. 
These absorption profiles are naturally interpreted as outflowing
material, in which blueshifted gas in front of the nucleus  
absorbs the background nuclear continuum emission.
Figure 22 presents detailed line profiles of HCN, HCO$^{+}$, and HNC at
J=4--3 and J=3--2, relative to the systemic velocity. 
HCO$^{+}$ displays a clear broad blueshifted absorption feature at
J=4--3 and J=3--2, whereas HCN and HNC do not.
By summing negative signals at the absorption features 
at $-$400 $\sim$ $-$215 (km s$^{-1}$) for HCO$^{+}$ J=4--3 and 
$-$850 $\sim$ $-$188 (km s$^{-1}$) for HCO$^{+}$ J=3--2 in Figure 22, 
we obtain $-$0.49 (Jy beam$^{-1}$ km s$^{-1}$) (5.4$\sigma$) and
$-$0.47 (Jy beam$^{-1}$ km s$^{-1}$) (4.9$\sigma$) in the integrated
intensity (moment 0) maps for HCO$^{+}$ J=4--3 and J=3--2, respectively. 
In the nearby well-studied ULIRG, Arp 220 (z=0.018), stronger
blueshifted absorption profiles were observed for HCO$^{+}$ than HCN at
J=4--3 and J=3--2 \citep{sak09,sco15,mar16} in a similar way to IRAS
15250$+$3609. 
HCO$^{+}$ may be more abundant than HCN and HNC in the outflowing gas in
these ULIRGs.  
 
In Figure 22, emission sub-peaks appear at 400--500 km s$^{-1}$ on the
redshifted (lower frequency) side of HCN J=4--3, HCO$^{+}$ J=4--3,
HCN J=3--2, and HCO$^{+}$ J=3--2 emission lines. 
We interpret that these sub-peaks originate in emission from  
redshifted outflow components, as opposed to other
weaker emission lines, because both the blueshifted absorption 
(for HCO$^{+}$ J=4--3 and J=3--2) and redshifted
emission lines show peaks at $\pm$400--500 km s$^{-1}$, relative to the
systemic velocity of IRAS 15250$+$3609 (Figure 22). 
The sub-peak emission, relative to the main emission component, is
stronger for HCO$^{+}$ than HCN.
This is naturally reproduced by outflowing material if HCO$^{+}$ is
more abundant than HCN there. 

The vibrationally excited HCN v$_{2}$=1f J=4--3 and J=3--2 emission
lines are located at the $\sim$400 km s$^{-1}$ redshifted (lower
frequency) side of HCO$^{+}$ J=4--3 and J=3--2 emission lines,
respectively \citep{sak10,ima13b,aal15b,mar16,ima16b,ima17}.   
However, we regard that the sub-peaks for HCO$^{+}$ are unlikely to
come solely from these HCN v$_{2}$=1f emission lines for the following
two reasons.
First, the sub-peaks are detected also at the 400--500 km s$^{-1}$
redshifted side of HCN J=4--3 and J=3--2 emission lines, which
cannot be explained by the HCN v$_{2}$=1f emission lines. 
Second, for the sub-peaks of HCO$^{+}$ J=4--3 and J=3--2 emission
lines, the observed peak frequencies are $\nu_{\rm obs}$ $\sim$ 337.55
GHz and 253.15 GHz \citep{ima16c}, respectively, which correspond to 
rest-frequencies with $\nu_{\rm rest}$ $\sim$ 356.18 GHz and 267.12 GHz
at the molecular-line-derived redshift of z=0.0552.
The vibrationally excited HCN v$_{2}$=1f emission lines at 
J=4--3 ($\nu_{\rm rest}$ = 356.26 GHz) and 
J=3--2 ($\nu_{\rm rest}$ = 267.20 GHz) are $\sim$0.08 GHz shifted 
to the higher frequency side of the sub-peaks of HCO$^{+}$ J=4--3 and
HCO$^{+}$ J=3--2.  
In another ULIRG IRAS 20551$-$4250 at z=0.0430, for which we argued the
detection of HCN v$_{2}$=1f J=4--3 and J=3--2 emission lines, their
redshifted frequencies agree (within $<<$0.08 GHz) with the
frequencies of the observed sub-peak emission for HCO$^{+}$ J=4--3 and
J=3--2 \citep{ima13b,ima16c,ima17}.  

The rest frequencies for the individual emission sub-peaks for HCN
J=4--3, HCO$^{+}$ J=4--3, HCN J=3--2, and HCO$^{+}$ J=3--2 are 
$\nu_{\rm rest}$ $\sim$ 354.02 GHz, 356.18 GHz, 265.49 GHz, and 267.12
GHz, respectively.
Based on the Splatalogue database (http://www.splatalogue.net), SO$_{2}$
emission lines are present at close to the frequencies of individual
sub-peaks, SO$_{2}$ 47(5,43)--48(2,46) ($\nu_{\rm rest}$ = 354.08 GHz), 
SO$_{2}$ 15(7,9)--16(6,10) ($\nu_{\rm rest}$ = 356.04 GHz), 
SO$_{2}$ 34(4,30)--34(3,31) ($\nu_{\rm rest}$ = 265.48 GHz), and 
SO$_{2}$ 63(6,58)--62(7,55) ($\nu_{\rm rest}$ = 267.19 GHz).
However, the SO$_{2}$ frequency is offset, relative to the sub-peak, to
the higher frequency side in some cases and lower frequency side in 
others, making it difficult to interpret that all four sub-peaks
originate solely from SO$_{2}$ emission. 

Summarizing, the broad negative absorption at the 400--500 km s$^{-1}$
blueshifted side and sub-peak emission at the 400--500 km s$^{-1}$
redshifted side of HCO$^{+}$ emission lines at J=4--3 and J=3--2
can naturally be explained by HCO$^{+}$ outflow. 
Narrow absorption features at the redshifted side of HCO$^{+}$
J=4--3 and J=3--2 lines, at the velocity of $\sim$250--300 km s$^{-1}$,
were also observed (Figure 22 and \citet{ima16c}). 
As the signals at the center of these absorption features are 
negative,
these features cannot be explained solely by self-absorption inside 
redshifted outflowing gas; absorption of background
continuum emission by foreground gas is required.   
A similar narrow absorption feature was detected at the velocity of
$\sim$250 km s$^{-1}$ in the OH line data at far-infrared 65--119
$\mu$m and was interpreted as inflow origin \citep{vei13,gon17}. 
Thus, we regard that the redshifted narrow absorption features of
HCO$^{+}$ J=4--3 and HCO$^{+}$ J=3--2 are due to inflowing gas.  

Regarding the blueshifted broad absorption, adopting the continuum
flux levels of $\sim$16.9 (mJy beam$^{-1}$) (Table 3) and $\sim$11 (mJ
beam$^{-1}$) \citep{ima16c}, the above estimated integrated intensities
of $-$0.49 (Jy beam$^{-1}$ km s$^{-1}$) (5.4$\sigma$) and $-$0.47 (Jy
beam$^{-1}$ km s$^{-1}$) (4.9$\sigma$) (paragraph 1 of this subsection)
provide the absorption equivalent widths to be 2.9 $\times$
10$^{6}$ (cm s$^{-1}$) and 4.3 $\times$ 10$^{6}$ (cm s$^{-1}$) for
HCO$^{+}$ J=4--3 and HCO$^{+}$ J=3--2, respectively.  
We then convert these absorption equivalent widths (EW) to the column
density, using the formula 
\begin{eqnarray}
EW = \frac{\lambda^{3} A_{ul} g_{u}}{8 \pi g_{l}} \times N 
\end{eqnarray}
\citep{ryb79,gon14}, where 
EW is in units of (cm s$^{-1}$), 
$\lambda$ is wavelength in (cm), 
A$_{ul}$ is the Einstein A coefficient for spontaneous emission from the
upper (u) level to the lower (l) level in units of (s$^{-1}$), 
g$_{u}$ and g$_{l}$ are the statistical weights at the upper and lower 
levels, respectively, and N is the column density in the lower
energy level in units of (cm$^{-2}$). The statistical weight 
of level J is 2J+1. A$_{ul}$ 
values for HCO$^{+}$ J=4--3 and J=3--2 are 35.7 
$\times$ 10$^{-4}$ (s$^{-1}$) and 14.5 $\times$ 10$^{-4}$ (s$^{-1}$),
respectively, based on data from the Cologne Database of Molecular
Spectroscopy (CDMS) \citep{mul05} via Splatalogue
(http://www.splatalogue.net).  
From these values, the column density of HCO$^{+}$ at J=3 and
J=2 in the blueshifted outflowing molecular gas is 2.7 $\times$
10$^{13}$ (cm$^{-2}$) and 3.8 $\times$ 10$^{13}$ (cm$^{-2}$),
respectively. 
Thus, the HCO$^{+}$ column density in the blueshifted outflowing gas in
front of the nuclear continuum-emitting energy source is at least $>$6
$\times$ 10$^{13}$ (cm$^{-2}$), because HCO$^{+}$ should be populated at
other J-levels than J=3 and J=2. 
Adopting the HCO$^{+}$-to-H$_{2}$ abundance ratio of $\sim$10$^{-8}$
\citep{gre09}, the molecular H$_{2}$ column density of the
blueshifted outflowing gas in IRAS 15250$+$3609 is estimated to be
N$_{\rm H2}$ $>$ 6 $\times$ 10$^{21}$ (cm$^{-2}$). 
In this equation, we do not take into account the emission within the
absorbing gas. 
The actual N$_{\rm H2}$ values must be even higher, because 
(1) emission by the approaching gas itself fills in the absorption
features and (2)  the population in levels other than J=3 and J=2 is
ignored.   

We next derive the molecular outflow rate (\.{M}$_{\rm outf}$) from the
redshifted emission sub-peak, based on the assumption that the sub-peak
originates from redshifted outflowing gas.
This is because blueshifted absorption features can be diluted by
emission within the same blueshifted outflowing molecular gas and thus
the molecular outflow rate can be underestimated.  
The integrated intensity of the redshifted emission sub-peaks of
HCO$^{+}$ J=4--3 and HCO$^{+}$ J=3--2 are 
2.0 (Jy beam$^{-1}$ km s$^{-1}$) (10.8$\sigma$) and 0.77 (Jy beam$^{-1}$
km s$^{-1}$) (9.1$\sigma$), respectively.  
Assuming that the blueshifted outflow gas has the same amount of emission,
we obtain HCO$^{+}$ J=4--3 and HCO$^{+}$ J=3--2 emission line
luminosities in outflowing gas to be 5.6 $\times$ 10$^{7}$ 
(K km s$^{-1}$ pc$^{2}$) and 3.9 $\times$ 10$^{7}$ (K km s$^{-1}$
pc$^{2}$), respectively, based on equation (3) of \citet{sol05}.
Assuming that the HCO$^{+}$ J=4--3 and HCO$^{+}$ J=3--2 emission lines
are thermalized 
\footnote{
If these emission lines are sub-thermally excited, 
the derived J=1--0 luminosity will increase and the estimated molecular
mass will be even higher.}
and optically thick, and that the relationship between 
dense molecular mass and HCO$^{+}$ J=1--0 luminosity is the same as that
between dense molecular mass and HCN J=1--0 luminosity, 
M$_{\rm dense}$ = 10 $\times$ HCN J=1--0 luminosity 
[M$_{\odot}$ (K km s$^{-1}$ pc$^{2}$)$^{-1}$] \citep{gao04a} 
\footnote{
The relation between dense molecular mass and HCO$^{+}$ J=1--0
luminosity has not been established in galaxies. 
As HCN J=1--0 and HCO$^{+}$ J=1--0 luminosities are usually
comparable within a factor of a few in the majority of galaxies
\citep{koh05,kri08,ima07a,ima09}, we make this assumption.}, 
the outflowing dense molecular mass would be M$_{\rm outf}$ $=$ 
5.6 $\times$ 10$^{8}$ M$_{\odot}$ and 3.9 $\times$ 10$^{8}$ M$_{\odot}$
from HCO$^{+}$ J=4--3 and HCO$^{+}$ J=3--2 data, respectively. 
Both M$_{\rm outf}$ values agree within $\sim$30\%. 
We adopt the value from HCO$^{+}$ J=3--2 data to be conservative.

From Figure 22b, the outflow velocity of V = 500 km s$^{-1}$ is adopted 
for IRAS 15250$+$3609. 
In Figure 23, the peak position of the blueshifted HCO$^{+}$ J=3--2
absorption and redshifted HCO$^{+}$ J=3--2 emission is displaced with
0$\farcs$14 or $\sim$160 pc in projected distance. 
Adopting the outflowing gas size of R $\sim$ 80 pc from the nucleus and 
the relation of \.{M}$_{\rm outf}$ = 3 $\times$ M$_{\rm outf}$ $\times$
V/R \citep{mai12,cic14}, we obtain a molecular outflow rate with 
\.{M}$_{\rm outf}$ $\sim$ 750 (M$_{\odot}$ yr$^{-1}$). 
If we adopt the formula of 
\.{M}$_{\rm outf}$ = M$_{\rm outf}$ $\times$ V/R \citep{gon17,vei17}, 
the outflow rate is a factor of 3 lower (i.e., $\sim$250 M$_{\odot}$
yr$^{-1}$.) 
Adopting the $\sim$50\% AGN contribution to the infrared luminosity of
IRAS 15250$+$3609 \citep{nar08,nar09,nar10}, the AGN luminosity is
L$_{\rm AGN}$ $\sim$ 1.9 $\times$ 10$^{45}$ erg s$^{-1}$. The estimated
molecular outflow rate in IRAS 15250$+$3609 follows the relation between
the AGN luminosity and molecular outflow rate established in other
AGN-containing ULIRGs \citep{cic14}. 
The molecular outflow kinetic power is P$_{\rm outf}$
$\equiv$ 0.5 $\times$ \.{M}$_{\rm outf}$ $\times$ V$^{2}$ $\sim$ 2.0--5.9 
$\times$ 10$^{43}$ (erg s$^{-1}$) or $\sim$1--3\% of the AGN luminosity.
The molecular outflow momentum rate is \.{P}$_{\rm outf}$ 
$\equiv$ \.{M}$_{\rm outf}$ $\times$ V $\sim$ 0.8--2.4 $\times$ 10$^{31}$ 
(kg m s$^{-2}$). 
This is $\sim$12--37 $\times$ L$_{\rm AGN}$/c, and supports the energy
conserving outflow scenario rather than the acceleration by radiation
pressure \citep{cic14}. 
Overall, the outflow properties of IRAS 15250$+$3609 are similar to 
those in other molecular-outflow-detected AGN-containing-ULIRGs
\citep{cic14,ima17}.  

\subsubsection{IRAS 12112$+$0305 NE} 

Broad negative absorption features also appear at the higher frequency
(blueshifted) side of the SiO J=6--5 emission line in the spectrum of
IRAS 12112$+$0305 NE in Figure 4g. 
\citet{tun15} observed the same line in the well-studied ULIRG, Arp 220
(z=0.018) and detected the P Cygni profile at the blueshifted side of the
SiO J=6--5 line; this was ascribed to SiO outflow.
Figure 24 plots the line profiles with respect to the systemic velocity
at z=0.0730 (V$_{\rm sys}$ = 21900 km s$^{-1}$) for the SiO J=6--5 
line. 
A textbook-type P Cygni profile, with redshifted emission and blueshifted 
absorption features, is seen.
Figure 24 also shows a negative narrow absorption feature on the
redshifted (lower frequency) side of the SiO J=6--5 emission line as
well, which may be due to (1) inflowing gas traced by the SiO 
J=6--5 line and/or 
(2) the P Cygni profile of the H$^{13}$CO$^{+}$ J=3--2
emission line.
However, given that the detection significance of this narrow absorption
feature in an integrated intensity map that we created was $<$3$\sigma$,
we do not discuss this feature further.  

The integrated intensity at the blueshifted SiO J=6--5 absorption
features at $-$190 $\sim$ $-$10 (km s$^{-1}$) in Figure 24 provides a
negative signal of $-$0.11 (Jy beam$^{-1}$ km s$^{-1}$) (4.6$\sigma$).
Its absolute value is smaller than the integrated intensity of the
redshifted emission part of SiO J=6--5 at $+$14 $\sim$ $+$195 (km
s$^{-1}$) with 0.18 (Jy beam$^{-1}$ km s$^{-1}$) (7.3$\sigma$), again
suggesting that the blueshifted SiO J=6--5 absorption feature is 
filled in by emission within the absorbing gas.
Adopting the continuum flux level of $\sim$4.6 (mJy beam$^{-1}$) at the
frequency of SiO J=6--5 (Table 3), we obtain an equivalent width of the
blueshifted SiO J=6--5 absorption feature of 2.4 $\times$ 10$^{6}$
(cm s$^{-1}$). 
Based on equation (1) and the Einstein coefficient of A$_{ul}$ =
9.1 $\times$ 10$^{-4}$ for SiO J=6--5 (http://www.splatalogue.net), 
we obtain SiO column density at J=5 in the blueshifted outflowing
gas in front of the continuum-emitting energy source of 
N(SiO J=5) = 3.6 $\times$ 10$^{13}$ cm$^{-2}$. 
SiO is well known to be a good tracer of shocked gas \citep{mar92}.
The SiO-to-H$_{2}$ abundance ratio in shocked gas is estimated to be
10$^{-10}$--10$^{-9}$ \citep{gar01,use06,gar10}. 
Adopting the highest value of 10$^{-9}$, we obtain the conservative
molecular H$_{2}$ column density in the outflow to be 
N$_{\rm H2}$ $\sim$ 3.6 $\times$ 10$^{22}$ (cm$^{-2}$). 
This is again a lower limit, because (1) the absorption feature in the 
blueshifted outflowing gas can be diluted by the emission from the same 
gas and (2) the SiO-to-H$_{2}$ abundance ratio can be lower than the
adopted conservative value of 10$^{-9}$.
For IRAS 12112$+$0305 NE, we did not estimate the molecular outflow rate, 
outflow kinetic power, or momentum rate, given the large ambiguity of
the conversion from the SiO J=6--5 emission line luminosity to H$_{2}$
molecular mass. 

\subsection{Molecular Line Flux Ratios}

\subsubsection{Observed HCN-to-HCO$^{+}$ Flux Ratios} 

The HCN-to-HCO$^{+}$ flux ratios of the observed (U)LIRGs at J=4--3
and J=3--2 are plotted in the ordinate of Figure 25. 
As previously seen at J=3--2 \citep{ima16c}, ULIRGs with
infrared-identified luminous AGNs show elevated HCN-to-HCO$^{+}$ flux
ratios at J=4--3, compared with the known starburst-classified galaxies,
NGC 1614, IRAS 12112$+$0305 SW, and IRAS 13509$+$0442.
The three ULIRGs with no obvious infrared AGN signatures, IRAS 12112$+$0305
NE, IRAS 22491$-$1808, and IRAS 20414$-$1651, also show elevated
HCN-to-HCO$^{+}$ flux ratios at J=4--3, as previously recognized at
J=3--2 \citep{ima16c}. 
These three ULIRGs show signatures of vibrationally excited HCN
v$_{2}$=1f J=3--2 emission lines, although the detection significance
is $<$3$\sigma$ for IRAS 20414$-$1651 \citep{ima16c}.  
In our new ALMA data, the signature of the HCN v$_{2}$=1f J=4--3
emission line is seen in IRAS 12112$+$0305 NE (Figures 4c and 19), which
can be taken as an AGN signature, because vibrational excitation of HCN
is most likely due to infrared radiative pumping by AGN-heated
hot dust emission ($\S$1).   
If we adopt the HCN v$_{2}$=1f J=4--3 emission line flux of 1.1 (Jy
beam$^{-1}$ km s$^{-1}$) estimated from the moment 0 map (3.0$\sigma$
detection, Table 16), the HCN v$_{2}$=1f to v=0 flux ratio at J=4--3
of $\sim$0.17 is even higher than that of IRAS 20551$-$4250
($\sim$0.04), the AGN-containing ULIRG with clearly detectable HCN
v$_{2}$=1f emission lines at J=4--3 and J=3--2, thanks to very small
observed molecular line widths \citep{ima13b,ima16b,ima17}.  
For IRAS 12112$+$0305 NE and IRAS 22491$-$1808, the HCN v$_{2}$=1f to
v=0 flux ratios at J=3--2, estimated in the same way as J=4--3, are 
both $\sim$0.06, which is also higher
than that of IRAS 20551$-$4250 ($\sim$0.04) \citep{ima16b,ima16c}. 
Although some emission component of the nearby bright HCO$^{+}$ v=0
line may contaminate the moment 0 maps of the HCN v$_{2}$=1f line, 
the HCN v$_{2}$=1f line is estimated to be emitted fairly 
strongly in IRAS 12112$+$0305 NE and IRAS 22491$-$1808. 
The HCN v$_{2}$=1f J=4--3 and J=3--2 emission lines can be produced
in regions where the infrared 14 $\mu$m photons are available to
vibrationally excite HCN to the v$_{2}$=1 level. 
In fact, the emission lines have been detected in Galactic stars and
star-forming regions \citep{ziu86,mil13,vea13,nag15}.
However, \citet{ima16b} estimated that the HCN v$_{2}$=1f J=4--3 to 
infrared luminosity ratio in IRAS 20551$-$4250 is more than a factor of
four higher than the Galactic active star-forming core W49A.
An AGN scenario is preferred, because the infrared radiative pumping is
much more efficient in an AGN than in a starburst ($\S$1).
These two ULIRGs may contain luminous AGNs that are too highly obscured
to be detectable in infrared spectra, but are detected in our
(sub)millimeter observations due to smaller dust extinction effects.
IRAS 12112$+$0305 NE shows a detectable CS J=7--6 emission line 
(Figure 4d).
The observed HCN J=4--3 to CS J=7--6 flux ratio for IRAS 12112$+$0305 NE
is estimated to be $\sim$5 (Tables 5 and 16), which is in the range
occupied by AGNs \citep{izu16}, supporting the AGN-important scenario
for IRAS 12112$+$0305 NE.  
We thus regard IRAS 12112$+$0305 NE and IRAS 22491$-$1808 as ULIRGs
containing infrared-elusive, but (sub)millimeter detectable, extremely
deeply buried AGNs. 
For IRAS 20414$-$1651, we need higher quality data to investigate if 
it belongs to this class of object.  
We have confirmed the trend of elevated HCN-to-HCO$^{+}$ flux ratios
in AGNs at J=3--2 and J=4--3 compared to starburst-dominated galaxies,
as previously argued at J=1--0 \citep{koh05,kri08,pri15}. 

Since HCN has a factor of $\sim$5 higher critical density than HCO$^{+}$ 
at the same J-transition under the same line opacity \citep{mei07,gre09}, 
HCN excitation can be more sub-thermal  (meaning that the excitation
temperature T$_{\rm ex}$ is much smaller than the molecular gas kinetic
temperature T$_{\rm kin}$) than HCO$^{+}$, particularly at higher-J
transitions. 
If HCN is only sub-thermally excited in starburst galaxies,
while its excitation is thermal (T$_{\rm ex}$ $\sim$ T$_{\rm kin}$) in
warm and highly dense molecular gas in the close vicinity of AGNs, then
this different excitation could partly explain the trend of higher
HCN-to-HCO$^{+}$ flux ratios in AGN-important galaxies than in
starburst galaxies.   
A comparison of the HCN J=4--3 to J=3--2 flux ratios and HCO$^{+}$
J=4--3 to J=3--2 flux ratios is shown in Figure 26.
The ratios of ``HCO$^{+}$ J=4--3-to-J=3--2 flux ratios''-to-``HCN
J=4--3-to-J=3--2 flux ratios'' are not systematically different between
starburst galaxies (objects A and K) and AGN-important ULIRGs (other
than A and K).
Excitation has a limited effect for the different HCN-to-HCO$^{+}$ flux
ratios between AGNs and starbursts. 

Another mechanism that could explain the difference in the
HCN-to-HCO$^{+}$ flux ratios is abundance. 
If HCN abundance, relative to HCO$^{+}$, is elevated in AGNs compared with
starbursts, then higher HCN-to-HCO$^{+}$ flux ratios in AGNs can
naturally be explained. 
The elevated HCN abundance in AGNs suggests that flux attenuation by
line opacity (not dust extinction) can be higher for HCN than
HCO$^{+}$.
In this case, (1) the {\it observed} HCN-to-HCO$^{+}$ flux ratios in
AGNs can be smaller than the {\it intrinsic} ratios, and (2) the excess
of the {\it observed} HCN-to-HCO$^{+}$ flux ratios could apparently
disappear in some HCN-abundance-enhanced AGNs in Figure 25.  

To investigate the abundance ratio of HCN and HCO$^{+}$, we
tabulate in Table 13 the HCN-to-H$^{13}$CN and
HCO$^{+}$-to-H$^{13}$CO$^{+}$ flux ratios at
J=3--2, based on our deep Cycle 4 isotopologue observations.  
The flux ratios of HCN-to-H$^{13}$CN and
HCO$^{+}$-to-H$^{13}$CO$^{+}$ at J=3--2 are expected to be comparable,
if all emission lines are optically thin and an intrinsic flux ratio of
main molecule to its isotopologue line at J=3--2 = $\alpha$ $\times$
main-to-isotopologue abundance ratio (= $^{12}$C-to-$^{13}$C ratio),
where $\alpha$ is comparable for HCN and HCO$^{+}$    
\footnote{
The most abundant isotopologue is generally more highly excited
than the low abundant isotopologue through collisions, due to a larger
line opacity ($\tau$) and resulting smaller effective critical density
($\propto$ 1/$\tau$).  
Thus, the $\alpha$ value is likely to be larger than unity. 
This excitation difference at J=3--2 between a main and isotopologue
line, and thereby the $\alpha$ value can be different between HCN and
HCO$^{+}$.   
If HCN abundance is higher than HCO$^{+}$ as expected ($\S$1), the
effective critical density of HCN will be reduced more than that of
HCO$^{+}$. 
It is likely that excitation to J=3--2 is promoted more for HCN and the
intrinsic HCN-to-H$^{13}$CN J=3--2 flux ratio is higher
than the intrinsic HCO$^{+}$-to-H$^{13}$CO$^{+}$ J=3--2 flux ratio. 
The estimated HCN J=3--2 flux attenuation derived from the comparison of 
the intrinsic and observed HCN-to-H$^{13}$CN J=3--2 flux ratio will
be even larger, strengthening our high HCN abundance scenario.
}. 
In Table 13, the observed HCN-to-H$^{13}$CN flux ratios are
substantially smaller than the
HCO$^{+}$-to-H$^{13}$CO$^{+}$ flux ratios in IRAS 08572$+$3915, IRAS
12112$+$0305 NE, and IRAS 22491$-$1808. 
Based on the reasonable assumption that isotopologue emission lines are
optically thin, it is suggested that the HCN J=3--2 flux attenuation by
line opacity is higher than HCO$^{+}$ J=3--2. 
Higher HCN abundance, relative to HCO$^{+}$, is a natural explanation.

For other sources, H$^{13}$CO$^{+}$ J=3--2 data were not taken, and only
shallow H$^{13}$CN J=3--2 line data were obtained in Cycle 2 or 3. 
Table 14 summarizes the HCN-to-H$^{13}$CN flux ratios at J=3--2 for
these sources.    
Only IRAS 15250$+$3609 shows a detectable H$^{13}$CN J=3--2 emission
line, and the HCN-to-H$^{13}$CN flux ratio at J=3--2 ($\sim$12) is 
as small as the above three ULIRGs, suggesting that the HCN J=3--2 flux
is significantly attenuated by line opacity.  
IRAS 15250$+$3609 shows a higher HCN-to-HCO$^{+}$ flux ratio than
starburst galaxies.  
It is possible that high HCN abundance is responsible for the ratio.

\subsubsection{Observed HCN-to-HNC Flux Ratios}

Regarding the HCN-to-HNC flux ratios in the abscissa of Figure 25,
AGN-hosting ULIRGs with high observed HCN-to-HCO$^{+}$ flux ratios show
a wide range of HCN-to-HNC flux ratios at J=3--2 and J=4--3. 
We see a trend that ULIRGs with optically identified Seyfert-type AGNs
(e.g., G: Superantennae, I: PKS 1345$+$12) distribute at the right side
with higher HCN-to-HNC flux ratios than those with optically elusive, but
infrared-identified buried AGNs (e.g., C: IRAS 20551$-$4250, D: IRAS
08572$+$3915, H: IRAS 15250$+$3609; J: IRAS 06035$-$7102, L: IRAS
12127$-$1412) \citep{vei99}.
The two ULIRGs with optically and infrared-elusive but (sub)millimeter
detectable extremely deeply buried AGNs (E: IRAS 12112$+$0305 NE and F: 
IRAS 22491$-$1808) distribute at the left side with small HCN-to-HNC
flux ratios. 

The HNC abundance is known to be relatively low in high radiation
density environments in close vicinity to luminous energy sources
(either AGNs and/or starbursts) \citep{sch92,hir98,gra14}.
In an optically-identifiable Seyfert-type AGN, the column density of the
obscuring material around a central AGN is modest, and the HNC
abundance can be smaller than that of HCN in a large fraction of
molecular gas volume.  
In an optically-elusive, but infrared-detectable buried AGN, the column
density of the obscuring material is expected to be higher than the
Seyfert-type optically-detectable AGN \citep{ima06a,ima08}.
It can be even higher in an optically- and infrared-elusive, but
(sub)millimeter-detectable, extremely deeply buried AGN.
The volume fraction of molecular gas with sufficient HNC abundance can
be higher in AGNs with a larger column of surrounding material, due to
shielding. 
The overall observed trend of the HCN-to-HNC flux ratios may be 
caused by the differences in the molecular column density around 
central AGNs. 
In the literature, low HCN-to-HNC flux ratios at J=3--2 were found in
Arp 220, Mrk 231, and NGC 4418 \citep{aal07a}, all which are (or may be) 
classified as containing buried AGNs surrounded by a large amount of
obscuring material 
\citep{dud97,spo01,ima04,arm07,dow07,vei09,gon13,aal15a,sak17}, 
supporting our scenario.   

Both HCN and HNC have a comparable critical density at the same
J-transitions under the same line opacity \citep{gre09}.
For IRAS 12112$+$0305 NE (object E in Figure 25) and IRAS
22491$-$1808 (object F in Figure 25), the HCN-to-H$^{13}$CN J=3--2
flux ratios are smaller than the HNC-to-HN$^{13}$C J=3--2 flux ratios
(Table 13), suggesting higher HCN J=3--2 line opacity and thereby higher
HCN abundance than HNC.  
For ULIRGs with HCN-to-HNC flux ratios larger than unity in Figure 25,
it is naively expected that HCN abundance is higher than HNC. 
In this case, the effective critical density ($\propto$1/$\tau$; $\tau$
is optical depth) of HCN is smaller than HNC, so that HCN can be more
easily excited by collision than HNC and that HCN J=4--3 to J=3--2 flux
ratios can be systematically higher than HNC J=4--3 to J=3--2 flux ratios.
The comparison of the J=4--3 to J=3--2 flux ratios between HCN and HNC 
in Figure 26 does not show such an expected trend.
Instead, the bulk of ULIRGs tend to show slightly higher J=4--3 to J=3--2 
flux ratios for HNC than HCN (filled symbols above the solid
dashed line). 

A possible mechanism to explain the higher J=4--3 to J=3--2 flux ratios
for HNC than HCN is infrared radiative pumping.
HCN and HNC can be vibrationally excited by absorbing infrared $\sim$14
$\mu$m and $\sim$21.5 $\mu$m photons, respectively, and through the
decay back to the vibrational ground level (v=0), the rotational
J-transition fluxes of HCN and HNC at v=0 could be enhanced, as compared to
collisional excitation alone.  
Since the number of infrared $\sim$21.5 $\mu$m photons is usually higher
than that of $\sim$14 $\mu$m photons in actual (U)LIRGs, HNC is more easily
vibrationally excited than HCN \citep{aal07a,ima16b,ima17}.
Since an AGN can emit mid-infrared 10--25 $\mu$m continuum 
from hot dust more efficiently than a starburst, this infrared 
radiative pumping has stronger effects in an AGN than in a starburst. 
In Figure 26, a higher J=4--3 to J=3--2 flux ratio for HNC than HCN is  
seen particularly in IRAS 12112$+$1305 NE (object E), which is diagnosed
to contain an extremely deeply buried AGN.
It is possible that infrared radiative pumping contributes to an increase
in the ratios for HNC in Figure 26, even though the detection of the HNC
v$_{2}$=1f emission lines is not clear partly due to contamination
by other nearby bright emission lines.  

In molecular gas around a central AGN, the infrared 21.5 $\mu$m to 14
$\mu$m continuum flux ratio at the outer part is expected to be higher
than the inner part, due to the reddening of the infrared continuum by
dust extinction. 
If HNC abundance increases substantially at the outer part of the
obscuring material due to shielding, then enhancement of the HNC flux,  
relative to the HCN flux via infrared radiative pumping can be
stronger there than in the inner part. 
This can partly work to decrease the HCN-to-HNC flux ratios in buried
AGNs surrounded by a large column density of obscuring molecular gas
more than those in optically identified Seyfert-type AGNs with a 
modest amount of surrounding obscuring material, as observed in Figure
25. 

\subsubsection{Intrinsic HCN-to-HCO$^{+}$ and HCN-to-HNC Flux Ratios}

If the enhanced HCN emission line fluxes in AGN-important ULIRGs are
due to elevated HCN abundance, then flux attenuation by line opacity is
expected to be higher for HCN than HCO$^{+}$ and HNC. 
In this case, the observed HCN-to-HCO$^{+}$ and HCN-to-HNC flux ratios
are smaller than the intrinsic values corrected for line opacities.
For ULIRGs with detected isotopologue H$^{13}$CN, H$^{13}$CO$^{+}$, and
HN$^{13}$C J=3--2 emission lines (IRAS 12112$+$0305 NE and 
IRAS 22491$-$1808), 
we can estimate the HCN-to-H$^{13}$CN,
HCO$^{+}$-to-H$^{13}$CO$^{+}$, and HNC-to-HN$^{13}$C flux ratios at
J=3--2.
As explained in $\S$5.2.1 (paragraph 4), if the observed flux ratios are
smaller than the $^{12}$C-to-$^{13}$C abundance ratio of 50--100
estimated for ULIRGs \citep{hen93a,hen93b,mar10,hen14,sli17}, we 
attribute the difference to the flux attenuation of the main molecule
(i.e., HCN, HCO$^{+}$, HNC) by line opacity.  
We derive the ratio of flux attenuation between HCN and HCO$^{+}$ (HNC)
and plot in Figure 25 the intrinsic HCN-to-HCO$^{+}$
and HCN-to-HNC flux ratios after line opacity correction
\footnote{
The {\it ratio} of flux attenuation between HCN and HCO$^{+}$ (HNC), and
thereby the correction factor in the HCN-to-HCO$^{+}$ (HCN-to-HNC) flux
ratio is independent of the adopted absolute value of the
$^{12}$C-to-$^{13}$C abundance ratio (between 50--100) and the $\alpha$
value in $\S$5.2.1 (paragraph 4).
}.
The same intrinsic flux ratios for AGN-containing ULIRG IRAS
20551$-$4205 \citep{ima16b,ima17} are also plotted in Figure 25. 
In the intrinsic flux ratios, AGNs with elevated HCN abundance and
starburst galaxies are separated more clearly than in the observed
flux ratios, making the distinction of AGNs and starbursts even more
solid.
Strictly speaking, the correction must be applied to starburst galaxies
(objects A, B, and K in Figure 25) in the same way as the
AGN-important HCN-flux-enhanced ULIRGs, but we have no isotopologue data
for these starburst galaxies.  
Starbursts show non-high HCN-to-HCO$^{+}$ flux ratios at
various J-transitions \citep{koh05,ima07a,kri08,ima09,pri15,izu16,ima16c}, 
including J=1--0 where unlike higher 
J-transitions, excitation of both HCN and HCO$^{+}$ is expected to be
close to thermal in typical starbursts. 
A much higher HCN abundance than HCO$^{+}$ is not required to explain 
the observed non-high HCN-to-HCO$^{+}$ flux ratios in starbursts. 
It is very unlikely that starburst galaxies will move to the upper-right
direction substantially after possible line opacity correction.

Our results suggest that in AGN-important ULIRGs, (1) HCN abundance is
high, (2) HCN flux is high, and (3) HCN line opacity is larger than
unity.
These three derived indications can be reconciled if molecular gas is
clumpy \citep{sol87}, where HCN flux increase is expected with
increased HCN abundance even under larger line opacity than unity,
because the surface area of line-emitting regions in individual clumps
can increase with increasing line opacity, providing a larger area
filling factor of line-emitting regions inside a molecular cloud
\citep{ima07a}. However, the indications are difficult to explain if
molecular gas spatially distributes smoothly, because in this geometry,
HCN flux saturates when HCN line opacity exceeds unity. We thus regard
that molecular gas in the observed AGN-important HCN-flux-excess ULIRGs
consists of a clumpy structure.   

In Figure 25, the observed HCN-to-HCO$^{+}$ and HCN-to-HNC flux ratios
at J=3--2 and J=4--3 are high in Superantennae (object G), which shows
by far the largest molecular line widths (Tables 4--7 and \citet{ima16c}).
According to the original clumpy molecular gas model in our Milky Way
galaxy \citep{sol87}, line opacity predominantly comes within individual 
clumps, with a minor contribution from foreground different clumps in a
molecular cloud.   
However, the molecular volume number density of (U)LIRGs is expected to be
higher than that in Galactic molecular clouds, in which case, obscuration
by foreground clumps may not be negligible. 
If the high {\it observed} HCN-to-HCO$^{+}$ and HCN-to-HNC flux ratios
in Superantennae are partly due to reduced HCN line opacity by
large line widths caused by high turbulence of clumps within molecular 
clouds, then it is indicated that obscuration by foreground clumps
within a molecular cloud contributes to flux attenuation to some 
degree in ULIRGs other than Superantennae. 
It is possible that in ULIRGs, molecular gas is in a clumpy form with a 
larger volume filling factor than Galactic molecular clouds and line
opacity comes both within individual clumps and from foreground 
clumps. 
If this scenario is the case, the line opacity by foreground clumps can
be high in less turbulent ULIRGs with small observed molecular line
widths.  
AGNs obscured by less turbulent molecular gas with elevated HCN
abundance may not necessarily show high {\it observed} HCN-to-HCO$^{+}$
flux ratios; thus, AGN identification may be missed based on the 
{\it observed} flux ratios.  
Evaluation of line opacity ratios among HCN, HCO$^{+}$, and HNC is
important particularly for AGN-important ULIRGs with relatively small
observed molecular line widths.  

\subsubsection{RADEX Calculations and Molecular Gas Physical Properties}

By assuming collisional excitation and using the large velocity gradient
(LVG) method in the widely used RADEX code \citep{van07} 
we attempted to constrain the molecular gas physical parameters, based
on the available J=4--3 and J=3--2 data of HCN, HCO$^{+}$, and HNC.  
The collisional rates are from \citet{sch05}.
Molecular line flux ratios are primarily determined by the following
three parameters: H$_{2}$ volume number density (n$_{\rm H2}$), H$_{2}$
kinetic temperature (T$_{\rm kin}$), and molecular column density
divided by line width (N$_{\rm mol}$/$\Delta$v).  
Here, the $\sim$3 K cosmic microwave background radiation is always
included, and $\Delta$v is derived from observed molecular line widths.
{\it At least three} independent observational data are necessary to
meaningfully constrain these three parameters, while we have only two
J-transitions.
We thus assume that the column densities of HCN, HCO$^{+}$, and HNC are 
1 $\times$ 10$^{16}$ cm$^{-2}$, because (1) N$_{\rm H}$ $\sim$ 10$^{24}$
cm$^{-2}$ is typically found in (U)LIRGs \citep{ric17} and (2) the
abundance ratios of HCN, HCO$^{+}$, and HNC, relative to H$_{2}$, in warm
molecular gas in (U)LIRGs are estimated to be $\sim$10$^{-8}$ \citep{gre09}.
Although the HCN abundance is estimated to be higher than HCO$^{+}$
and HNC in many of our ULIRG samples ($\S$5.2.1--5.2.3), we adopted the
same value as above, so as not to increase the free parameters relative
to the limited number of observational constraints.  

Figure 27 presents the J=4--3 to J=3--2 flux ratios of HCN, HCO$^{+}$,
and HNC, as a function of kinetic temperature (T$_{\rm kin}$ [K]) and
H$_{2}$ volume number density (n$_{\rm H2}$ [cm$^{-3}$]), where line
widths of 400 km s$^{-1}$ and 1000 km s$^{-1}$ are used to represent all
galaxies (except for Superantennae) and Superantennae, respectively. 
The ratios of J=4--3 to J=3--2 flux in (Jy km s$^{-1}$) in the observed
(U)LIRGs are 0.6--1.3, 0.7--1.5, and 0.8--1.5 for HCN, HCO$^{+}$, and HNC,
respectively (Table 11). 
In Figures 27a, b, c, assuming  a molecular gas kinetic temperature of
T$_{\rm kin}$ = 30--80 K, a volume number density of n$_{\rm H2}$ = 
10$^{4}$--10$^{7}$ cm$^{-3}$, 10$^{3}$--10$^{7}$ cm$^{-3}$, and
10$^{5}$--10$^{7}$ cm$^{-3}$ can reproduce the observed flux ratios of
HCN, HCO$^{+}$, and HNC, respectively, except for
a small fraction of sources with very low HCN J=4--3 to J=3--2 flux
ratios ($\sim$0.6).  
For the bulk of galaxies where the flux ratios are close to or larger than
unity, a volume number density of at least n$_{\rm H2}$ = 10$^{5}$--10$^{6}$
cm$^{-3}$ is needed.  
Figures 27d,e,f are RADEX calculations for the very large molecular line
width of Superantennae ($\Delta$v $\sim$ 1000 km s$^{-1}$).  
For a given kinetic temperature and volume number density, the flux
ratios tend to be slightly lower than in Figures 27a,b,c. 
Assuming the same molecular gas kinetic temperature range of 
T$_{\rm kin}$ = 30--80 K, 
a volume number density of n$_{\rm H2}$ = 10$^{5}$--10$^{6}$ cm$^{-3}$
is required to reproduce the J=4--3 to J=3--2 flux ratios of HCN,
HCO$^{+}$, and HNC for Superantennae. 
For ULIRGs with n$_{\rm H2}$ $\sim$ 10$^{5-6}$ cm$^{-3}$, in the case of
$\sim$100 pc-scale nuclear molecular size (Table 10 and \citet{sco15}),
the column density becomes N$_{\rm H2}$ $\sim$ 3 $\times$ 10$^{25-26}$
cm$^{-2}$ (i.e., heavily Compton thick), if the dense gas fully occupies
the volume of molecular clouds.  
If molecular gas is clumpy and the volume filling factor is
$\sim$1--10$\%$, the N$_{\rm H2}$ column density approaches $\sim$3
$\times$ 10$^{24}$ cm$^{-2}$ (i.e., mildly Compton thick), as observed
from X-ray observations of ULIRGs \citep{ten15,ric17,oda17}. 
The volume filling factor will be even smaller, if the nuclear molecular
size is substantially larger than $\sim$100 pc.


\subsection{Relation between Molecular Line and Infrared Luminosity}

Since stars are formed in dense molecular gas, it is expected that dense
molecular gas mass and star-formation rate are correlated. 
The correlation between infrared luminosities (= star-formation rate
indicator) and emission line luminosities of HCN and HCO$^{+}$ (= dense
molecular mass tracers) has been investigated and observationally
confirmed in various types of galaxies and Galactic star-forming regions
\citep{gao04a,wu05,eva06,gra08,ma13,zha14,liu16,tan17}.

Table 15 summarizes the luminosities of selected bright
molecular emission lines. 
Figure 28 plots the comparison of HCN and HCO$^{+}$ emission line
luminosity at J=4--3 and J=3--2 with infrared luminosity. 
For the spatially extended LIRG, NGC 1614 (the leftmost circle in
all figures), HCN and HCO$^{+}$ luminosities are derived from 
Gaussian fits of the spatially integrated spectra, but diffuse
emission with a spatial extent larger than the maximum recoverable scale
of our ALMA data can be missed.
For other ULIRGs, the luminosities are calculated from Gaussian fits 
of the spectra within the beam size, because (1) ULIRGs are usually
dominated by nuclear compact energy sources with $\lesssim$500 pc
\citep{soi00} and (2) the dense gas tracers, J=4--3 and J=3--2 emission
lines of HCN and HCO$^{+}$, are also estimated to mostly come from
nuclear compact regions (Table 10). 
We thus regard that the bulk of dense molecular gas emission is covered 
in our ALMA measurements for our sample 
\footnote{
For a small fraction of ULIRGs, the synthesized beam size is much 
smaller than 1$''$ and the probed physical scale is significantly
smaller than $\sim$1 kpc. 
For them, we measured HCN and HCO$^{+}$ fluxes at J=4--3 and J=3--2
using Gaussian fits in spectra taken with a 1$''$ diameter circular
aperture, and confirmed that the estimated fluxes in the 1$''$ aperture 
spectra agree within $<$30\% to those in the beam sized spectra.
}.

At J=4--3, the distribution of our ULIRG sample tends to be at the
upper-left side of the relation, namely the infrared to molecular line
luminosity ratios are slightly higher than the best fit lines of
\citet{tan17} for both HCN and HCO$^{+}$.  
The excess infrared emission is sometimes interpreted by significant AGN
contributions to the infrared luminosities \citep{eva06}.
However, if dust and dense molecular gas coexist around luminous AGNs in
(U)LIRGs, not only the infrared dust emission luminosity but also the
molecular line luminosity of dense gas tracers will increase.  
If we ignore possible differences in the abundance and excitation
conditions between AGNs and star-formation, the infrared to dense
molecular line luminosity ratios are not expected to change much in the
presence of luminous buried AGNs.  
If the HCN abundance increases in molecular gas in the close vicinity of
a buried AGN and/or if HCN rotational excitation is higher in an AGN, the 
infrared to HCN J=4--3 luminosity ratio could decrease, rather than
increase, in a buried AGN. 

At J=3--2, the displacement toward the upper-left side is smaller than
J=4--3. 
This can be explained if the J=4--3 emission lines of HCN and HCO$^{+}$ 
are sub-thermally excited. 
One would think that in the case of sub-thermal J=4--3 excitation, 
the J=4--3 to J=3--2 flux ratios for HCN would be smaller than
HCO$^{+}$, because the critical density of HCN is a factor of $\sim$5
higher than HCO$^{+}$ under the same line opacity.
However, the higher HCN line opacity decreases the effective HCN
critical density with $\propto$1/$\tau$; thus, it may be that the
J=4--3 excitation of HCN and HCO$^{+}$ are similarly sub-thermal. 
Unlike HCN and HNC, the infrared radiative pumping rate per given column
density differs only slightly between HCN and HCO$^{+}$
\citep{ima16b,ima17}, so that the J=4--3 to J=3--2 flux ratios are not 
expected to change differently between HCN and HCO$^{+}$ by infrared
radiative pumping. 

\subsection{Double-Peaked Emission Lines in IRAS 12112$+$0305 NE and 
IRAS 20414$-$1651}

IRAS 12112$+$0305 NE displays double-peaked emission line profiles for
HCN J=4--3, HCO$^{+}$ J=4--3, H$^{13}$CN J=3--2, CS J=5--4, and possibly
CS J=7--6 (Figure 4c,d,g,h), as previously observed for HCN J=3--2 and
HCO$^{+}$ J=3--2 \citep{ima16c}.
The central dips of HCN and HCO$^{+}$ are considerably stronger at
J=4--3 (Figure 4c) than at J=3--2 \citep{ima16c}.  
Such double-peaked emission line profiles are not clear for HNC at
J=3--2 and J=4--3 (Figure 4a,e).
One possibility to produce these double-peaked profiles is
self-absorption by foreground molecular gas \citep{aal15b}. 
If the molecular gas concentration in a galaxy nucleus is very high, the 
molecular gas spatial distribution can be better approximated by a
smooth distribution \citep{dow93,sco15}, rather than the widely accepted
clumpy distribution \citep{sol87}.  
In this scenario, molecular emission at systemic velocities is
selectively absorbed by foreground molecular gas,
because this velocity component is the most abundant. 
This effect is strong for molecules with high abundance. 

The central dip of HCO$^{+}$ J=4--3 looks stronger than that of HCN
J=4--3 (Figure 4c), which suggests that self-absorption is stronger
for HCO$^{+}$.
This seems inconsistent with the higher HCN abundance than HCO$^{+}$
derived from our ALMA isotopologue observations ($\S$5.2.1).  
The Einstein A coefficients for HCN J=4--3 and HCO$^{+}$ J=4--3 are 
20.6 $\times$ 10$^{-4}$ and 35.7 $\times$ 10$^{-4}$, respectively 
(http://www.splatalogue.net).
The Einstein B coefficient from J=3 to J=4 are related to the A
coefficient from J=4 to J=3, in the form of B$_{3-4}$ $\propto$
$\lambda^{3}$ $\times$ A$_{\rm 4-3}$, so that the B$_{3-4}$ coefficient 
for HCO$^{+}$ J=4--3 is a factor of $\sim$1.7 higher than that of HCN
J=4--3. 
Thus, the photon absorption rate at J=3 of HCO$^{+}$ could be comparable
to that of HCN, if the HCO$^{+}$ abundance is only a factor of $\sim$2
smaller than the HCN abundance (Table 13).
The central dip of HCO$^{+}$ J=4--3 could be as strong as that of HCN
J=4--3, but it is difficult to be much stronger in IRAS 12112$+$0305 NE.

To further test this scenario, we fit the data of HCN J=4--3 and
HCO$^{+}$ J=4--3 that were not strongly affected by the central dips
(Figure 4c) using a single Gaussian function.
The resulting single Gaussian fits are shown as curved dotted 
lines in Figure 17, and the estimated Gaussian fluxes are 18$\pm$17 and
19$\pm$8 (Jy km s$^{-1}$) for HCN J=4--3 and HCO$^{+}$ J=4--3,
respectively (Tables 5 and 6).
The uncertainty is large because only data at the tail of the Gaussian
profiles can be used for the fit, which precludes further quantitatively 
detailed discussion of whether the central dips of HCN J=4--3 and
HCO$^{+}$ J=4--3 can be explained by the self-absorption model in smoothly
distributed molecular gas, under the constraint of higher abundance of
HCN than HCO$^{+}$. 

However, the apparently much stronger central dips at J=4--3 than at
J=3--2 \citep{ima16c} are not easily reconciled with this self-absorption
scenario.
The Einstein A coefficients for HCN J=3--2 and HCO$^{+}$ J=3--2 are 8.4
$\times$ 10$^{-4}$ (s$^{-1}$) and 14.5 $\times$ 10$^{-4}$ (s$^{-1}$),
respectively. 
Since the wavelength at J=3--2 is longer than that at J=4--3, the
Einstein B coefficients ($\propto$ $\lambda^{3}$ $\times$ A) are
comparable within 5\% for J=2 to J=3 and J=3 to J=4, suggesting that
the strength of the central dips by self-absorption is also expected
to be roughly comparable for J=3--2 and J=4--3 of HCN and HCO$^{+}$.
In molecular gas around AGNs and starbursts, CS abundance is usually
predicted to be much lower than HCN and HCO$^{+}$ \citep{har13}, and yet
the CS J=5--4 emission line shows a strong central dip. 
Most importantly, the isotopologue H$^{13}$CN J=3--2 emission line,
which is believed to be optically thin, also displays a clear central
dip. 
It seems difficult to explain the double-peaked emission line profiles
of IRAS 12112$+$0305 NE seen in several lines with largely different line
opacities, solely by the self-absorption model in
smoothly distributed molecular gas.  
In IRAS 12112$+$0305 NE, (1) the flux of HNC J=4--3 is higher than 
those of HCN J=4--3 and HCO$^{+}$ J=4--3 (Tables 5, 6, and 7), and 
(2) the flux of HNC J=3--2 (Table 4) is also as high as those of HCN
J=3--2 and HCO$^{+}$ J=3--2 \citep{ima16c}, suggesting
that the HNC abundance is at least comparable to the HCN and HCO$^{+}$
abundances.   
In galaxies, CO abundance is usually more than 3--4 orders of magnitude
higher than HCN and HCO$^{+}$ \citep{mei05,har13}. 
Although strong central dips are expected for these high abundance
molecular lines, HNC J=4--3, HNC J=3--2, and CO J=1--0, they show
single-peaked emission profiles (Figures 4, 16, and 17, and
\citet{eva02}).

Another scenario is that the observed double-peaked emission line
profiles come from emission from molecular gas in rotating disks
\citep{sco17}.  
The velocity separations of the double-peaked emission lines are
$\sim$300 km s$^{-1}$ for HCN J=4--3 and HCO$^{+}$ J=4--3 (Figure 17),
which are comparable to those for HCN J=3--2 and HCO$^{+}$ J=3--2
\citep{ima16c}. 
The intensity-weighted mean velocity (moment 1) maps for
HCN J=4--3 and HCO$^{+}$ J=4--3 of IRAS 12112$+$0305 NE in Figure 20 show
rotational motion, with a velocity difference of $\sim$300 km s$^{-1}$. 
The observed double-peaked emission line profiles of HCN J=4--3 and
HCO$^{+}$ J=4--3 can largely be explained by rotation. 
The same argument was applied to HCN J=3--2 and HCO$^{+}$ J=3--2
\citep{ima16c}.
This rotating disk model can at least explain the double-peaked
emission profile of the optically thin H$^{13}$CN J=3--2 line more
naturally.   
However, the single-peaked emission line profiles of HNC J=4--3, HNC
J=3--2, and CO J=1--0 need to be explained. 
Since the critical density of CO J=1--0 is much smaller than those of
HCN, HCO$^{+}$, and HNC, the CO J=1--0 emission is expected to be much
more spatially extended.
For HNC, as explained in $\S$5.2.2, its abundance can be low at the
inner part close to the central buried energy source, but can be high at
the outer part, due to shielding.  
In fact, at the AGN-dominant nucleus of the nearby well-studied Seyfert
2 galaxy, NGC 1068, while HCN J=3--2 and HCO$^{+}$ J=3--2 emission lines
are clearly detected and their spatial distribution is very similar
\citep{ima16a,ima18}, HNC J=3--2 emission is much weaker (Imanishi et
al. in preparation).
If (a) the bulk of the CO J=1--0 emission line comes from
spatially extended lower density molecular gas and (b) HNC J=4--3 and
HNC J=3--2 emission lines also originate in outer regions than HCN and
HCO$^{+}$ J=4--3 and J=3--2 lines, then the dynamics of CO J=1--0, HNC
J=4--3, and HNC J=3--2 may not be dominated by rotation as strongly as
the inner {HCN and HCO$^{+}$ emitting dense molecular gas, 
which may be a reason for the smoother single Gaussian emission
line profiles.  
The stronger double-peaked profiles of HCN and HCO$^{+}$ at J=4--3
than at J=3--2 could also be explained by
rotation, because the J=4--3 emission lines come from more central regions
dominated by strong rotation than the J=3--2 emission lines, due to
a higher critical density at J=4--3.
The HNC-emitting molecular gas in the nuclear region of IRAS
12112$+$0305 NE should have a volume number density as high as the inner
HCN- and HCO$^{+}$-emitting molecular gases, because the estimated
volume number densities are comparable among HCN, HCO$^{+}$, and HNC
(Figure 27).     
It is likely that the HNC emitting molecular gas distributes more
outside than the HCN and HCO$^{+}$ emitting one, but is still much more
inside than the CO J=1--0 emitting lower density molecular gas.
The CH$_{3}$OH line also displays a single Gaussian profile (Figures 4e
and 31), but this is a shock tracer ($\S$5.5) whose spatial distribution
and line profile can be largely different from those of the nuclear dense
molecular gas close to the buried energy source probed with the HCN and
HCO$^{+}$ J=4--3 and J=3--2 emission lines.

The other ULIRG, IRAS 20414$-$1651, also displays clear double-peaked
emission line profiles for HCN J=4--3, HCO$^{+}$ J=4--3, and HNC J=4--3
(Figures 12 and 17), as well as HCN J=3--2 and HCO$^{+}$ J=3--2
\citep{ima16c}. 
A similar double-peaked emission line profile is also seen in CS 7--6
(Figures 12 and 31). 
For the same reasons as applied to IRAS 12112$+$0305 NE, the
double-peaked profiles of these molecular lines with various optical
depths can be reasonably reproduced by rotation, if these lines trace
similar dense molecular gas.

\subsection{Shock-Origin Molecular Emission Lines}

Other serendipitously detected emission lines are evident in the
spectra of some observed ULIRGs, including CH$_{3}$OH, SiO, SO, and 
SO$_{2}$, which were also detected in several nearby bright (U)LIRGs in
the literature 
\citep{mar11,cos11,mei14,ala15,cos15,ima16b,ima17,sai17,ued17,sli17}.  

Fairly strong CH$_{3}$OH emission was detected in the nuclear region of
IRAS 12112$+$0305 NE in Figure 4e. 
The gas-phase reaction is inefficient to form a large amount of
CH$_{3}$OH \citep{lee96}. Strong CH$_{3}$OH emission is interpreted
to originate in shocked regions by a slow shock (v $<$ 10--15 km
s$^{-1}$), shielded from intense UV radiation, because a shock can
liberate CH$_{3}$OH produced in the ice mantle at the surface of dust
grains to the gas phase \citep{meie05,gar10,mei12,sai17,ued17}.   
A buried AGN surrounded by a large column density of gas and dust can
explain strong CH$_{3}$OH emission if a shock occurs at the UV-shielded
region of the surrounding material. 
SiO J=6--5 emission is also detected in IRAS 12112$+$0305 NE (Figure 4g).
It is widely argued that a fast shock (v $>$ 15--20 km s$^{-1}$) can
destroy the cores of dust grains and release SiO to the gas phase,
explaining strong SiO emission in the (sub)millimeter wavelength range
\citep{mar92,gar01,use06,gar10,mei12}. 
It is possible that strong shocks are created by molecular outflow at
the nucleus of IRAS 12112$+$0305 NE.  

SO emission is detected in the spectrum of IRAS 22491$-$1808 
(Figure 5c).
As almost all of the sulfur is believed to be depleted onto dust grains
\citep{wak11}, shocks are thought to liberate sulfur-bearing species
into the gas phase \citep{mar03,mar05}.  
Another sulfur-bearing species, H$_{2}$S \citep{ima14b}, and another
shock tracer, SiO (Figure 5c), were also detected. 
IRAS 22491$-$1808 may contain some shock activity.

These shock tracers were also detected in other ULIRGs in our sample.
H$_{2}$S was detected in IRAS 08572$+$3915 \citep{ima14b}.
The ULIRG IRAS 20551$-$4250 shows several shock tracers: H$_{2}$S, 
SO$_{2}$, CH$_{3}$OH, SO, and HNCO \citep{ima13b,ima16b,ima17}.
The other shock tracer CH$_{3}$CN \citep{cod09} was detected in IRAS
12127$-$1412 \citep{ima14b} and IRAS 20551$-$4250 \citep{ima13b}.
Our ALMA data show that shock phenomena are common in the nearby ULIRG 
population. 

\section{Summary} 

We presented our ALMA Cycle 2, 3, and 4 observational results of eleven 
(U)LIRGs (IRAS 08572$+$3915, Superantennae, IRAS 12112$+$0305, IRAS
22491$-$1808, NGC 1614, IRAS 12127$-$1412, IRAS 15250$+$3609, PKS
1345$+$12, IRAS 06035$-$7102, IRAS 13509$+$0442, and IRAS 20414$-$1651) 
at J=3--2 of HNC and at J=4--3 of HCN, HCO$^{+}$, and HNC.
All of these (U)LIRGs have available ALMA data of HCN and HCO$^{+}$ at
J=3--2 \citep{ima16c}.
HCN, HCO$^{+}$, and HNC J=4--3 data had been taken in ALMA Cycle 0 for some
sources. 
By adding our new ALMA data to these existing data, we now have J=3--2
and J=4--3 data of HCN, HCO$^{+}$, HNC for most of these
eleven (U)LIRGs.  
These multiple rotational J-transition molecular line data make it
possible to investigate molecular gas physical properties and the
physical reason for the observed molecular line flux ratios.
Several interesting molecular line features were detected in selected
(U)LIRGs. 
We summarize our findings in the following.

\begin{enumerate}

\item The targeted HCN, HCO$^{+}$, and HNC emission lines were detected 
in all of the observed main (U)LIRG's nuclei. 
In IRAS 12112$+$0305, in addition to the brighter main north-eastern
nucleus (IRAS 12112$+$0305 NE), the HCO$^{+}$ J=4--3 emission line was 
detected in the fainter south-western nucleus (IRAS 12112$+$0305 SW).

\item We detected the signatures of the vibrationally excited
(v$_{2}$=1f) HCN J=4--3 emission line in the
infrared-starburst-classified ULIRG, IRAS 12112$+$0305 NE, which had
previously shown a detectable HCN v$_{2}$=1f J=3--2 emission line. 
The HCN vibrational excitation is believed to originate in infrared
radiative pumping by absorbing infrared 14 $\mu$m photons, 
most likely coming from AGN-heated hot dust emission. 
The v$_{2}$=1f to v=0 flux ratios of HCN at J=3--2 and J=4--3 in 
IRAS 12112$+$0305 NE could be even higher than those of IRAS
20551$-$4250, the AGN-containing ULIRG with clearly detected HCN
v$_{2}$=1f J=3--2 and J=4--3 emission lines.
IRAS 12112$+$0305 NE may be classified as a ULIRG, which contains an
infrared-elusive, but (sub)millimeter-detectable, extremely deeply
buried AGN. 

\item We confirmed that ULIRGs with infrared- or
(sub)millimeter-identified luminous obscured AGNs tend to  
show significantly higher HCN-to-HCO$^{+}$ flux ratios at J=3--2 and
J=4--3 than galaxies that are classified as starburst-dominated, with no
AGN signatures at all (i.e., NGC 1614, IRAS 12112$+$0305 SW, and
IRAS 13509$+$0442). 

\item HCN-to-HNC flux ratios at J=3--2 and J=4--3 vary among 
AGN-classified ULIRGs with elevated HCN-to-HCO$^{+}$ flux ratios. 
The HCN-to-HNC flux ratios tend to be higher in optically identified
Seyfert-type AGNs than in optically elusive infrared- or
(sub)millimeter-detectable buried AGNs.
The latter class is thought to contain a larger amount of molecular gas
and dust around a central AGN compared with the former class. 
We interpret that HNC abundance is low at the strongly AGN-affected
inner part of the surrounding obscuring material, but high at the outer
part, due to shielding.
In optically elusive buried AGNs surrounded by a large amount of
molecular gas and dust, HNC lines can be strongly emitted from the
outer high-HNC-abundance regions.

\item From isotopologue H$^{13}$CN, H$^{13}$CO$^{+}$, and HN$^{13}$C
J=3--2 observations for selected ULIRGs, classified as
AGN-important due to high HCN-to-HCO$^{+}$ J=3--2 and J=4--3 flux
ratios, larger flux attenuation by line opacity was estimated for HCN
J=3--2 than HCO$^{+}$ J=3--2 and HNC J=3--2.  
Higher HCN abundance is a natural explanation. 
AGN-important galaxies deviate from the distribution of starburst
galaxies even further if line-opacity-corrected intrinsic
HCN-to-HCO$^{+}$ and HCN-to-HNC flux ratios are used rather than observed
flux ratios, making our molecular line flux method even more solid. 

\item We interpret that elevated HCN abundance is primarily responsible
for the high HCN-to-HCO$^{+}$ flux ratios in AGN-important ULIRGs. 
This scenario and the significant flux attenuation of HCN J=3--2 can be
reconciled if molecular gas in ULIRGs distributes in a clumpy form, as
was widely argued \citep{sol87}, where line opacity largely comes within
individual clumps. 

\item The HCO$^{+}$ J=4--3-to-J=3--2 flux ratios are not systematically
higher than HCN J=4--3-to-J=3--2 flux ratios in our observed (U)LIRGs, 
despite the fact that HCN has a factor of $\sim$5 larger critical
density than HCO$^{+}$ at the same J-transition under the same line
opacity. 
Excitation has a limited effect for the higher HCN-to-HCO$^{+}$ flux
ratios observed in AGN-important ULIRGs than in starburst galaxies.

\item From the J=4--3 and J=3--2 fluxes of HCN, HCO$^{+}$, and HNC, we
constrained the molecular gas kinetic temperature and volume number
density, using the RADEX code and by assuming column densities of
these molecules to be 1 $\times$ 10$^{16}$ cm$^{-2}$. 
For the bulk of ULIRGs whose J=4--3 to J=3--2 flux ratios 
of HCN, HCO$^{+}$, and HNC are larger than or close to unity,
dense (n$_{\rm H2}$ $>$ 10$^{5}$--10$^{6}$ cm$^{-3}$) molecular gas is
needed, if the kinetic temperature is in the range of 30--80 K.

\item A possible signature of the vibrationally excited HNC v$_{2}$=1f 
J=4--3 emission line is seen in IRAS 15250$+$3609. 
However, for other (U)LIRGs, the weakness of this line and contamination by
other nearby bright emission lines preclude quantitatively reliable
estimates of the HNC v$_{2}$=1f emission line fluxes at J=3--2 and
J=4--3.   

\item A broad blueshifted absorption feature was detected in the
HCO$^{+}$ J=4--3 line in IRAS 15250$+$3609, as had previously been seen
in the HCO$^{+}$ J=3--2 line.  
When combined with the detection of the redshifted emission sub-peaks,
molecular outflow activity is a natural explanation.  
The profiles were weaker for HCN and HNC than HCO$^{+}$, suggesting 
enhanced HCO$^{+}$ abundance in the outflow of IRAS 15250$+$3609. 
We estimated a molecular outflow mass M$_{\rm outf}$ $\sim$ 4 $\times$
10$^{8}$ M$_{\odot}$, a molecular outflow rate \.{M}$_{\rm outf}$ $\sim$
250--750 (M$_{\odot}$ yr$^{-1}$), a molecular outflow kinetic power 
P$_{\rm outf}$ $\sim$ 3\% of AGN luminosity (L$_{\rm AGN}$), and 
a molecular outflow momentum rate \.{P}$_{\rm outf}$ $\sim$ 37 
$\times$ L$_{\rm AGN}$/c in IRAS 15250$+$3609.
These properties can be explained by energy-conserving outflow rather
than radiation pressure-driven outflow.

\item IRAS 12112$+$0305 NE also shows the P Cygni profile in the 
SiO J=6--5 line.
Given that SiO is a good fast shock tracer, the presence of
shock-important molecular outflow is suggested in IRAS 12112$+$0305 NE. 

\item Our (U)LIRG sample roughly follows the previously established
relation between the infrared luminosity and molecular line 
luminosity of HCN and HCO$^{+}$ at J=4--3 and J=3--2.

\item Clear double-peaked emission line profiles are detected for 
HCN J=4--3 and HCO$^{+}$ J=4--3 in IRAS 12112$+$0305 NE and IRAS
20414$-$1651, as previously seen in the HCN J=3--2 and HCO$^{+}$
J=3--2 lines. 
Given that similar double-peaked emission line profiles are seen in
other molecular lines with expected smaller line opacities, 
we interpret that rotation can largely explain the profiles.

\item Several shock tracers, such as CH$_{3}$OH, SiO, SO, and SO$_{2}$
emission lines, are detected fairly strongly in several ULIRGs,
suggesting ubiquitous shock phenomena in these ULIRGs.  

\end{enumerate}

\acknowledgments 
 
We thank the anonymous referee for his/her valuable comments which  
helped improve the clarity of this manuscript.
We are grateful to Dr. K. Saigo, Y. Ao, A. Kawamura, and F. Egusa for
their supports regarding ALMA data reduction. 
M.I. was supported by JSPS KAKENHI Grant Number 23540273,15K05030 and
the ALMA Japan Research Grant of the NAOJ Chile Observatory,
NAOJ-ALMA-0001, 0023, 0072.
This paper made use of the following ALMA data:
ADS/JAO.ALMA\#2013.1.00032.S, 2015.1.00027.S, and 2016.1.00051.S. 
ALMA is a partnership of ESO (representing its member states), NSF (USA) 
and NINS (Japan), together with NRC (Canada), NSC and ASIAA
(Taiwan), and KASI (Republic of Korea), in cooperation with the Republic
of Chile. The Joint ALMA Observatory is operated by ESO, AUI/NRAO, and
NAOJ. 
Data analysis was in part carried out on the open use data analysis
computer system at the Astronomy Data Center, ADC, of the National
Astronomical Observatory of Japan. 
This research has made use of NASA's Astrophysics Data System and the
NASA/IPAC Extragalactic Database (NED) which is operated by the Jet
Propulsion Laboratory, California Institute of Technology, under
contract with the National Aeronautics and Space Administration. 




\begin{deluxetable}{lccrrrrcll}
\tabletypesize{\scriptsize}
\tablecaption{Basic Properties of Observed (Ultra)luminous Infrared Galaxies 
\label{tbl-1}} 
\tablewidth{0pt}
\tablehead{
\colhead{Object} & \colhead{Redshift} & \colhead{Physical scale} & 
\colhead{f$_{\rm 12}$}   & 
\colhead{f$_{\rm 25}$}   & 
\colhead{f$_{\rm 60}$}   & 
\colhead{f$_{\rm 100}$}  & 
\colhead{log L$_{\rm IR}$} & 
\colhead{Optical} & \colhead{IR/submm} 
\\
\colhead{} & \colhead{} & \colhead{[kpc/$''$]}  & \colhead{[Jy]}
& \colhead{[Jy]} & \colhead{[Jy]} & \colhead{[Jy]}  &
\colhead{[L$_{\odot}$]} & \colhead{Class} & \colhead{AGN} \\  
\colhead{(1)} & \colhead{(2)} & \colhead{(3)} & \colhead{(4)} & 
\colhead{(5)} & \colhead{(6)} & \colhead{(7)} & \colhead{(8)} & 
\colhead{(9)} & \colhead{(10)} 
}
\startdata
IRAS 08572$+$3915 & 0.0580 & 1.1 & 0.32 & 1.70 & 7.43  & 4.59  & 12.1 &
LI$^{a}$(Sy2$^{b}$) & Y$^{g,h,i,j,k,l,m,n}$ \\   
Superantennae (IRAS 19254$-$7245) & 0.0617 & 1.2 & 0.22 & 1.24 & 5.48 &
5.79 & 12.1 & Sy2 $^{c}$ & Y$^{o,p}$ \\    
IRAS 12112$+$0305 & 0.0730 & 1.4 & 0.12 & 0.51 & 8.50 & 9.98 & 12.3 & LI
$^{a}$ (Sy2 $^{b}$) & Y$^{q}$ \\     
IRAS 22491$-$1808 & 0.0776 \tablenotemark{A} & 1.5 & 0.05 & 0.55 & 5.44
& 4.45 & 12.2 & HII$^{a,b}$ & Y$^{q}$ \\  
NGC 1614 & 0.0160 & 0.32 & 1.38 & 7.50 & 32.12 & 34.32 & 11.7 & HII
$^{d,e}$ (Cp$^{b}$) & N \\ 
IRAS 12127$-$1412 & 0.1332 \tablenotemark{A} & 2.3 & $<$0.13 & 0.24 &
1.54 & 1.13 & 12.2 & LI $^{a}$ (HII $^{b}$) & Y$^{i,k,m,n,p}$ \\
IRAS 15250$+$3609 & 0.0552 \tablenotemark{A} & 1.1 & 0.16 & 1.31 & 7.10
& 5.93 & 12.0 & LI $^{d}$ (Cp$^{b}$) & Y$^{m,n,p,r,s,t}$ \\
PKS 1345$+$12 (IRAS 13451$+$1232) & 0.1215 \tablenotemark{A} & 2.2 &
0.14 & 0.67 & 1.92 & 2.06 & 12.3 & Sy2 $^{a,b}$ & Y$^{m,n,p}$ \\
\enddata

\tablenotetext{A}{
Redshifts are based on ALMA molecular line data \citep{ima16c}.
}
\tablecomments{
Col.(1): Object name. 
Col.(2): Redshift. 
Col.(3): Physical scale in [kpc arcsec$^{-1}$]. 
Col.(4)--(7): f$_{12}$, f$_{25}$, f$_{60}$, and f$_{100}$ are 
{\it IRAS} fluxes at 12 $\mu$m, 25 $\mu$m, 60 $\mu$m, and 100 $\mu$m,
respectively, taken from \citet{kim98} or \citet{san03} or the IRAS
Faint Source Catalog (FSC).  
Col.(8): Decimal logarithm of infrared (8$-$1000 $\mu$m) luminosity
in units of solar luminosity (L$_{\odot}$), calculated with
$L_{\rm IR} = 2.1 \times 10^{39} \times$ D(Mpc)$^{2}$
$\times$ (13.48 $\times$ $f_{12}$ + 5.16 $\times$ $f_{25}$ +
$2.58 \times f_{60} + f_{100}$) [ergs s$^{-1}$] \citep{sam96}.
Col.(9): Optical spectroscopic classification.
``Sy2'', ``LI'', ``HII'', and ``Cp'' refer to Seyfert 2, LINER,
HII-region, and starburst$+$AGN composite, respectively. 
$^{a}$: \citet{vei99}.
$^{b}$: \citet{yua10}.
$^{c}$: \citet{mir91}.
$^{d}$: \citet{vei95}.
$^{e}$: \citet{kew01}.
$^{f}$: \citet{duc97}.
Col.(9): AGN signature in infrared and/or (sub)millimeter spectra.
``Y'' means that AGN signatures are detected, while sources with 
no such detectable signatures are marked as ``N''.
$^{g}$: \citet{dud97}.
$^{h}$: \citet{imd00}.
$^{i}$: \citet{ima06a}.
$^{j}$: \citet{spo06}.
$^{k}$: \citet{ima07b}.
$^{l}$: \citet{arm07}.
$^{m}$: \citet{vei09}.
$^{n}$: \citet{nar10}.
$^{o}$: \citet{ris03}.
$^{p}$: \citet{ima10}.
$^{q}$: \citet{ima16c}.
$^{r}$: \citet{spo02}.
$^{s}$: \citet{ima11b}.
$^{t}$: \citet{sti13}.
$^{u}$: \citet{dar07}.
$^{v}$: \citet{far09}.
}

\end{deluxetable}

\begin{deluxetable}{lllccc|ccc}
\tabletypesize{\scriptsize}
\tablecaption{Log of ALMA Cycle 2, 3, and 4 Observations \label{tbl-2}} 
\tablewidth{0pt}
\tablehead{
\colhead{Object} & \colhead{Line} & \colhead{Date} & \colhead{Antenna} & 
\colhead{Baseline} & \colhead{Integration} & \multicolumn{3}{c}{Calibrator} \\ 
\colhead{} & \colhead{} & \colhead{[UT]} & \colhead{Number} & \colhead{[m]} &
\colhead{[min]} & \colhead{Bandpass} & \colhead{Flux} & \colhead{Phase}  \\
\colhead{(1)} & \colhead{(2)} & \colhead{(3)} & \colhead{(4)} &
\colhead{(5)} & \colhead{(6)} & \colhead{(7)}  & \colhead{(8)} & \colhead{(9)}
}
\startdata 
IRAS 08572$+$3915 & HNC J=3--2 & 2015 Jun 14 & 39 & 21--784 & 12 &
J0750$+$1231 & J0854$+$2006 & J0927$+$3902 \\
& J=3--2 isotopologue & 2016 Oct 13 & 40 & 19--1808 & 49 & J0854$+$2006
& J0854$+$2006 & J0916$+$3854  \\  
& & 2016 Oct 14 & 42 & 19--2535 & 49 & J0854$+$2006 & J0750$+$1231 &
J0824$+$3916 \\  
& & 2016 Oct 15 & 41 & 19--1809 & 49 & J0854$+$2006 & J0854$+$2006 &
J0916$+$3854 \\  \hline
Superantennae & HNC J=3--2 & 2015 May 15 & 37 & 23--558 & 9.3 &
J2056$-$4714 & Ceres & J1933$-$6942 \\ 
 & HCN/HCO$^{+}$ J=4--3 & 2015 May 13 & 36 & 21--558 & 5.2 &
J1924$-$2914 & Ceres & J1933$-$6942 \\ 
 & HNC J=4--3 & 2015 May 14 & 37 & 21--558 & 5.2 & J1924$-$2914 & Ceres
& J1933$-$6942 \\ \hline  
IRAS 12112$+$0305 & HNC J=3--2 & 2015 May 14 & 37 & 23--558 & 5.2 &
J1058$+$0133 & Ganymede & J1229$+$0203  \\
\enddata

\tablecomments{ 
Col.(1): Object name. 
Col.(2): Observed molecular line. 
Col.(3): Observing date in UT. 
Col.(4): Number of antennas used for observations. 
Col.(5): Baseline length in meter. Minimum and maximum baseline lengths are 
shown.  
Col.(6): Net on source integration time in min.
Cols.(7), (8), and (9): Bandpass, flux, and phase calibrator for the 
target source, respectively.
}

\end{deluxetable}

\begin{deluxetable}{clrccl}
\tabletypesize{\scriptsize}
\tablecaption{Continuum Emission \label{tbl-3}}
\tablewidth{0pt}
\tablehead{
\colhead{Object} & \colhead{Frequency} & \colhead{Flux} & 
\colhead{Peak Coordinate} & \colhead{rms} & \colhead{Synthesized beam} \\
\colhead{} & \colhead{[GHz]} & \colhead{[mJy/beam]} & 
\colhead{(RA,DEC)J2000} & \colhead{[mJy/beam]} & 
\colhead{[$''$ $\times$ $''$] ($^{\circ}$)} \\  
\colhead{(1)} & \colhead{(2)} & \colhead{(3)}  & \colhead{(4)}  &
\colhead{(5)} & \colhead{(6)} 
}
\startdata 
08572$+$3915 & 32b: 241.8--245.4, 256.1--259.8 & 1.8 (23$\sigma$) &
(09 00 25.36, $+$39 03 54.1) & 0.079 & 0.90$\times$0.51 (9$^{\circ}$) \\  
 & iso32: 228.8--232.5, 243.7--247.4 & 1.4 (44$\sigma$) & (09 00 25.38,
$+$39 03 54.1) & 0.031 & 0.39$\times$0.18 (16$^{\circ}$) \\ 
Superantennae & 32b: 240.9--244.6, 255.3--259.0 & 4.4 (38$\sigma$) & (19
31 21.45, $-$72 39 21.5) & 0.12 & 1.0$\times$0.61 (59$^{\circ}$) \\ 
              & 43a: 333.0--336.8, 345.1--348.8 & 7.7 (29$\sigma$) & (19
31 21.43, $-$72 39 21.5) & 0.26 & 0.74$\times$0.41 (66$^{\circ}$) \\ 
              & 43b: 328.8--332.4, 340.7--344.5 & 9.2 (44$\sigma$) & (19
31 21.43, $-$72 39 21.6) & 0.21 & 0.61$\times$0.50 (12$^{\circ}$) \\ 
12112$+$0305 NE & 32b: 238.3--242.0, 252.5--256.2 & 7.9 (58$\sigma$)
& (12 13 46.06, $+$02 48 41.5) & 0.14 & 0.81$\times$0.56 (66$^{\circ}$)\\
  & 43a: 317.3--321.0, 329.4--333.2 &  16.0 (46$\sigma$) & (12 13 46.05,
$+$02 48 41.5) & 0.35 & 1.4$\times$0.56 (71$^{\circ}$) \\   
  & 43b: 336.9--340.8, 349.1--352.7 & 13.3 (39$\sigma$) & (12 13 46.06,
$+$02 48 41.6) & 0.34 & 1.0$\times$0.65 (81$^{\circ}$) \\
\enddata

\tablenotetext{A}{All LSB data were flagged.}

\tablecomments{
Col.(1): Object.
Col.(2): Frequency range in [GHz] used for continuum extraction.
Frequencies at obvious emission lines and absorption features are removed.
We denote continuum data taken with ``HNC J=3--2'', ``J=4--3 of HCN and
HCO$^{+}$'', ``HNC J=4--3'', and ``J=3--2 of H$^{13}$CN,
H$^{13}$CO$^{+}$, HN$^{13}$C'' as 32b, 43a, 43b, and iso32,
respectively. 
Col.(3): Flux in [mJy beam$^{-1}$] at the emission peak.
Value at the highest flux pixel (0$\farcs$03--0$\farcs$1 pixel$^{-1}$)
is extracted. 
The detection significance relative to the rms noise is shown in
parentheses. 
Possible systematic uncertainties, coming from the absolute flux
calibration uncertainty of ALMA and choice of frequency range for
the continuum level determination, are not included.  
Col.(4): Coordinate of the continuum emission peak in J2000.
Col.(5): The rms noise level (1$\sigma$) in [mJy beam$^{-1}$], derived
from the standard deviation of sky signals in each continuum map.
Col.(6): Synthesized beam in [arcsec $\times$ arcsec] and position angle
in [degree]. 
The position angle is 0$^{\circ}$ along the north--south direction
and increases in the counterclockwise direction. 
}

\end{deluxetable}

\begin{deluxetable}{l|lcl|cccc}
\tabletypesize{\scriptsize}
\tablecaption{Flux of HNC J=3--2 Emission Line \label{tbl-4}} 
\tablewidth{0pt}
\tablehead{
\colhead{Object} & \multicolumn{3}{c}{Integrated intensity (moment 0) map} & 
\multicolumn{4}{c}{Gaussian line fit} \\  
\colhead{} & \colhead{Peak} &
\colhead{rms} & \colhead{Beam} & \colhead{Velocity}
& \colhead{Peak} & \colhead{FWHM} & \colhead{Flux} \\ 
\colhead{} & \multicolumn{2}{c}{[Jy beam$^{-1}$ km s$^{-1}$]} & 
\colhead{[$''$ $\times$ $''$] ($^{\circ}$)} &
\colhead{[km s$^{-1}$]} & \colhead{[mJy]} & \colhead{[km s$^{-1}$]} &
\colhead{[Jy km s$^{-1}$]} \\  
\colhead{(1)} & \colhead{(2)} & \colhead{(3)} & \colhead{(4)} & 
\colhead{(5)} & \colhead{(6)} & \colhead{(7)} & \colhead{(8)} 
}
\startdata 
IRAS 08572$+$3915 & 0.59 (7.5$\sigma$) & 0.079 & 0.87$\times$0.49
($-$172$^{\circ}$) & 17497$\pm$12 & 2.6$\pm$0.2 & 265$\pm$28 &
0.71$\pm$0.10 \\      
Superantennae & 1.5 (9.4$\sigma$) & 0.16 & 0.97$\times$0.59
(59$^{\circ}$) & 18493$\pm$24 & 2.6$\pm$0.2 & 637$\pm$55 & 1.6$\pm$0.2
\\      
IRAS 12112$+$0305 NE & 7.1 (45$\sigma$) & 0.16 & 0.79$\times$0.54
(66$^{\circ}$) & 21796$\pm$4 & 22$\pm$1 & 362$\pm$11 & 7.8$\pm$0.3 \\     
IRAS 12112$+$0305 SW & 0.47 (3.7$\sigma$) & 0.13& 0.79$\times$0.54
(66$^{\circ}$) & 21980$\pm$32 & 1.5$\pm$0.3 & 321$\pm$70 & 0.47$\pm$0.14
\\       
IRAS 22491$-$1808 & 5.4 (38$\sigma$) & 0.14 & 0.40$\times$0.35
(79$^{\circ}$) & 23300$\pm$5 & 17$\pm$1 & 352$\pm$12 & 6.0$\pm$0.3 \\      
NGC 1614 N-peak & 0.23 (1.8$\sigma$) & 0.13 & 1.0$\times$0.38
($-$68$^{\circ}$) & 4833$\pm$7 & 3.8$\pm$0.6 & 71$\pm$13 & 0.28$\pm$0.07
\\       
NGC 1614 S-peak & 0.29 (2.3$\sigma$) & 0.13 & 1.0$\times$0.38
($-$68$^{\circ}$) & 4697$\pm$34 & 1.7$\pm$0.5 & 274$\pm$182 &
0.49$\pm$0.36 \\       
NGC 1614 2$\farcs$5 radius & --- & --- & --- & 4808 (fix) & 9.5$\pm$2.0 &
229$\pm$73 & 2.3$\pm$0.9 \\
\enddata

\tablecomments{ 
Col.(1): Object.
Col.(2): Integrated intensity of the HNC J=3--2 emission line 
($\nu_{\rm rest}$=271.981 GHz) in [Jy beam$^{-1}$ km s$^{-1}$] at the 
emission peak. 
Detection significance relative to the rms noise (1$\sigma$) in the 
moment 0 map is shown in parentheses. 
Possible systematic uncertainty is not included. 
Col.(3): The rms noise (1$\sigma$) level in the moment 0 map in 
[Jy beam$^{-1}$ km s$^{-1}$], derived from the standard deviation 
of sky signals in each moment 0 map. 
Col.(4): Beam size in [arcsec $\times$ arcsec] and position angle in
[degree]. Position angle is 0$^{\circ}$ along the north-south direction, 
and increases counterclockwise. 
Cols.(5)--(8): Gaussian fits of emission lines in the spectra at the 
continuum peak position, within the beam size. 
For NGC 1614, the fit in a spatially integrated spectrum is also
shown. 
Col.(5): Optical LSR velocity (v$_{\rm opt}$) of emission peak in 
[km s$^{-1}$].  
Col.(6): Peak flux in [mJy]. 
Col.(7): Observed full width at half maximum (FWHM) in [km s$^{-1}$].
Col.(8): Flux in [Jy km s$^{-1}$]. The observed FWHM in 
[km s$^{-1}$] in column 7 is divided by ($1+z$) to obtain the
intrinsic FWHM in [km s$^{-1}$]. 
}

\end{deluxetable}


\begin{deluxetable}{l|lcl|cccc}
\tabletypesize{\scriptsize}
\tablecaption{Flux of HCN J=4--3 Emission Line \label{tbl-5}} 
\tablewidth{0pt}
\tablehead{
\colhead{Object} & \multicolumn{3}{c}{Integrated intensity (moment 0) map} & 
\multicolumn{4}{c}{Gaussian line fit} \\  
\colhead{} & \colhead{Peak} &
\colhead{rms} & \colhead{Beam} & \colhead{Velocity}
& \colhead{Peak} & \colhead{FWHM} & \colhead{Flux} \\ 
\colhead{} & \multicolumn{2}{c}{[Jy beam$^{-1}$ km s$^{-1}$]} & 
\colhead{[$''$ $\times$ $''$] ($^{\circ}$)} &
\colhead{[km s$^{-1}$]} & \colhead{[mJy]} & \colhead{[km s$^{-1}$]} &
\colhead{[Jy km s$^{-1}$]} \\  
\colhead{(1)} & \colhead{(2)} & \colhead{(3)} & \colhead{(4)} & 
\colhead{(5)} & \colhead{(6)} & \colhead{(7)} & \colhead{(8)} 
}
\startdata 
Superantennae & 8.6 (4.3$\sigma$) & 2.0 & 0.76$\times$0.43 (66$^{\circ}$)
& 18569$\pm$18 & 10$\pm$1 & 957$\pm$47 & 9.7$\pm$0.6 \\      
IRAS 12112$+$0305NE & 6.0 (22$\sigma$) & 0.27 & 1.4$\times$0.56
(71$^{\circ}$) & 21688$\pm$29,21976$\pm$12 & 12$\pm$1, 17$\pm$2 &
315$\pm$67,164$\pm$28 & 6.6$\pm$1.0 \\  
 & & & & 21818$\pm$6 \tablenotemark{A} & 66$\pm$59 \tablenotemark{A} &
279$\pm$41 \tablenotemark{A} & 18$\pm$17 \tablenotemark{A} \\
IRAS 12112$+$0305SW & $<$0.63 ($<$3$\sigma$) & 0.21& 1.4$\times$0.56
(71$^{\circ}$) & --- & --- & --- & --- \\  
IRAS 15250$+$3609 & 6.3 (26$\sigma$)\tablenotemark{B} & 0.25
\tablenotemark{B} &
0.88$\times$0.54(26$^{\circ}$)\tablenotemark{B} & 16561$\pm$5
\tablenotemark{B} & 25$\pm$1 \tablenotemark{B} & 269$\pm$12
\tablenotemark{B} & 6.8$\pm$0.4 \tablenotemark{B} \\ 
 & 6.8 (22$\sigma$)\tablenotemark{C} & 0.32 \tablenotemark{C}
& 0.88$\times$0.54(26$^{\circ}$)\tablenotemark{C} & 16561$\pm$4,17004$\pm$16 \tablenotemark{C} & 25$\pm$1, 4.7$\pm$1.1 \tablenotemark{C}
& 268$\pm$9,117$\pm$43\tablenotemark{C} & 7.3$\pm$0.4
\tablenotemark{C} \\ 
PKS 1345$+$12 & 1.5 (7.8$\sigma$) & 0.19 & 0.69$\times$0.58
(42$^{\circ}$) & 36458$\pm$27 & 3.6$\pm$0.4 & 510$\pm$60 & 1.7$\pm$0.3 \\     
IRAS 13509$+$0442 & 0.68 (10$\sigma$) & 0.067 & 0.70$\times$0.51
(48$^{\circ}$) & 40924$\pm$13 & 3.1$\pm$0.3 & 268$\pm$34 & 0.78$\pm$0.12 \\
\enddata

\tablenotetext{A}{Single Gaussian fit using data not strongly affected
by the central dip.
Data at 21340--21620 km s$^{-1}$ and 22010--22150 km s$^{-1}$ in Figure
17 are used for the fit.
}

\tablenotetext{B}{Main emission component only.
}

\tablenotetext{C}{Main emission component and sub-peak emission
component are included (moment 0 map) or fit simultaneously (Gaussian fit).
}

\tablecomments{ 
Col.(1): Object.
Col.(2): Integrated intensity of the HCN J=4--3 emission line 
($\nu_{\rm rest}$=354.505 GHz) in [Jy beam$^{-1}$ km s$^{-1}$] at the 
emission peak. 
Detection significance relative to the rms noise (1$\sigma$) in the 
moment 0 map is shown in parentheses. 
Possible systematic uncertainty is not included. 
Col.(3): The rms noise (1$\sigma$) level in the moment 0 map in 
[Jy beam$^{-1}$ km s$^{-1}$], derived from the standard deviation 
of sky signals in each moment 0 map. 
Col.(4): Beam size in [arcsec $\times$ arcsec] and position angle in
[degree]. Position angle is 0$^{\circ}$ along the north-south direction, 
and increases counterclockwise. 
Cols.(5)--(8): Gaussian fits of emission lines in the spectra at the 
continuum peak position, within the beam size. 
Col.(5): Optical LSR velocity (v$_{\rm opt}$) of emission peak in 
[km s$^{-1}$].  
Col.(6): Peak flux in [mJy]. 
Col.(7): Observed FWHM in [km s$^{-1}$].
Col.(8): Flux in [Jy km s$^{-1}$]. The observed FWHM in 
[km s$^{-1}$] in column 7 is divided by ($1+z$) to obtain the
intrinsic FWHM in [km s$^{-1}$]. 
}

\end{deluxetable}


\begin{deluxetable}{l|lcl|cccc}
\tabletypesize{\scriptsize}
\tablecaption{Flux of HCO$^{+}$ J=4--3 Emission Line \label{tbl-6}} 
\tablewidth{0pt}
\tablehead{
\colhead{Object} & \multicolumn{3}{c}{Integrated intensity (moment 0) map} & 
\multicolumn{4}{c}{Gaussian line fit} \\  
\colhead{} & \colhead{Peak} &
\colhead{rms} & \colhead{Beam} & \colhead{Velocity}
& \colhead{Peak} & \colhead{FWHM} & \colhead{Flux} \\ 
\colhead{} & \multicolumn{2}{c}{[Jy beam$^{-1}$ km s$^{-1}$]} & 
\colhead{[$''$ $\times$ $''$] ($^{\circ}$)} &
\colhead{[km s$^{-1}$]} & \colhead{[mJy]} & \colhead{[km s$^{-1}$]} &
\colhead{[Jy km s$^{-1}$]} \\  
\colhead{(1)} & \colhead{(2)} & \colhead{(3)} & \colhead{(4)} & 
\colhead{(5)} & \colhead{(6)} & \colhead{(7)} & \colhead{(8)} 
}
\startdata 
Superantennae & 5.3 (6.6$\sigma$) & 0.80 & 0.75$\times$0.42
(67$^{\circ}$) & 18565$\pm$18 & 7.3$\pm$0.3 & 827$\pm$46 & 6.1$\pm$0.4 \\      
IRAS 12112$+$0305NE & 3.9 (11$\sigma$) & 0.38 & 1.4$\times$0.54
(71$^{\circ}$) & 21638$\pm$16,21988$\pm$8 & 9.7$\pm$1.4,16$\pm$2 &
203$\pm$43,178$\pm$28 & 4.8$\pm$0.7 \\      
 & & & & 21818$\pm$6 \tablenotemark{A} & 69$\pm$29 \tablenotemark{A} & 
278$\pm$27 \tablenotemark{A} & 19$\pm$8 \tablenotemark{A} \\
IRAS 12112$+$0305SW & 1.4 (4.2$\sigma$) & 0.34& 1.4$\times$0.55
(71$^{\circ}$) & 21873$\pm$29 & 5.9$\pm$1.6 & 271$\pm$80 & 1.6$\pm$0.6
\\ 
IRAS 15250$+$3609 & 2.2 (16$\sigma$)\tablenotemark{B} &
0.14\tablenotemark{B} & 
0.86$\times$0.48 (30$^{\circ}$)\tablenotemark{B} & 16568$\pm$12
\tablenotemark{B} & 14$\pm$2 \tablenotemark{B} & 156$\pm$28
\tablenotemark{B} & 2.3$\pm$0.5 \tablenotemark{B} \\ 
& 3.9 (12$\sigma$)\tablenotemark{C} & 0.32\tablenotemark{C} &
0.86$\times$0.48 (30$^{\circ}$)\tablenotemark{C}  
& 16567$\pm$7,17055$\pm$21\tablenotemark{C} & 14$\pm$1, 7.9$\pm$1.2
\tablenotemark{C} & 157$\pm$16,233$\pm$56\tablenotemark{C} & 4.1$\pm$0.6
\tablenotemark{C} \\  
PKS 1345$+$12 & 2.6 (13$\sigma$) & 0.21 & 0.68$\times$0.58
(42$^{\circ}$) & 36474$\pm$17 & 6.9$\pm$0.5 & 479$\pm$36 & 3.2$\pm$0.3 \\      
IRAS 13509$+$0442 & 0.80 (10$\sigma$) & 0.080 & 0.69$\times$0.50
(47$^{\circ}$) & 40941$\pm$13 & 3.5$\pm$0.3 & 292$\pm$28 & 0.95$\pm$0.13 \\
\enddata

\tablenotetext{A}{Single Gaussian fit using data not strongly affected
by the central dip. 
Data at 21500--21620 km s$^{-1}$ and 22000--22130 km s$^{-1}$ in Figure
17 are used. 
}

\tablenotetext{B}{Main emission component only.
}

\tablenotetext{C}{Main emission component and sub-peak emission
component are included (moment 0 map) or fit simultaneously (Gaussian fit).
}

\tablecomments{ 
Col.(1): Object.
Col.(2): Integrated intensity of the HCO$^{+}$ J=4--3 emission line 
($\nu_{\rm rest}$=356.734 GHz) in [Jy beam$^{-1}$ km s$^{-1}$] at the 
emission peak. 
Detection significance relative to the rms noise (1$\sigma$) in the 
moment 0 map is shown in parentheses. 
Possible systematic uncertainty is not included. 
Col.(3): The rms noise (1$\sigma$) level in the moment 0 map in 
[Jy beam$^{-1}$ km s$^{-1}$], derived from the standard deviation 
of sky signals in each moment 0 map. 
Col.(4): Beam size in [arcsec $\times$ arcsec] and position angle in
[degree]. Position angle is 0$^{\circ}$ along the north-south direction, 
and increases counterclockwise. 
Cols.(5)--(8): Gaussian fits of emission lines in the spectra at the 
continuum peak position, within the beam size. 
Col.(5): Optical LSR velocity (v$_{\rm opt}$) of emission peak in 
[km s$^{-1}$]. 
Col.(6): Peak flux in [mJy]. 
Col.(7): Observed FWHM in [km s$^{-1}$].
Col.(8): Flux in [Jy km s$^{-1}$]. The observed FWHM in 
[km s$^{-1}$] in column 7 is divided by ($1+z$) to obtain the
intrinsic FWHM in [km s$^{-1}$]. 
}

\end{deluxetable}


\begin{deluxetable}{l|lcl|cccc}
\tabletypesize{\scriptsize}
\tablecaption{Flux of HNC J=4--3 Emission Line \label{tbl-7}} 
\tablewidth{0pt}
\tablehead{
\colhead{Object} & \multicolumn{3}{c}{Integrated intensity (moment 0) map} & 
\multicolumn{4}{c}{Gaussian line fit} \\  
\colhead{} & \colhead{Peak} &
\colhead{rms} & \colhead{Beam} & \colhead{Velocity}
& \colhead{Peak} & \colhead{FWHM} & \colhead{Flux} \\ 
\colhead{} & \multicolumn{2}{c}{[Jy beam$^{-1}$ km s$^{-1}$]} & 
\colhead{[$''$ $\times$ $''$] ($^{\circ}$)} &
\colhead{[km s$^{-1}$]} & \colhead{[mJy]} & \colhead{[km s$^{-1}$]} &
\colhead{[Jy km s$^{-1}$]} \\  
\colhead{(1)} & \colhead{(2)} & \colhead{(3)} & \colhead{(4)} & 
\colhead{(5)} & \colhead{(6)} & \colhead{(7)} & \colhead{(8)} 
}
\startdata 
Superantennae & 1.1 (6.6$\sigma$) & 0.16 & 0.60$\times$0.49
(11$^{\circ}$) & 18504$\pm$37 & 2.6$\pm$0.3 & 546$\pm$85 & 1.4$\pm$0.3
\\      
IRAS 12112$+$0305 NE & 9.6 (26$\sigma$) & 0.37 & 1.1$\times$0.68
(81$^{\circ}$) &  
21797$\pm$5 & 26$\pm$1 & 409$\pm$11 & 11$\pm$1 \\     
IRAS 12112$+$0305 SW & $<$0.76 ($<$3$\sigma$) & 0.25 & 
1.1$\times$0.68 (81$^{\circ}$) & --- & --- & --- & --- \\ 
IRAS 15250$+$3609 & 9.6 (32$\sigma$) & 0.30 & 1.1$\times$0.55
($-$26$^{\circ}$) & 16589$\pm$4 & 39$\pm$1 & 253$\pm$8 & 10$\pm$1 \\ 
IRAS 13509$+$0442 & 0.72 (6.5$\sigma$) & 0.11 & 0.66$\times$0.59
($-$41$^{\circ}$) & 40867$\pm$15 & 3.5$\pm$0.4 & 242$\pm$39 &
0.80$\pm$0.16 \\
\enddata

\tablecomments{ 
Col.(1): Object.
Col.(2): Integrated intensity of the HNC J=4--3 emission line 
($\nu_{\rm rest}$=362.630 GHz) in [Jy beam$^{-1}$ km s$^{-1}$] at the 
emission peak. 
Detection significance relative to the rms noise (1$\sigma$) in the 
moment 0 map is shown in parentheses. 
Possible systematic uncertainty is not included. 
Col.(3): The rms noise (1$\sigma$) level in the moment 0 map in 
[Jy beam$^{-1}$ km s$^{-1}$], derived from the standard deviation 
of sky signals in each moment 0 map. 
Col.(4): Beam size in [arcsec $\times$ arcsec] and position angle in
[degree]. Position angle is 0$^{\circ}$ along the north-south direction, 
and increases counterclockwise. 
Cols.(5)--(8): Gaussian fits of emission lines in the spectra at the 
continuum peak position, within the beam size. 
Col.(5): Optical LSR velocity (v$_{\rm opt}$) of emission peak in 
[km s$^{-1}$]. 
Col.(6): Peak flux in [mJy]. 
Col.(7): Observed FWHM in [km s$^{-1}$].
Col.(8): Flux in [Jy km s$^{-1}$]. The observed FWHM in 
[km s$^{-1}$] in column 7 is divided by ($1+z$) to obtain the
intrinsic FWHM in [km s$^{-1}$]. 
}

\end{deluxetable}


\begin{turnpage}
\begin{deluxetable}{lc|lcl|cccc}
\tabletypesize{\scriptsize}
\tablecaption{Flux of Isotopologue Emission Lines Based on Cycle 4 Data
\label{tbl-8}} 
\tablewidth{0pt}
\tablehead{
\colhead{Object} & \colhead{Line} & 
\multicolumn{3}{c}{Integrated intensity (moment 0) map} & 
\multicolumn{4}{c}{Gaussian line fit} \\  
\colhead{} & \colhead{} & \colhead{Peak} &
\colhead{rms} & \colhead{Beam} & \colhead{Velocity}
& \colhead{Peak} & \colhead{FWHM} & \colhead{Flux} \\ 
\colhead{} & \colhead{} 
& \multicolumn{2}{c}{[Jy beam$^{-1}$ km s$^{-1}$]} &  
\colhead{[$''$ $\times$ $''$] ($^{\circ}$)} &
\colhead{[km s$^{-1}$]} & \colhead{[mJy]} & \colhead{[km s$^{-1}$]} &
\colhead{[Jy km s$^{-1}$]} \\  
\colhead{(1)} & \colhead{(2)} & \colhead{(3)} & \colhead{(4)} & 
\colhead{(5)} & \colhead{(6)} & \colhead{(7)} & \colhead{(8)} &
\colhead{(9)} 
} 
\startdata 
08572$+$3915 & H$^{13}$CN J=3--2 & 0.11 (2.9$\sigma$) & 0.039 &
0.37$\times$0.18 (17$^{\circ}$) & 17524$\pm$29 & 0.34$\pm$0.05 &
393$\pm$87 & 0.14$\pm$0.04 \\       
 & H$^{13}$CO$^{+}$ J=3--2 & $<$0.094 ($<$3$\sigma$) \tablenotemark{A} 
& 0.031 \tablenotemark{A} & 0.37$\times$0.18 (15$^{\circ}$) & 17450$\pm$90
&0.20$\pm$0.09 & 235 (fix) & 0.047$\pm$0.021 \\  
 & HN$^{13}$C J=3--2 & $<$0.067 ($<$3$\sigma$) & 0.022 &
0.37$\times$0.18 (15$^{\circ}$) & --- & --- & --- & --- \\       
12112$+$0305 NE & H$^{13}$CN J=3--2 & 0.56 (12$\sigma$) &
0.049 & 0.22$\times$0.20 (39$^{\circ}$) & 21687$\pm$6, 21969$\pm$5 &
1.6$\pm$0.1, 1.8$\pm$0.1 & 237$\pm$17, 125$\pm$11 & 0.6$\pm$0.1 \\    
 & H$^{13}$CO$^{+}$ J=3--2 & 0.060 (2.6$\sigma$)
\tablenotemark{B} & 0.023 \tablenotemark{B} & 0.22$\times$0.19
(38$^{\circ}$) & 22003$\pm$31 \tablenotemark{C} & 
0.43 (fix) \tablenotemark{C} & 131 (fix) \tablenotemark{C} &
0.060 \tablenotemark{C} (0.16) \tablenotemark{D} \\    
 & HN$^{13}$C J=3--2 & 0.32 (7.0$\sigma$) &
0.046 & 0.22$\times$0.19 (38$^{\circ}$) & 21793$\pm$9 &
1.3$\pm$0.1 & 305$\pm$29 & 0.38$\pm$0.04 \\    
12112$+$0305 SW & H$^{13}$CN J=3--2 & $<$0.11 ($<$3$\sigma$) &
0.033 & 0.22$\times$0.20 (39$^{\circ}$) & --- & --- & --- & --- \\    
 & H$^{13}$CO$^{+}$ J=3--2 & $<$0.086 ($<$3$\sigma$) & 0.029 &
0.22$\times$0.19 (38$^{\circ}$) & --- & --- & --- & --- \\
\enddata
\end{deluxetable}
\end{turnpage}

\begin{deluxetable}{lcccc}
\tabletypesize{\scriptsize}
\tablecaption{Intrinsic Continuum Emission Size after Deconvolution
  \label{tbl-9}} 
\tablewidth{0pt}
\tablehead{
\colhead{Object} & \colhead{continuum}  & \colhead{apparent size}  &
\colhead{beam}  & \colhead{physical size} \\
\colhead{} & \colhead{} & 
\colhead{[mas $\times$ mas] ($^{\circ}$)} & 
\colhead{[arcsec $\times$ arcsec]} &  \colhead{[pc $\times$ pc]} \\ 
\colhead{(1)} & \colhead{(2)} & \colhead{(3)} & \colhead{(4)} & \colhead{(5)} 
}
\startdata
IRAS 08572$+$3915 & 32b & 903$\pm$98 $\times$ 468$\pm$46 (17$\pm7) $&
0.90 $\times$ 0.51 & 1000 $\times$ 520 \\ 
                  & iso32 & 196$\pm$15 $\times$ 157$\pm$18 (58$\pm$18) &
0.39 $\times$ 0.18 & 220 $\times$ 170 \\ \hline
Superantennae & 32b & 504$\pm$53 $\times$ 476$\pm$45 (48$\pm$87) & 1.0
$\times$ 0.61 & 590 $\times$ 560 \\
              & 43a & 513$\pm$68 $\times$ 334$\pm$141 (131$\pm$24) &
0.74 $\times$ 0.41 & 600 $\times$ 390 \\
              & 43b & 548$\pm$39 $\times$ 279$\pm$53 (143$\pm$7) & 0.61
$\times$ 0.50 & 640 $\times$ 330 \\ \hline
IRAS 12112$+$0305 NE & 32b & 269$\pm$42 $\times$ 215$\pm$37 (88$\pm$47)
& 0.81 $\times$ 0.56 & 370 $\times$ 290 \\
                     & 43a & 341$\pm$90 $\times$ 163$\pm$114
(124$\pm$32) & 1.4 $\times$ 0.56 & 470 $\times$ 220 \\
                     & 43b & 365$\pm$96 $\times$ 84$\pm$64 (75$\pm$18) &
1.0 $\times$ 0.65 & 500 $\times$ 120 \\
\enddata

\tablecomments{
Col.(1): Object name. 
Col.(2): Continuum band.
Col.(3): Intrinsic continuum emission size in [mas], after deconvolution
using the CASA task ``imfit''. 
The position angle in [degree] is shown in parentheses.
Col.(4): Synthesized beam size in [arcsec].
Col.(5): Intrinsic continuum emission size in physical scale in [pc].
No value is shown for IRAS 13509$+$0442 NE, because its redshift is
unknown. 
}

\end{deluxetable}

\begin{deluxetable}{lcccc}
\tabletypesize{\scriptsize}
\tablecaption{Intrinsic Size of Molecular Emission Lines after 
Deconvolution \label{tbl-10}} 
\tablewidth{0pt}
\tablehead{
\colhead{Object} & \colhead{HNC J=3--2} & \colhead{HCN J=4--3} &
\colhead{HCO$^{+}$ J=4--3} & \colhead{HNC J=4--3} \\
\colhead{} & \colhead{[mas $\times$ mas] ($^{\circ}$)} & 
\colhead{[mas $\times$ mas] ($^{\circ}$)} & 
\colhead{[mas $\times$ mas] ($^{\circ}$)} & 
\colhead{[mas $\times$ mas] ($^{\circ}$)} \\   
\colhead{(1)} & \colhead{(2)} & \colhead{(3)} & \colhead{(4)} &
\colhead{(5)}  
}
\startdata
IRAS 08572$+$3915 & 781$\pm$276 $\times$ 180$\pm$170 &
could not deconvolve \tablenotemark{A} & could not deconvolve
\tablenotemark{A} & --- \\
Superantennae & 440$\pm$179 $\times$ 80$\pm$329  &
$<$490 $\times$ $<$150 & could not deconvolve & 1210$\pm$330 $\times$
270$\pm$130 \\
IRAS 12112$+$0305 NE & 266$\pm$49 $\times$ 222$\pm$46 & 631$\pm$158
$\times$ 196$\pm$82 & could not deconvolve & $<$440 $\times$ $<$180 \\
IRAS 12112$+$0305 SW & could not deconvolve & signal too weak & point
source & point source \\
IRAS 22491$-$1808 & 313$\pm$37 $\times$ 273$\pm$40 & could not
deconvolve \tablenotemark{A} &  
could not deconvolve \tablenotemark{A} & could not deconvolve
\tablenotemark{A} \\
NGC 1614 & 2260$\pm$670 $\times$ 1430$\pm$410 \tablenotemark{B} & 
1860$\pm$140 $\times$ 710$\pm$290 \tablenotemark{A} & 
1854$\pm$42 $\times$ 1028$\pm$54 \tablenotemark{A} &
1330$\pm$330 $\times$ 870$\pm$560 \tablenotemark{A} \\
IRAS 12127$-$1412 & could not deconvolve & could not deconvolve
\tablenotemark{A} & could not deconvolve \tablenotemark{A} & 
could not deconvolve \tablenotemark{A} \\
IRAS 15250$+$3609 & 249$\pm$120 $\times$ 197$\pm$53 & 466$\pm$85
$\times$ 64$\pm$84 & $<$480 $\times$ $<$86 & could not deconvolve \\
\enddata

\tablenotetext{A}{Derived from Cycle 0 data presented by
\citet{ima13a,ima14b}.
Due to limited data quality and the large synthesized beam size of Cycle 0
data, compared to Cycle 2, 3, and 4 data, constraints on the 
intrinsic emission size after deconvolution are not strong in many cases.  
}

\tablenotetext{B}{Measured at the southern HNC J=3--2 emission peak in 
Figure 14. 
}

\tablecomments{
Col.(1): Object name. 
Cols.(2), (3), (4), and (5): Intrinsic emission size of HNC J=3--2, HCN
J=4--3, HCO$^{+}$ J=4--3, and HNC J=4--3 in [mas], respectively, after
deconvolution using the CASA task ``imfit''. 
Position angle is 0$^{\circ}$ along the north-south direction, 
and increases counterclockwise. }

\end{deluxetable}

\begin{deluxetable}{lccc}
\tablecaption{Ratio of J=4--3 to J=3--2 Flux in (Jy km s$^{-1}$) for 
HCN, HCO$^{+}$, and HNC \label{tbl-11}}  
\tablewidth{0pt}
\tablehead{
\colhead{Object} & \colhead{HCN} & \colhead{HCO$^{+}$} & \colhead{HNC} \\
\colhead{(1)} & \colhead{(2)} & \colhead{(3)} & \colhead{(4)} 
}
\startdata 
IRAS 08572$+$3915 & 1.3$\pm$0.2 & 1.4$\pm$0.2 & --- \\
Superantennae & 0.9$\pm$0.2 & 0.8$\pm$0.1 & 0.9$\pm$0.2 \\
IRAS 12112$+$0305 NE & 0.8$\pm$0.2 & 1.0$\pm$0.2 & 1.4$\pm$0.2 \\     
IRAS 12112$+$0305 SW \tablenotemark{n} & $<$1.6 & 1.2$\pm$0.5 & $<$2.3 \\     
IRAS 22491$-$1808 & 0.9$\pm$0.2 &  1.5$\pm$0.2 & 1.2$\pm$0.2 \\      
NGC 1614 \tablenotemark{n} & 0.6$\pm$0.2 & 1.0$\pm$0.2 & 0.8$\pm$0.4 \\ 
IRAS 12127$-$1412 & 0.9$\pm$0.2 & 0.8$\pm$0.2 & 1.1$\pm$0.3 \\
IRAS 15250$+$3609 & 1.3$\pm$0.2 & 1.2$\pm$0.3 & 1.5$\pm$0.2 \\
PKS 1345$+$12 & 0.6$\pm$0.1 & 0.8$\pm$0.1 & --- \\
IRAS 13509$+$0442 \tablenotemark{n} & 0.7$\pm$0.1 & 0.7$\pm$0.1 &
0.9$\pm$0.2 \\  
IRAS 20414$-$1651 & 1.3$\pm$0.3 & 0.8$\pm$0.1 & --- 

\enddata

\tablenotetext{n}{No detectable AGN signatures are seen in the optical, 
infrared, and (sub)millimeter spectra (Table 1).} 

\tablecomments{ 
Col.(1): Object.
Cols.(2), (3), and (4): Ratio of J=4--3 to J=3--2 flux in (Jy km
s$^{-1}$) for HCN, HCO$^{+}$, and HNC, respectively. 
Possible absolute flux calibration uncertainty with maximum $\sim$10\%
in individual ALMA band 6 and 7 data are included.
}

\end{deluxetable}

\begin{deluxetable}{lcccc}
\tabletypesize{\small}
\tablecaption{Ratio of HCN-to-HCO$^{+}$ and HCN-to-HNC Flux in 
(Jy km s$^{-1}$) at J=4--3 and J=3--2 \label{tbl-12}} 
\tablewidth{0pt}
\tablehead{
\colhead{Object} & \multicolumn{2}{c}{HCN/HCO$^{+}$} &
\multicolumn{2}{c}{HCN/HNC} \\ 
\colhead{} & \colhead{J=4--3} & \colhead{J=3--2} & \colhead{J=4--3} &
\colhead{J=3--2} \\ 
\colhead{(1)} & \colhead{(2)} & \colhead{(3)} & \colhead{(4)} & \colhead{(5)} 
}
\startdata 
IRAS 08572$+$3915 & 0.8$\pm$0.1 & 0.9$\pm$0.1 & ---  & 3.6$\pm$0.6 \\
Superantennae & 1.6$\pm$0.3 & 1.4$\pm$0.2 & 6.8$\pm$1.8 & 6.4$\pm$1.0 \\
IRAS 12112$+$0305 NE & 1.4$\pm$0.3 & 1.8$\pm$0.3 & 0.6$\pm$0.1 &
1.0$\pm$0.2 \\      
IRAS 12112$+$0305 SW \tablenotemark{n} & $<$1.6 & 0.5$\pm$0.2 & --- &
1.4$\pm$0.6 \\      
IRAS 22491$-$1808 & 1.4$\pm$0.1 & 2.3$\pm$0.2 & 1.1$\pm$0.1 &
1.4$\pm$0.2 \\       
NGC 1614 \tablenotemark{n} & 0.18$\pm$0.06 & 0.33$\pm$0.11 & 1.6$\pm$0.7
& 2.2$\pm$1.1 \\ 
IRAS 12127$-$1412 & 1.5$\pm$0.5 & 1.3$\pm$0.2 & 1.3$\pm$0.4 & 1.6$\pm$0.3 \\
IRAS 15250$+$3609 & 3.0$\pm$0.7 & 2.6$\pm$0.4 & 0.7$\pm$0.1 & 0.8$\pm$0.1 \\
\enddata

\tablenotetext{n}{No detectable AGN signatures are seen in the
optical, infrared, and (sub)millimeter spectra (Table 1).}

\tablecomments{ 
Col.(1): Object. 
Col.(2): Ratio of HCN-to-HCO$^{+}$ flux in (Jy km s$^{-1}$) at J=4--3.
Col.(3): Ratio of HCN-to-HCO$^{+}$ flux in (Jy km s$^{-1}$) at J=3--2.
Col.(4): Ratio of HCN-to-HNC flux in (Jy km s$^{-1}$) at J=4--3.
Col.(5): Ratio of HCN-to-HNC flux in (Jy km s$^{-1}$) at J=3--2.
Since HCN and HCO$^{+}$ line data were taken simultaneously at J=3--2 or
J=4--3, only statistical uncertainties for Gaussian fits are considered 
in Cols. (2) and (3), except Superantennae, for which possible additional
$\sim$10\% systematic uncertainty is added, because the extremely broad
molecular emission line profiles make continuum determination ambiguous
\citep{ima16c}.
HCN and HNC line data were taken separately.
For the HCN-to-HNC flux ratios in Cols. (4) and (5), possible $\sim$10\%
absolute flux calibration uncertainty is added.
}

\end{deluxetable}

\begin{deluxetable}{lccc}
\tabletypesize{\scriptsize}
\tablecaption{Ratio of Main to Isotopologue Emission Line Flux in 
[Jy km s$^{-1}$] Based on ALMA Cycle 4 Data \label{tbl-13}}  
\tablewidth{0pt}
\tablehead{
\colhead{Object} & \colhead{HCN-to-H$^{13}$CN J=3--2} &
\colhead{HCO$^{+}$-to-H$^{13}$CO$^{+}$ J=3--2} & 
\colhead{HNC-to-HN$^{13}$C J=3--2}  \\
\colhead{(1)} & \colhead{(2)} & \colhead{(3)} & \colhead{(4)} 
}
\startdata 
IRAS 08572$+$3915 & 19 & $>$31 & $>$10 \\
IRAS 12112$+$0305 NE & 14 & 77 (29 \tablenotemark{A}) & 20 \\
IRAS 12112$+$0305 SW & $>$6 & $>$15 & $>$4 \\
IRAS 22491$-$1808 & 8 & 11 & 31  

\enddata

\tablenotetext{A}{We assume that H$^{13}$CO$^{+}$ J=3--2 has the same 
emission line profile as H$^{13}$CN J=3--2, and that the blue component
of the H$^{13}$CO$^{+}$ J=3--2 emission is overlapped with the stronger
SiO J=6--5 emission line in Figure 18.}

\tablecomments{ 
Col.(1): Object.
Col.(2): Ratio of HCN J=3--2 to H$^{13}$CN J=3--2 flux in (Jy km
s$^{-1}$).
HCN J=3--2 flux is taken from \citet{ima16c}.
Col.(3): Ratio of HCO$^{+}$ J=3--2 to H$^{13}$CO$^{+}$ J=3--2 flux in
(Jy km s$^{-1}$). 
HCO$^{+}$ J=3--2 flux is taken from \citet{ima16c}.
Col.(4): Ratio of HNC J=3--2 to HN$^{13}$C J=3--2 flux in (Jy km s$^{-1}$).
}

\end{deluxetable}

\begin{deluxetable}{lccc}
\tabletypesize{\small}
\tablecaption{Ratio of HCN J=3--2 to H$^{13}$CN J=3--2 Flux in 
[Jy km s$^{-1}$] Based on ALMA Cycle 2 and 3 Data \label{tbl-14}} 
\tablewidth{0pt}
\tablehead{
\colhead{Object} & \colhead{HCN J=3--2} & 
\colhead{H$^{13}$CN J=3--2} & \colhead{HCN/H$^{13}$CN J=3--2}  \\ 
\colhead{(1)} & \colhead{(2)} & \colhead{(3)} & \colhead{(4)} 
}
\startdata 
Superantennae & 10.6$\pm$0.3 & $<$0.21 & $>$51 \\
NGC 1614 & 0.87$\pm$2.0, 0.85$\pm$0.22 & $<$0.25
\tablenotemark{A} & $>$3.5 \tablenotemark{A} \\
IRAS 12127$-$1412 & 1.3$\pm$0.1 & $<$0.14 & $>$9 \\ 
IRAS 15250$+$3609 & 5.2$\pm$0.3 & 0.43$\pm$0.16 & 12 \\
PKS 1345$+$12 & 2.7$\pm$0.3 & $<$0.13 & $>$21 \\
IRAS 06035$-$7102 & 4.9$\pm$0.2 & $<$0.24 & $>$20 \\
IRAS 13509$+$0442 & 1.1$\pm$0.2 & $<$0.12 & $>$9 

\enddata

\tablenotetext{A}{Both at the northern and southern HCN J=3--2 emission
peaks defined in \citet{ima16c}. 
}

\tablecomments{ 
Col.(1): Object.
Col.(2): HCN J=3--2 flux in (Jy km s$^{-1}$) taken from \citet{ima16c}.
Col.(3): H$^{13}$CN J=3--2 flux in (Jy km s$^{-1}$) derived from 
integrated intensity (moment 0) maps in ALMA Cycle 2 or 3 data (Table 16).
For undetected sources, 3$\sigma$ upper limits are provided.
Col.(4): Ratio of HCN J=3--2 to H$^{13}$CN J=3--2 flux in 
(Jy km s$^{-1}$). 
}

\end{deluxetable}


\begin{deluxetable}{lcccc}
\tabletypesize{\scriptsize}
\tablecaption{Luminosity of Selected Bright Molecular Emission Lines 
\label{tbl-15}}
\tablewidth{0pt}
\tablehead{
\colhead{Object} & \colhead{HNC J=3--2} & 
\colhead{HCN J=4--3} & 
\colhead{HCO$^{+}$ J=4--3} &
\colhead{HNC J=4--3} \\
\colhead{(1)} & \colhead{(2)} & \colhead{(3)} & \colhead{(4)} & \colhead{(5)} 
}
\startdata
IRAS 08572$+$3915 & 1.2$\pm$0.2 & 7.3$\pm$0.7 \tablenotemark{A} &
9.7$\pm$0.7 \tablenotemark{A} & --- \\ 
 & 1.9$\pm$0.3 & 5.1$\pm$0.5 \tablenotemark{A} &
6.6$\pm$0.5 \tablenotemark{A} & --- \\
Superantennae & 3.2$\pm$0.4 & 25$\pm$2 & 16$\pm$1 & 3.7$\pm$0.8 \\
 & 4.9$\pm$0.6 & 18$\pm$1 & 11$\pm$1 & 2.4$\pm$0.5 \\
IRAS 12112$+$0305 NE & 22$\pm$1 & 24$\pm$4 & 18$\pm$3 & 37$\pm$4 \\
 & 34$\pm$1 & 17$\pm$3 & 12$\pm$2 & 24$\pm$2 \\
IRAS 12112$+$0305 SW & 1.3$\pm$0.4 & non-detection & 5.9$\pm$2.2 &
non-detection \\
 & 2.0$\pm$0.6 & non-detection & 4.0$\pm$1.5 &
non-detection \\
\enddata

\tablenotetext{A}{Taken from \citet{ima14b}.}

\tablenotetext{B}{Taken from \citet{ima13a}, measured with a 3$''$ 
$\times$ 3$''$ square aperture.}

\tablenotetext{C}{Nuclear main emission line component only, 
after excluding the outflow-origin sub-peak emission component.}

\tablecomments{
Col.(1): Object.
Cols.(2)--(5): Luminosity of HNC J=3--2, HCN J=4--3, HCO$^{+}$ J=4--3,
and HNC J=4--3, in units of (10$^{4}$ L$_{\odot}$) calculated with 
Equation (1) of \citet{sol05} (first row) and in units of (10$^{7}$ K km
s$^{-1}$ pc$^{2}$) calculated with Equation (3) of \citet{sol05} (second
row).
}

\end{deluxetable}


\begin{figure}
\begin{center}
\vspace*{-0.9cm}
\includegraphics[angle=0,scale=.406]{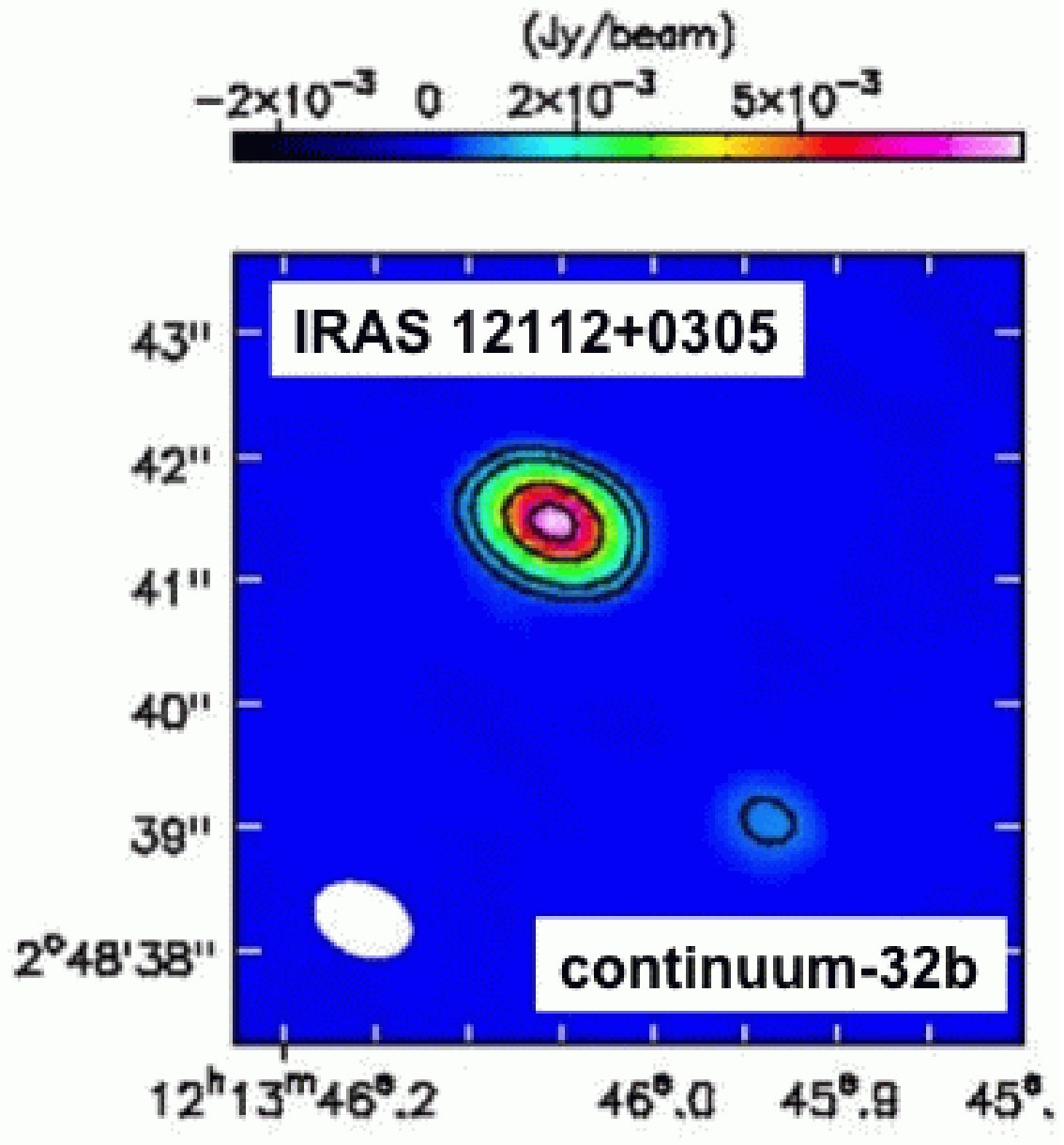} 
\includegraphics[angle=0,scale=.406]{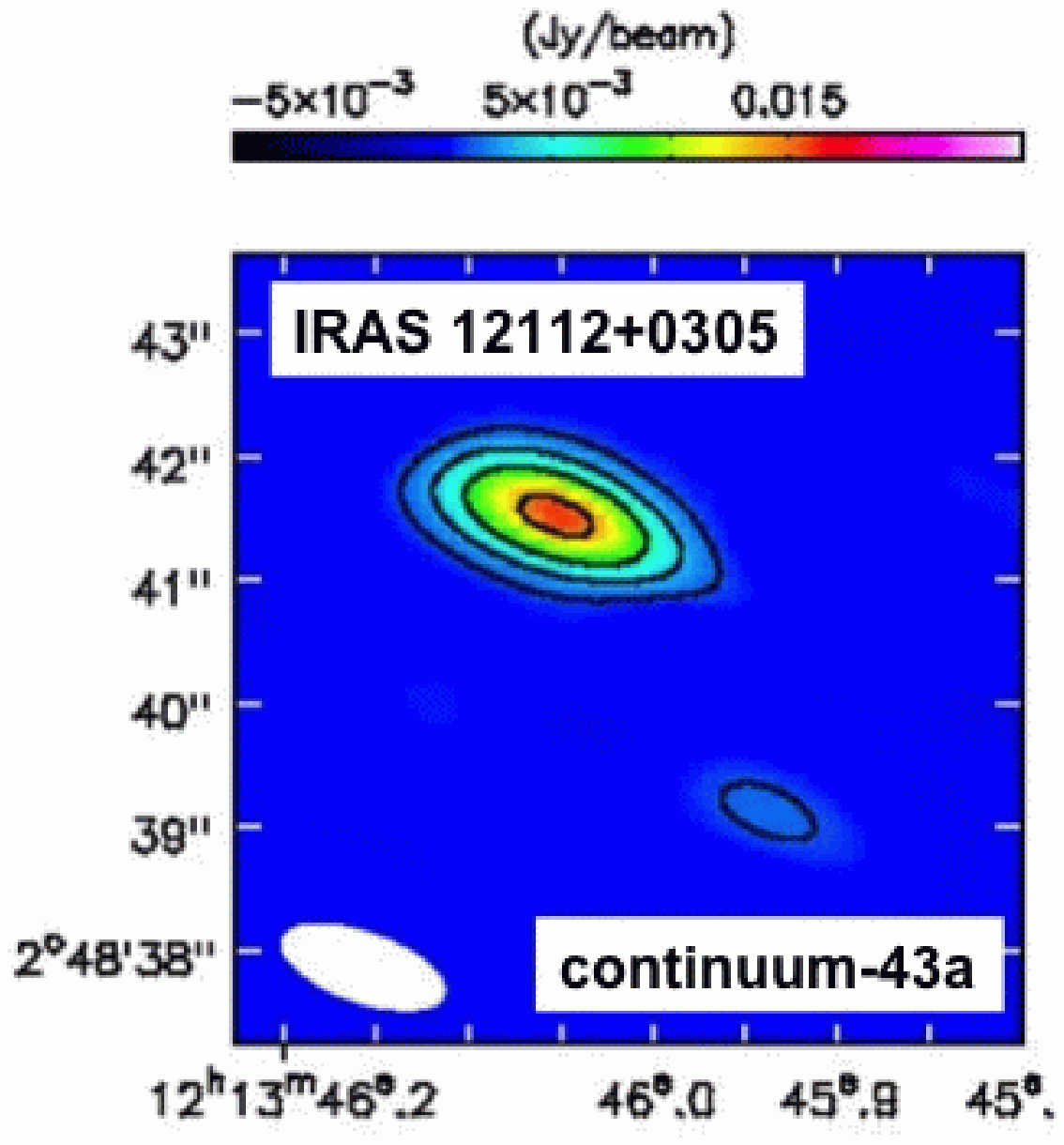}  
\includegraphics[angle=0,scale=.406]{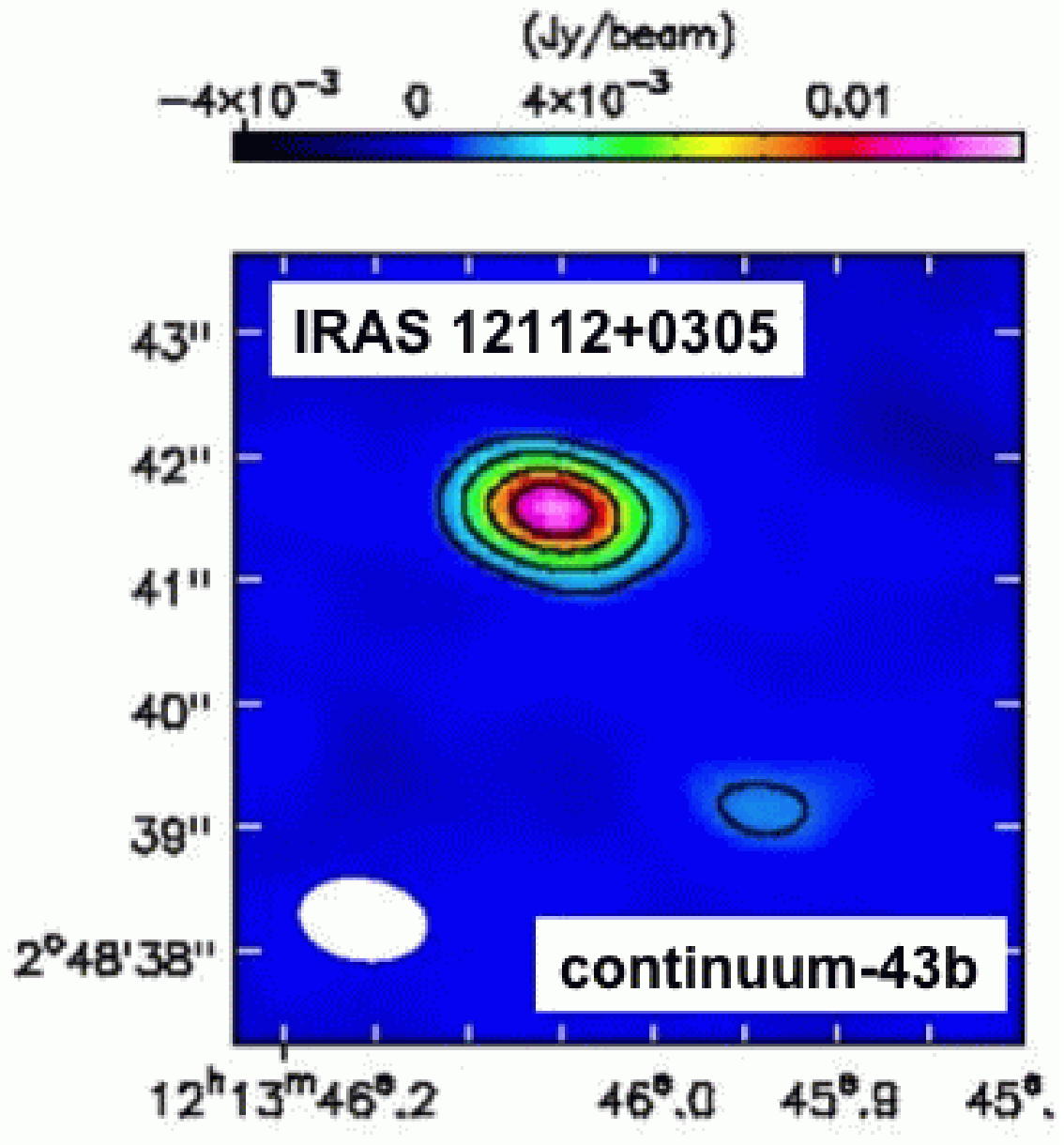} \\
\vspace{-1.5cm}
\includegraphics[angle=0,scale=.406]{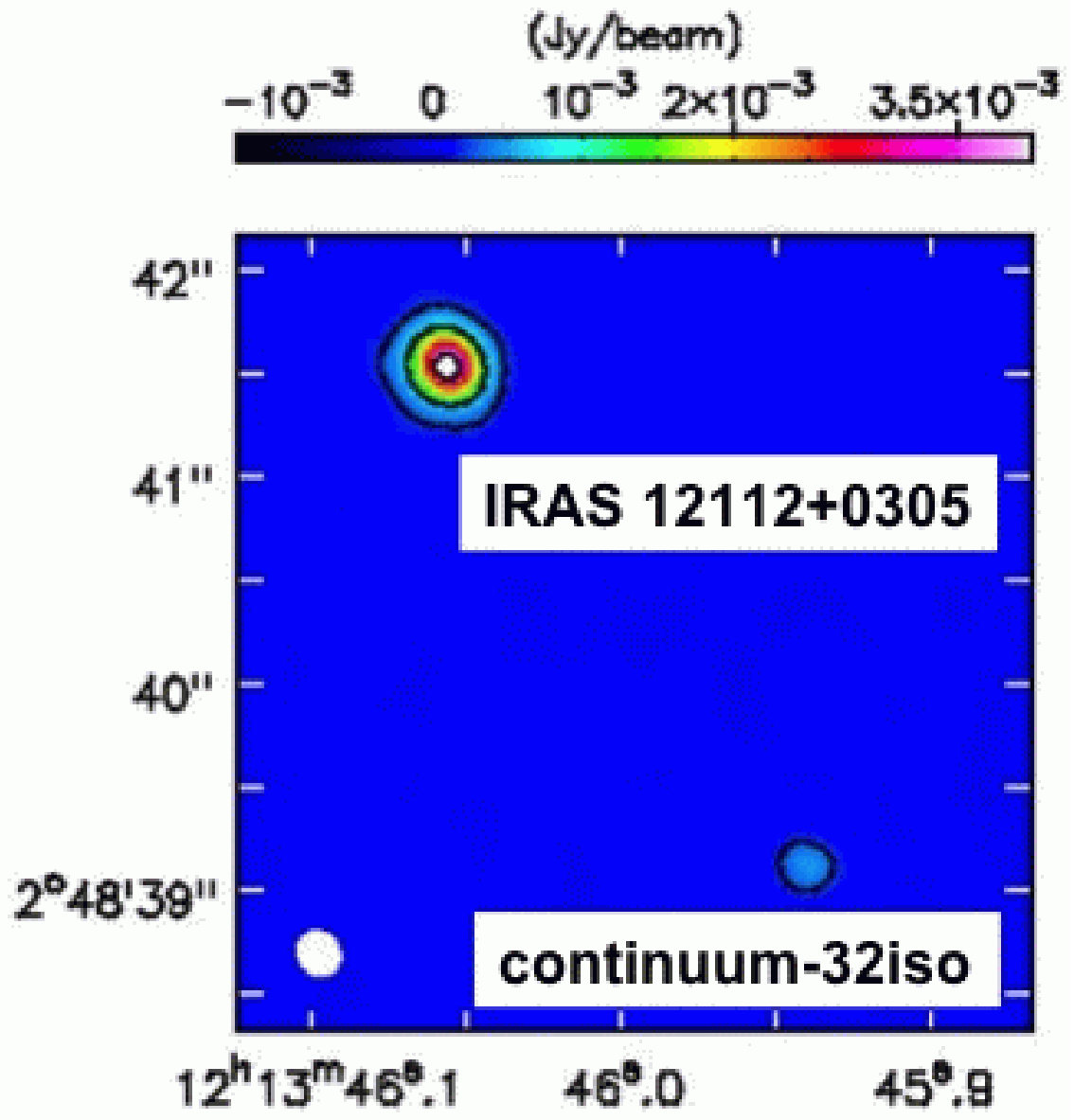} 
\includegraphics[angle=0,scale=.406]{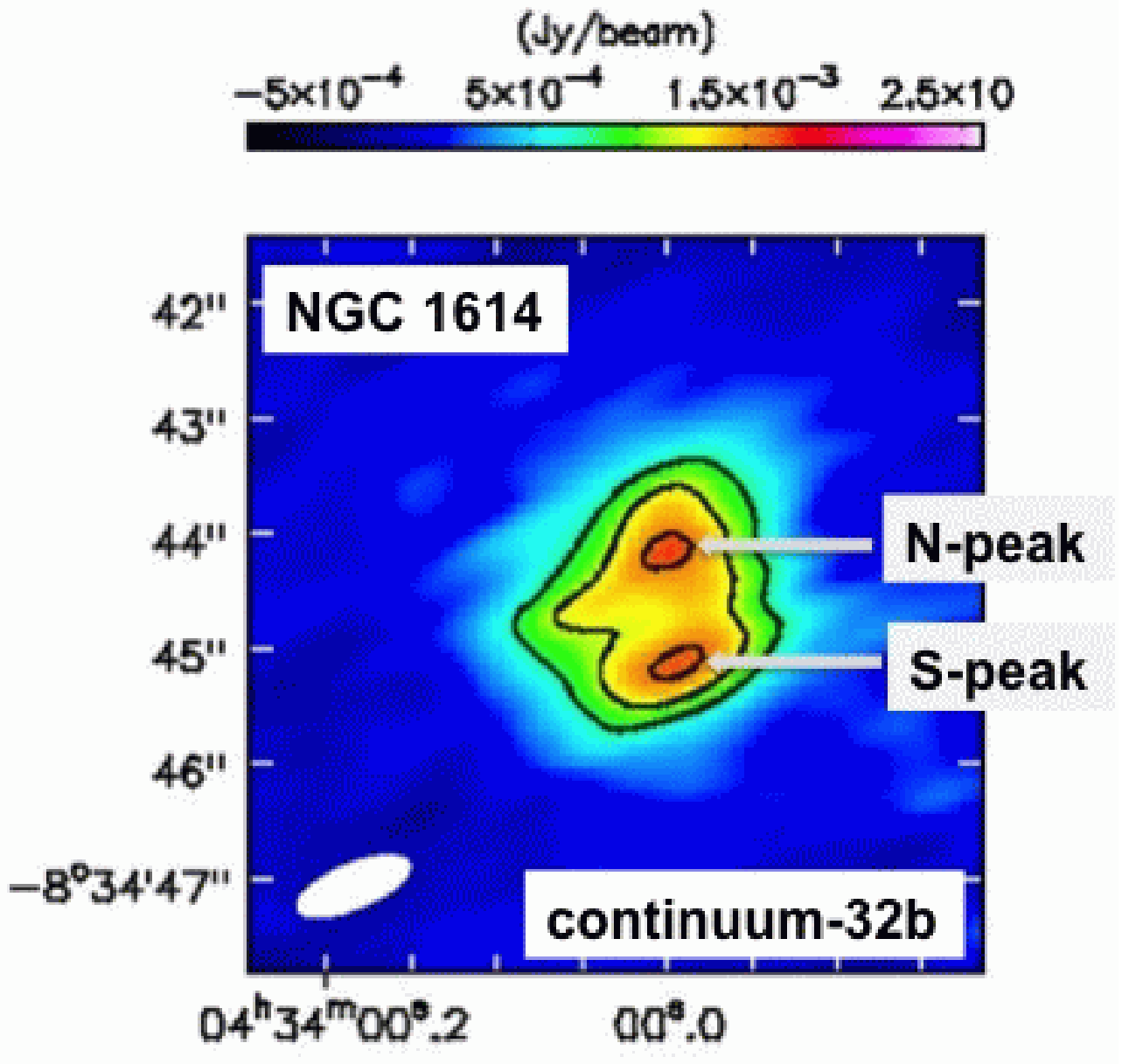} 
\includegraphics[angle=0,scale=.406]{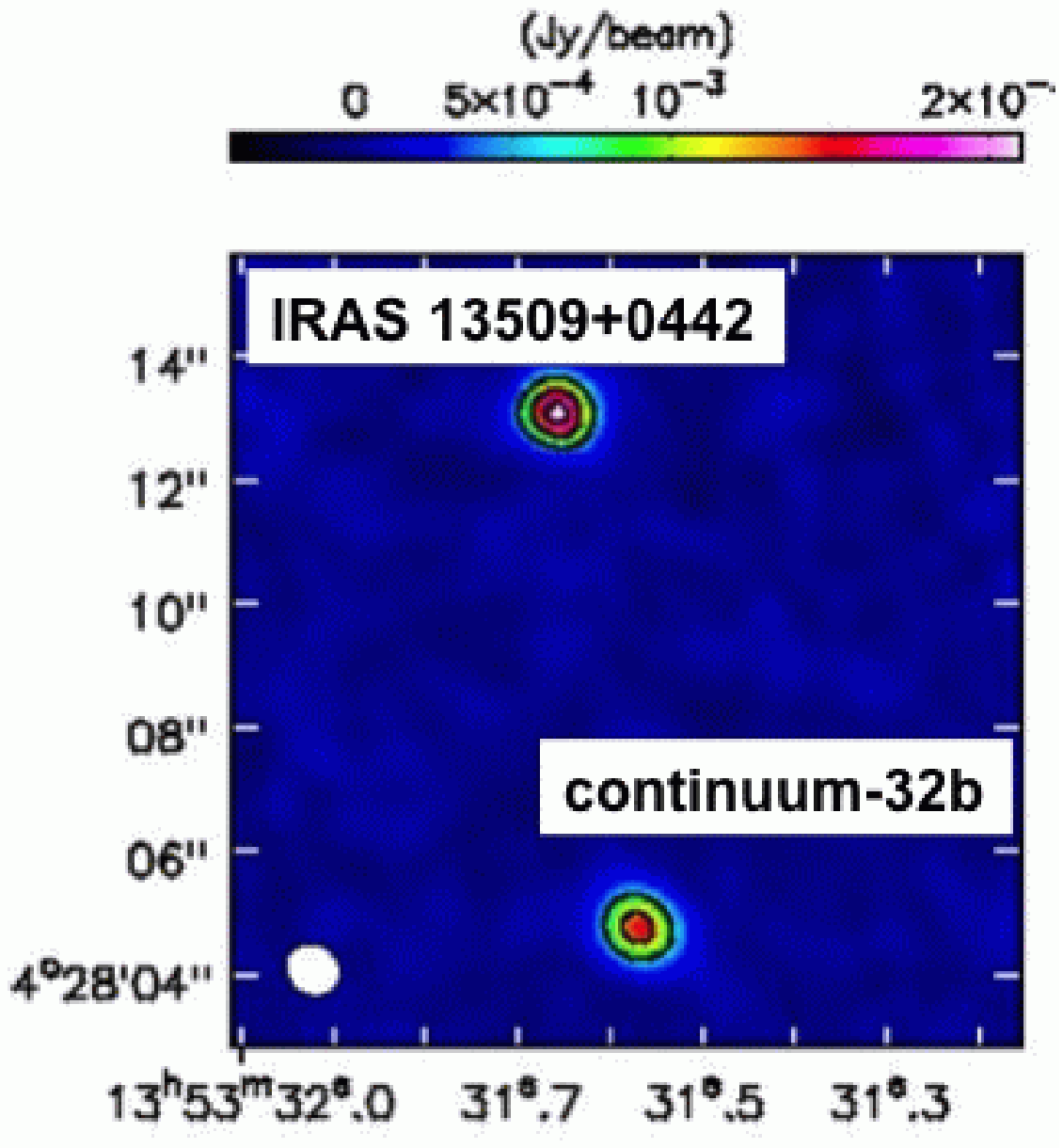} \\
\vspace{-1.5cm}
\includegraphics[angle=0,scale=.406]{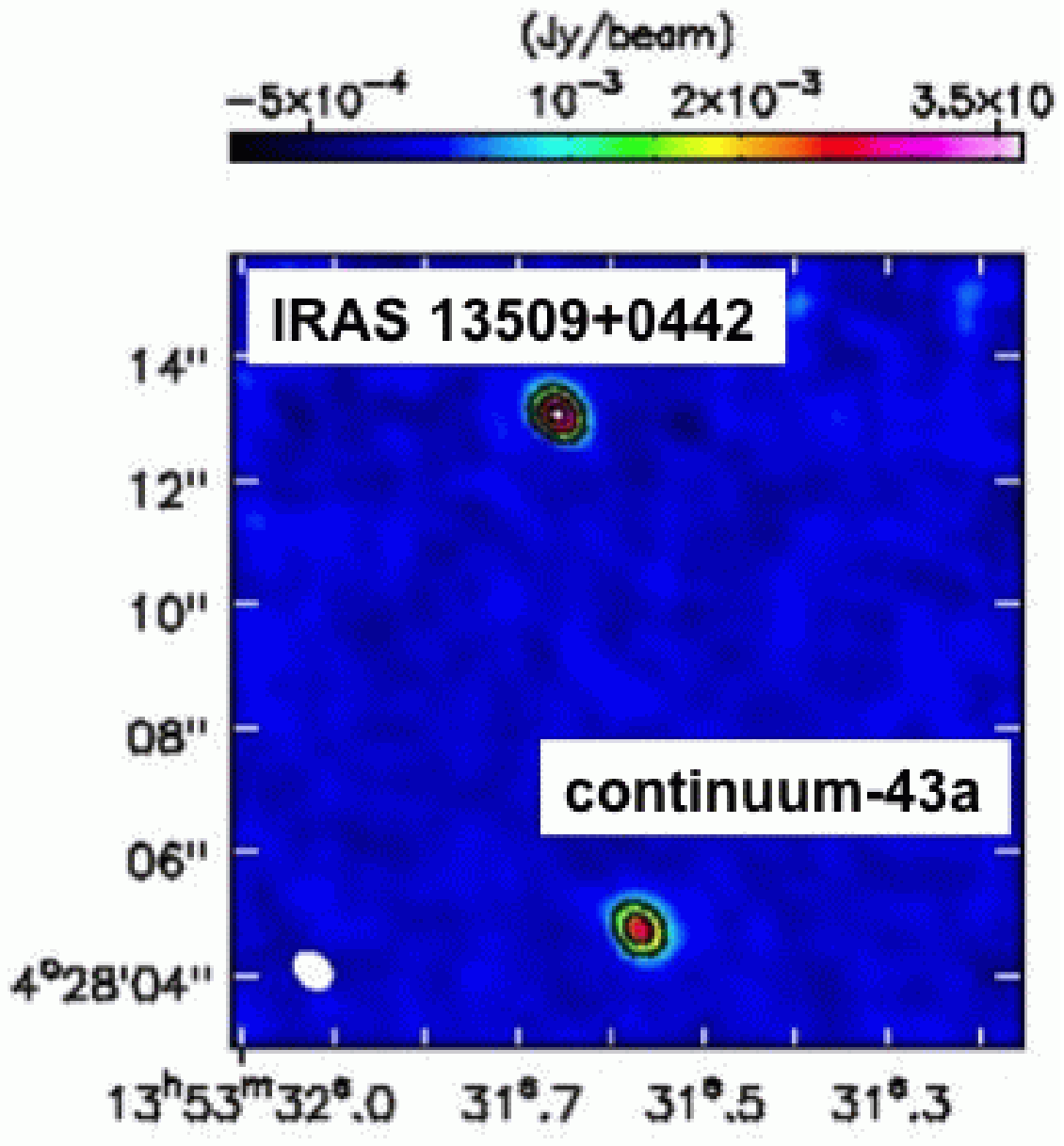} 
\includegraphics[angle=0,scale=.406]{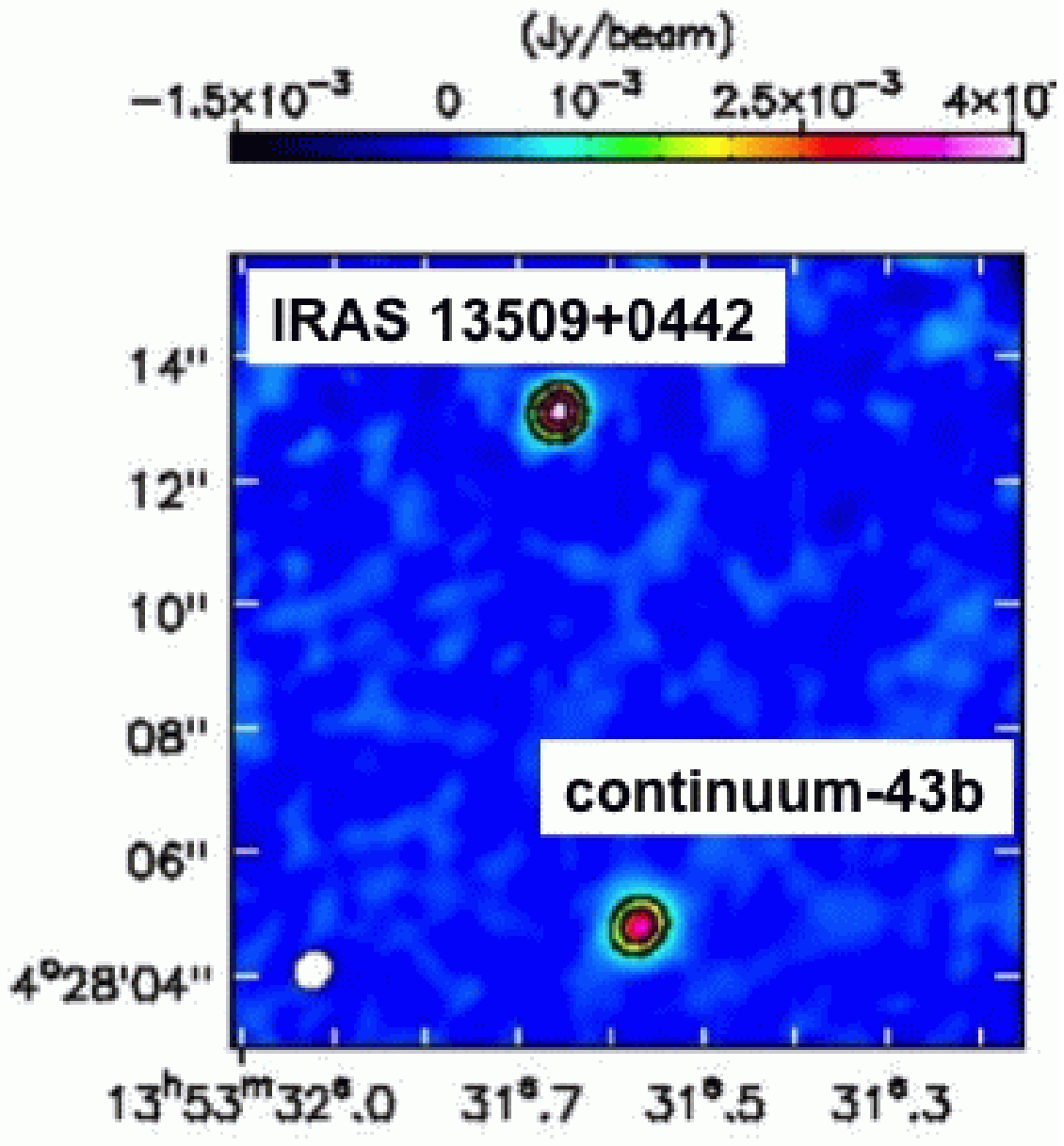} \\
\end{center}
\caption{
Continuum maps of selected (U)LIRGs with some morphological features. 
The abscissa and ordinate are R.A. (J2000) and decl. (J2000),
respectively.
We denote continuum data taken with ``HNC J=3--2'', 
``J=4--3 of HCN and HCO$^{+}$'', ``HNC J=4--3'', 
and ``J=3--2 of H$^{13}$CN, H$^{13}$CO$^{+}$, and HN$^{13}$C'', 
as 32b, 43a, 43b, and 32iso, respectively. 
The contours are 
5$\sigma$, 10$\sigma$, 30$\sigma$, 50$\sigma$ for continuum-32b of IRAS
12112$+$0305, 
4$\sigma$, 10$\sigma$, 20$\sigma$, 40$\sigma$ for continuum-43a of IRAS
12112$+$0305,
4$\sigma$, 10$\sigma$, 20$\sigma$, 30$\sigma$ for continuum-43b of IRAS
12112$+$0305,
5$\sigma$, 20$\sigma$, 80$\sigma$ for continuum-32iso of IRAS 12112$+$0305, 
5$\sigma$, 6$\sigma$, 8$\sigma$ for continuum-32b of NGC 1614, 
10$\sigma$, 20$\sigma$, 30$\sigma$ for continuum-32b, -43a, and -43b of
IRAS 13509$+$0442.
For IRAS 12112$+$0305, both the north-eastern (NE) and south-western (SW)
nuclei are detected.
For IRAS 13509$+$0442, the southern source is the ULIRG IRAS
13509$+$0442, and the north-eastern (NE) one is a
serendipitously detected bright continuum emitting source. 
The projected physical separation of the two nuclei of IRAS
12112$+$0305 is $\sim$4 kpc and that of the two continuum peaks of NGC
1614 is $\sim$0.3 kpc.
Beam sizes are shown as filled circles in the lower-left region.
}
\end{figure}


\begin{figure}
\begin{center}
\includegraphics[angle=0,scale=.3]{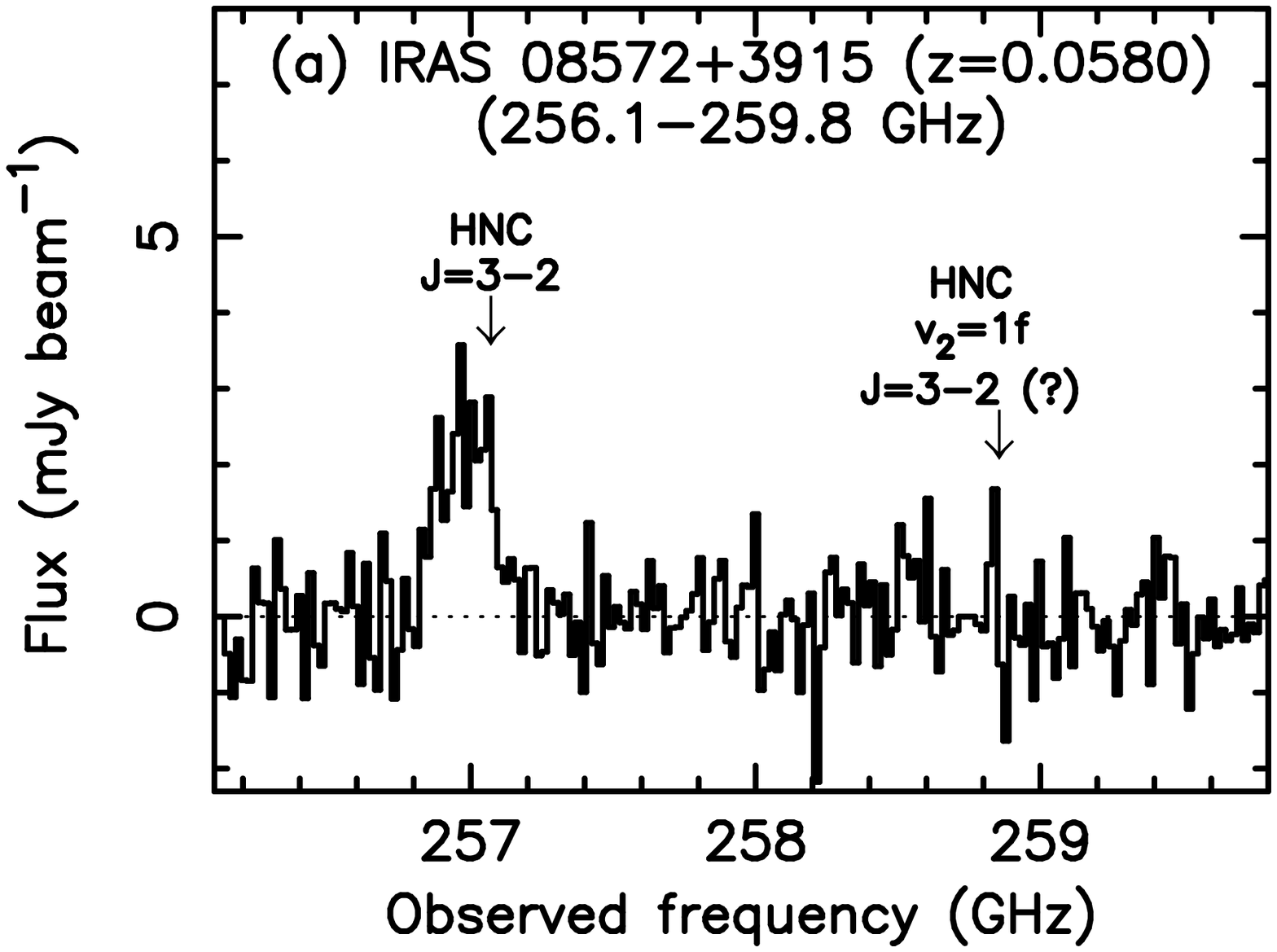} 
\includegraphics[angle=0,scale=.3]{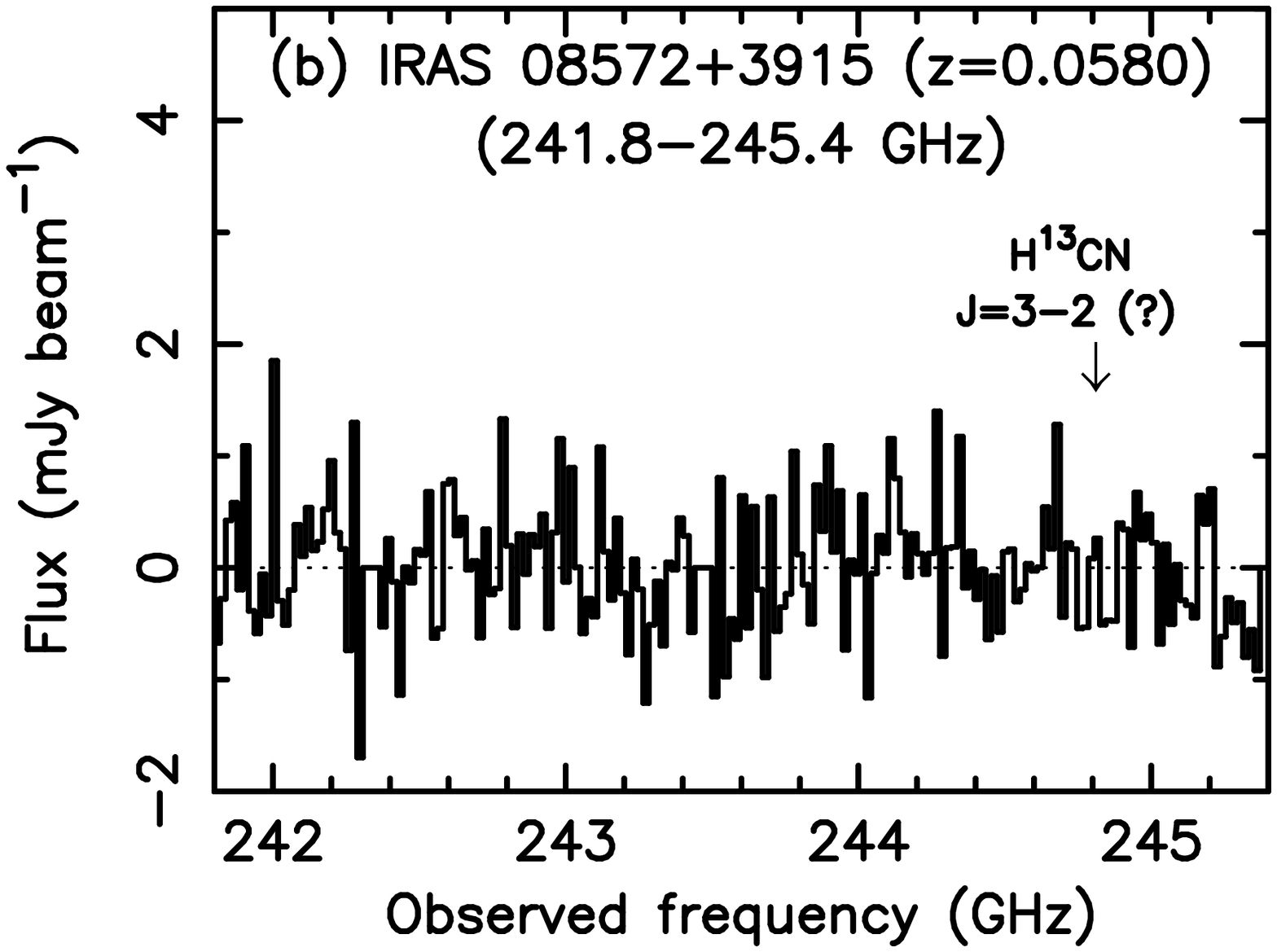} 
\includegraphics[angle=0,scale=.3]{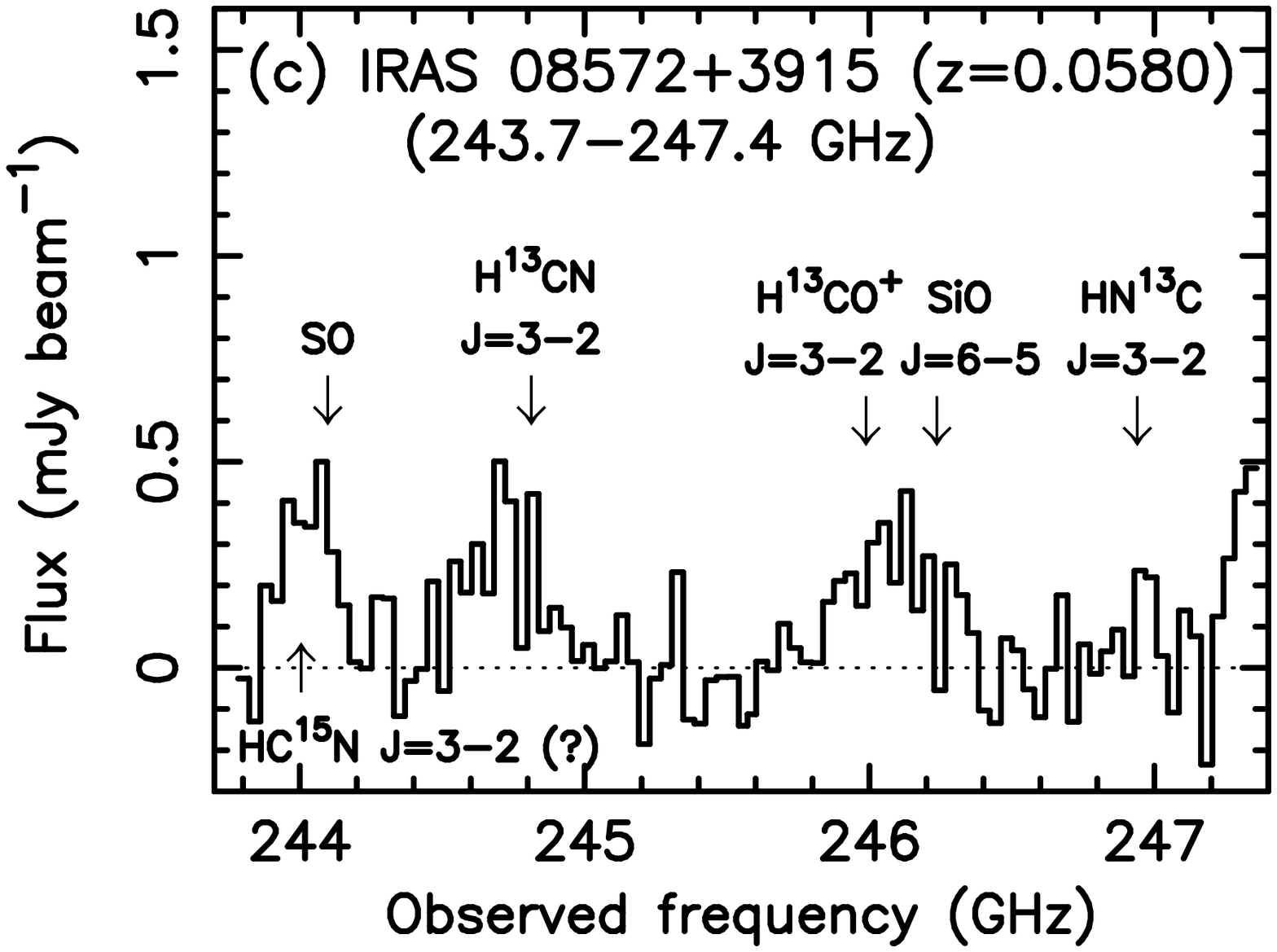} \\
\includegraphics[angle=0,scale=.3]{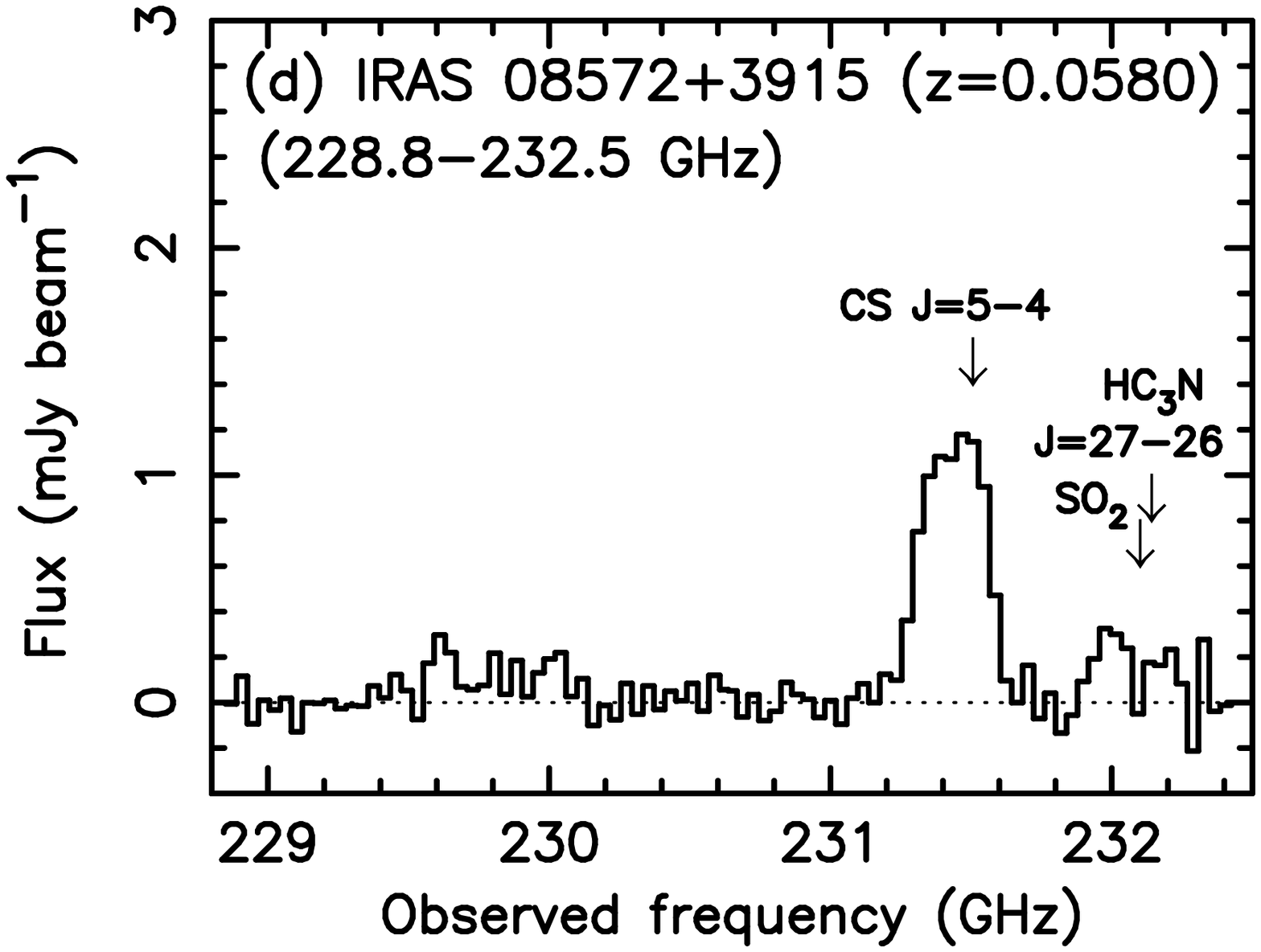} 
\end{center}
\caption{ALMA spectra of IRAS 08572$+$3915.
In (c), a downward arrow and upward arrow are shown for 
SO 6(6)--5(5) ($\nu_{\rm rest}$=258.256 GHz) and 
HC$^{15}$N J=3--2 ($\nu_{\rm rest}$=258.157 GHz), respectively.
In (d), a downward arrow is shown for 
SO$_{2}$ 10(3,7)--10(2,8) ($\nu_{\rm rest}$=245.563 GHz).
}
\end{figure}


\begin{figure}
\begin{center}
\vspace{-0.5cm}
\includegraphics[angle=0,scale=.3]{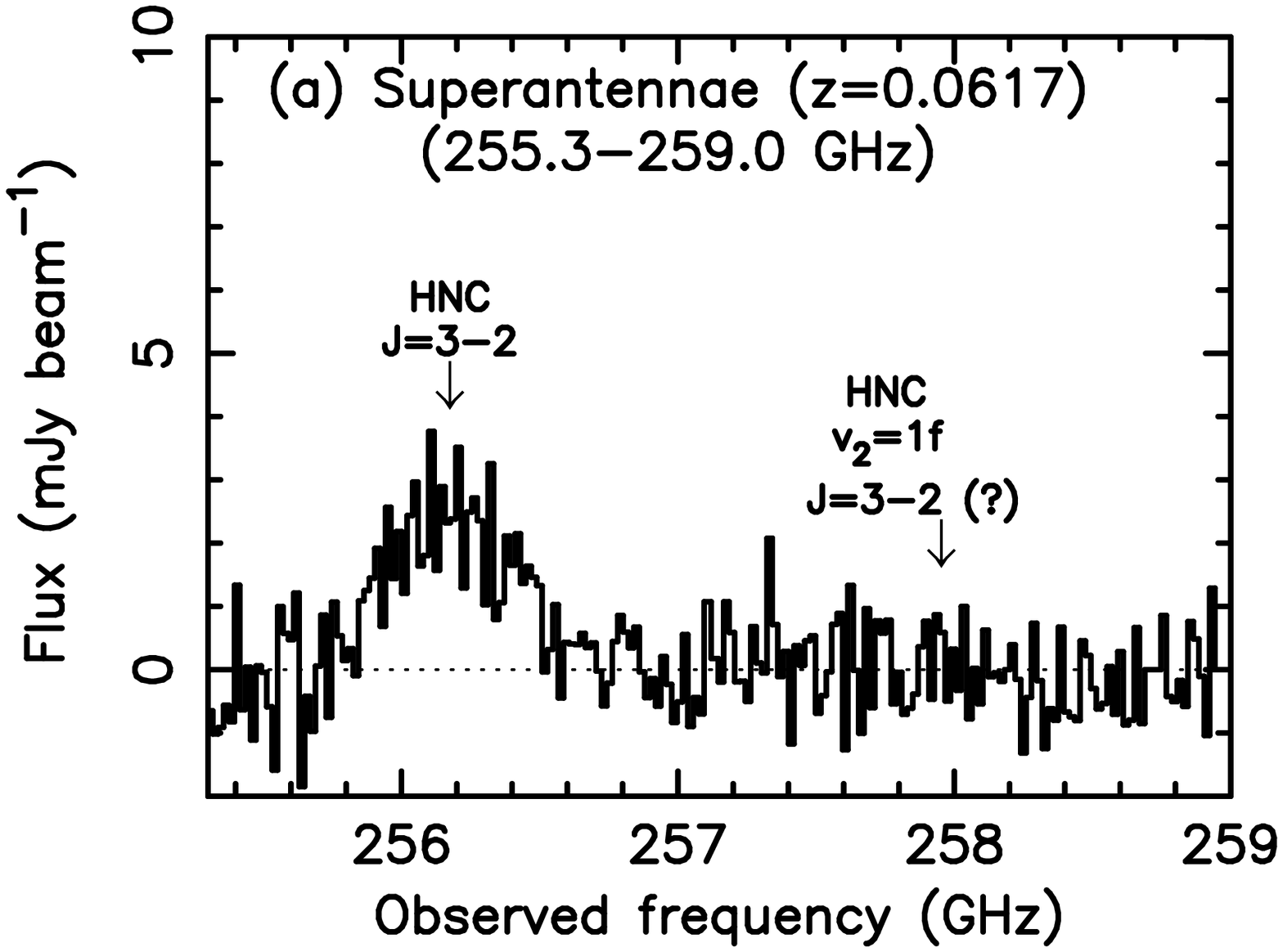} 
\includegraphics[angle=0,scale=.3]{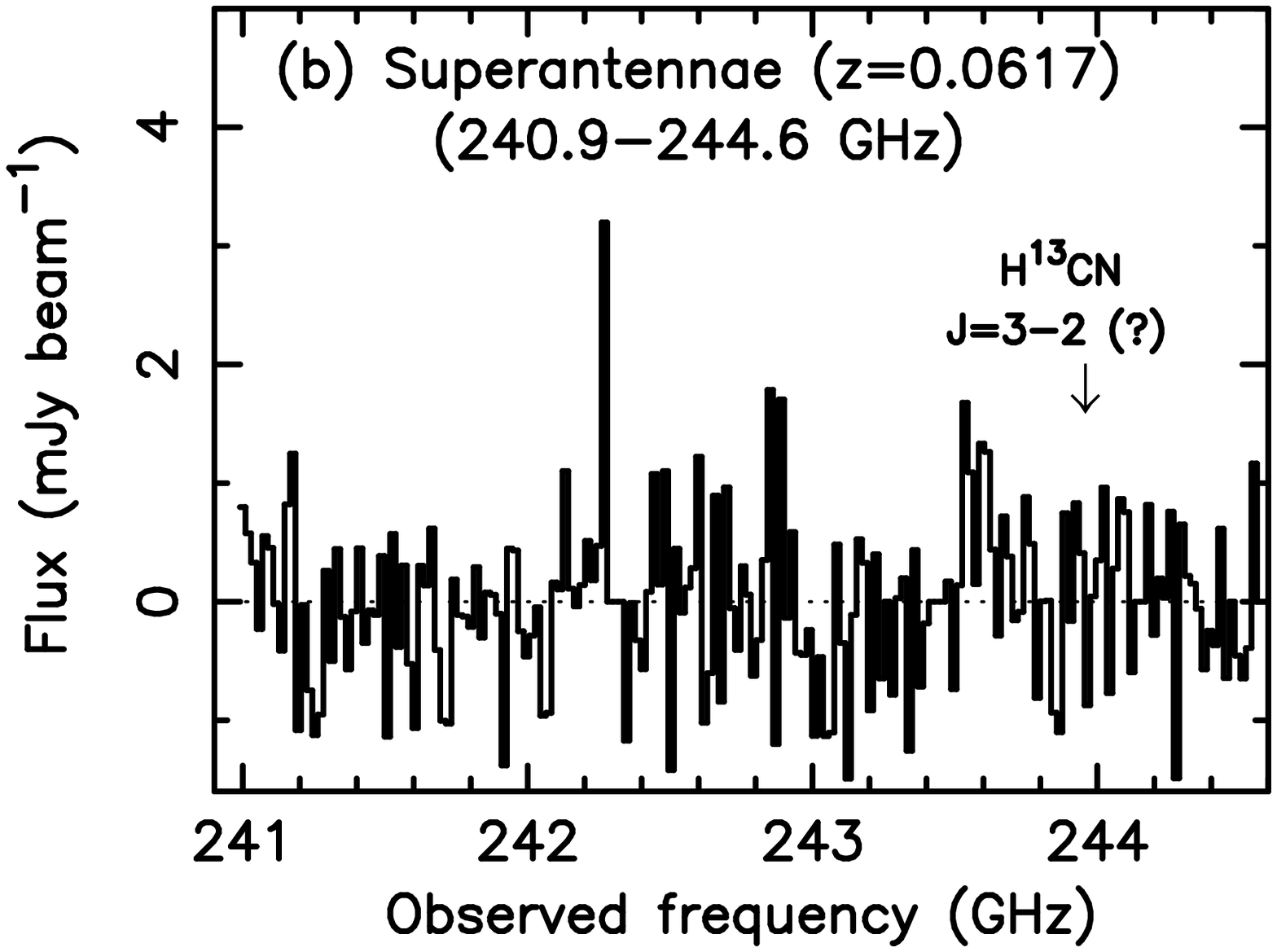} 
\includegraphics[angle=0,scale=.3]{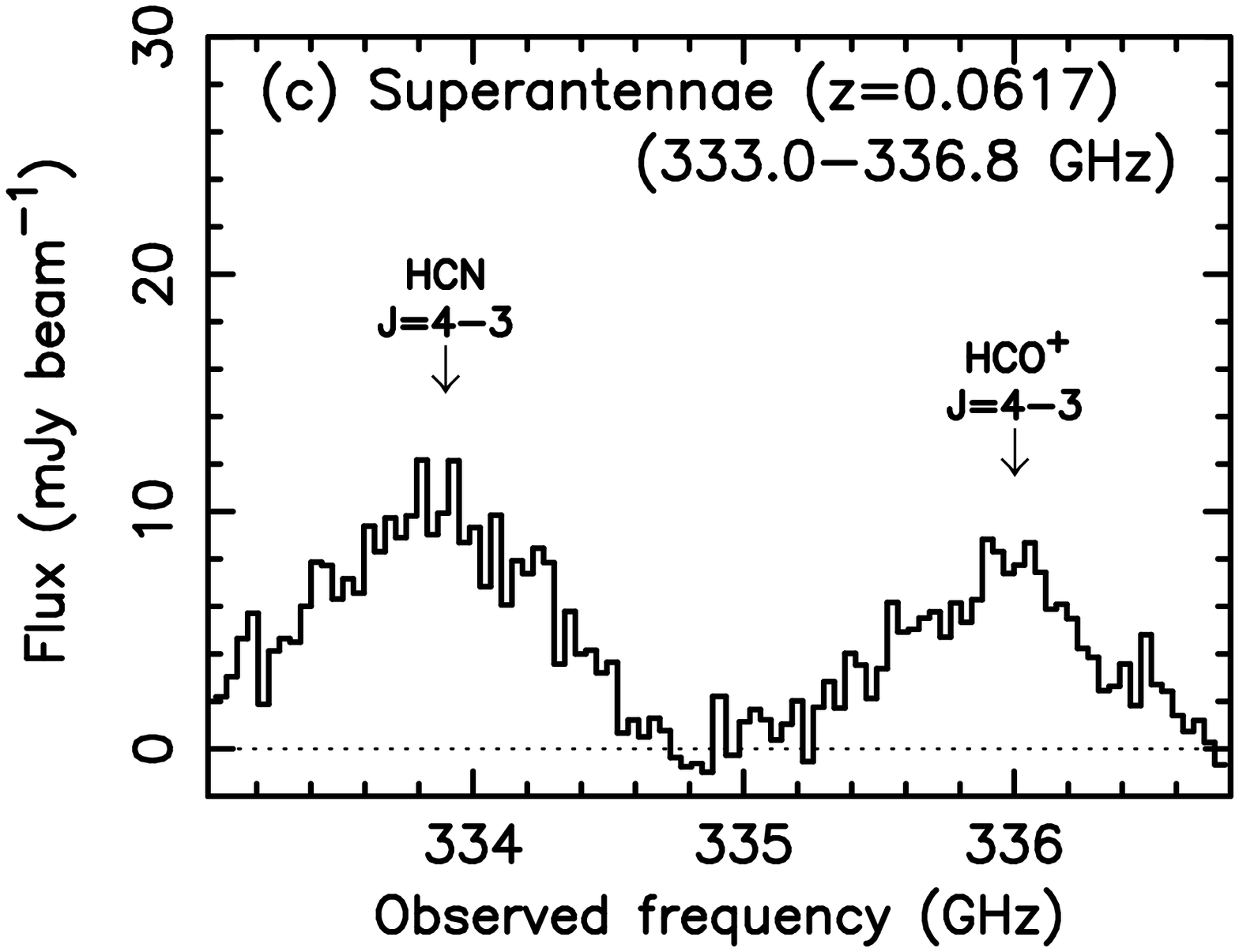} \\
\vspace{-0.5cm}
\includegraphics[angle=0,scale=.3]{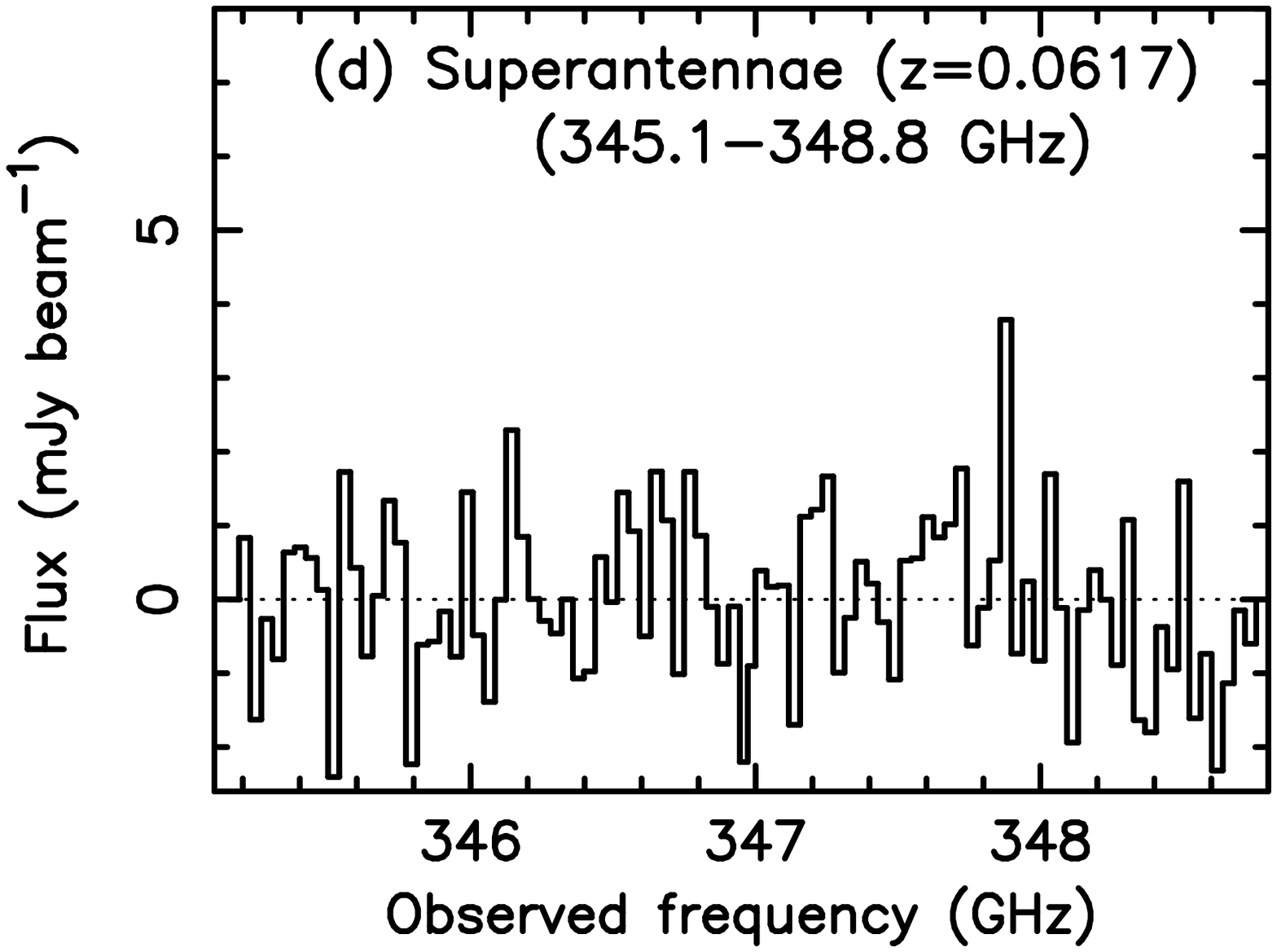} 
\includegraphics[angle=0,scale=.3]{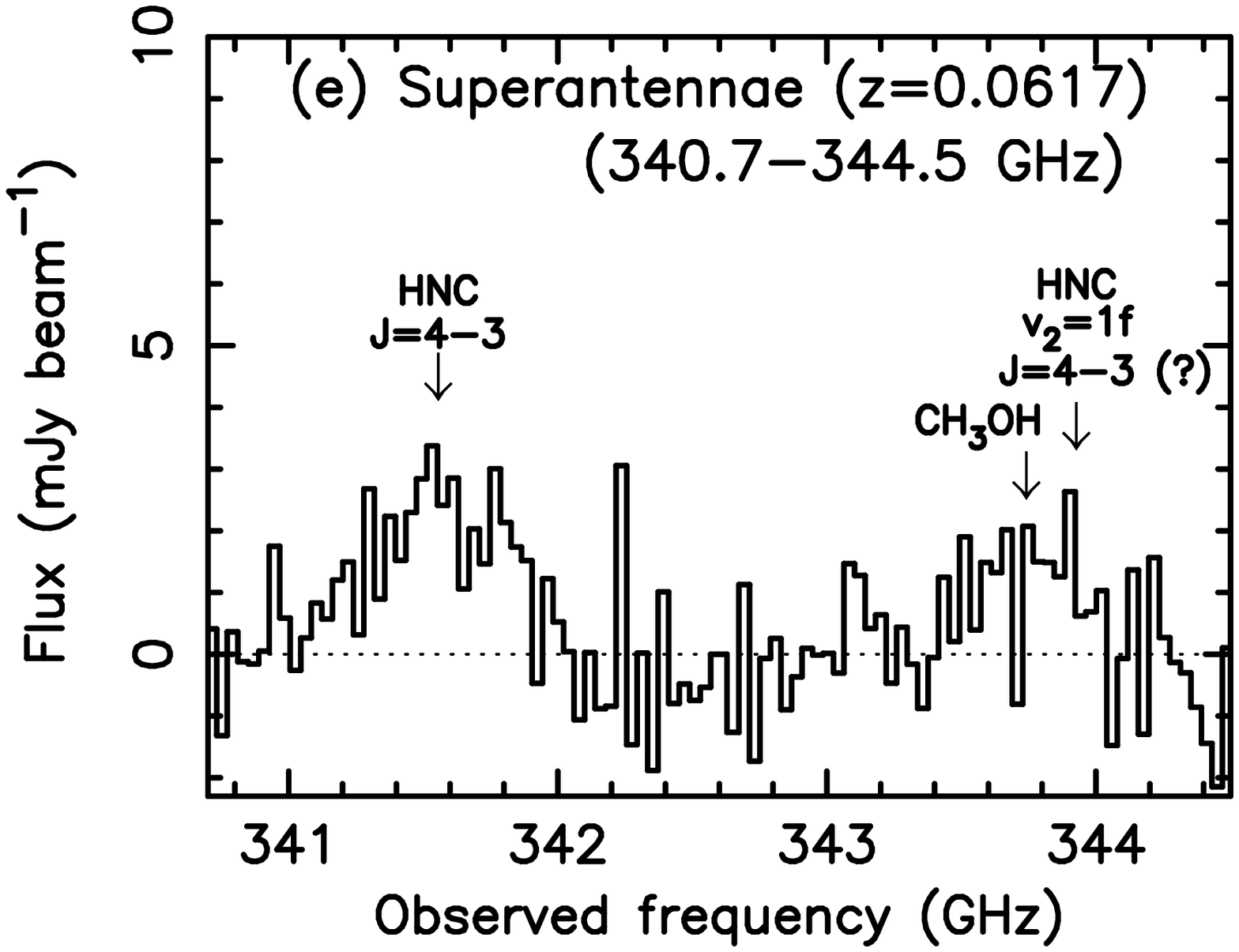} 
\includegraphics[angle=0,scale=.3]{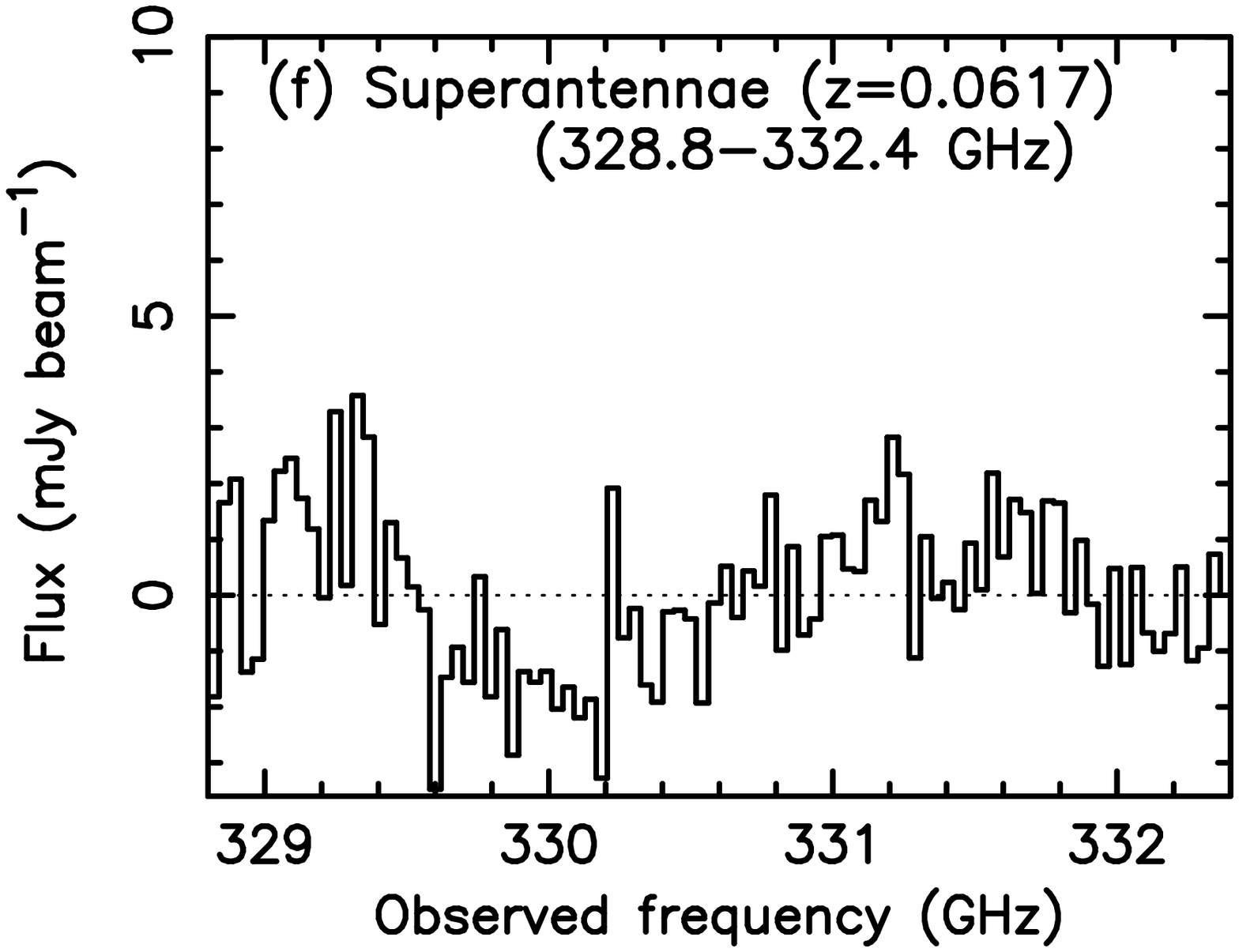} \\
\end{center}
\caption{ALMA spectra of Superantennae.
In (e), a downward arrow is shown for 
CH$_{3}$OH 16(2,14)--16($-$1,16) ($\nu_{\rm rest}$=364.859 GHz).}
\end{figure}


\begin{figure}
\begin{center}
\includegraphics[angle=0,scale=.3]{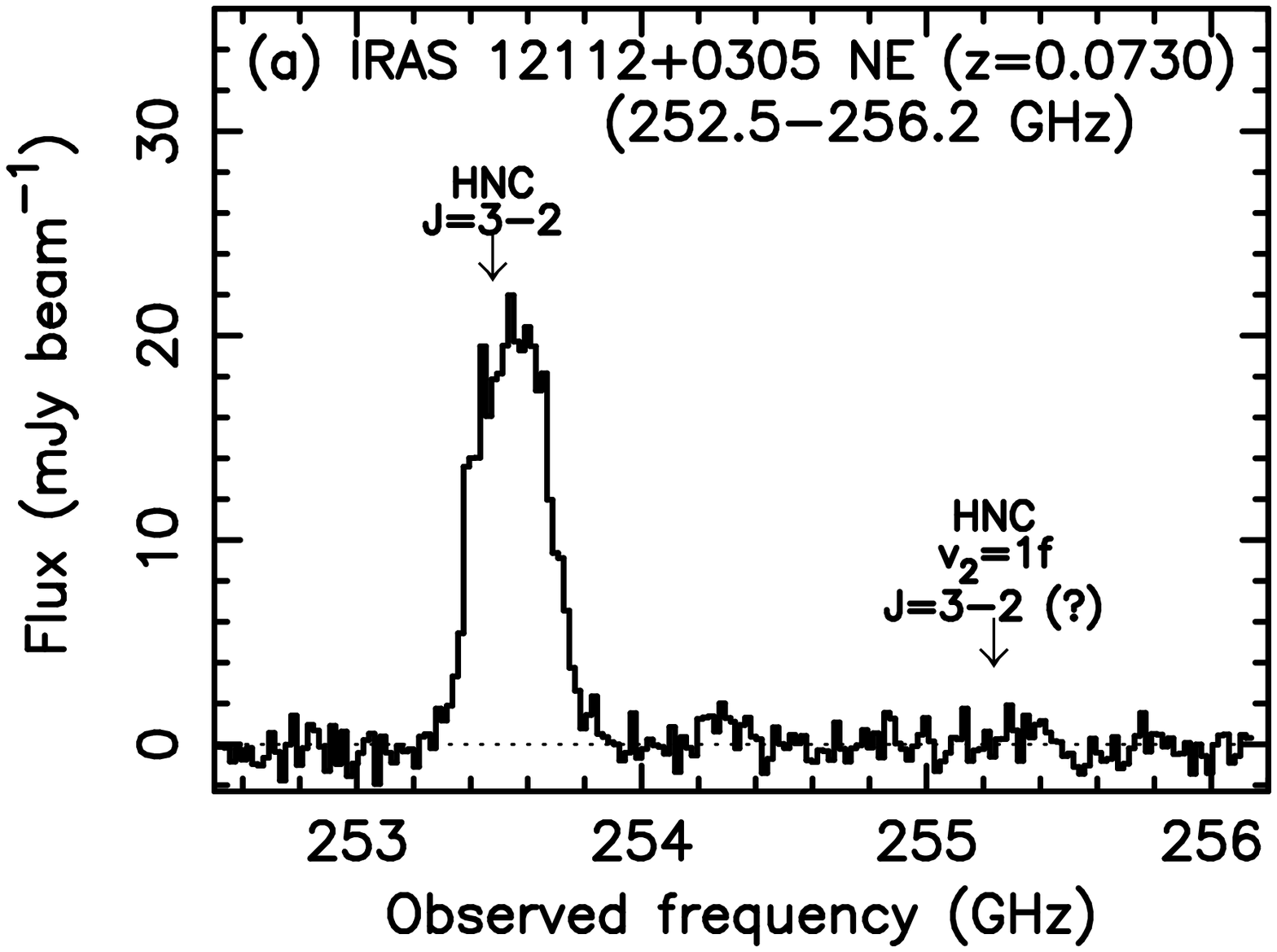} 
\includegraphics[angle=0,scale=.3]{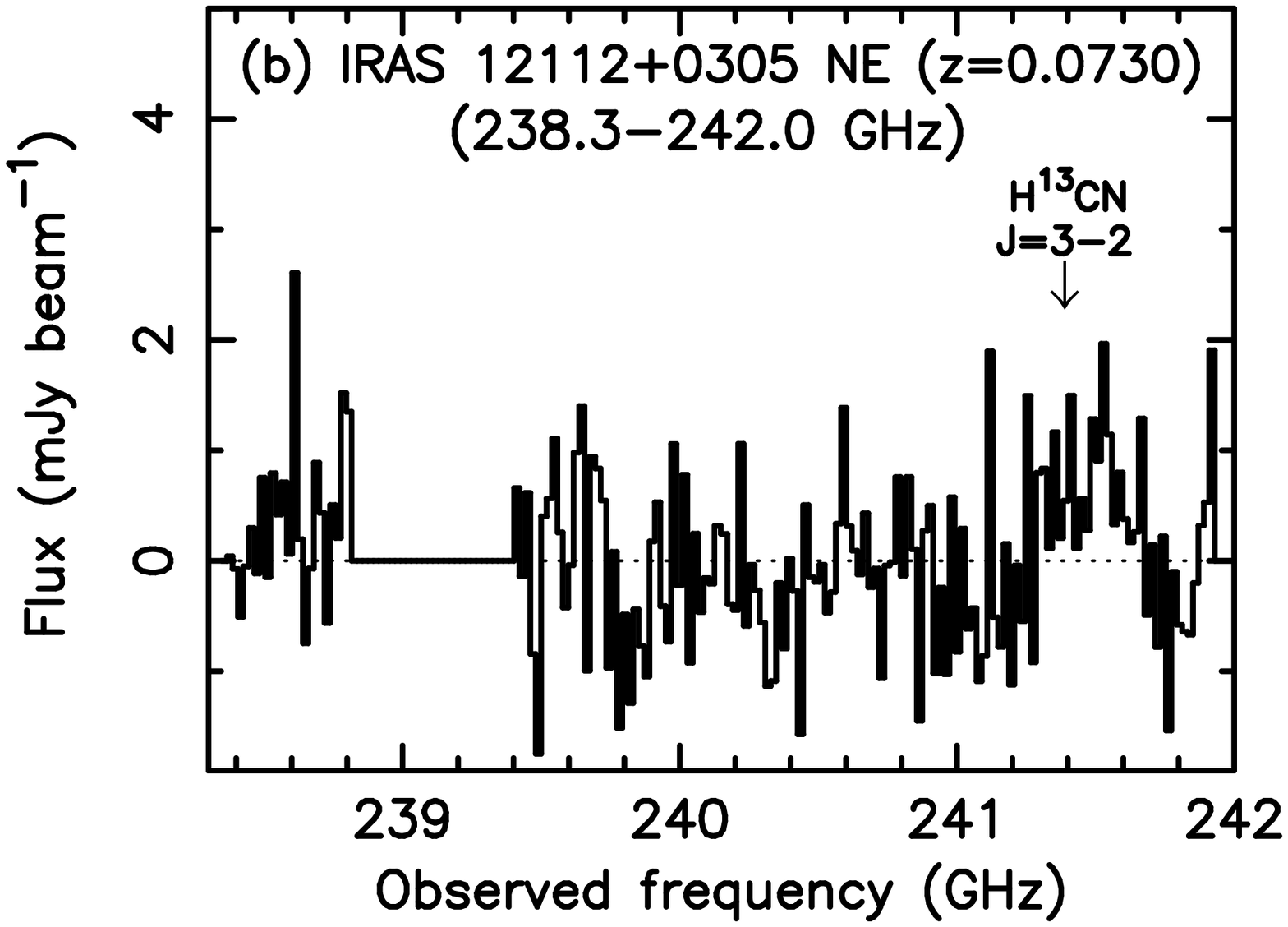} 
\includegraphics[angle=0,scale=.3]{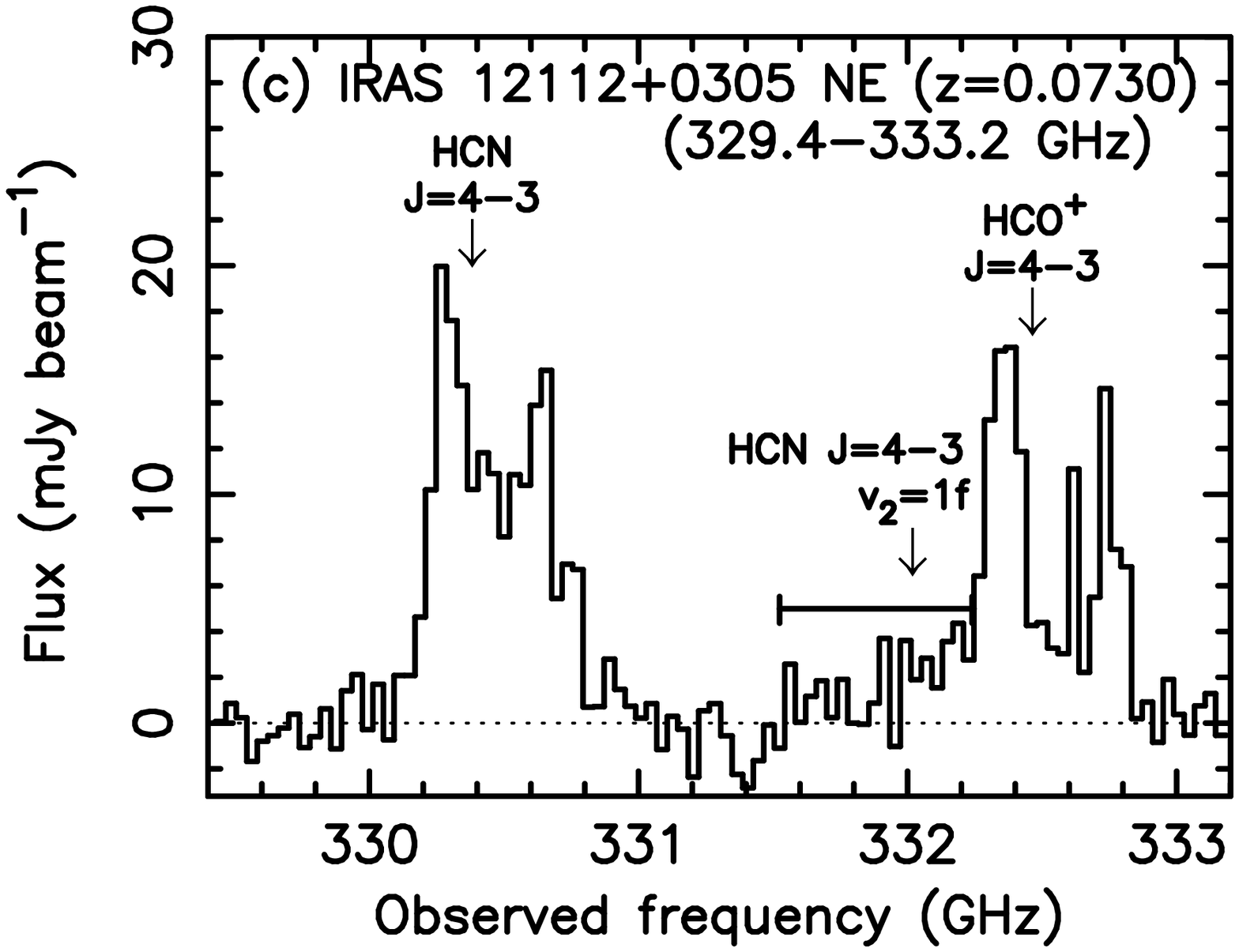} \\
\vspace{-0.5cm}
\includegraphics[angle=0,scale=.3]{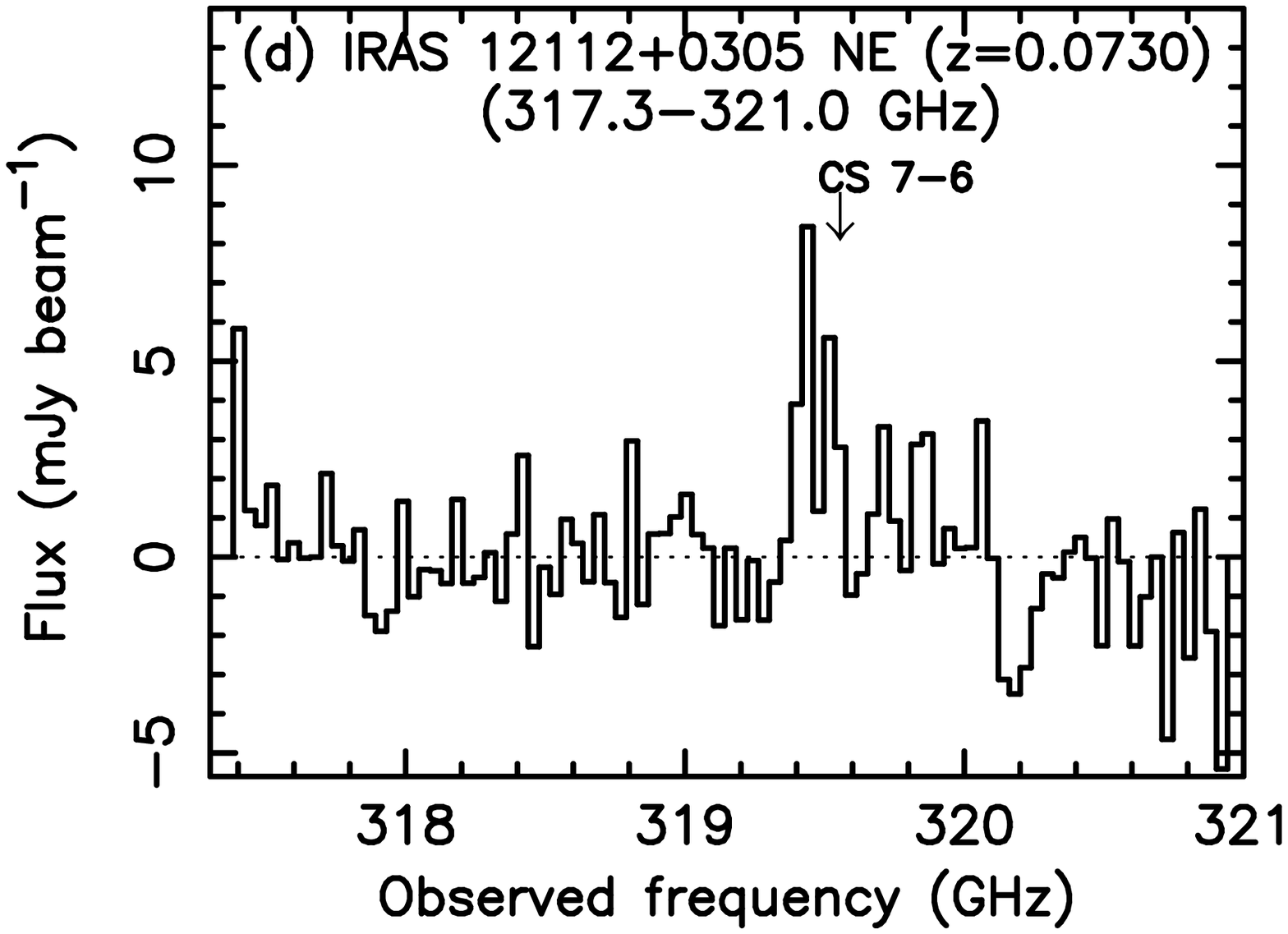} 
\includegraphics[angle=0,scale=.3]{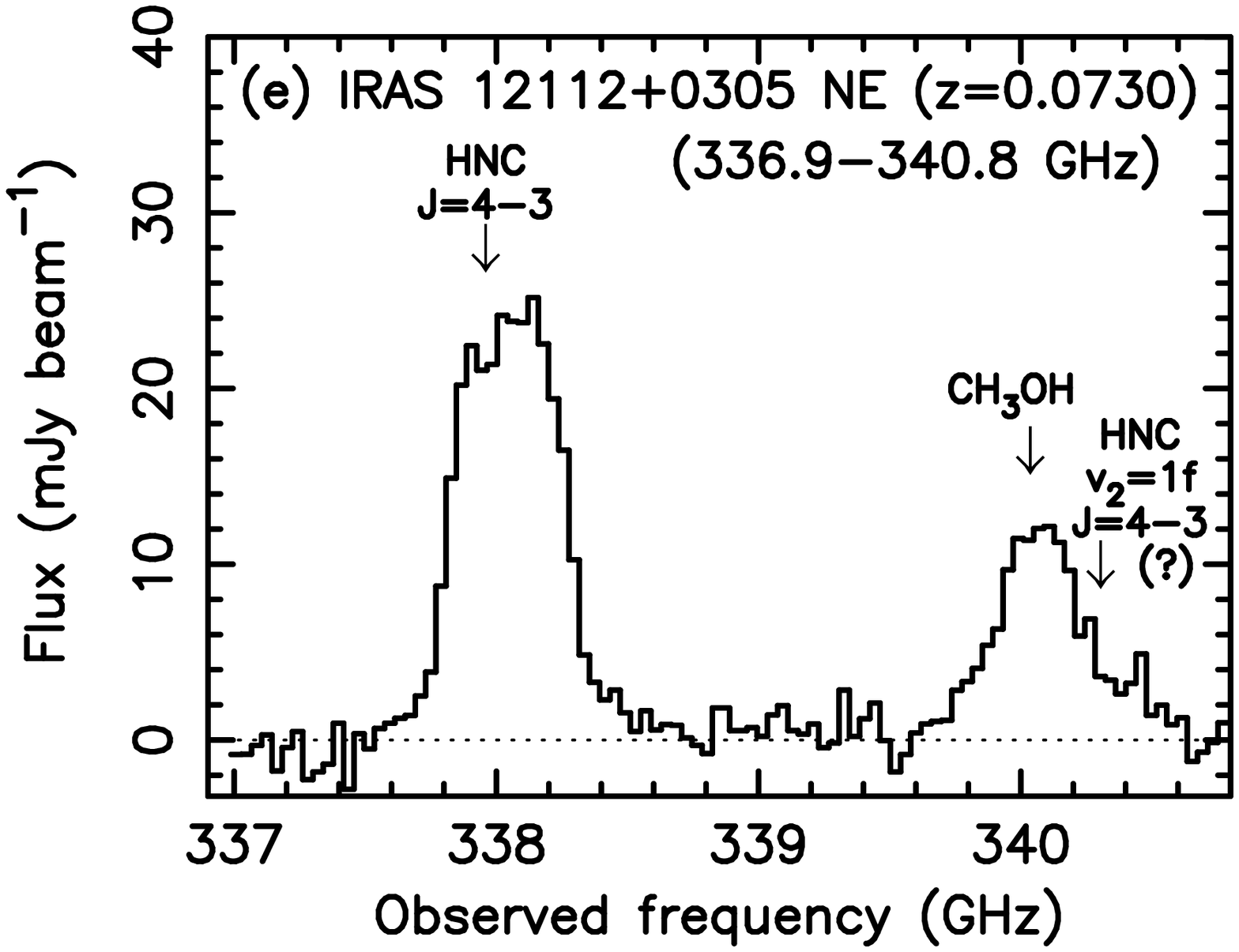} 
\includegraphics[angle=0,scale=.3]{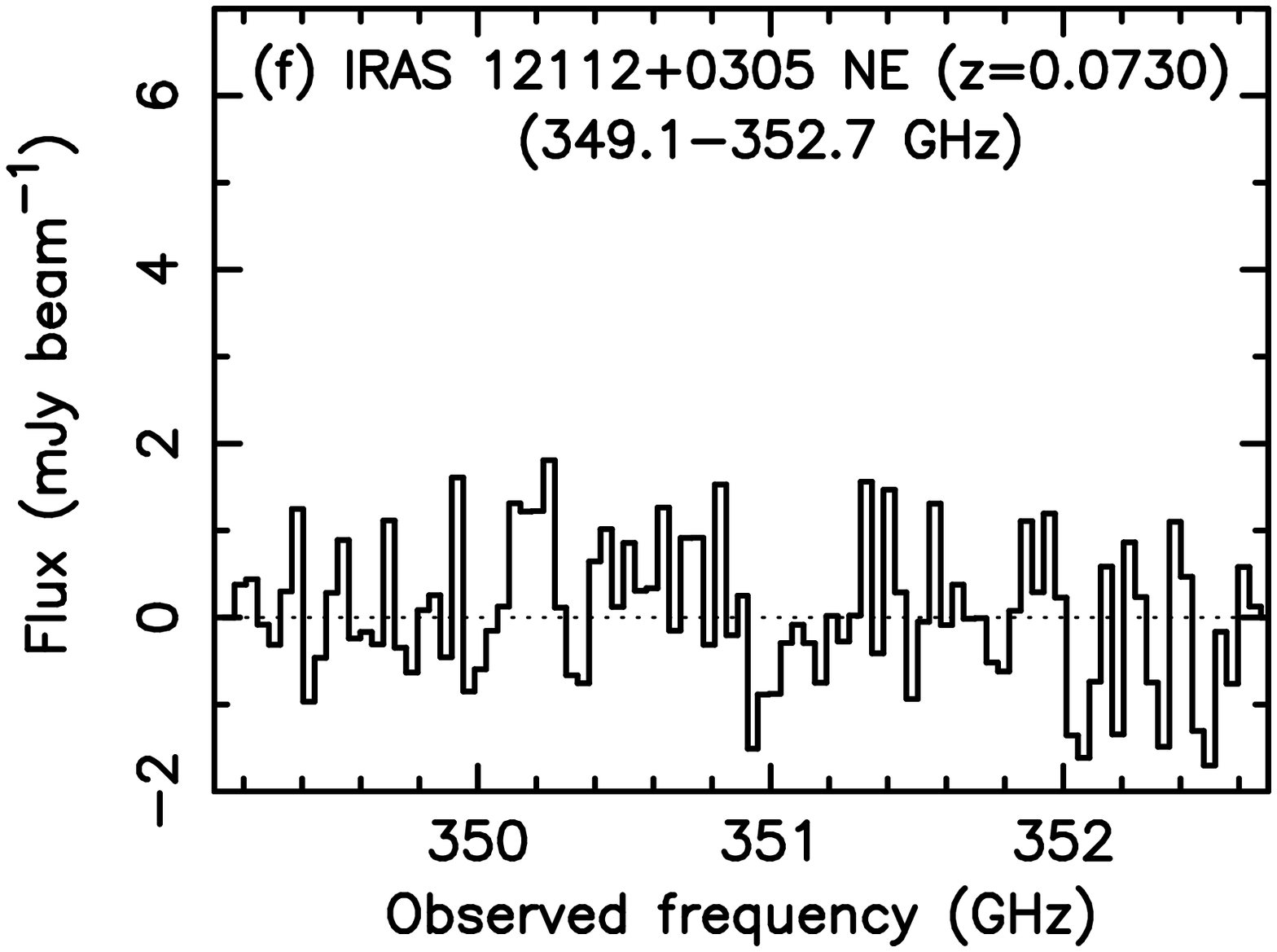} \\
\vspace{-0.5cm}
\includegraphics[angle=0,scale=.3]{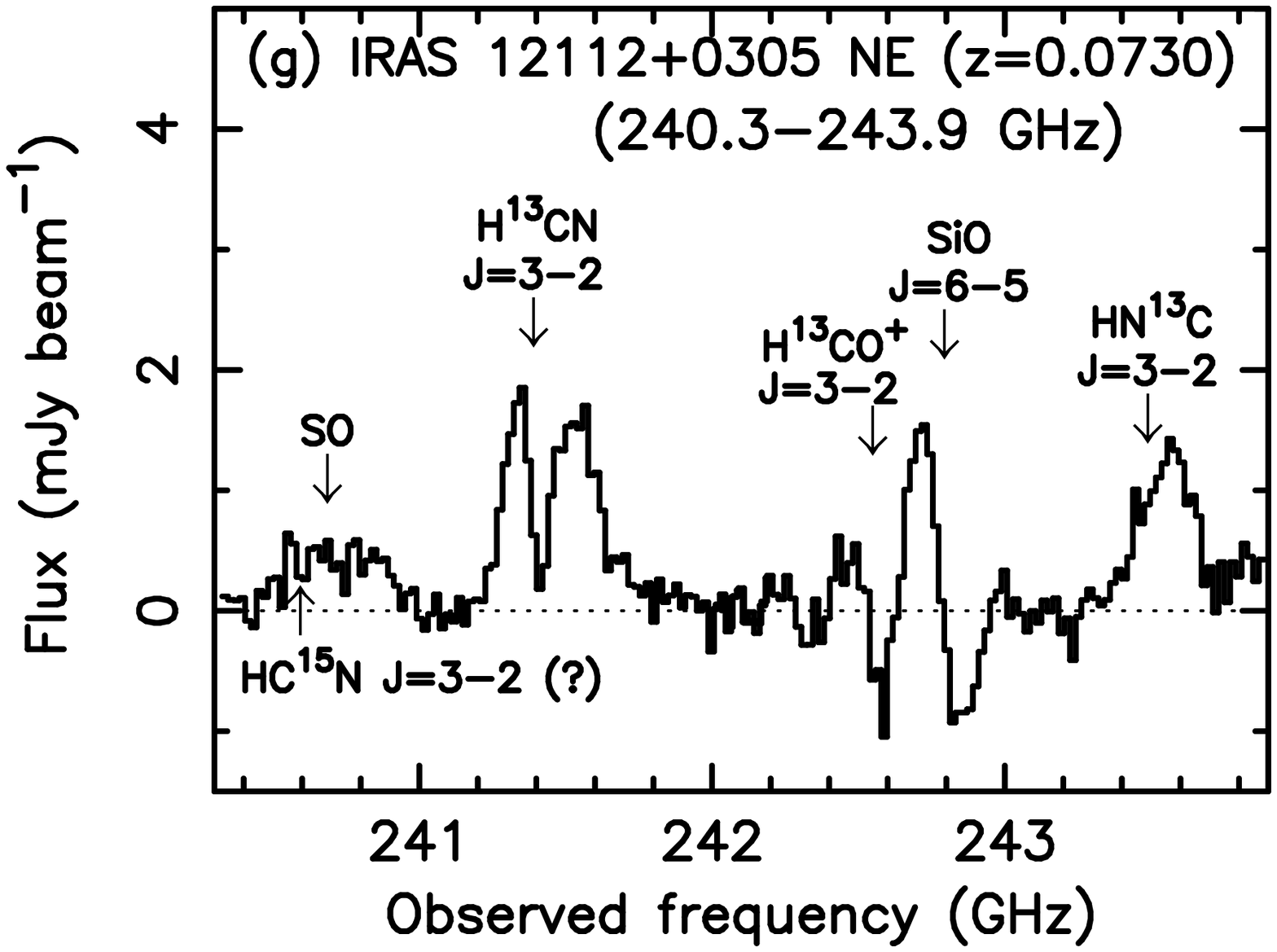} 
\includegraphics[angle=0,scale=.3]{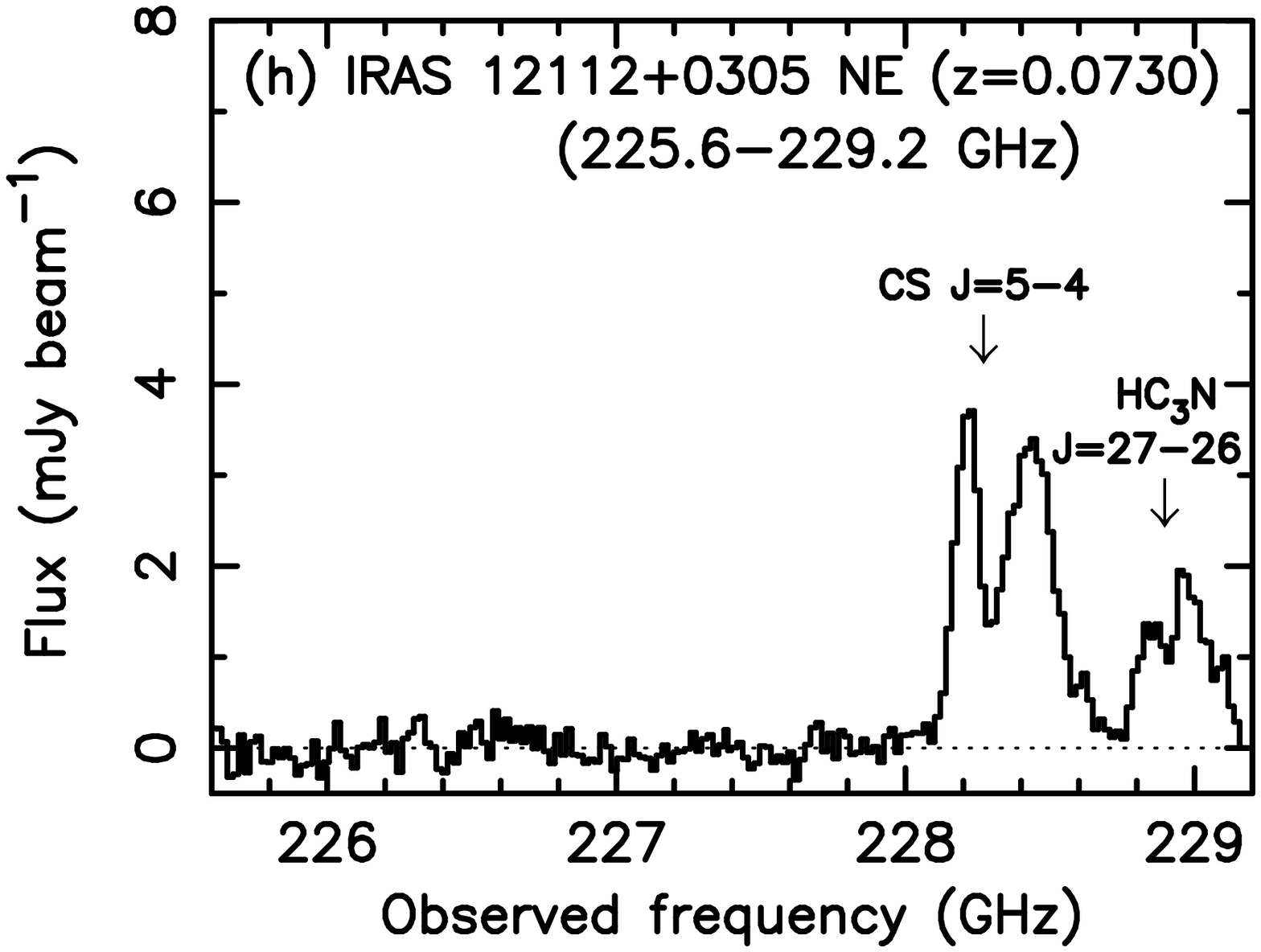} 
\includegraphics[angle=0,scale=.3]{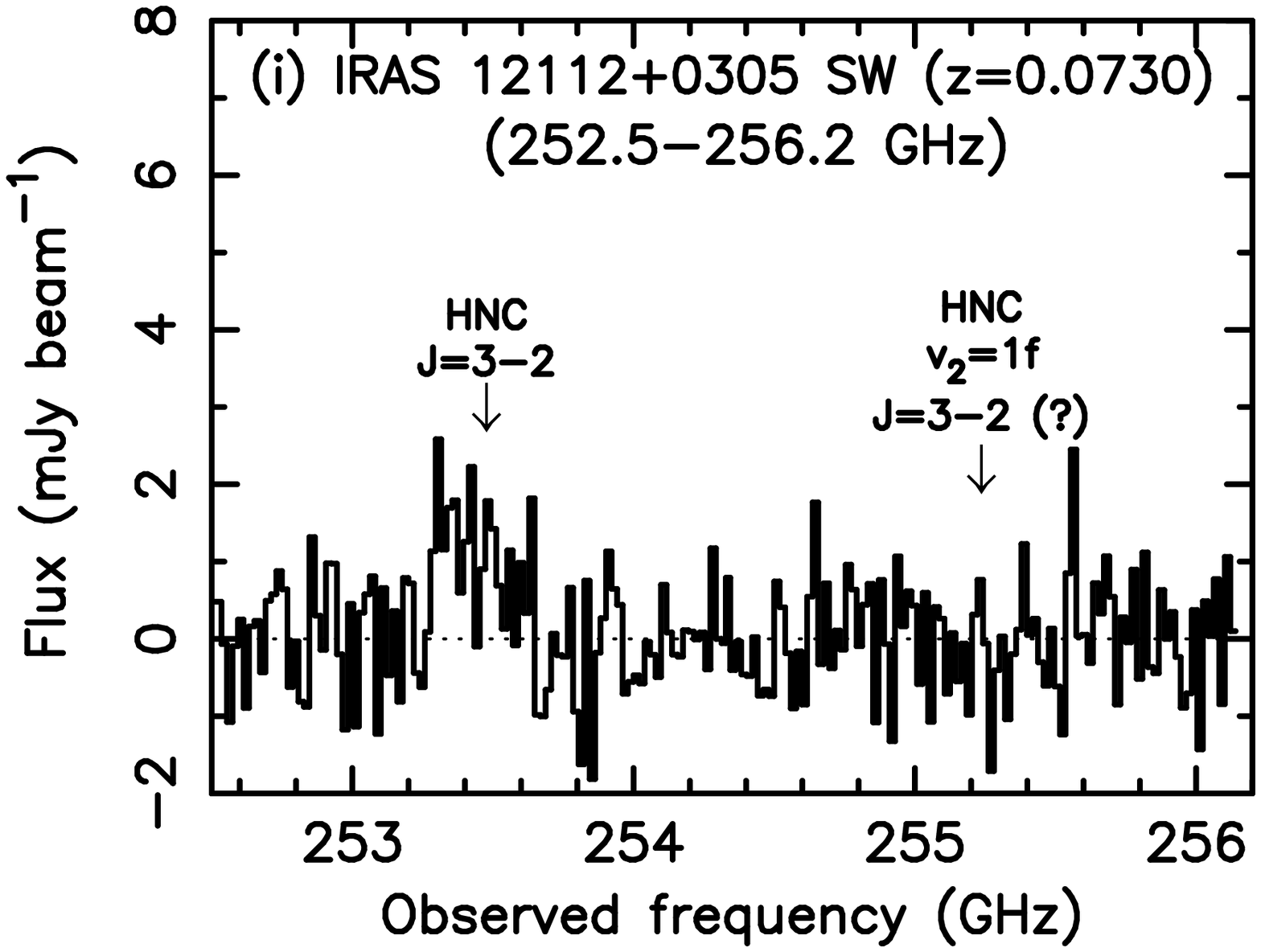} 
\includegraphics[angle=0,scale=.3]{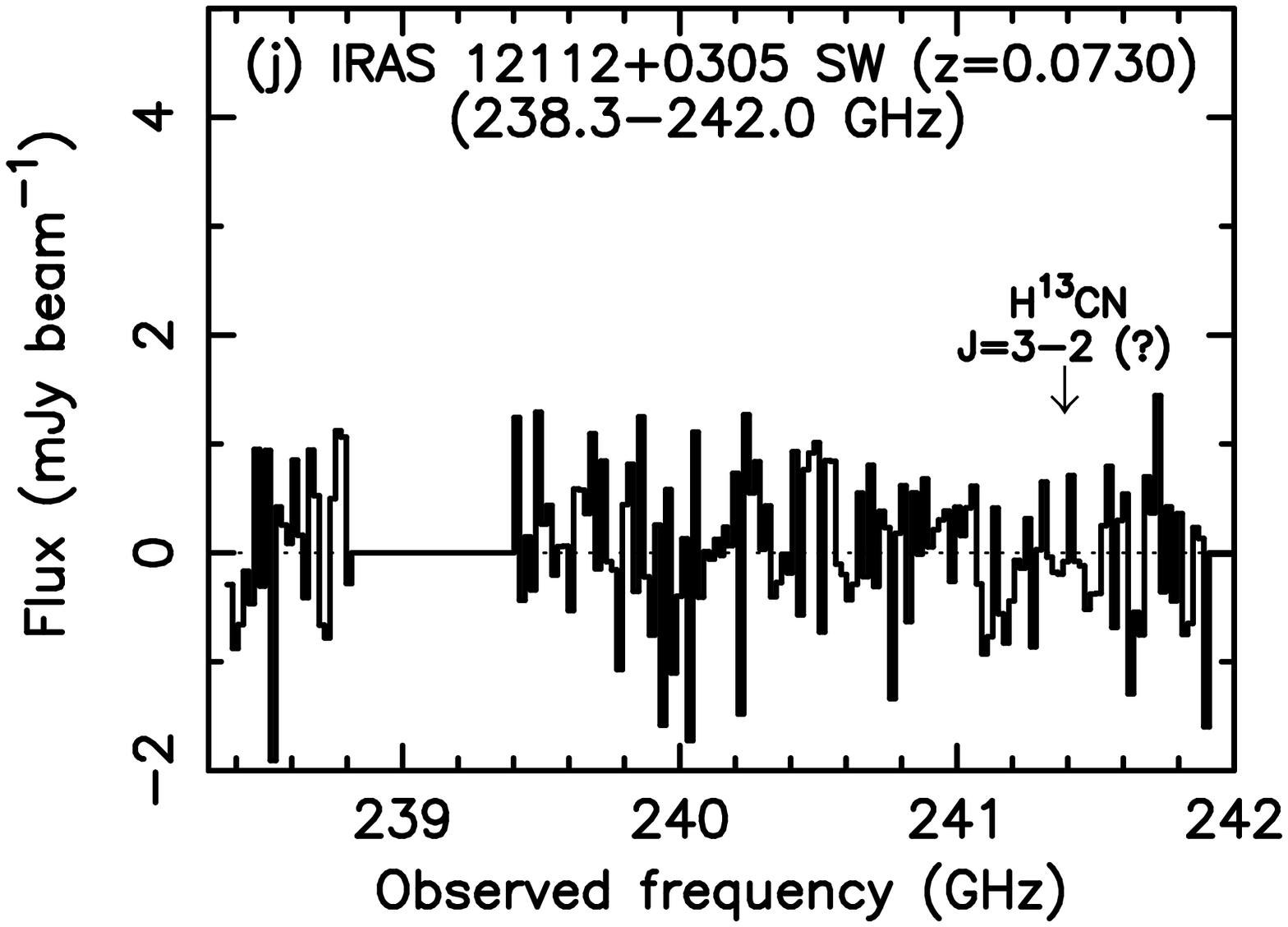} 
\includegraphics[angle=0,scale=.3]{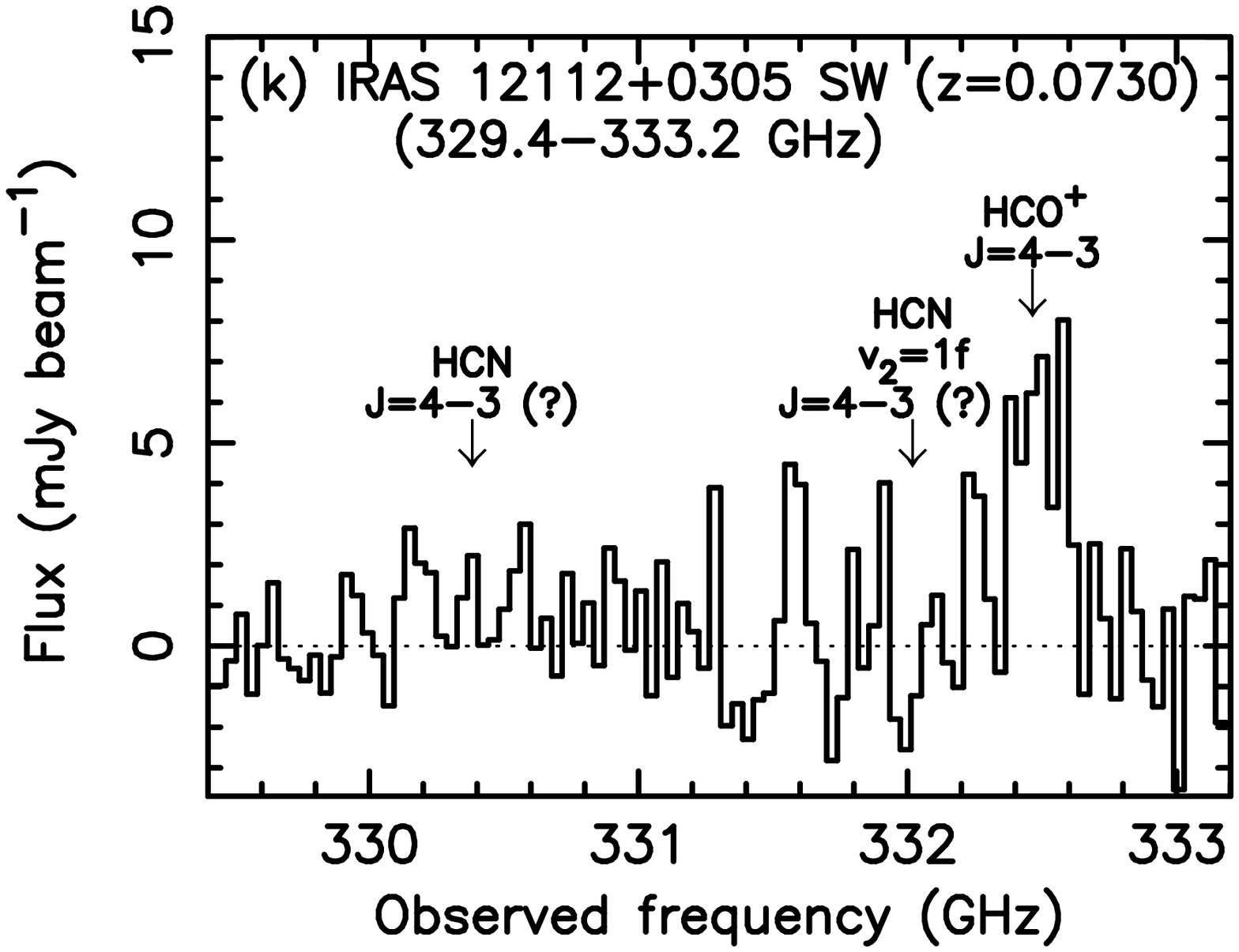} 
\includegraphics[angle=0,scale=.3]{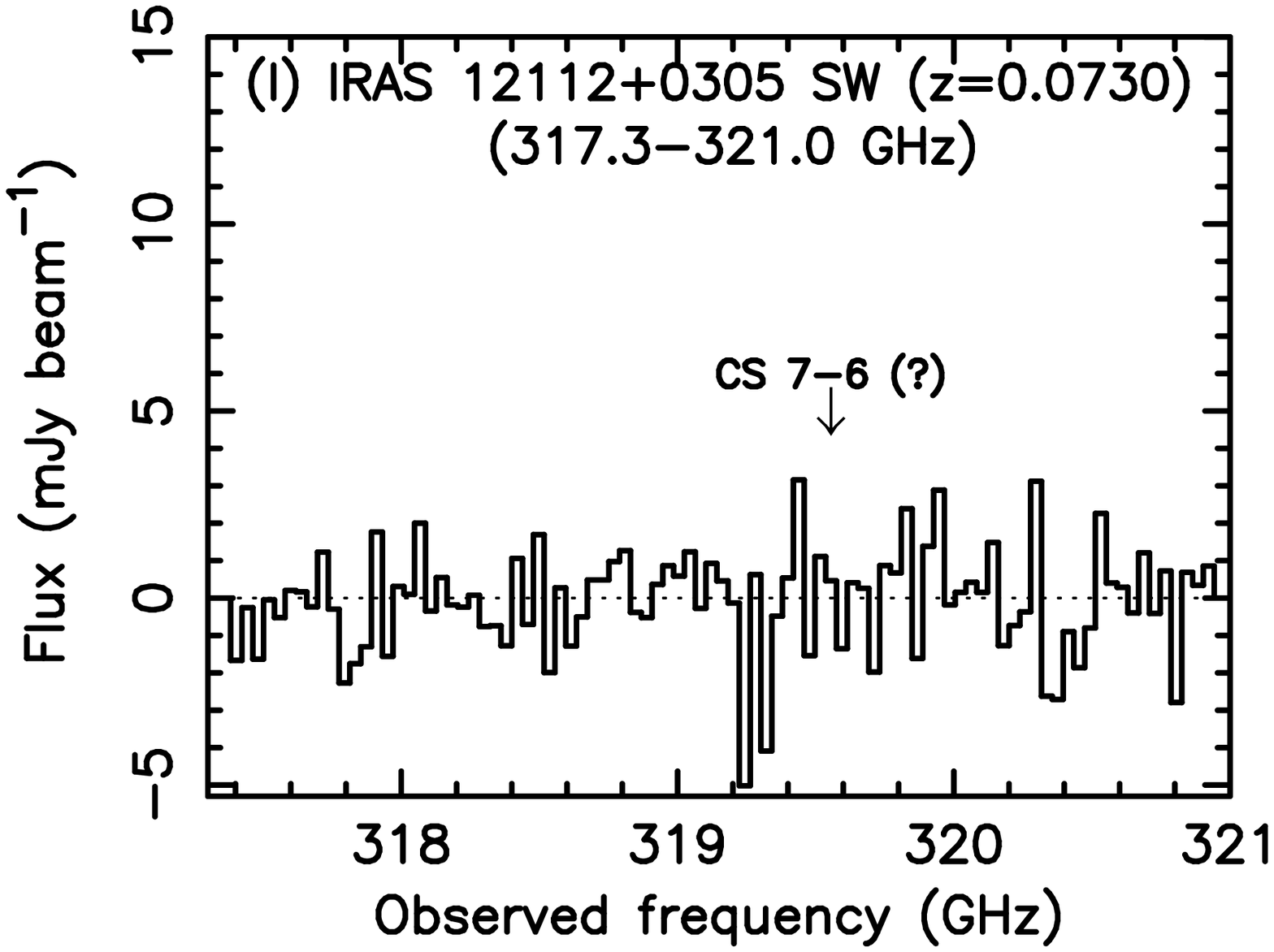} \\
\vspace{-0.5cm}
\includegraphics[angle=0,scale=.3]{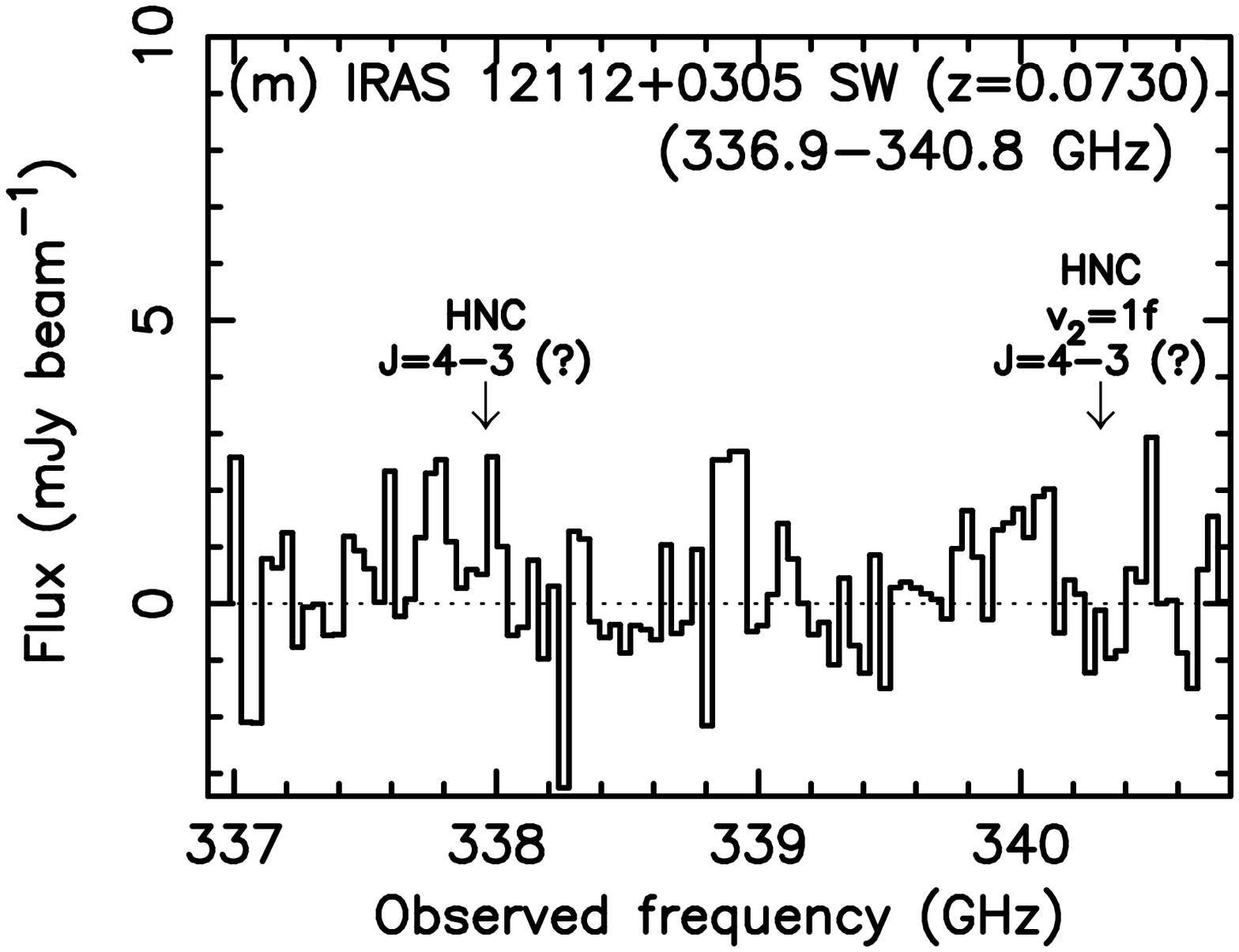} 
\includegraphics[angle=0,scale=.3]{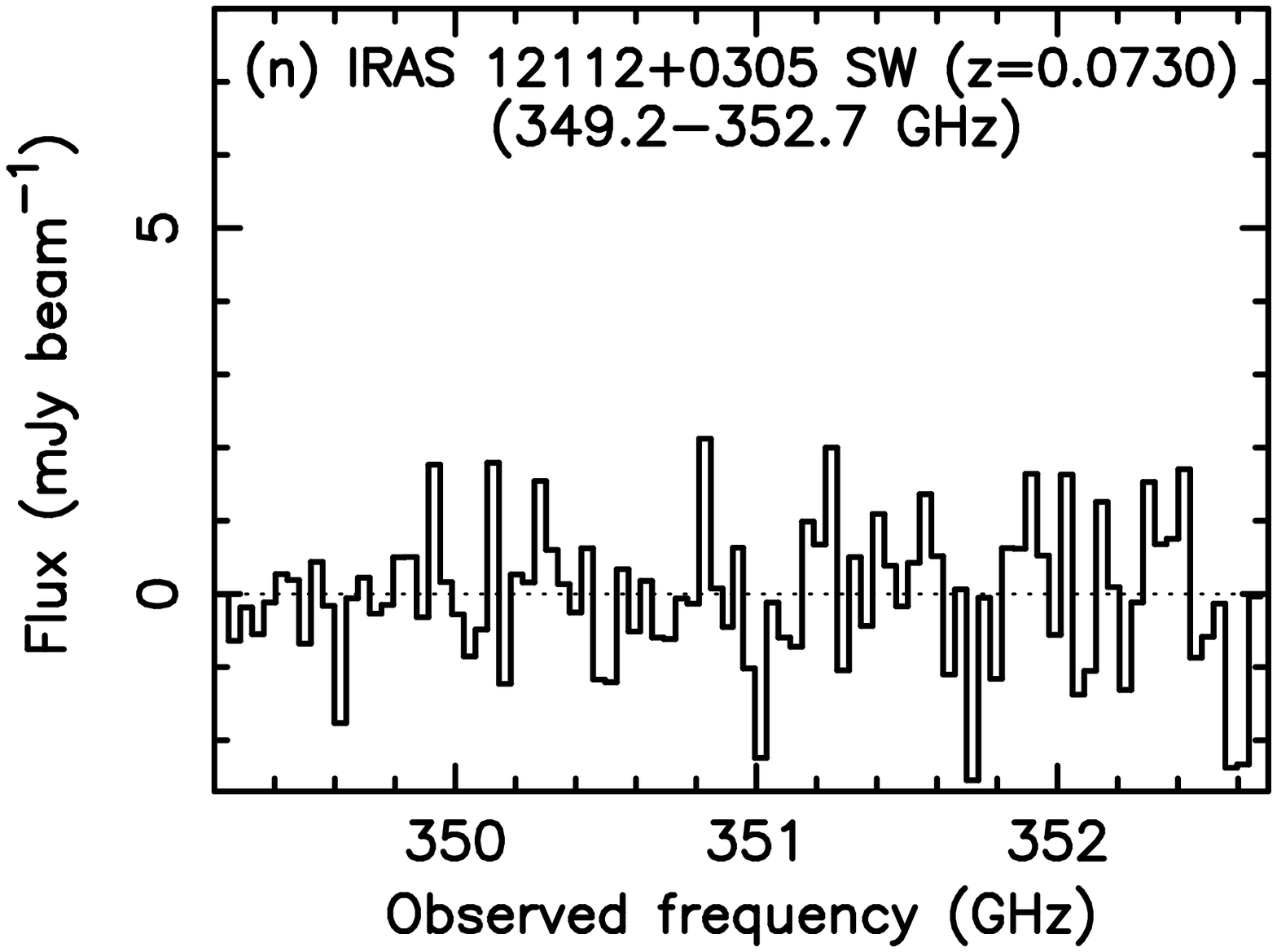} 
\includegraphics[angle=0,scale=.3]{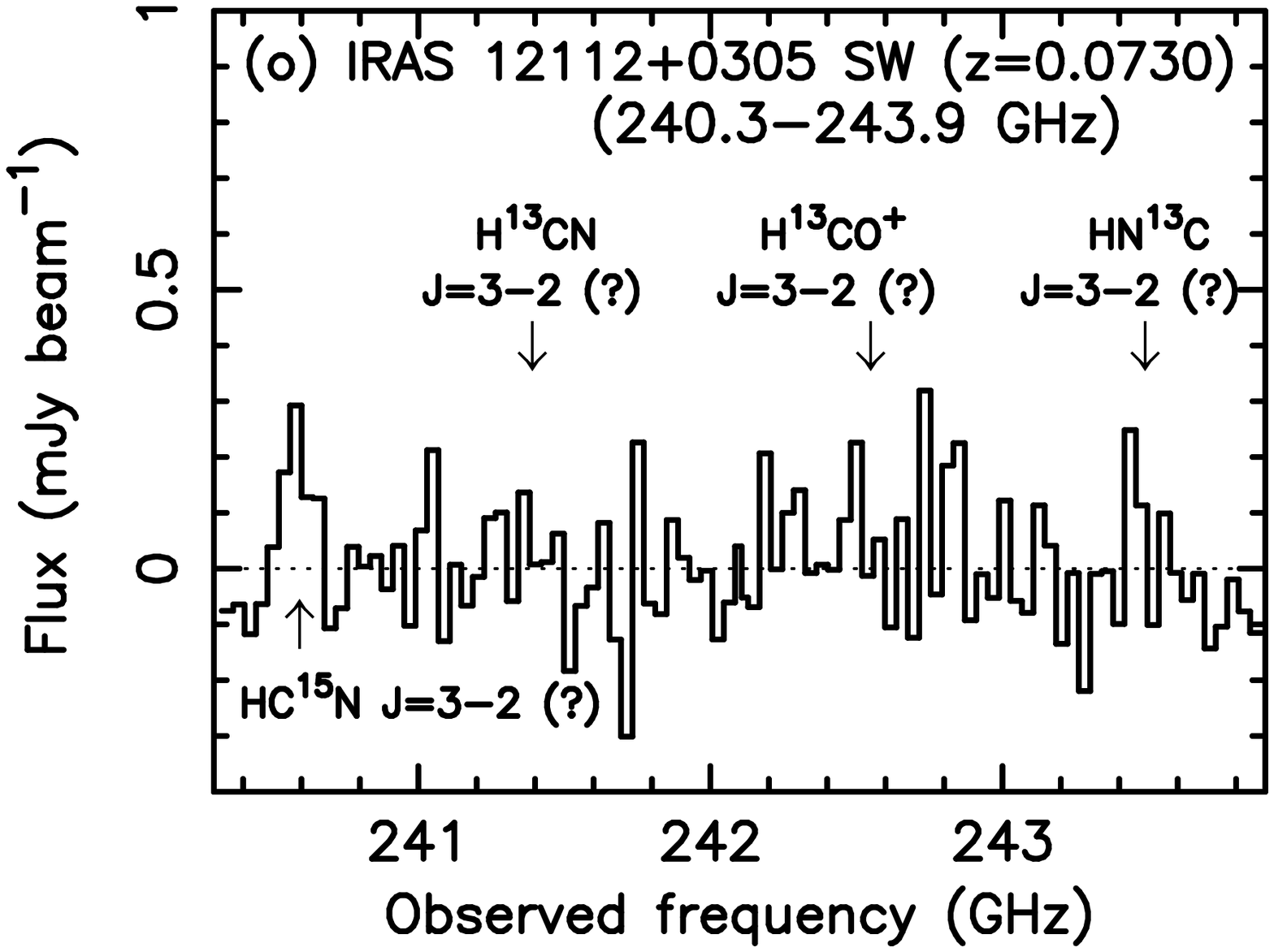} \\
\end{center}
\end{figure}

\clearpage

\begin{figure}
\begin{center}
\includegraphics[angle=0,scale=.41]{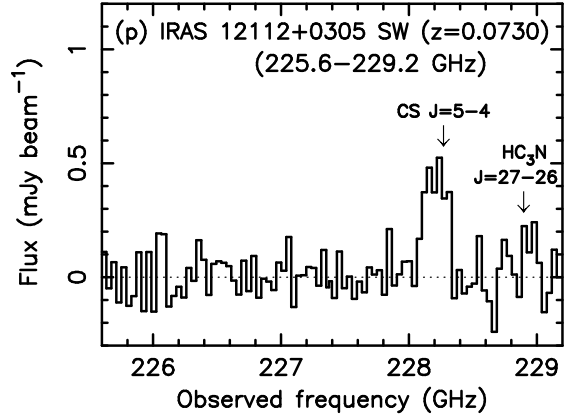} 
\end{center}
\caption{ALMA spectra of IRAS 12112$+$0305.
Those of the north-eastern nucleus (IRAS 12112$+$0305 NE) are shown
first, and those of the south-western nucleus (IRAS 12112$+$0305 SW) are
shown later. 
In (e), a downward arrow is shown for 
CH$_{3}$OH 16(2,14)--16($-$1,16) ($\nu_{\rm rest}$=364.859 GHz).
The expected frequency of SO 6(6)--5(5) ($\nu_{\rm rest}$=258.256 GHz) 
is shown with a downward arrow in (g), and that of HC$^{15}$N J=3--2 
($\nu_{\rm rest}$=258.157 GHz) is shown with an upward arrow in (g) and (o).
In (b) and (j), data at $\nu_{\rm obs}$ $\sim$ 239 GHz were flagged in 
the pipeline processed data, due to the influence of Earth's
atmospheric features.   
}
\end{figure}


\begin{figure}
\begin{center}
\includegraphics[angle=0,scale=.41]{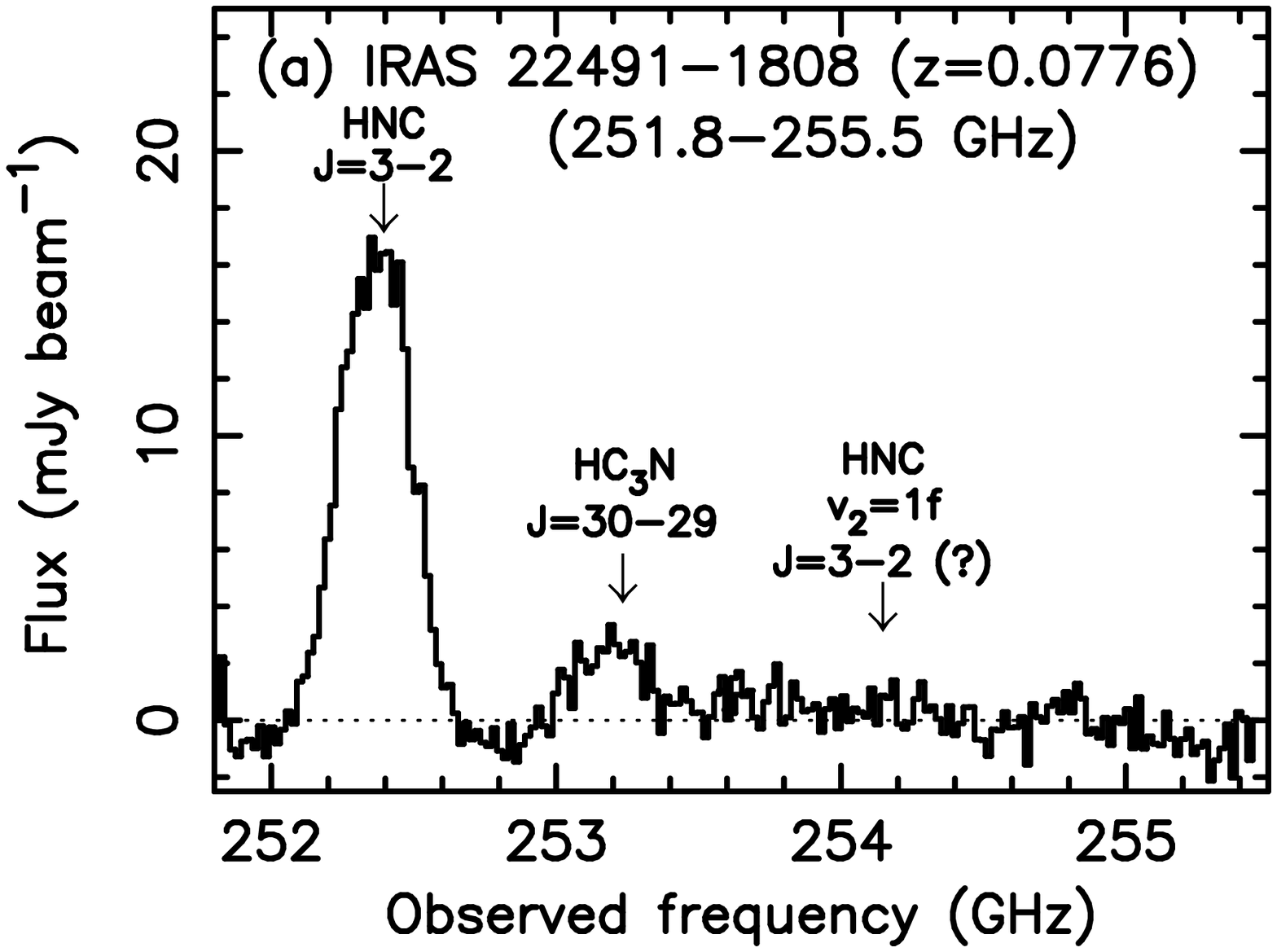} 
\includegraphics[angle=0,scale=.41]{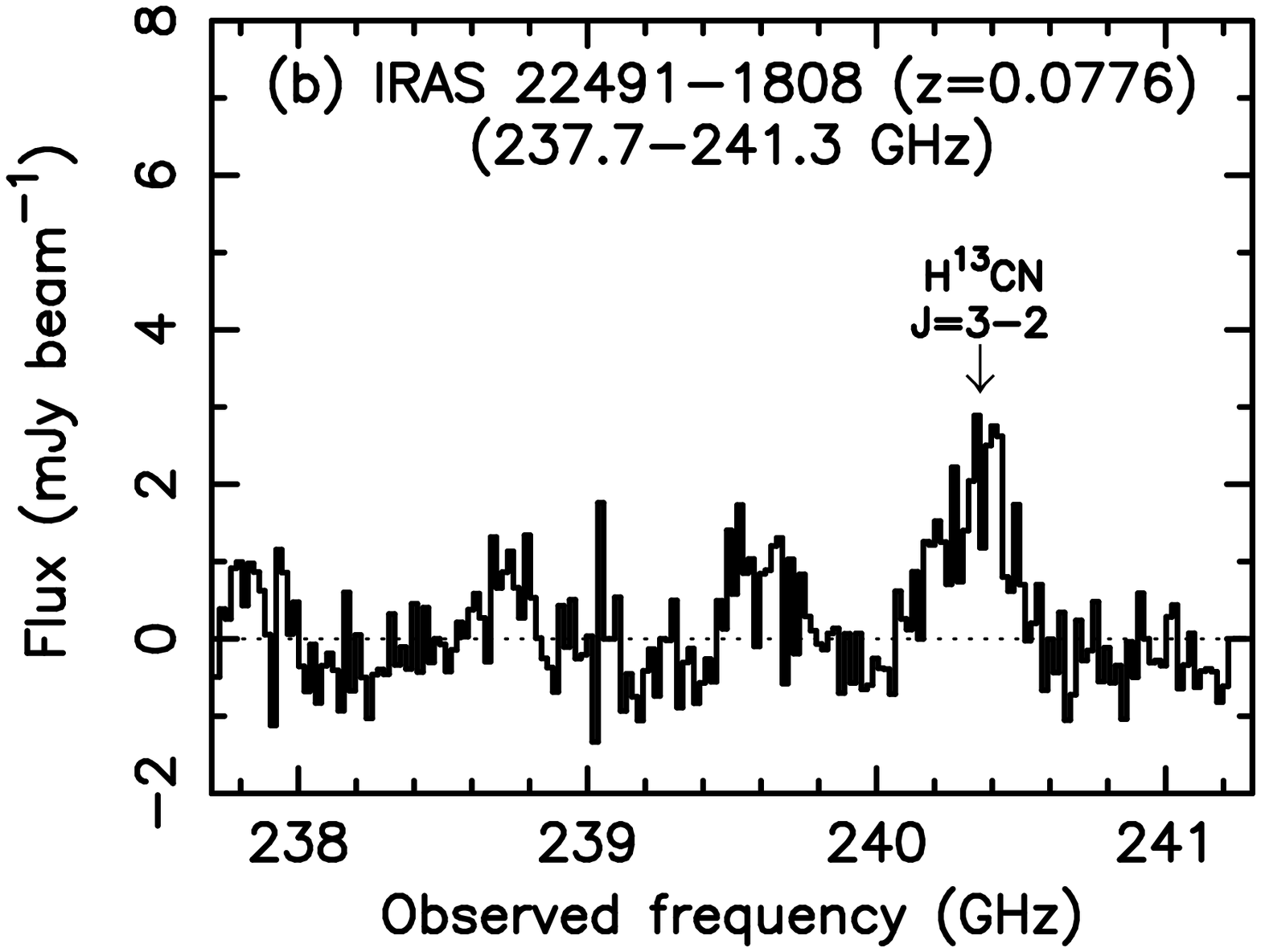} \\
\vspace{-0.5cm}
\includegraphics[angle=0,scale=.41]{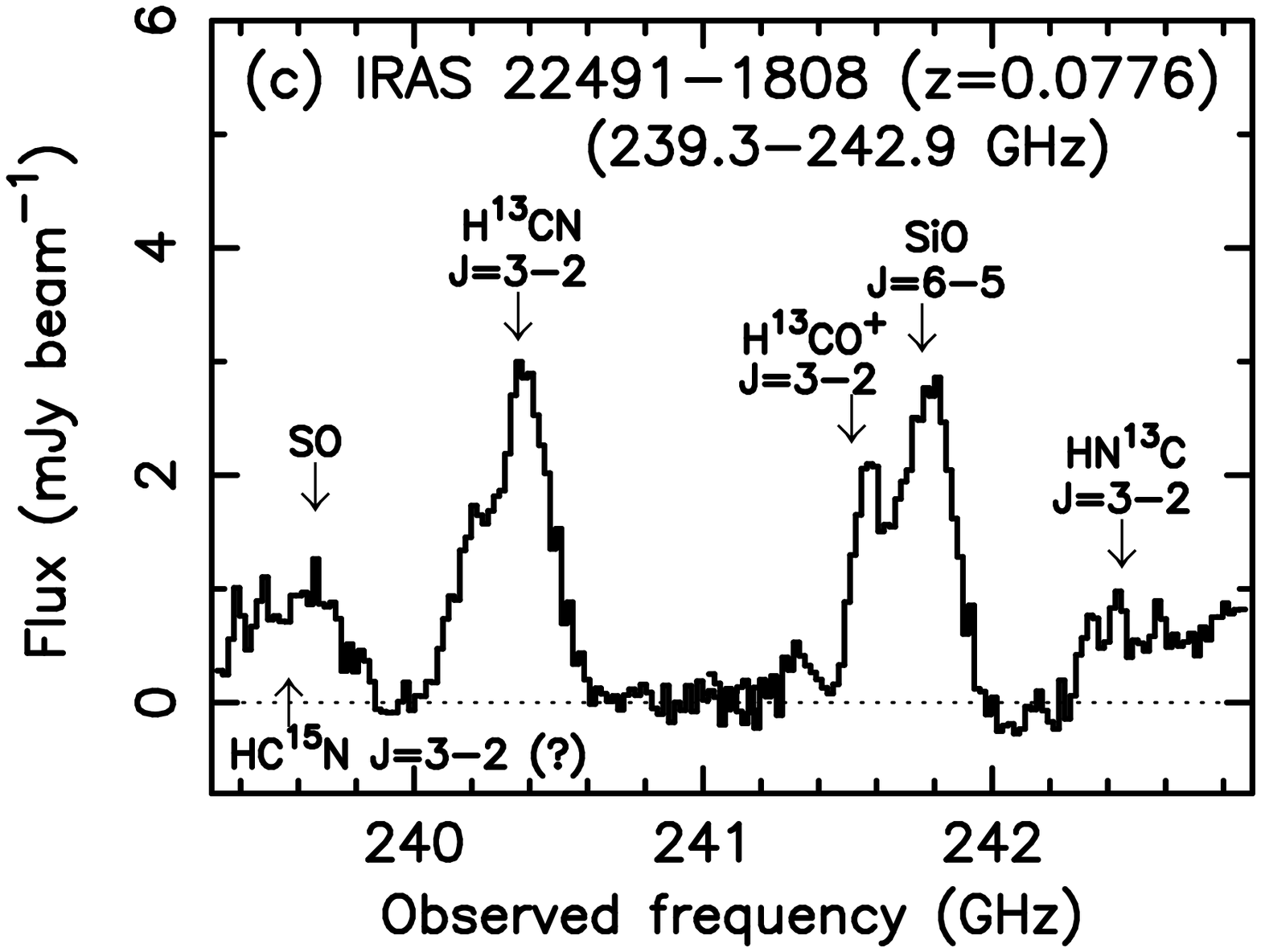} 
\includegraphics[angle=0,scale=.41]{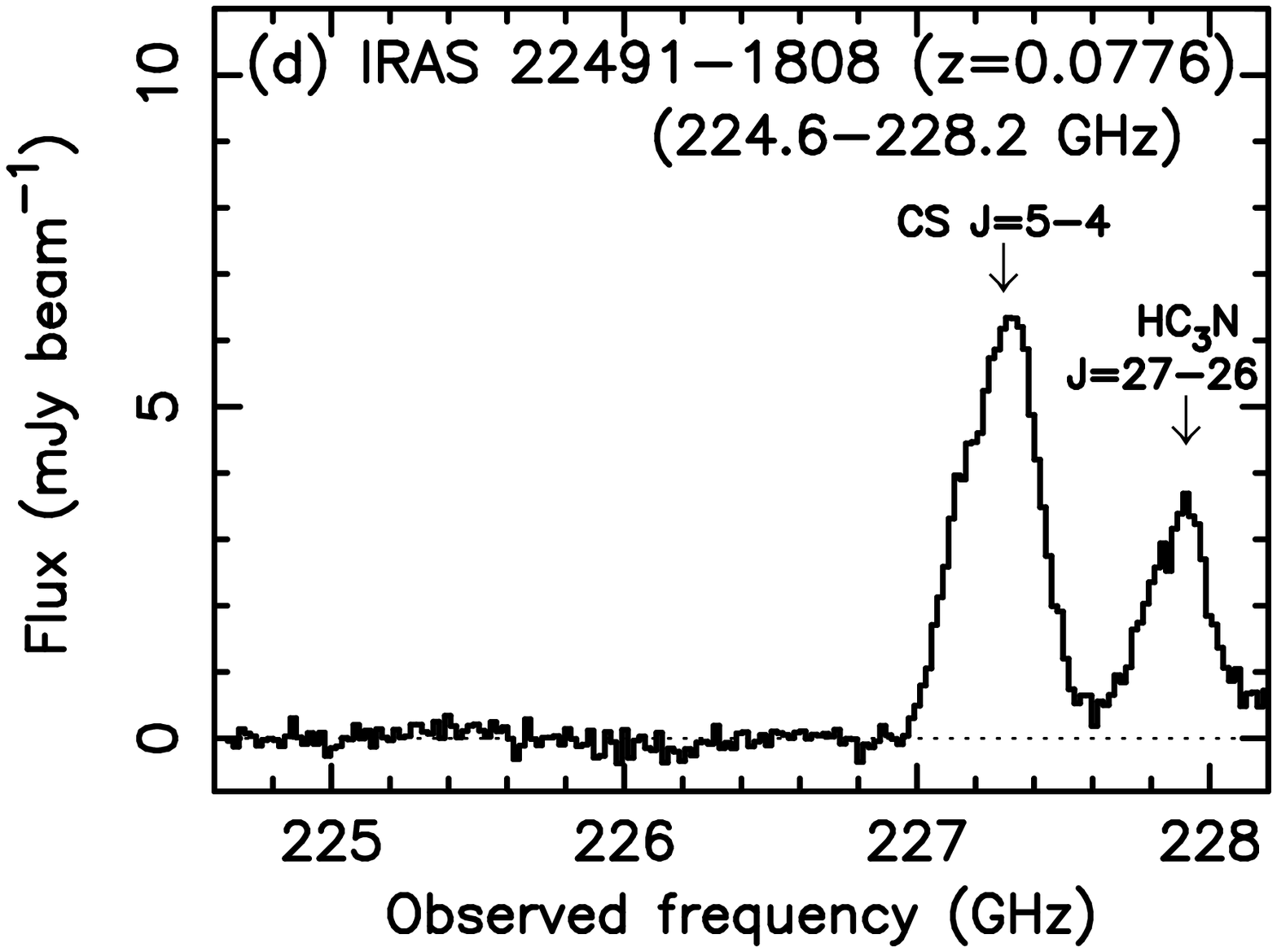} \\
\end{center}
\caption{ALMA spectra of IRAS 22491$-$1808.
In (c), a downward arrow and upward arrow are shown for 
SO 6(6)--5(5) ($\nu_{\rm rest}$=258.256 GHz) and 
HC$^{15}$N J=3--2 ($\nu_{\rm rest}$=258.157 GHz), respectively.
}
\end{figure}


\begin{figure}
\begin{center}
\includegraphics[angle=0,scale=.3]{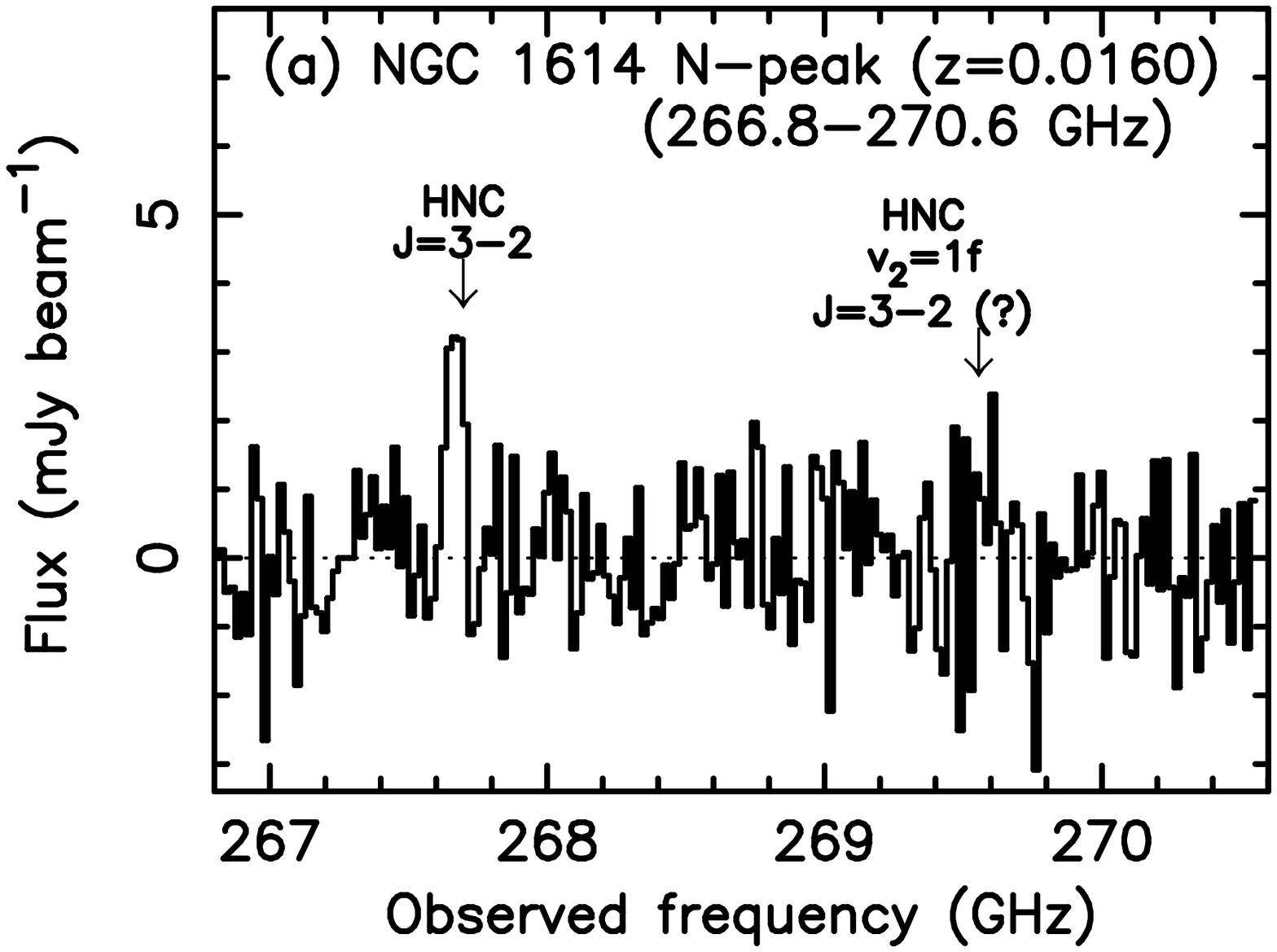}  
\includegraphics[angle=0,scale=.3]{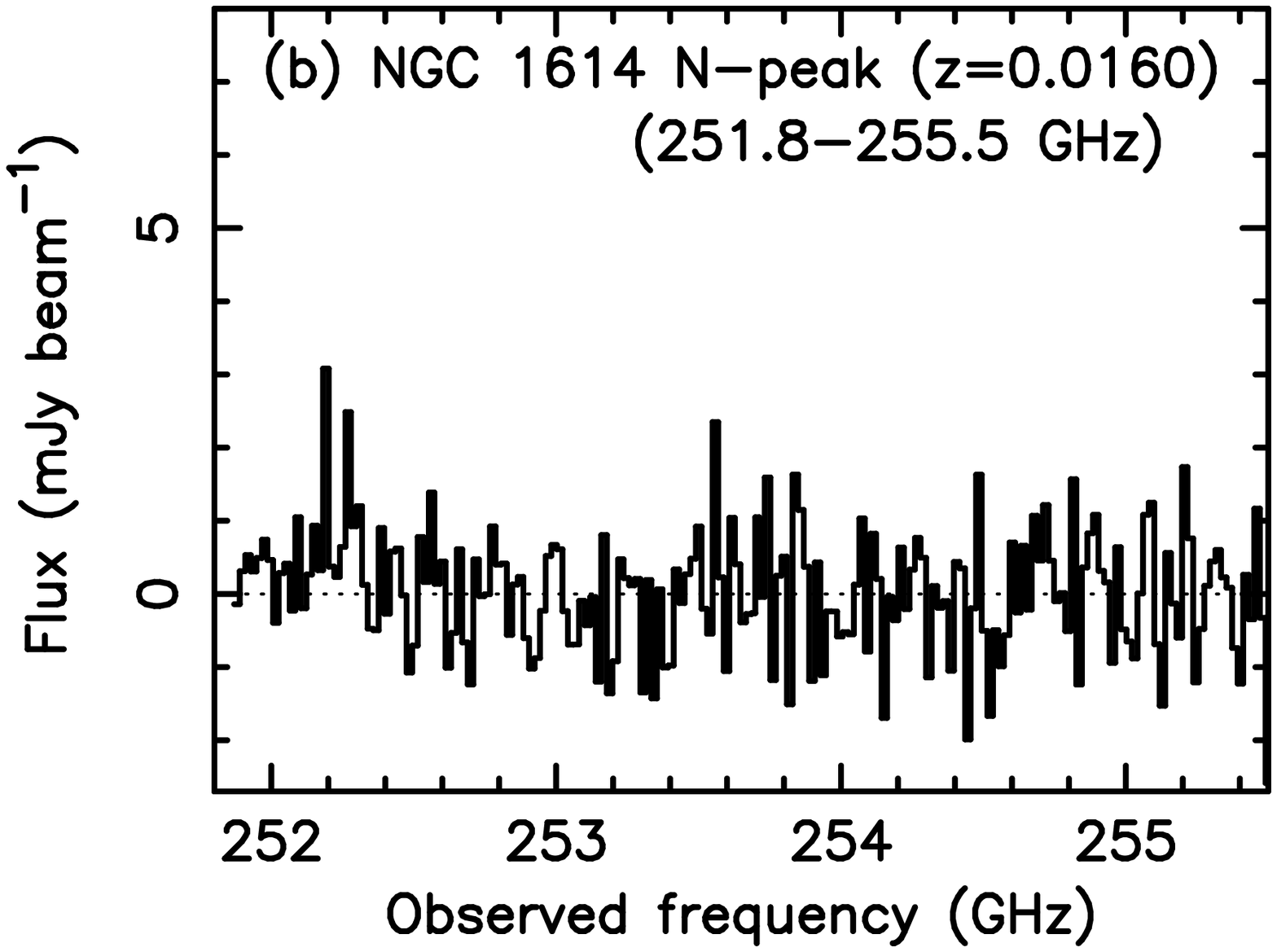}    
\includegraphics[angle=0,scale=.3]{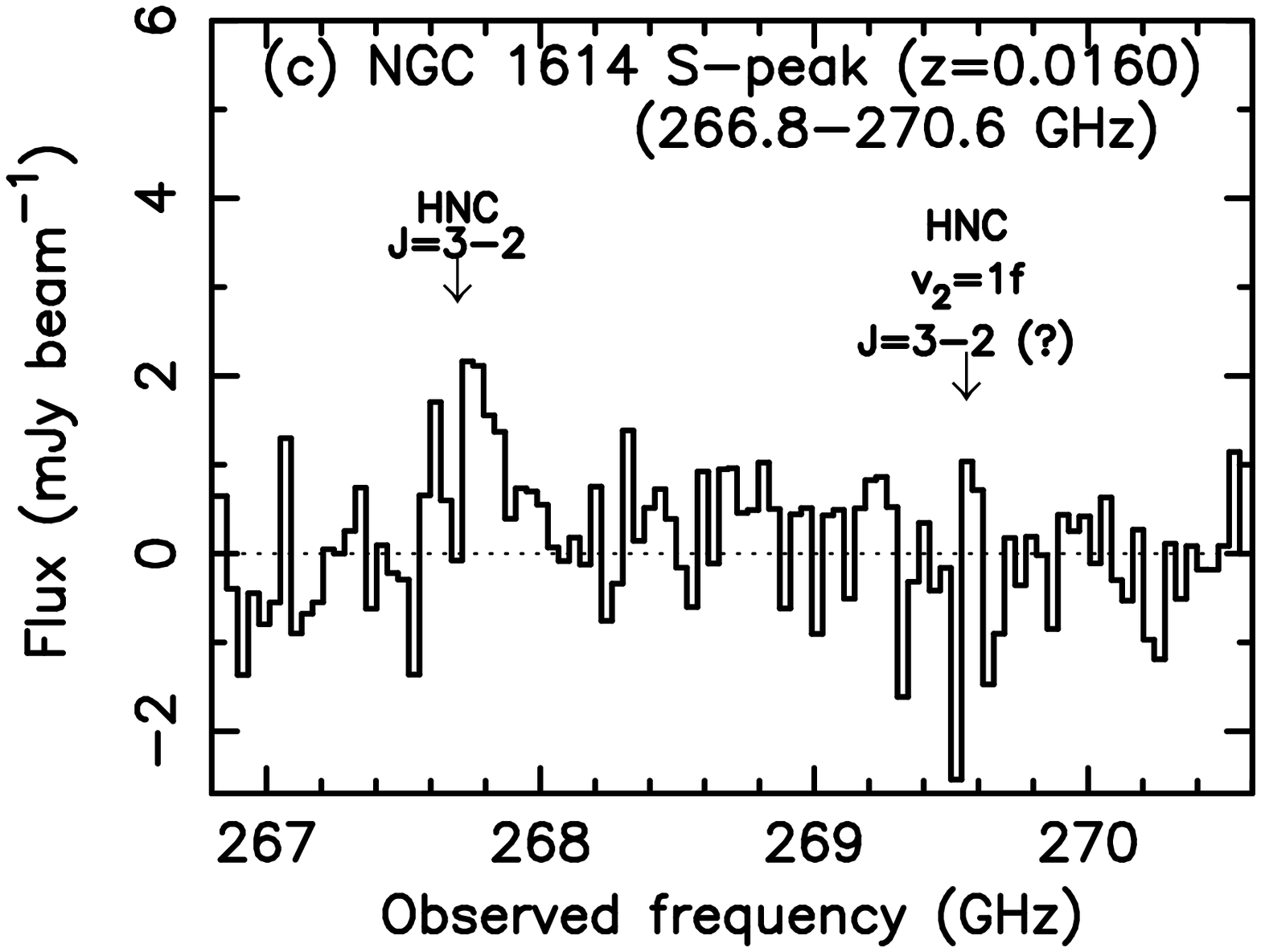} \\ 
\vspace{-0.5cm}
\includegraphics[angle=0,scale=.3]{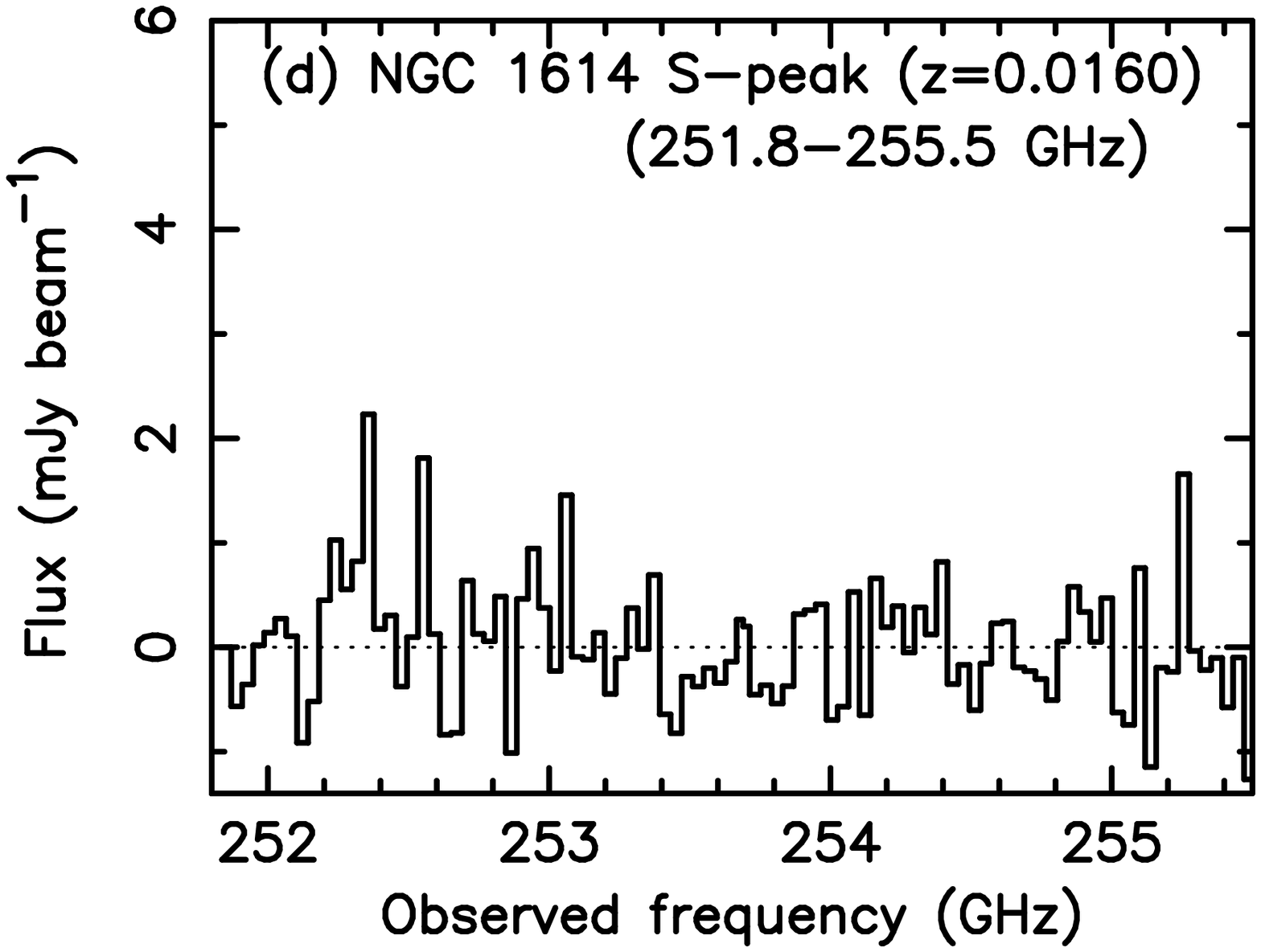} 
\includegraphics[angle=0,scale=.3]{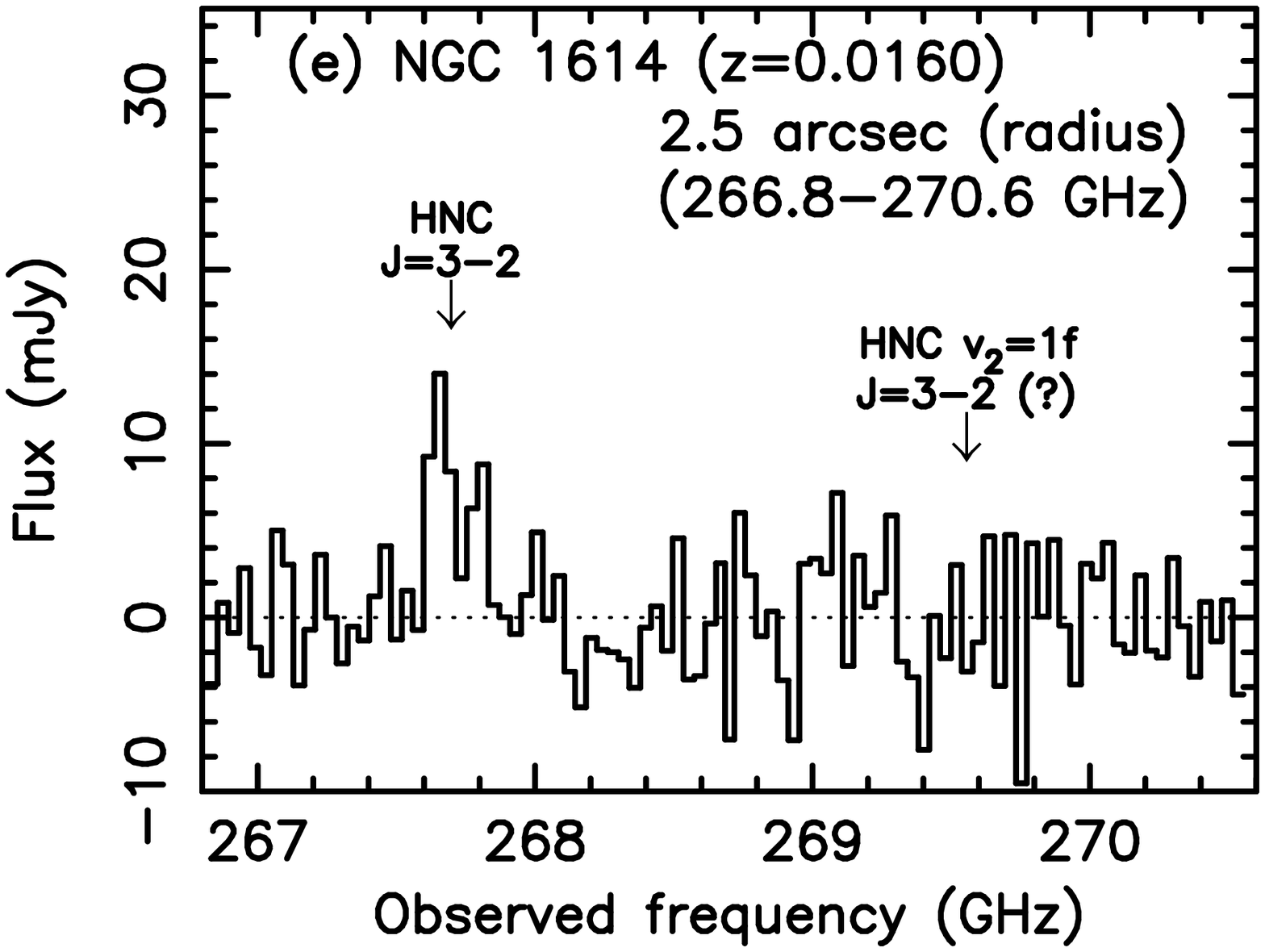}  
\includegraphics[angle=0,scale=.3]{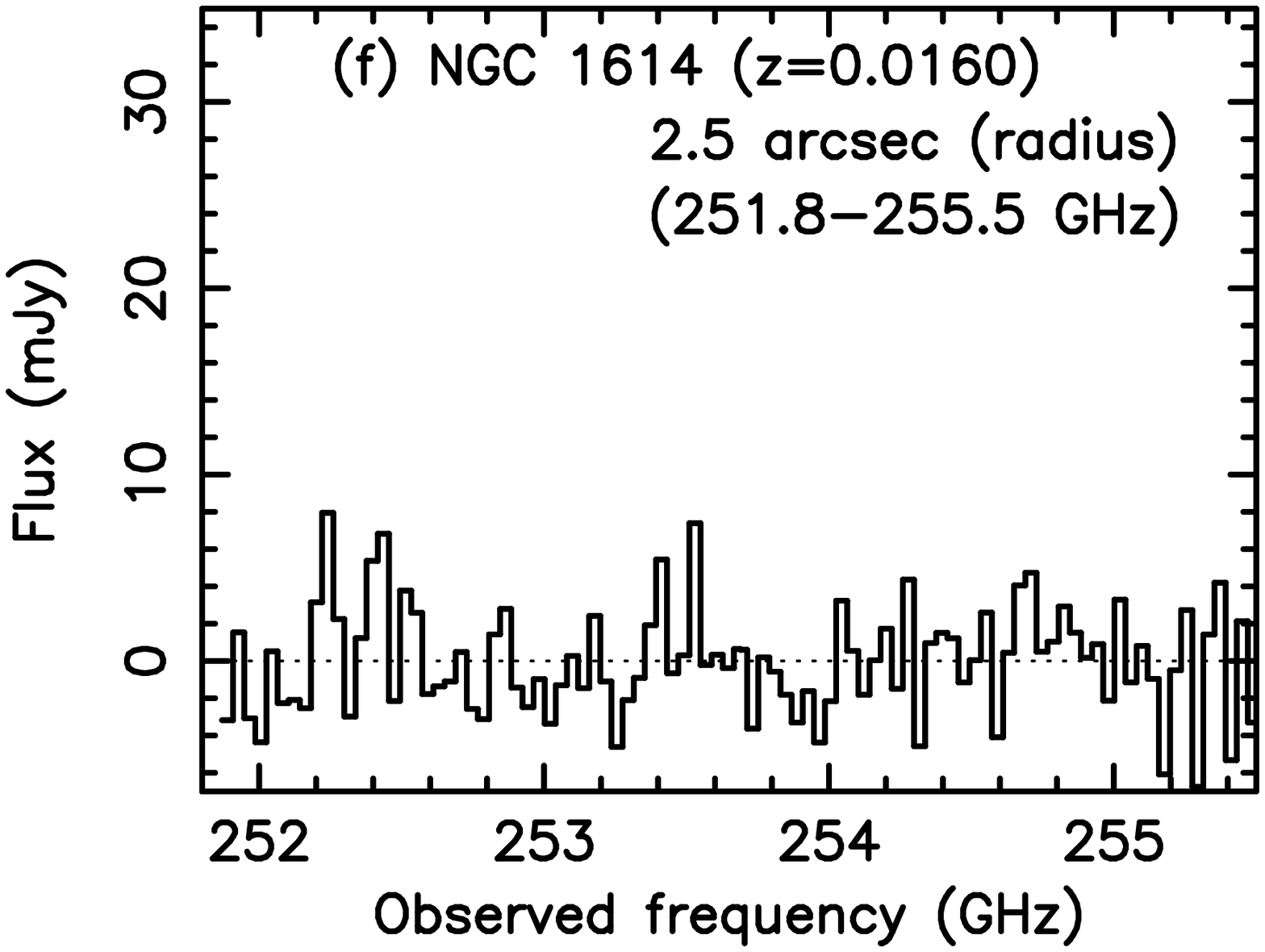} \\   
\end{center}
\caption{ALMA spectra of NGC 1614. 
In addition to the spectra at the continuum peak positions within the
beam size, spatially integrated spectra are shown.} 
\end{figure}


\begin{figure}
\begin{center}
\includegraphics[angle=0,scale=.41]{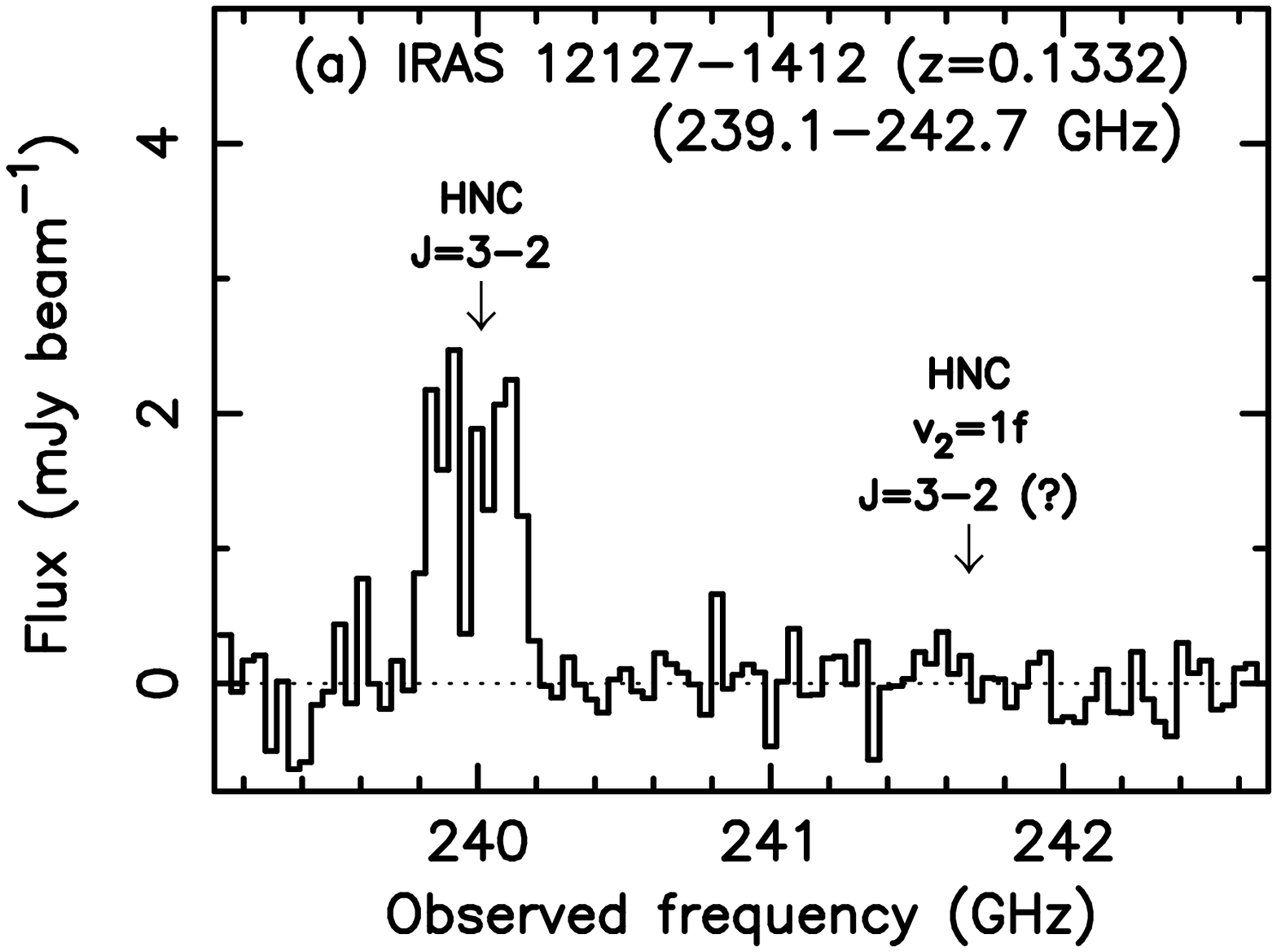}    
\includegraphics[angle=0,scale=.41]{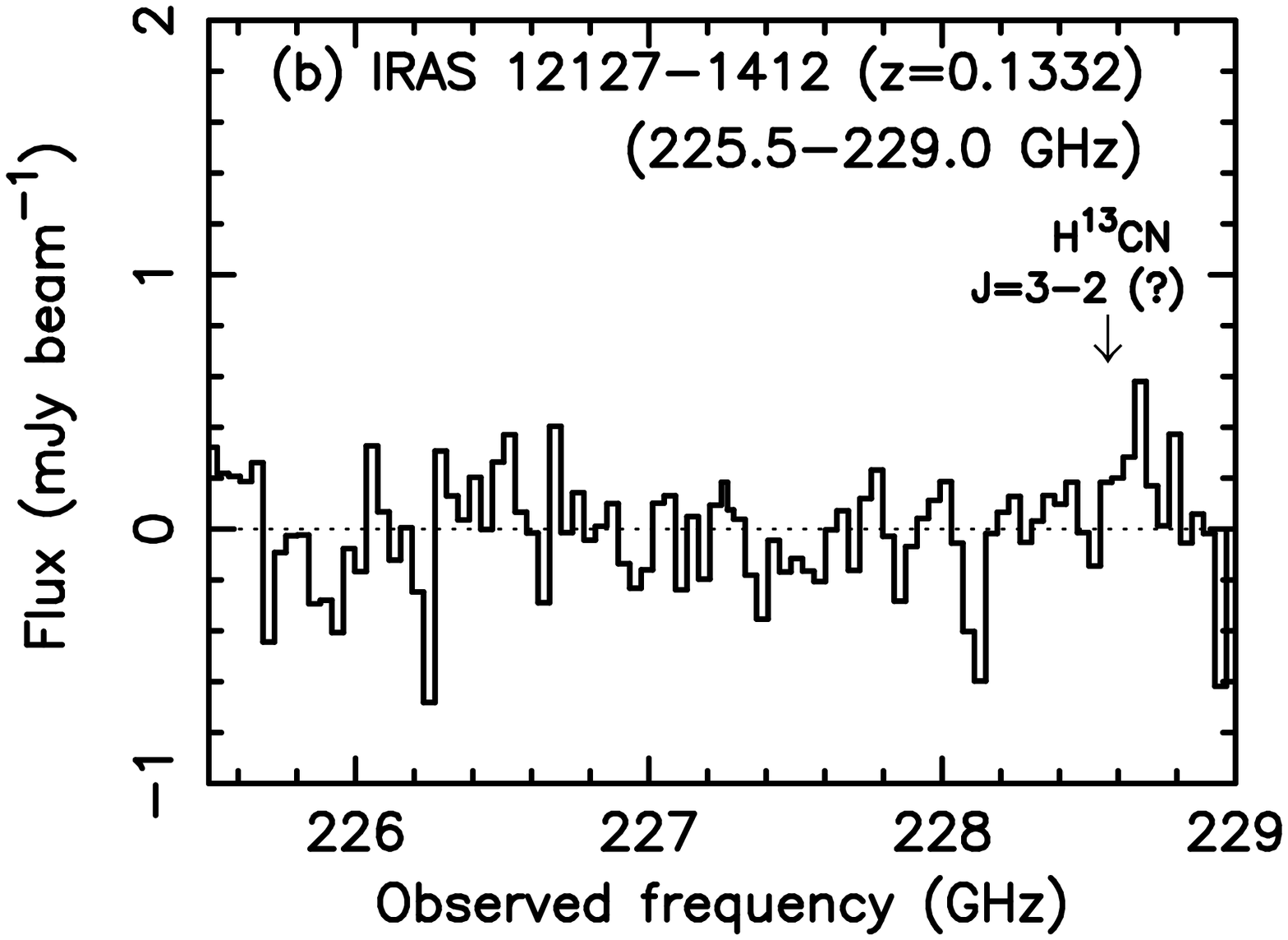}    
\end{center}
\caption{ALMA spectra of IRAS 12127$-$1412.}
\end{figure}


\begin{figure}
\begin{center}
\includegraphics[angle=0,scale=.3]{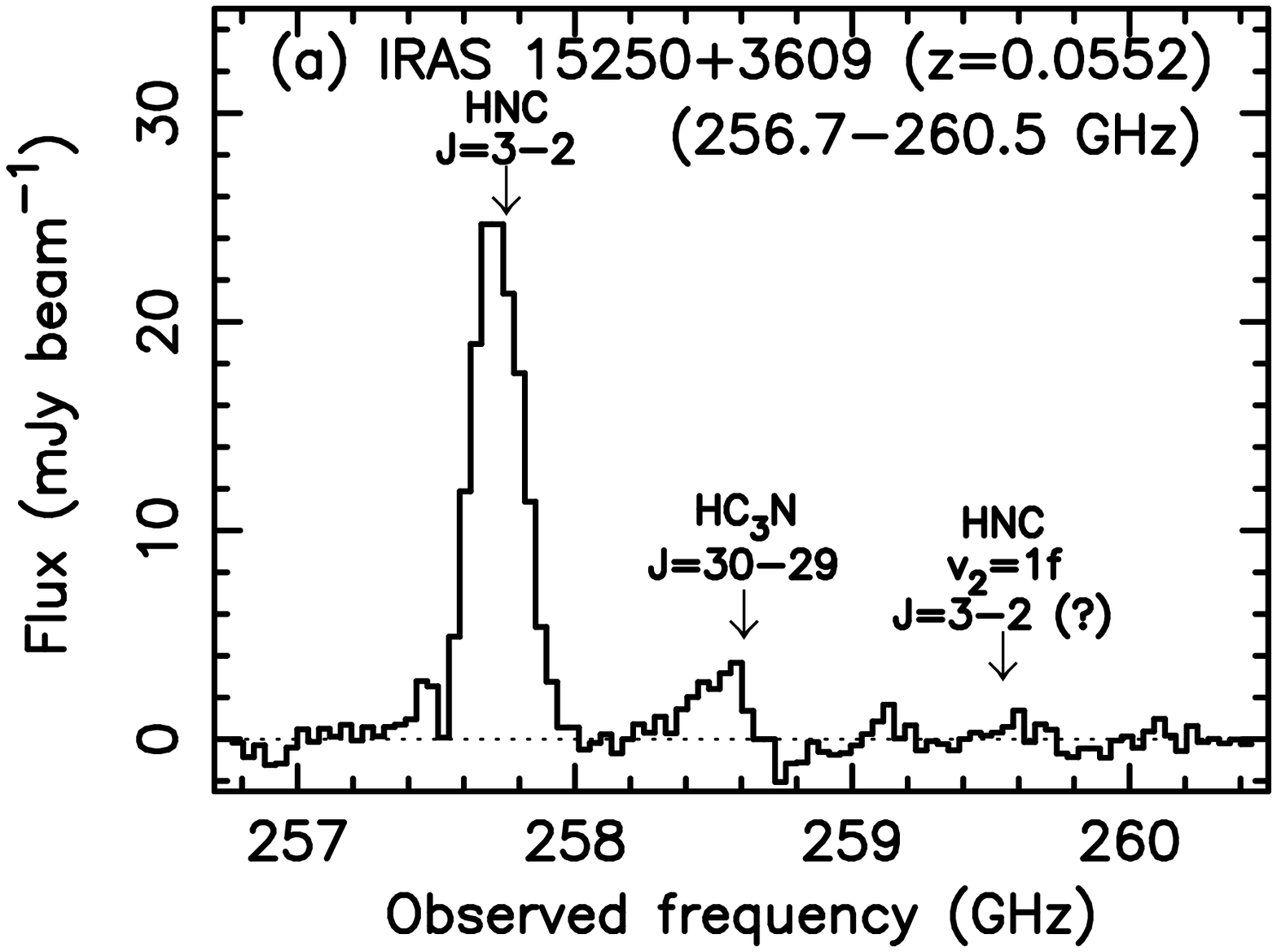}  
\includegraphics[angle=0,scale=.3]{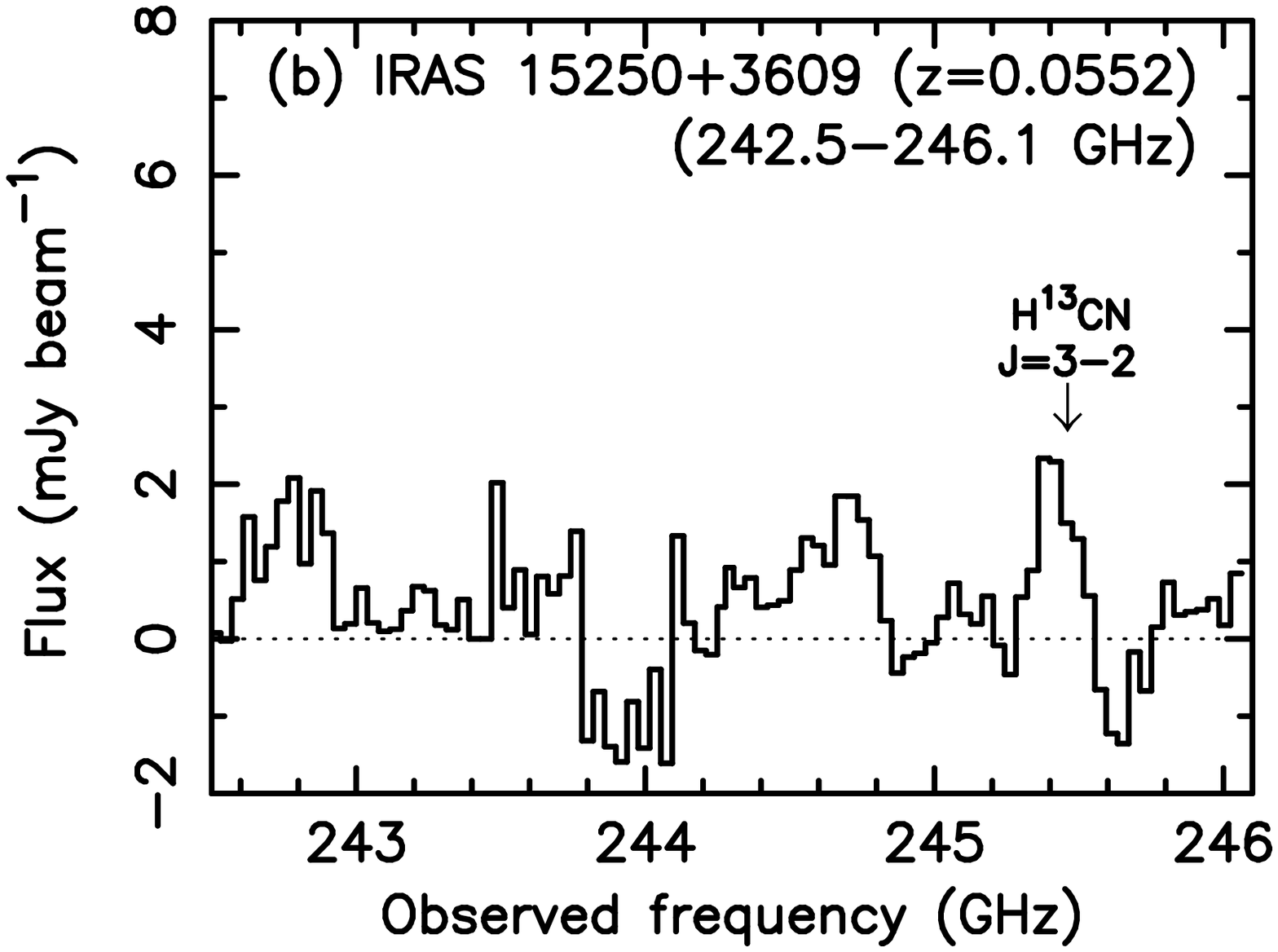}    
\includegraphics[angle=0,scale=.3]{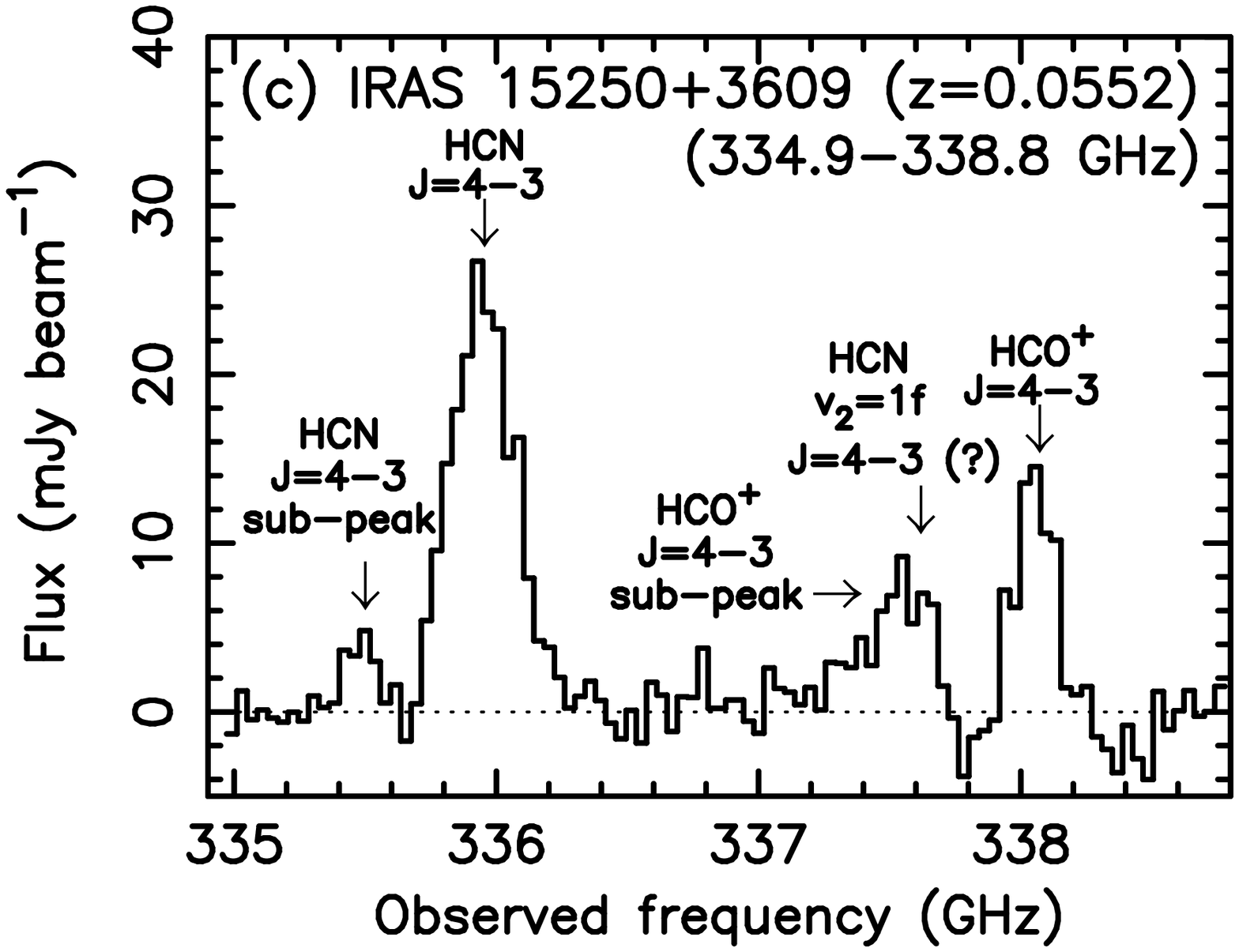} \\
\includegraphics[angle=0,scale=.3]{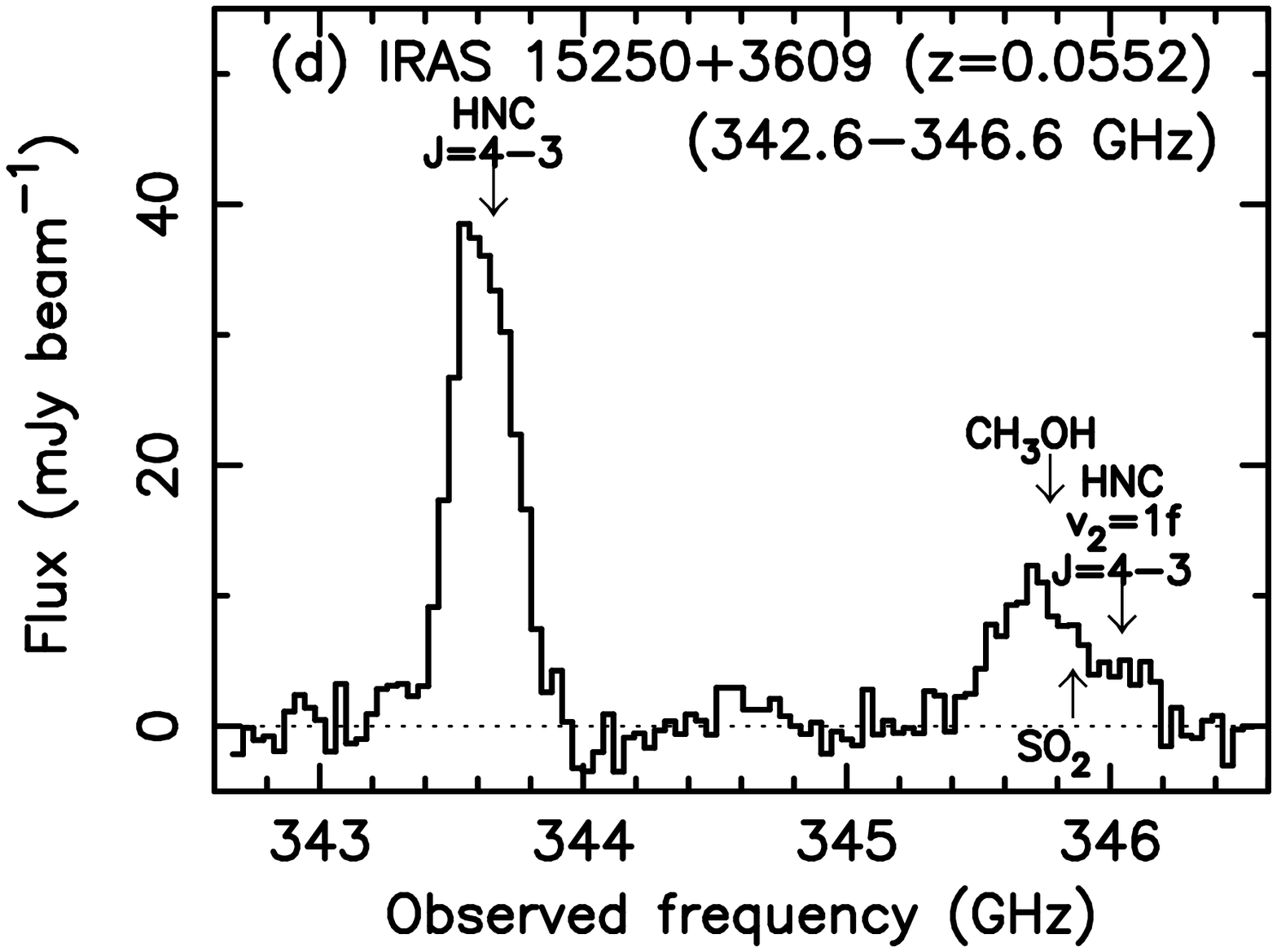} 
\includegraphics[angle=0,scale=.3]{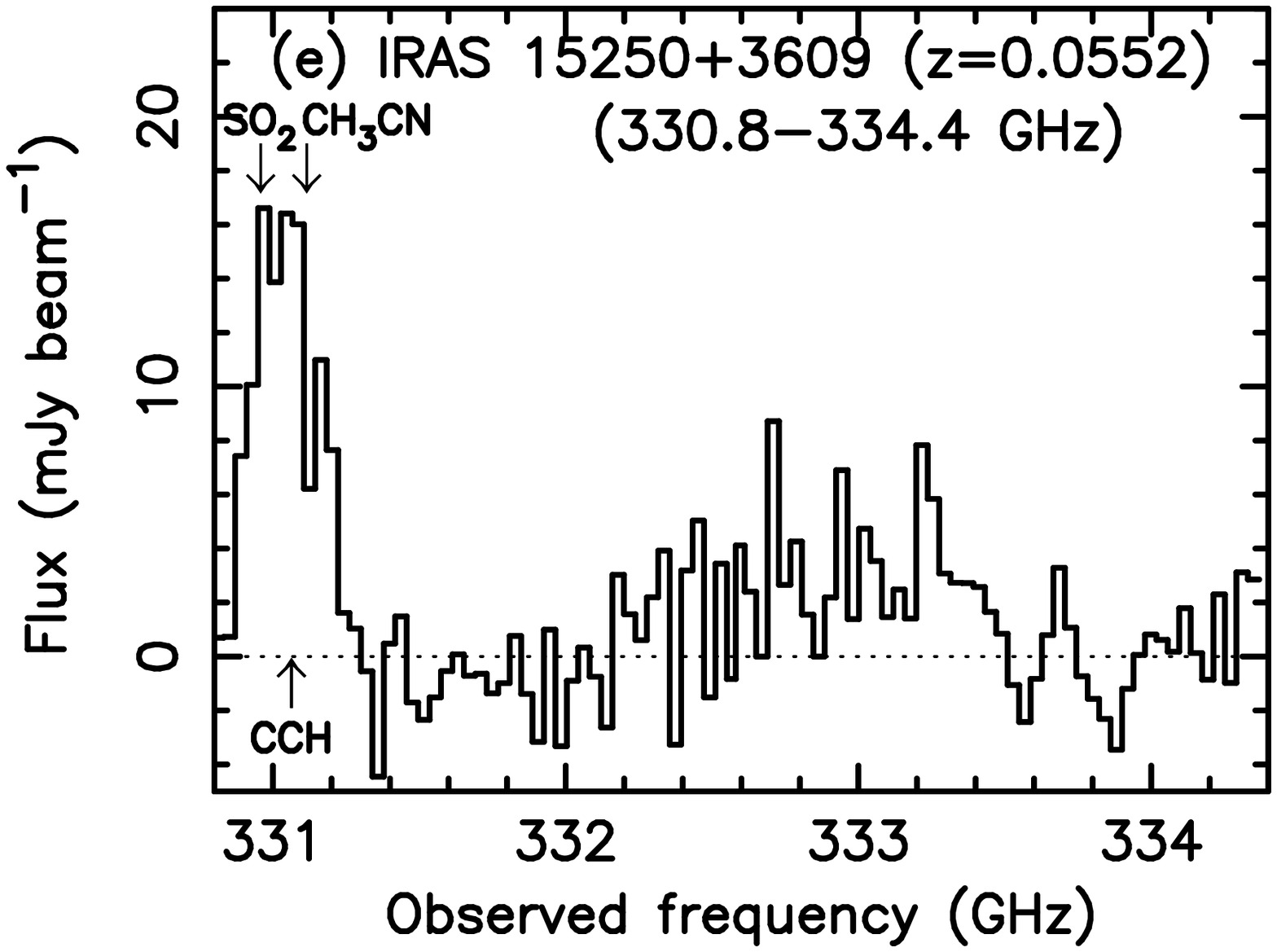} 
\end{center}
\caption{
ALMA spectra of IRAS 15250$+$3609.
In (d), a downward arrow and upward arrow are shown for 
CH$_{3}$OH 16(2,14)--16($-$1,16) ($\nu_{\rm rest}$=364.859 GHz) and
SO$_{2}$ 25(9,17)--26(8,18) ($\nu_{\rm rest}$=364.950 GHz), respectively.
The HNC v$_{2}$=1f J=4--3 line may be contaminated by the
CH$_{3}$OH and SO$_{2}$ lines.
In (e), downward arrows are shown for SO$_{2}$ 31(10,22)--32(9,23)
($\nu_{\rm rest}$=349.227 GHz) and CH$_{3}$CN 19(3)--18(3) 
($\nu_{\rm rest}$=349.393 GHz), and upward arrows are shown for 
CCH N=4--3, J=9/2--7/2, F=5--4 ($\nu_{\rm rest}$=349.337 GHz) and 
CCH N=4--3, J=9/2--7/2, F=4--3 ($\nu_{\rm rest}$=349.339 GHz).    
}
\end{figure}


\begin{figure}
\begin{center}
\includegraphics[angle=0,scale=.3]{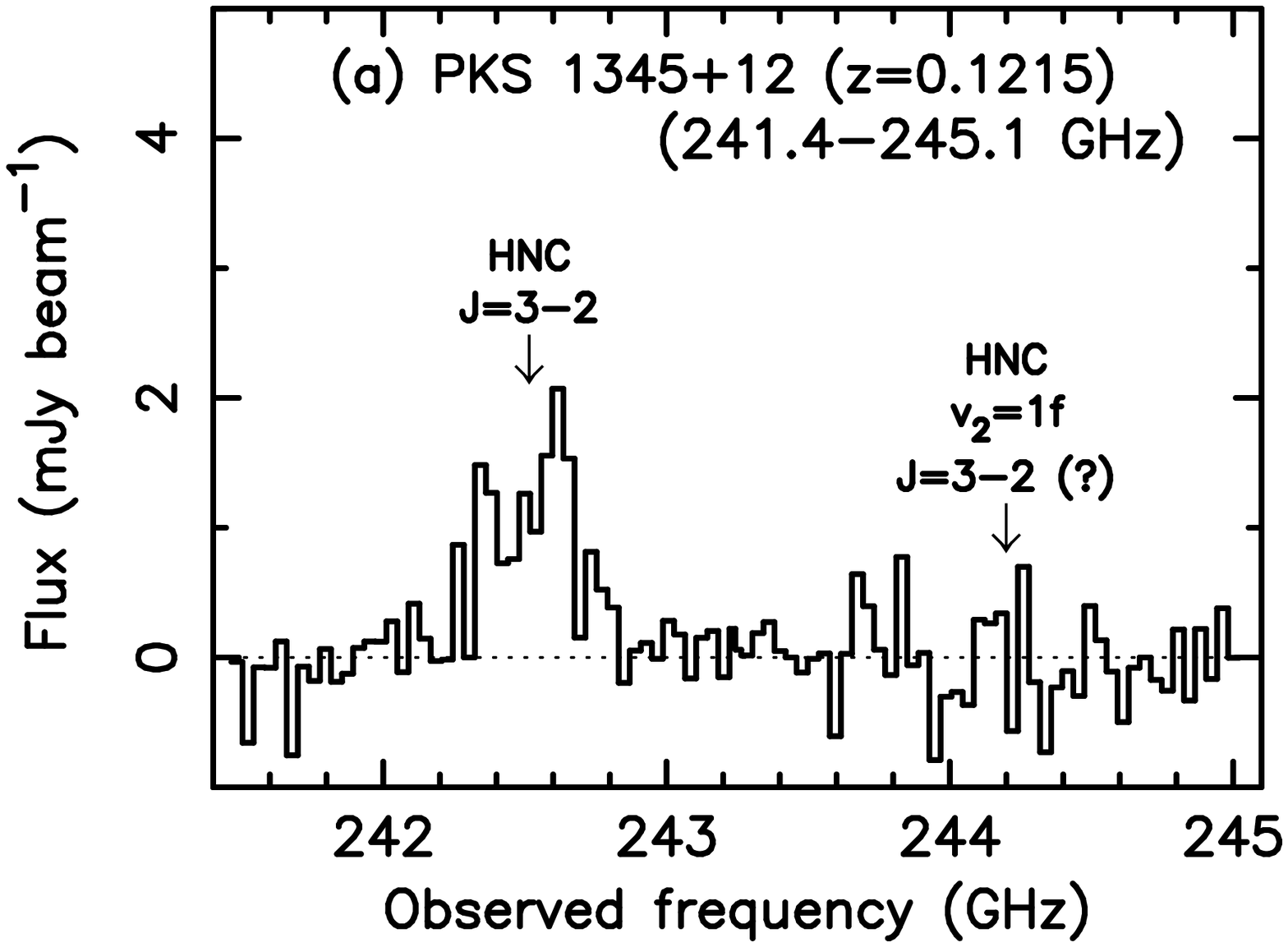}  
\includegraphics[angle=0,scale=.3]{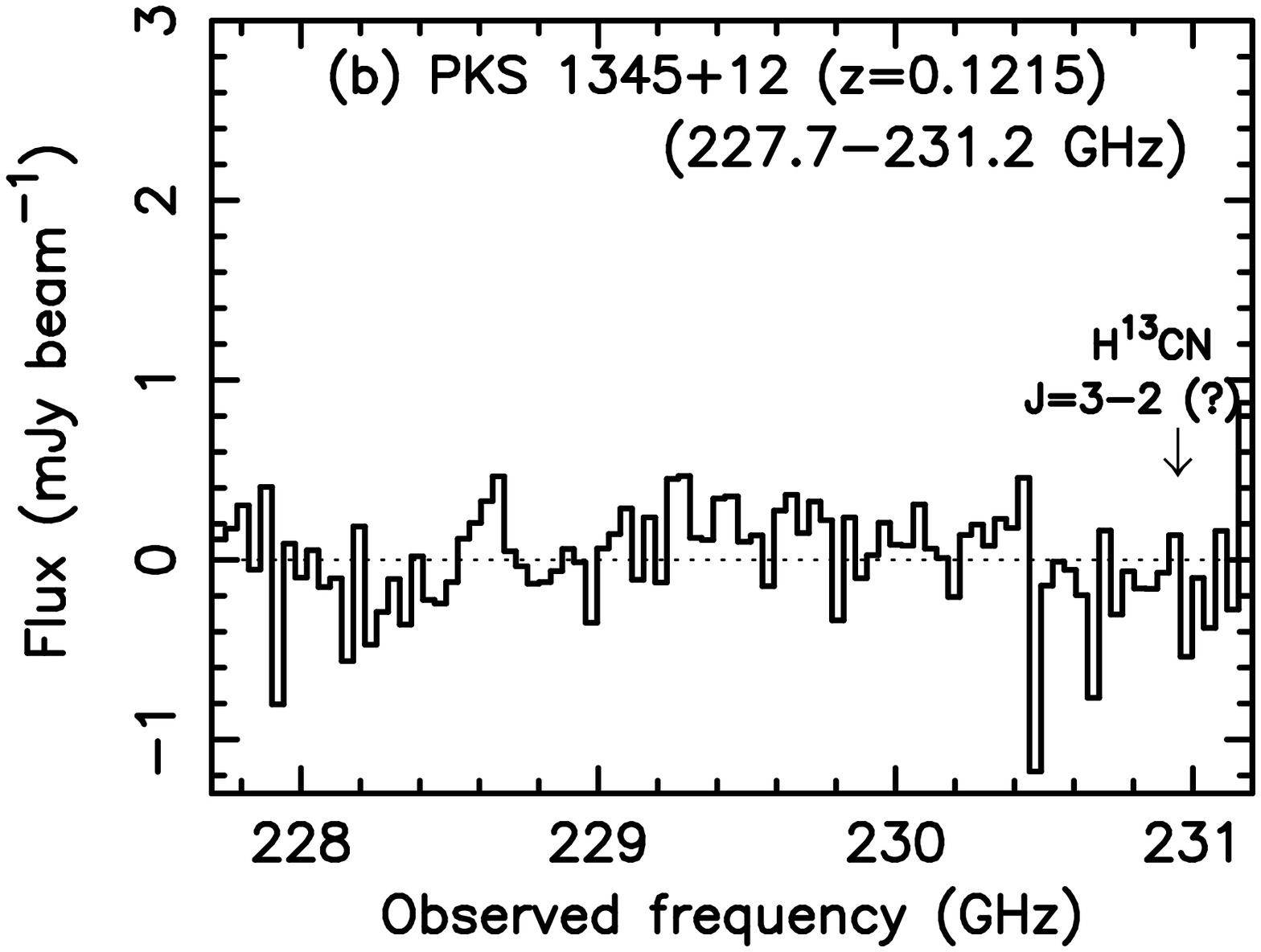}   
\includegraphics[angle=0,scale=.3]{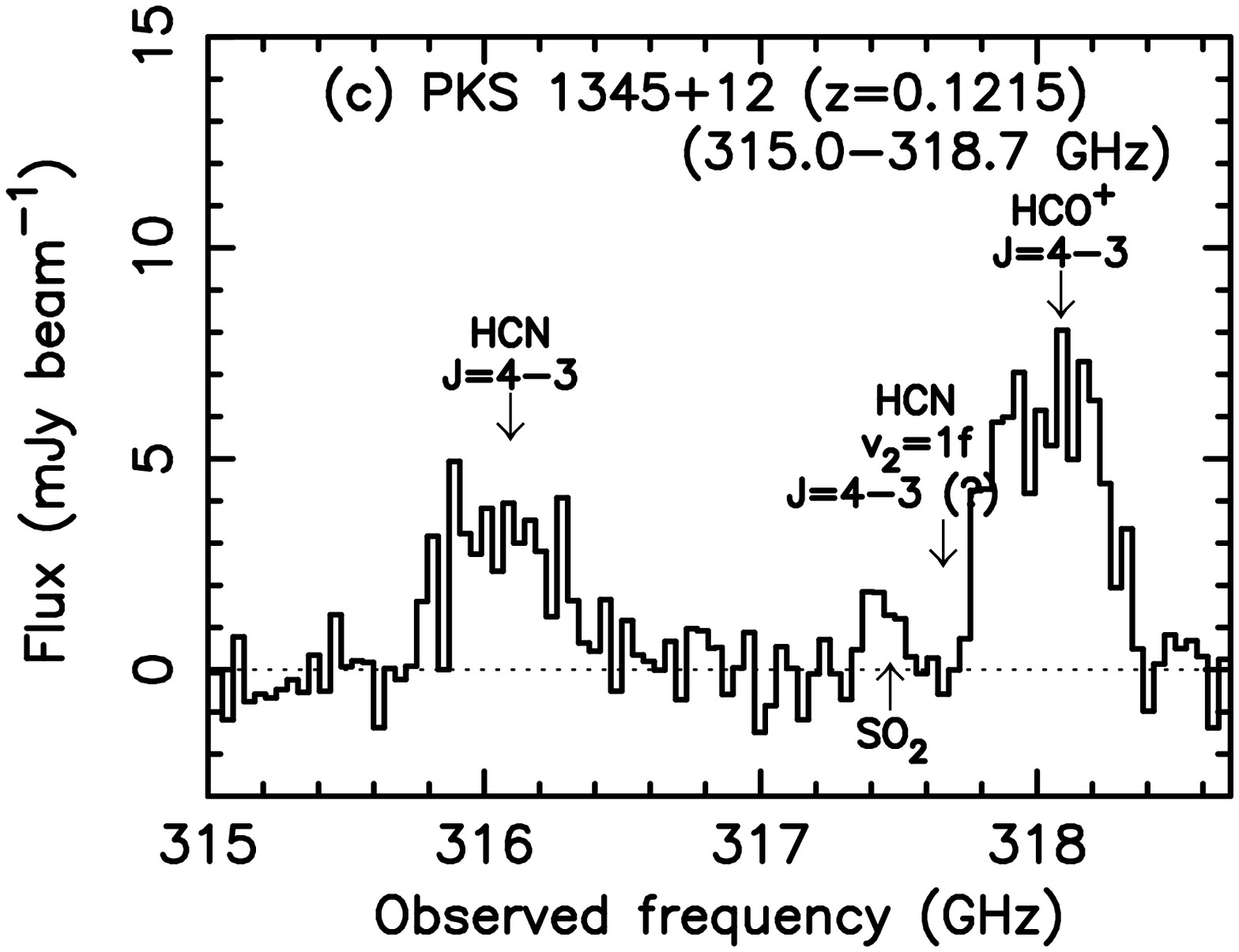}  \\
\vspace{-0.6cm}
\includegraphics[angle=0,scale=.3]{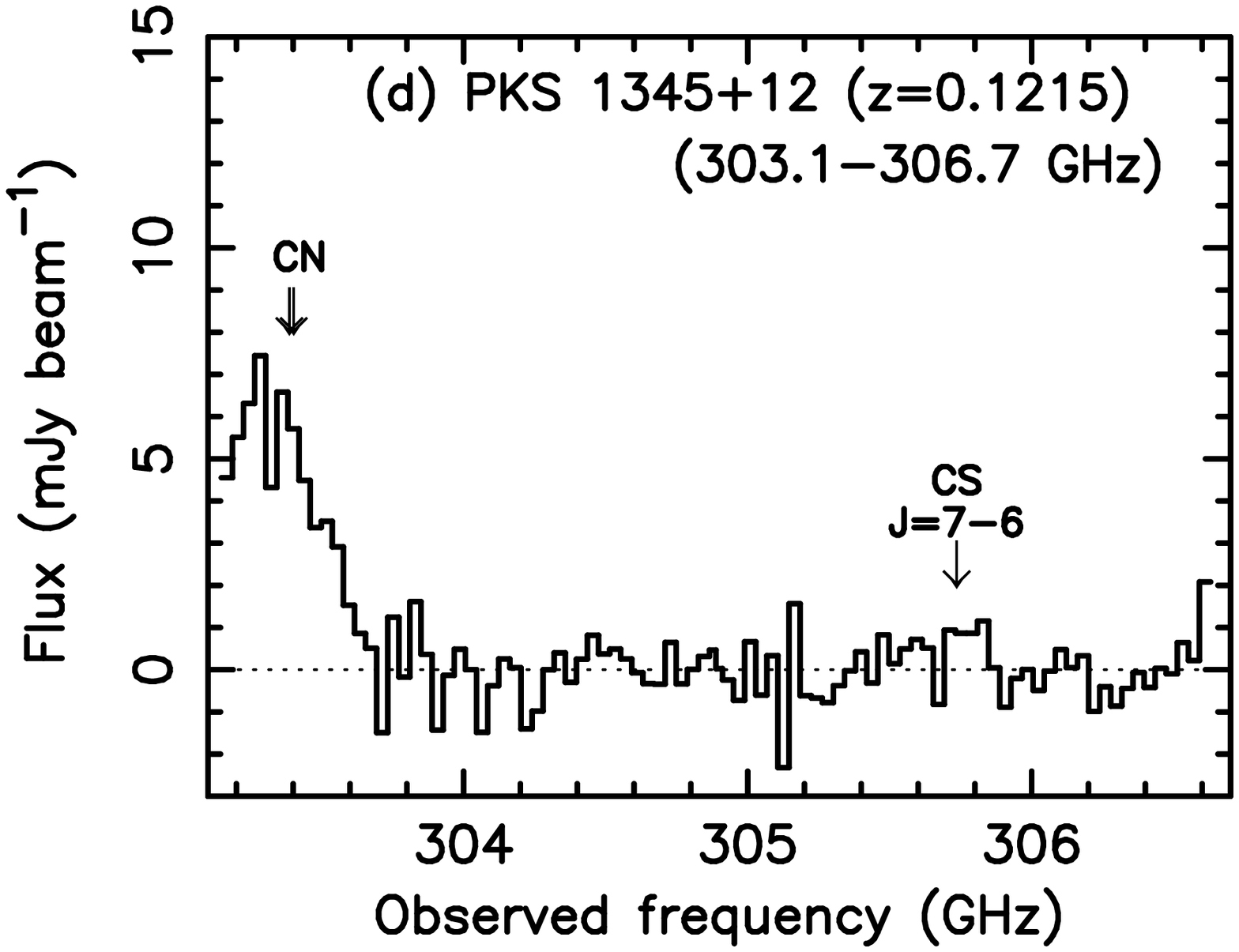} 
\end{center}
\caption{
ALMA spectra of PKS 1345$+$12.
In (c), an upward arrow is shown for SO$_{2}$ 15(7,9)--16(6,10)
($\nu_{\rm rest}$=356.041 GHz). 
In (d), downward arrows are shown  for CN N=3--2, J=7/2--5/2 
($\nu_{\rm rest}$=340.248--340.265 GHz).
}
\end{figure}

\begin{figure}
\vspace{-0.8cm}
\begin{center}
\includegraphics[angle=0,scale=.41]{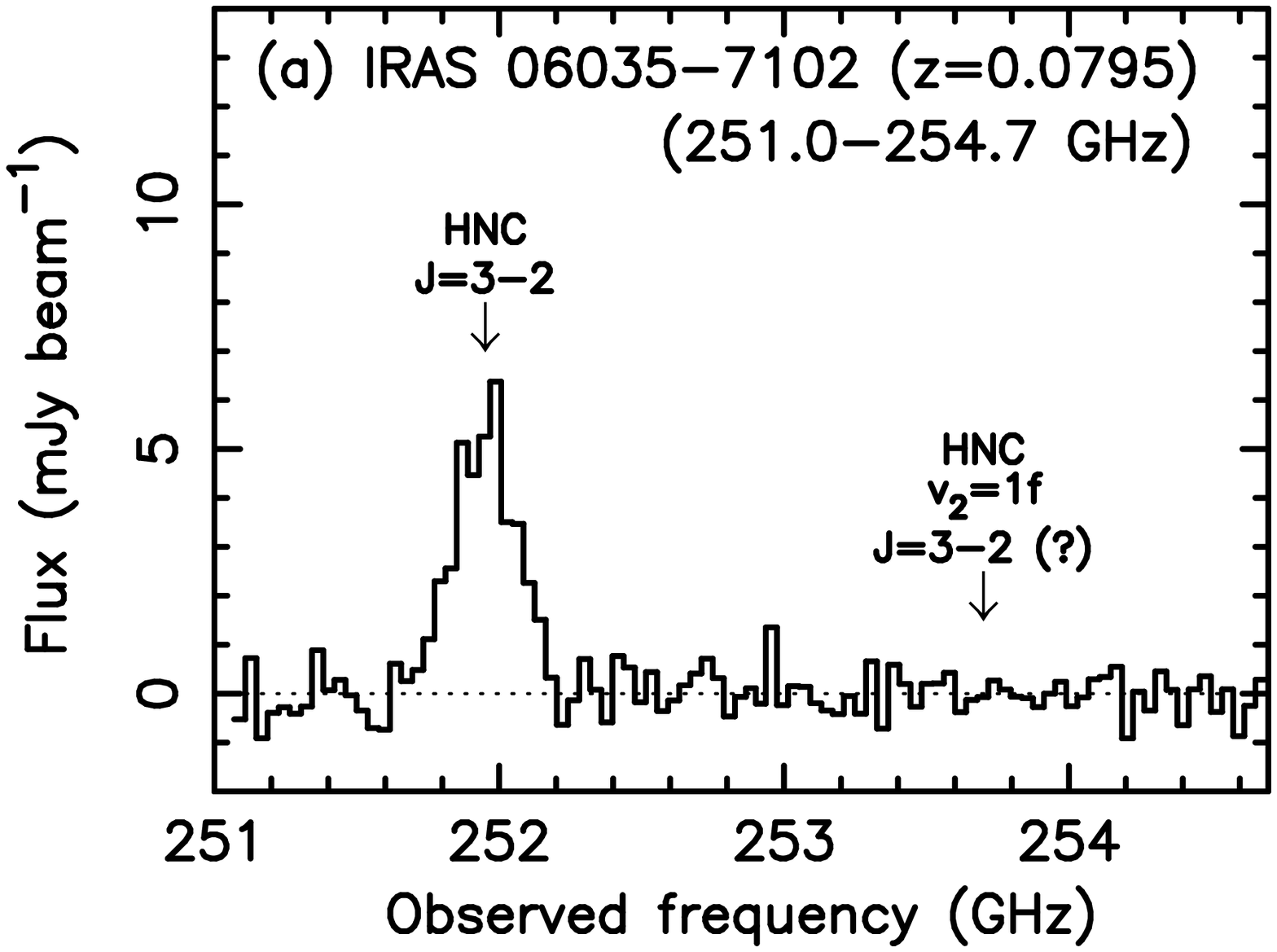}  
\includegraphics[angle=0,scale=.41]{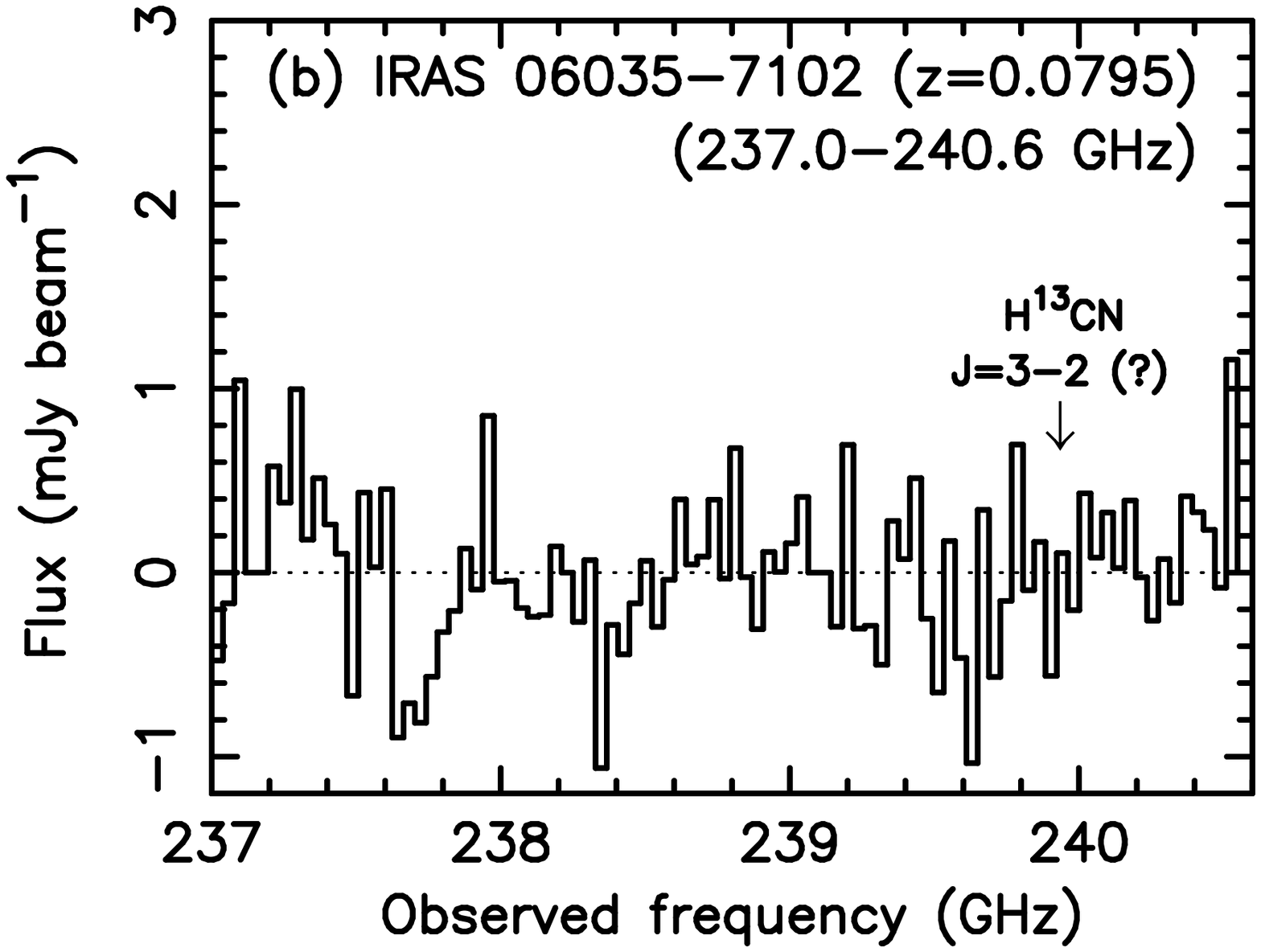} 
\end{center}
\caption{
ALMA spectra of IRAS 06035$-$7102.}
\end{figure}


\begin{figure}
\begin{center}
\includegraphics[angle=0,scale=.3]{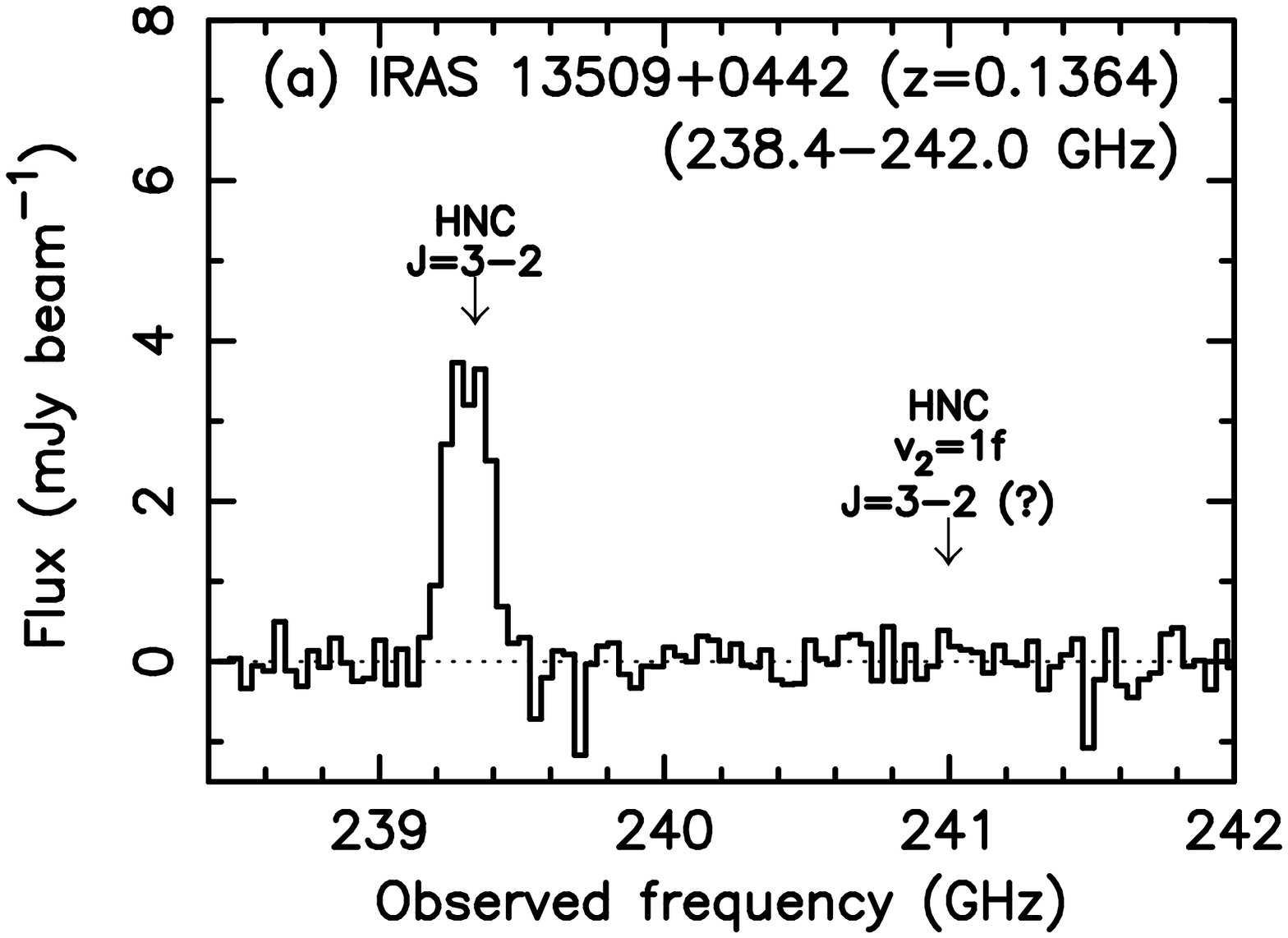} 
\includegraphics[angle=0,scale=.3]{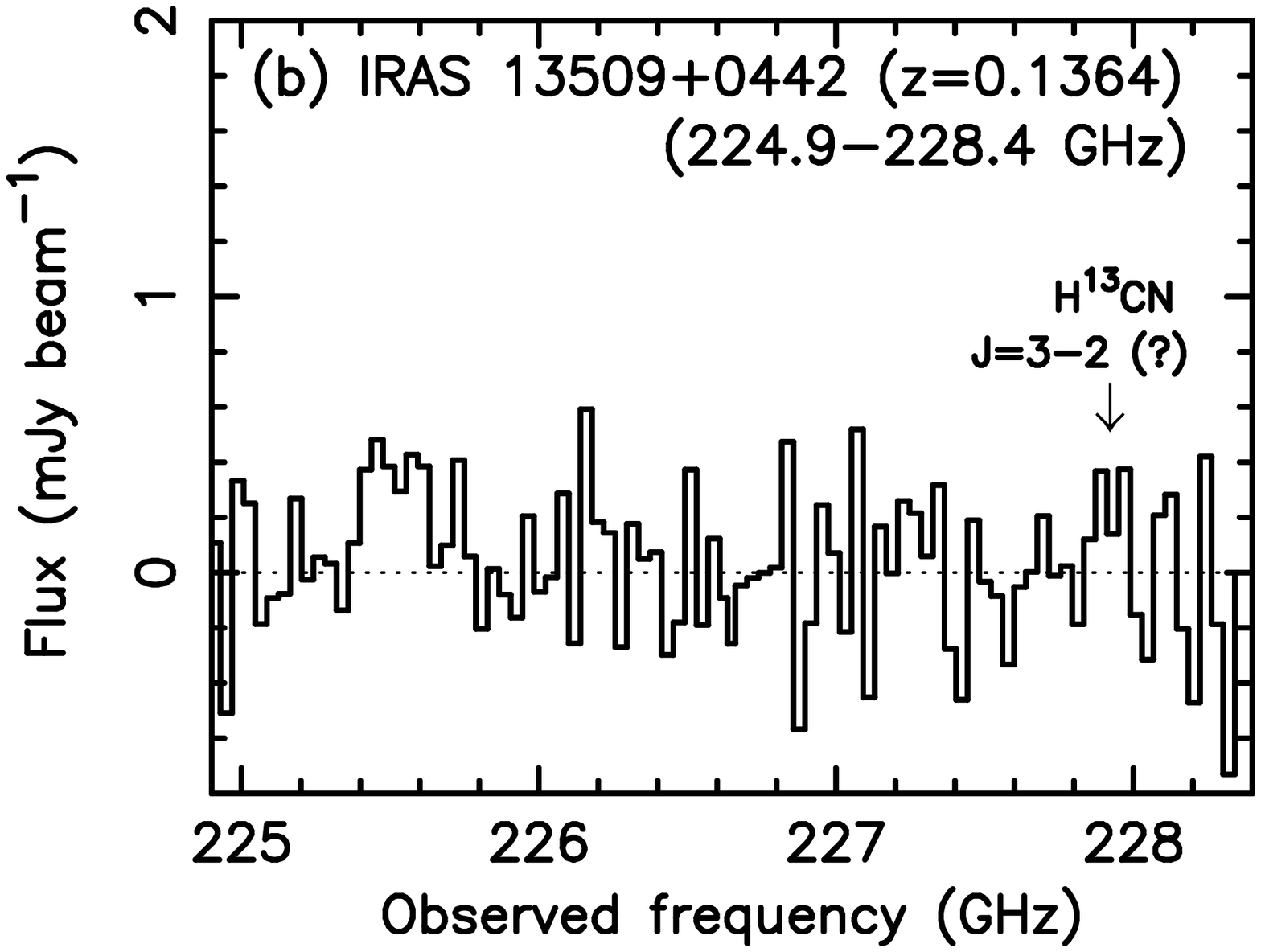} 
\includegraphics[angle=0,scale=.3]{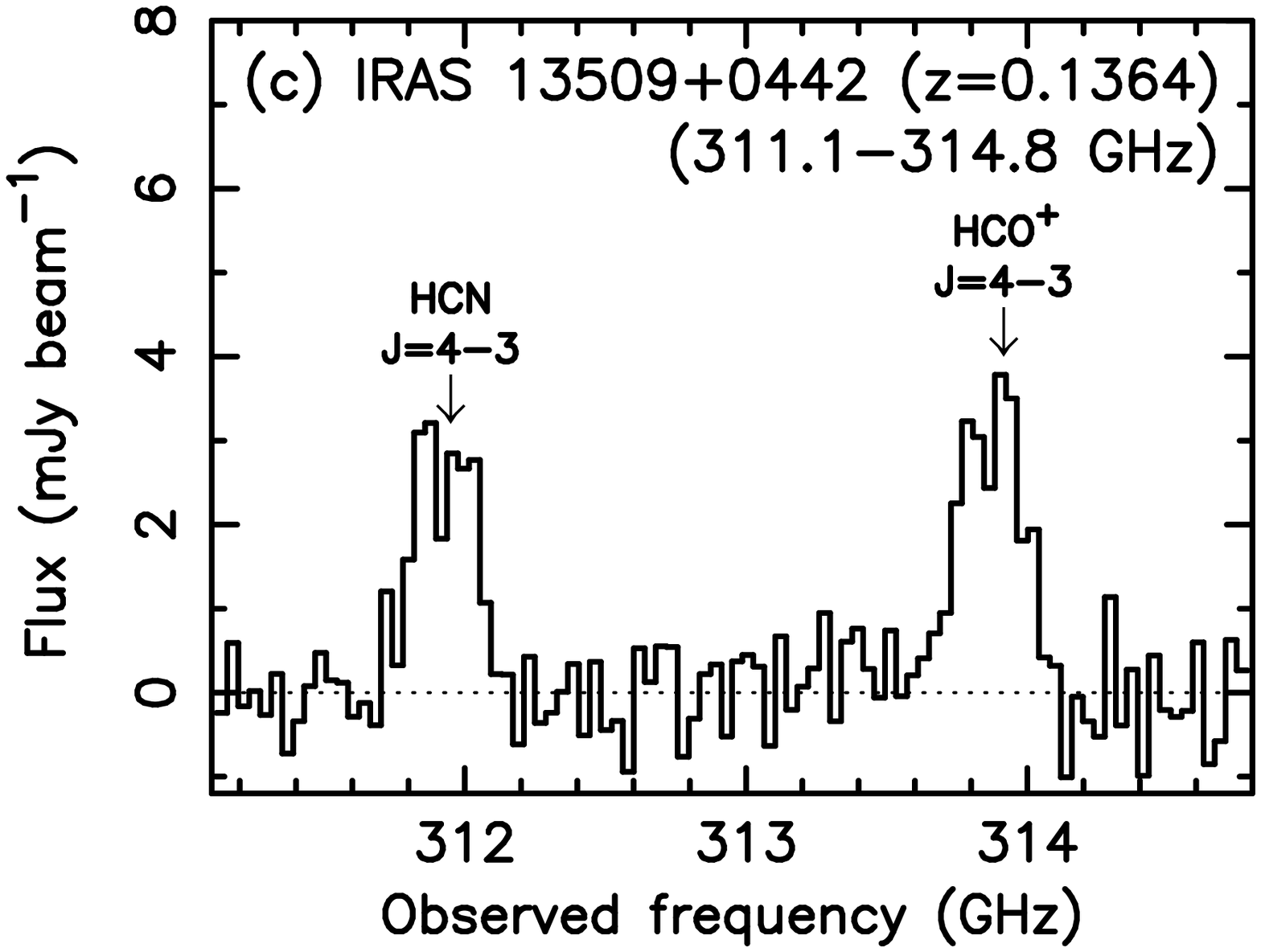} \\
\vspace{-0.5cm} 
\includegraphics[angle=0,scale=.3]{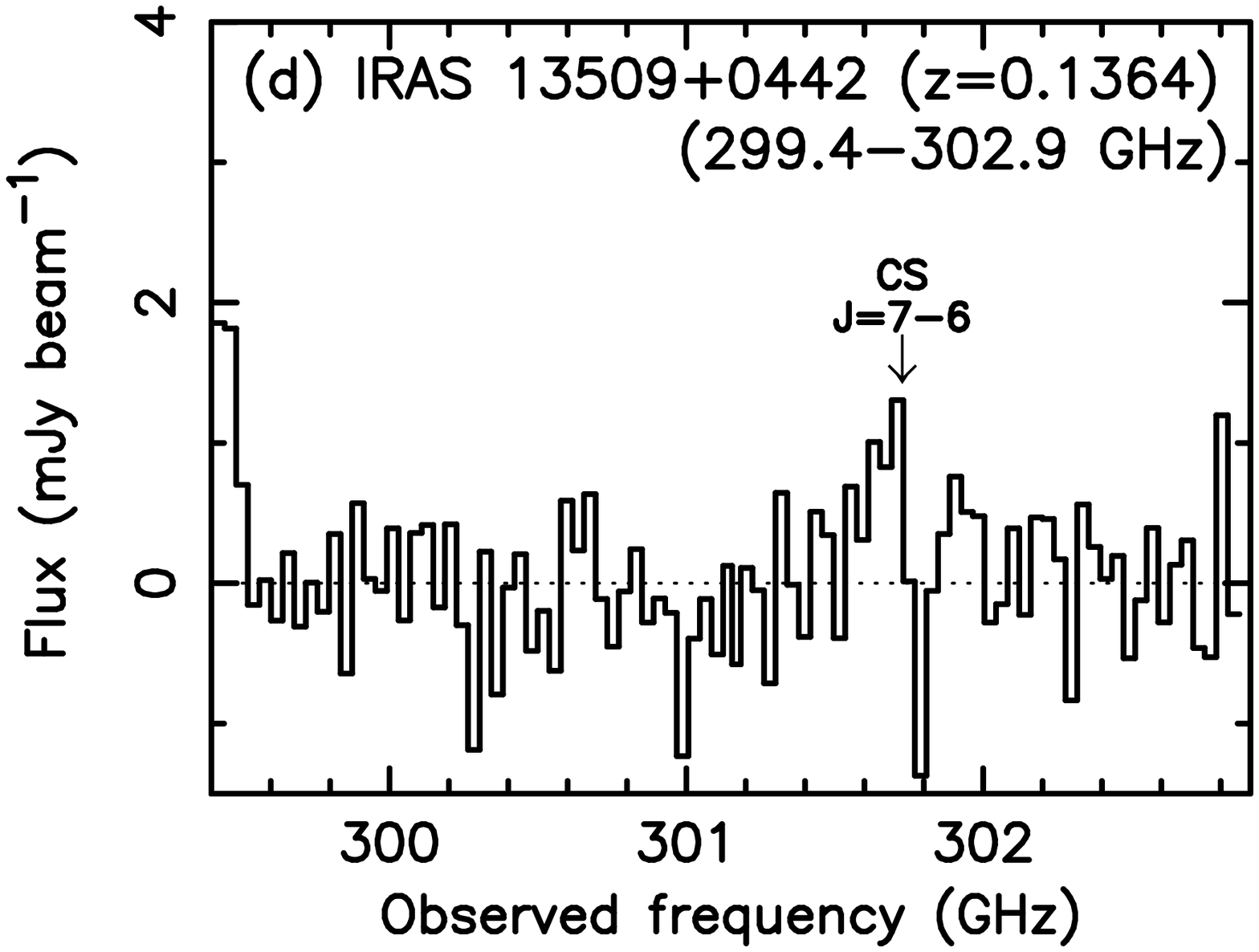}    
\includegraphics[angle=0,scale=.3]{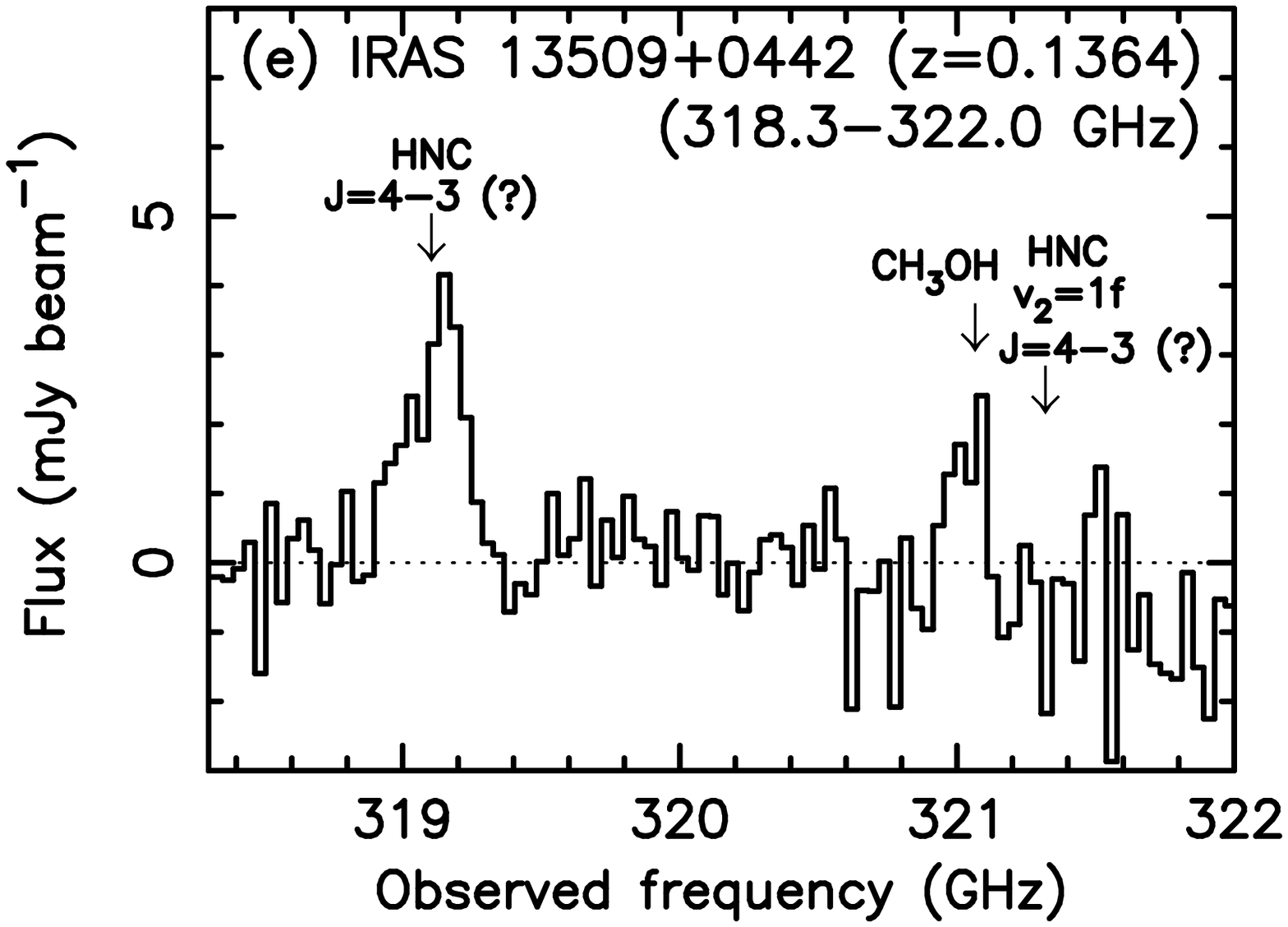}
\includegraphics[angle=0,scale=.3]{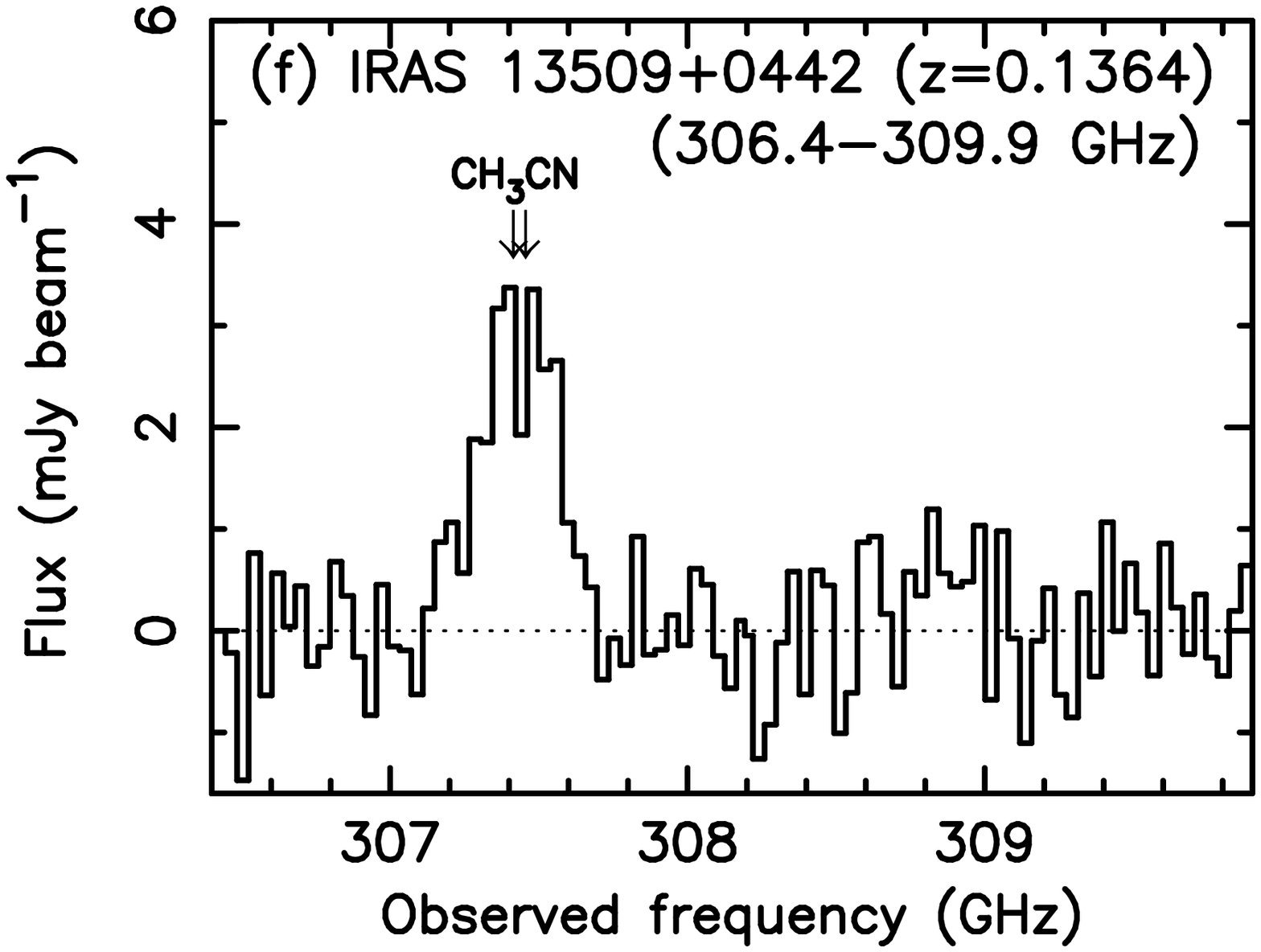}  \\   
\includegraphics[angle=0,scale=.3]{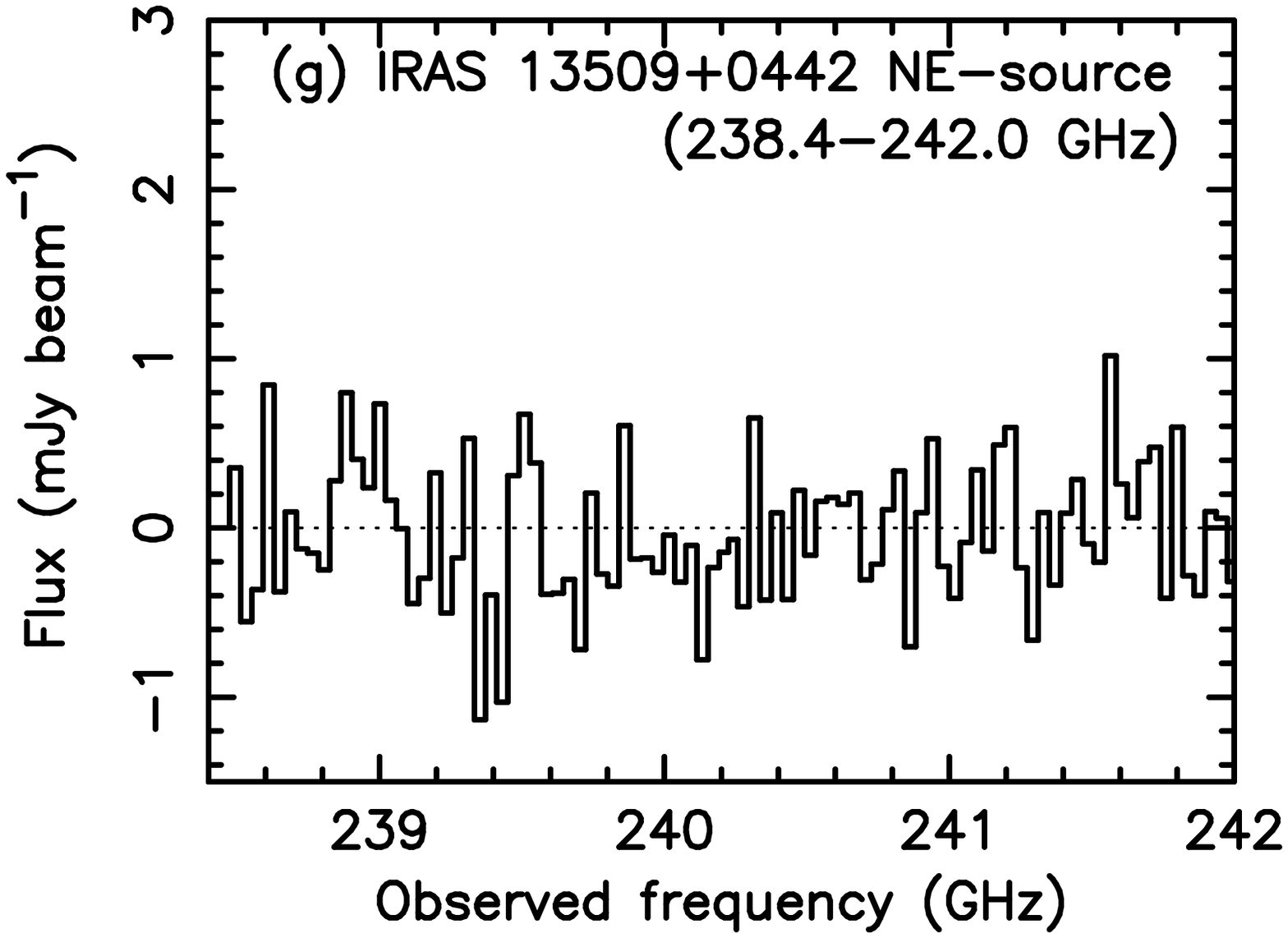} 
\includegraphics[angle=0,scale=.3]{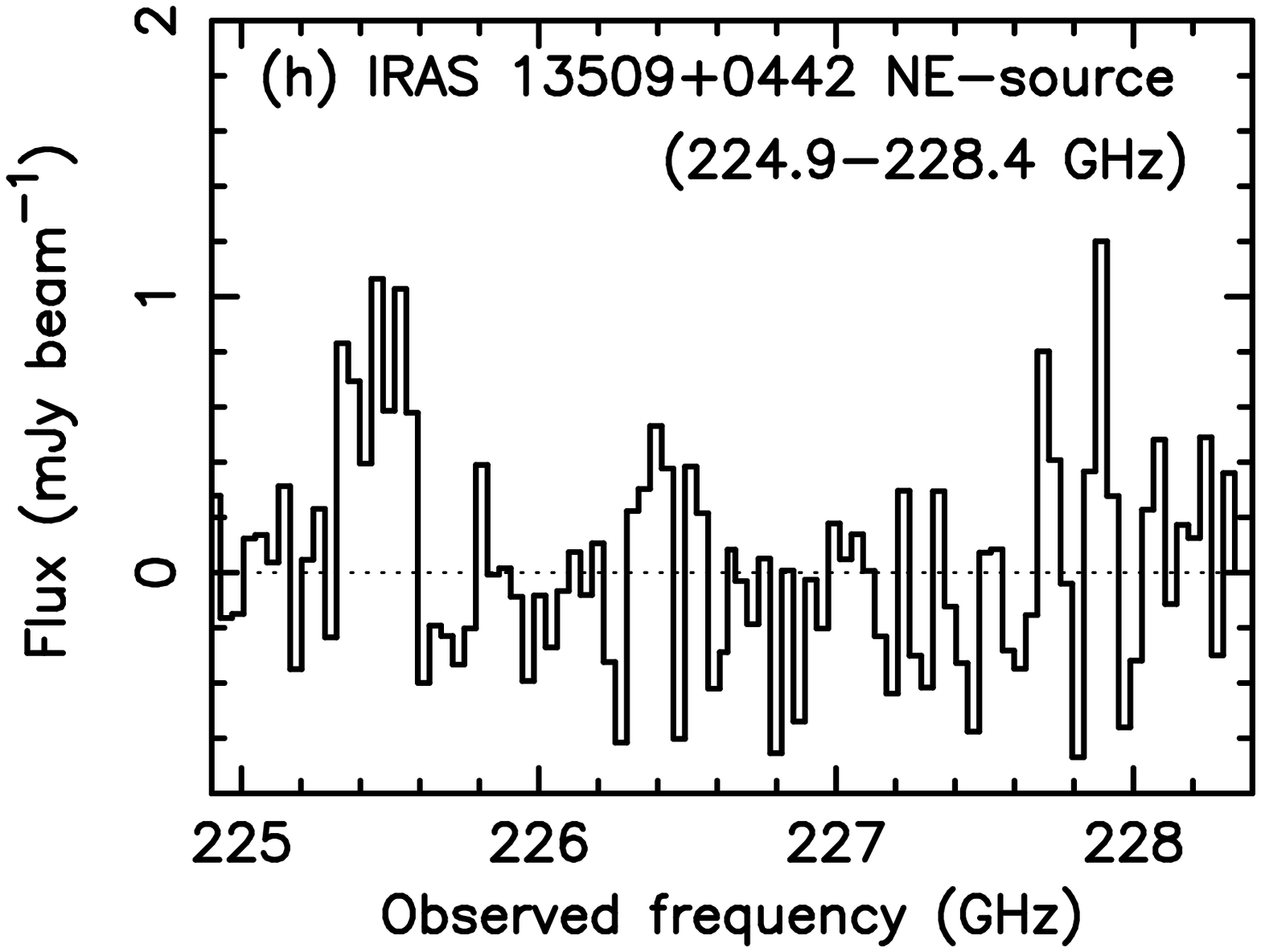} 
\includegraphics[angle=0,scale=.3]{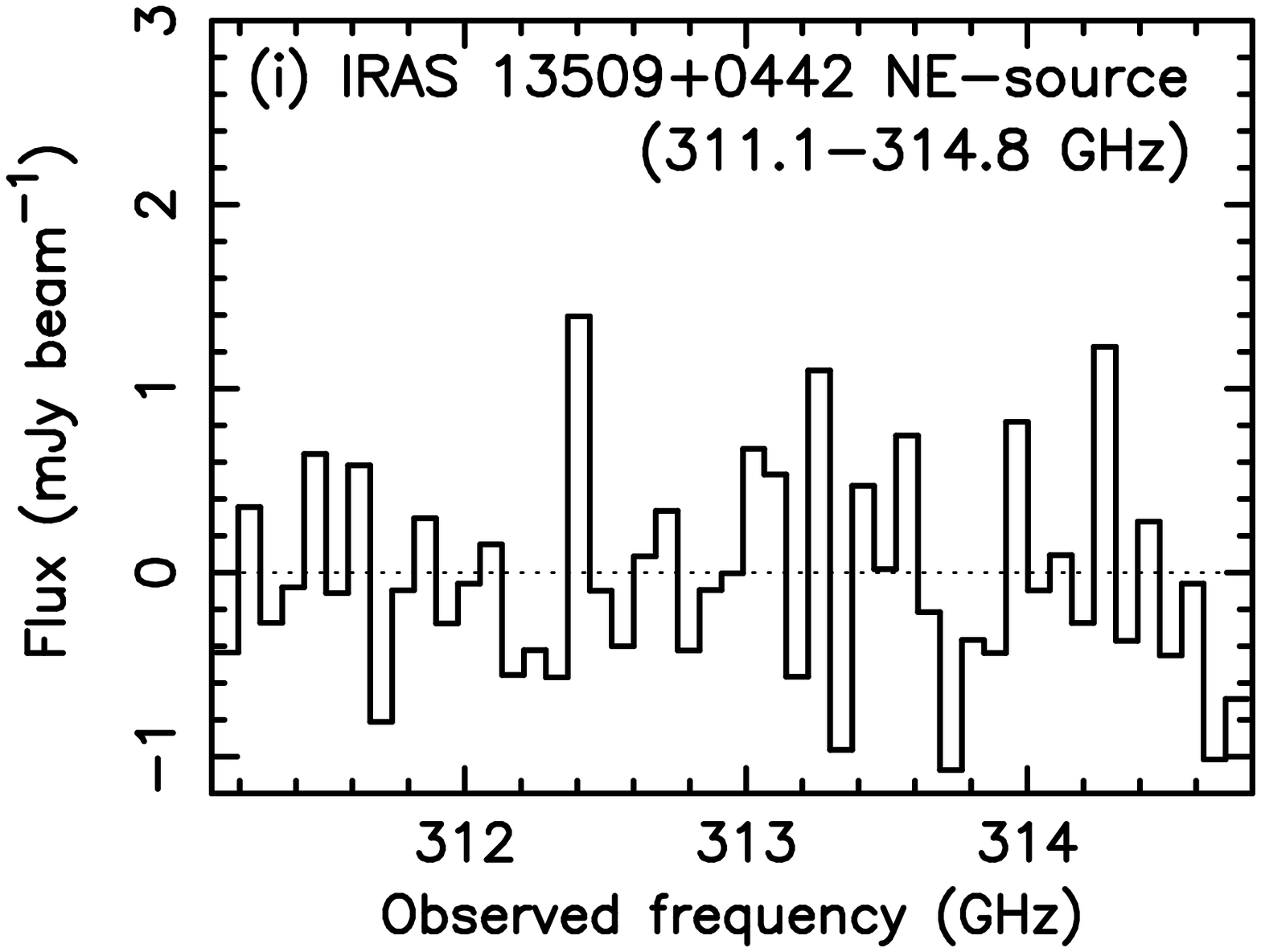} \\ 
\end{center}
\end{figure}

\begin{figure}
\begin{center}
\includegraphics[angle=0,scale=.3]{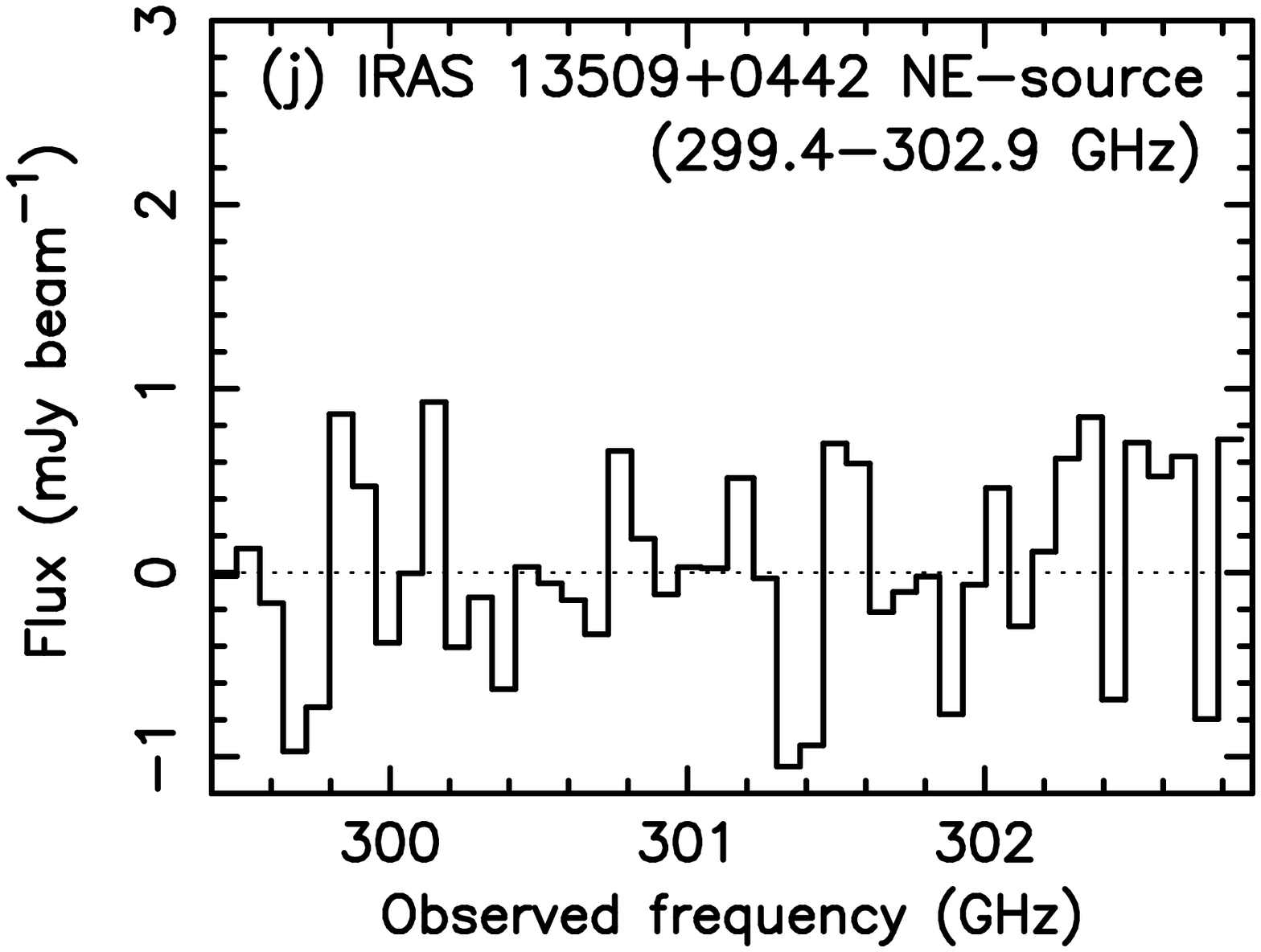}    
\includegraphics[angle=0,scale=.3]{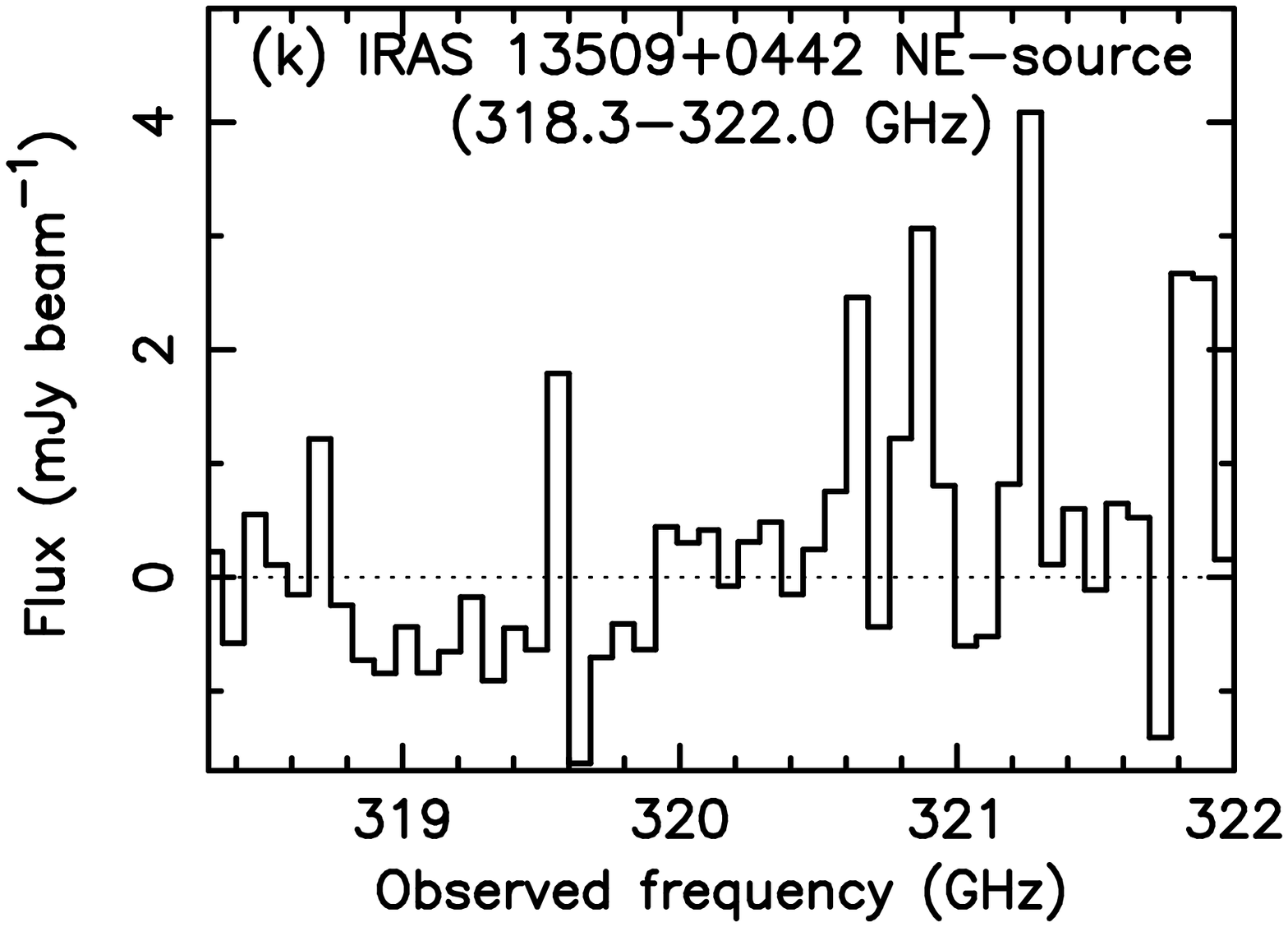}  
\includegraphics[angle=0,scale=.3]{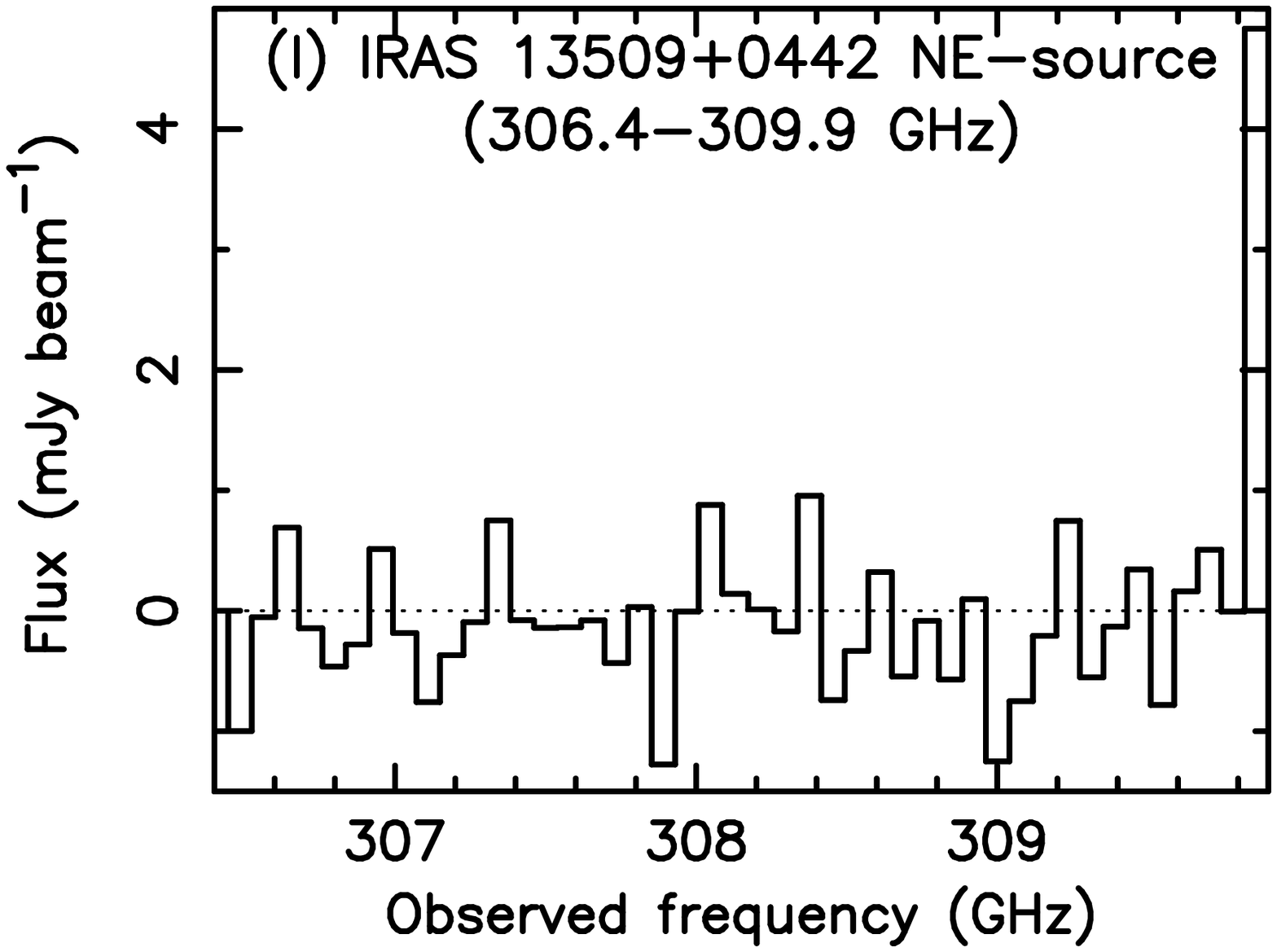} \\   
\end{center}
\caption{
ALMA spectra of IRAS 13509$+$0442.
Those of the ULIRG IRAS 13509$+$0442 itself are shown
first, and those of the north-eastern bright continuum emitting source 
(IRAS 13509$+$0442 NE) are shown later. 
In (e), a downward arrow is shown for CH$_{3}$OH 16(2,14)--16($-$1,16) 
($\nu_{\rm rest}$=364.859 GHz). 
In (f), downward arrows are shown for 
CH$_{3}$CN 19(4)--18(4) ($\nu_{\rm rest}$=349.346 GHz) and 
CH$_{3}$CN 19(3)--18(3) ($\nu_{\rm rest}$=349.393 GHz).
}
\end{figure}


\begin{figure}
\begin{center}
\vspace{-0.5cm}
\includegraphics[angle=0,scale=.41]{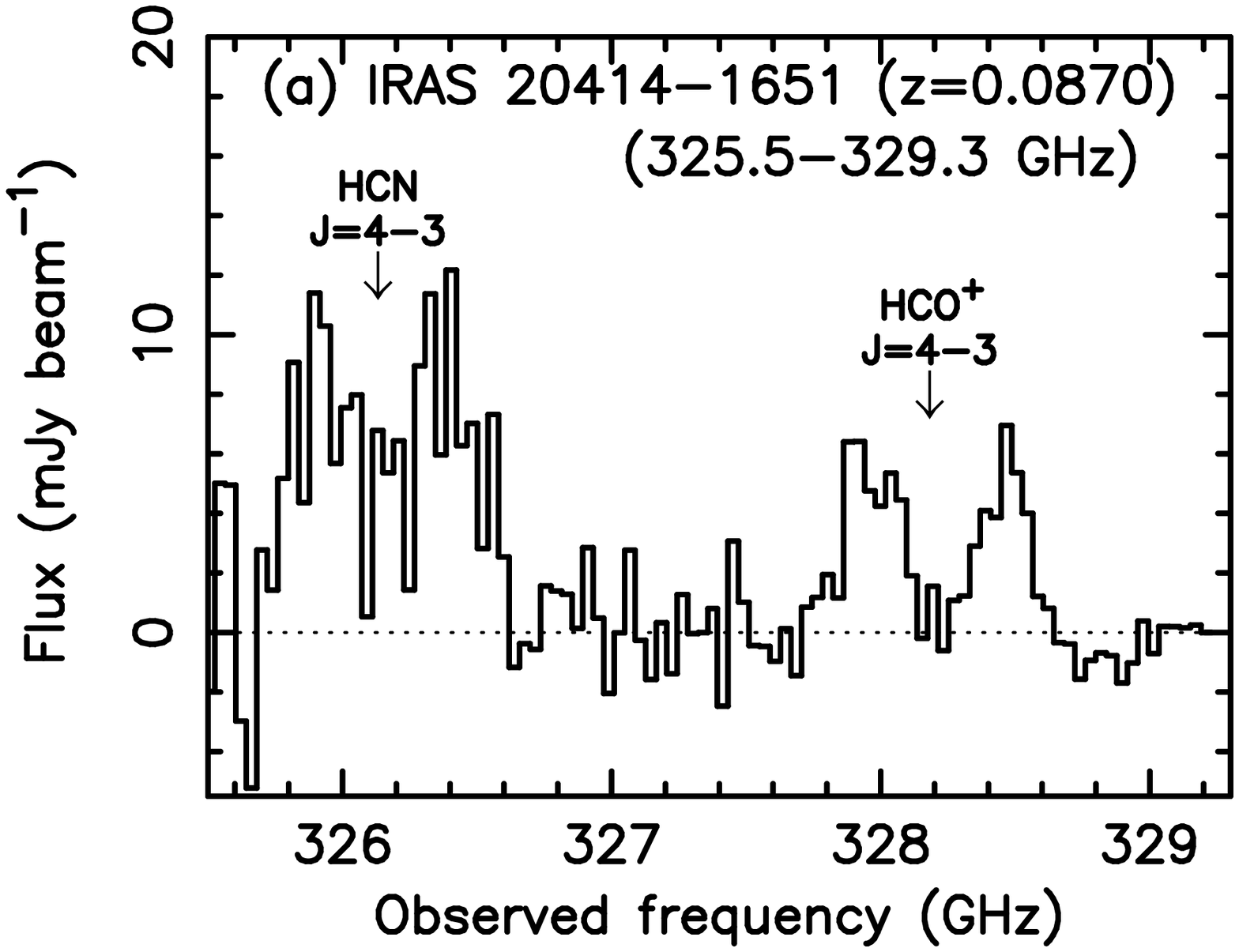}  
\includegraphics[angle=0,scale=.41]{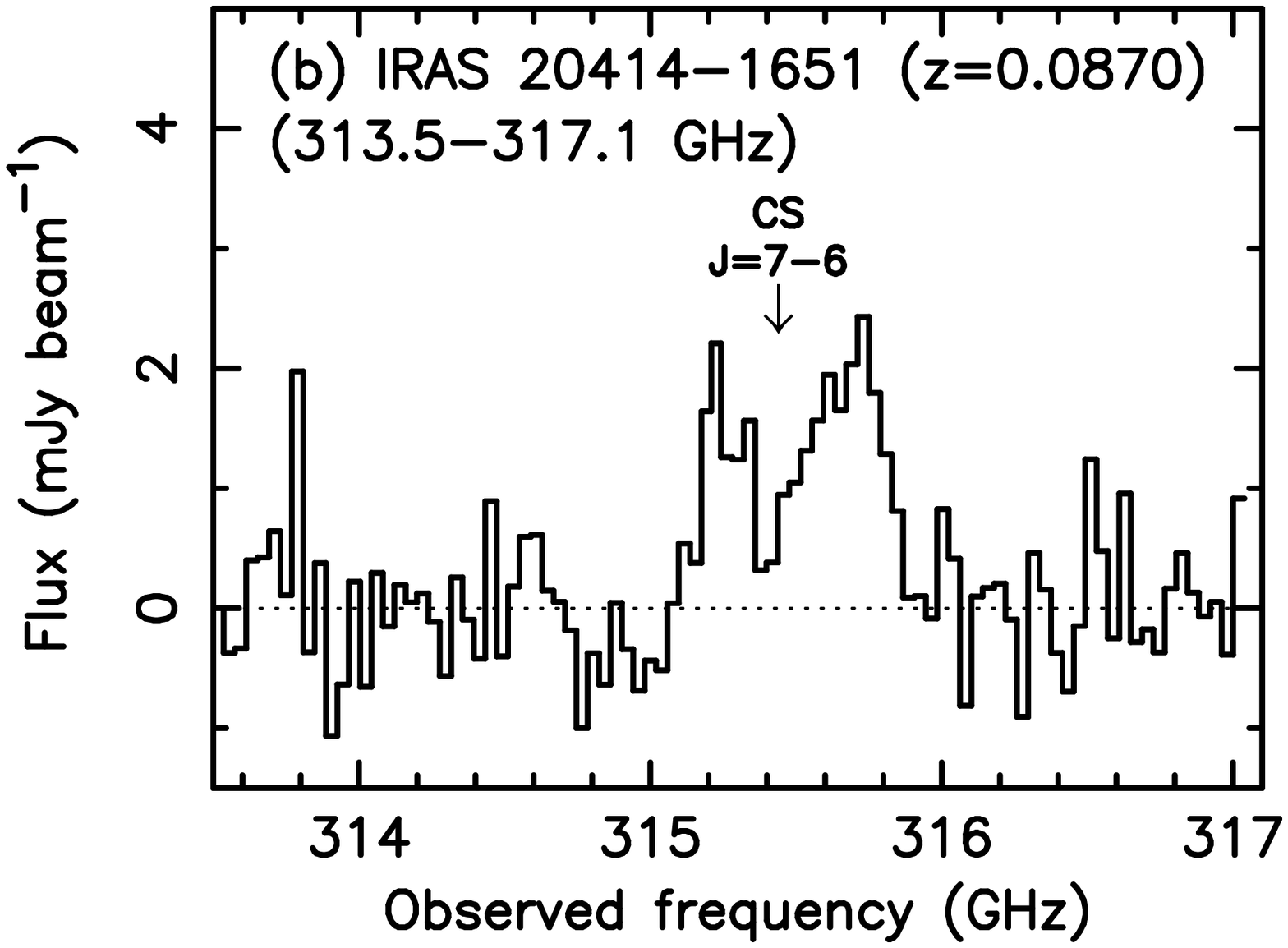}    
\includegraphics[angle=0,scale=.41]{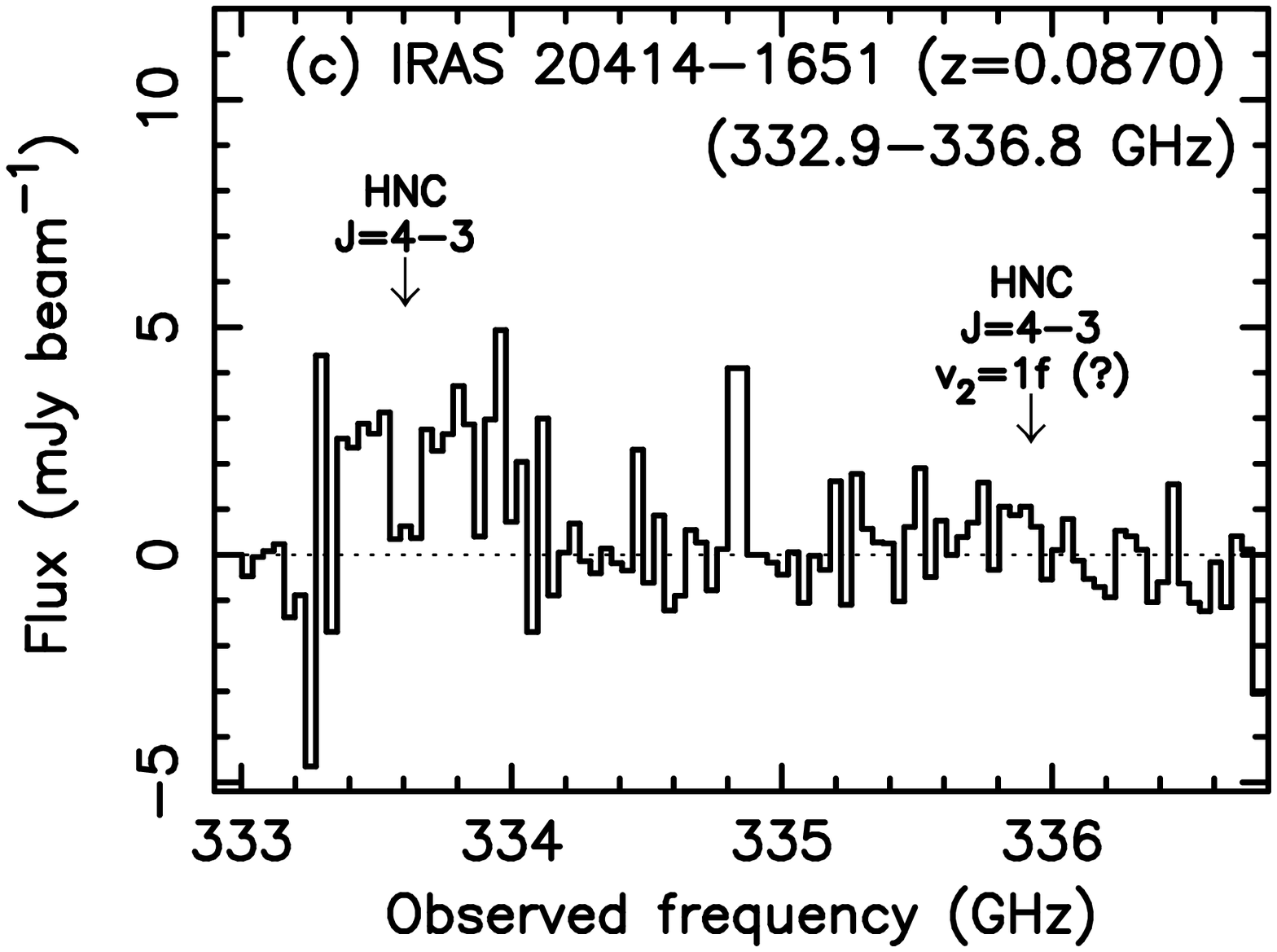}  
\includegraphics[angle=0,scale=.41]{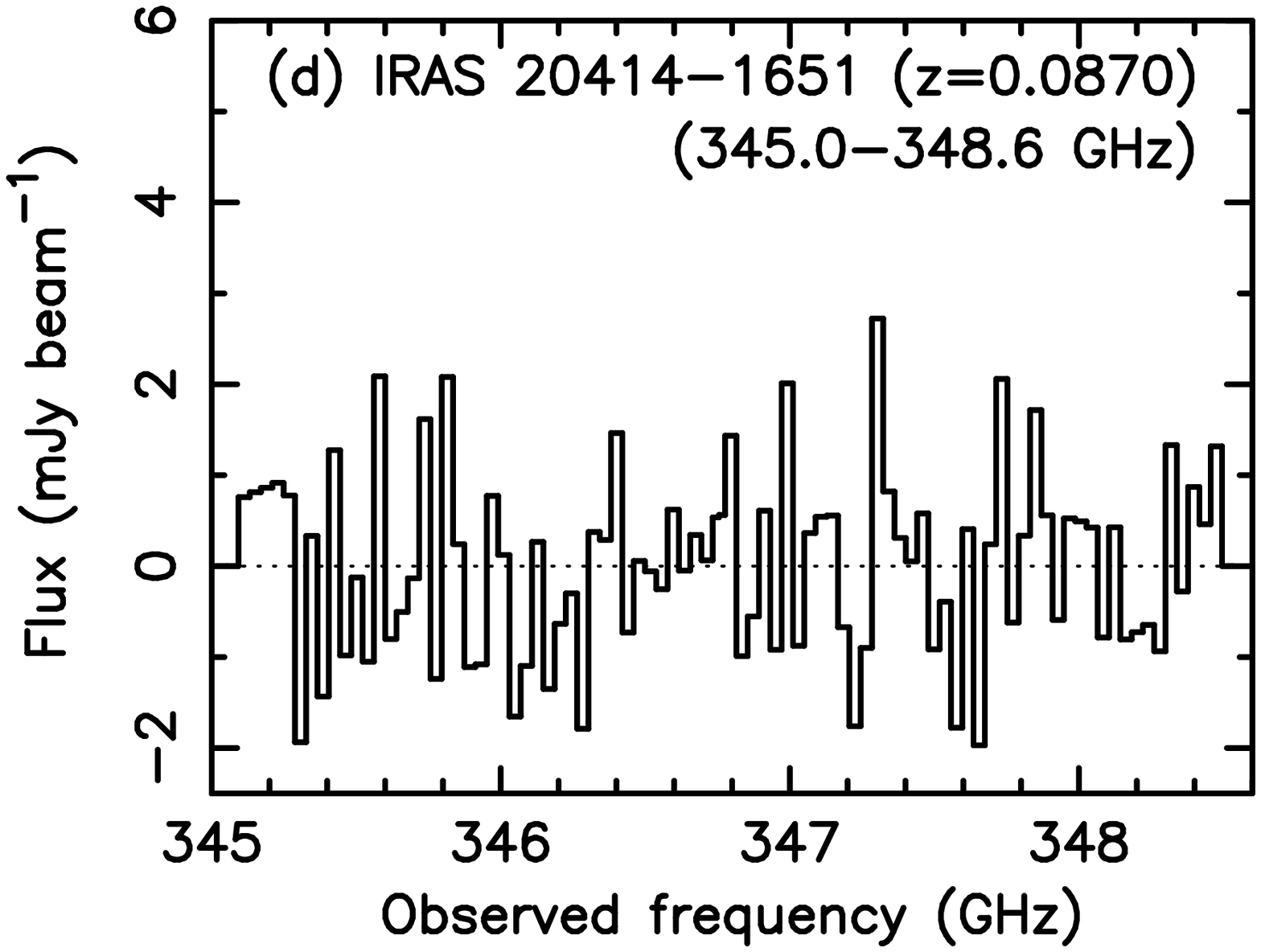} \\   
\end{center}
\caption{
ALMA spectra of IRAS 20414$-$1651. 
}
\end{figure}

\begin{figure}
\begin{center}
\includegraphics[angle=0,scale=.41]{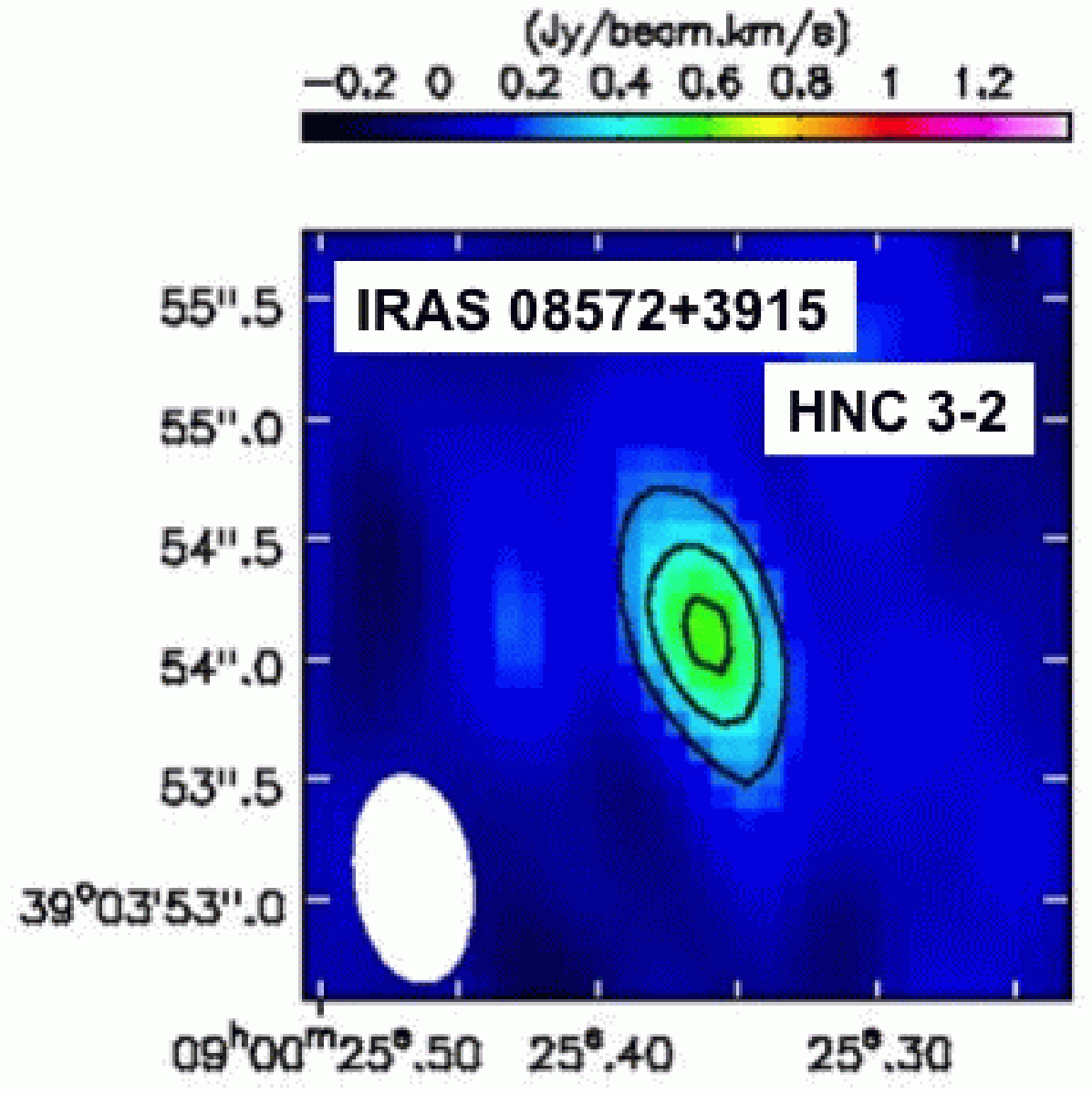} 
\includegraphics[angle=0,scale=.41]{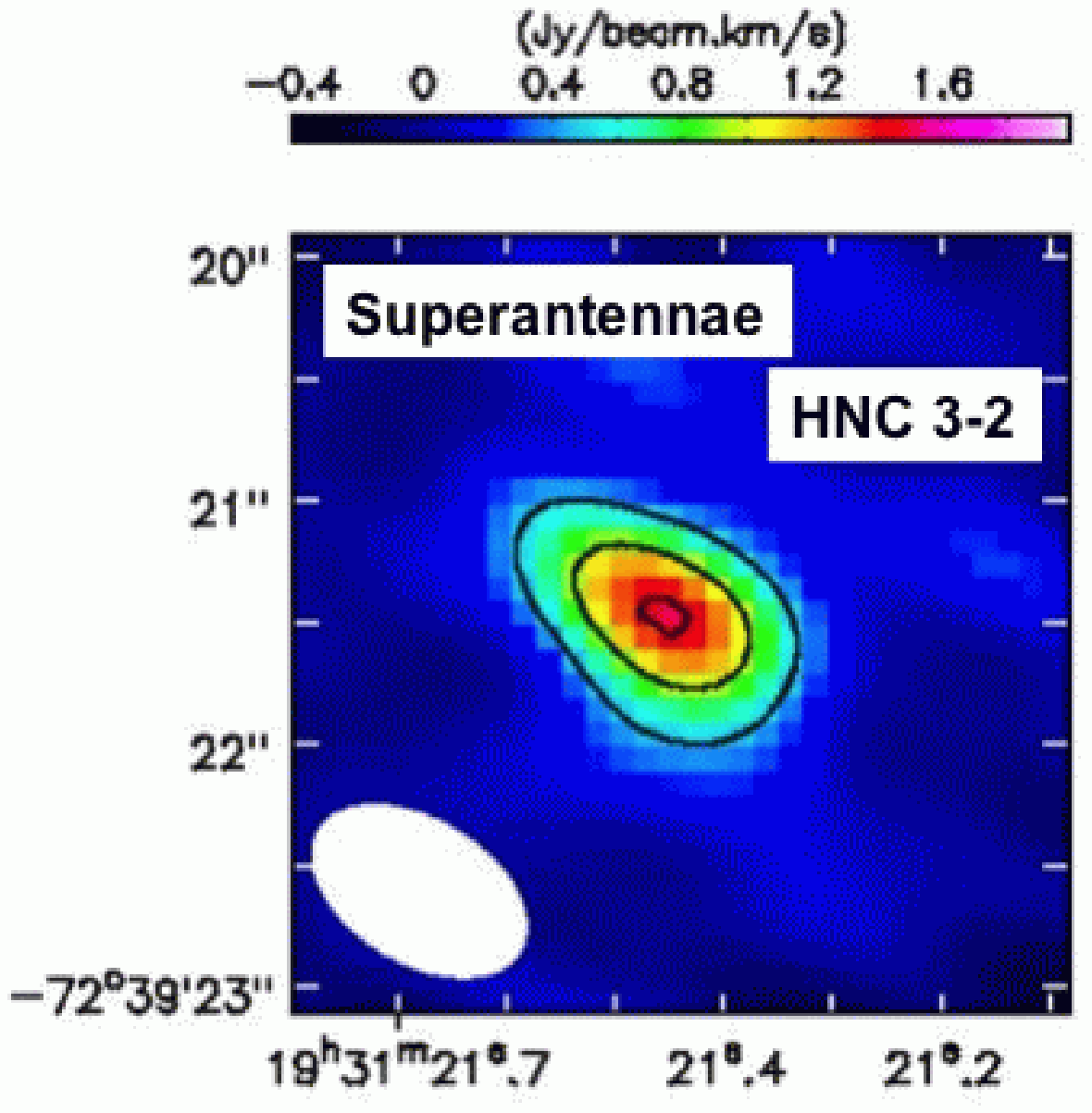} 
\includegraphics[angle=0,scale=.41]{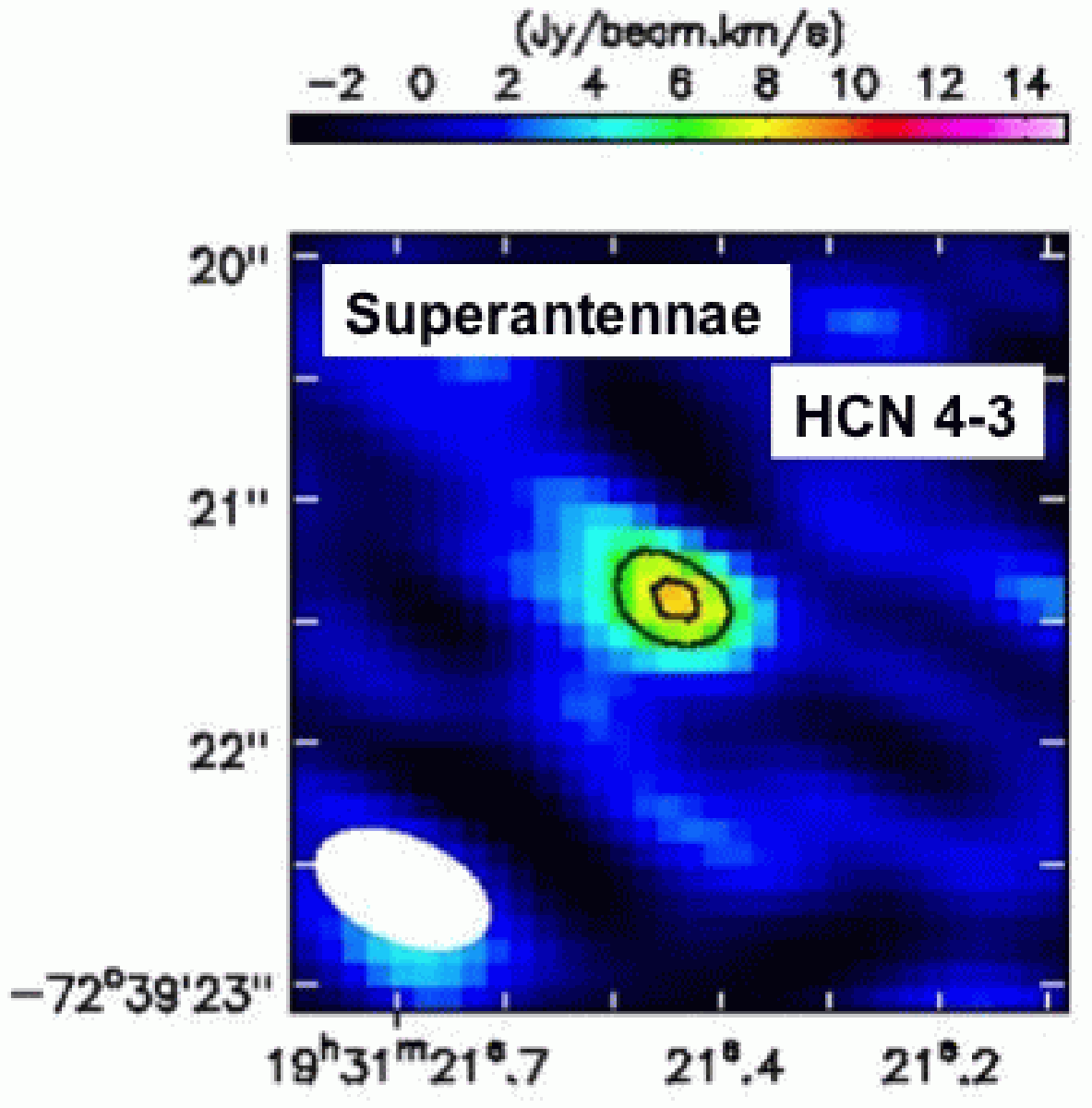} \\
\vspace{-1.3cm}
\includegraphics[angle=0,scale=.41]{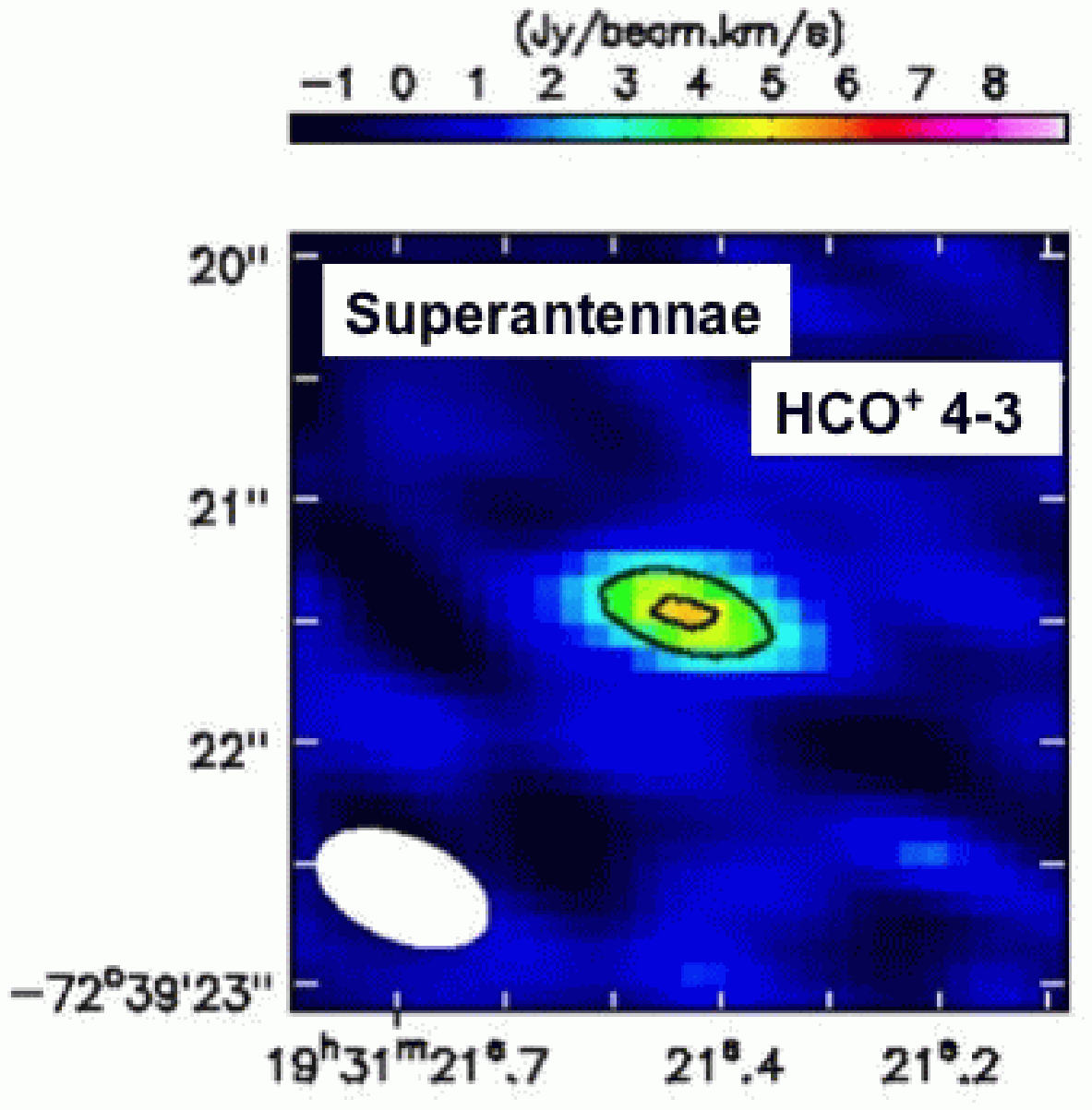} 
\includegraphics[angle=0,scale=.41]{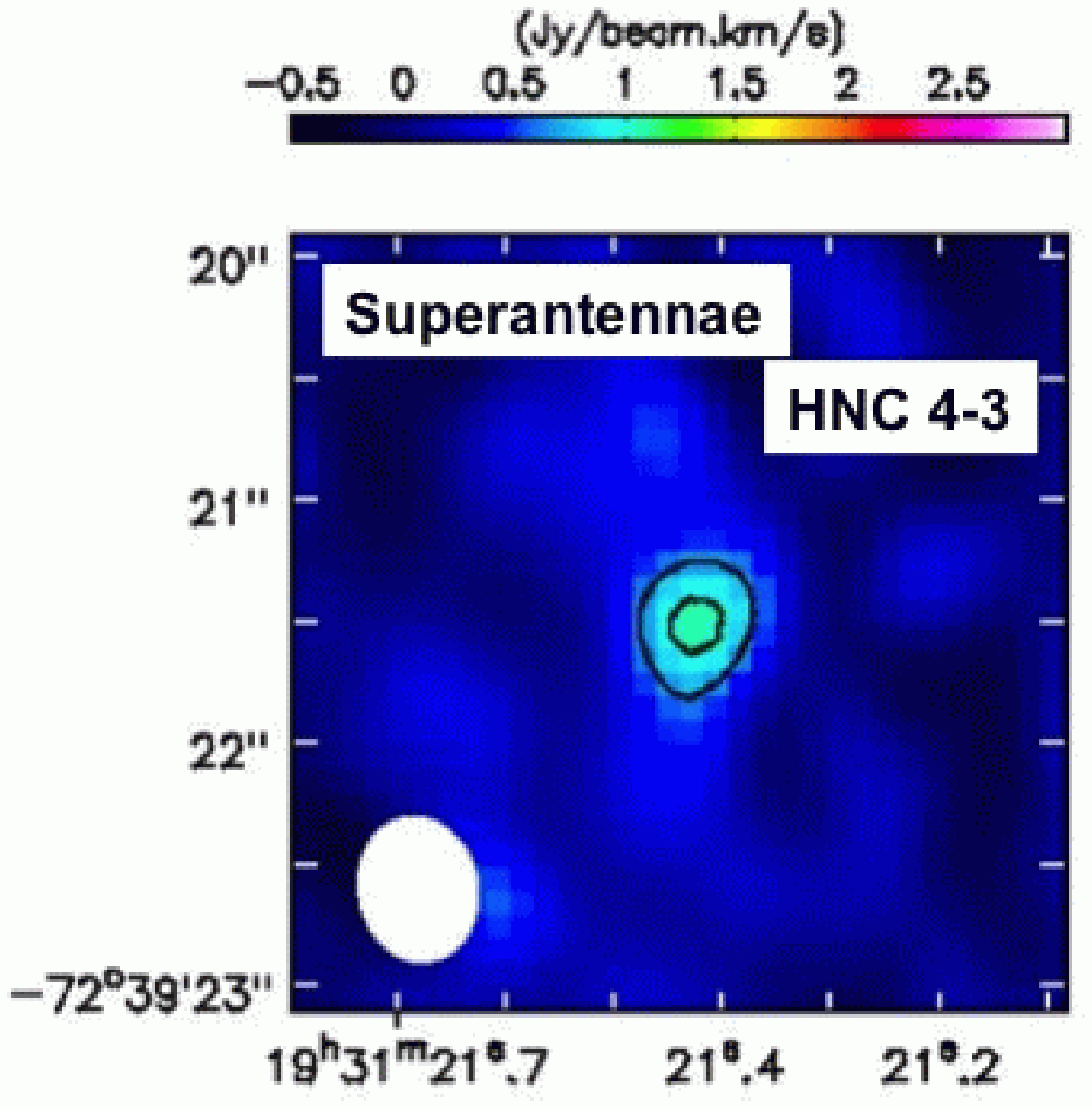} 
\includegraphics[angle=0,scale=.41]{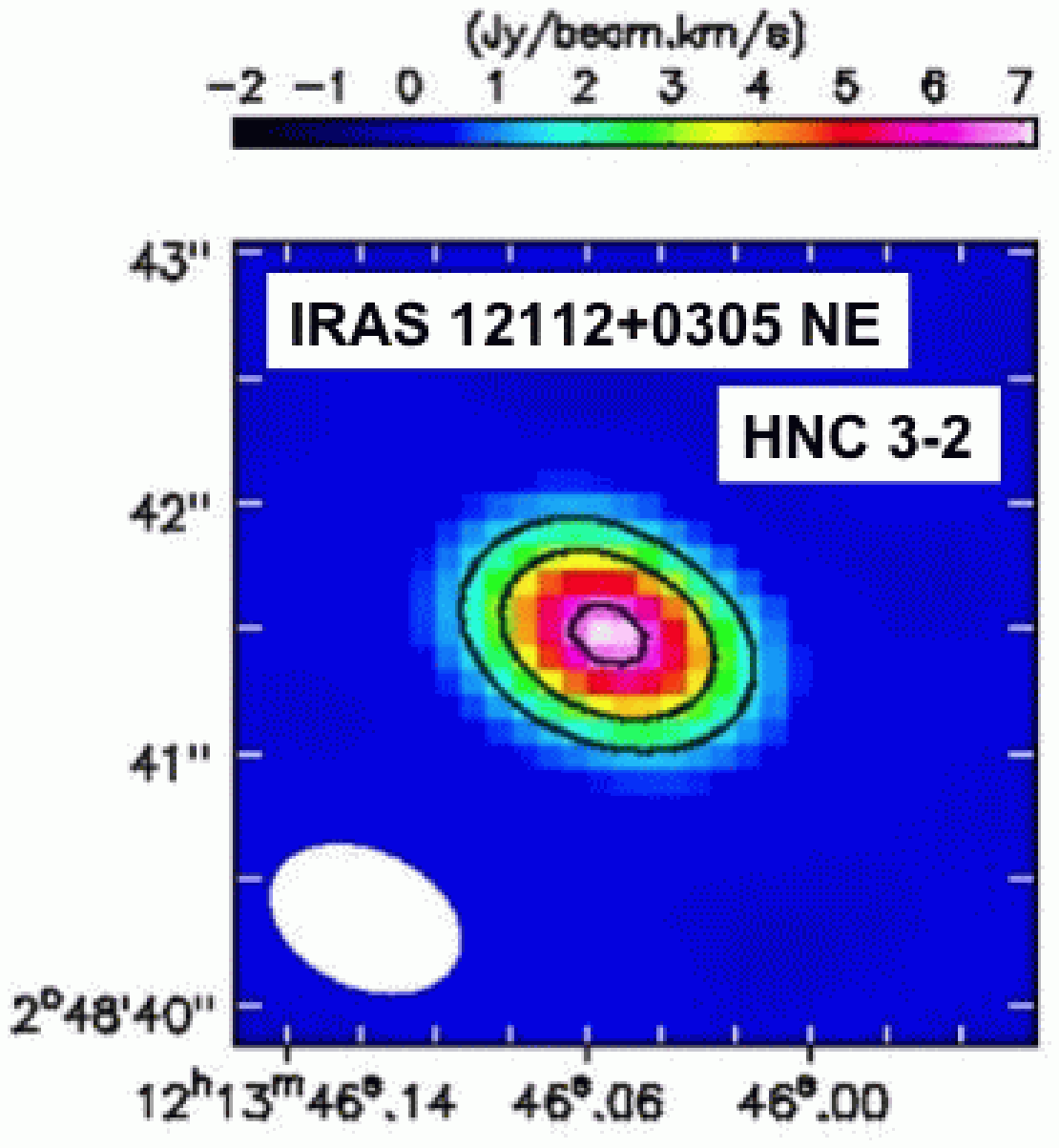} \\
\vspace{-1.3cm}
\includegraphics[angle=0,scale=.41]{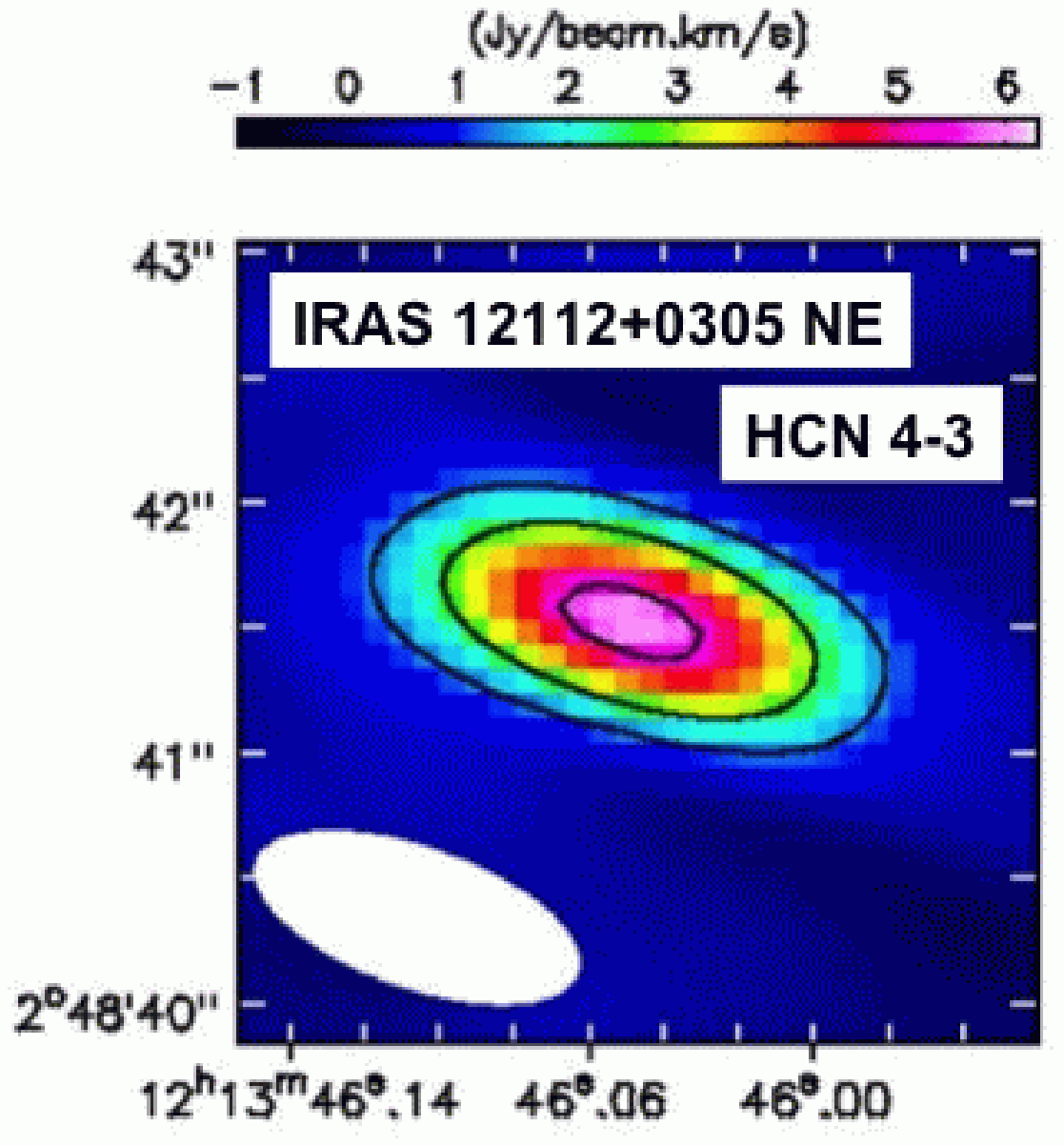} 
\includegraphics[angle=0,scale=.41]{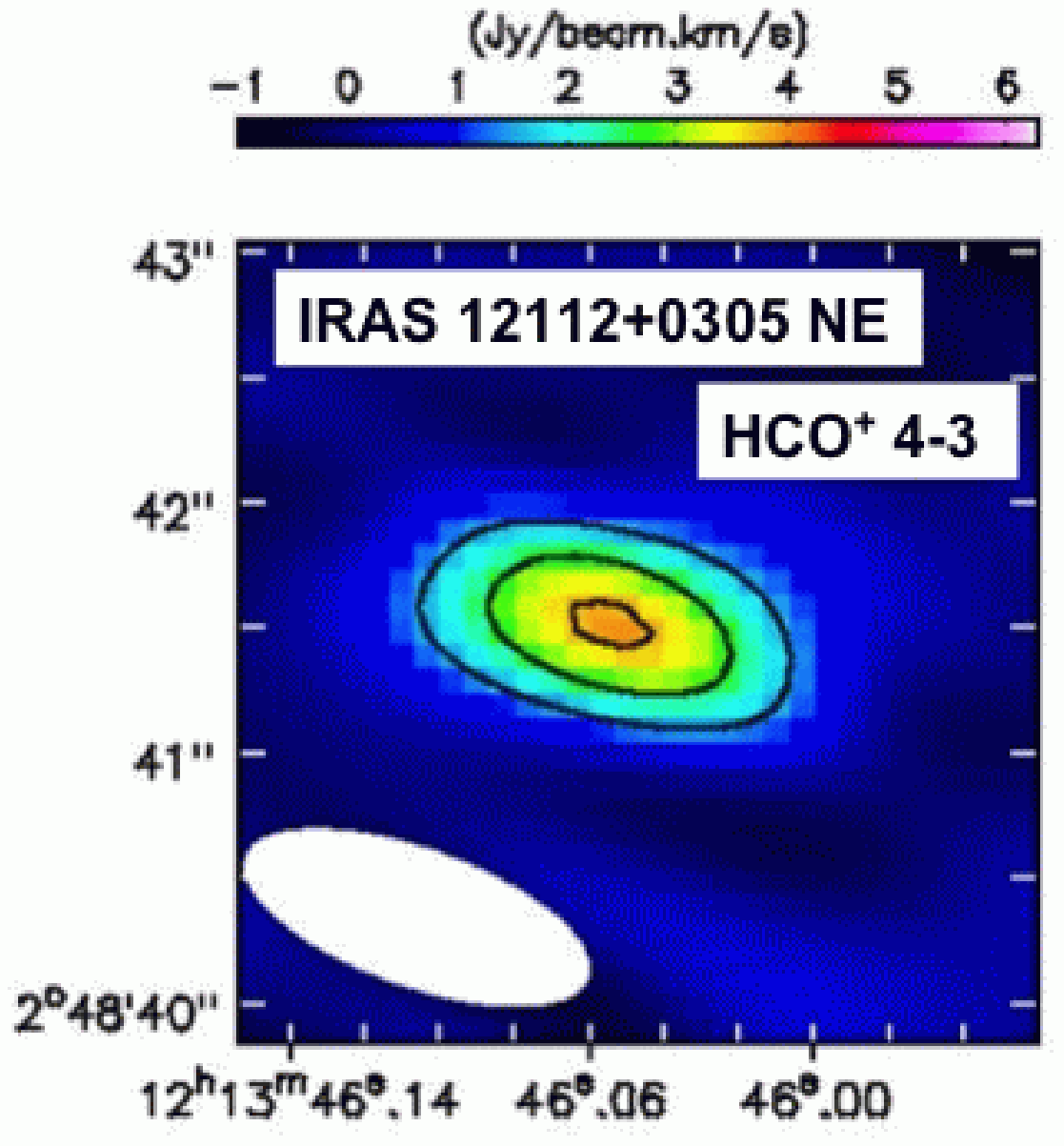} 
\includegraphics[angle=0,scale=.41]{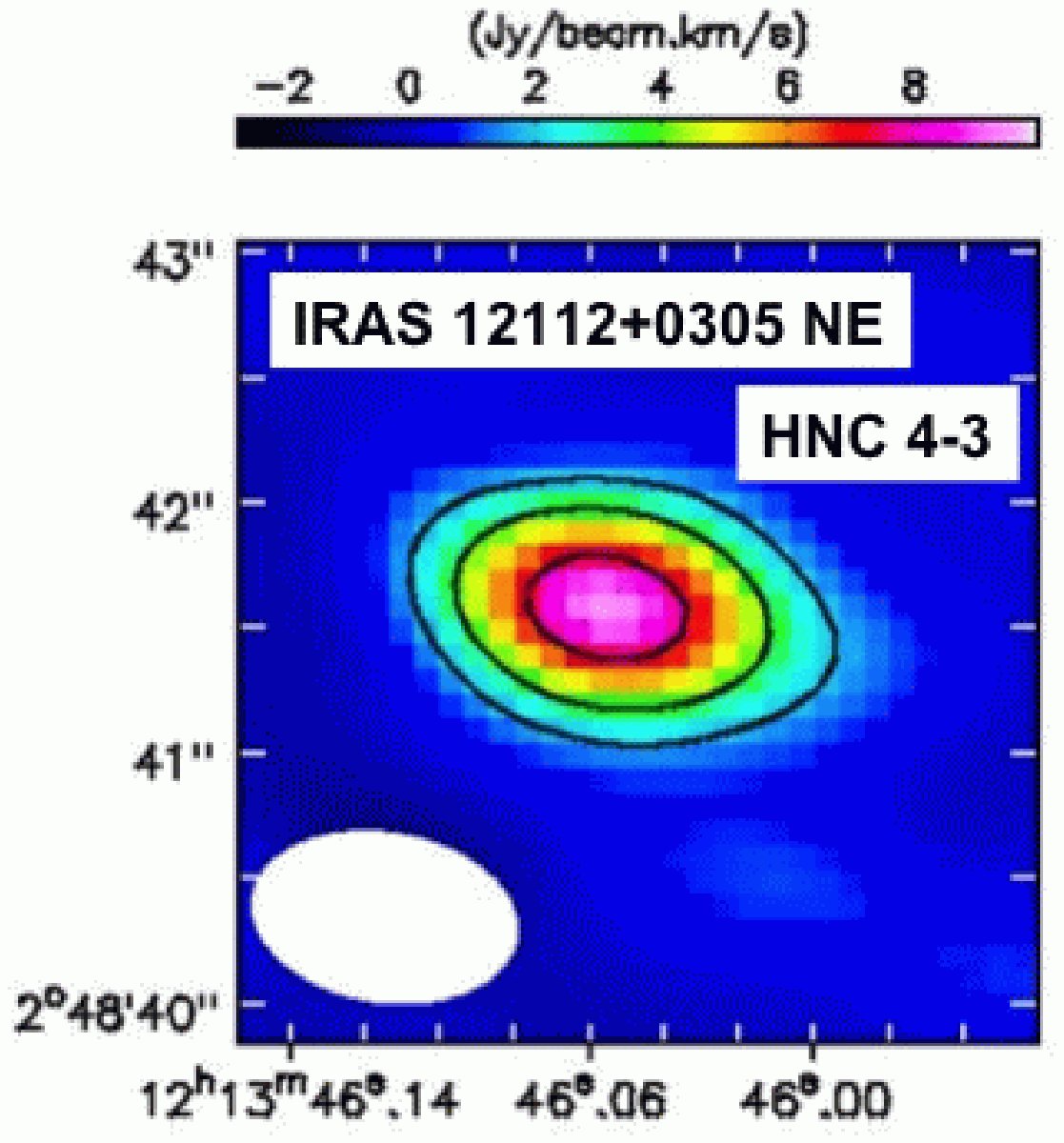} \\
\vspace{-1.3cm}
\includegraphics[angle=0,scale=.41]{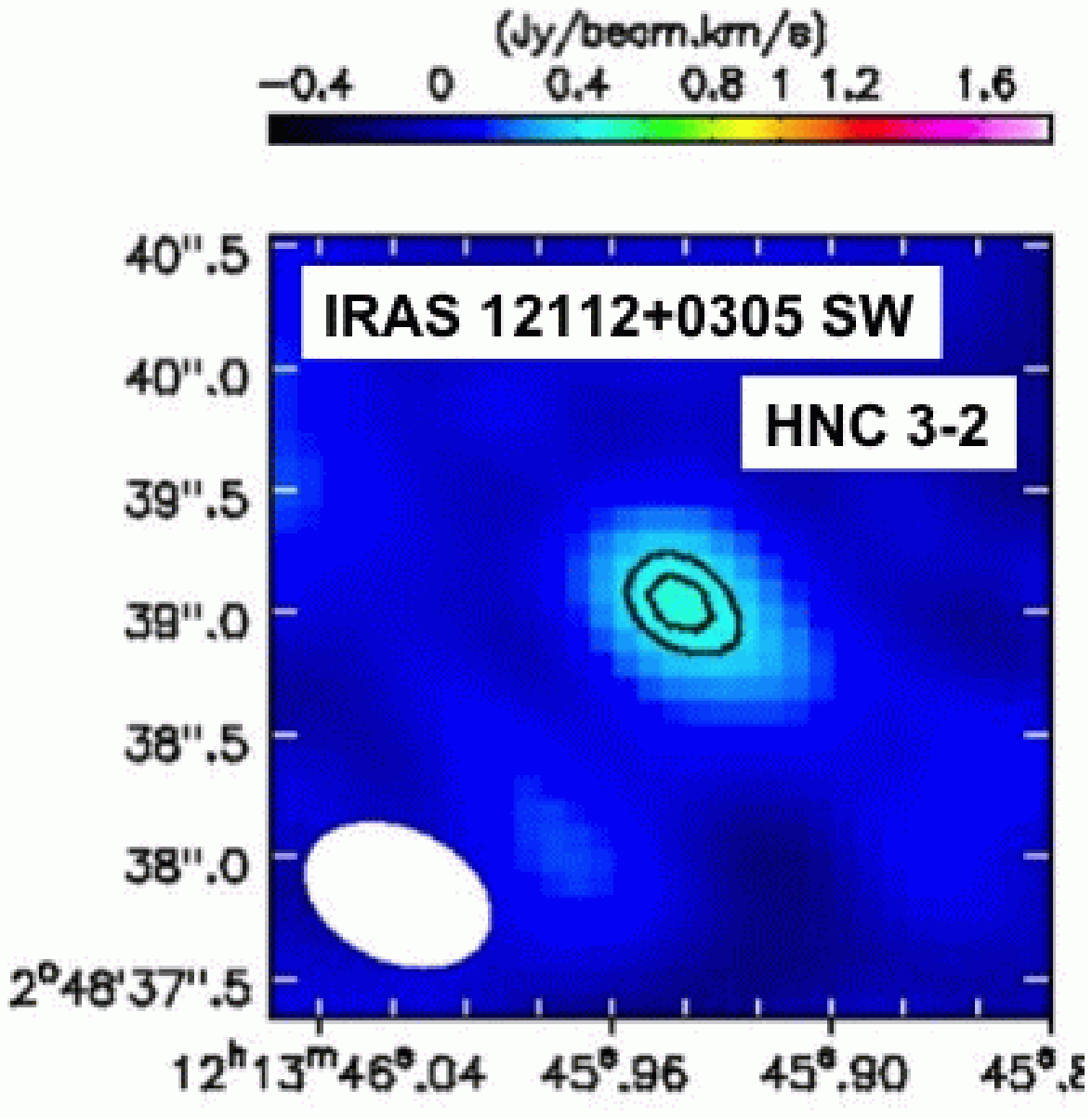} 
\includegraphics[angle=0,scale=.41]{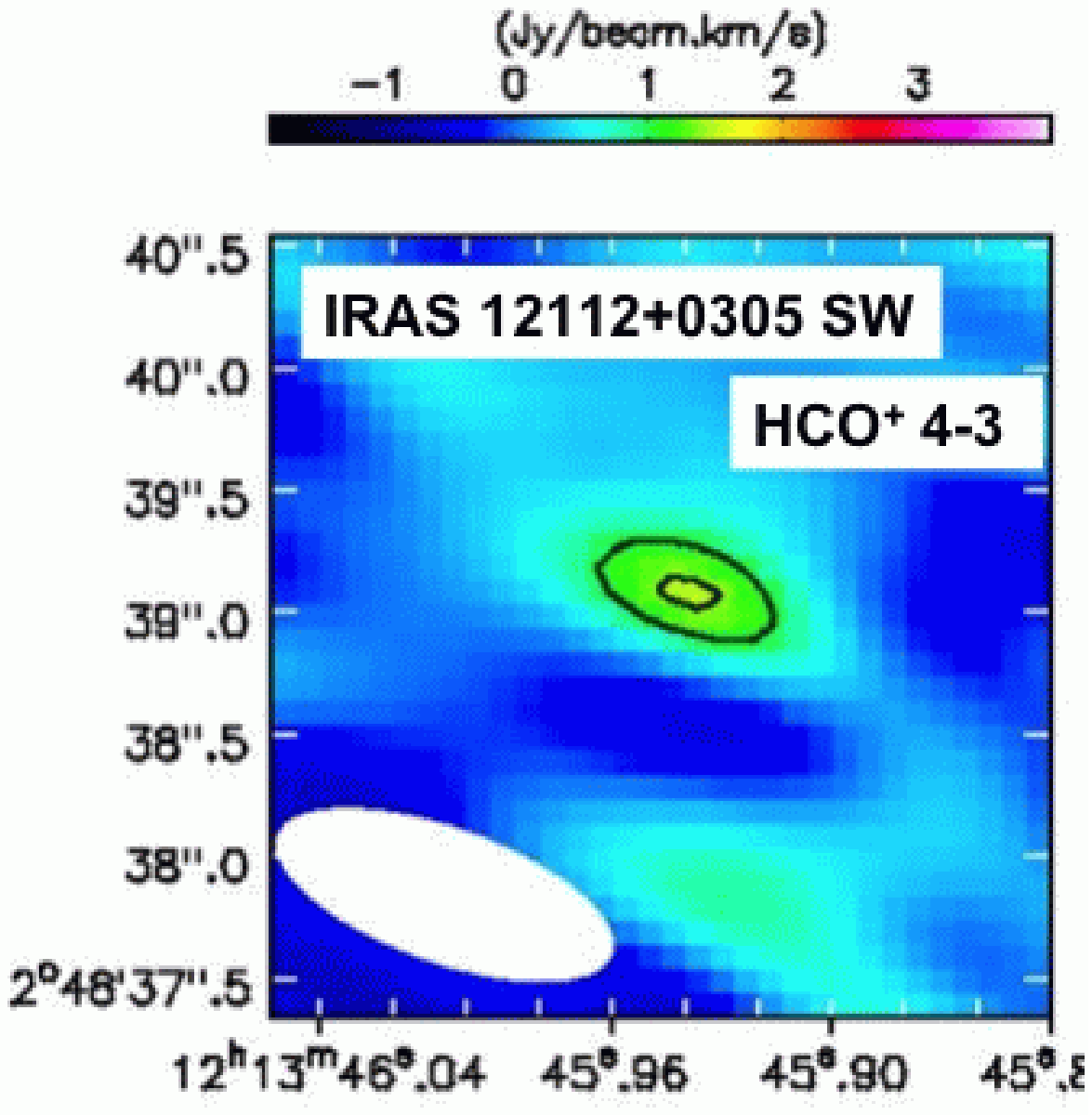} 
\includegraphics[angle=0,scale=.41]{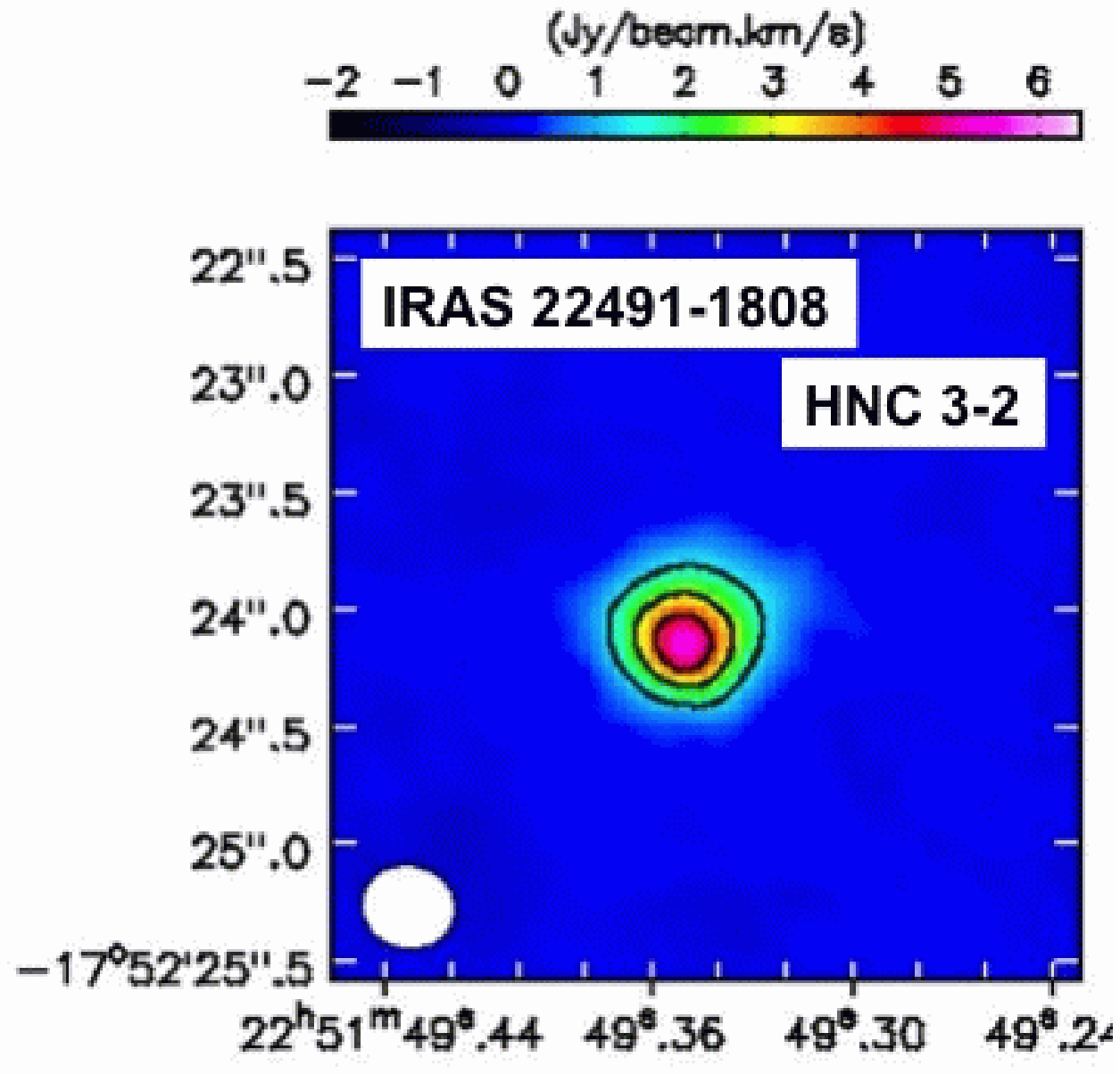} \\
\end{center}
\end{figure}

\clearpage

\begin{figure}
\begin{center}
\includegraphics[angle=0,scale=.41]{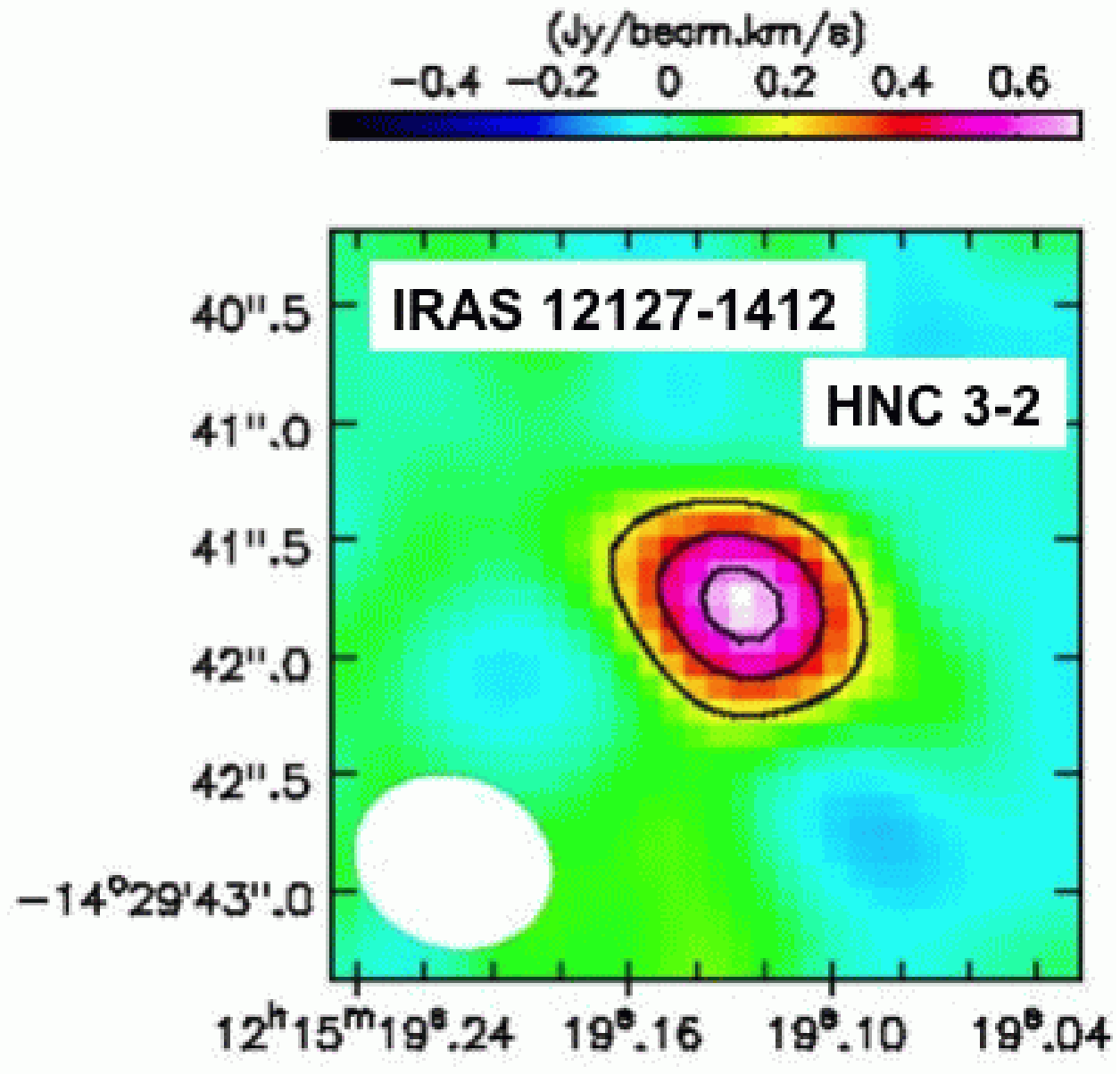} 
\includegraphics[angle=0,scale=.41]{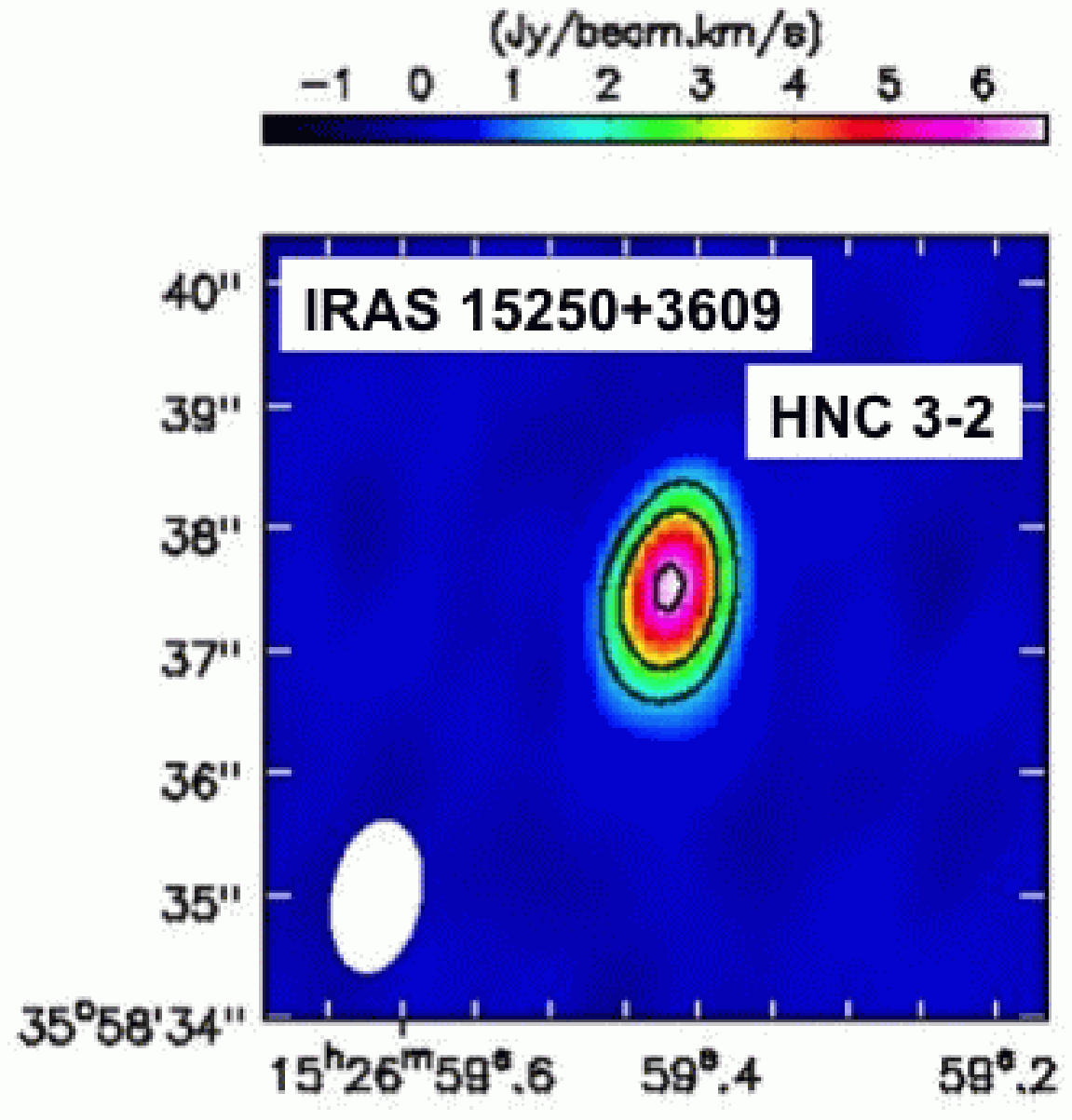} 
\includegraphics[angle=0,scale=.41]{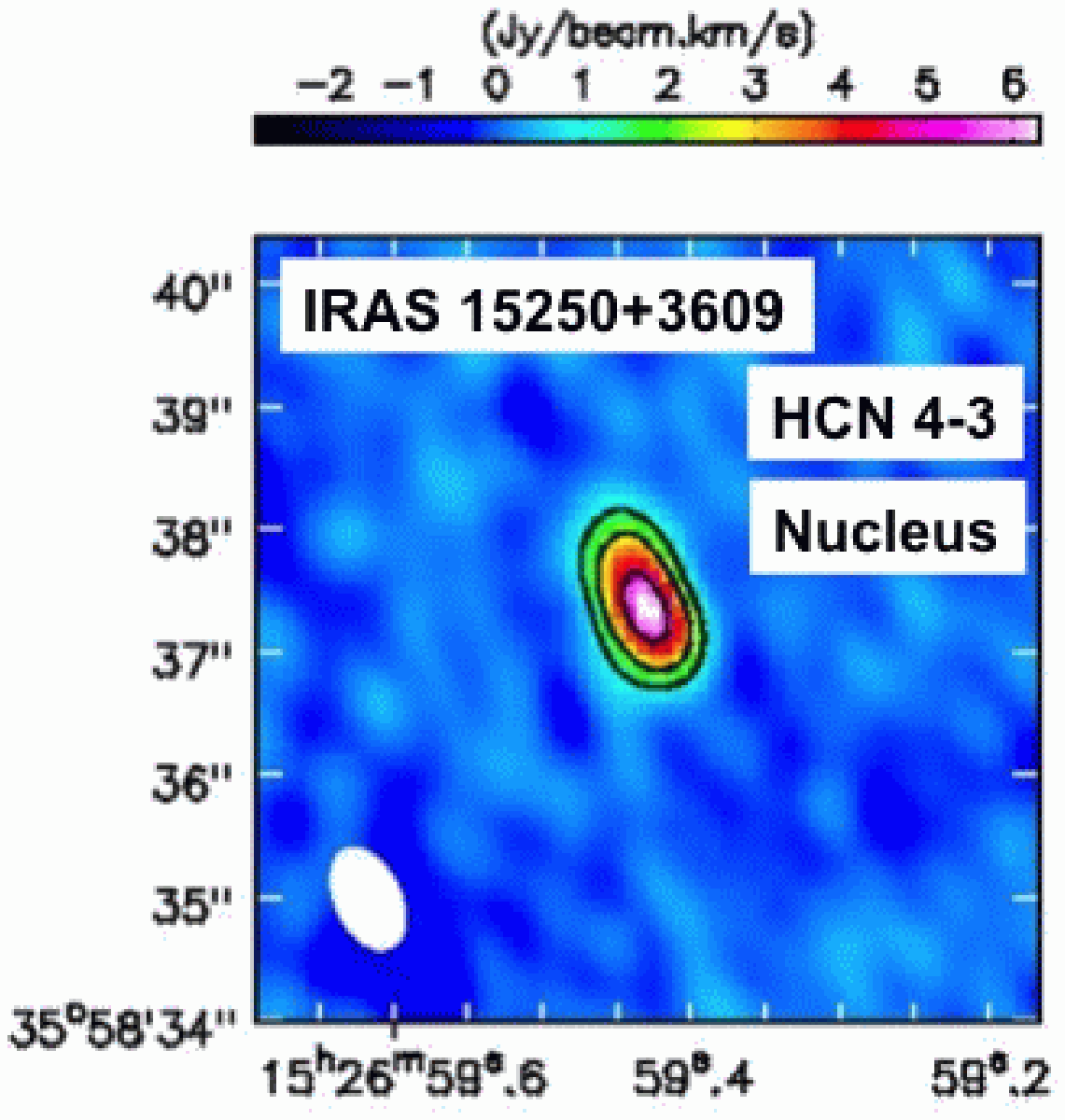} \\
\vspace{-1.3cm}
\includegraphics[angle=0,scale=.41]{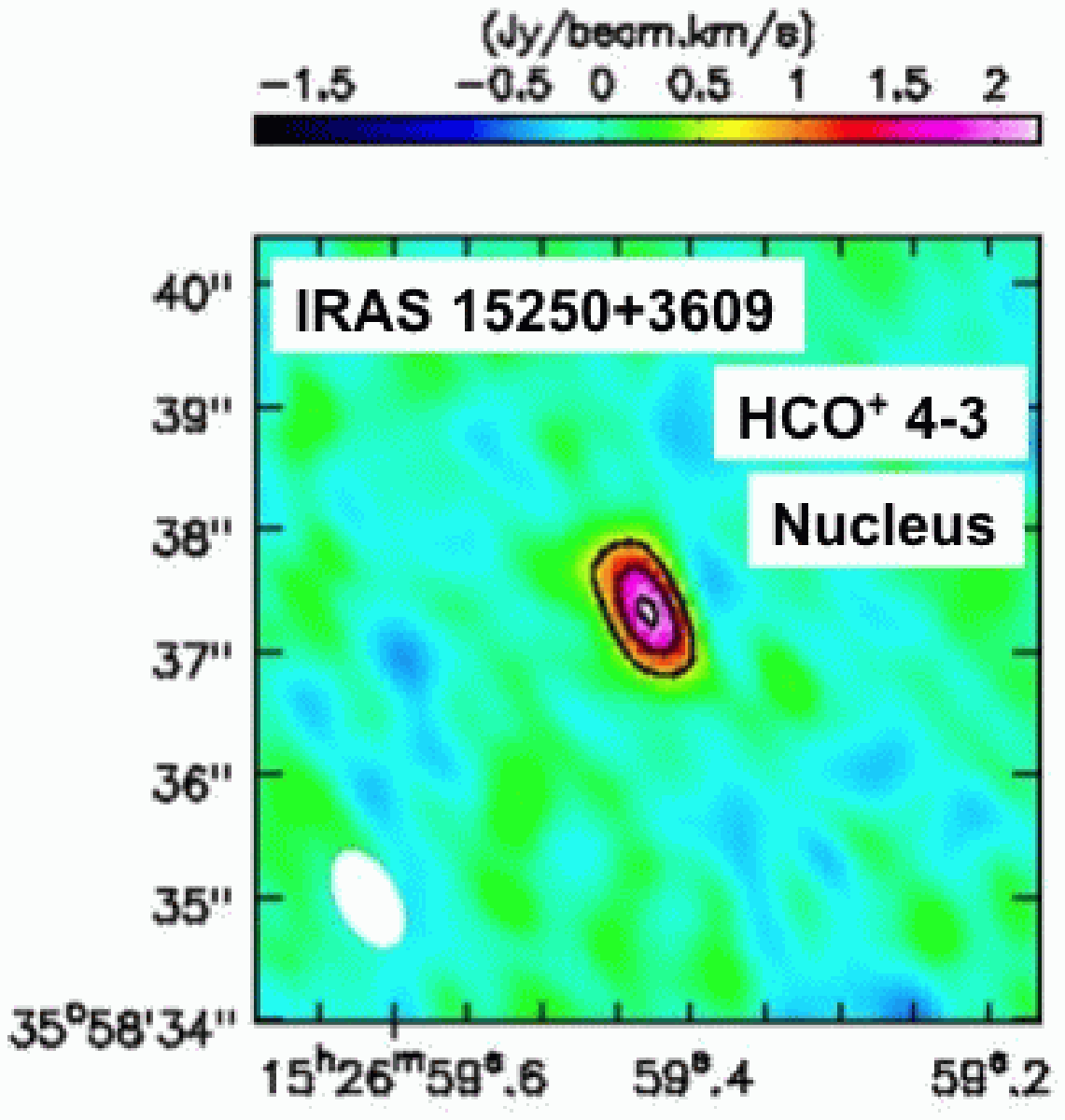} 
\includegraphics[angle=0,scale=.41]{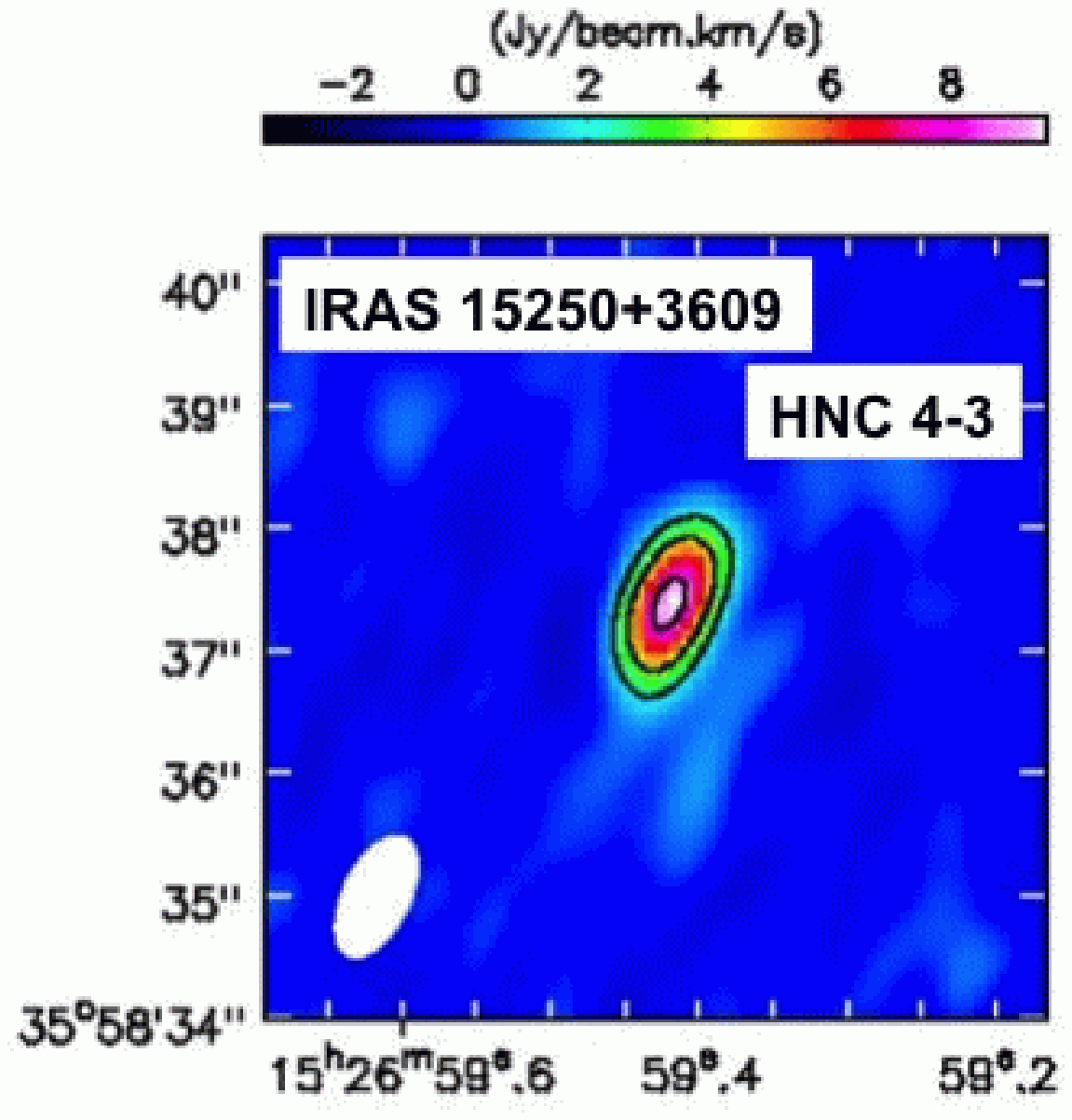} 
\includegraphics[angle=0,scale=.41]{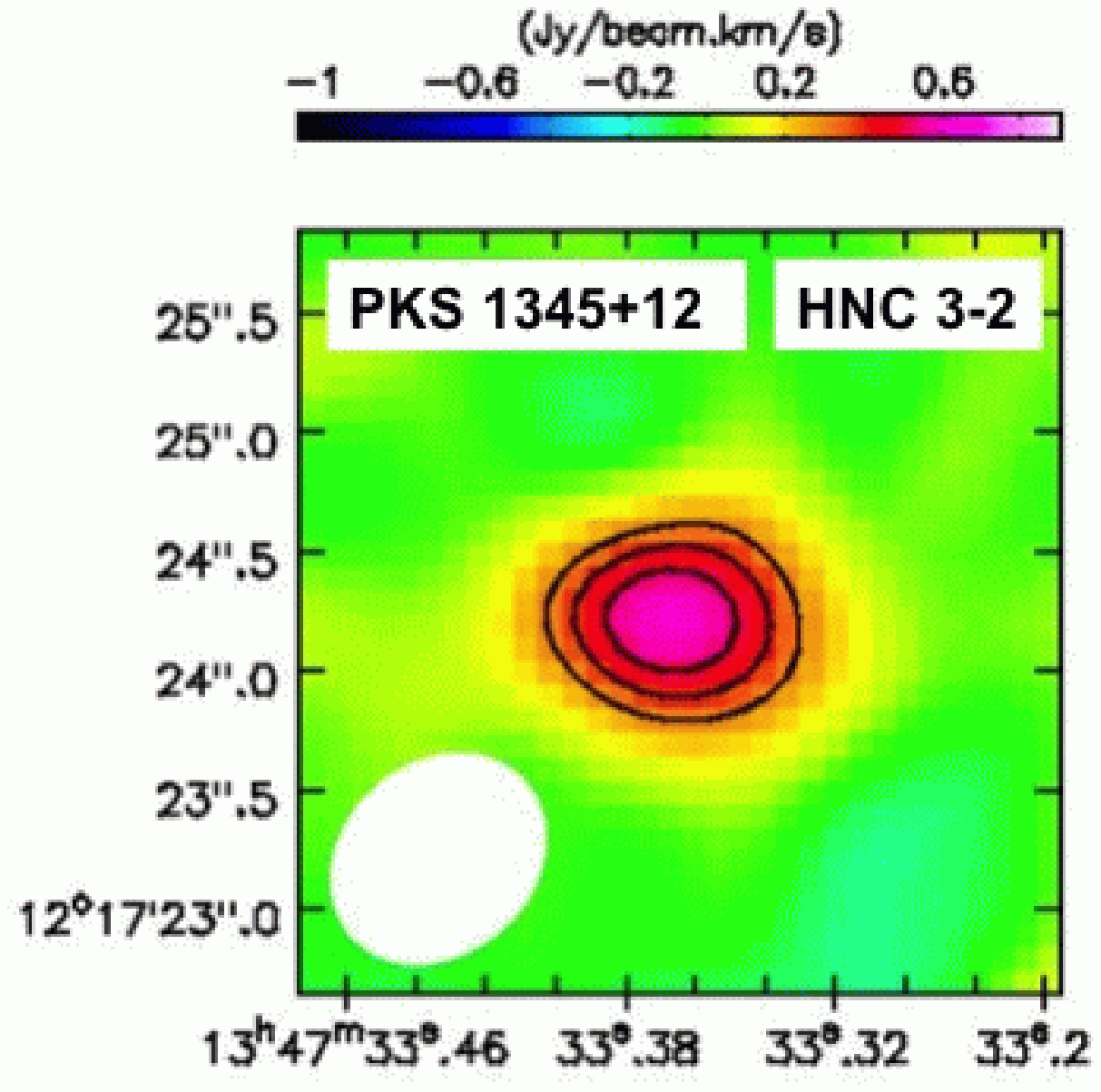} \\ 
\vspace{-1.3cm}
\includegraphics[angle=0,scale=.41]{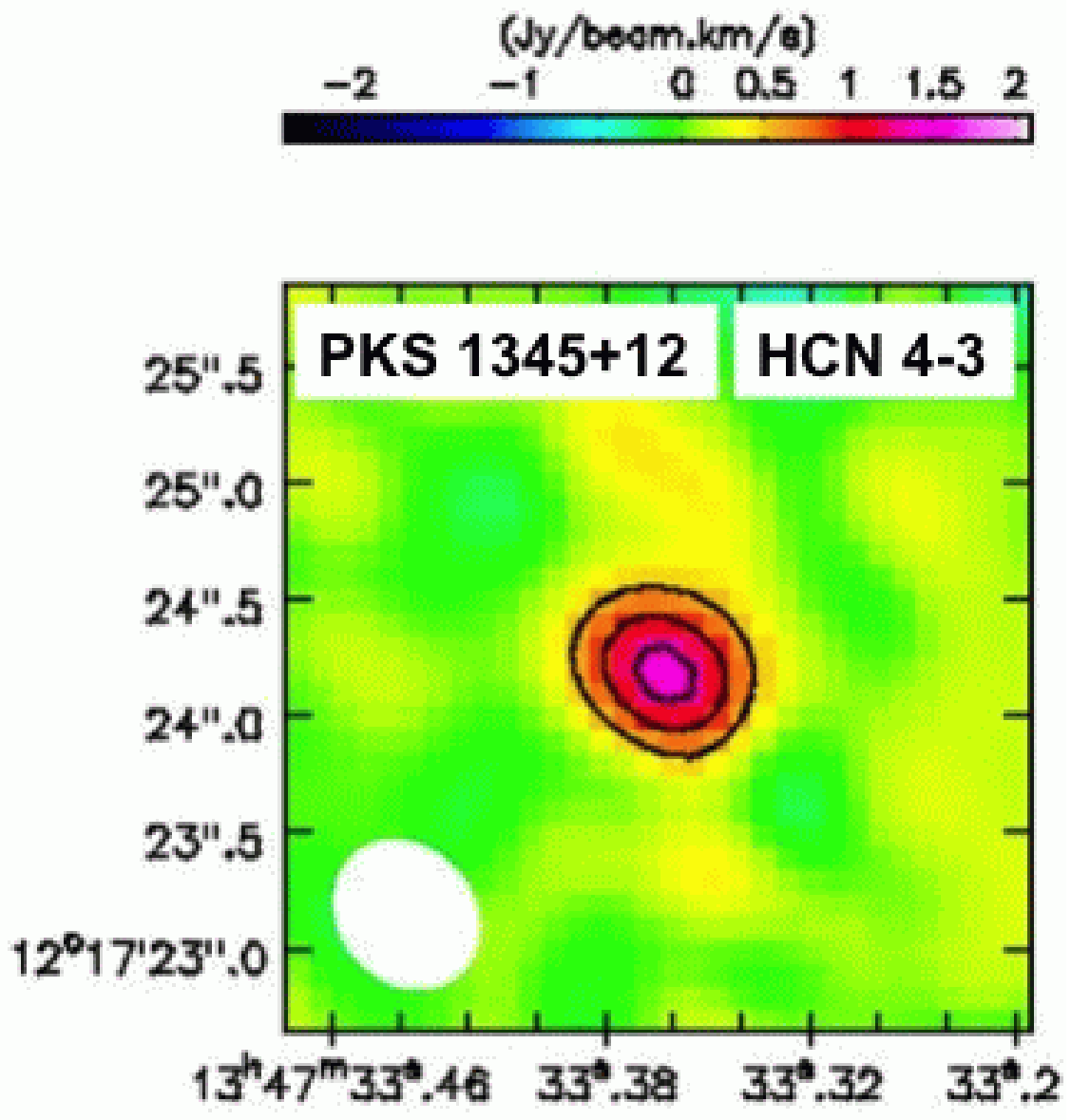} 
\includegraphics[angle=0,scale=.41]{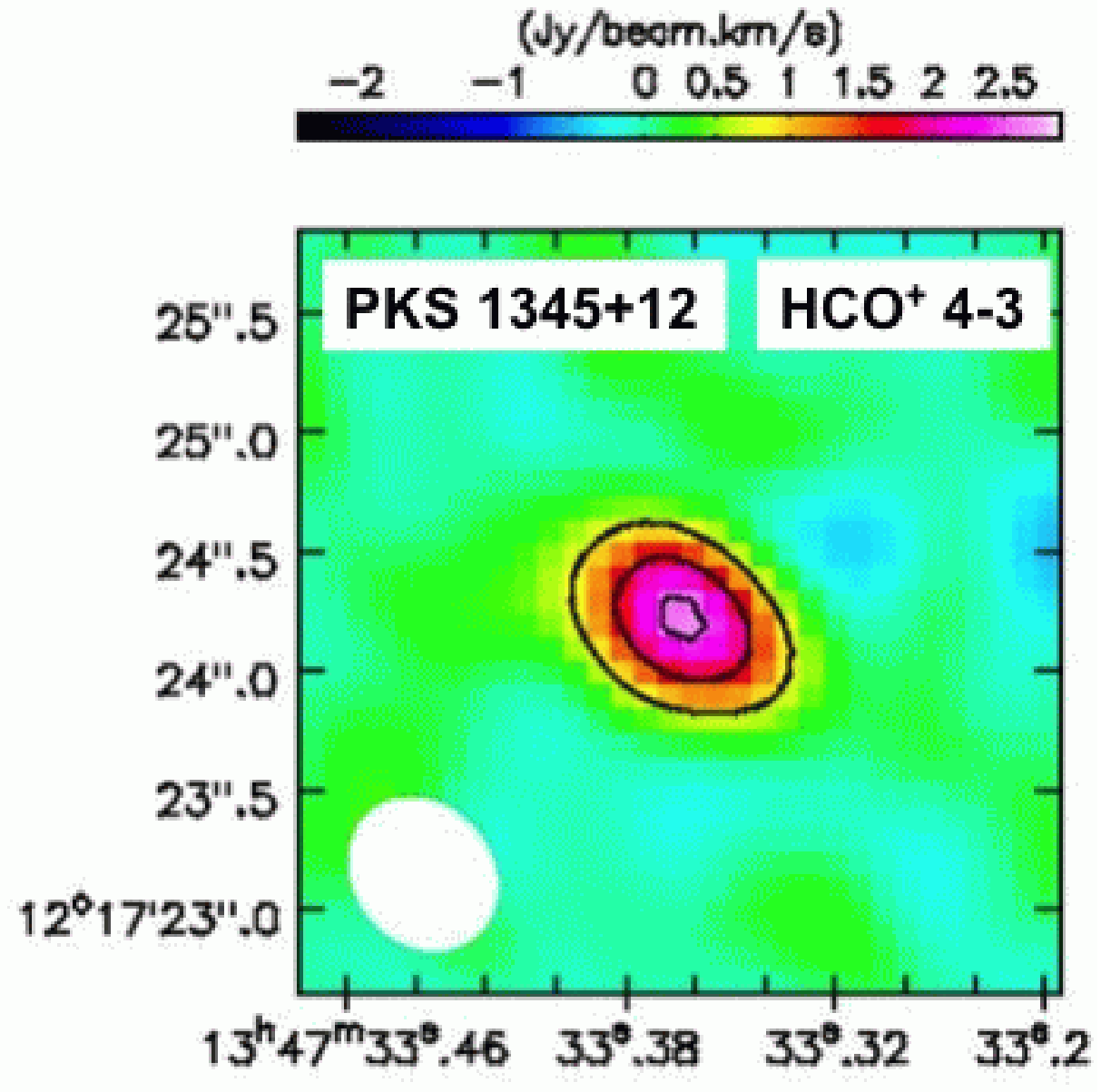} 
\includegraphics[angle=0,scale=.41]{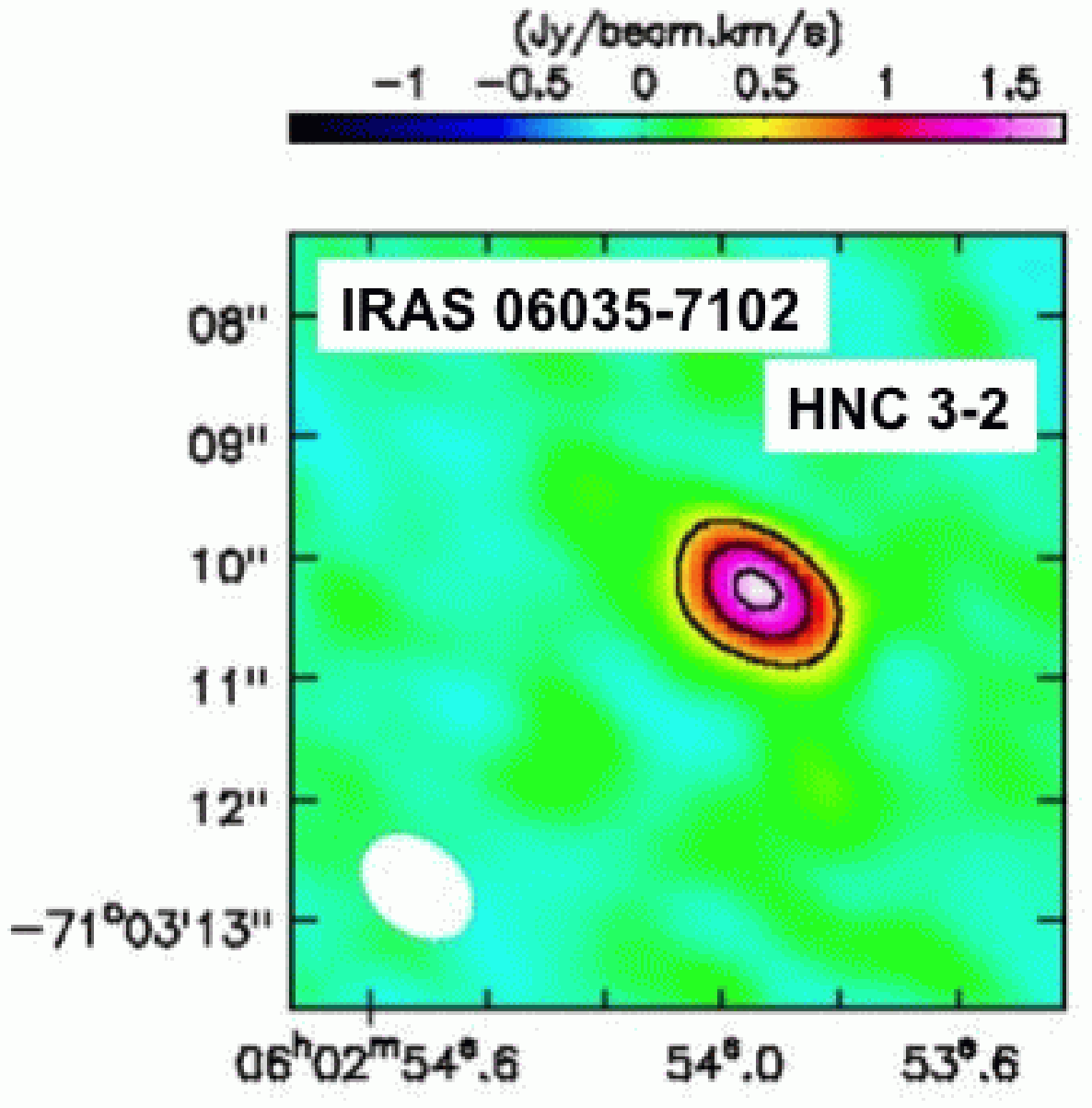} \\
\vspace{-1.3cm}
\includegraphics[angle=0,scale=.41]{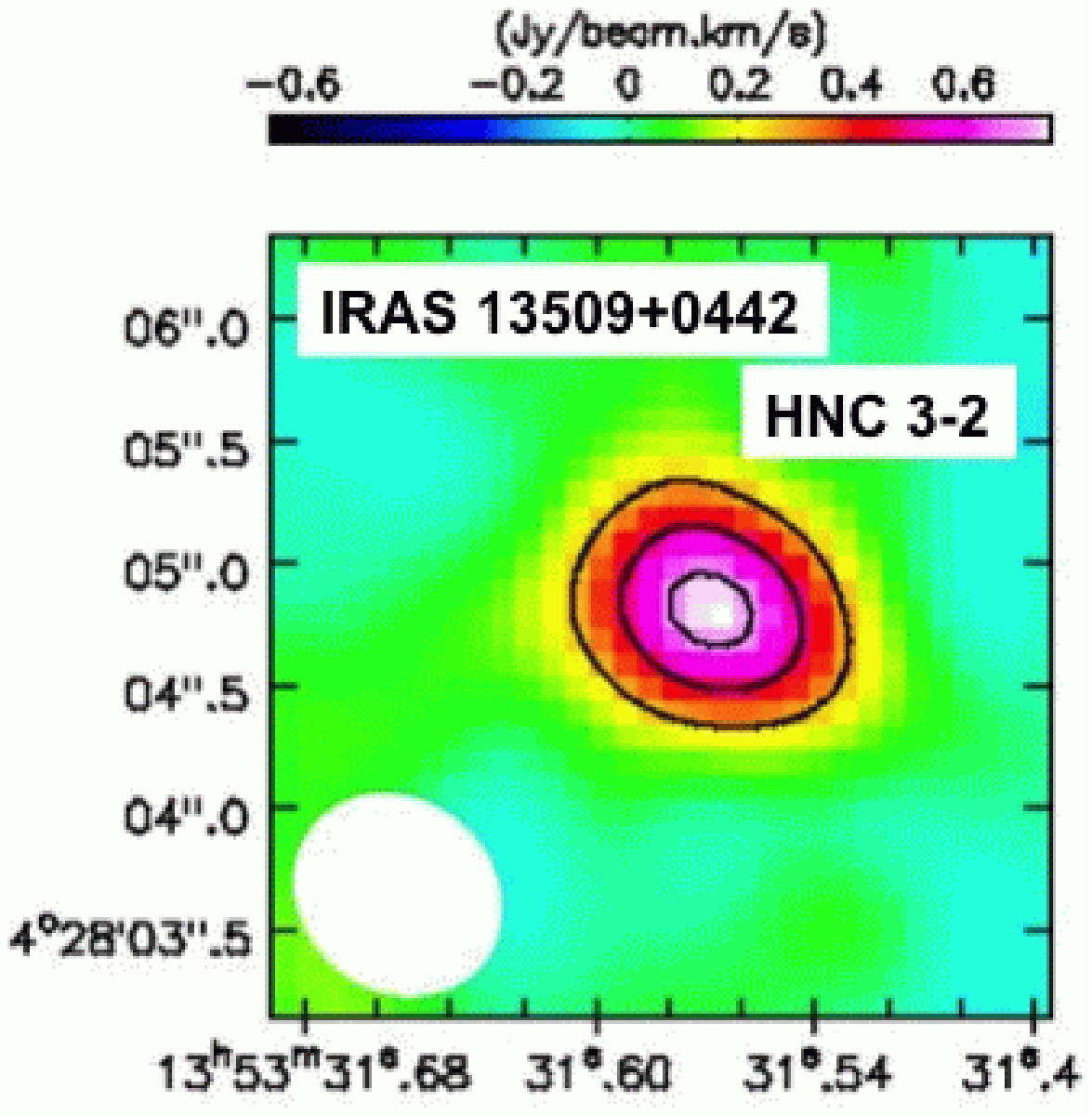} 
\includegraphics[angle=0,scale=.41]{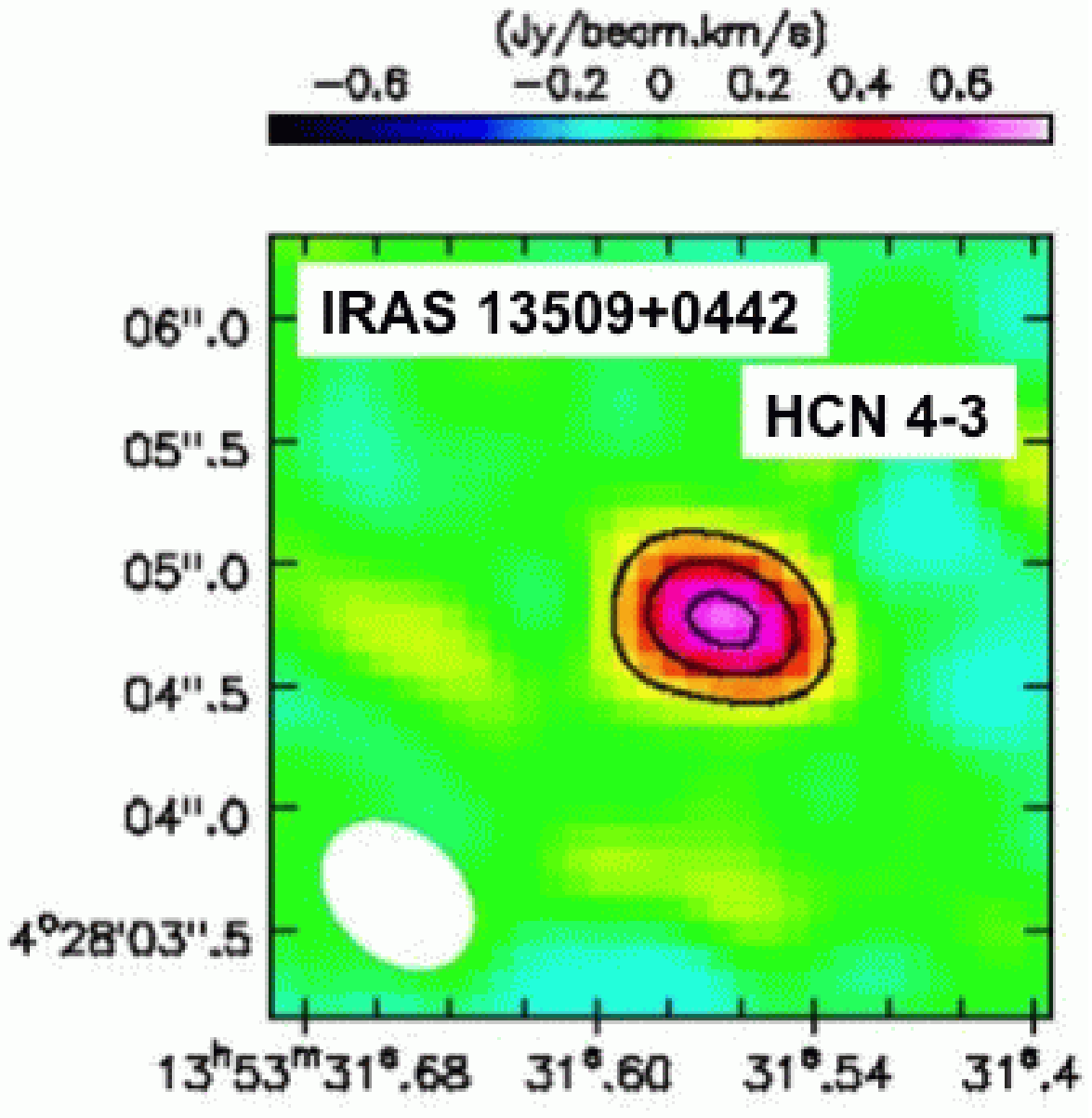}  
\includegraphics[angle=0,scale=.41]{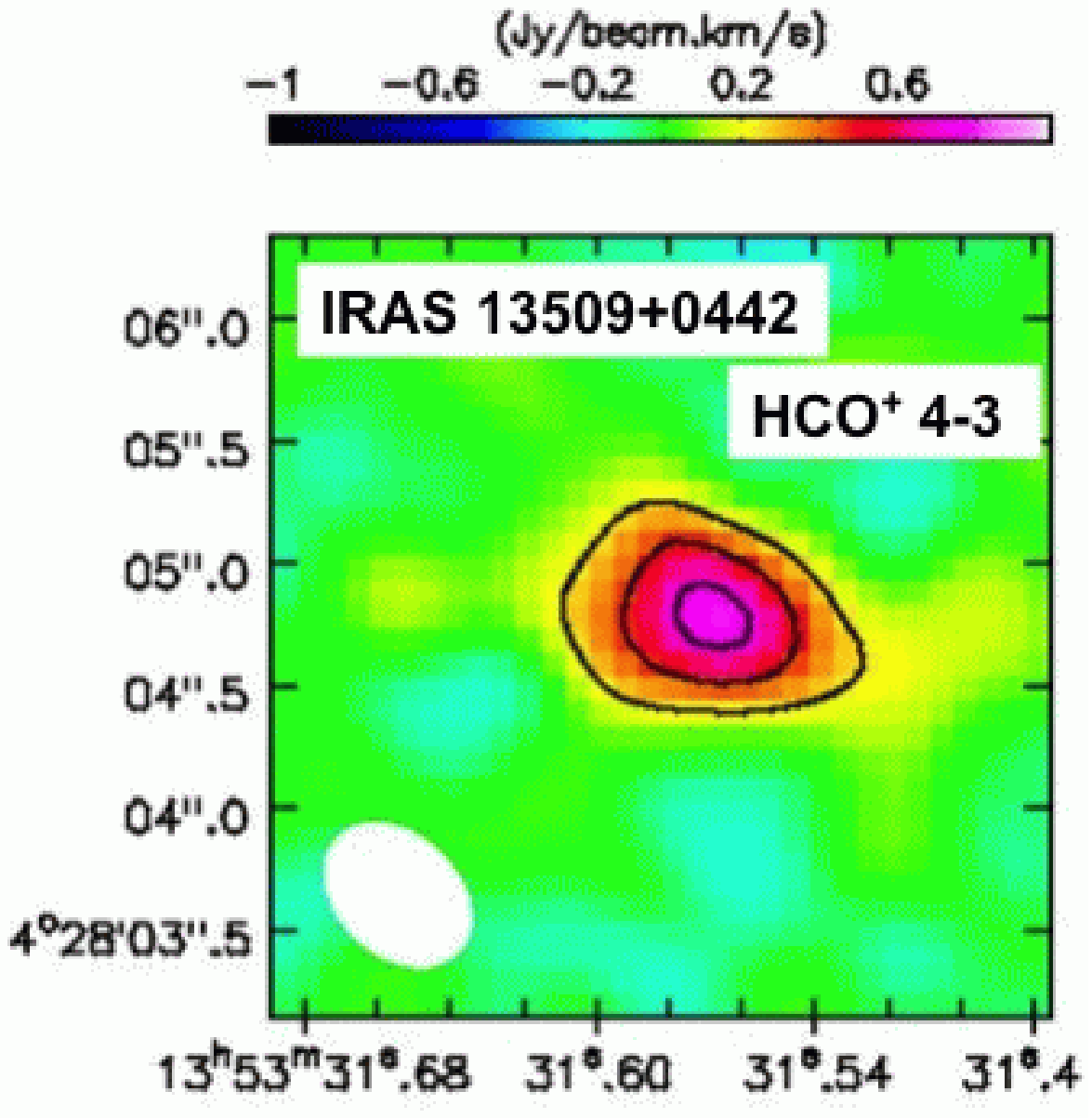} \\
\end{center}
\end{figure}

\clearpage

\begin{figure}
\begin{center}
\includegraphics[angle=0,scale=.41]{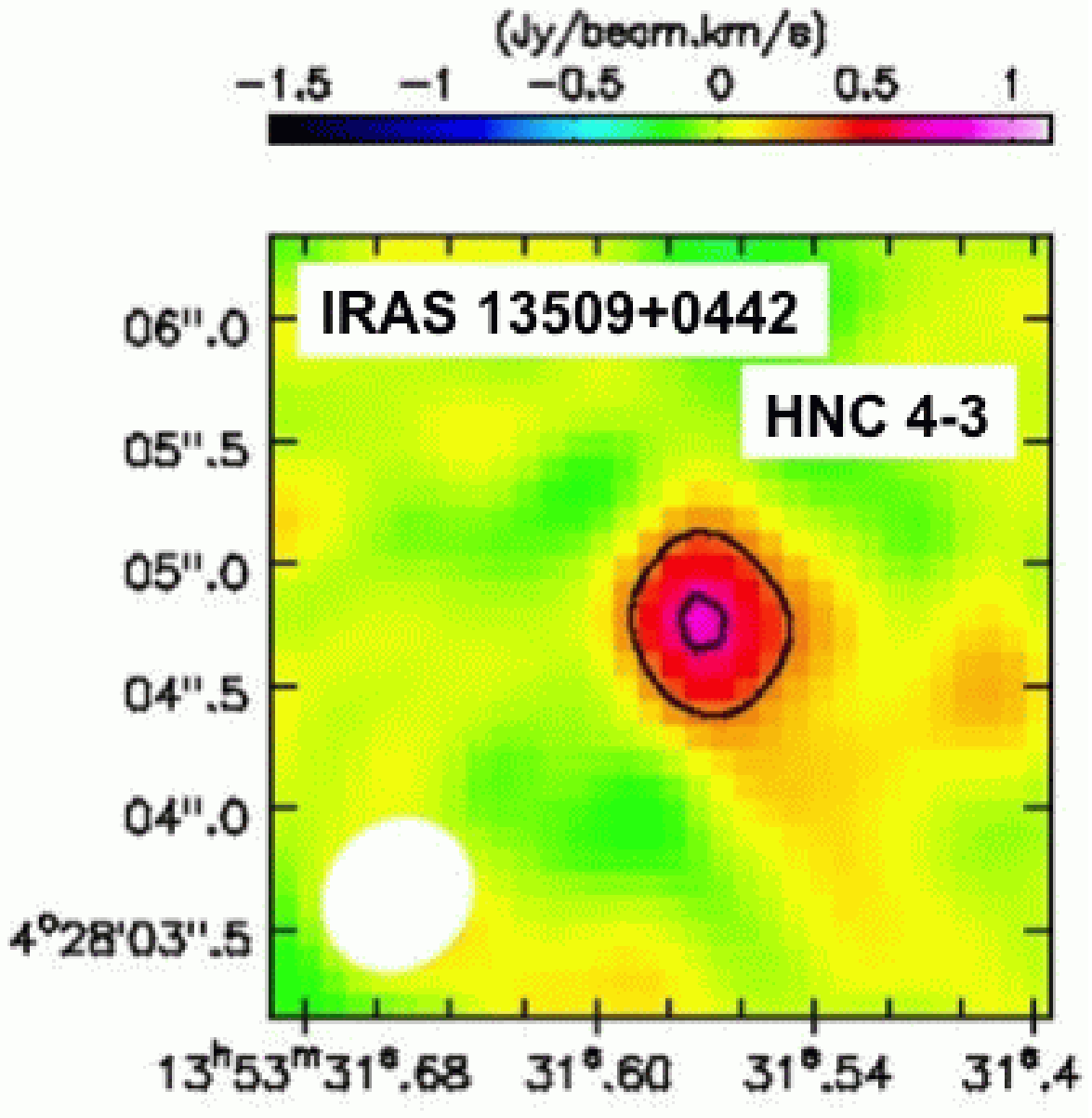}  
\includegraphics[angle=0,scale=.381]{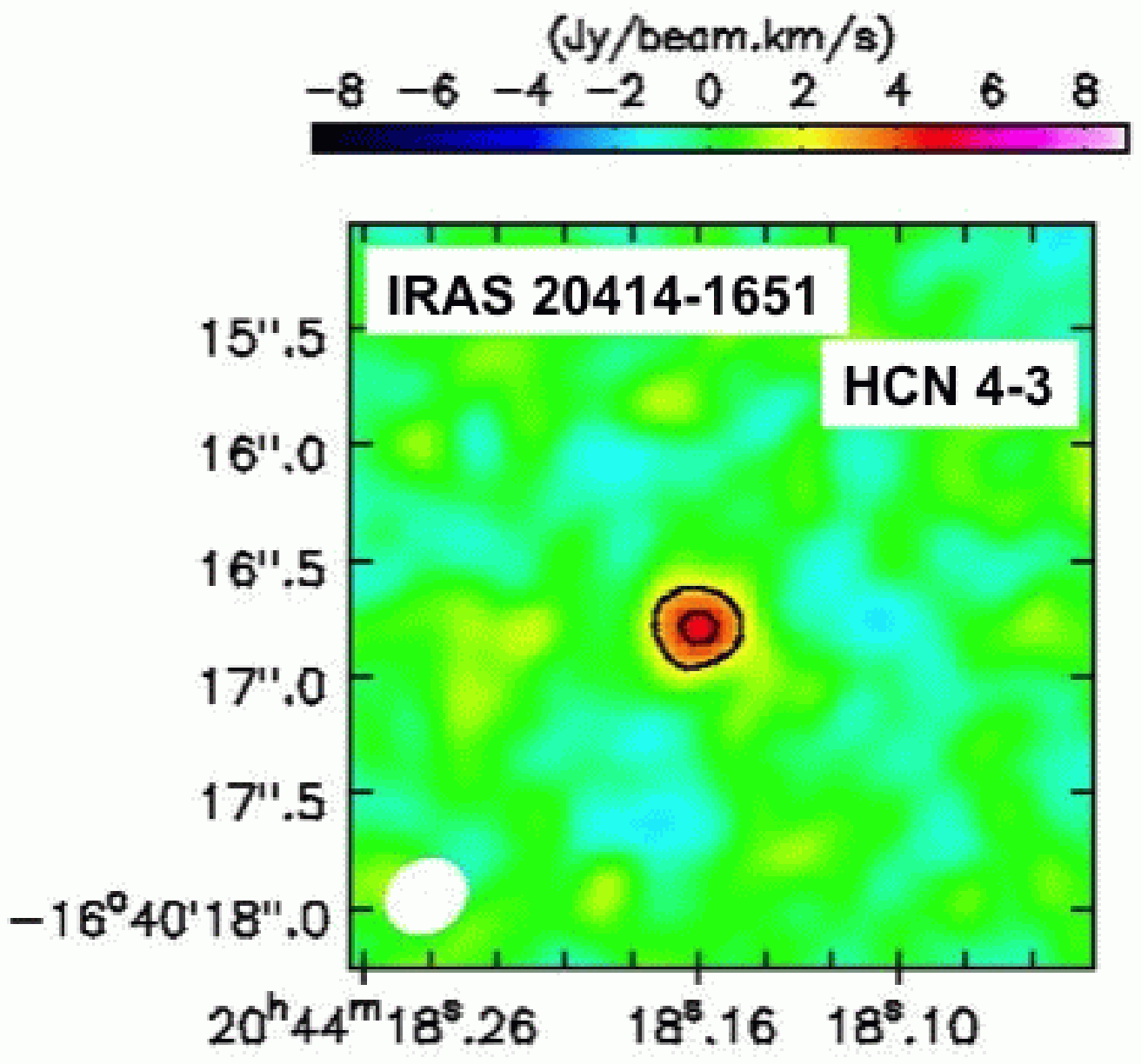}
\includegraphics[angle=0,scale=.381]{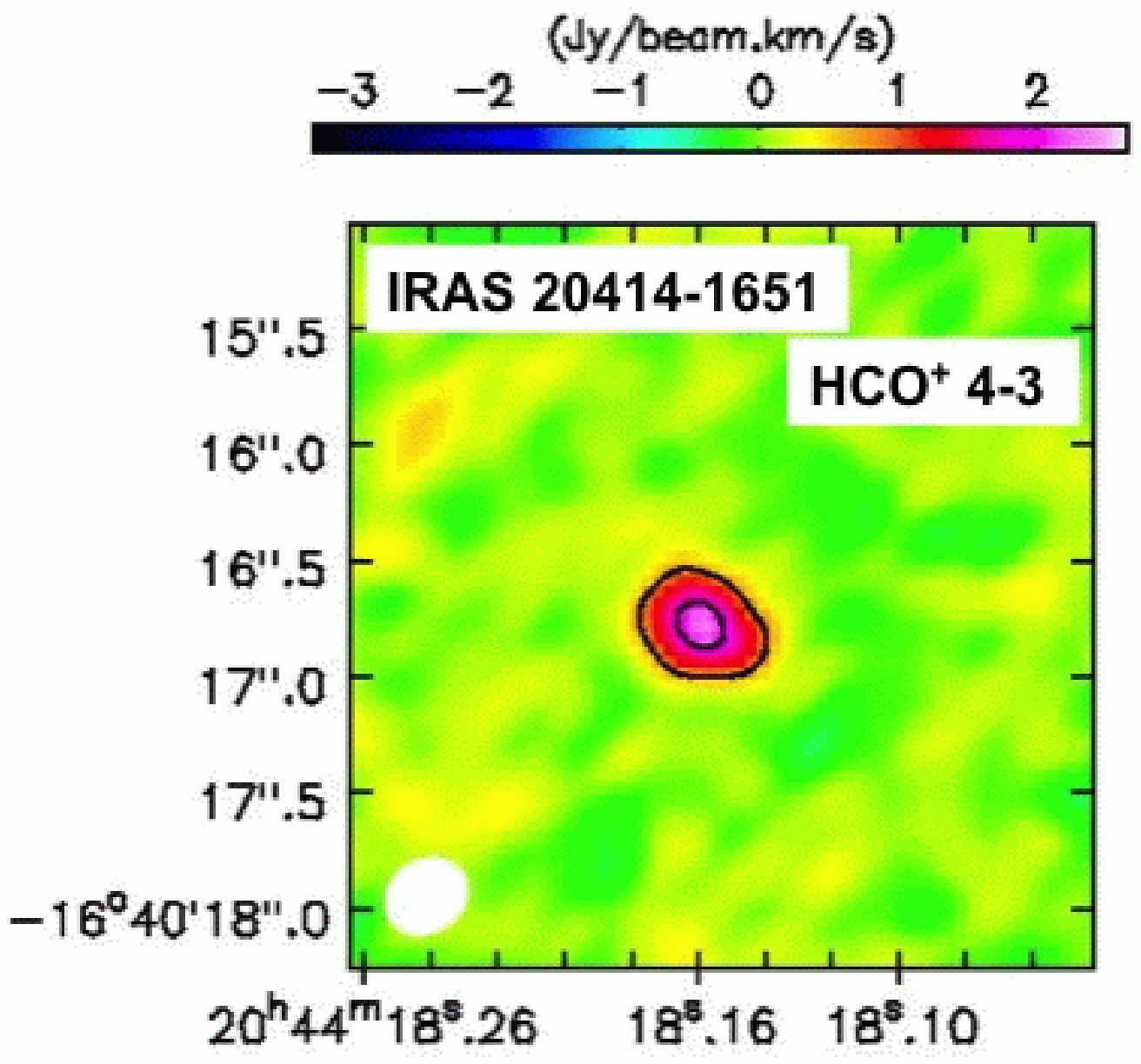} \\ 
\vspace{-1.3cm}
\includegraphics[angle=0,scale=.381]{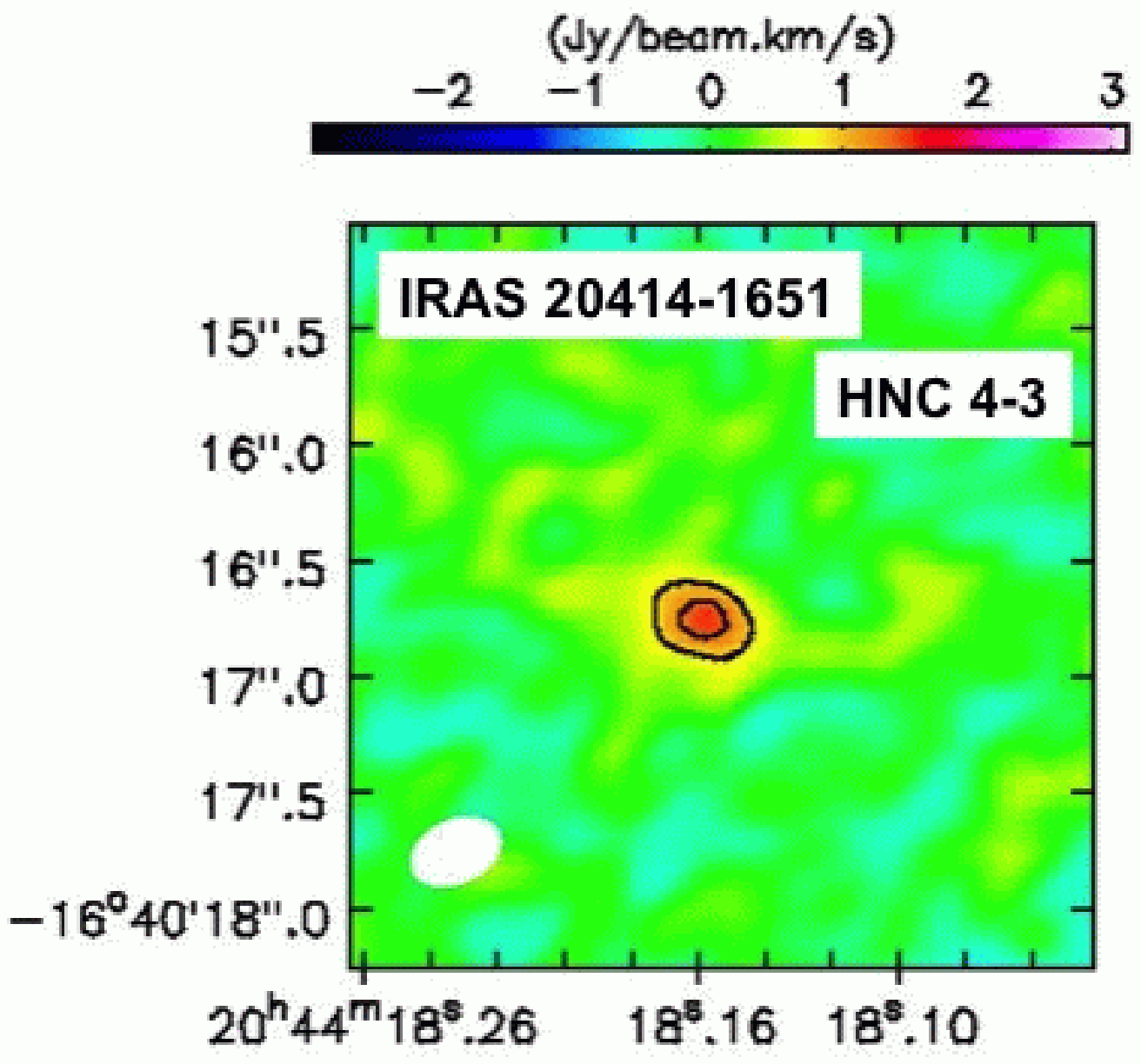}  
\end{center}
\caption{
Integrated intensity (moment 0) maps of the detected bright HNC J=3--2,
HCN J=4--3, HCO$^{+}$ J=4--3, and HNC J=4--3 emission lines. 
The abscissa and ordinate are R.A. (J2000) and decl. (J2000),
respectively. 
The contours represent 
3$\sigma$, 5$\sigma$, 7$\sigma$ for IRAS 08572$+$3915 HNC J=3--2; 
3$\sigma$, 6$\sigma$, 9$\sigma$ for Superantennae HNC J=3--2; 
3$\sigma$, 4$\sigma$ for Superantennae HCN J=4--3; 
4$\sigma$, 6$\sigma$ for Superantennae HCO$^{+}$ J=4--3; 
4$\sigma$, 6$\sigma$ for Superantennae HNC J=4--3;
10$\sigma$, 20$\sigma$, 40$\sigma$ for IRAS 12112$+$0305 NE HNC J=3--2;
5$\sigma$, 10$\sigma$, 20$\sigma$ for IRAS 12112$+$0305 NE HCN J=4--3;
4$\sigma$, 7$\sigma$, 10$\sigma$ for IRAS 12112$+$0305 NE HCO$^{+}$ J=4--3;
5$\sigma$, 10$\sigma$, 20$\sigma$ for IRAS 12112$+$0305 NE HNC J=4--3;
3$\sigma$, 3.5$\sigma$ for IRAS 12112$+$0305 SW HNC J=3--2;
3$\sigma$, 4$\sigma$ for IRAS 12112$+$0305 SW HCO$^{+}$ J=4--3;
10$\sigma$, 20$\sigma$, 30$\sigma$ for IRAS 22491$-$1808 HNC J=3--2; 
3$\sigma$, 6$\sigma$, 9$\sigma$ for IRAS 12127$-$1412 HNC J=3--2;
8$\sigma$, 16$\sigma$, 32$\sigma$ for IRAS 15250$+$3609 HNC J=3--2; 
5$\sigma$, 10$\sigma$, 20$\sigma$ for IRAS 15250$+$3609 HCN J=4--3; 
5$\sigma$, 10$\sigma$, 15$\sigma$ for IRAS 15250$+$3609 HCO$^{+}$ J=4--3; 
7$\sigma$, 14$\sigma$, 28$\sigma$ for IRAS 15250$+$3609 HNC J=4--3;
3$\sigma$, 4$\sigma$, 5$\sigma$ for PKS 1345$+$12 HNC J=3--2; 
3$\sigma$, 5$\sigma$, 7$\sigma$ for PKS 1345$+$12 HCN J=4--3; 
4$\sigma$, 8$\sigma$, 12$\sigma$ for PKS 1345$+$12 HCO$^{+}$ J=4--3; 
4$\sigma$, 8$\sigma$, 12$\sigma$ for IRAS 06035$-$7102 HNC J=3--2; 
4$\sigma$, 7$\sigma$, 10$\sigma$ for IRAS 13509$+$0442 HNC J=3--2;
3$\sigma$, 6$\sigma$, 9$\sigma$ for IRAS 13509$+$0442 HCN J=4--3;
3$\sigma$, 6$\sigma$, 9$\sigma$ for IRAS 13509$+$0442 HCO$^{+}$ J=4--3;
3$\sigma$, 6$\sigma$ for IRAS 13509$+$0442 HNC J=4--3;
4$\sigma$, 7$\sigma$ for IRAS 20414$-$1651 HCN J=4--3; 
4$\sigma$, 10$\sigma$ for IRAS 20414$-$1651 HCO$^{+}$ J=4--3; and
4$\sigma$, 6$\sigma$ for IRAS 20414$-$1651 HNC J=4--3.
IRAS 15250$+$3609 displays an outflow-origin sub-peak emission
component, in addition to the nuclear main emission component, for HCN
J=4--3 and HCO$^{+}$ J=4--3 ($\S$5.1.1). The moment 0 maps are for the
nuclear main emission components only. 
Beam sizes are shown as filled circles in the lower-left region.
}
\end{figure}

\begin{figure}
\begin{center}
\includegraphics[angle=0,scale=.64]{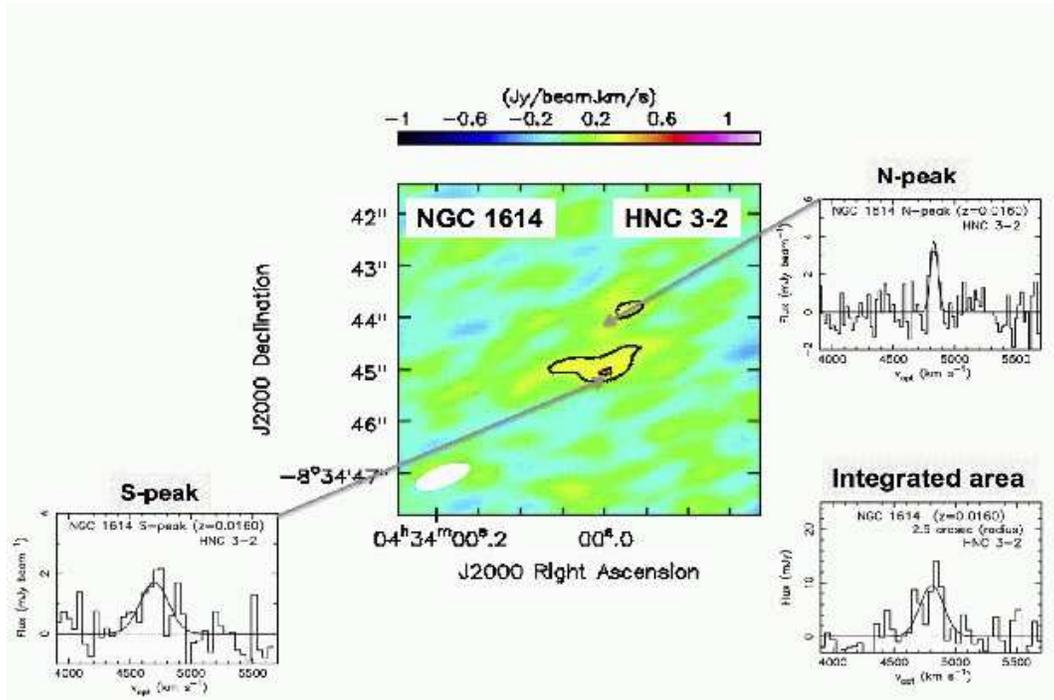} 
\caption{
Integrated intensity (moment 0) map of HNC J=3--2 emission line
of NGC 1614. 
The abscissa and ordinate are R.A. (J2000) and decl. (J2000),
respectively. 
The contours are 2$\sigma$ and 2.8$\sigma$.
A beam size is shown as the filled circle in the lower-left region.
Gaussian fits of HNC J=3--2 emission line in the spectra at the N- and
S-peaks in the continuum emission map (Figure 1), within the beam size, and
in a spatially integrated spectrum with a 2$\farcs$5 radius
circular region around (04$^{h}$33$^{m}$59.99$^{s}$,
$-$08$^{\circ}$34$'$44.70$''$)J2000, are shown.   
The abscissa is optical LSR velocity in (km s$^{-1}$). 
The ordinate is flux in (mJy beam$^{-1}$) for the spectra at the peaks
and flux in (mJy) for the spatially integrated spectrum. 
}
\end{center}
\end{figure}

\begin{figure}
\begin{center}
\includegraphics[angle=0,scale=.41]{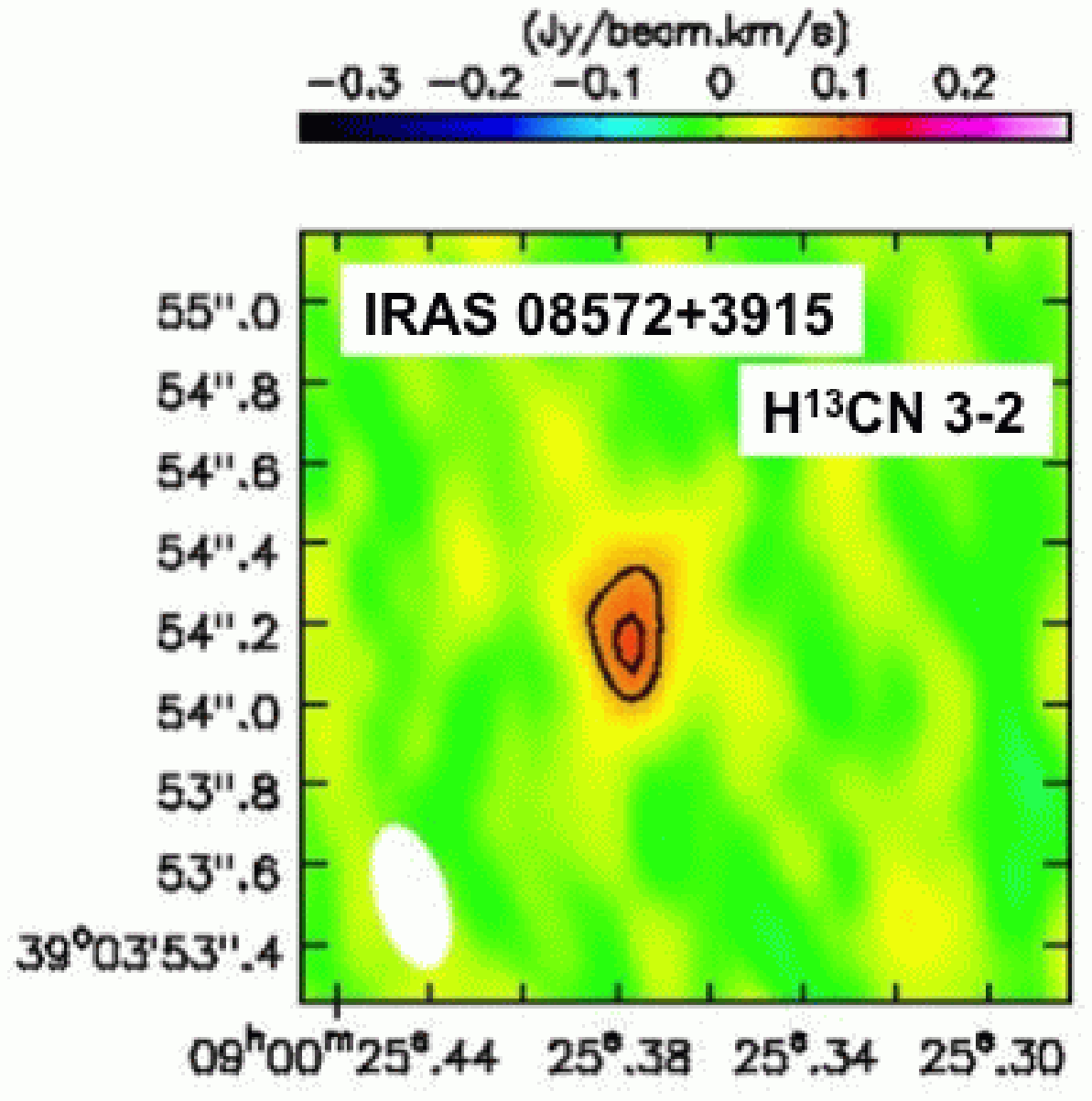} 
\includegraphics[angle=0,scale=.41]{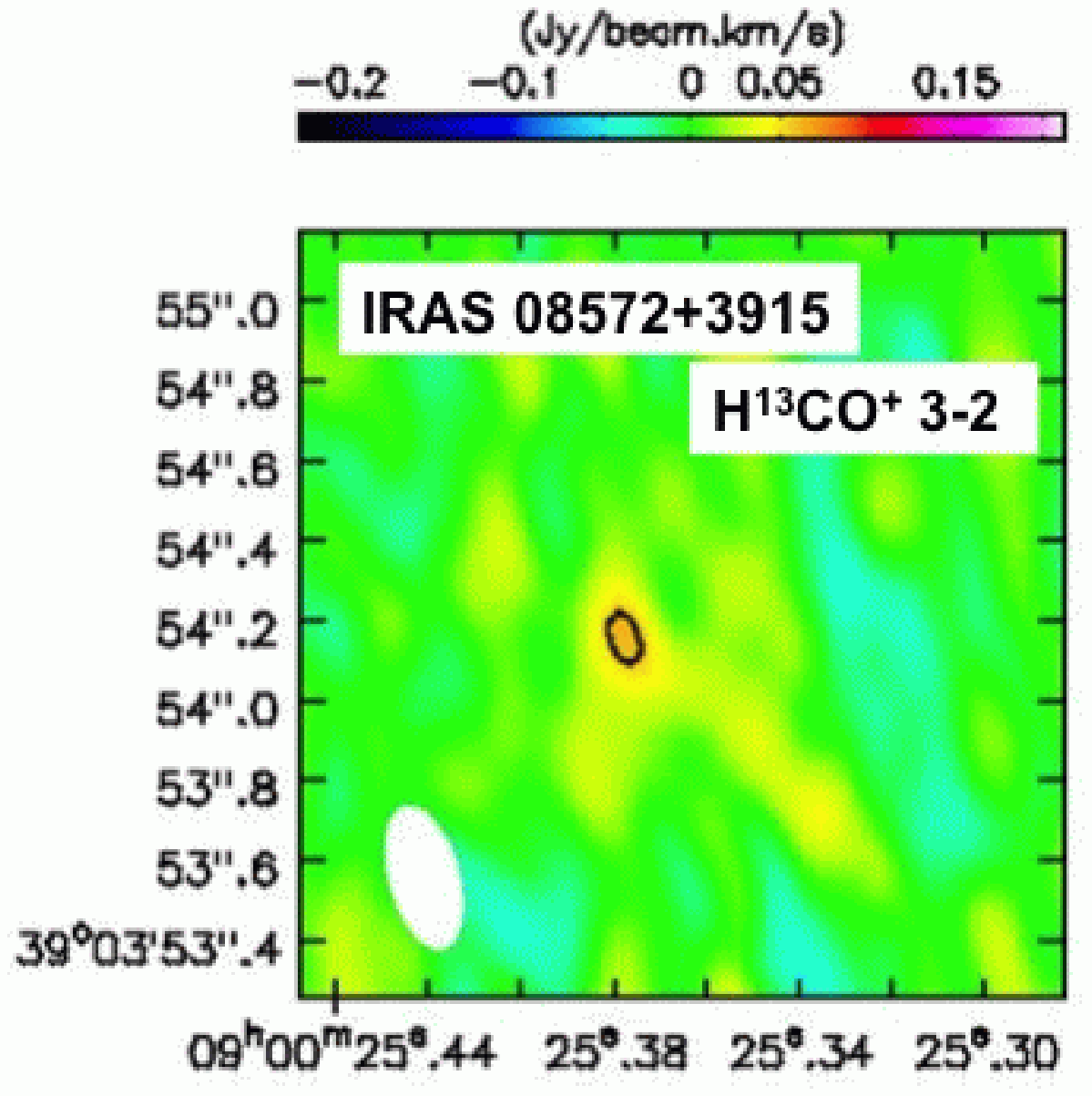} 
\includegraphics[angle=0,scale=.41]{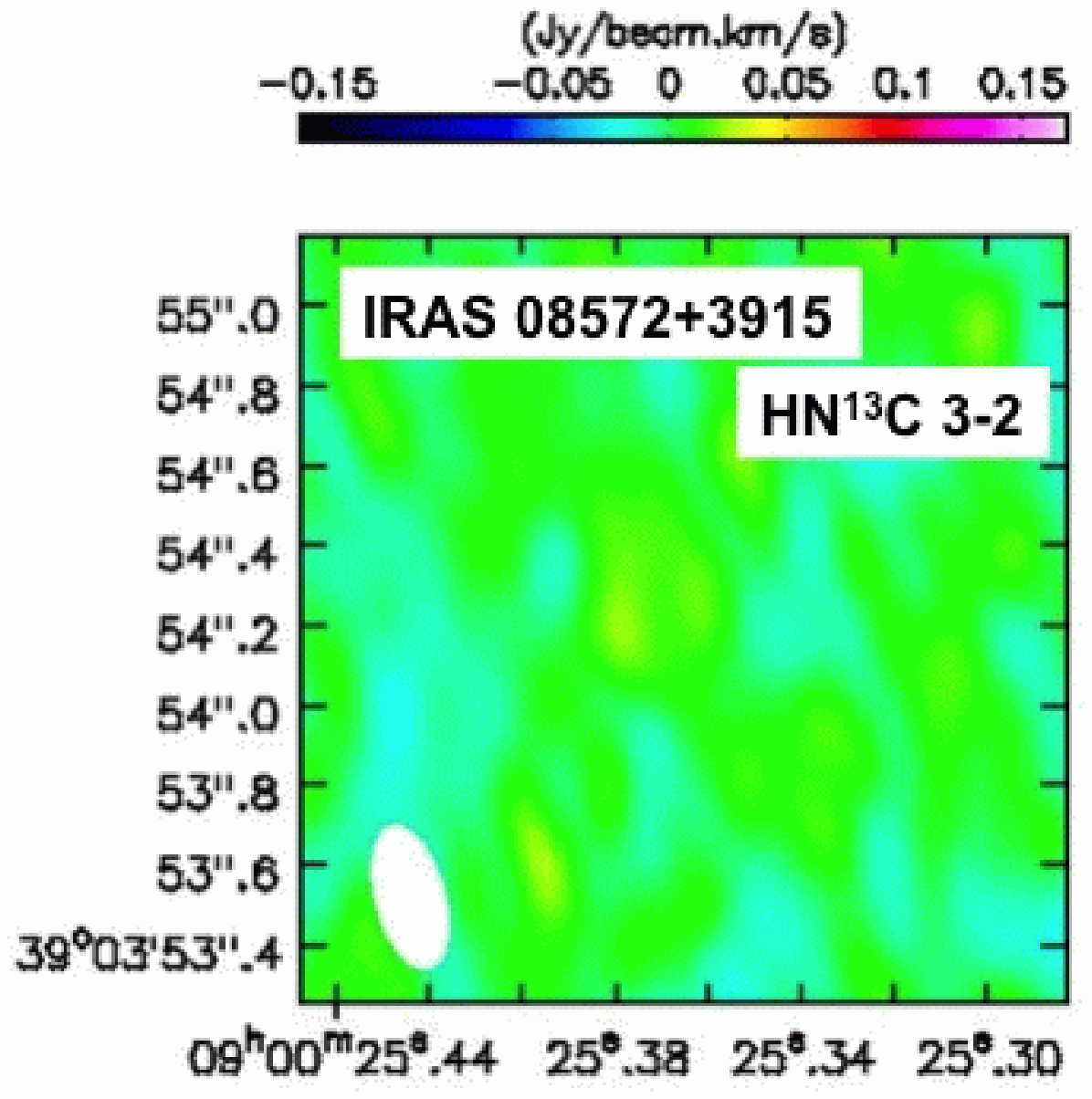} \\
\vspace{-1.2cm}
\includegraphics[angle=0,scale=.41]{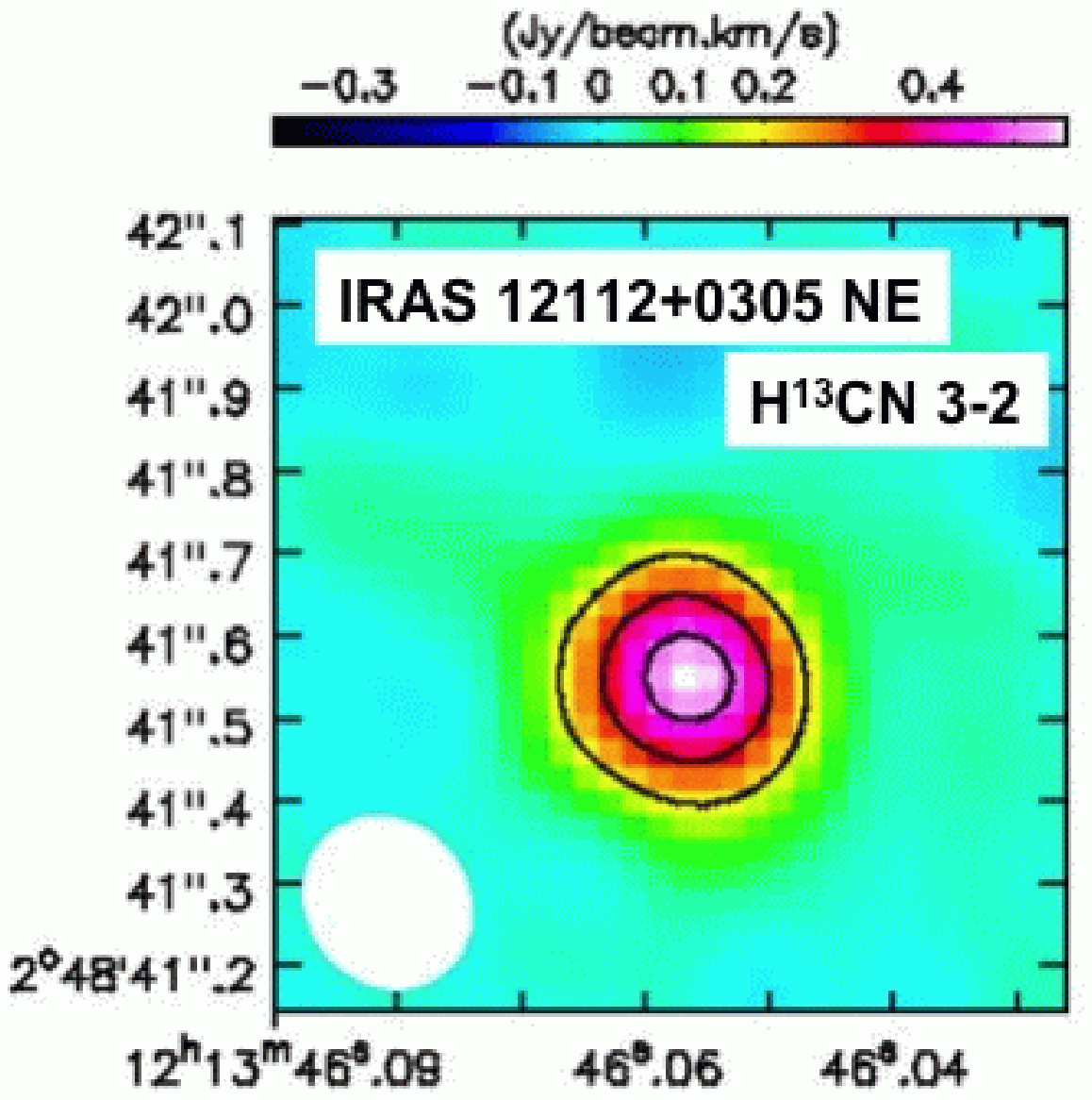} 
\includegraphics[angle=0,scale=.41]{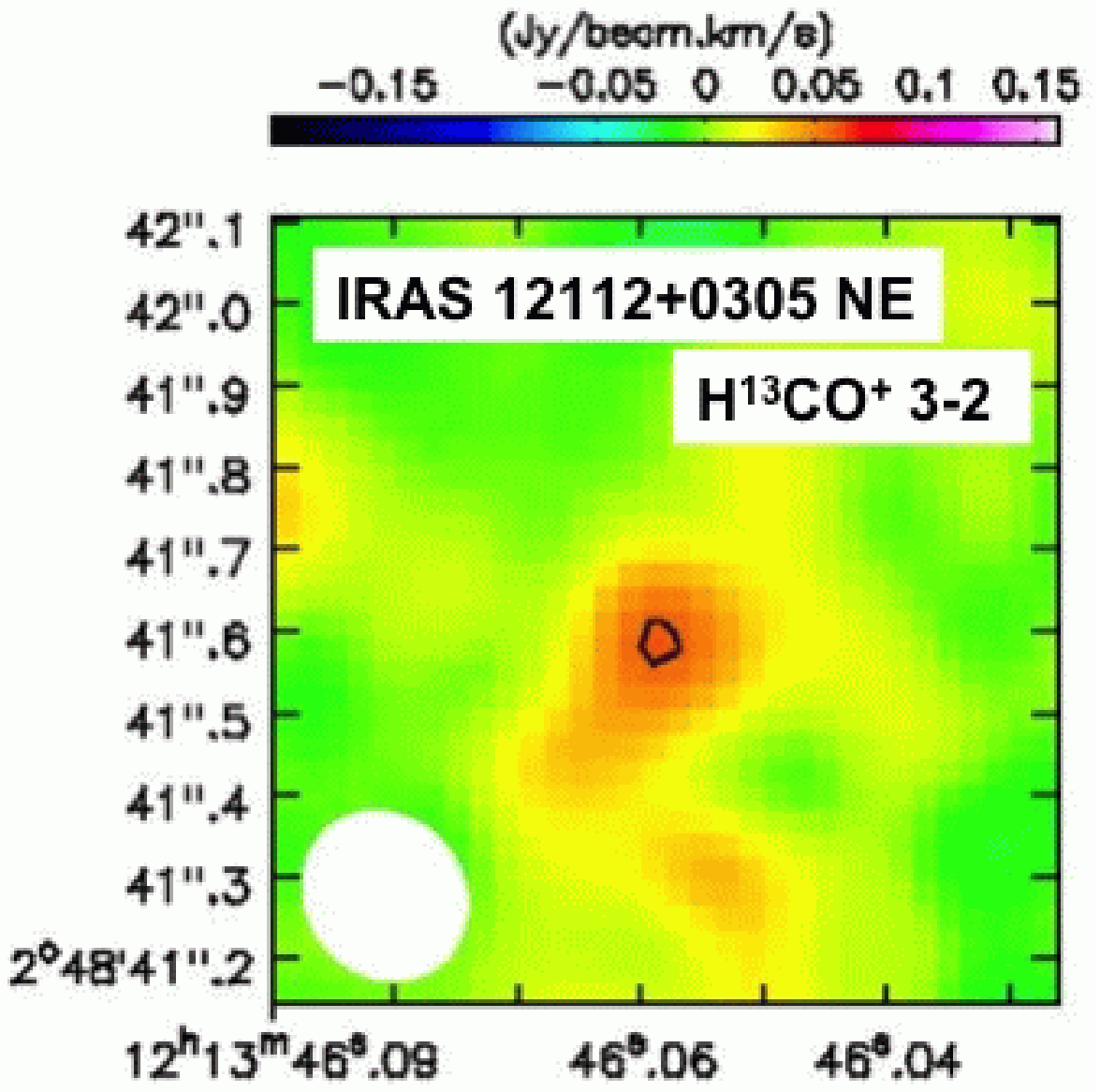} 
\includegraphics[angle=0,scale=.41]{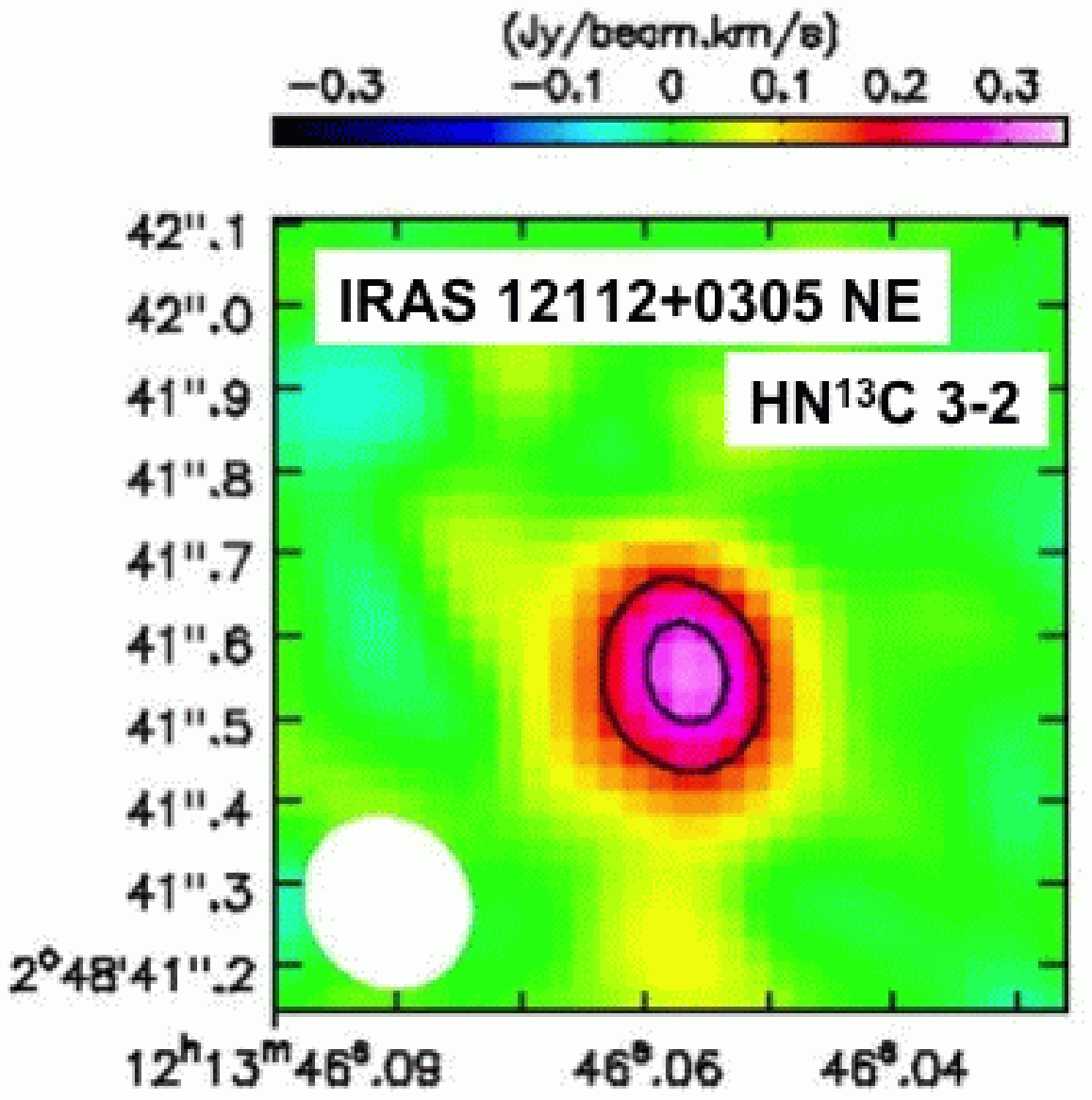} \\
\vspace{-1.2cm}
\includegraphics[angle=0,scale=.41]{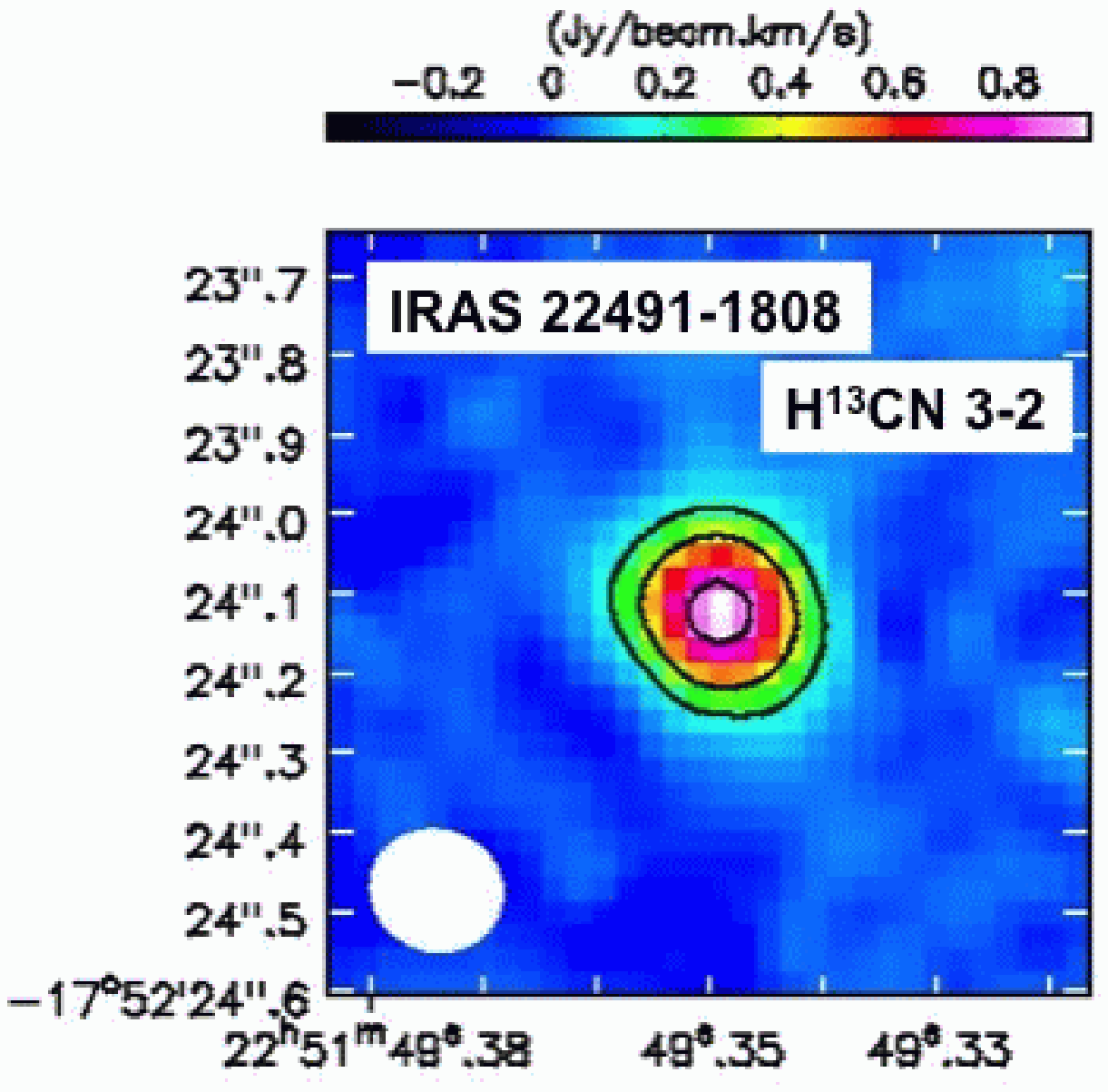} 
\includegraphics[angle=0,scale=.41]{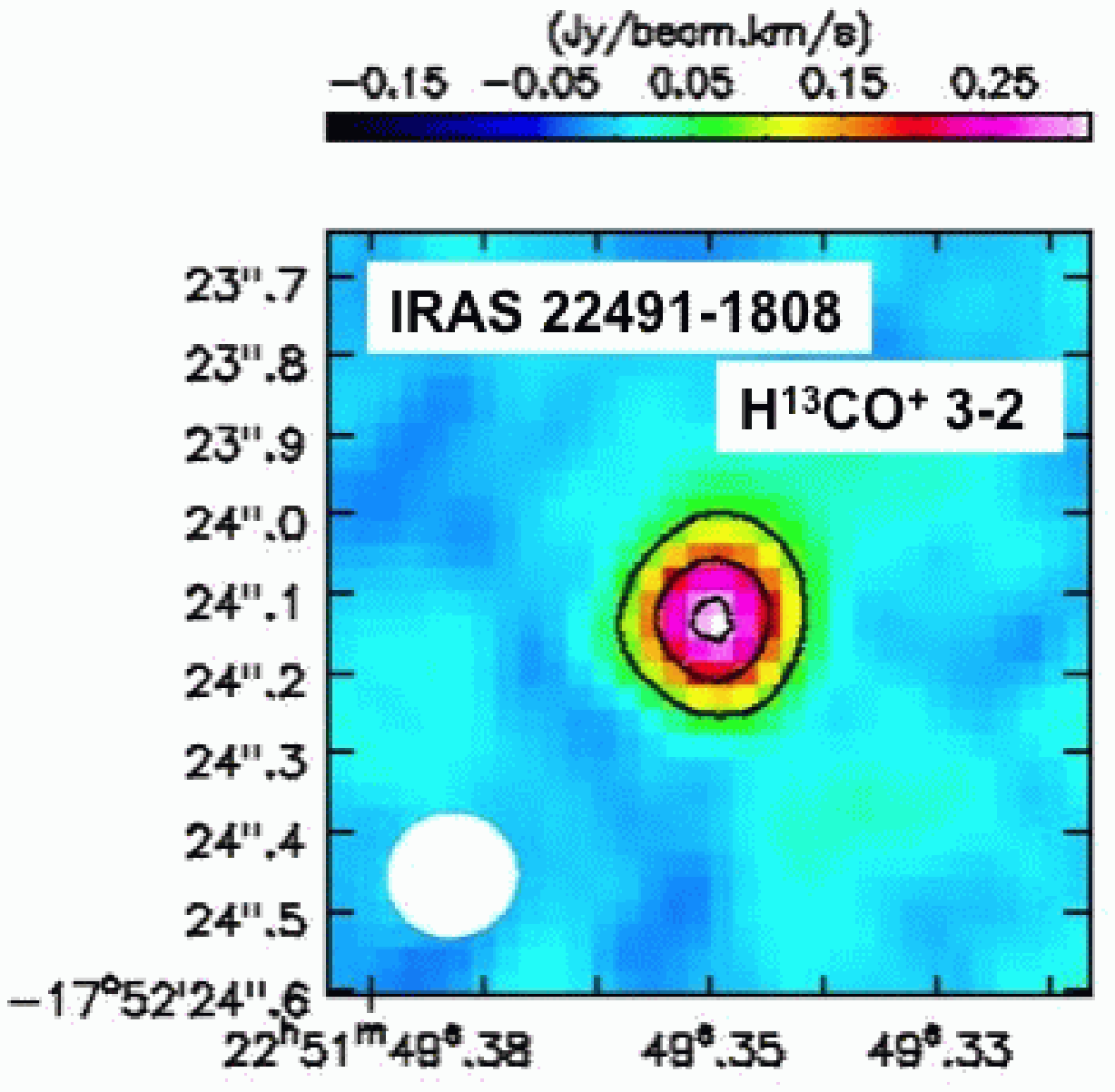} 
\includegraphics[angle=0,scale=.41]{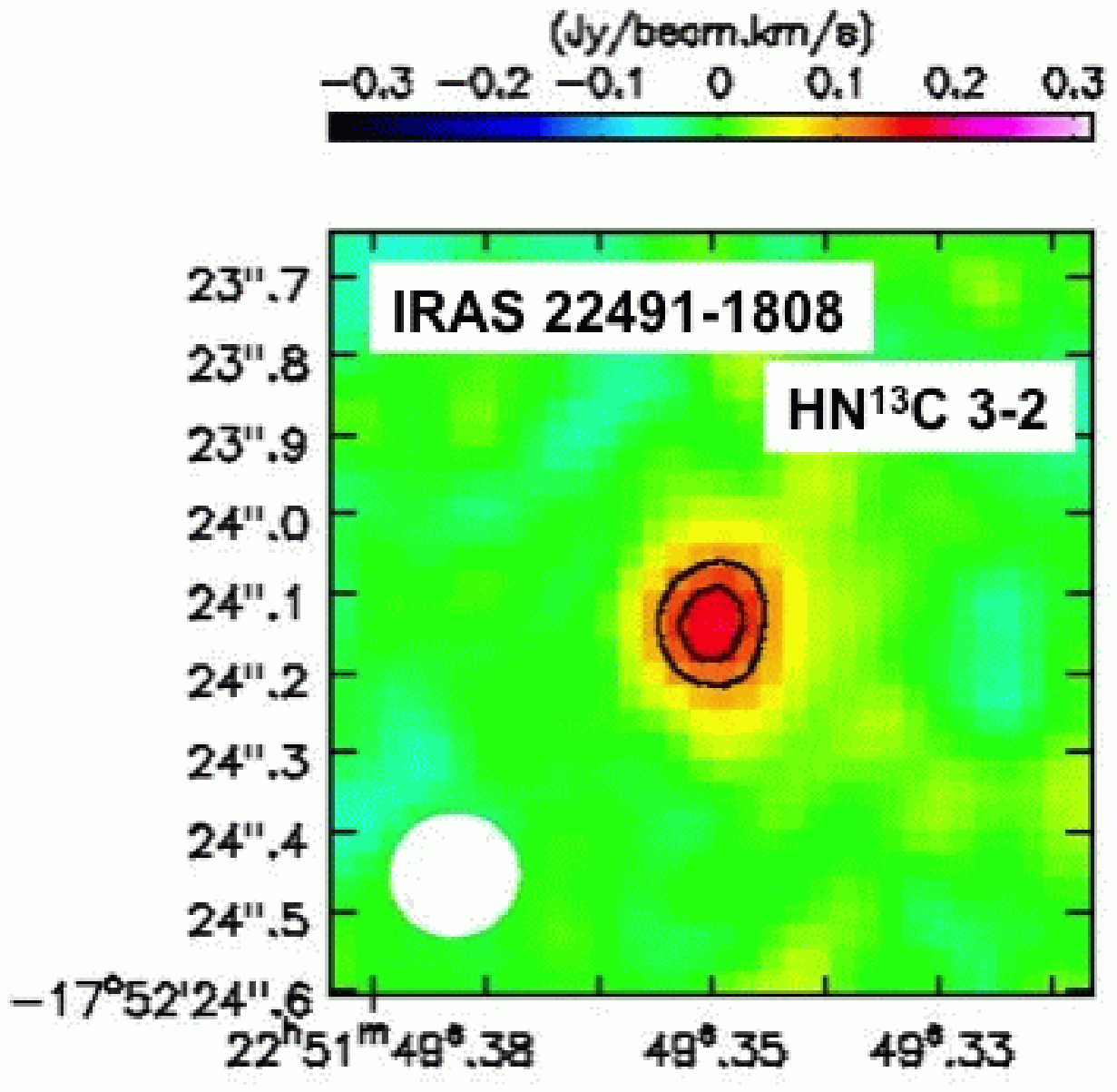} \\
\end{center}
\caption{
Integrated intensity (moment 0) maps of the isotopologue 
H$^{13}$CN J=3--2, H$^{13}$CO$^{+}$ J=3--2, and HN$^{13}$C J=3--2
emission lines taken in Cycle 4 observations.
The abscissa and ordinate are R.A. (J2000) and decl. (J2000),
respectively. 
The contours represent 
2$\sigma$, 2.7$\sigma$ for IRAS 08572$+$3915  H$^{13}$CN J=3--2,
1.8$\sigma$ for IRAS 08572$+$3915 H$^{13}$CO$^{+}$ J=3--2,
no contours with $>$1.5$\sigma$ for IRAS 08572$+$3915 HN$^{13}$C J=3--2,
4$\sigma$, 7$\sigma$, 10$\sigma$ for IRAS 12112$+$0305 NE H$^{13}$CN J=3--2,
2.5$\sigma$ for IRAS 12112$+$0305 NE H$^{13}$CO$^{+}$ J=3--2,
4$\sigma$, 6$\sigma$ for IRAS 12112$+$0305 NE HN$^{13}$C J=3--2,
4$\sigma$, 8$\sigma$, 16$\sigma$ for IRAS 22491$-$1808 H$^{13}$CN J=3--2,
3$\sigma$, 7$\sigma$, 11$\sigma$ for IRAS 22491$-$1808 H$^{13}$CO$^{+}$ J=3--2,
and 
3$\sigma$, 4$\sigma$ for IRAS 22491$-$1808 HN$^{13}$C J=3--2.
Beam sizes are shown as filled circles in the lower-left region.
}
\end{figure}

\begin{figure}
\begin{center}
\includegraphics[angle=0,scale=.273]{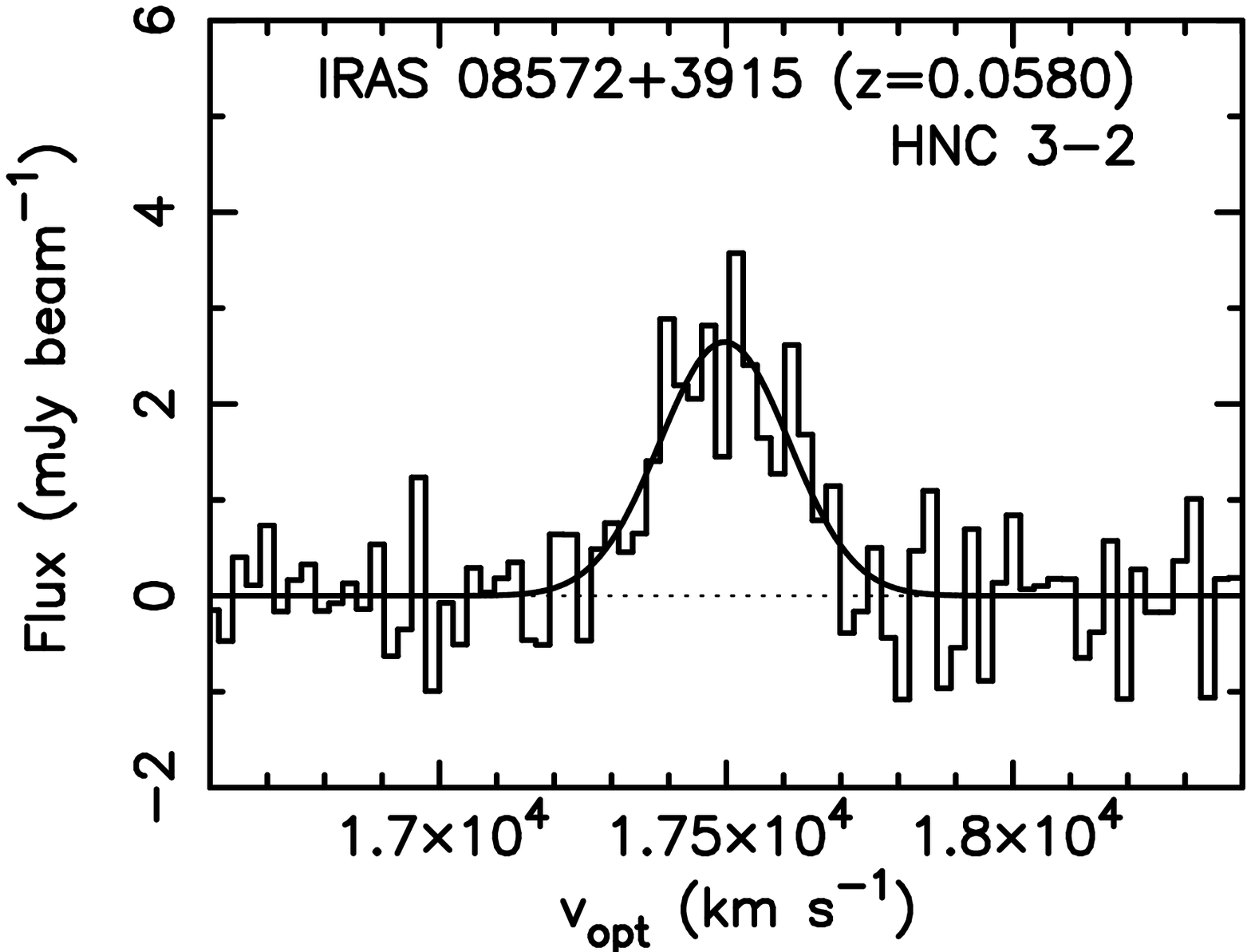} 
\includegraphics[angle=0,scale=.273]{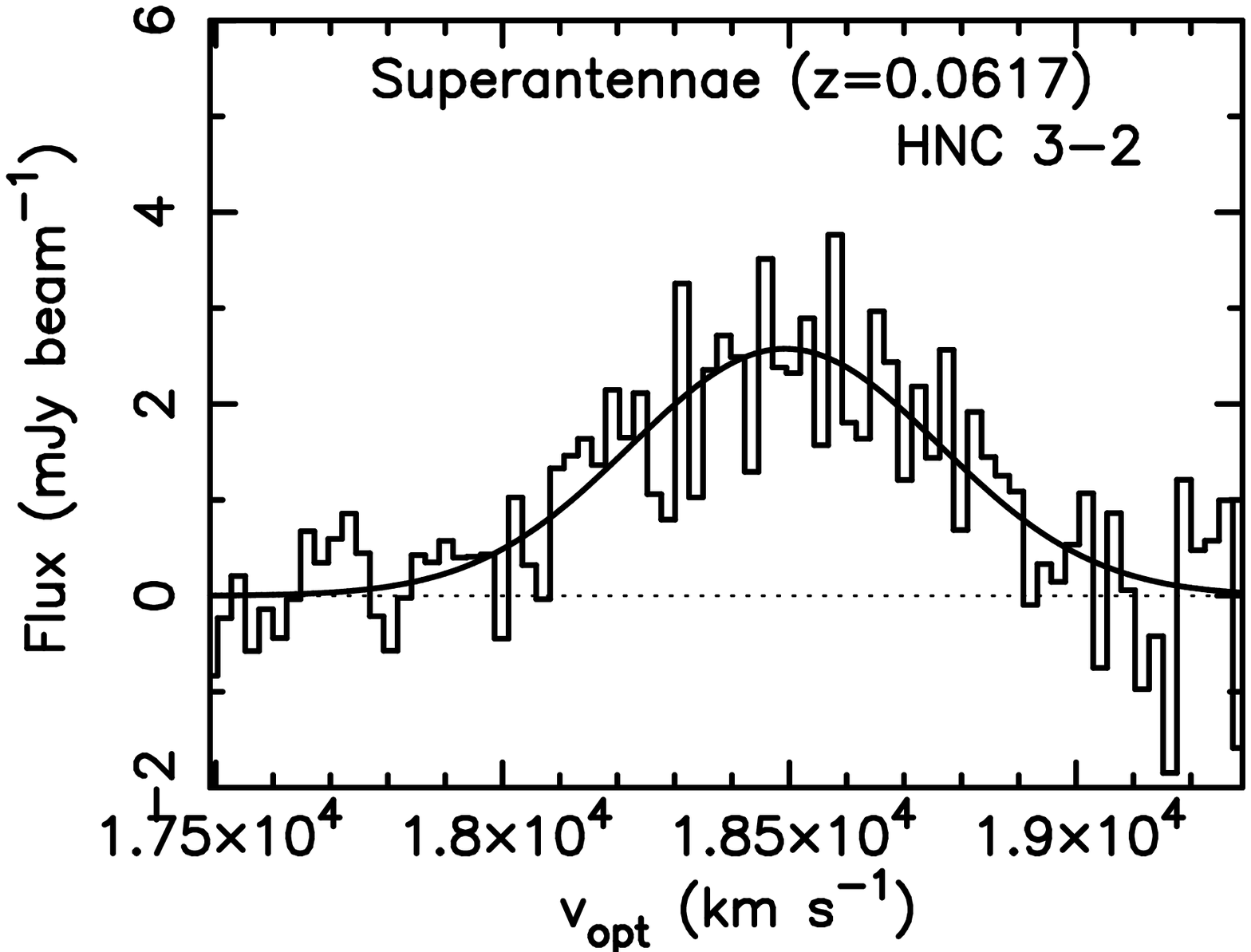} 
\includegraphics[angle=0,scale=.273]{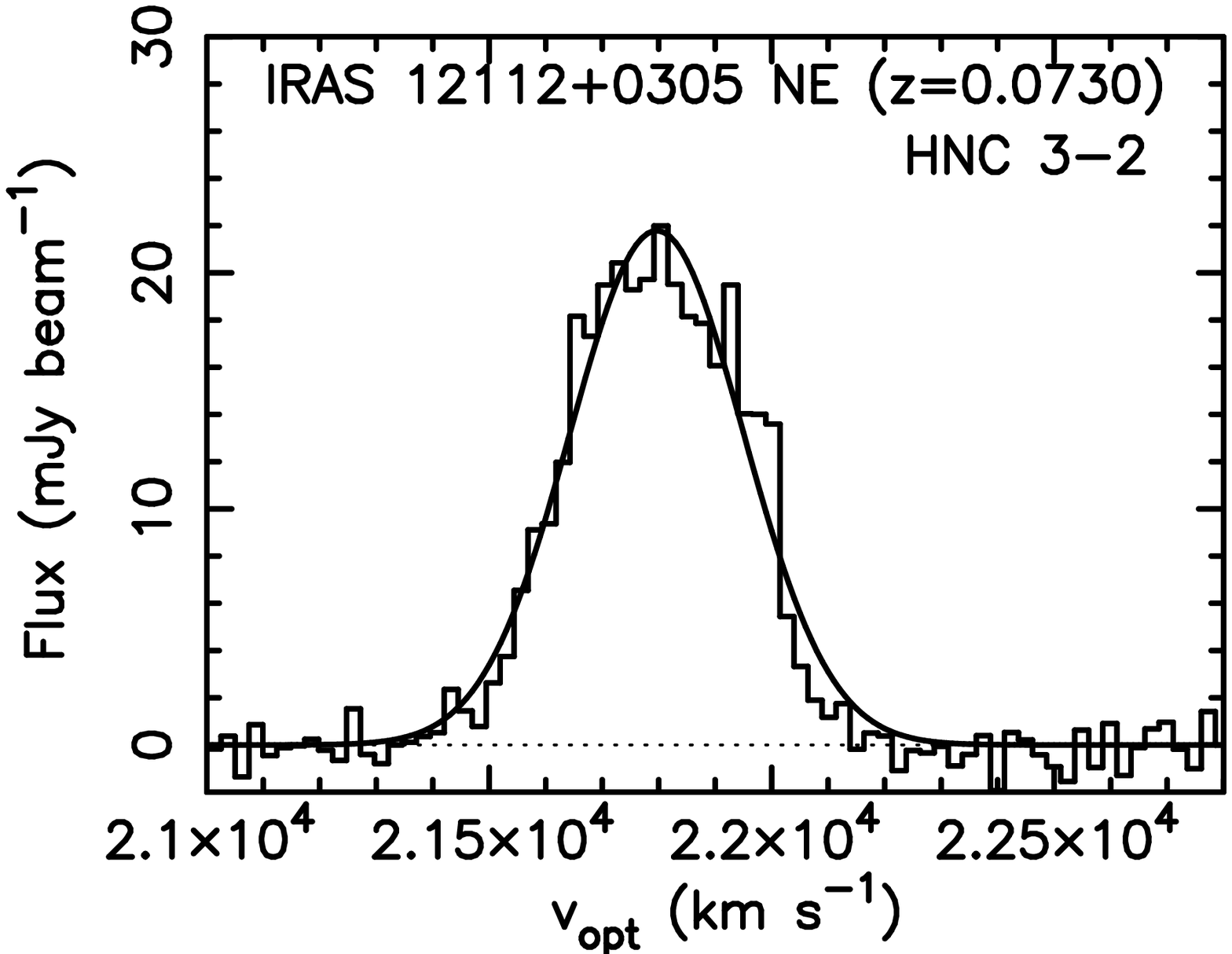} \\
\includegraphics[angle=0,scale=.273]{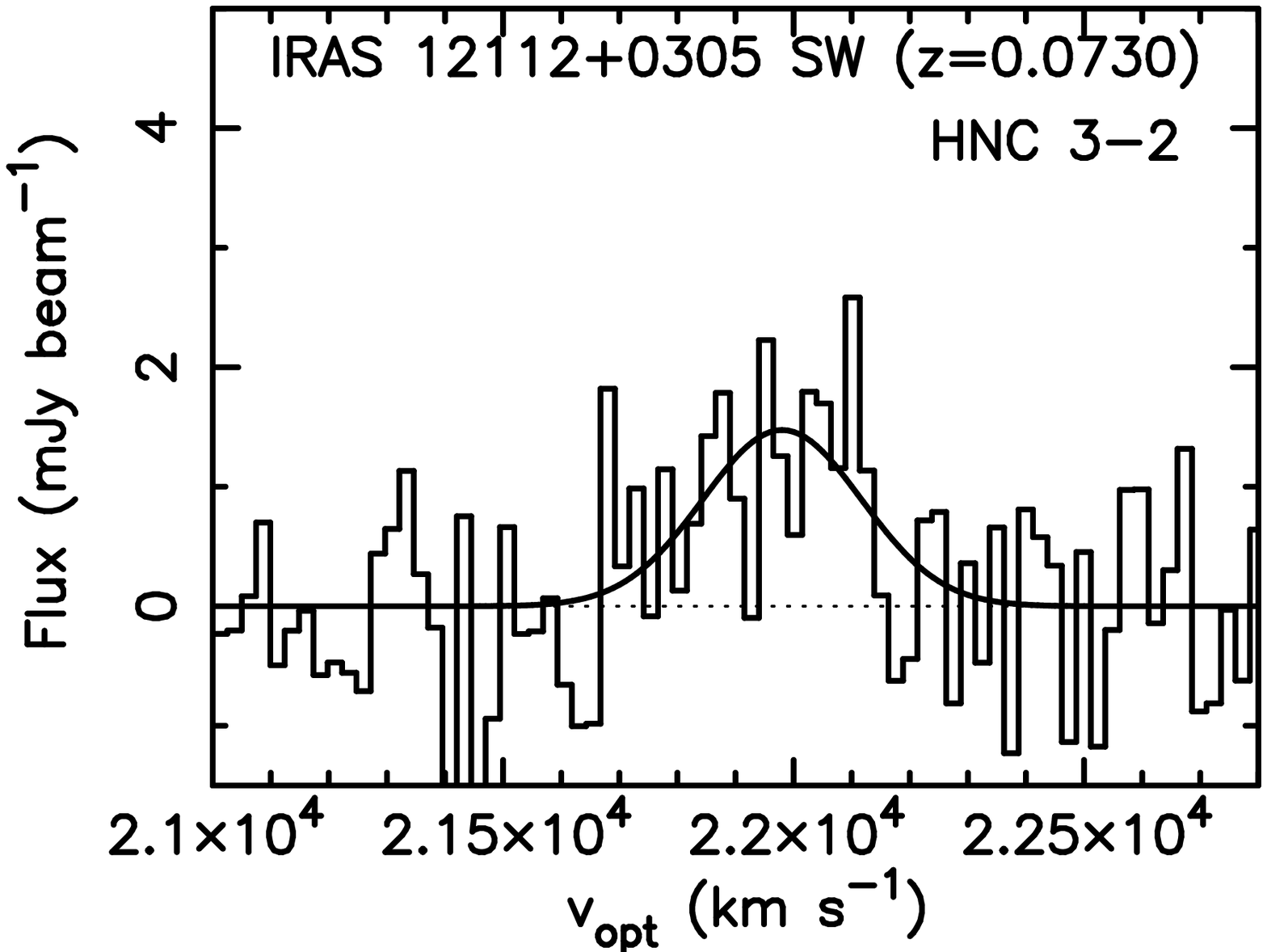} 
\includegraphics[angle=0,scale=.273]{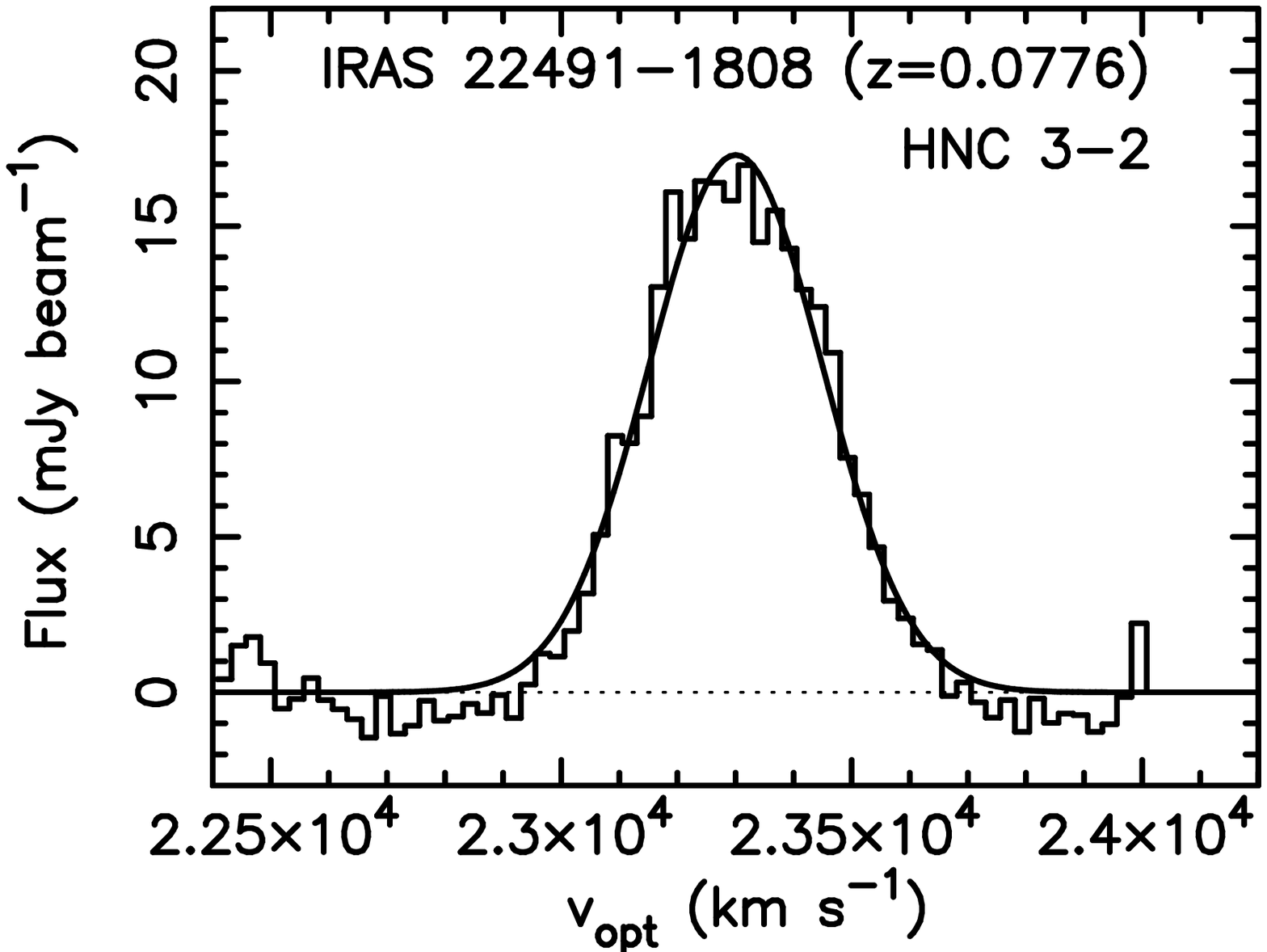} 
\includegraphics[angle=0,scale=.273]{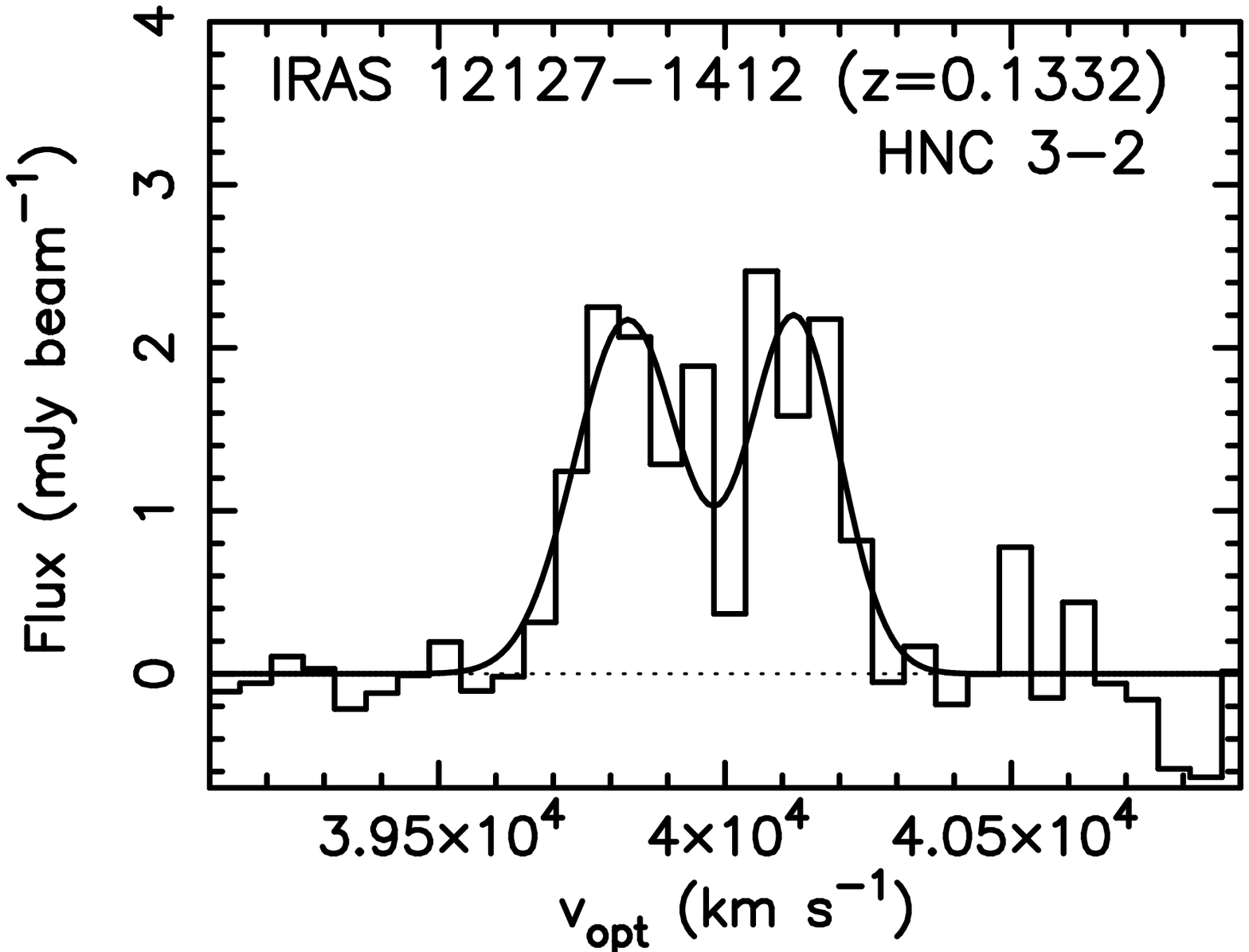} \\
\includegraphics[angle=0,scale=.273]{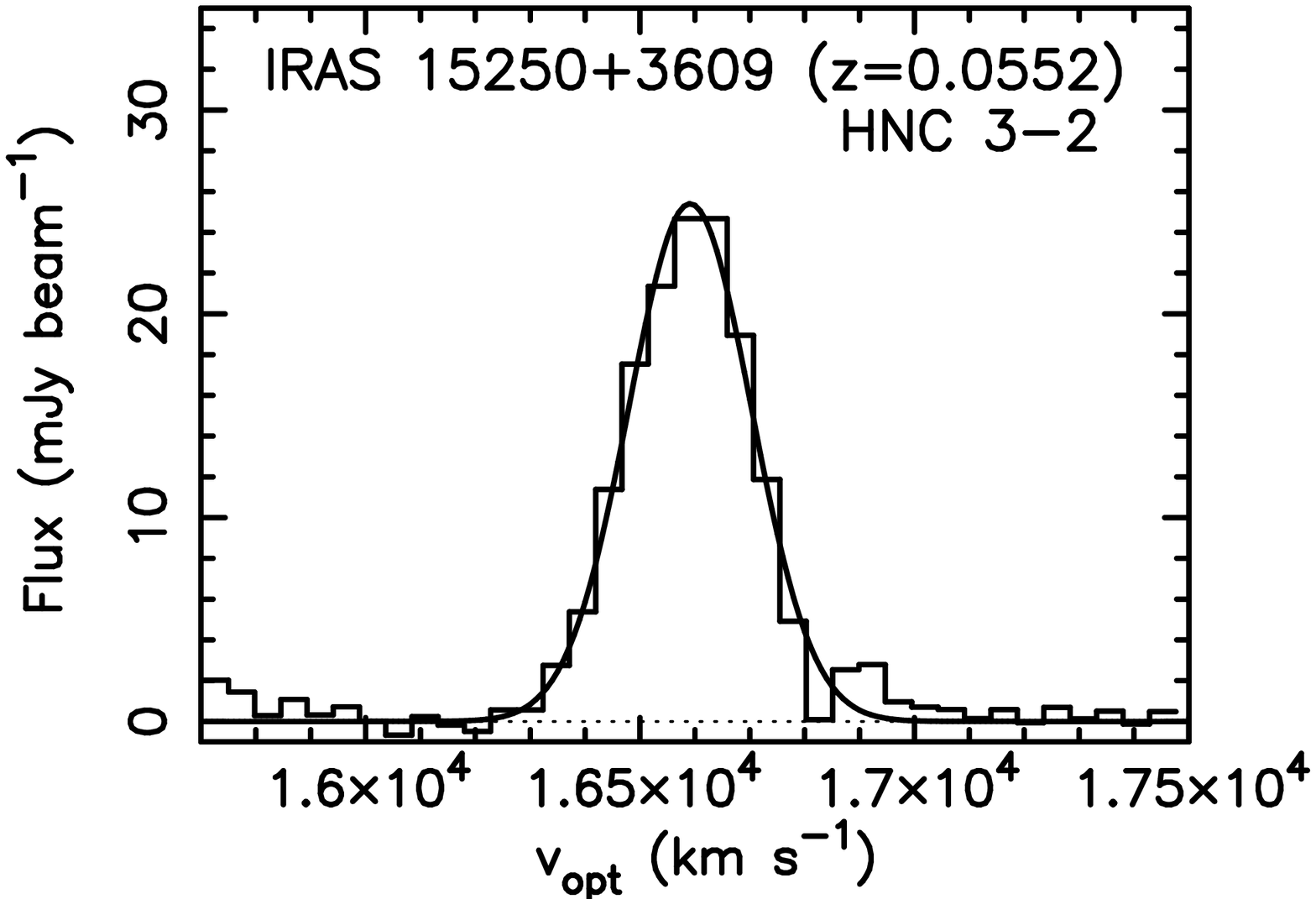} 
\includegraphics[angle=0,scale=.273]{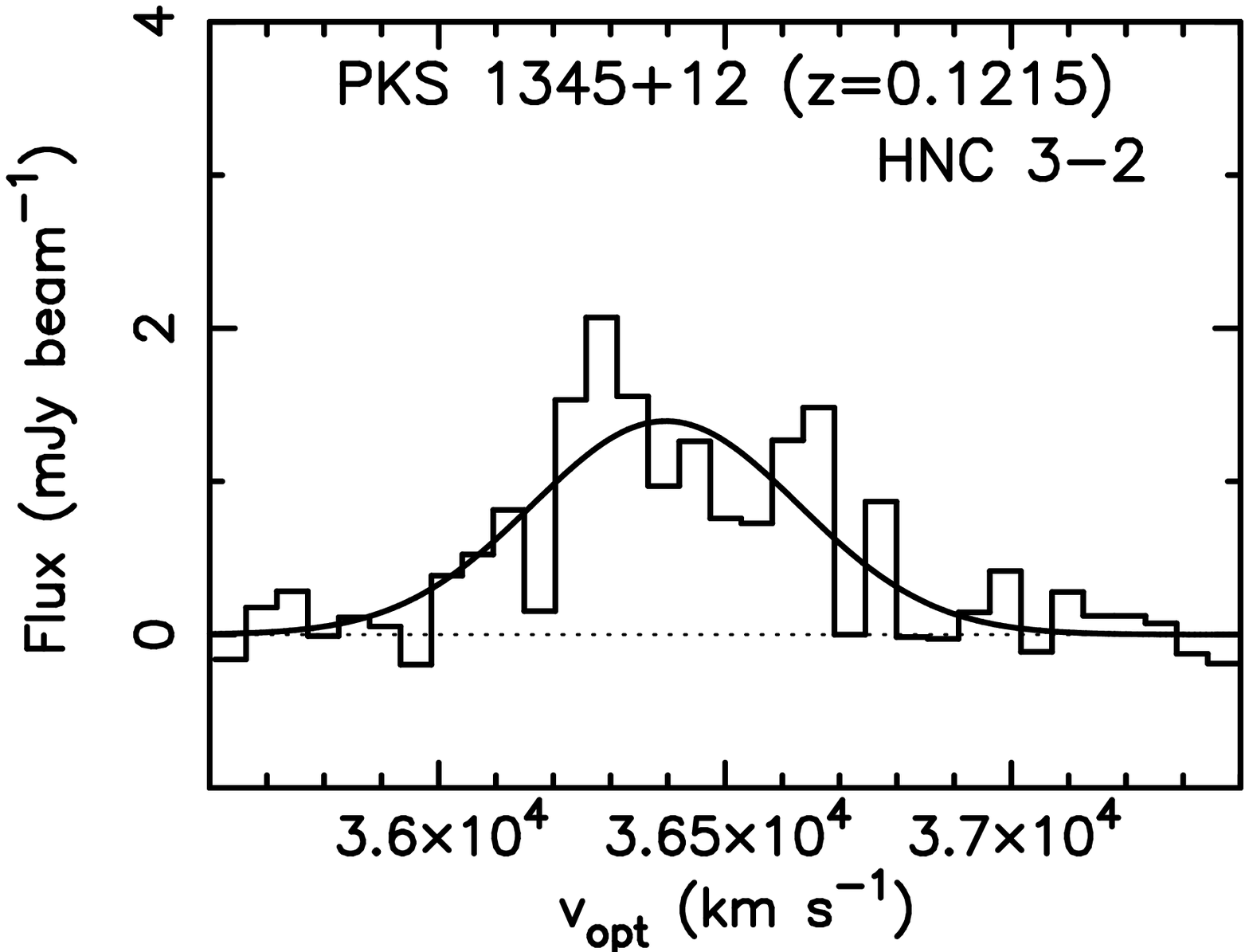} 
\includegraphics[angle=0,scale=.273]{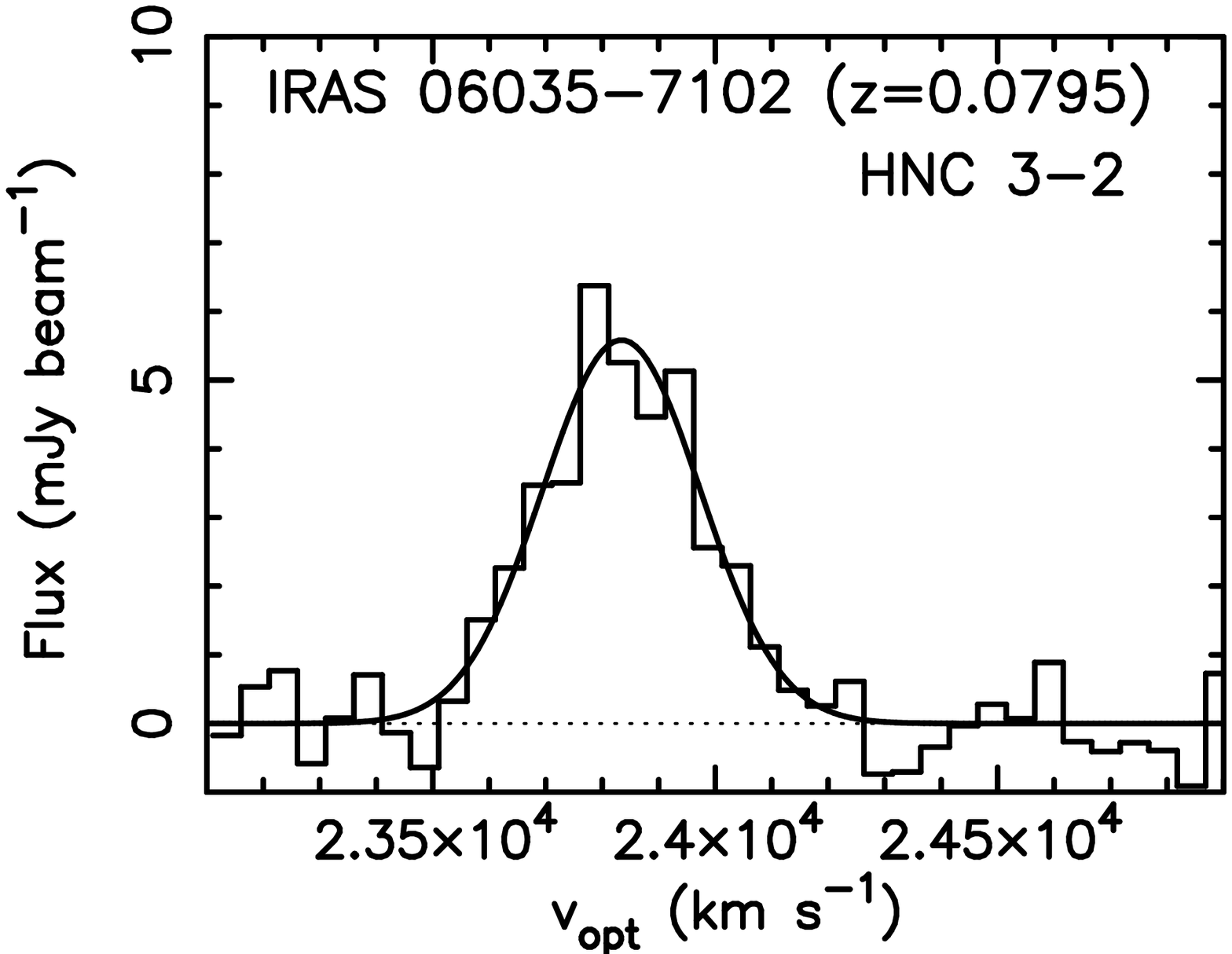} \\
\includegraphics[angle=0,scale=.273]{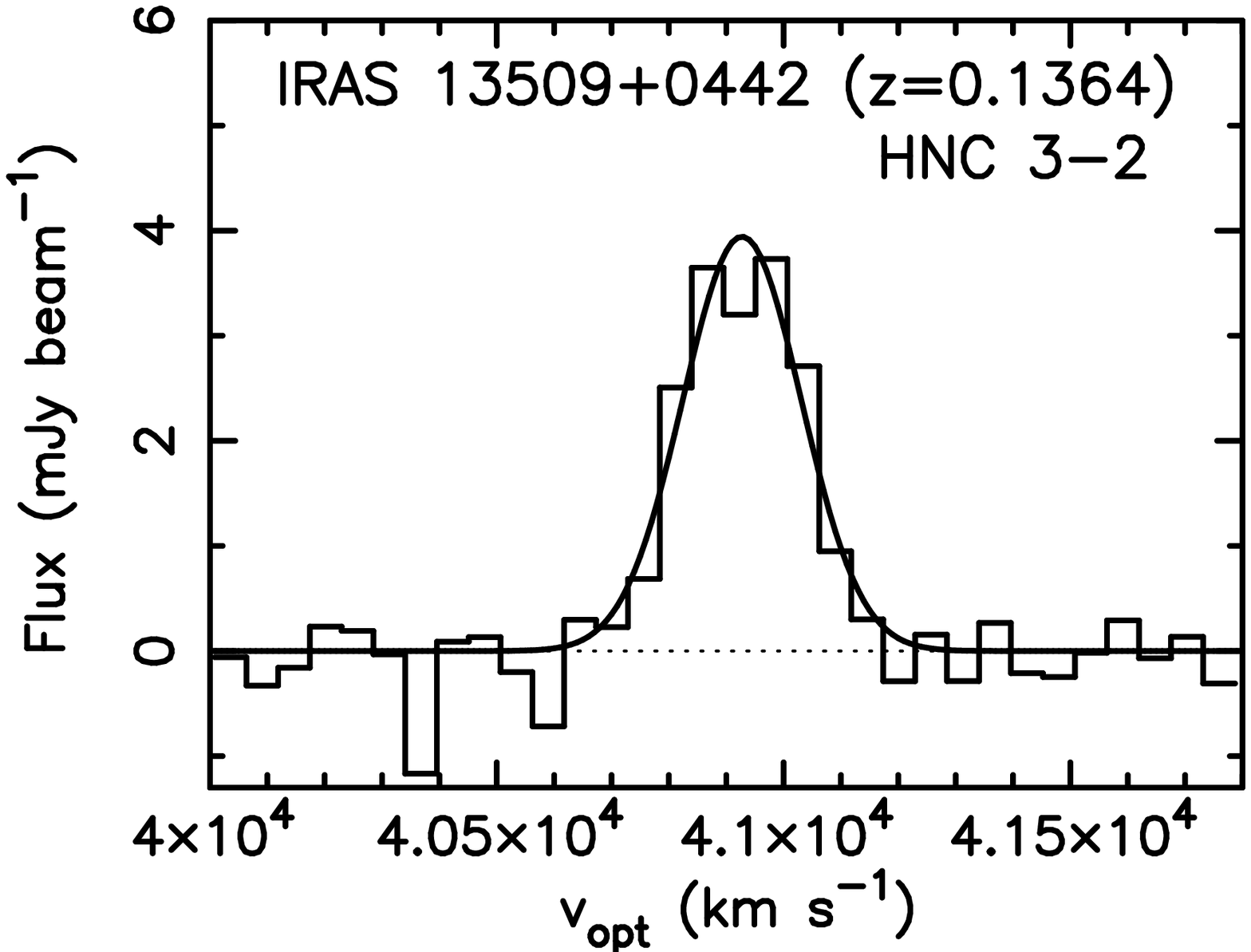} 
\end{center}
\caption{
Gaussian fits to the detected HNC J=3--2 emission lines in band 6
spectra at ULIRG's nuclei within the beam size.    
The abscissa is optical LSR velocity in (km s$^{-1}$) and the ordinate
is flux in (mJy beam$^{-1}$). 
}
\end{figure}

\begin{figure}
\begin{center}
\includegraphics[angle=0,scale=.273]{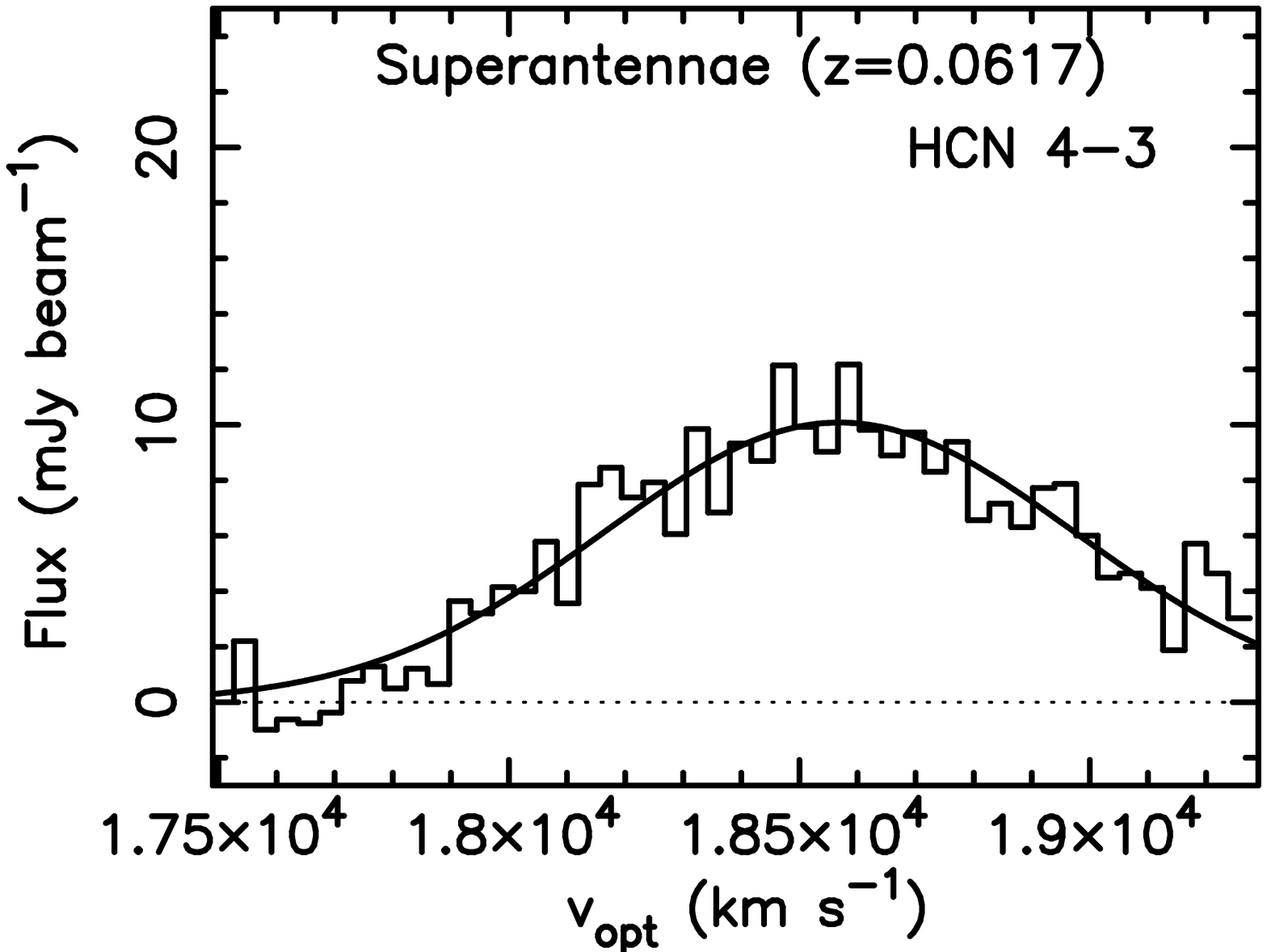}  
\includegraphics[angle=0,scale=.273]{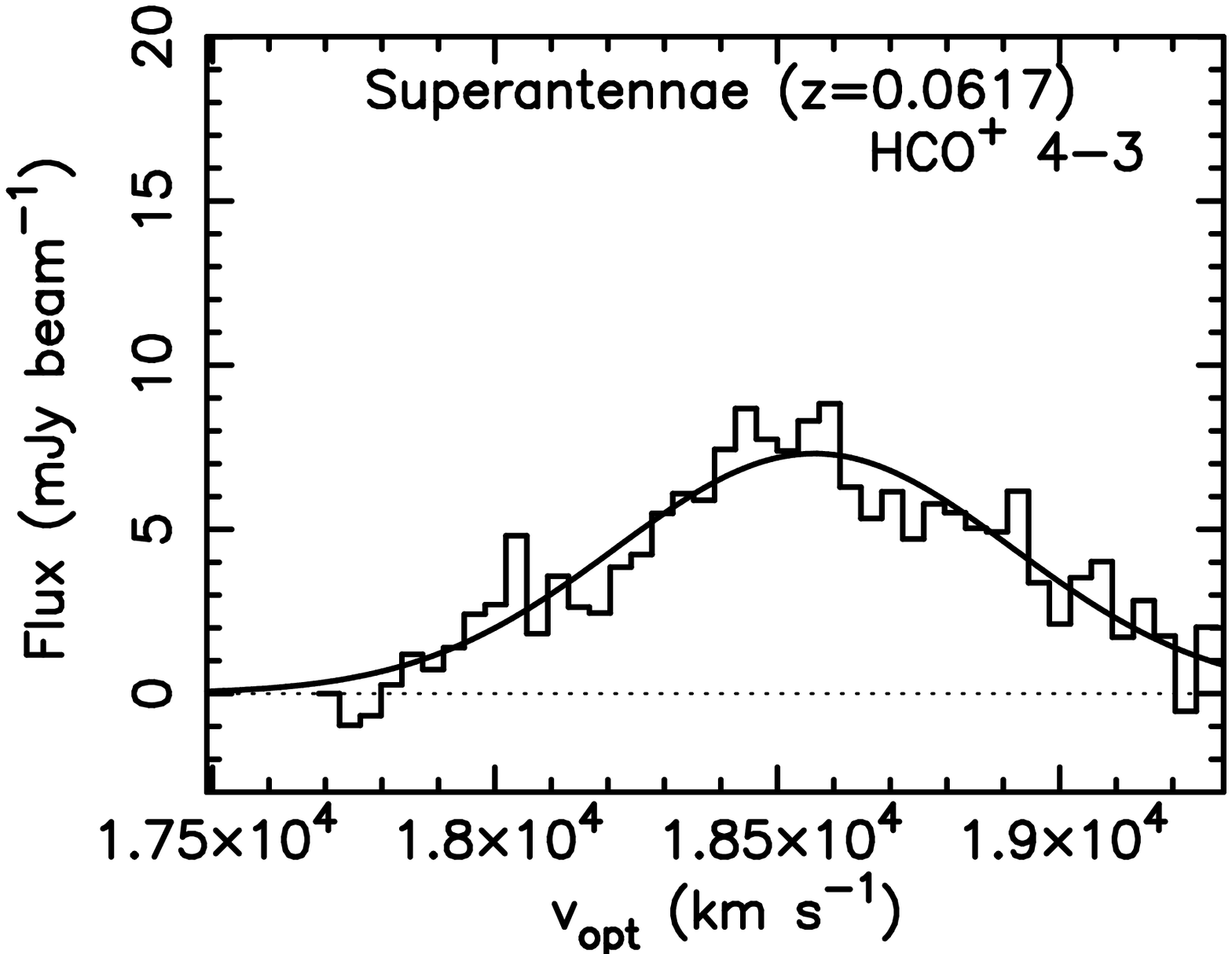} 
\includegraphics[angle=0,scale=.273]{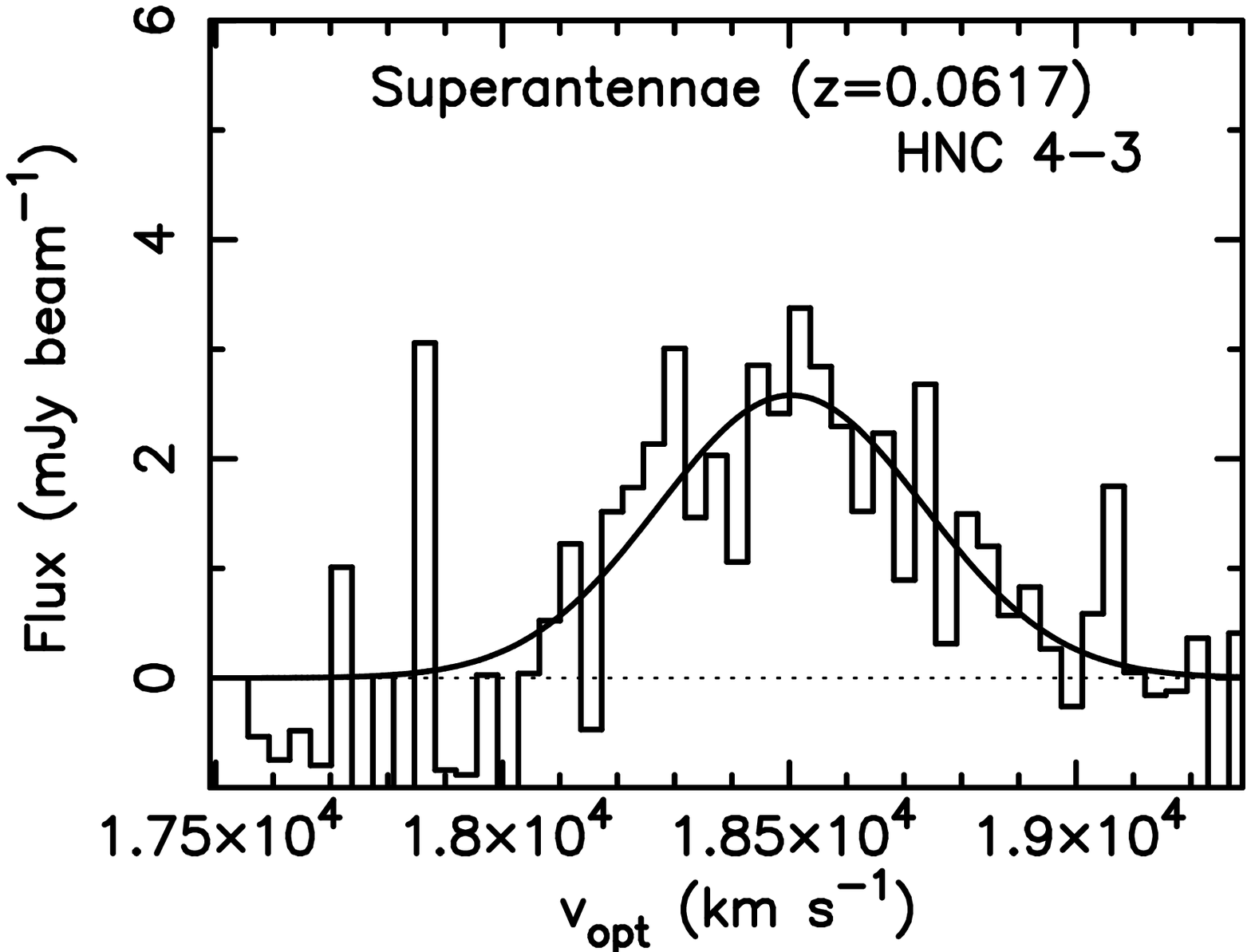} \\
\includegraphics[angle=0,scale=.273]{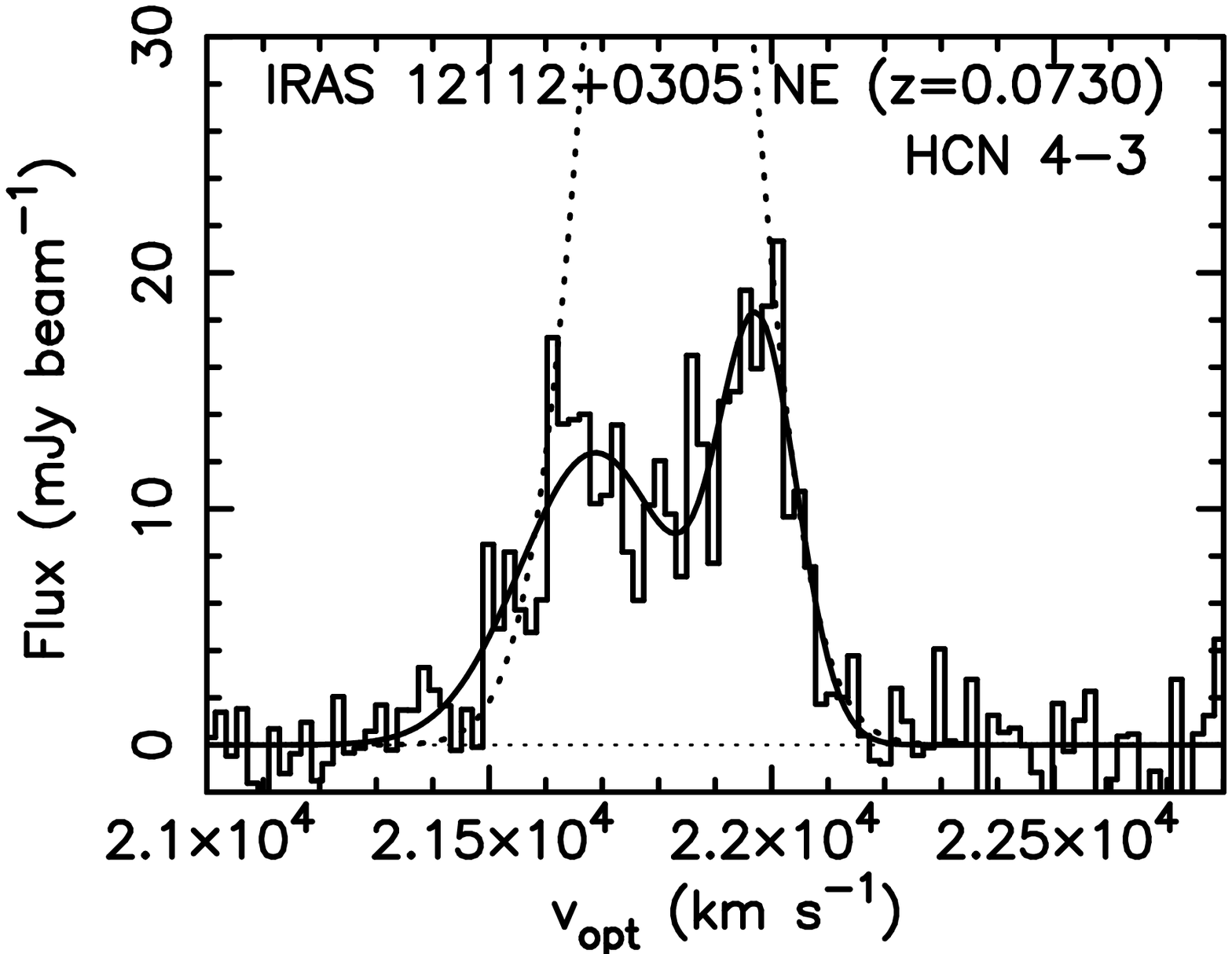} 
\includegraphics[angle=0,scale=.273]{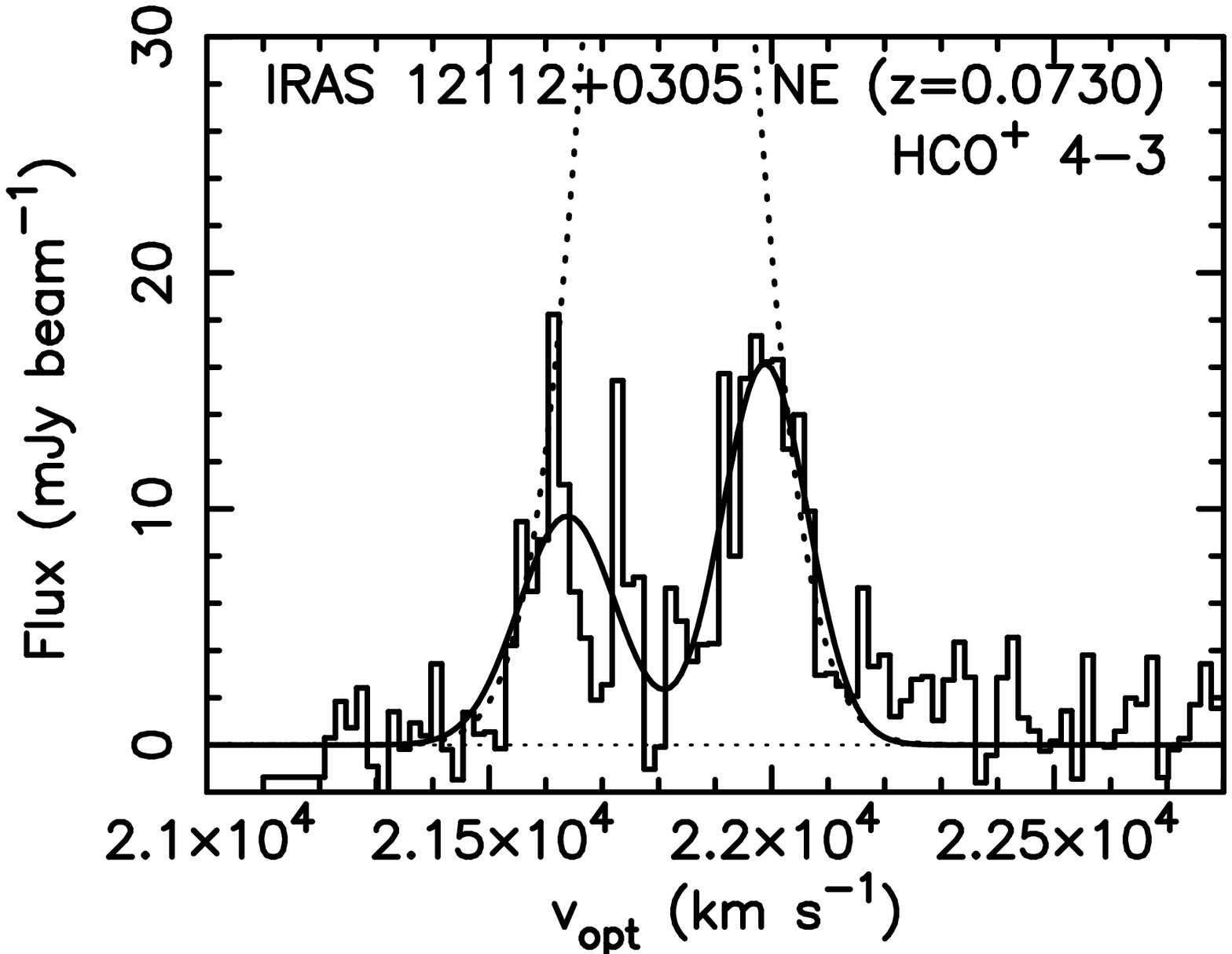} 
\includegraphics[angle=0,scale=.273]{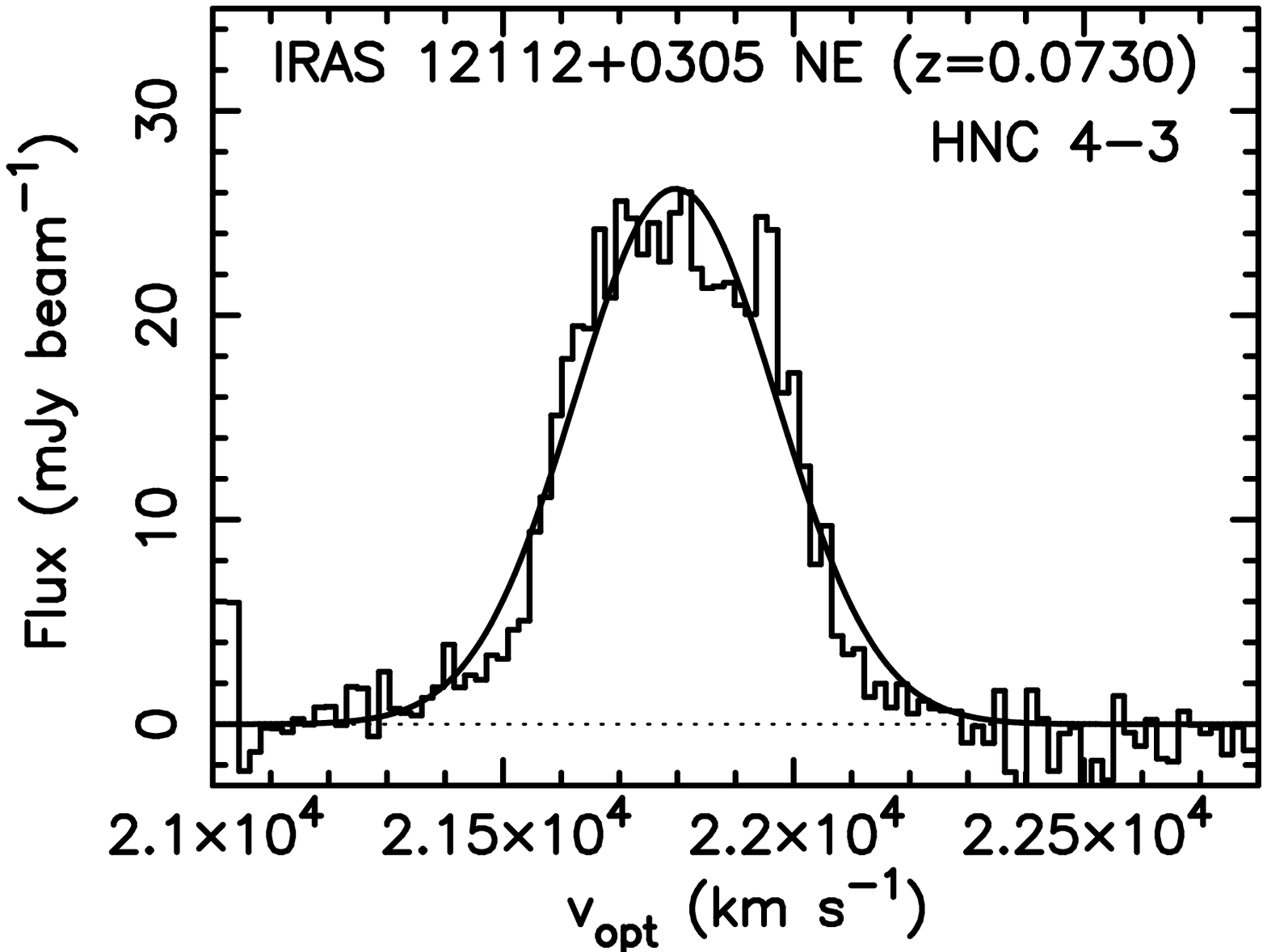} \\
\includegraphics[angle=0,scale=.273]{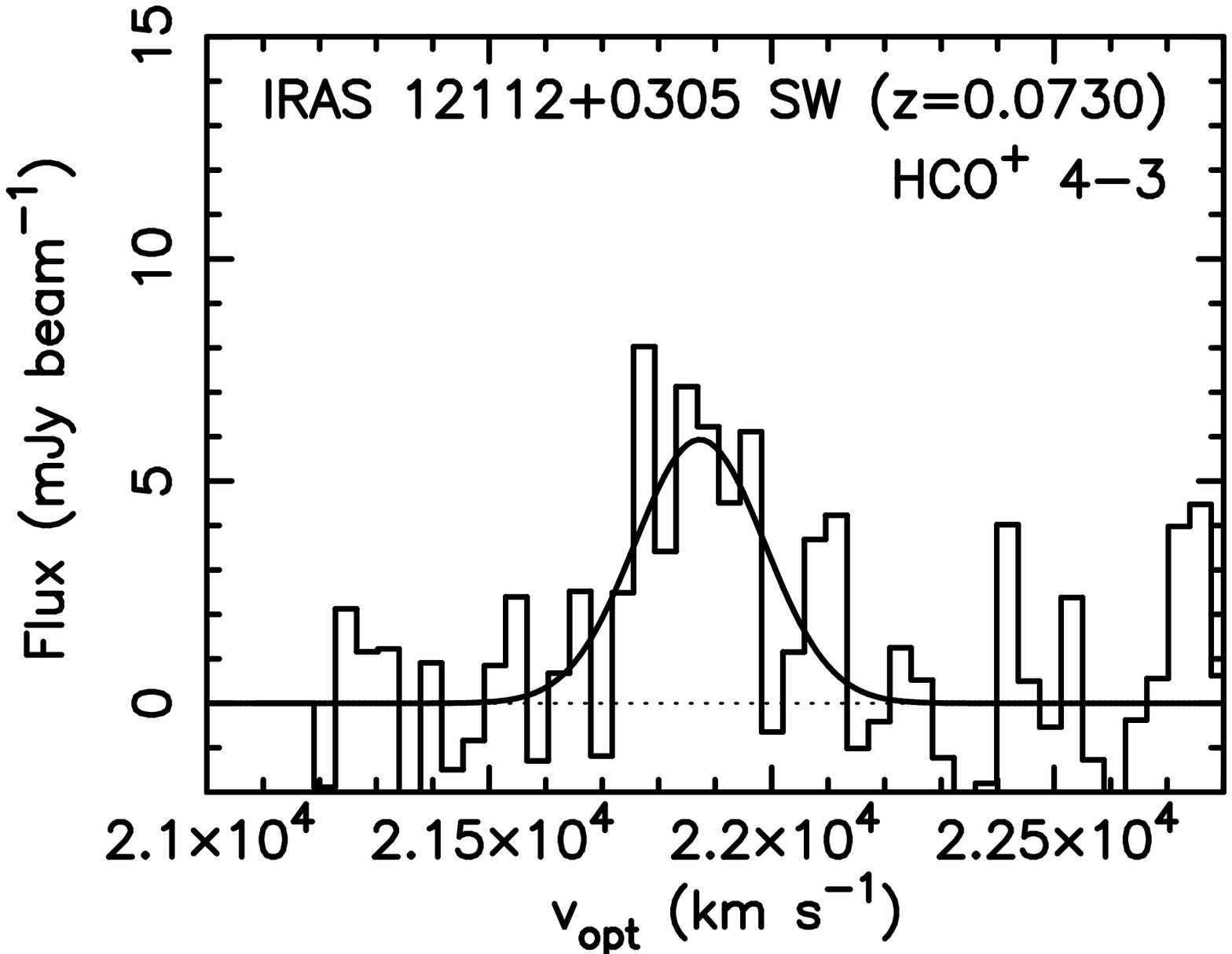} 
\includegraphics[angle=0,scale=.273]{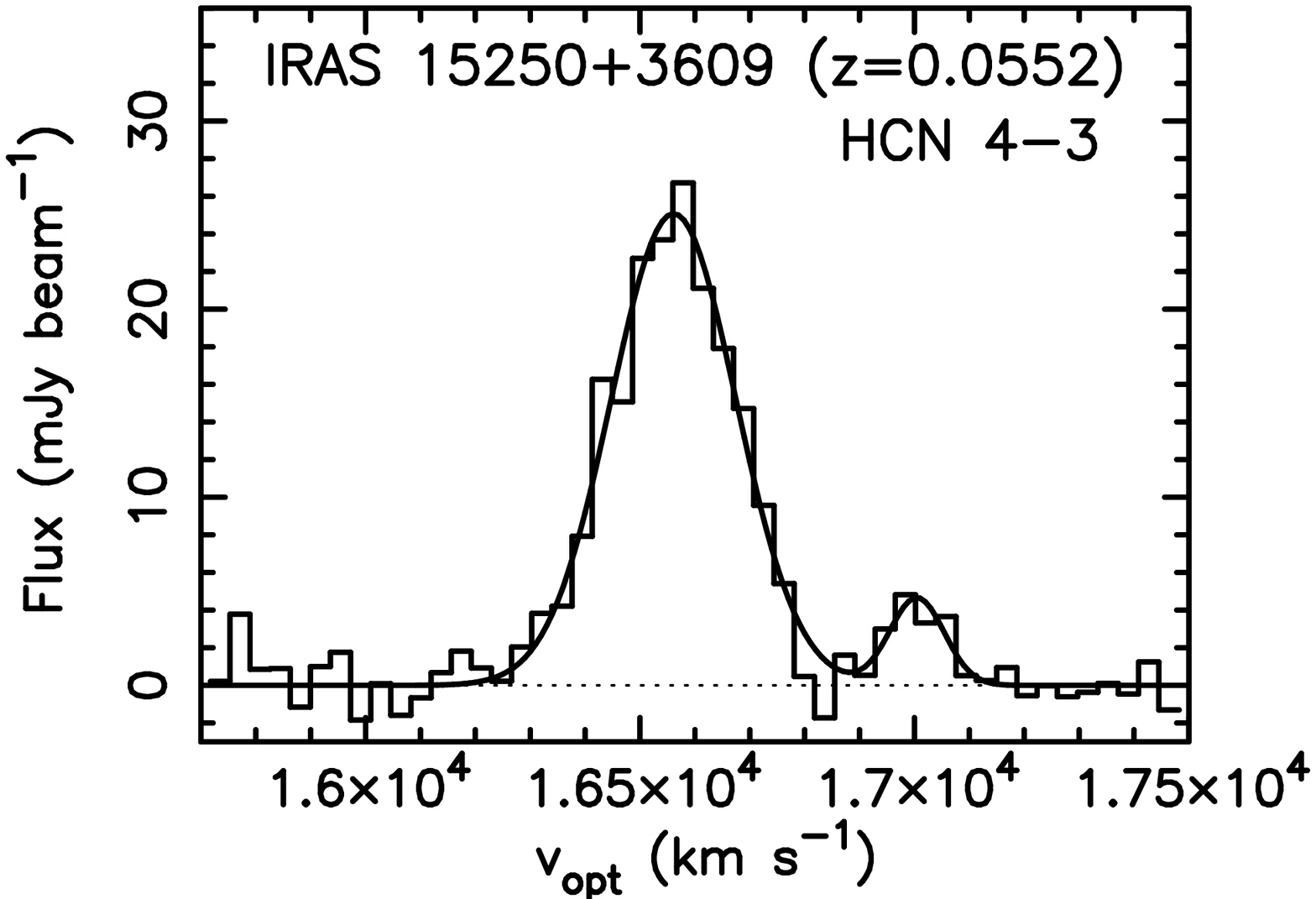} 
\includegraphics[angle=0,scale=.273]{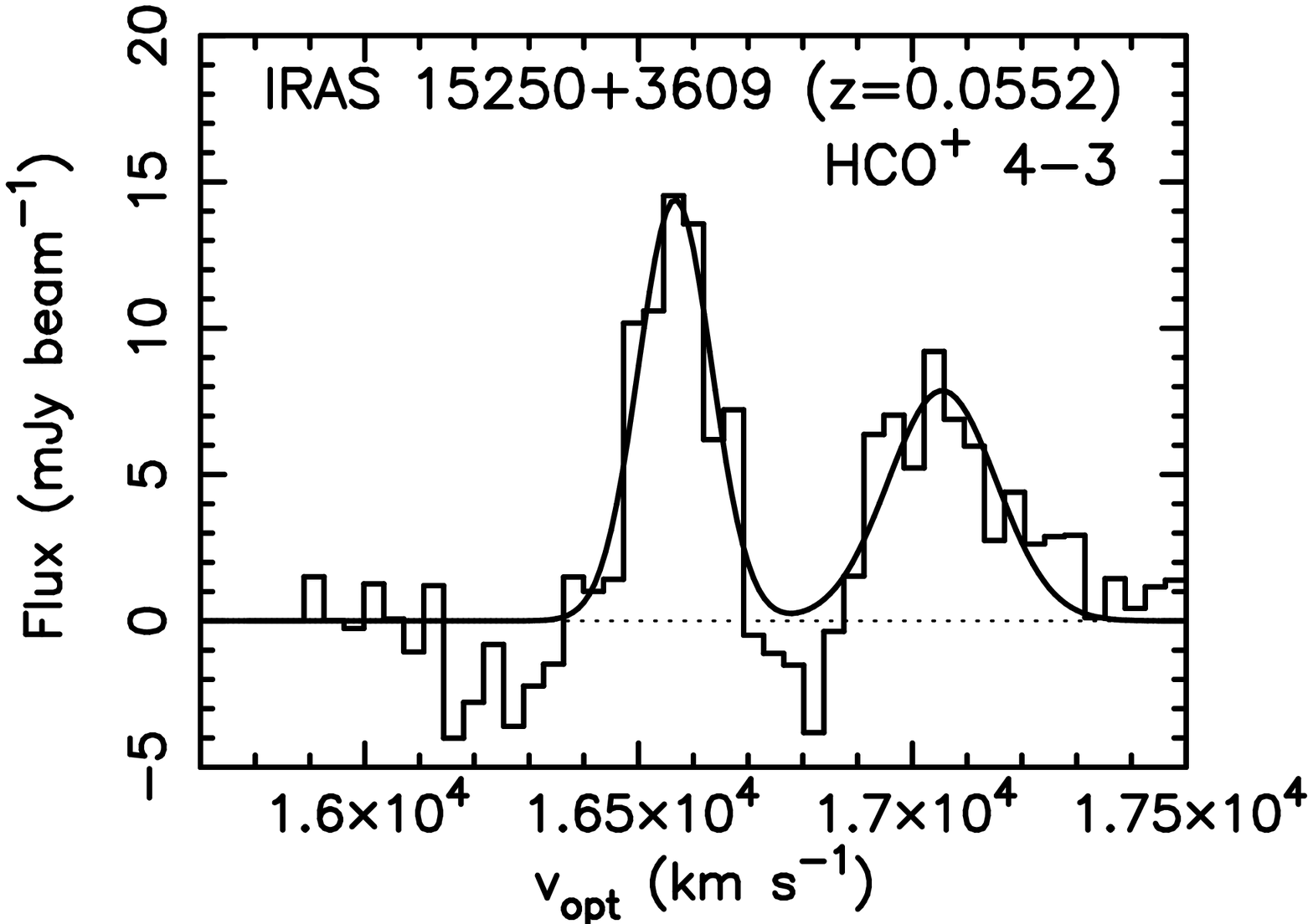} \\
\includegraphics[angle=0,scale=.273]{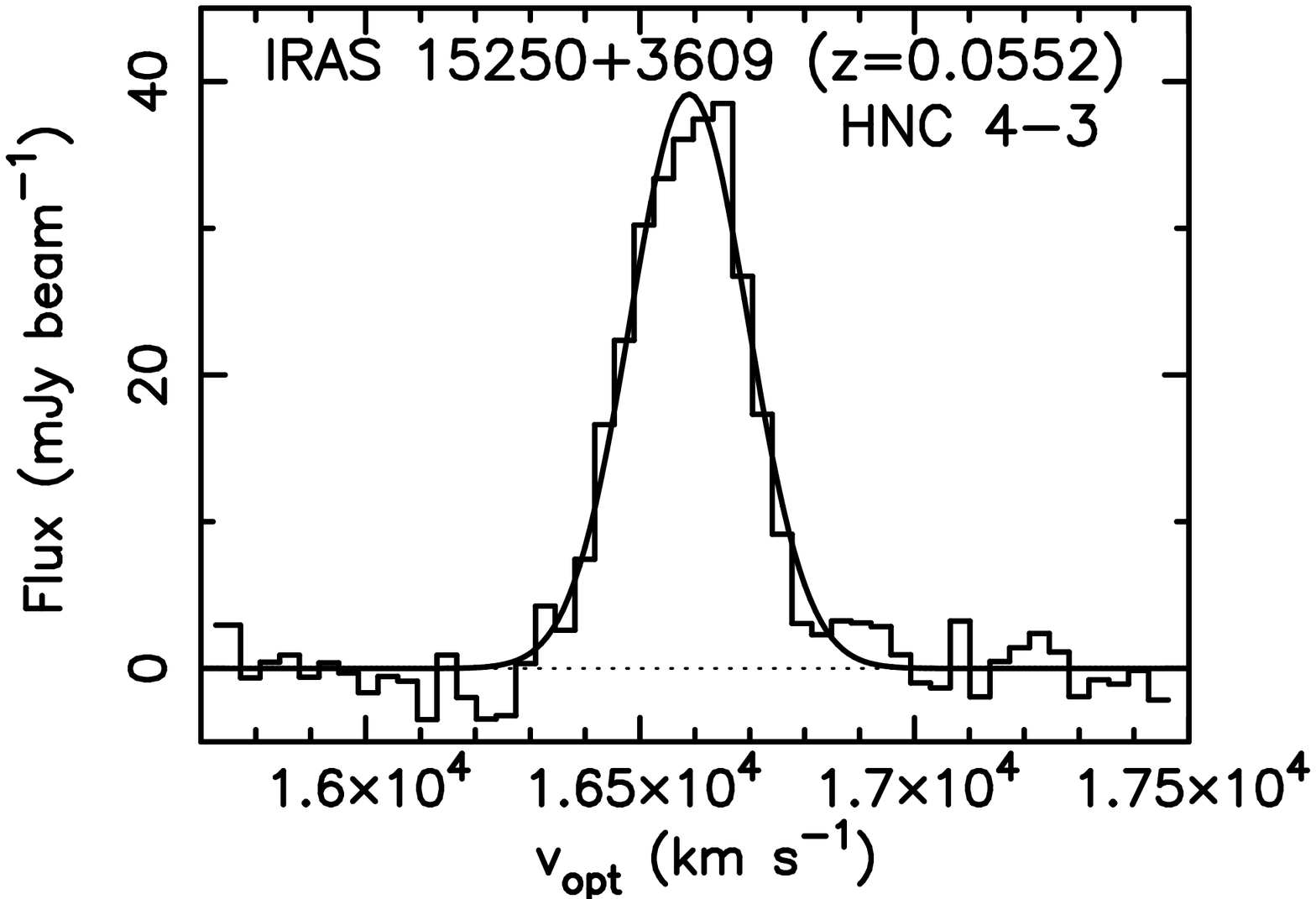} 
\includegraphics[angle=0,scale=.273]{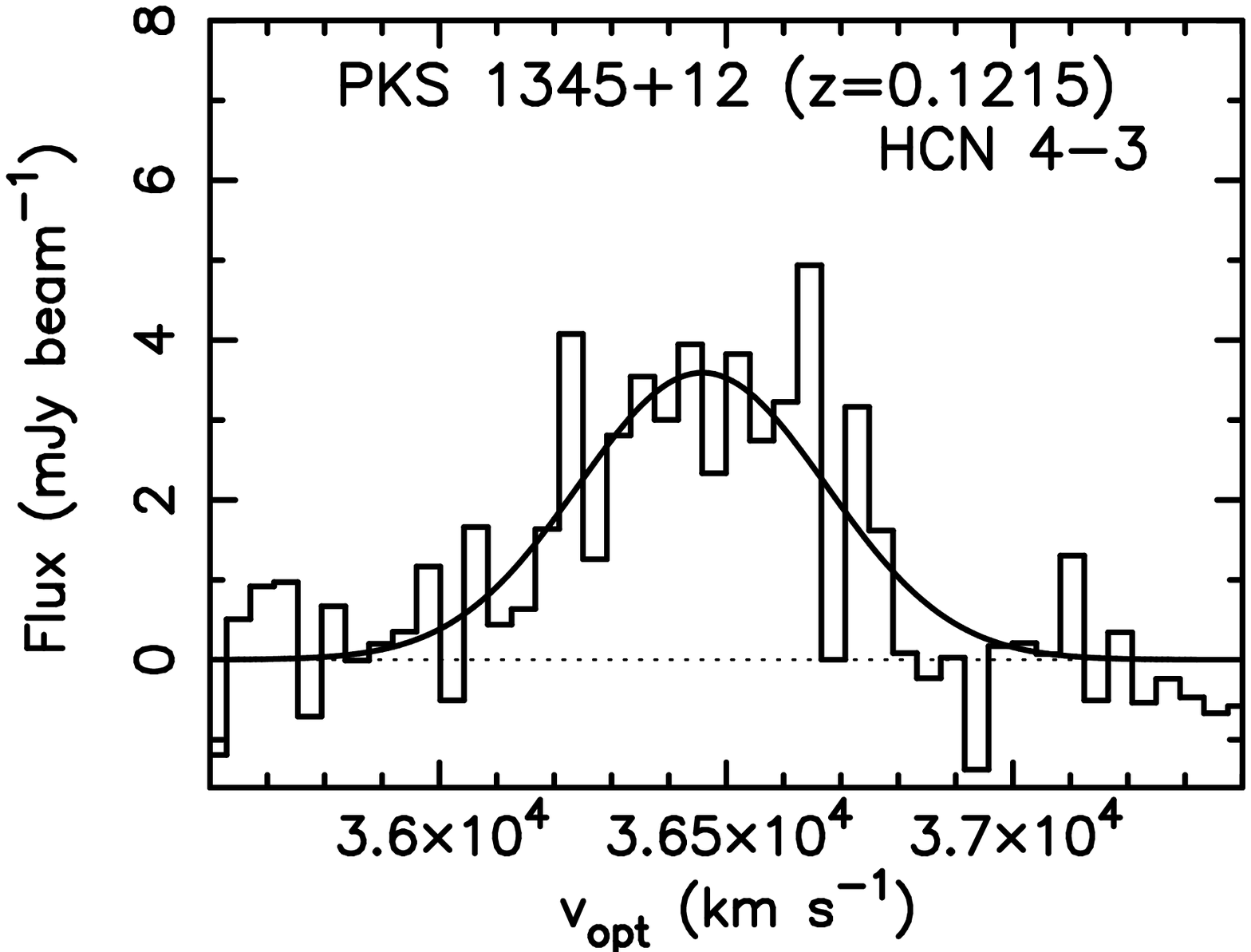} 
\includegraphics[angle=0,scale=.273]{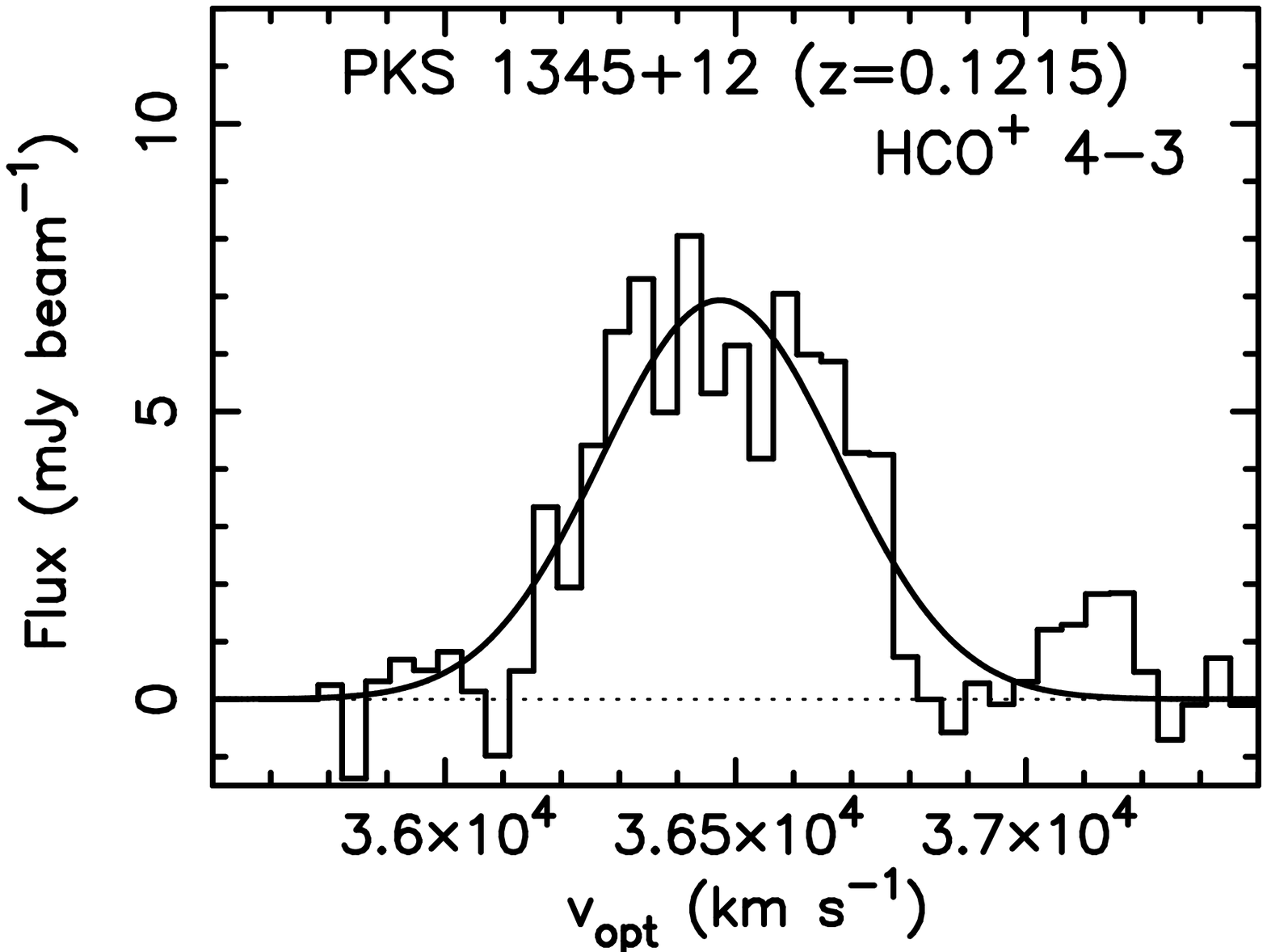} \\
\includegraphics[angle=0,scale=.273]{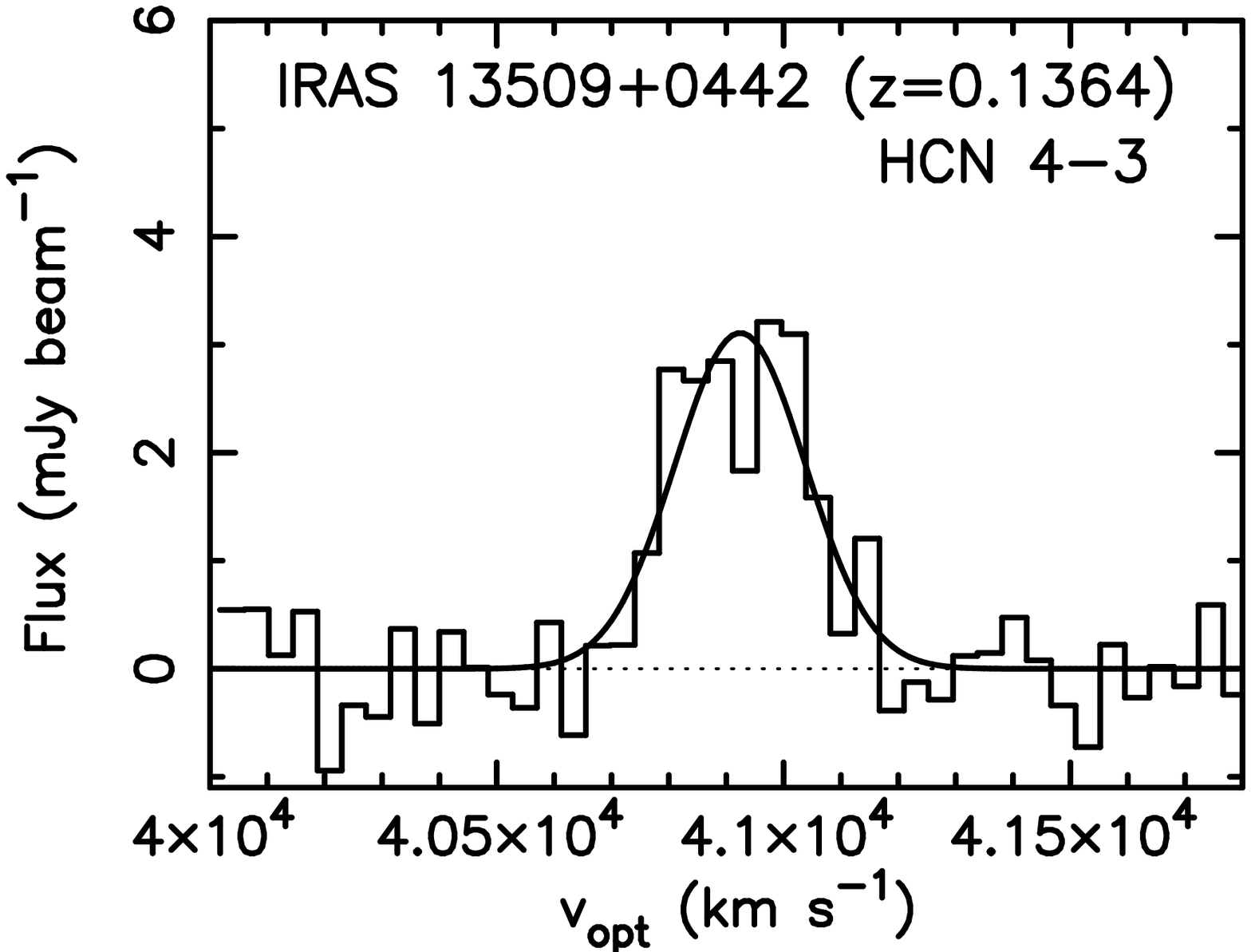}  
\includegraphics[angle=0,scale=.273]{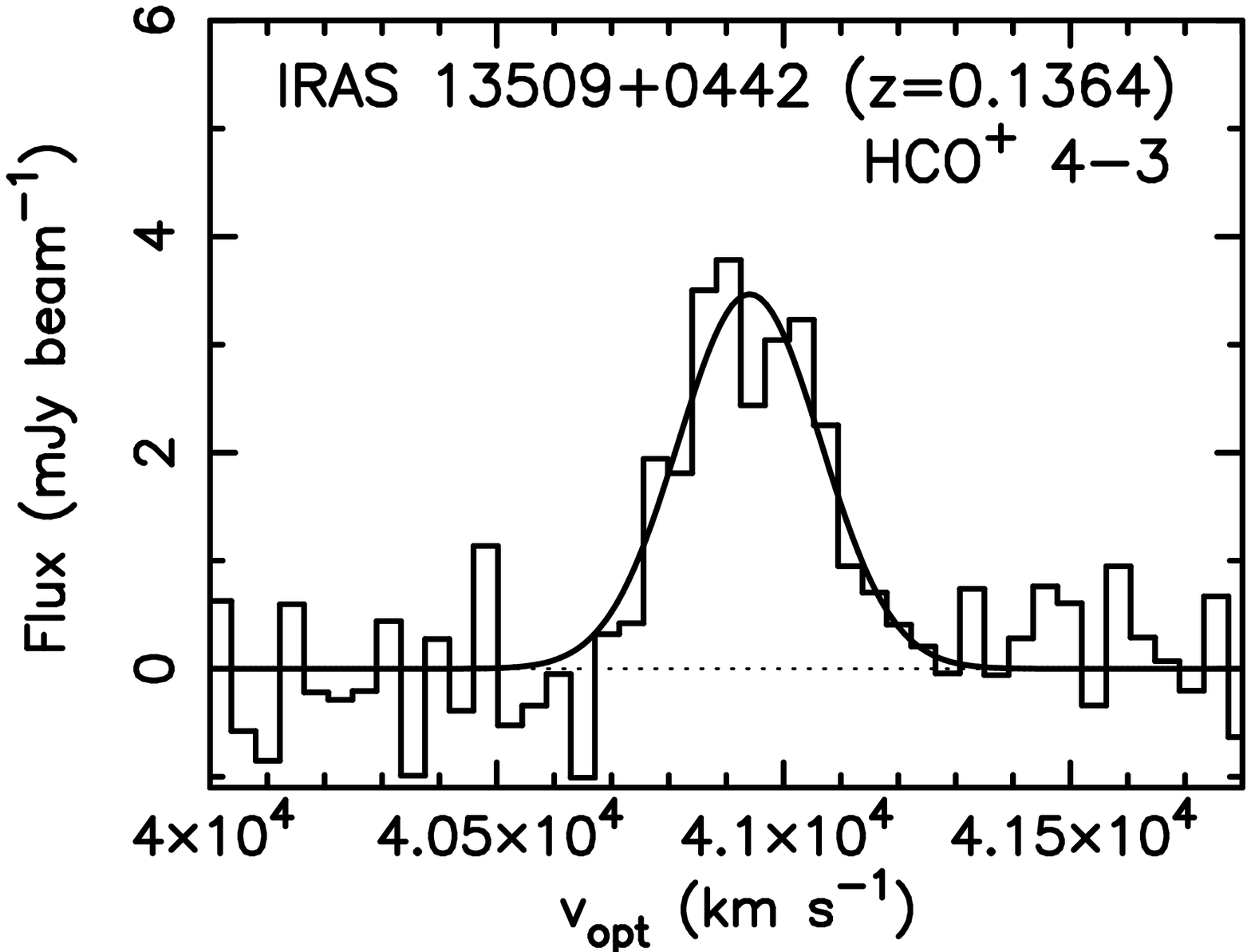}    
\includegraphics[angle=0,scale=.273]{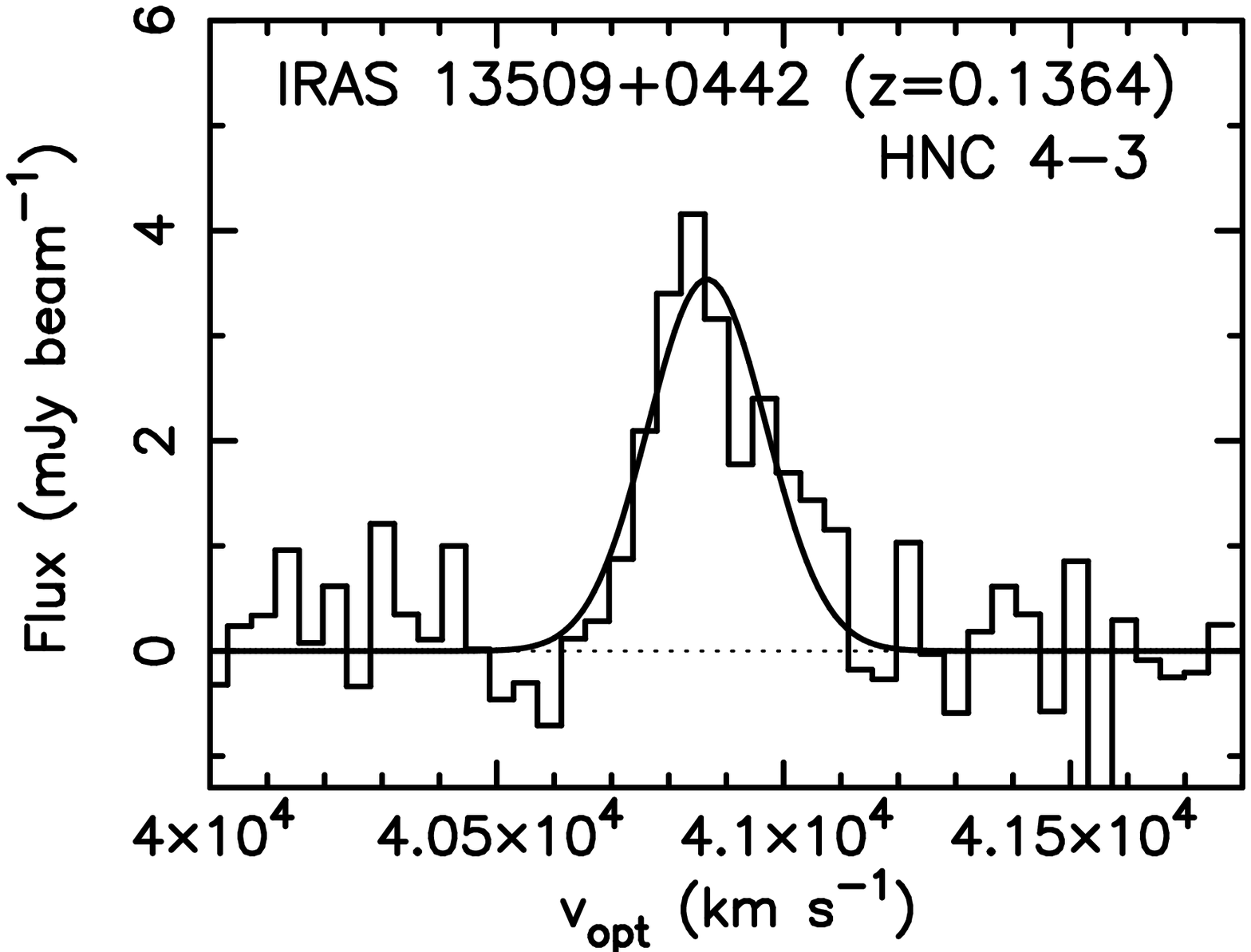} \\ 
\end{center}
\end{figure}

\clearpage

\begin{figure}
\begin{center}
\includegraphics[angle=0,scale=.273]{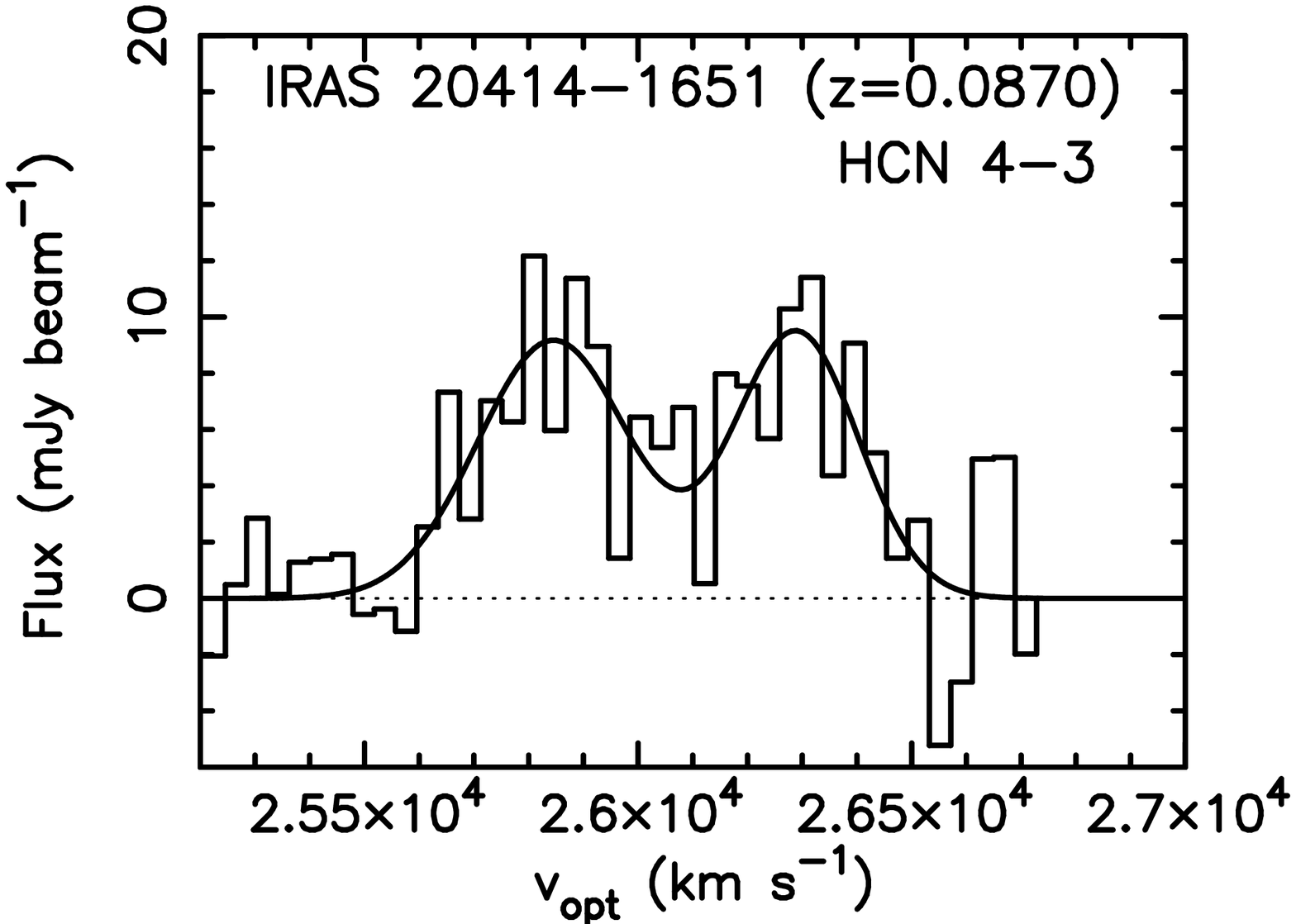}  
\includegraphics[angle=0,scale=.273]{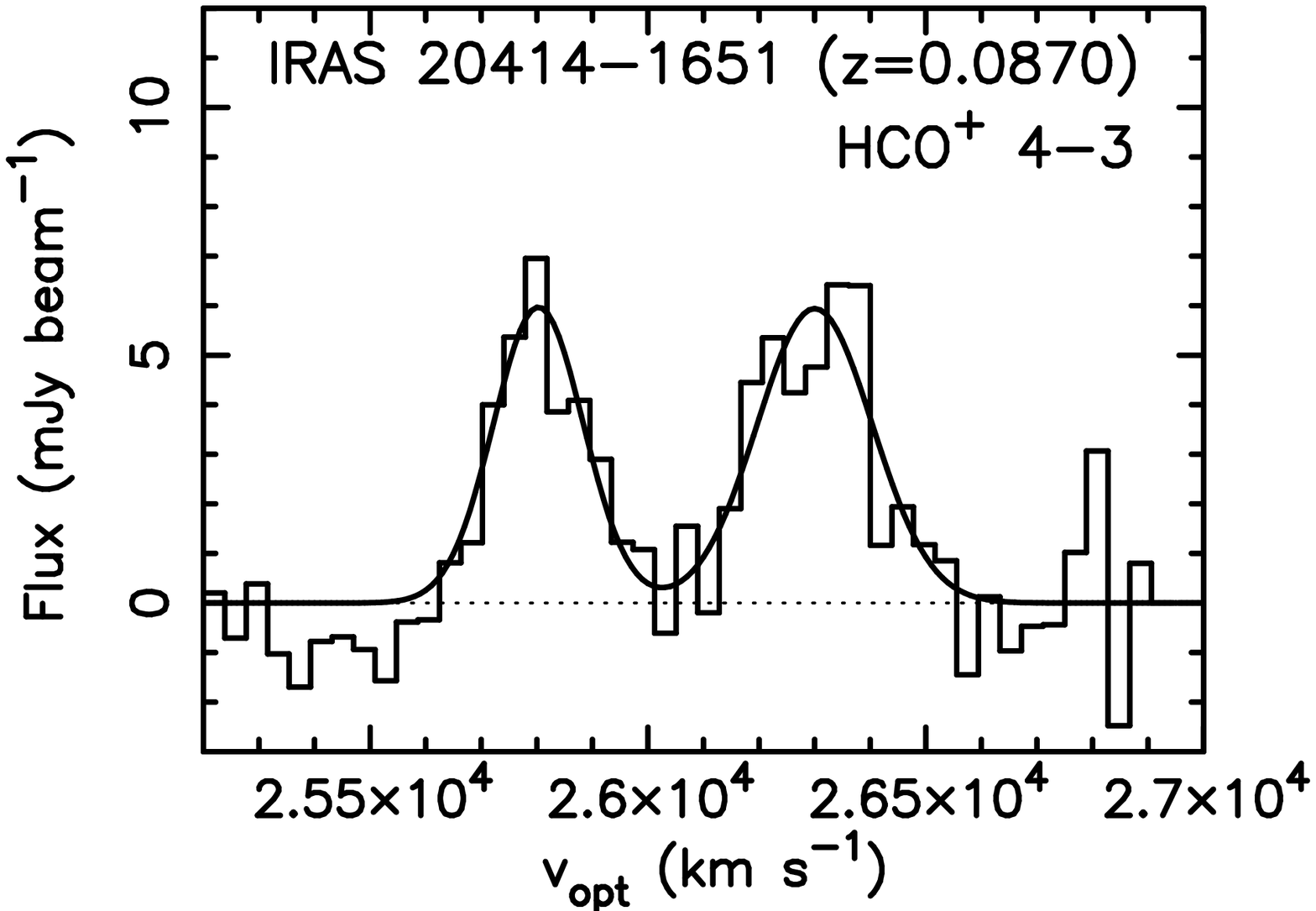} 
\includegraphics[angle=0,scale=.273]{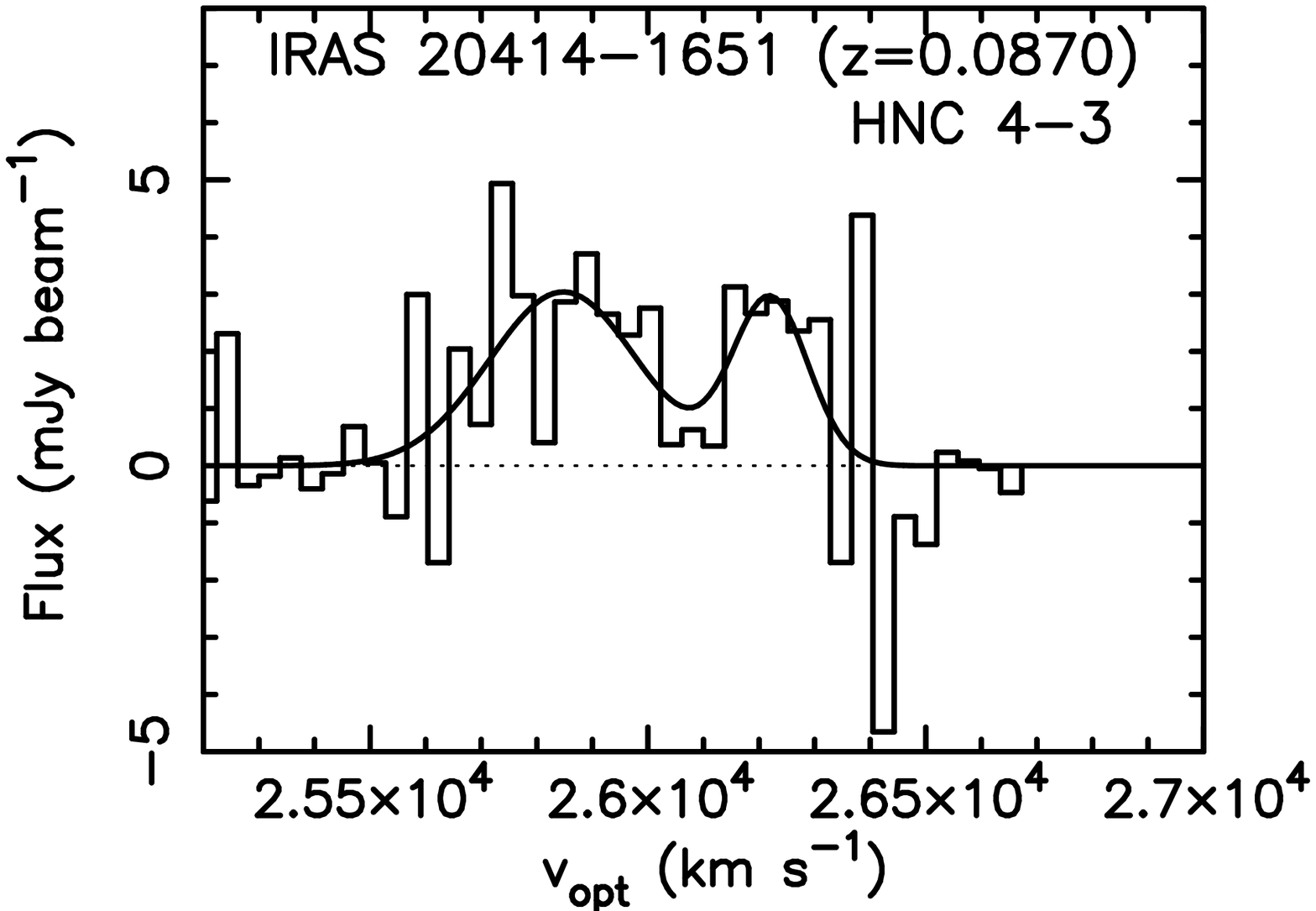} \\
\end{center}
\caption{
Gaussian fits to the detected HCN, HCO$^{+}$, and HNC J=4--3 emission
lines in band 7 spectra at ULIRG's nuclei within the beam size.
The abscissa is optical LSR velocity in (km s$^{-1}$) and the ordinate
is flux in (mJy beam$^{-1}$). 
For HCN J=4--3 and HCO$^{+}$ J=4--3 of IRAS 12112$+$0305 NE, the dotted
curved lines are single Gaussian fits using data not strongly
affected by the central dips. 
For HCN J=4--3, data at 21340--21620 km s$^{-1}$ and 22010--22150 km
s$^{-1}$ are used for the fit.
For HCO$^{+}$ J=4--3, data at 21500--21620 km s$^{-1}$ and 22000--22130
km s$^{-1}$ are used. 
}
\end{figure}

\begin{figure}
\begin{center}
\includegraphics[angle=0,scale=.273]{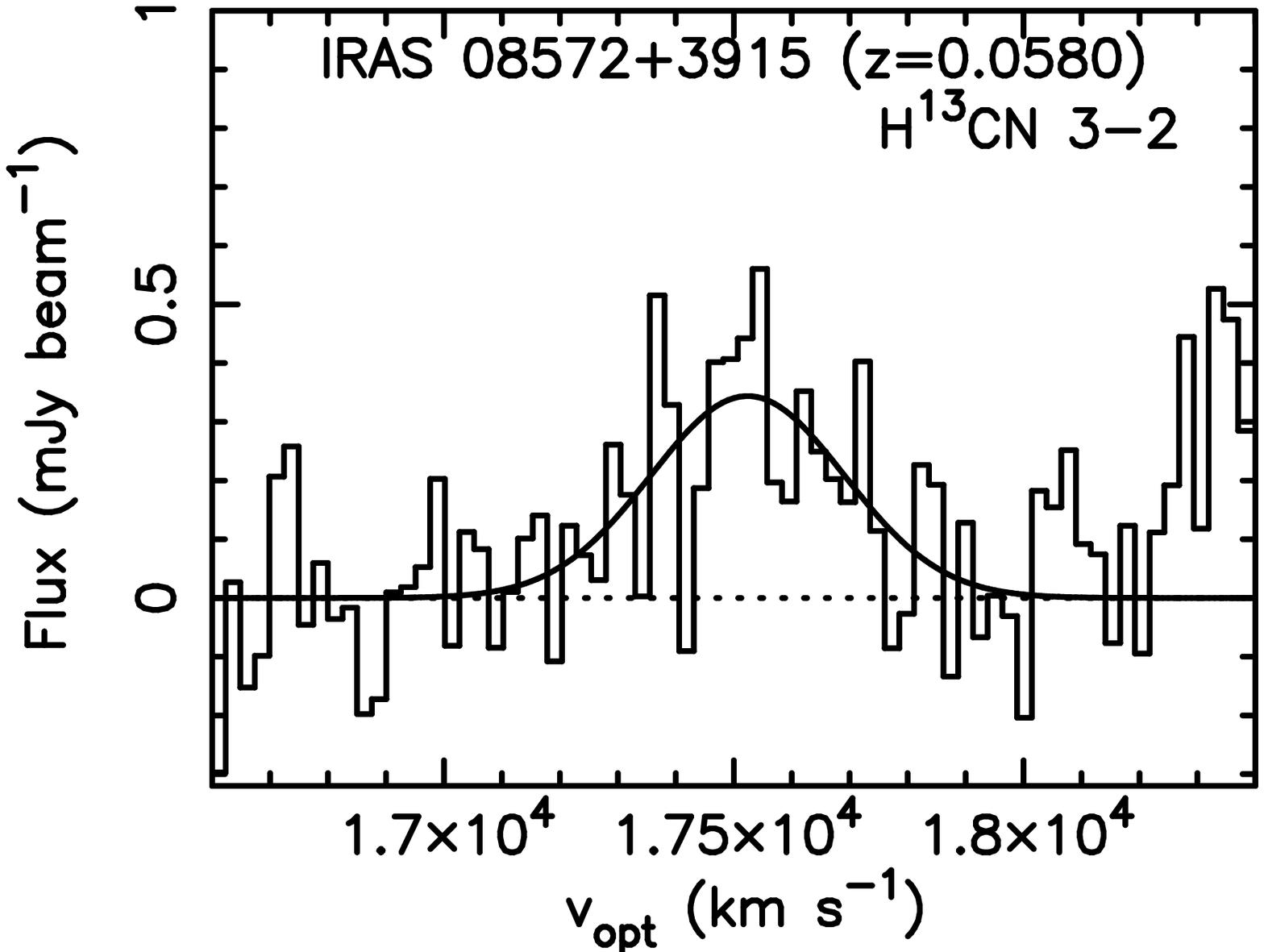}
\includegraphics[angle=0,scale=.273]{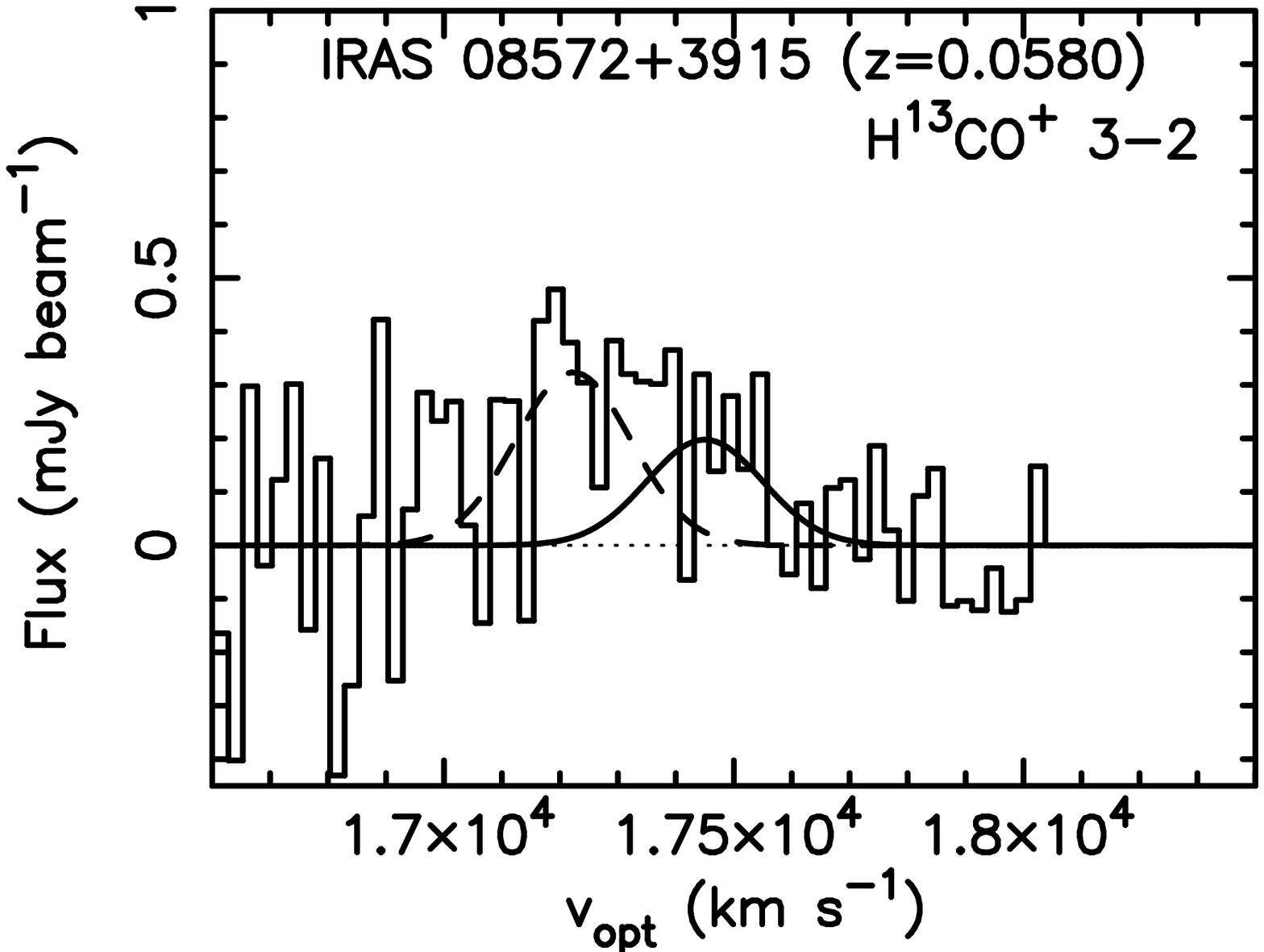} \\
\includegraphics[angle=0,scale=.273]{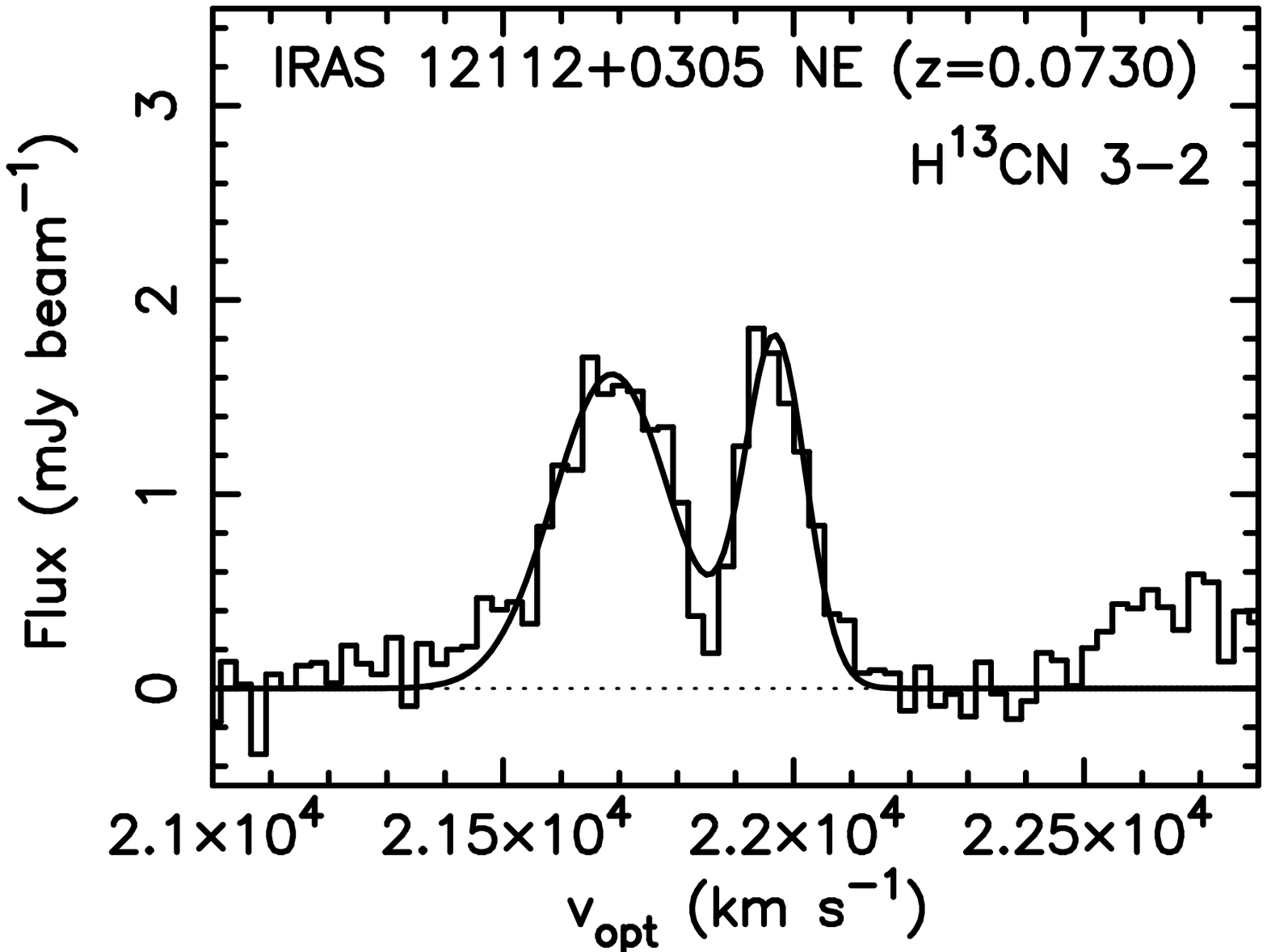}
\includegraphics[angle=0,scale=.273]{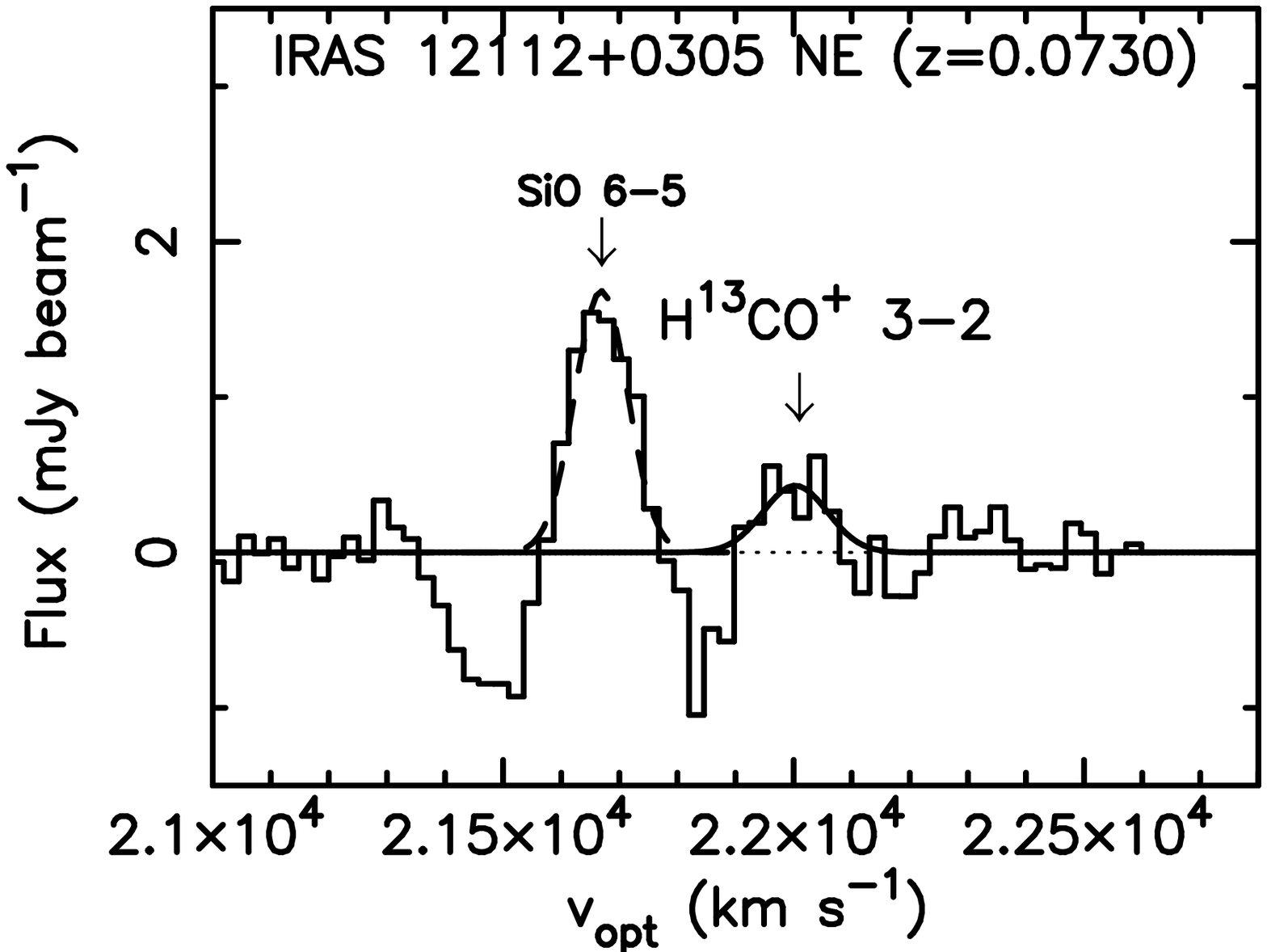}
\includegraphics[angle=0,scale=.273]{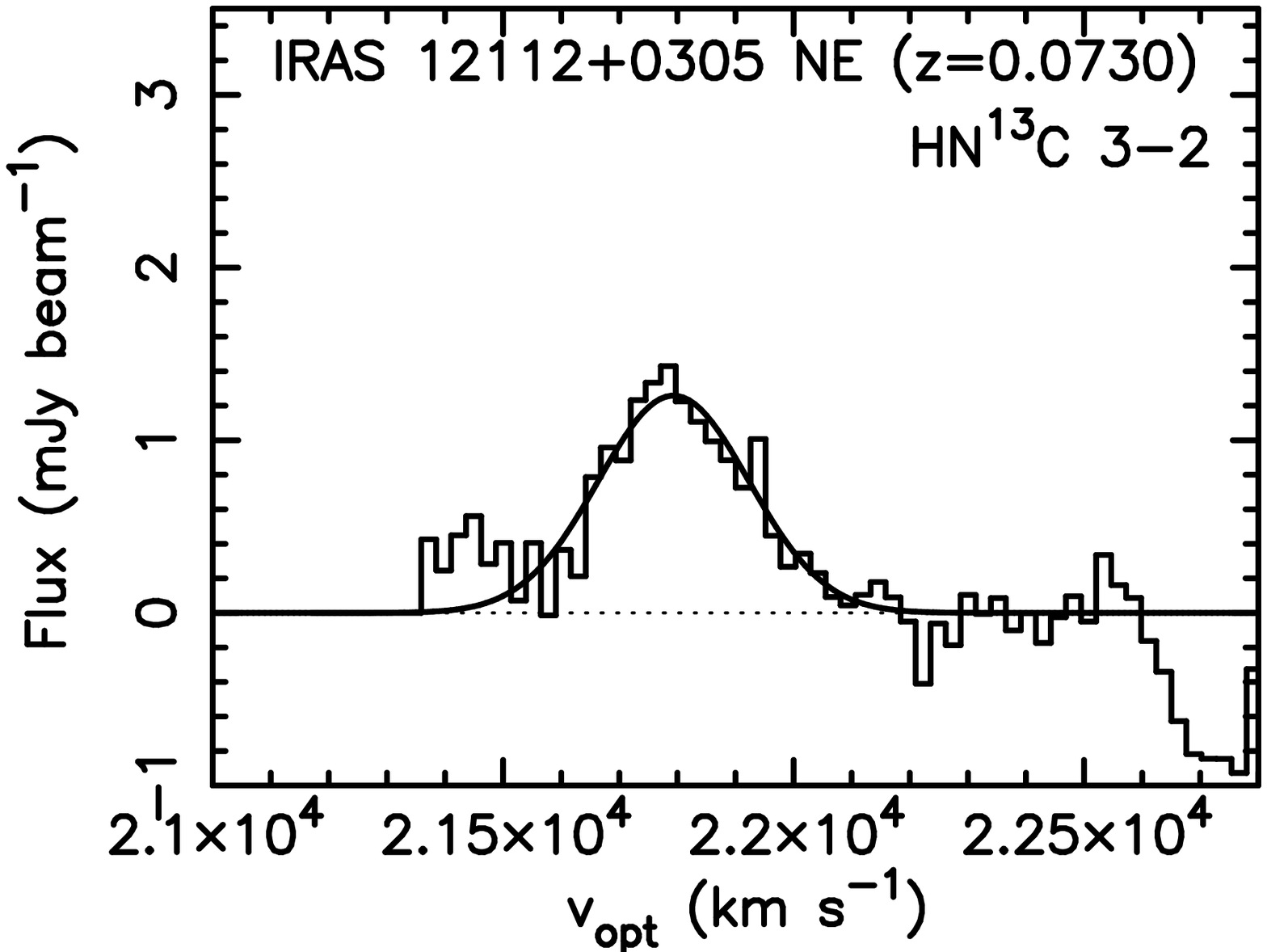} \\
\includegraphics[angle=0,scale=.273]{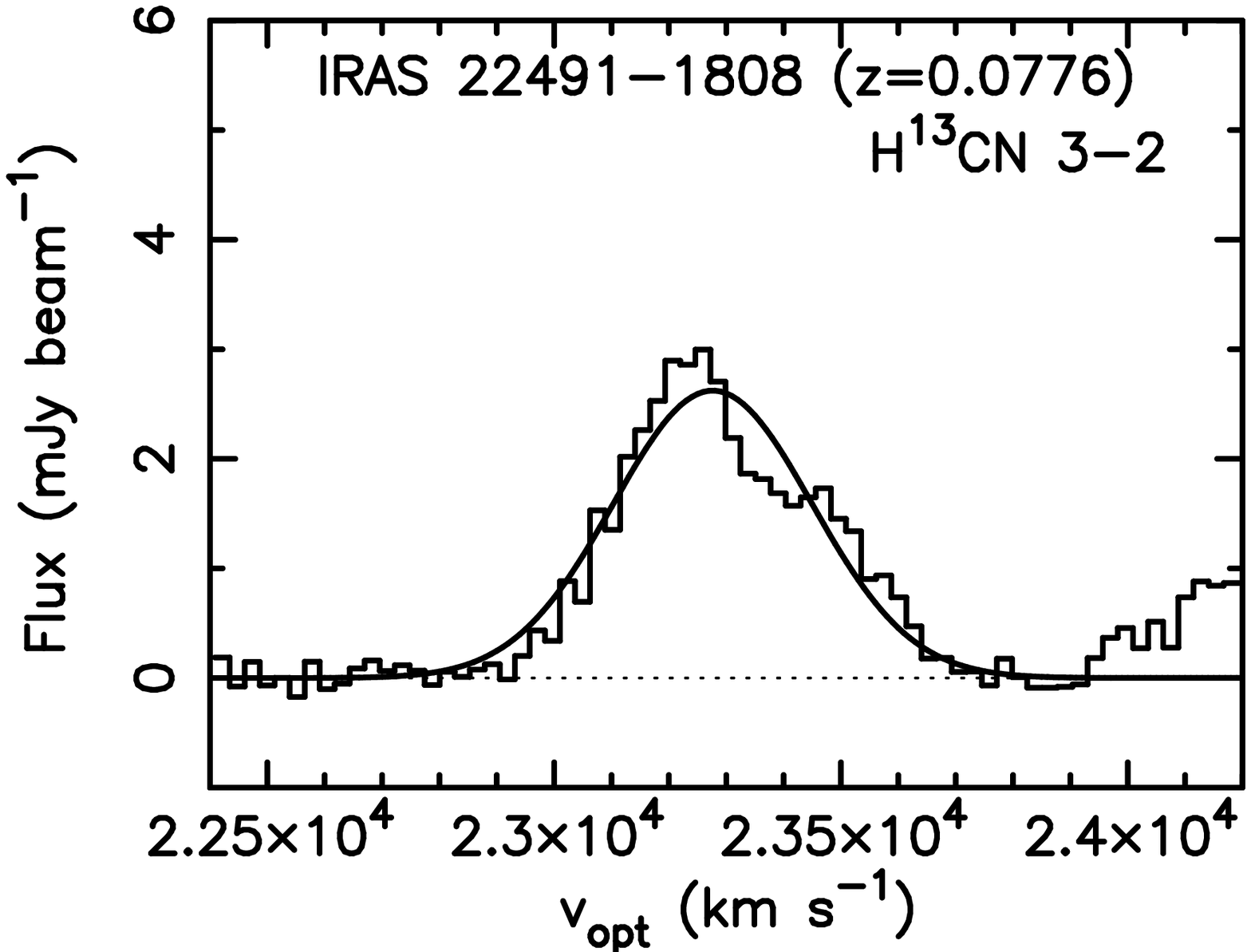}  
\includegraphics[angle=0,scale=.273]{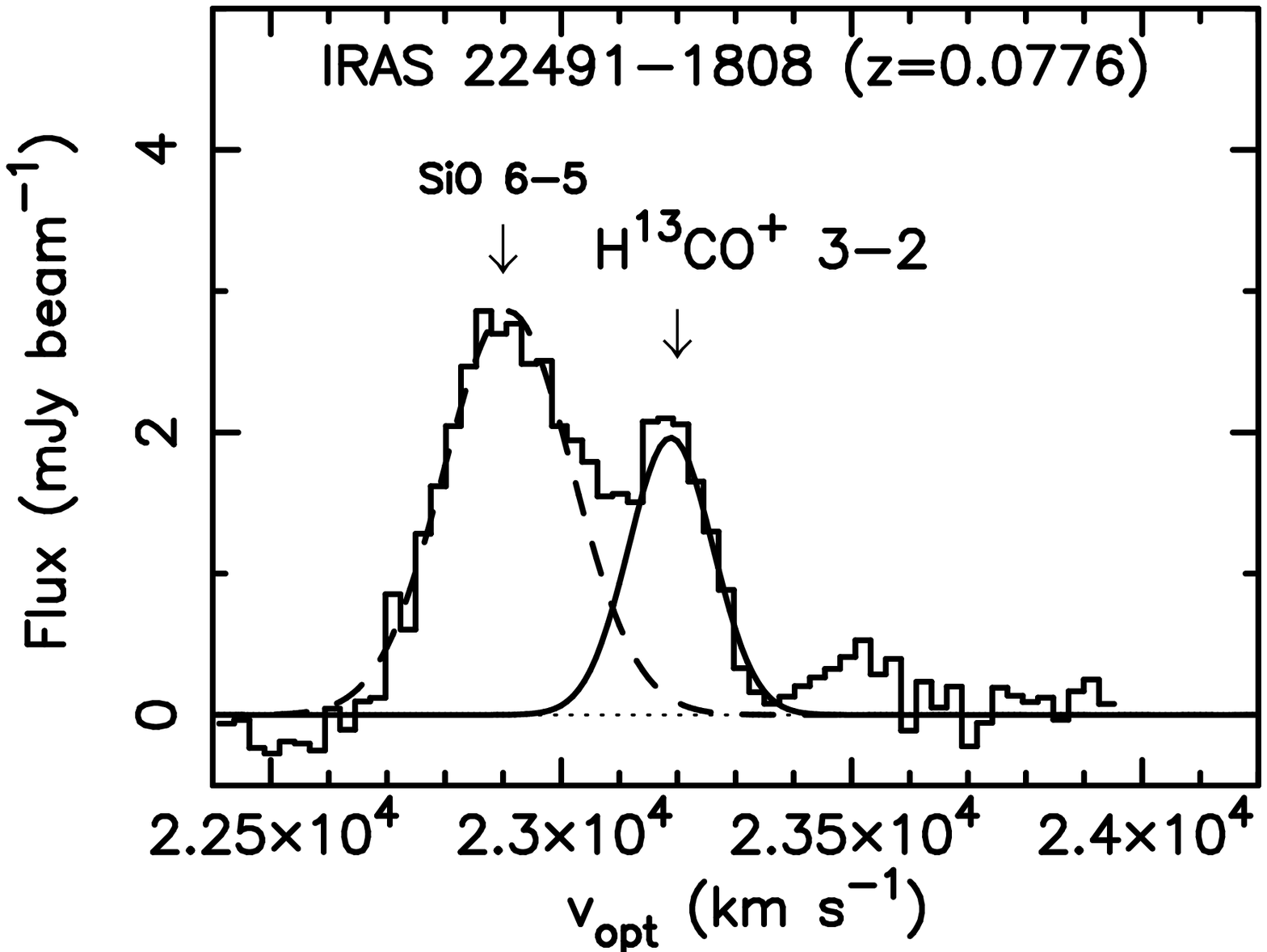}  
\includegraphics[angle=0,scale=.273]{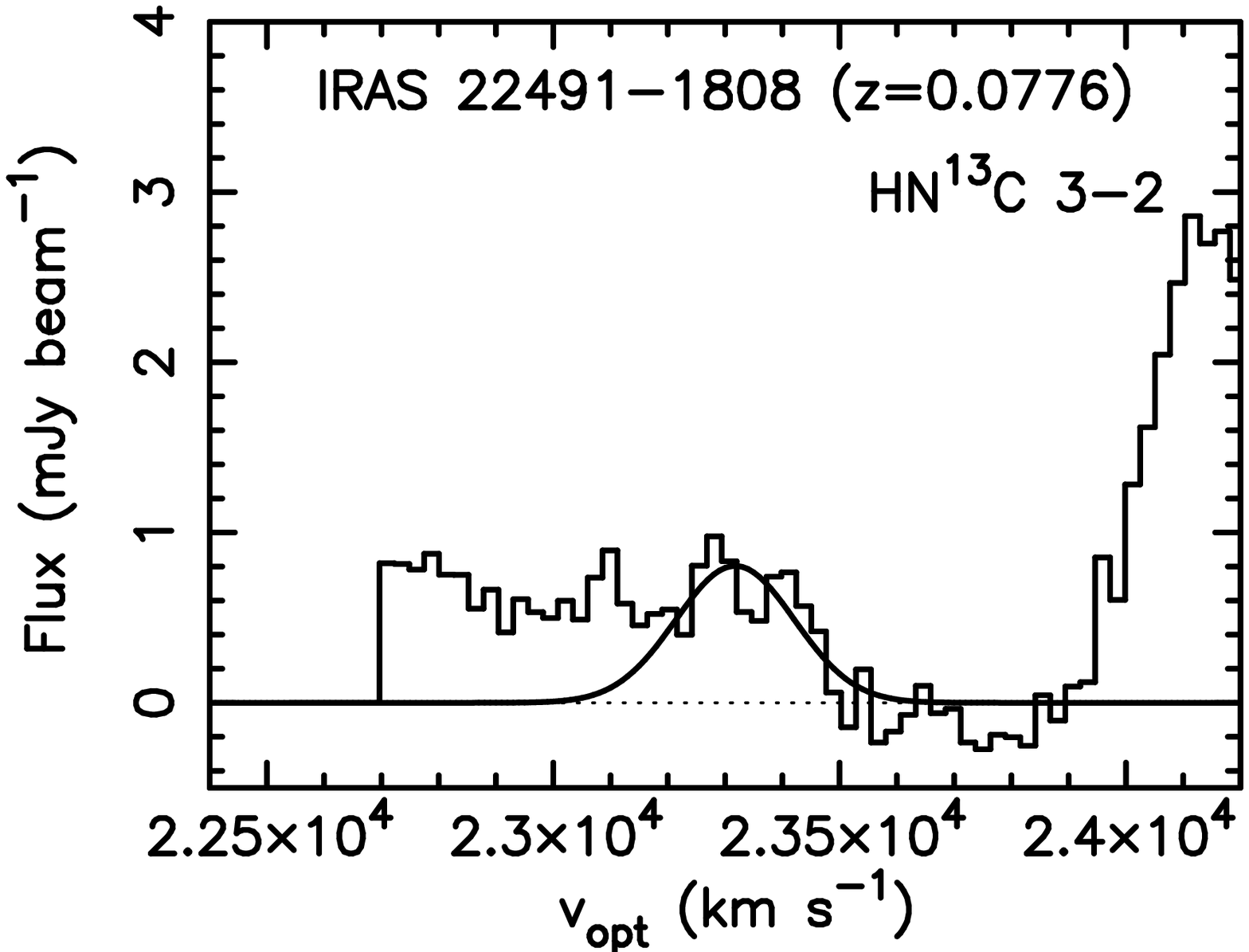} \\ 
\end{center}
\caption{
Gaussian fits to the detected isotopologue H$^{13}$CN, H$^{13}$CO$^{+}$,
and HN$^{13}$C J=3--2 emission lines in Cycle 4 deep spectra, within the
beam size, at the nuclei of IRAS 08572$+$3915, IRAS 12112$+$0305 NE, and
IRAS 22491$-$1808.
The abscissa is optical LSR velocity in (km s$^{-1}$) and the ordinate
is flux in (mJy beam$^{-1}$). 
The HN$^{13}$C J=3--2 emission line is not clearly detected in IRAS
08572$+$3915. 
The best fit Gaussian is shown as a solid curved line.
For the H$^{13}$CN J=3--2 emission line of IRAS 12112$+$0305 NE, two
Gaussians are used for the fit. 
For all sources, we simultaneously fit spectrally-overlapped
H$^{13}$CO$^{+}$ J=3--2 and SiO J=6--5 emission lines with two Gaussian
components.  
The fits for the SiO J=6--5 lines are shown as dashed curved lines.
For HN$^{13}$C J=3--2 of IRAS 22491$-$1808, only data at v$_{\rm opt}$
$>$ 23229 km s$^{-1}$ are used for the Gaussian fit.
}
\end{figure}

\begin{figure}
\begin{center}
\includegraphics[angle=0,scale=.63]{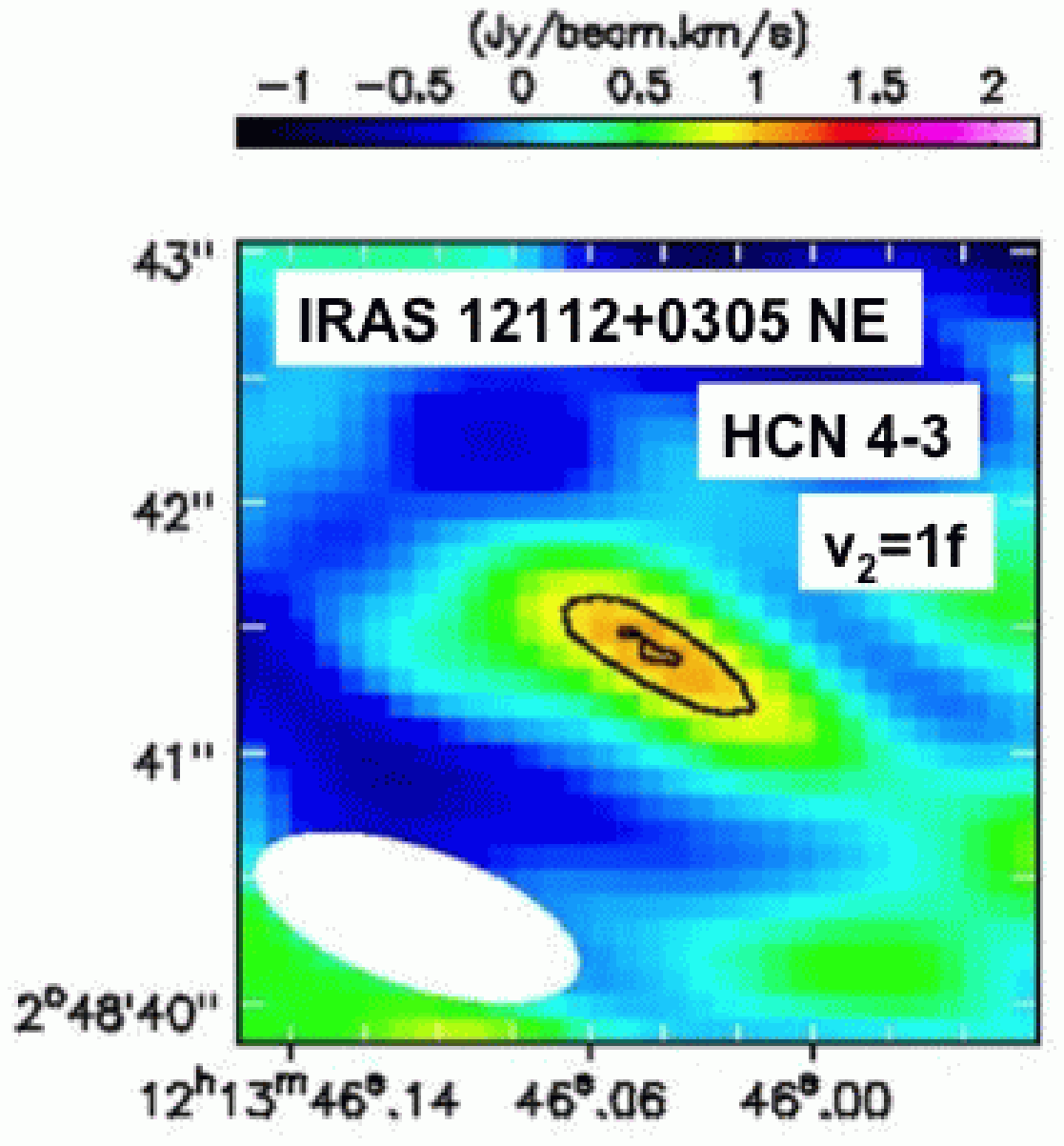}  
\includegraphics[angle=0,scale=.63]{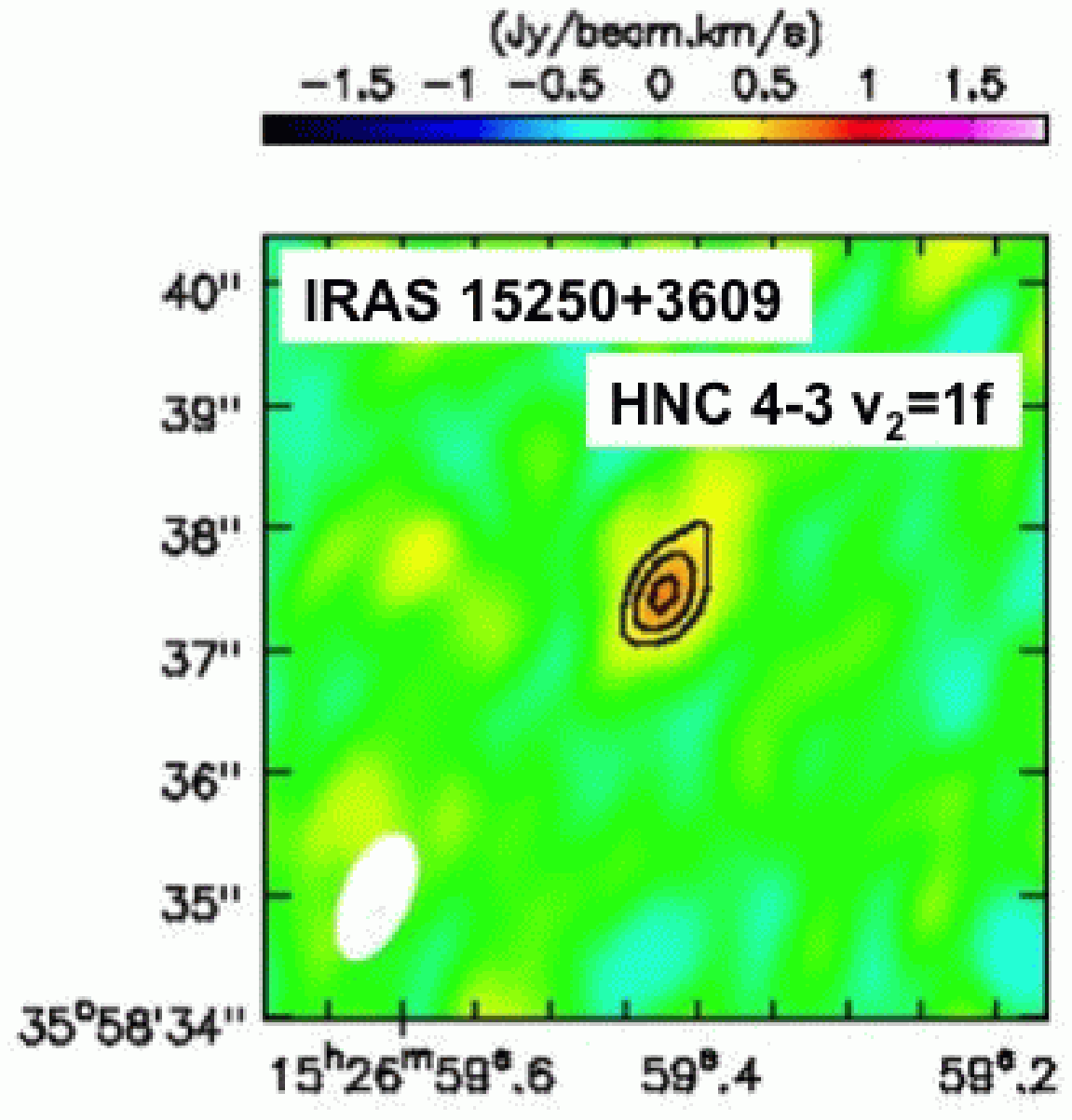}  
\caption{Integrated intensity (moment 0) maps of the
vibrationally excited HCN v$_{2}$=1f J=4--3 emission line for 
IRAS 12112$+$0305 NE and HNC v$_{2}$=1f J=4--3 emission line for IRAS 
15250$+$3609. 
The abscissa and ordinate are R.A. (J2000) and decl. (J2000),
respectively. 
The contours are 2.5$\sigma$ and 3$\sigma$ for HCN v$_{2}$=1f J=4--3
of IRAS 12112$+$0305 NE and 3$\sigma$, 4$\sigma$, and 5$\sigma$ for 
HNC v$_{2}$=1f J=4--3 of IRAS 15250$+$3609. 
Beam sizes are shown as filled circles in the lower-left region.
}
\end{center}
\end{figure}

\begin{figure}
\begin{center}
\includegraphics[angle=0,scale=.41]{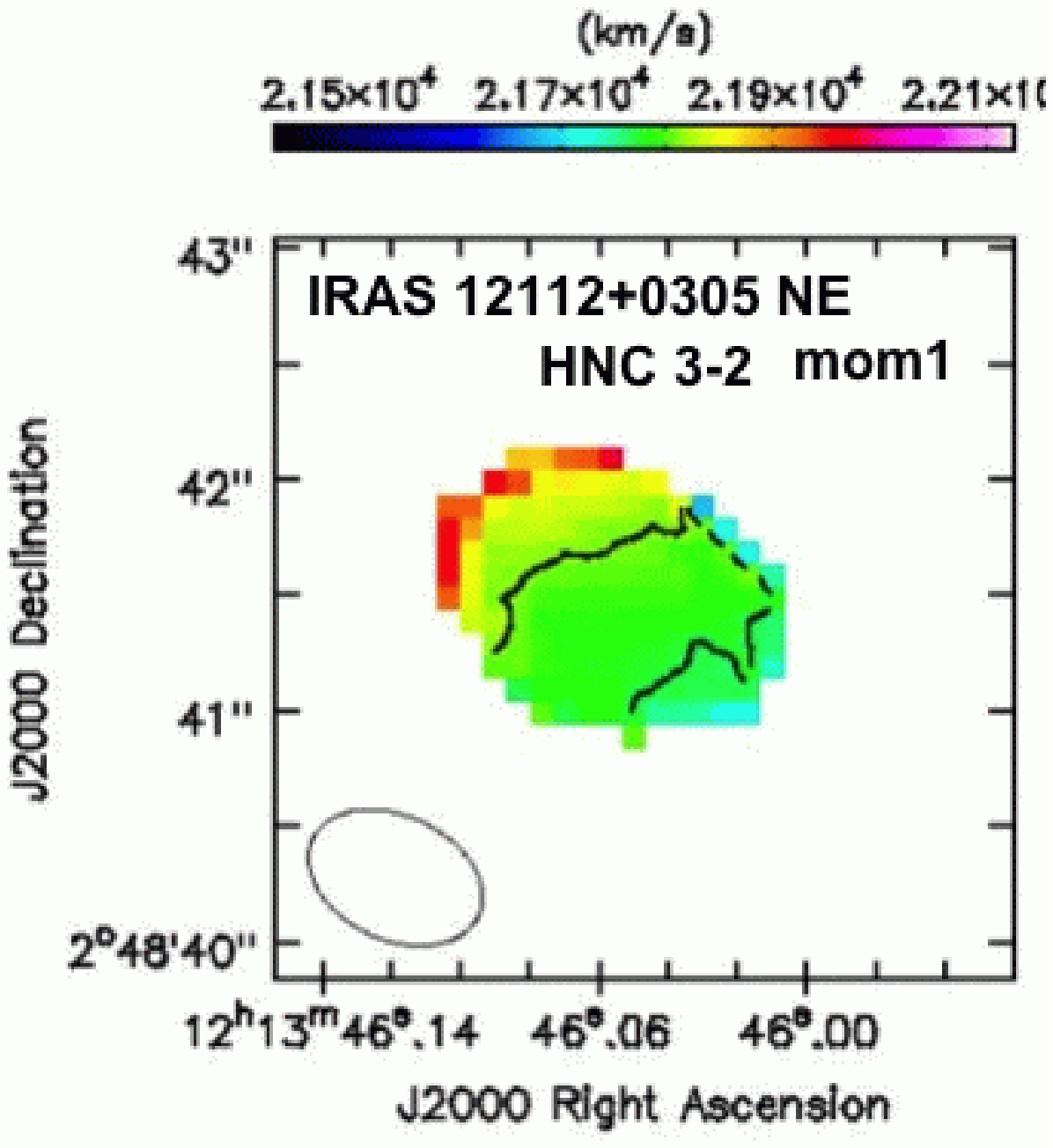}  
\includegraphics[angle=0,scale=.41]{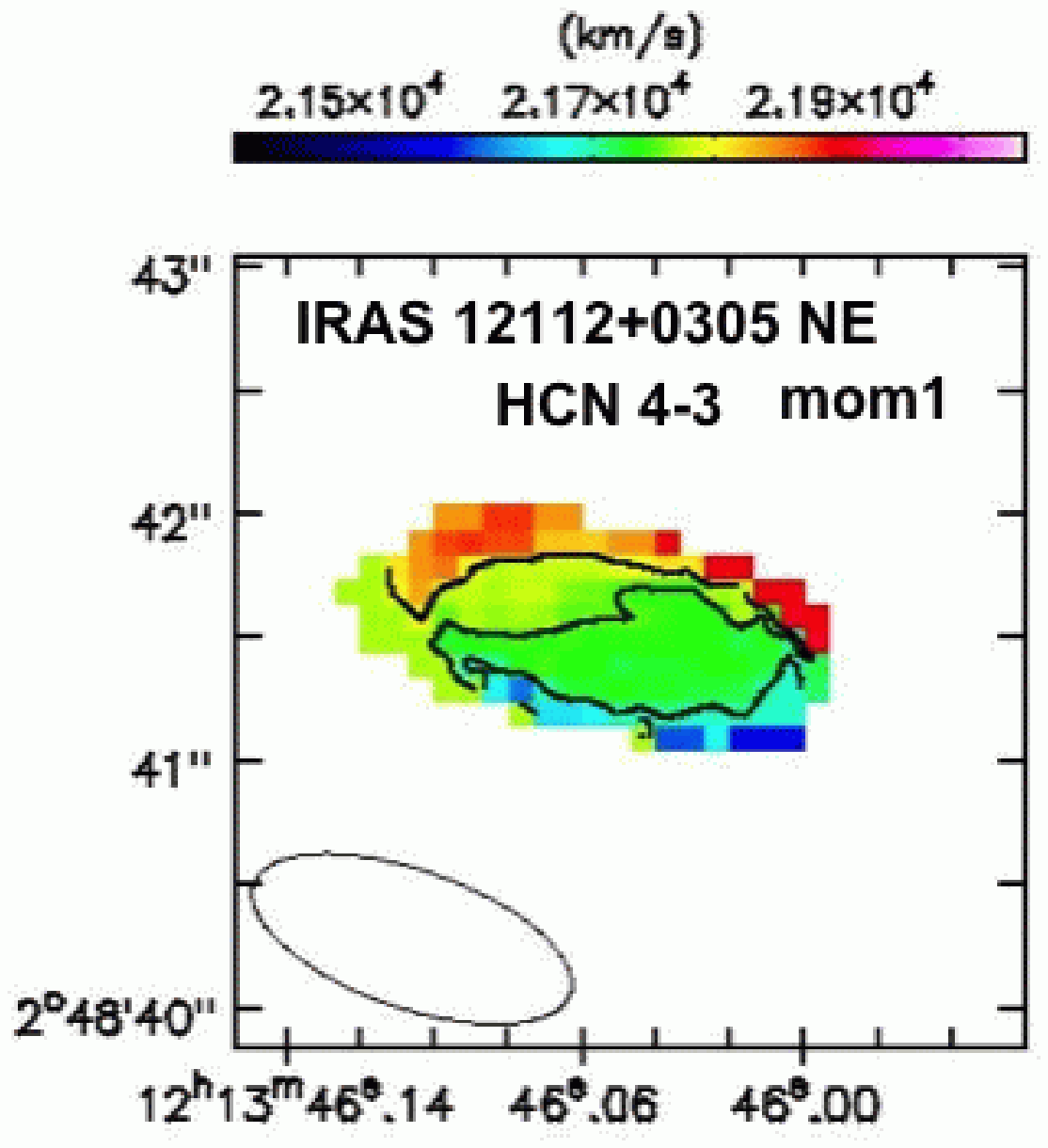}  
\includegraphics[angle=0,scale=.41]{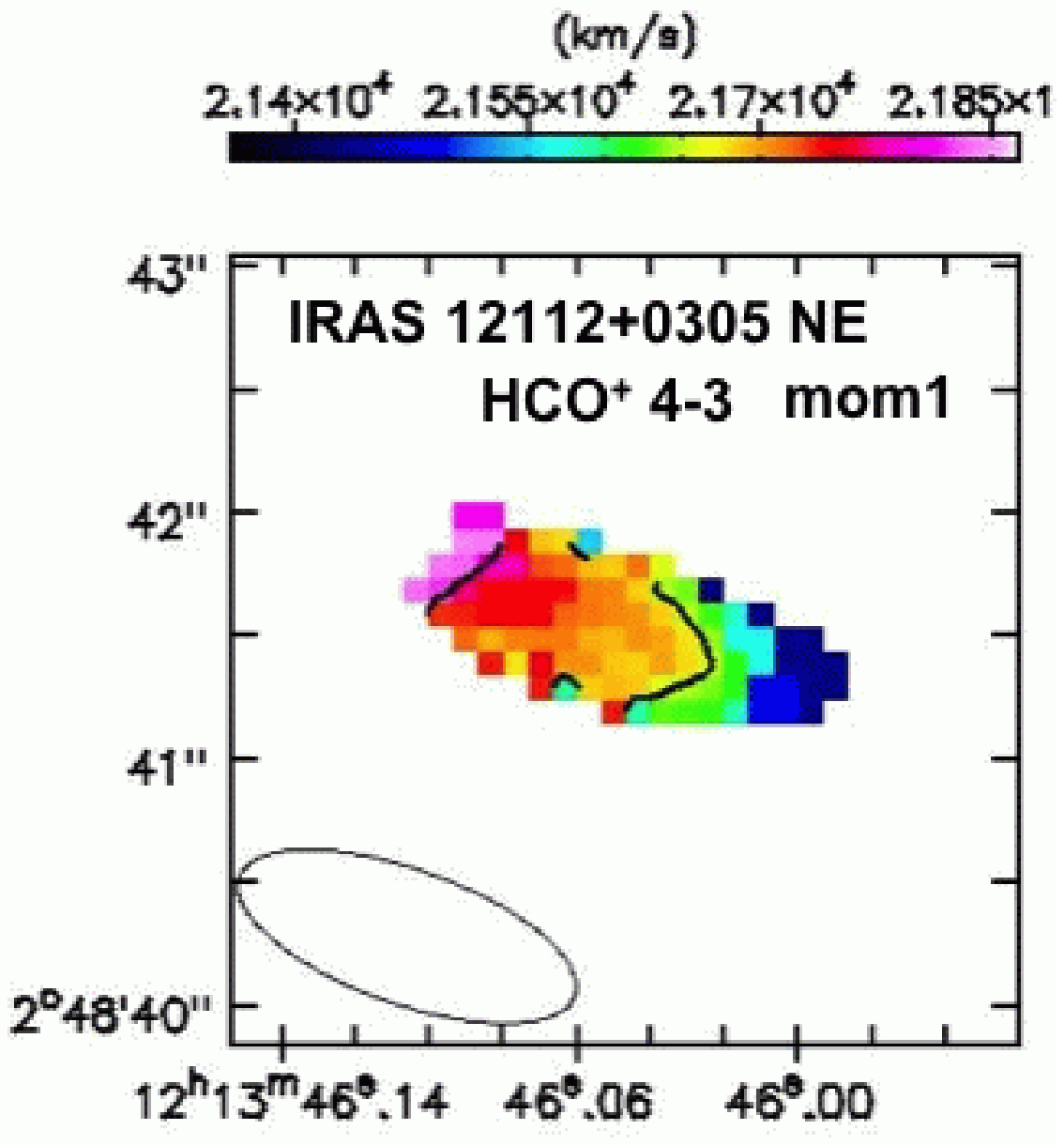} \\ 
\vspace{-1.3cm}
\includegraphics[angle=0,scale=.41]{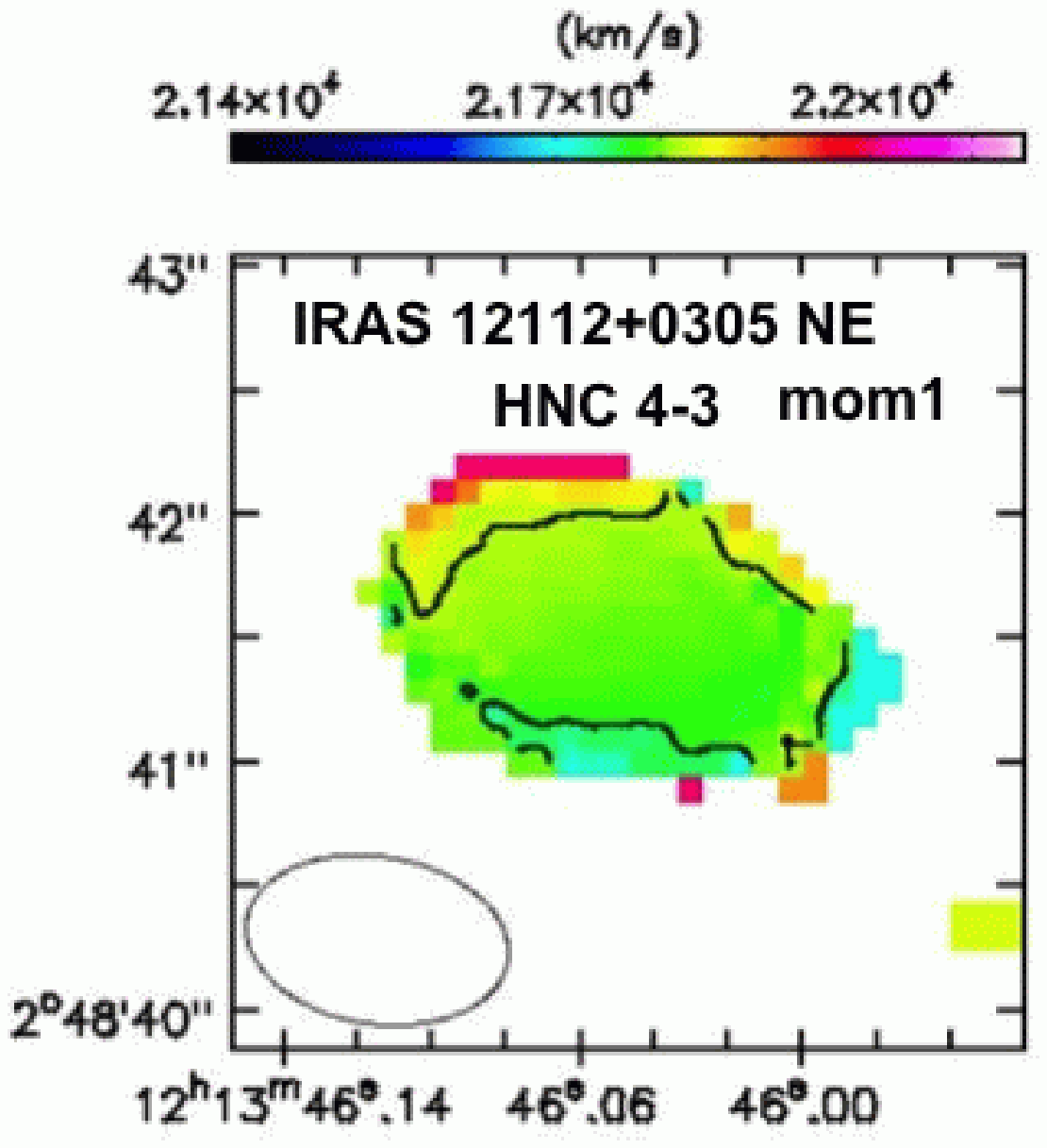}   
\includegraphics[angle=0,scale=.41]{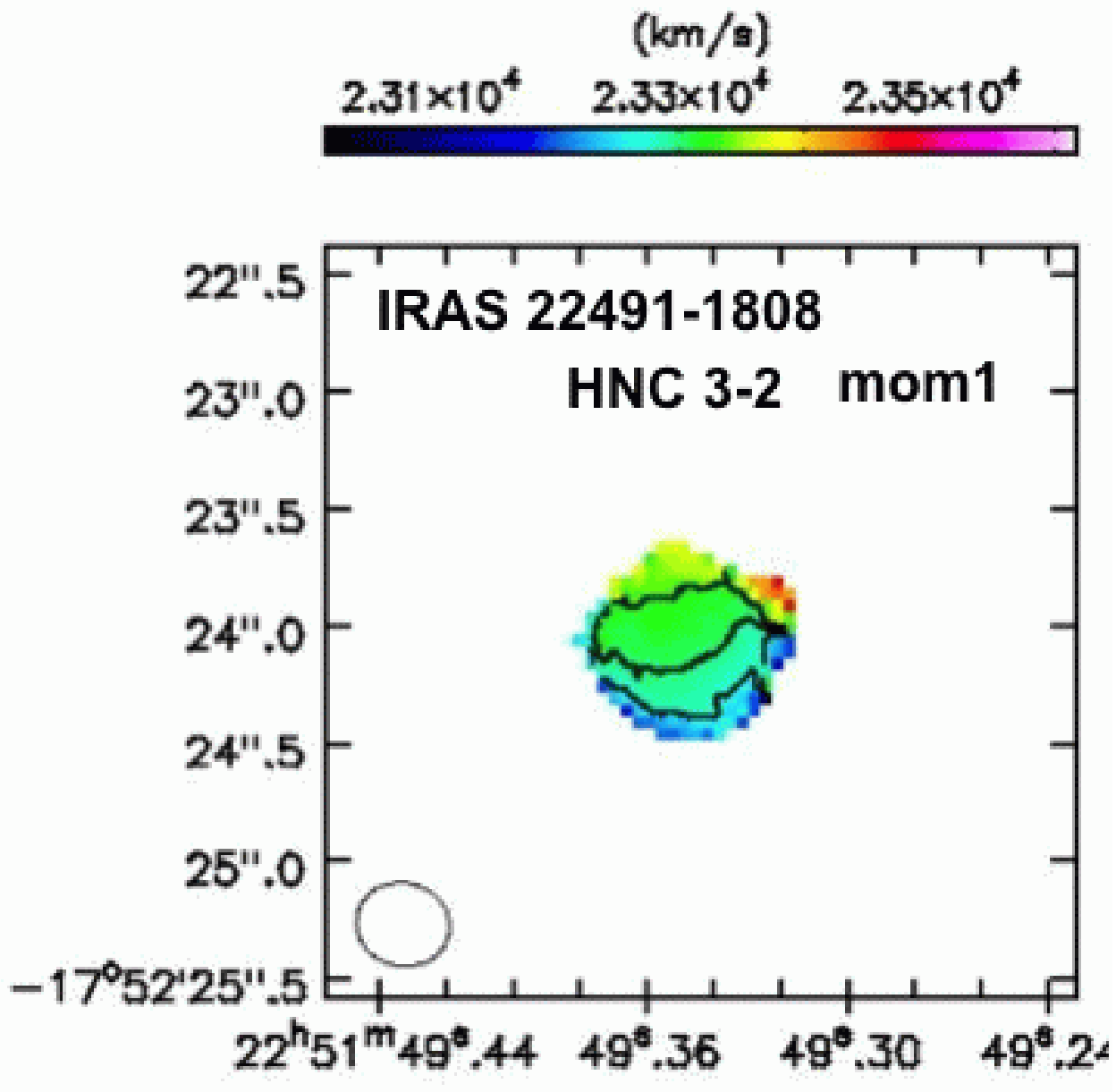} 
\includegraphics[angle=0,scale=.41]{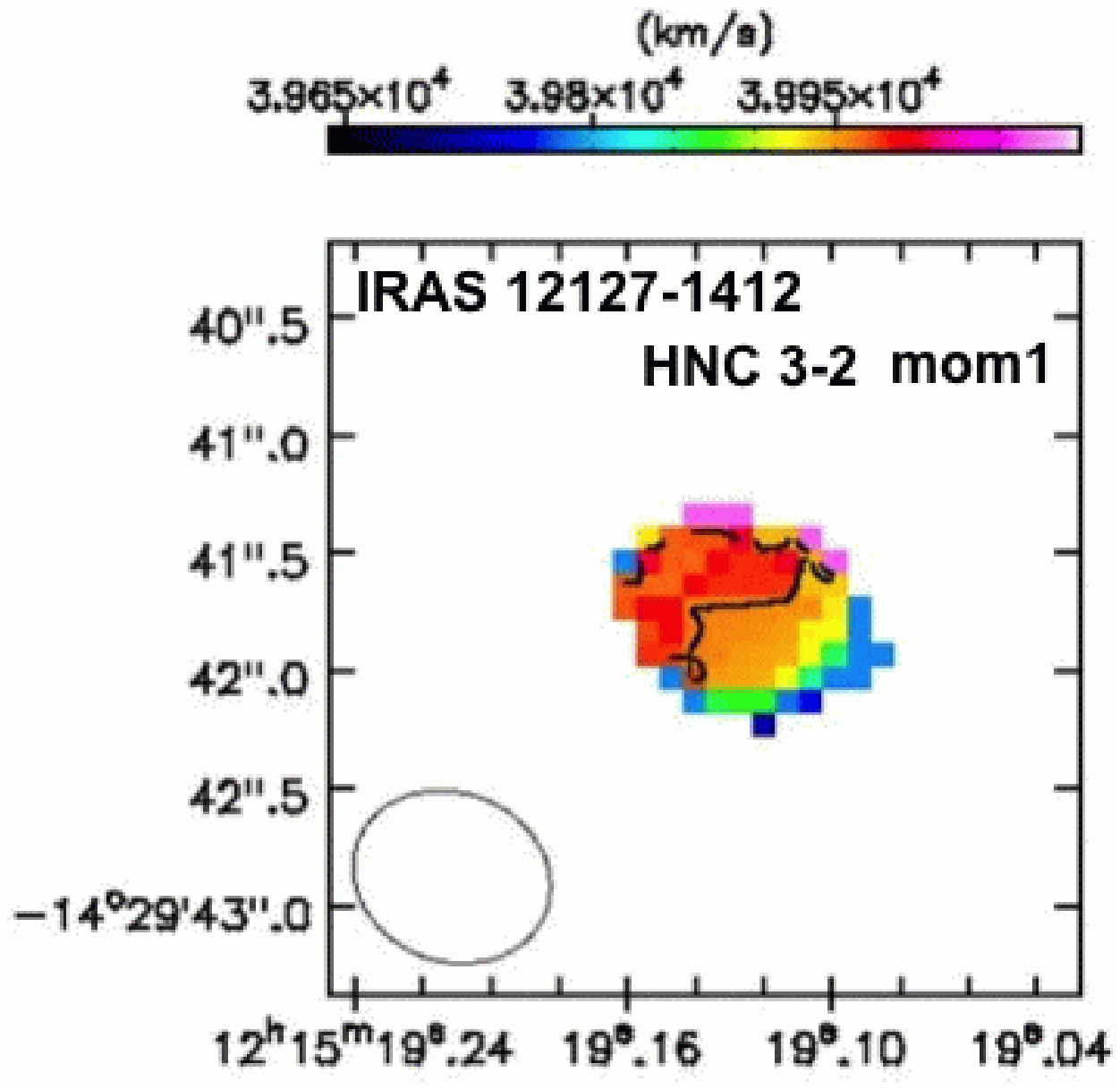} \\
\vspace{-1.3cm}
\includegraphics[angle=0,scale=.41]{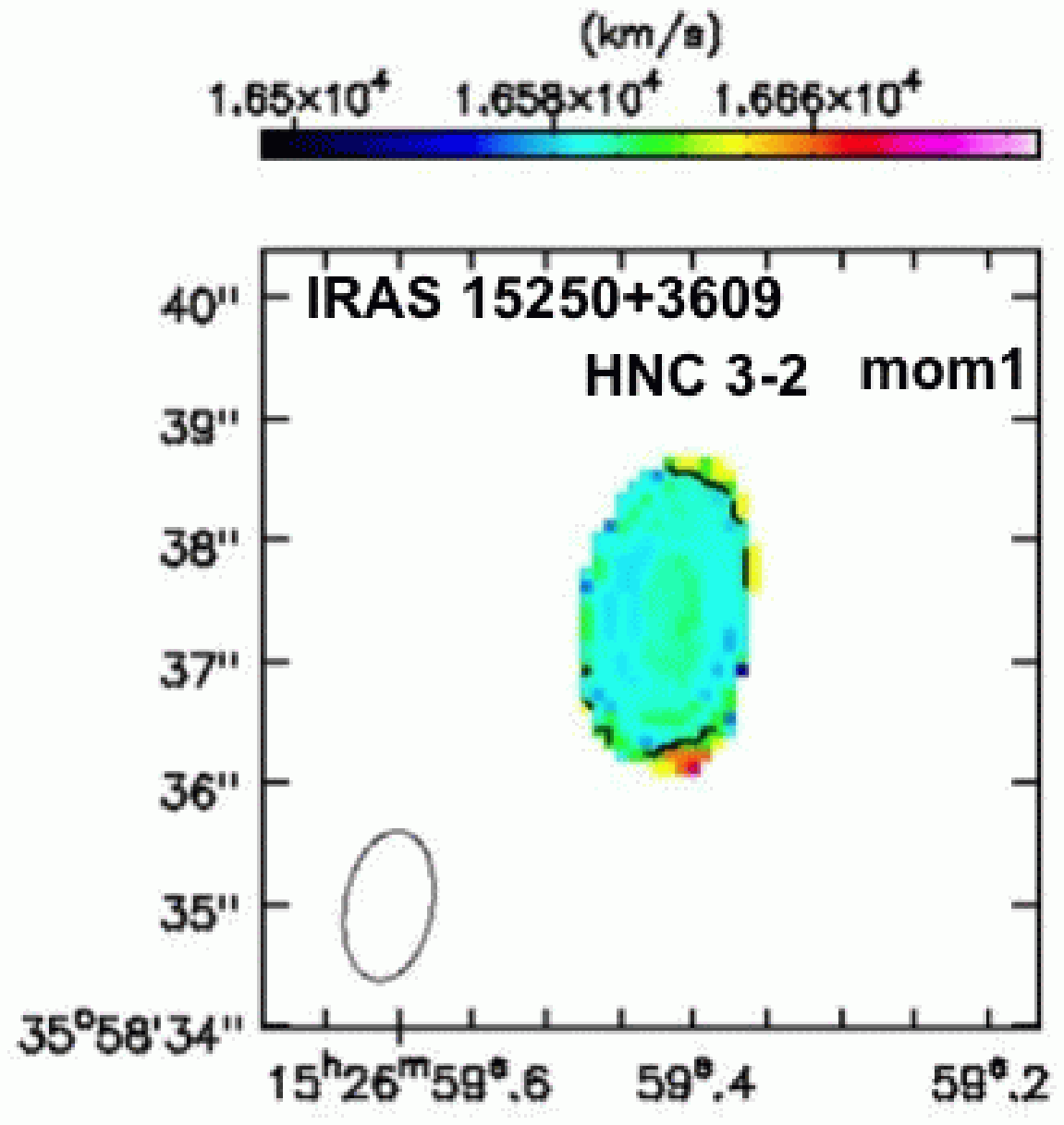}
\includegraphics[angle=0,scale=.41]{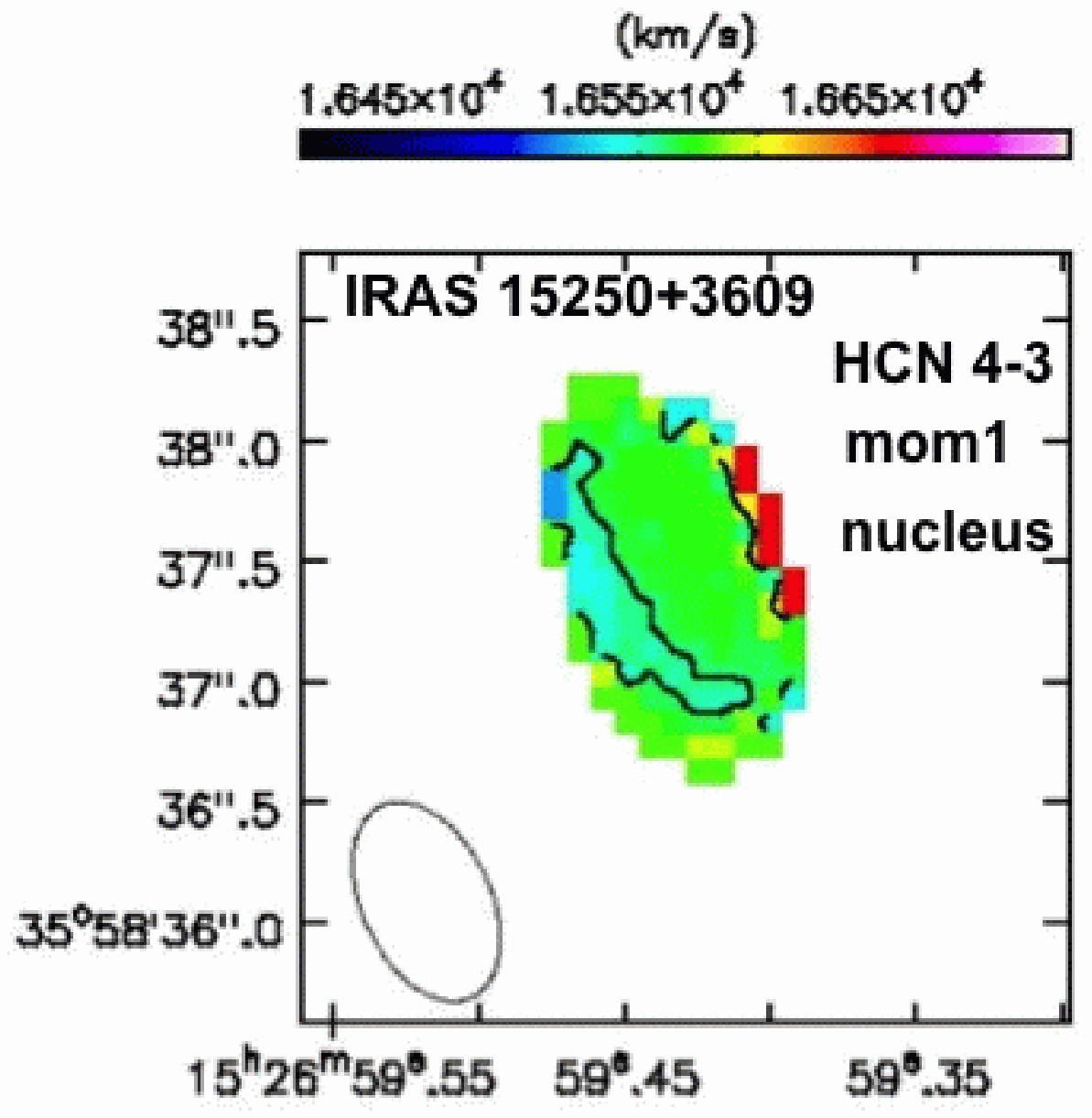}
\includegraphics[angle=0,scale=.41]{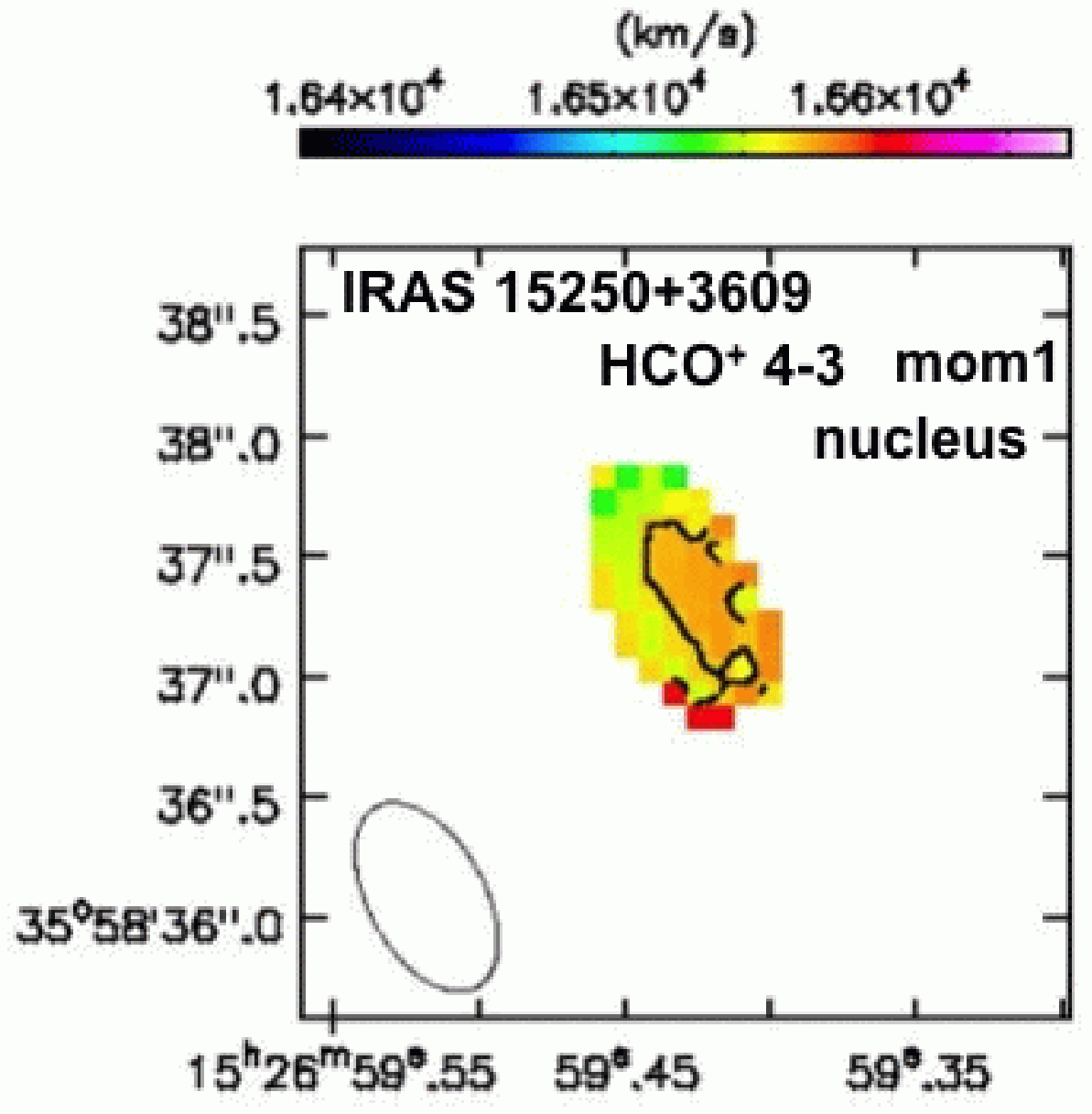} \\
\vspace{-1.3cm}
\includegraphics[angle=0,scale=.41]{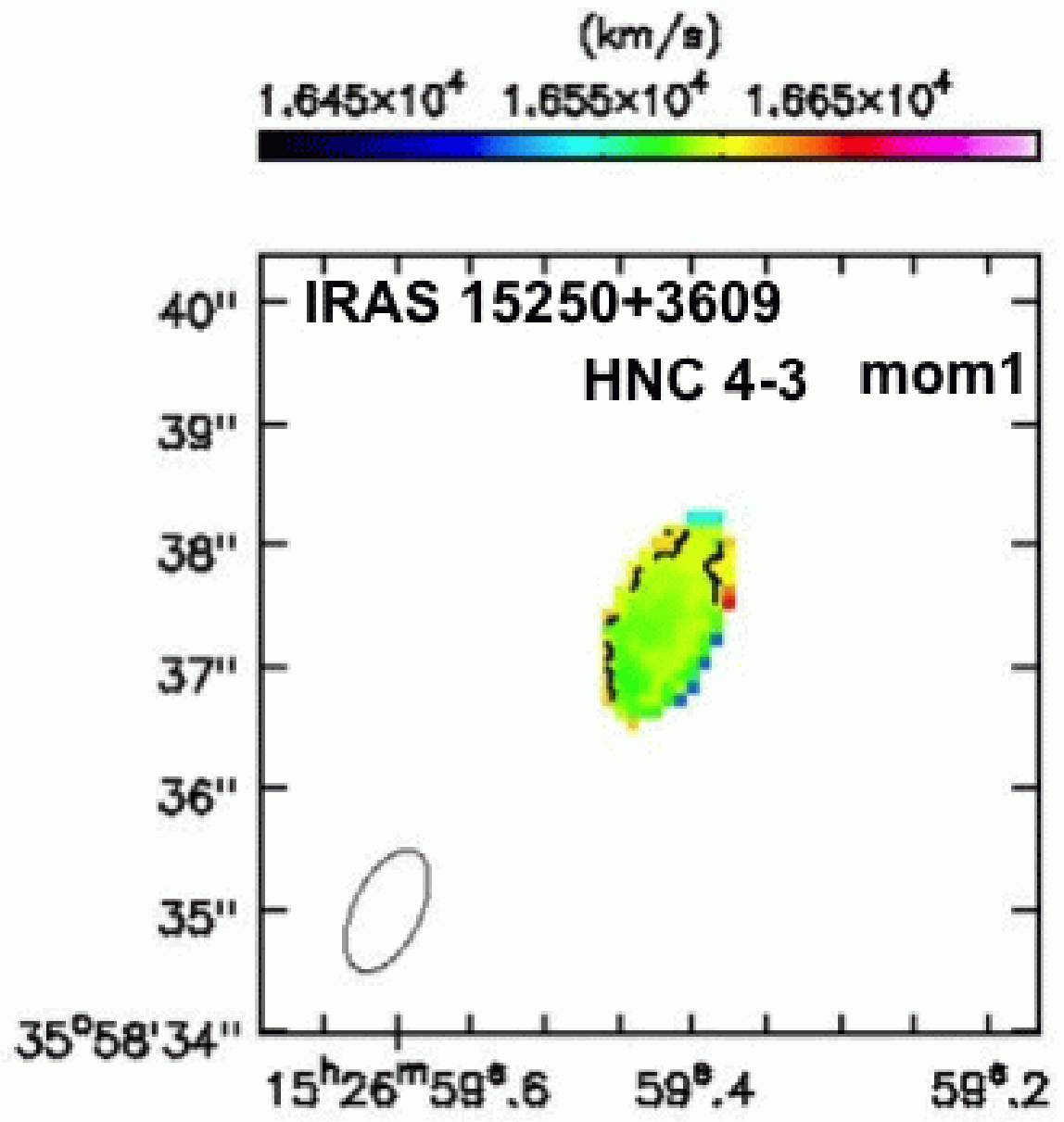} 
\includegraphics[angle=0,scale=.41]{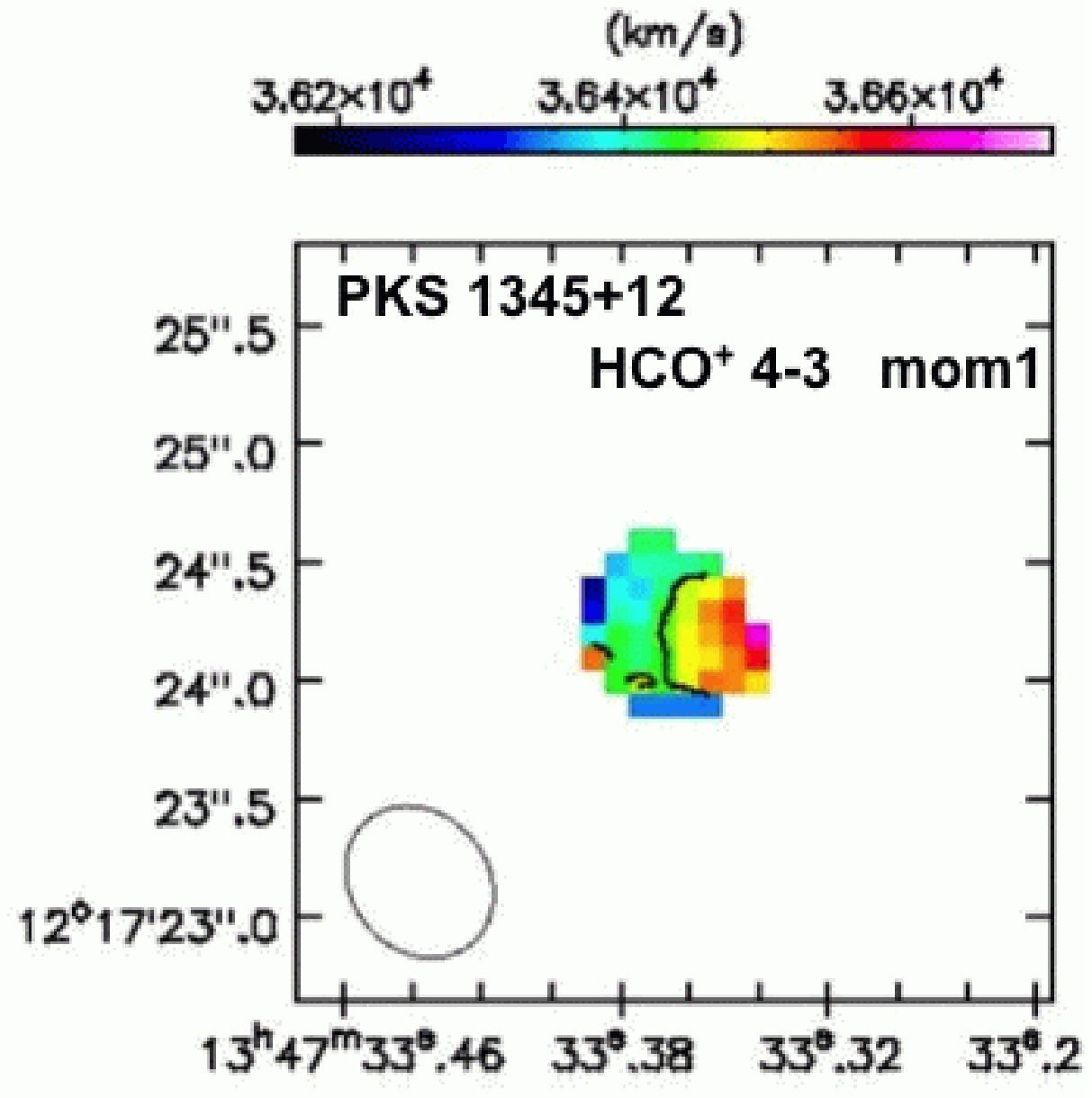}
\includegraphics[angle=0,scale=.41]{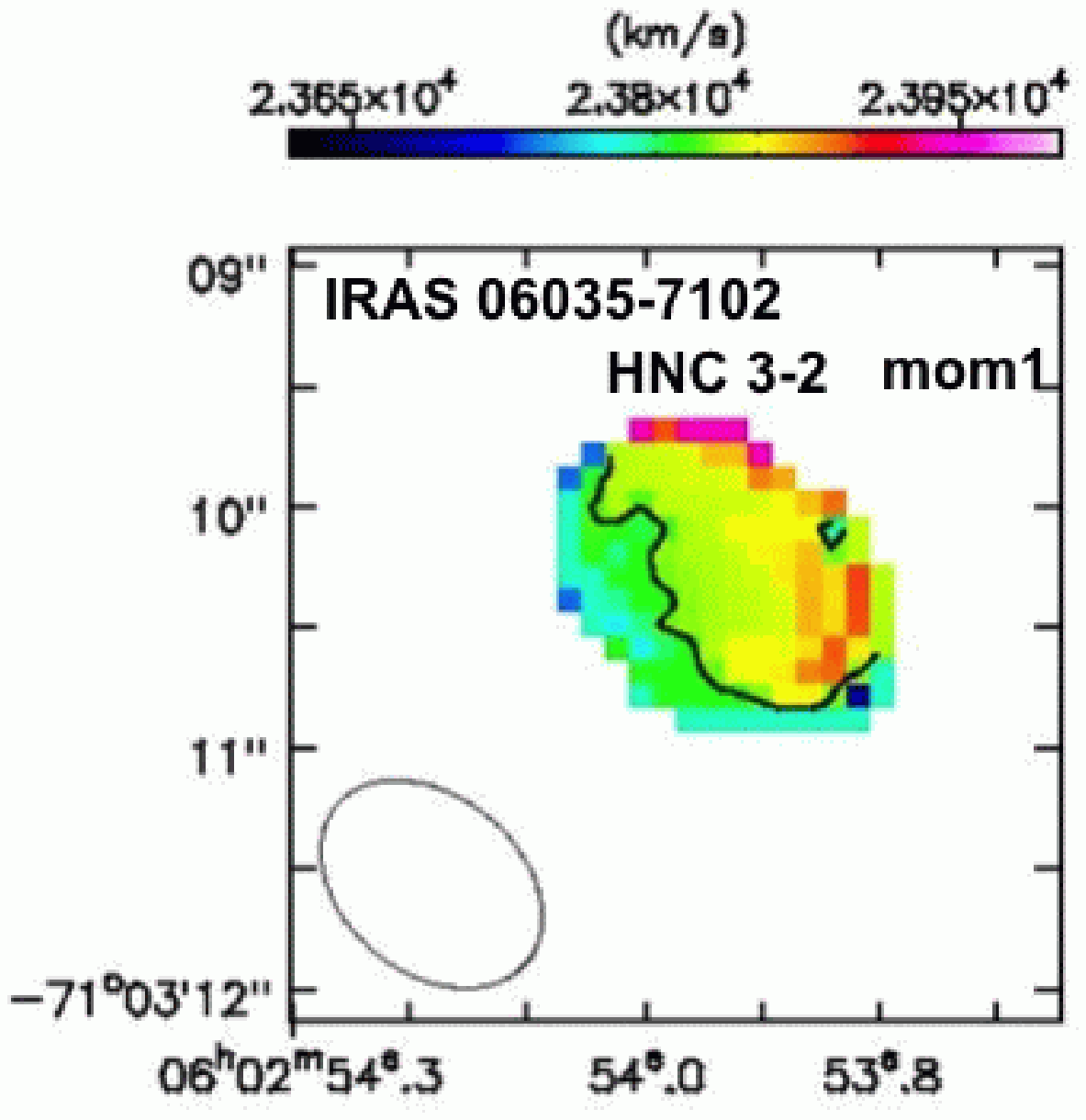} 
\end{center}
\end{figure}

\clearpage

\begin{figure}
\begin{center}
\includegraphics[angle=0,scale=.41]{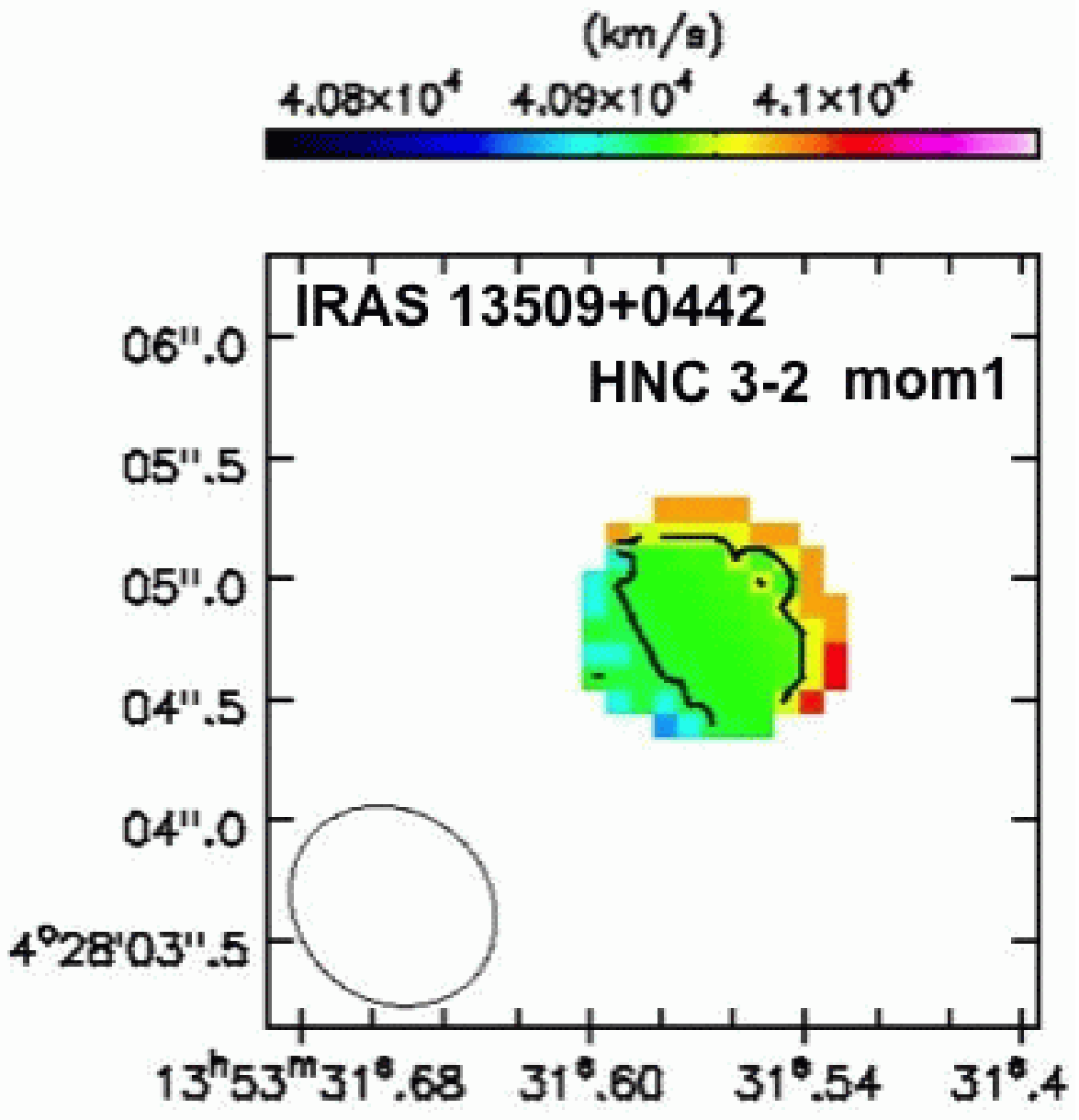} 
\includegraphics[angle=0,scale=.41]{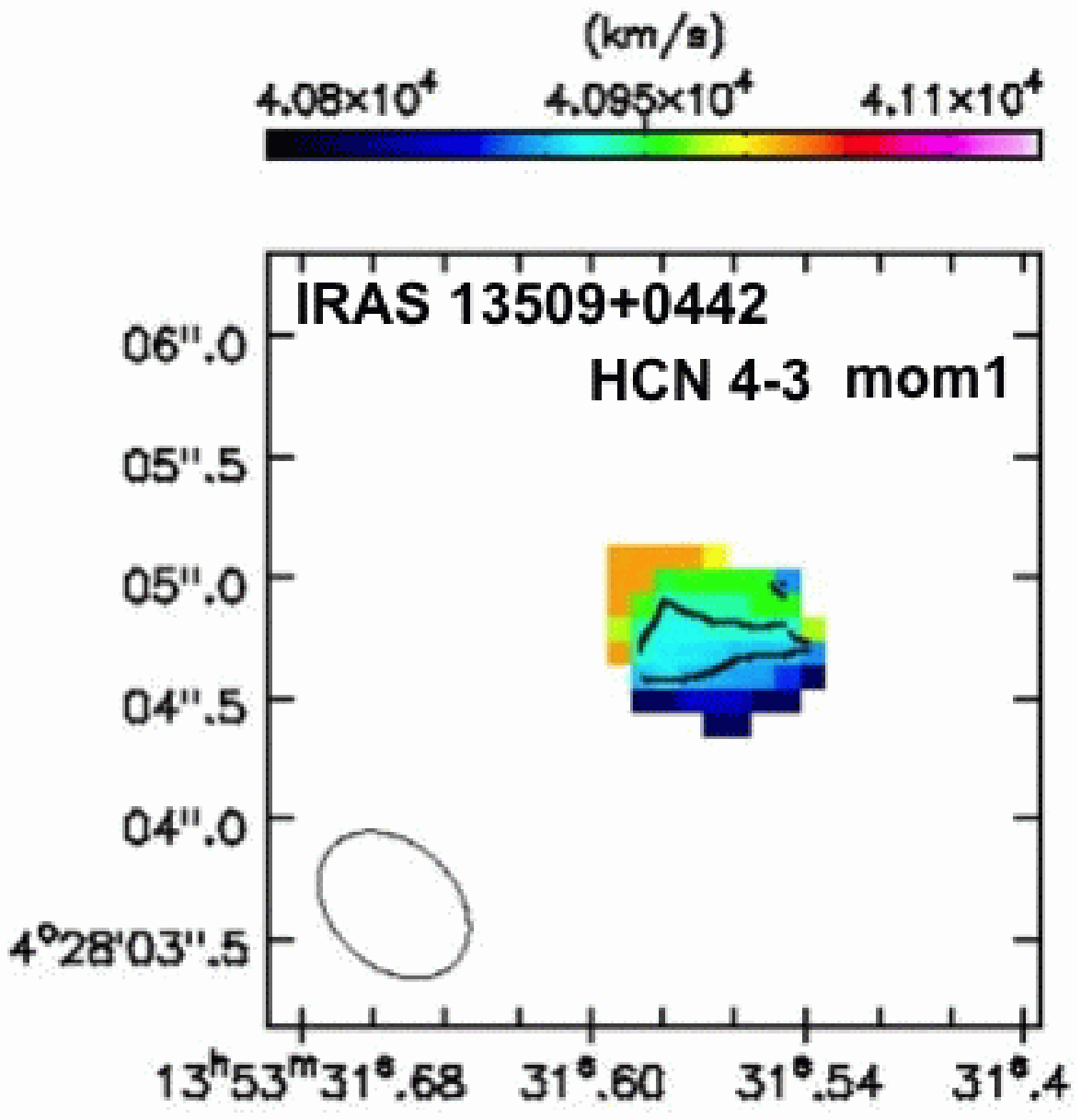} 
\includegraphics[angle=0,scale=.41]{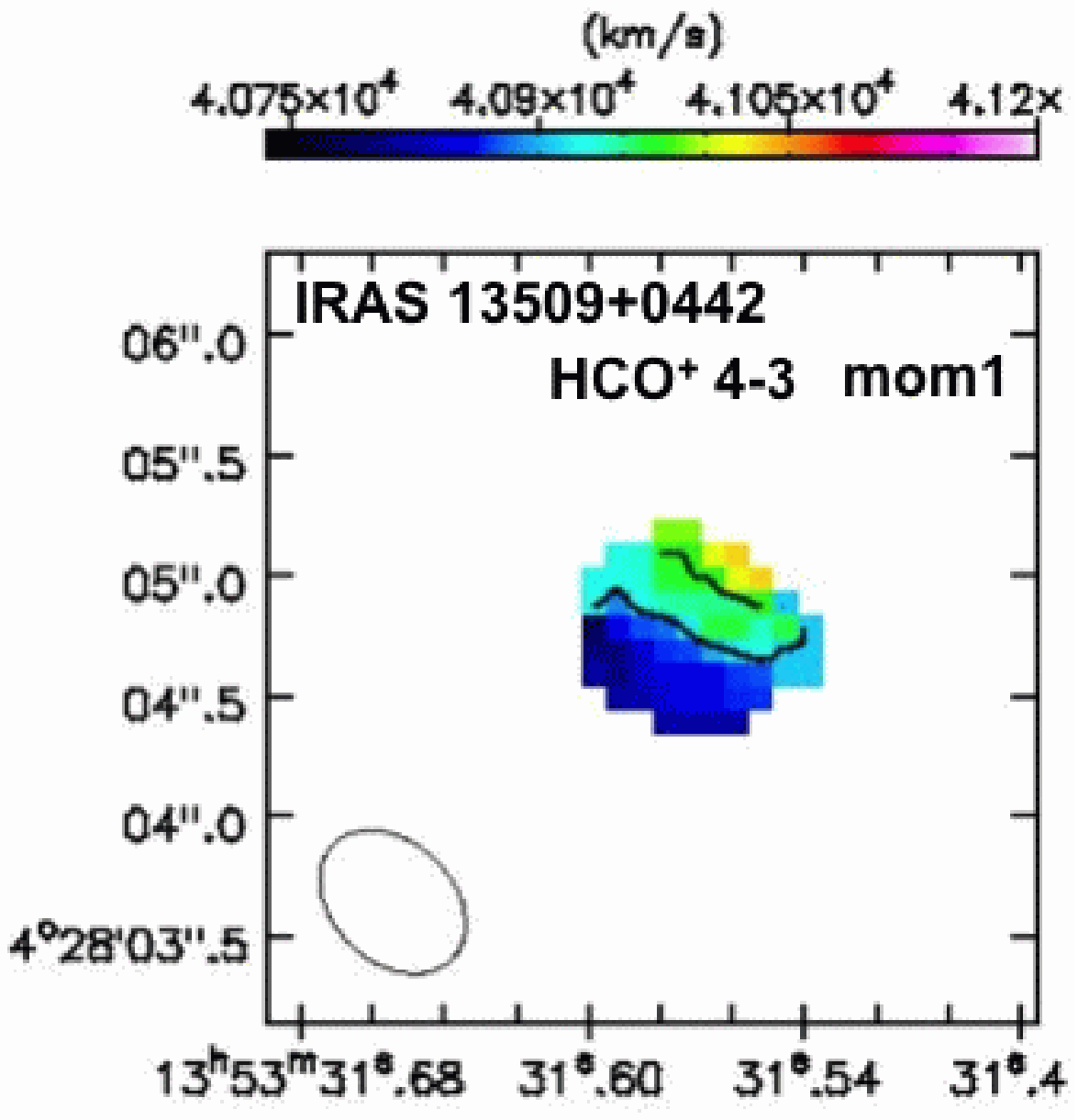} \\
\vspace{-1.3cm}
\includegraphics[angle=0,scale=.39]{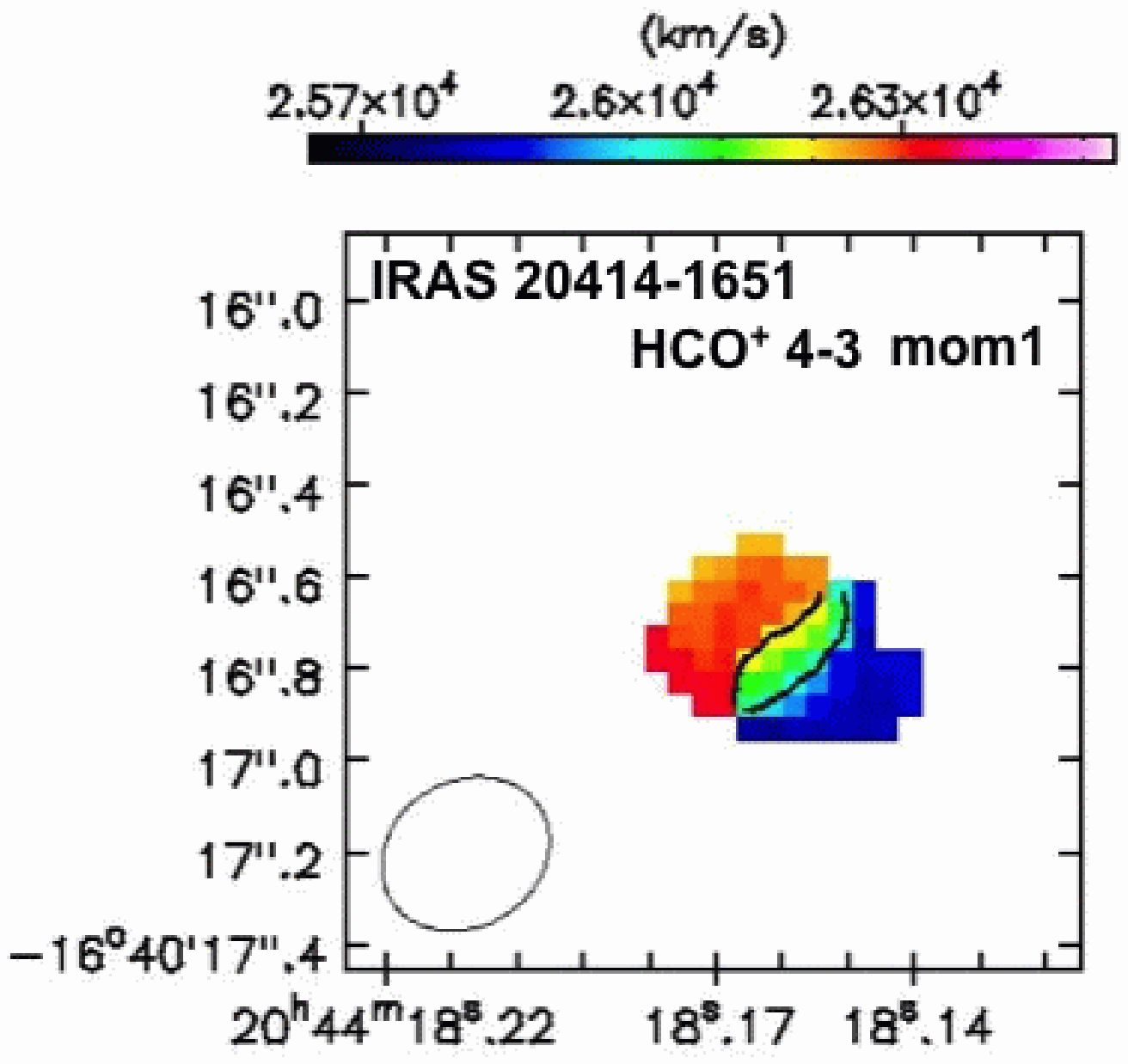} 
\caption{
Intensity-weighted mean velocity (moment 1) maps for emission lines with
$>$10$\sigma$ detection in their integrated intensity (moment 0) maps.
The abscissa and ordinate are R.A. (J2000) and decl. (J2000),
respectively. 
The contours represent 
21770, 21810 km s$^{-1}$ for IRAS 12112$+$0305 NE HNC J=3--2,
21700, 21750, 21800 km s$^{-1}$ for IRAS 12112$+$0305 NE HCN J=4--3,
21650, 21800 km s$^{-1}$ for IRAS 12112$+$0305 NE HCO$^{+}$ J=4--3, 
21765, 21820 km s$^{-1}$ for IRAS 12112$+$0305 NE HNC J=4--3, 
23270, 23300, 23330 km s$^{-1}$ for IRAS 22491$-$1808 HNC J=3--2, 
39960, 40000 km s$^{-1}$ for IRAS 12127$-$1412 HNC J=3--2,
16610 km s$^{-1}$ for IRAS 15250$+$3609 HNC J=3--2,
16560, 16600 km s$^{-1}$ for IRAS 15250$+$3609 HCN J=4--3,
16570 km s$^{-1}$ for IRAS 15250$+$3609 HCO$^{+}$ J=4--3,
16600 km s$^{-1}$ for IRAS 15250$+$3609 HNC J=4--3,
36450 km s$^{-1}$ for PKS 1345$+$12 HCO$^{+}$ J=4--3,
23820 km s$^{-1}$ for IRAS 06035$-$7102 HNC J=3--2, 
40910, 40950 km s$^{-1}$ for IRAS 13509$+$0442 HNC J=3--2, 
40900, 40930 km s$^{-1}$ for IRAS 13509$+$0442 HCN J=4--3, 
40910, 40970 km s$^{-1}$ for IRAS 13509$+$0442 HCO$^{+}$ J=4--3, 
and 
26000, 26200 km s$^{-1}$ for IRAS 20414$-$1651 HCO$^{+}$ J=4--3.
For HCN J=4--3 and HCO$^{+}$ J=4--3 emission of IRAS 15250$+$3609, 
only the main nuclear emission components (see $\S$5.1.1) are used to
create the moment 1 maps.
Beam sizes are shown as open circles at the lower-left part.
We applied an appropriate cut-off to prevent the resulting maps
from being dominated by noise.
}
\end{center}
\end{figure}


\begin{figure}
\begin{center}
\includegraphics[angle=0,scale=.41]{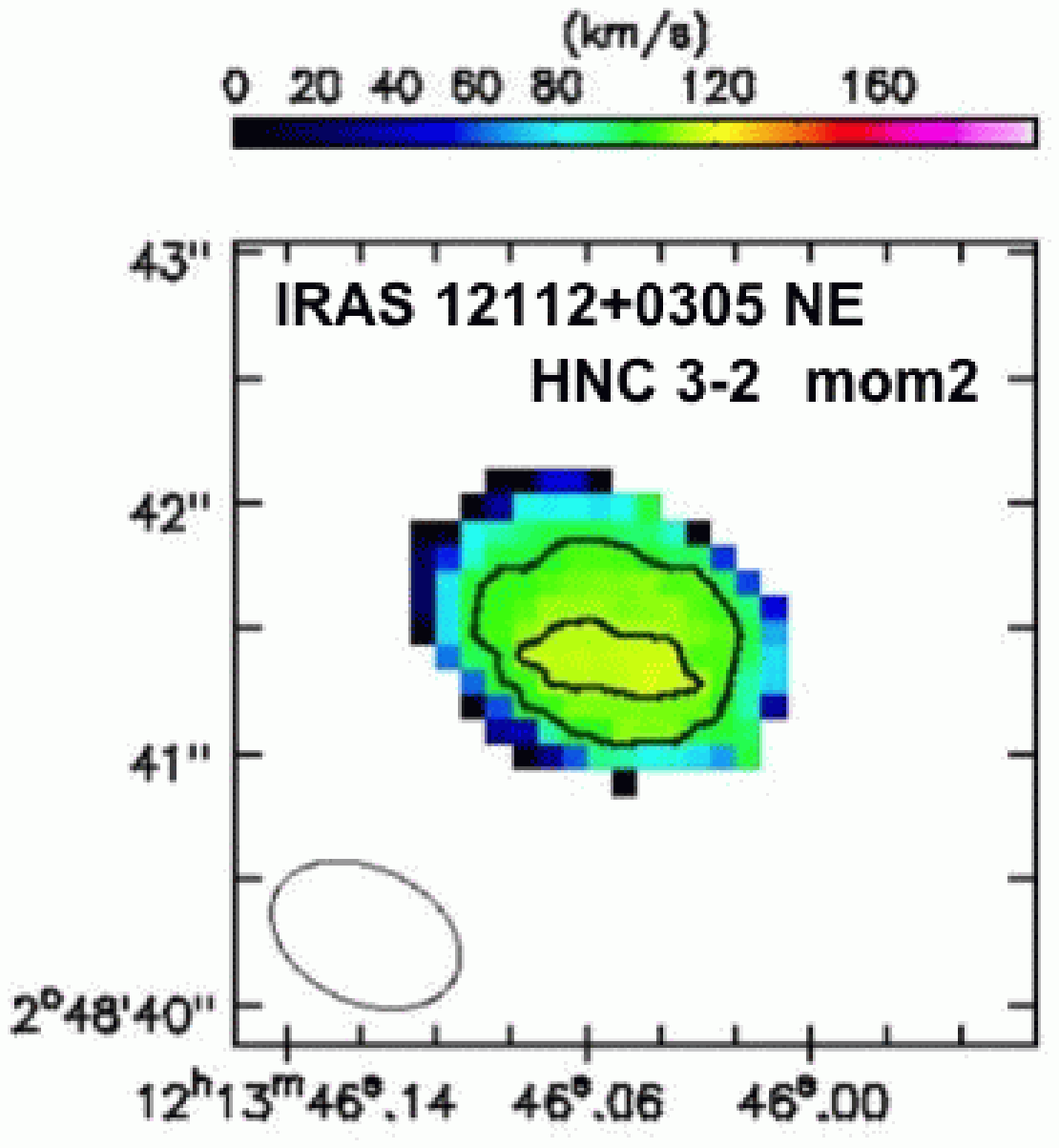} 
\includegraphics[angle=0,scale=.41]{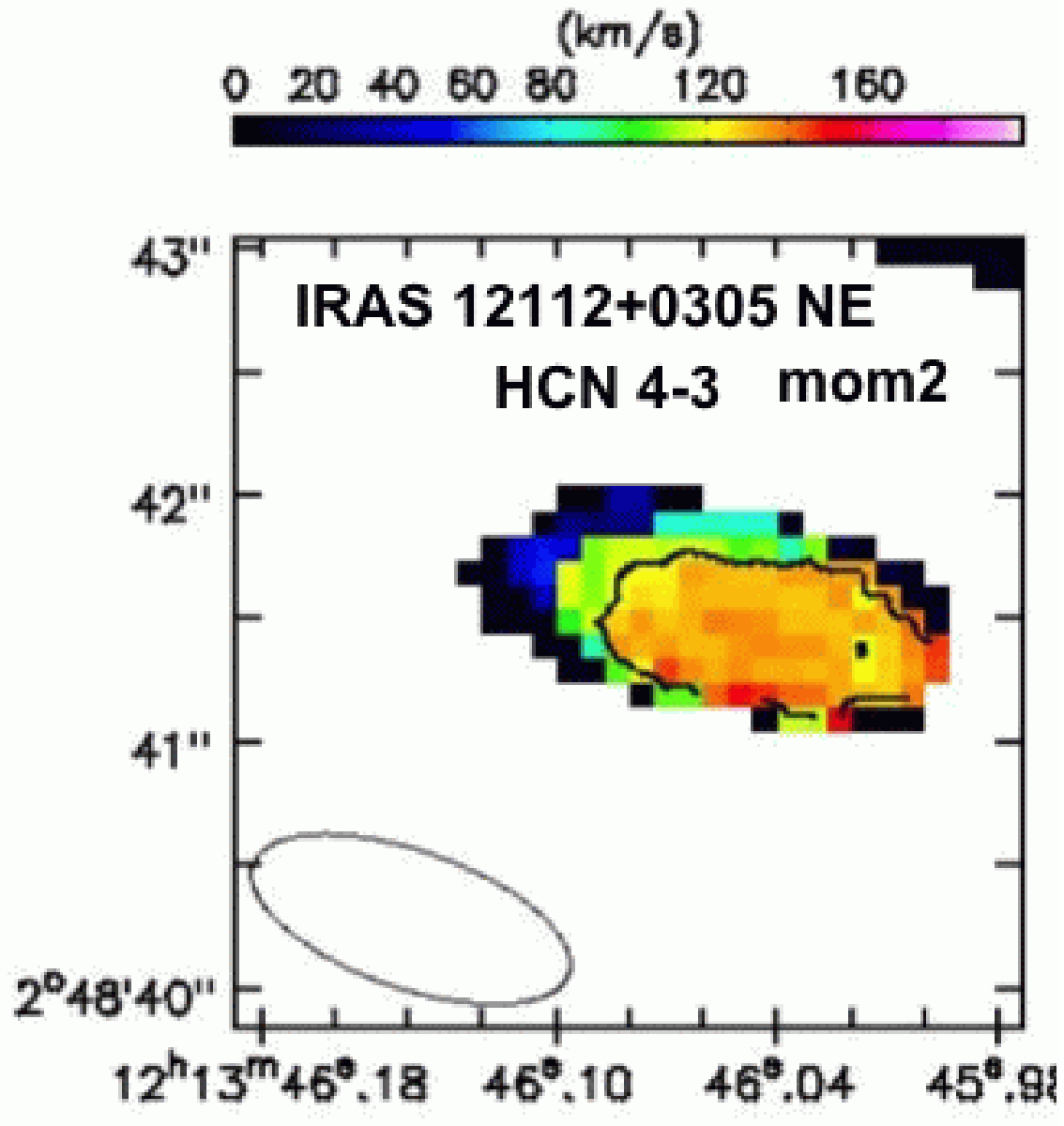}  
\includegraphics[angle=0,scale=.41]{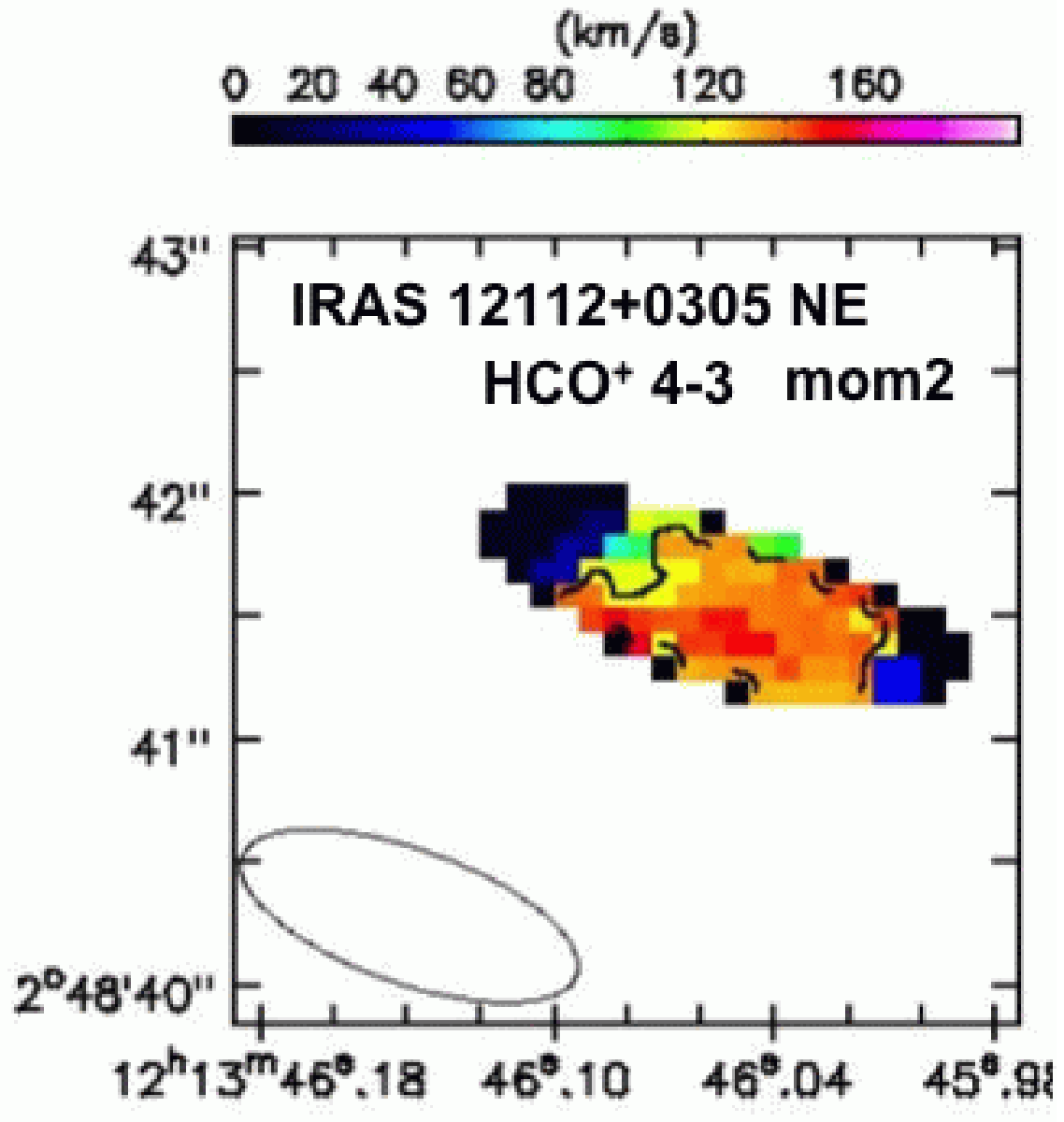} \\ 
\vspace{-1.3cm}
\includegraphics[angle=0,scale=.41]{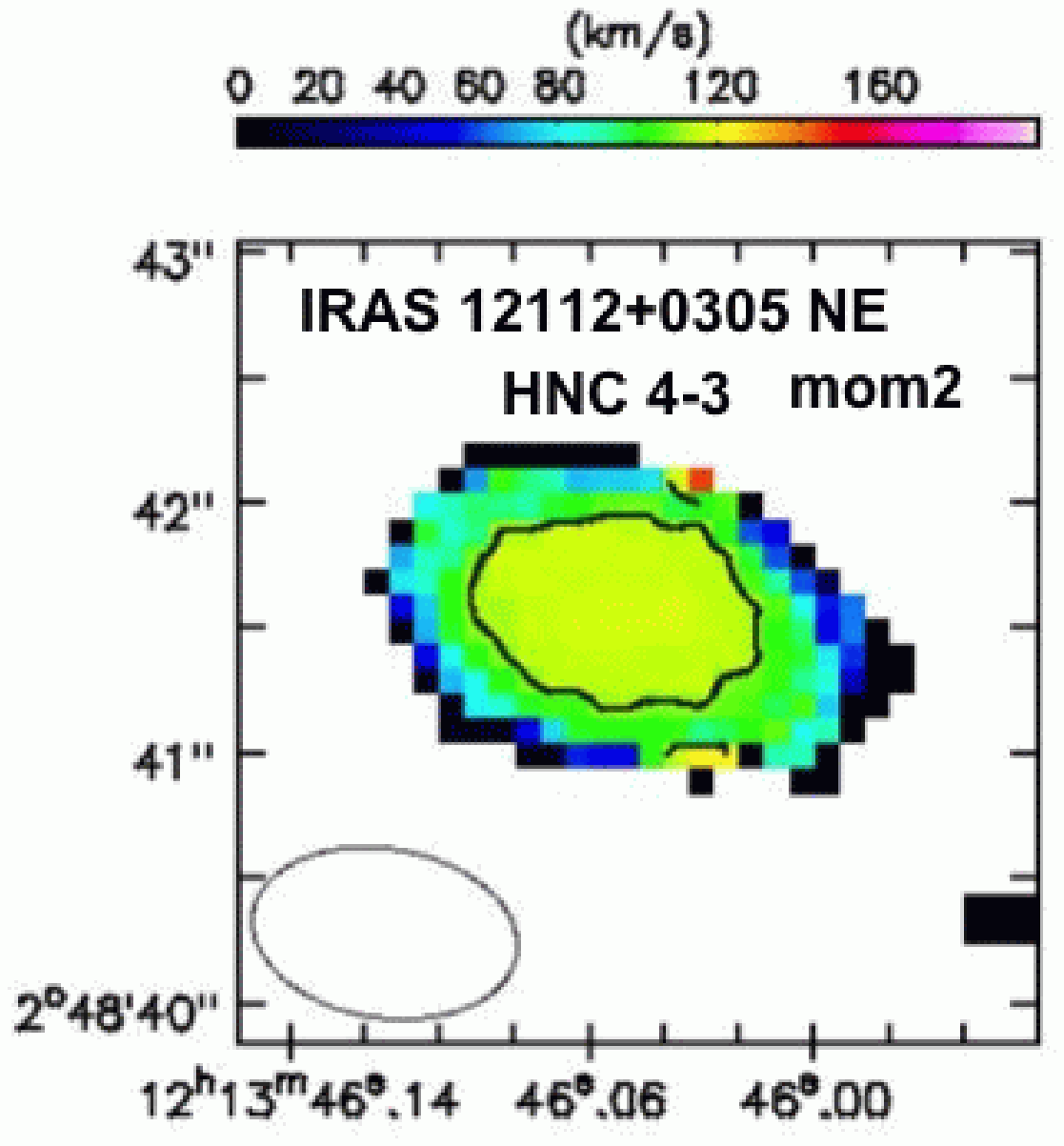}  
\includegraphics[angle=0,scale=.41]{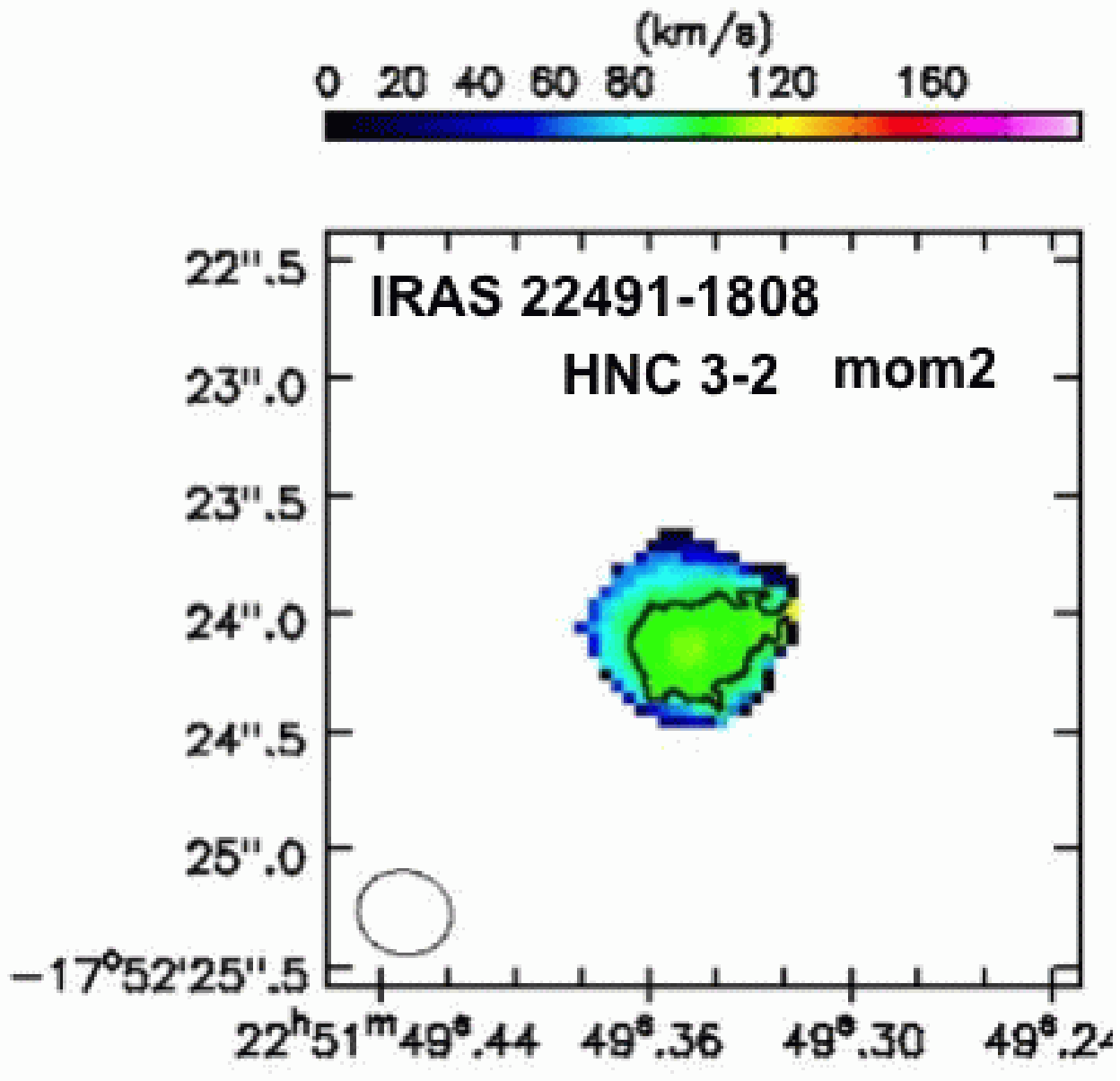} 
\includegraphics[angle=0,scale=.41]{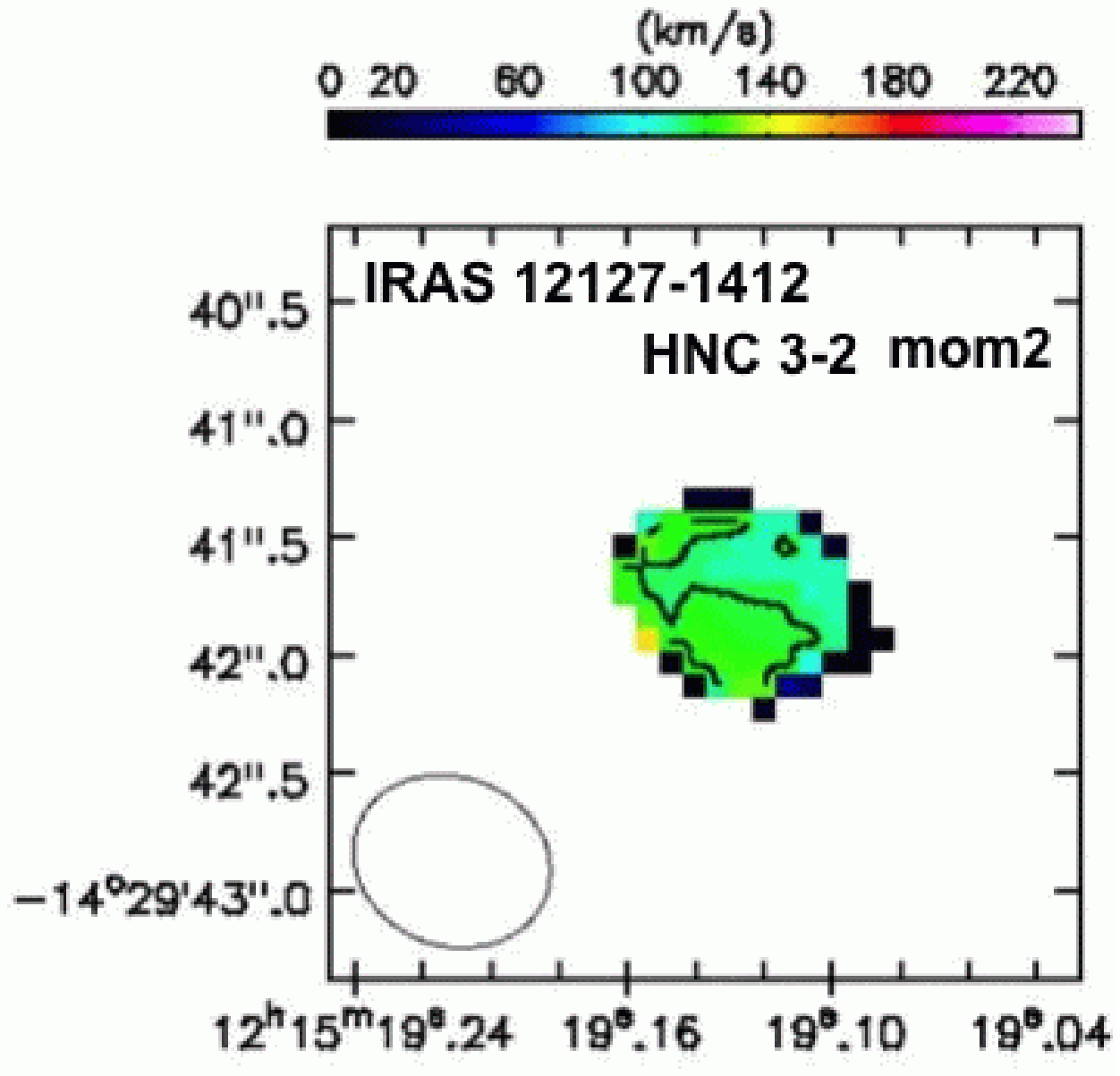} \\  
\vspace{-1.3cm}
\includegraphics[angle=0,scale=.41]{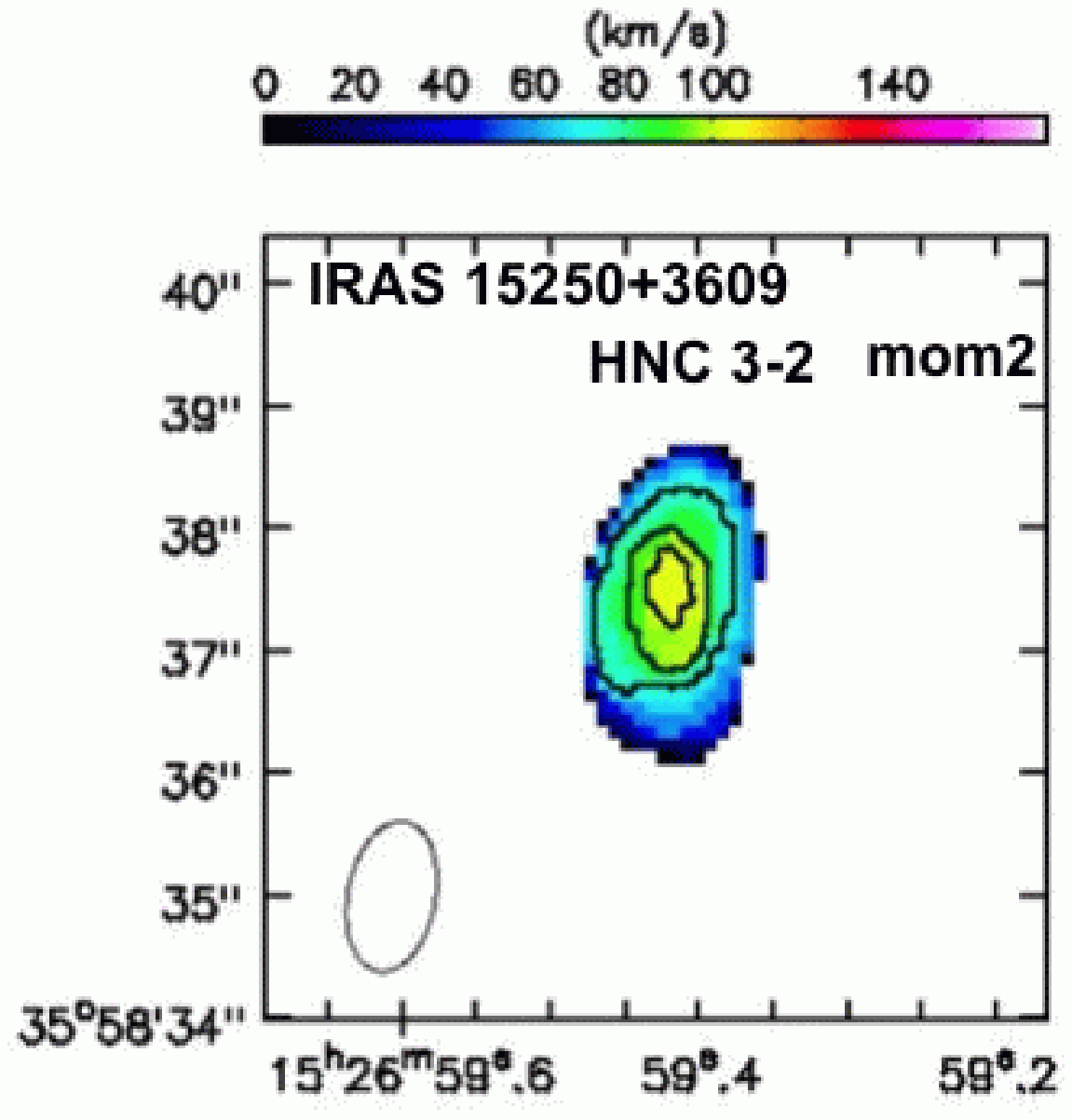} 
\includegraphics[angle=0,scale=.41]{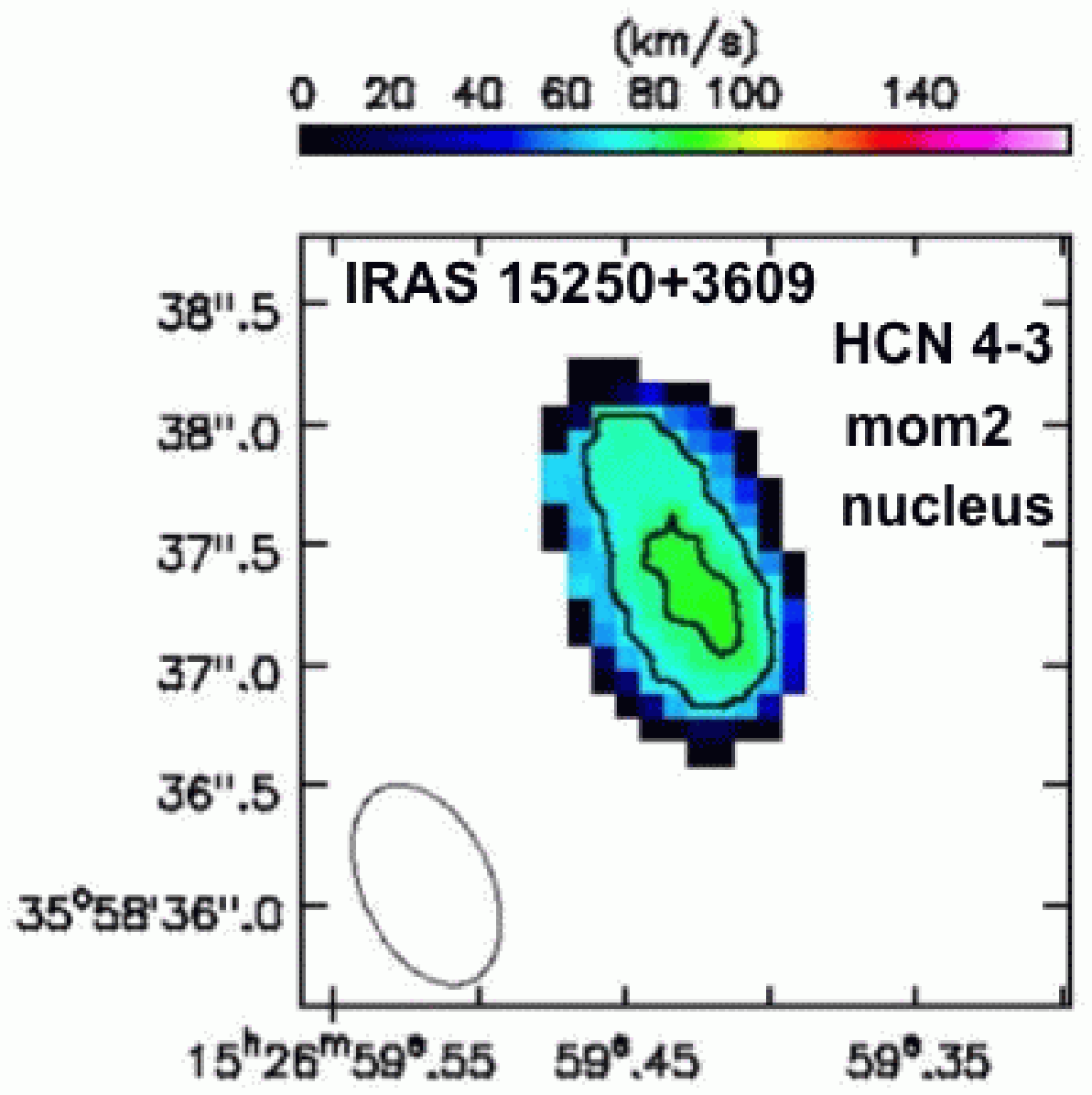}  
\includegraphics[angle=0,scale=.41]{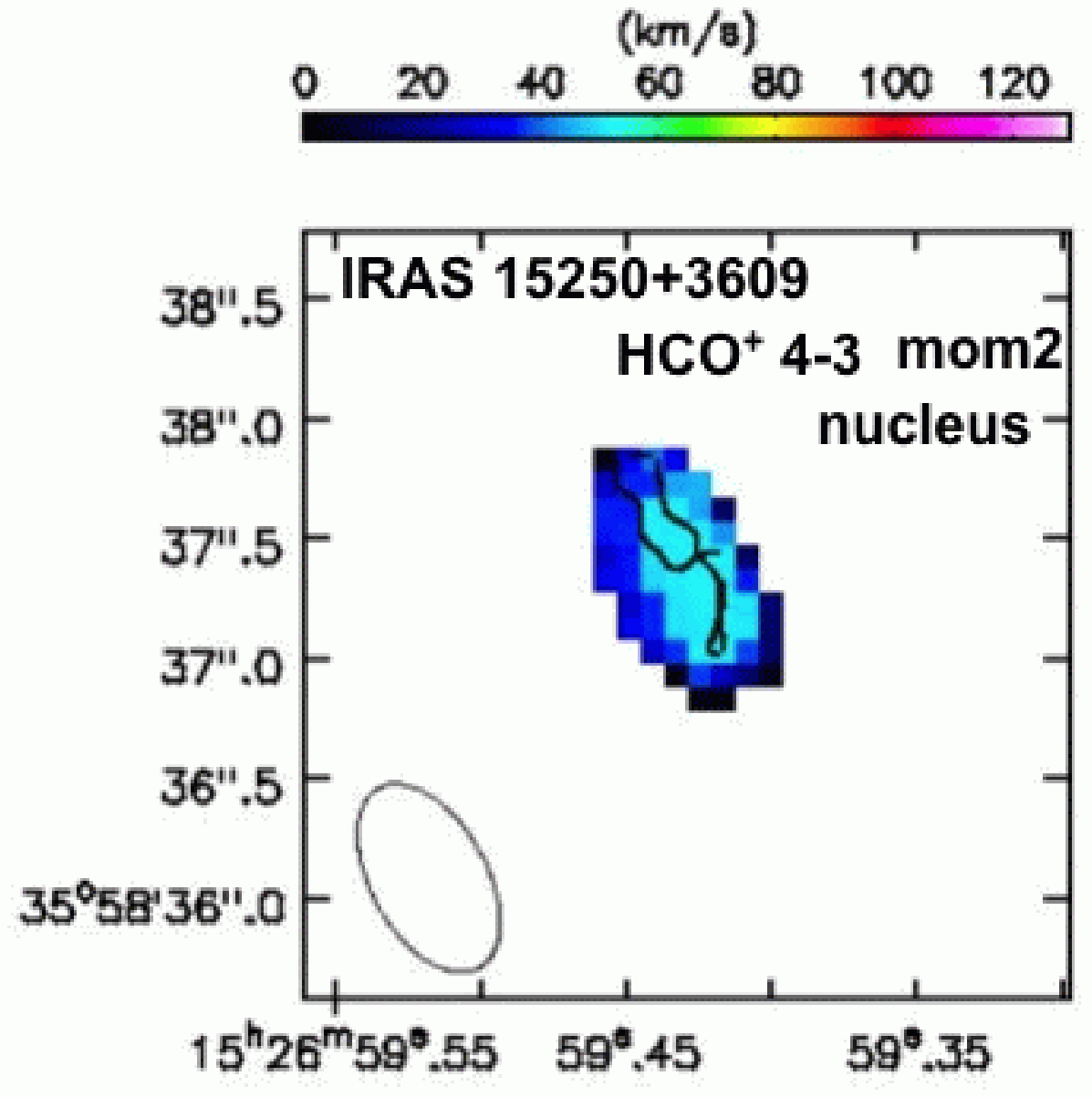} \\ 
\vspace{-1.3cm}
\includegraphics[angle=0,scale=.41]{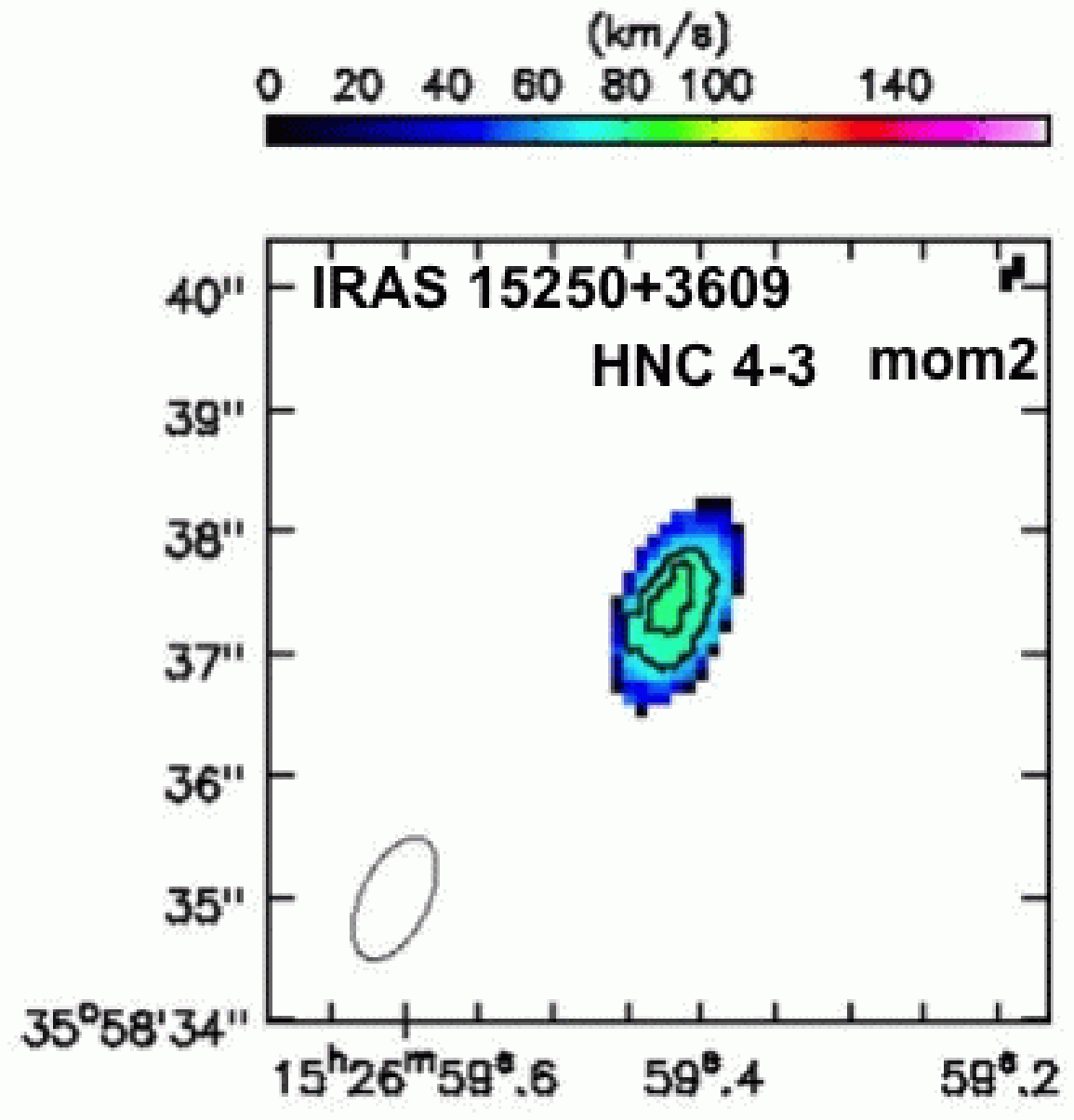} 
\includegraphics[angle=0,scale=.41]{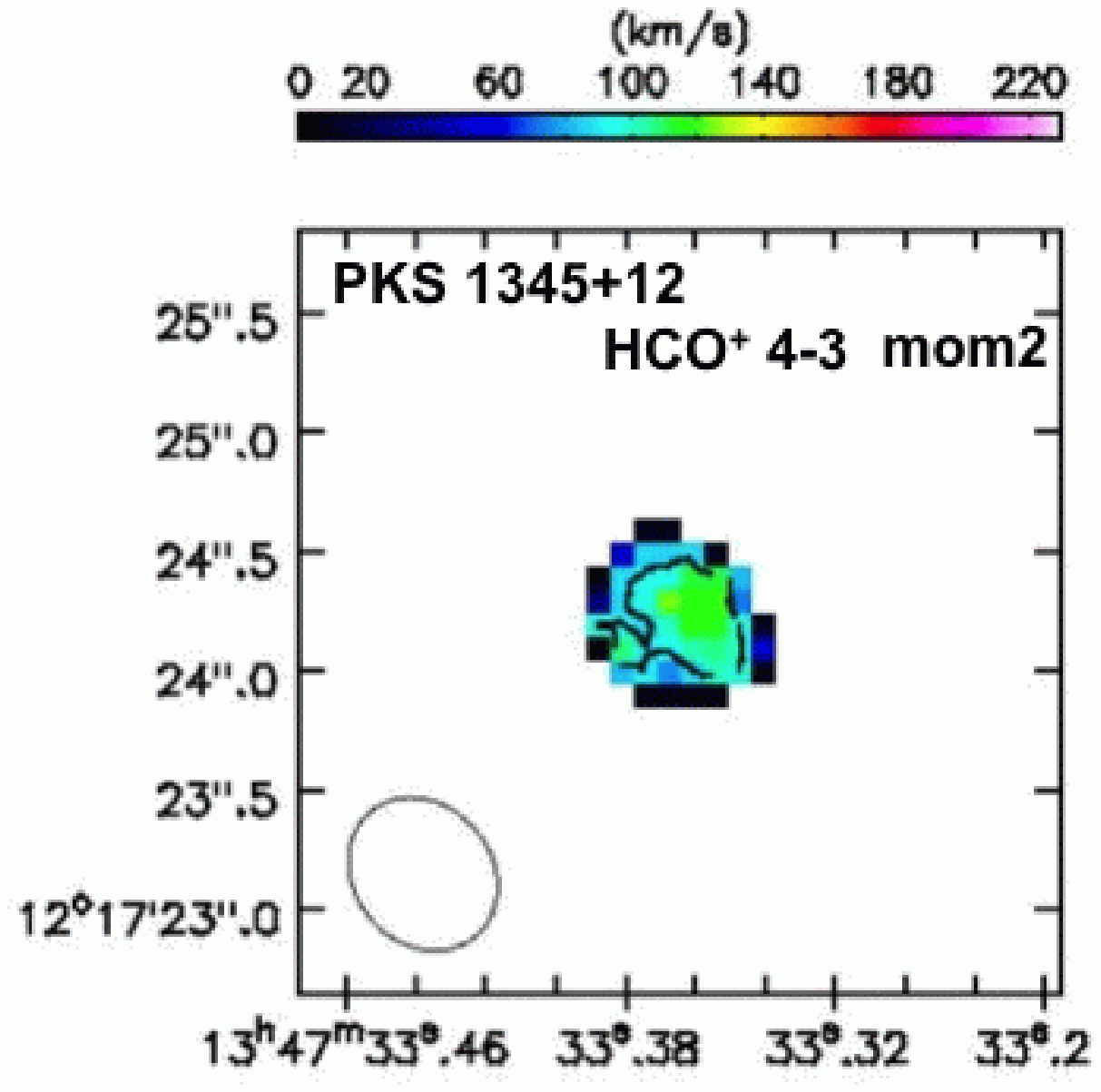}  
\includegraphics[angle=0,scale=.41]{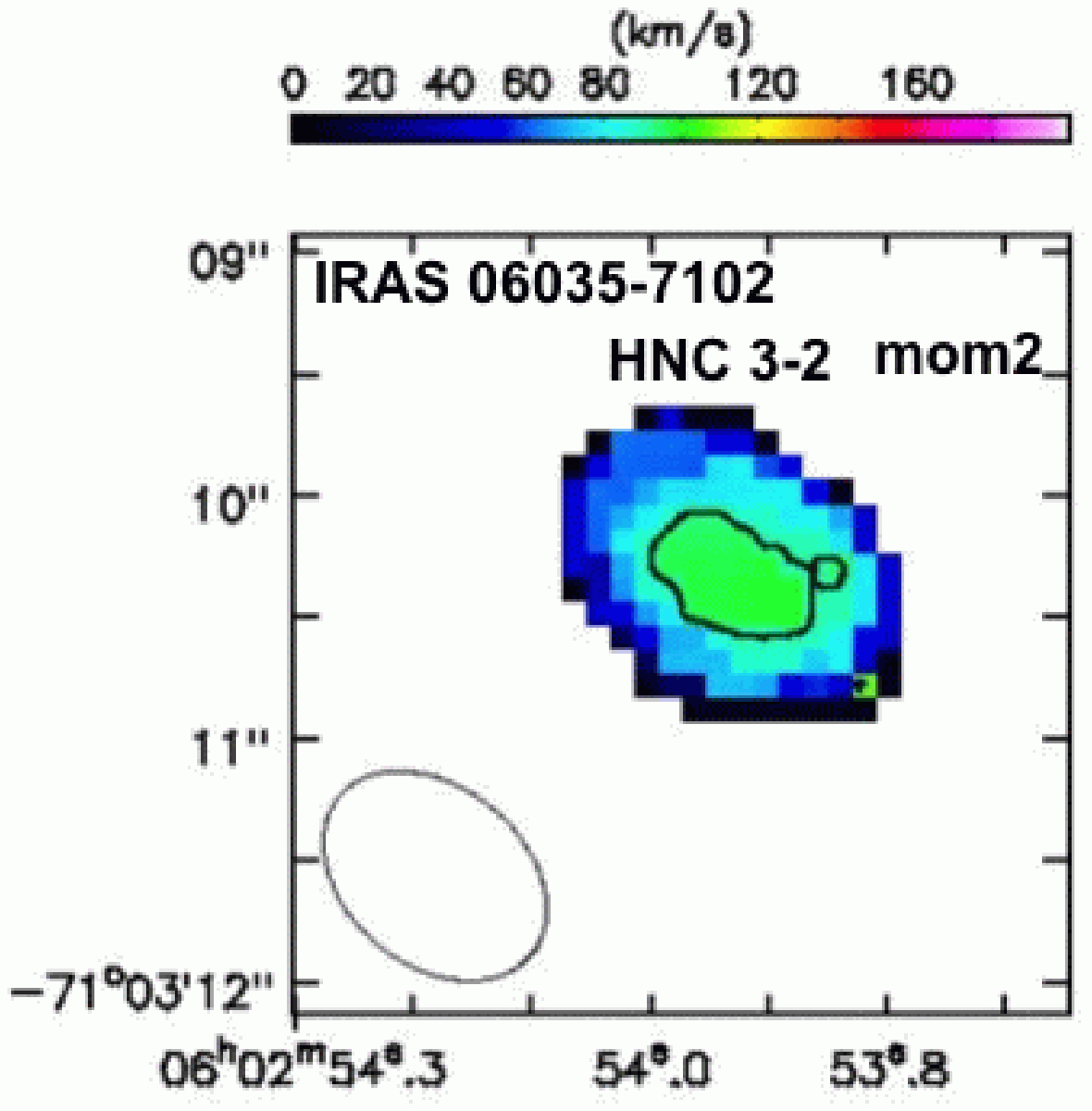} \\
\end{center}
\end{figure}

\clearpage

\begin{figure}
\begin{center}
\includegraphics[angle=0,scale=.41]{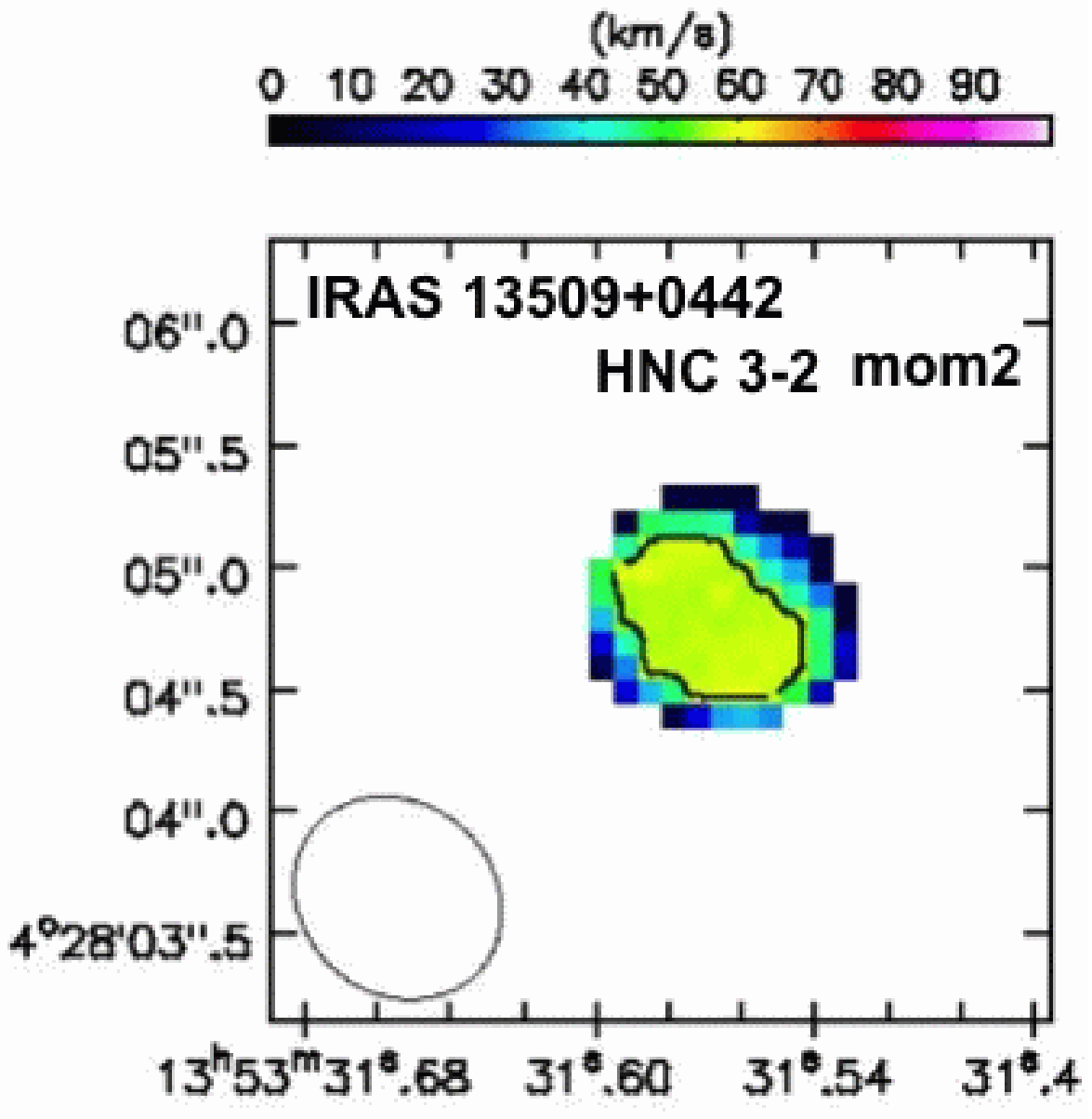} 
\includegraphics[angle=0,scale=.41]{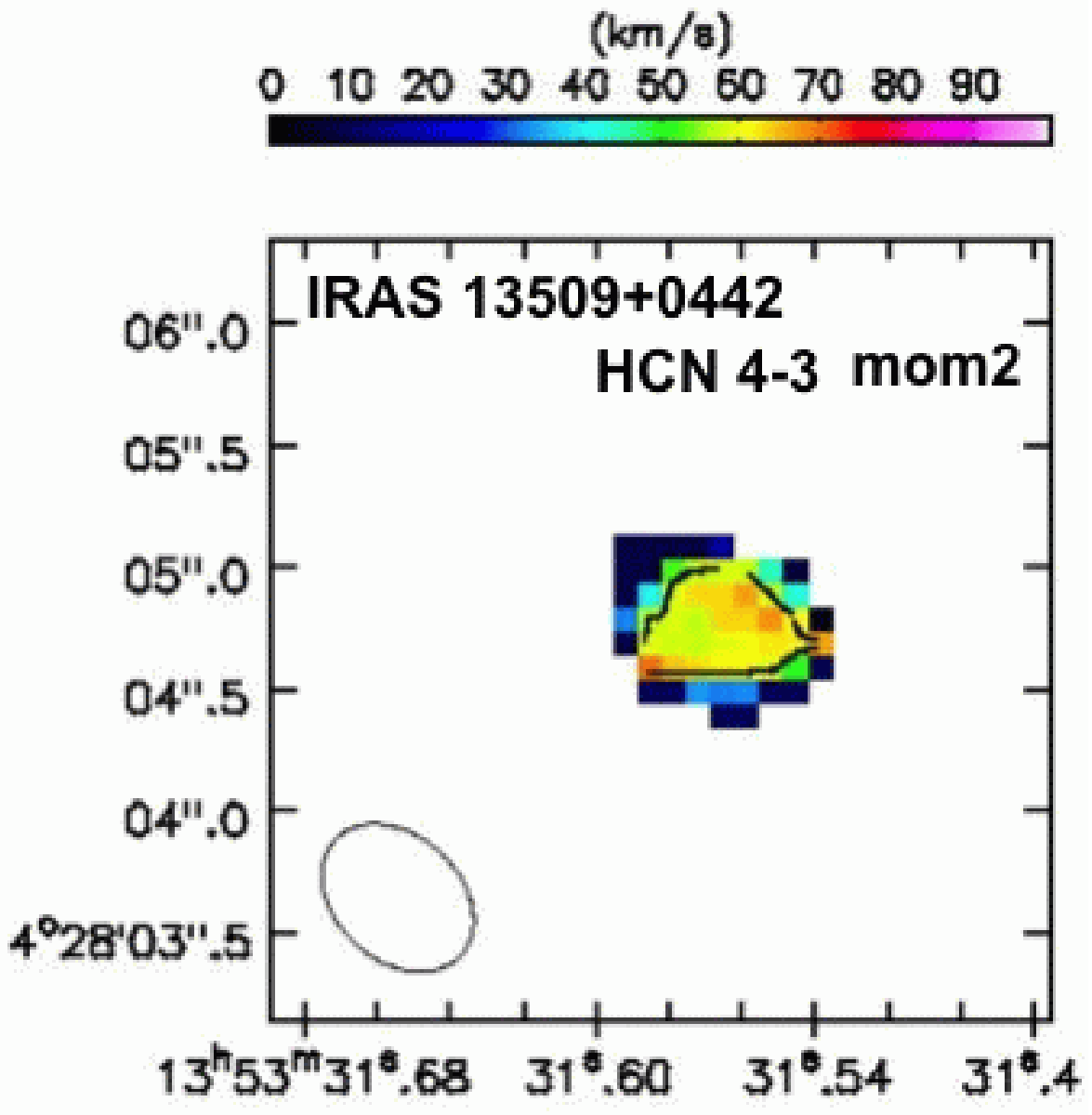} 
\includegraphics[angle=0,scale=.41]{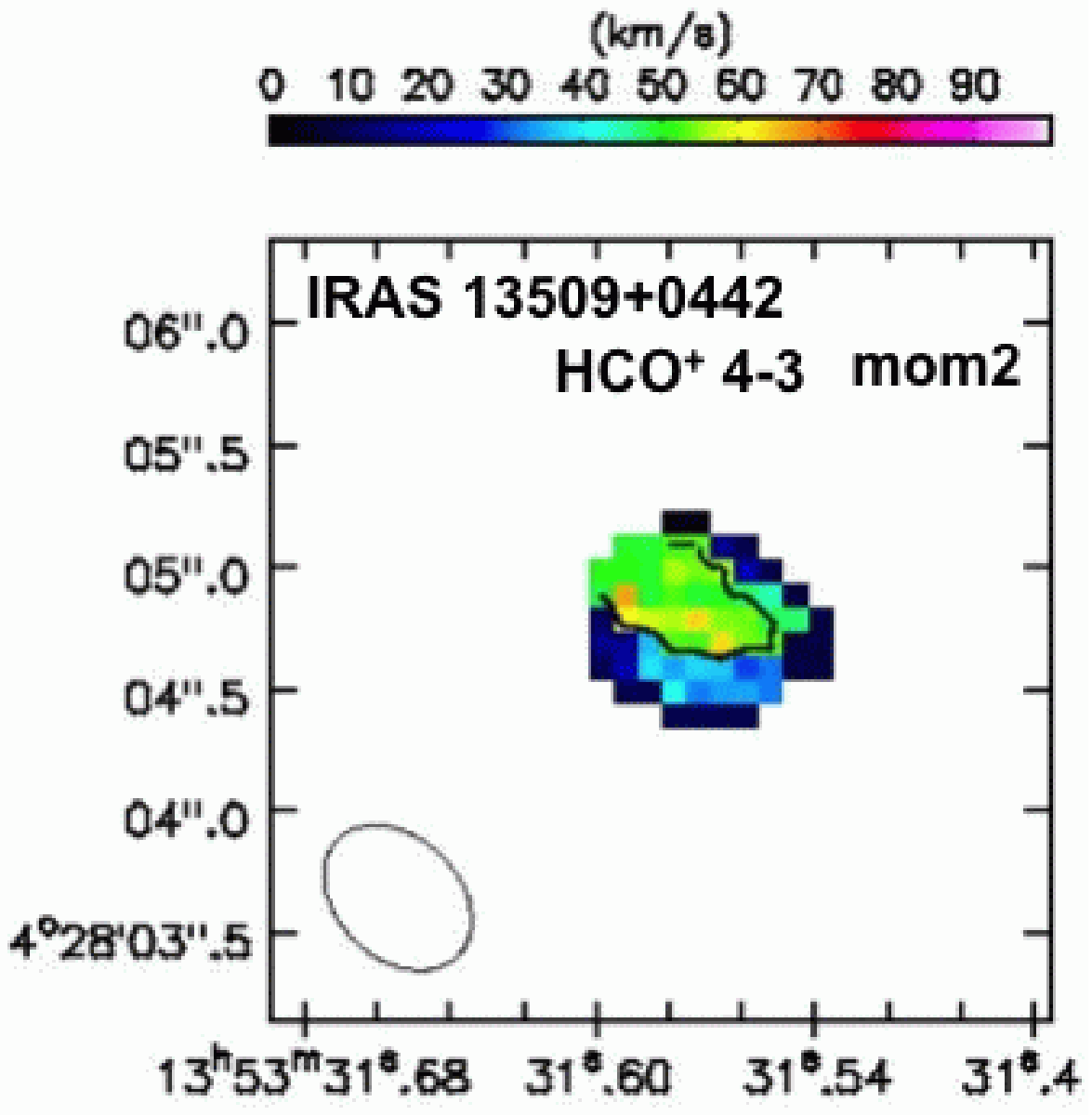} \\
\vspace{-1.3cm}
\includegraphics[angle=0,scale=.39]{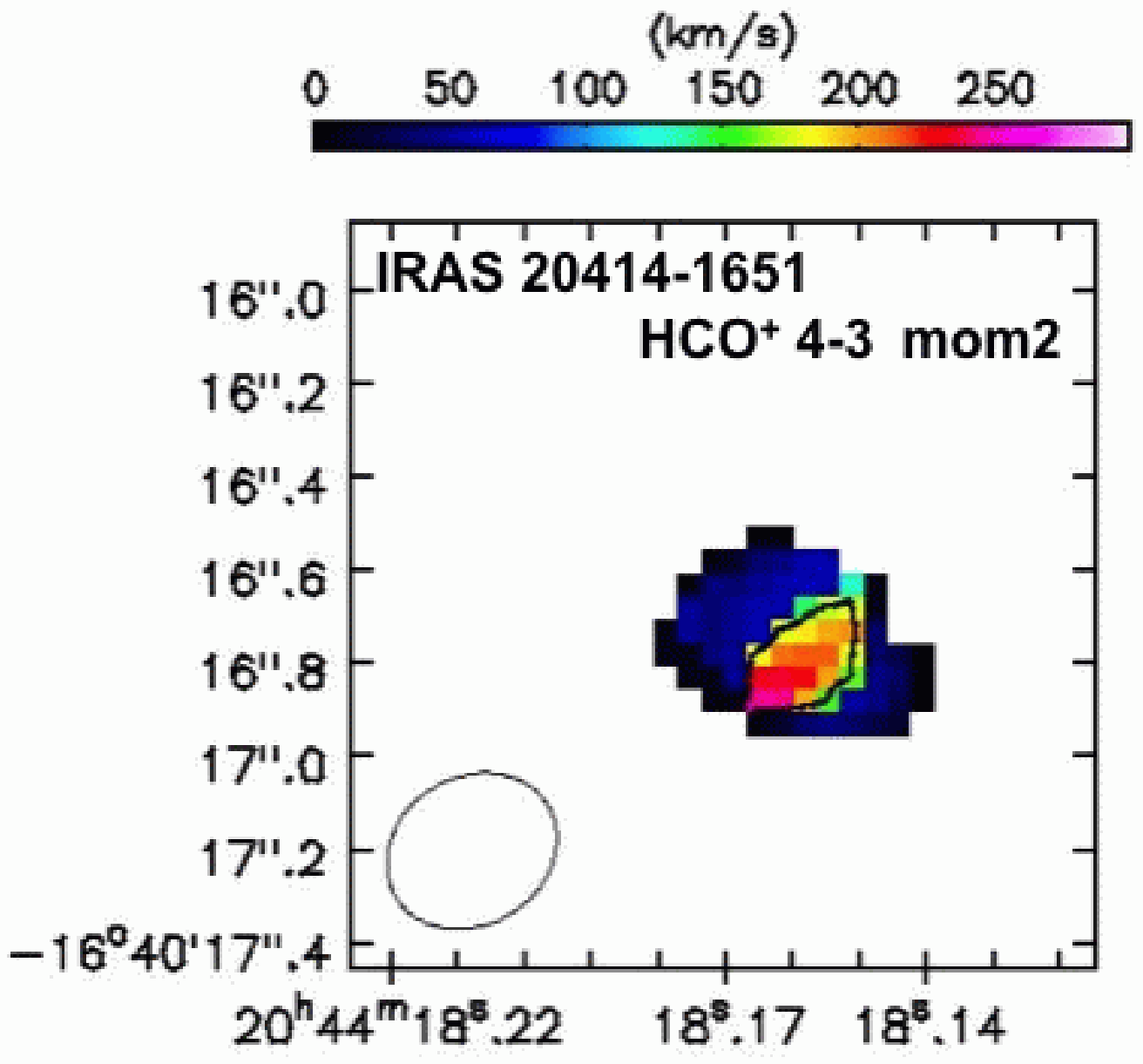} \\
\end{center}
\caption{
Intensity-weighted velocity dispersion (moment 2) maps for emission lines with
$>$10$\sigma$ detection in their integrated intensity (moment 0) maps.
The abscissa and ordinate are R.A. (J2000) and decl. (J2000),
respectively. 
The contours represent 
100, 112 km s$^{-1}$ for IRAS 12112$+$0305 NE HNC J=3--2,
120 km s$^{-1}$ for IRAS 12112$+$0305 NE HCN J=4--3,
120 km s$^{-1}$ for IRAS 12112$+$0305 NE HCO$^{+}$ J=4--3, 
110 km s$^{-1}$ for IRAS 12112$+$0305 NE HNC J=4--3, 
100 km s$^{-1}$ for IRAS 22491$-$1808 HNC J=3--2,
116 km s$^{-1}$ for IRAS 12127$-$1412 HNC J=3--2,
74, 88, 102 km s$^{-1}$ for IRAS 15250$+$3609 HNC J=3--2,
70, 84 km s$^{-1}$ for IRAS 15250$+$3609 HCN J=4--3,
60 km s$^{-1}$ for IRAS 15250$+$3609 HCO$^{+}$ J=4--3,
73, 80 km s$^{-1}$ for IRAS 15250$+$3609 HNC J=4--3,
120 km s$^{-1}$ for PKS 1345$+$12 HCO$^{+}$ J=4--3,
90 km s$^{-1}$ for IRAS 06035$-$7102 HNC J=3--2, 
54 km s$^{-1}$ for IRAS 13509$+$0442 HNC J=3--2, 
56 km s$^{-1}$ for IRAS 13509$+$0442 HCN J=4--3, 
48 km s$^{-1}$ for IRAS 13509$+$0442 HCO$^{+}$ J=4--3, 
and 
160 km s$^{-1}$ for IRAS 20414$-$1651 HCO$^{+}$ J=4--3.
For HCN J=4--3 and HCO$^{+}$ J=4--3 emission of IRAS 15250$+$3609, 
only the main nuclear emission components (see $\S$5.1.1) are used to
create the moment 2 maps. 
Beam sizes are shown as open circles at the lower-left part.
An appropriate cut-off was chosen for these moment 2 maps.
}
\end{figure}

\begin{figure}
\begin{center}
\includegraphics[angle=0,scale=.3]{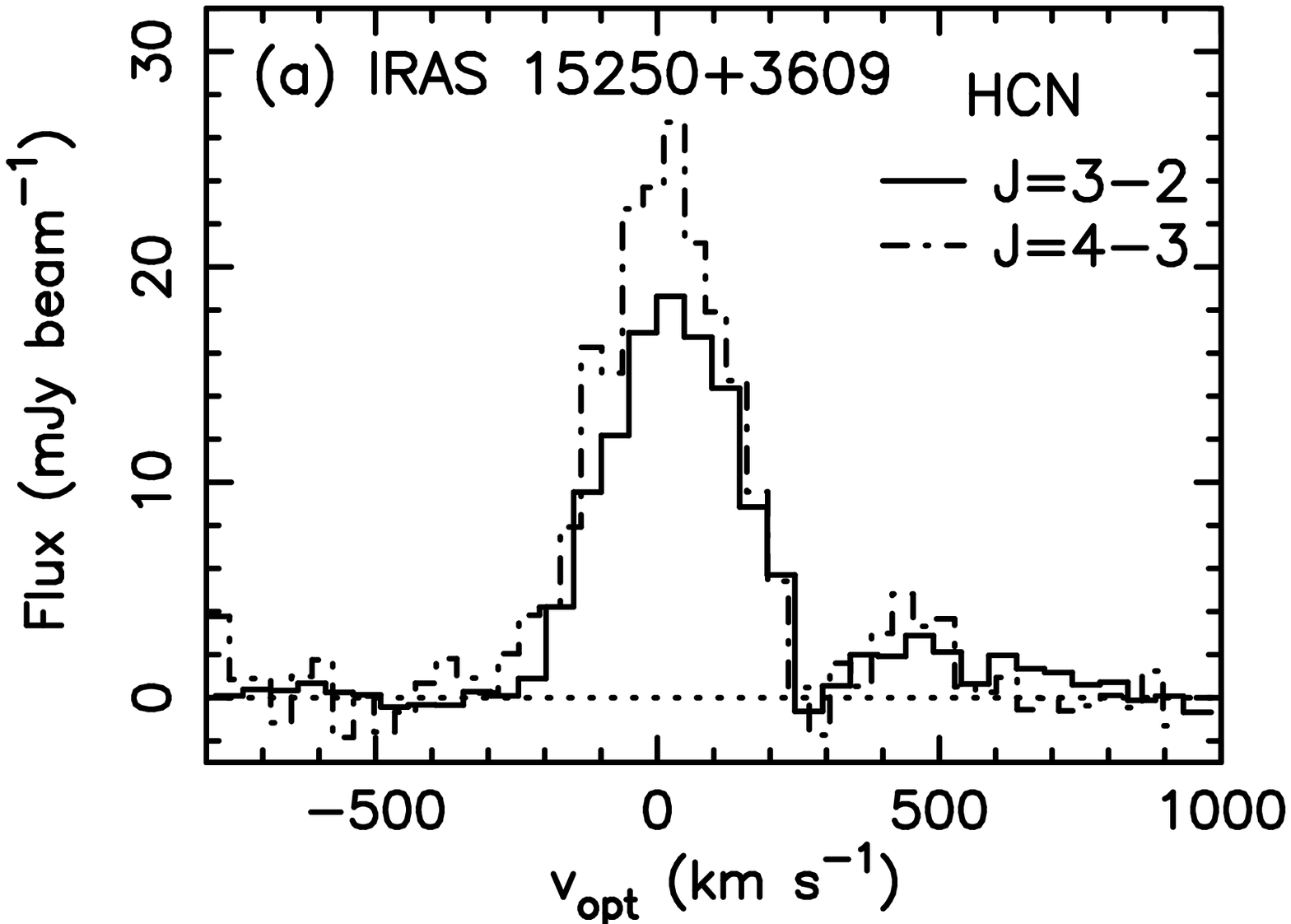}  
\includegraphics[angle=0,scale=.3]{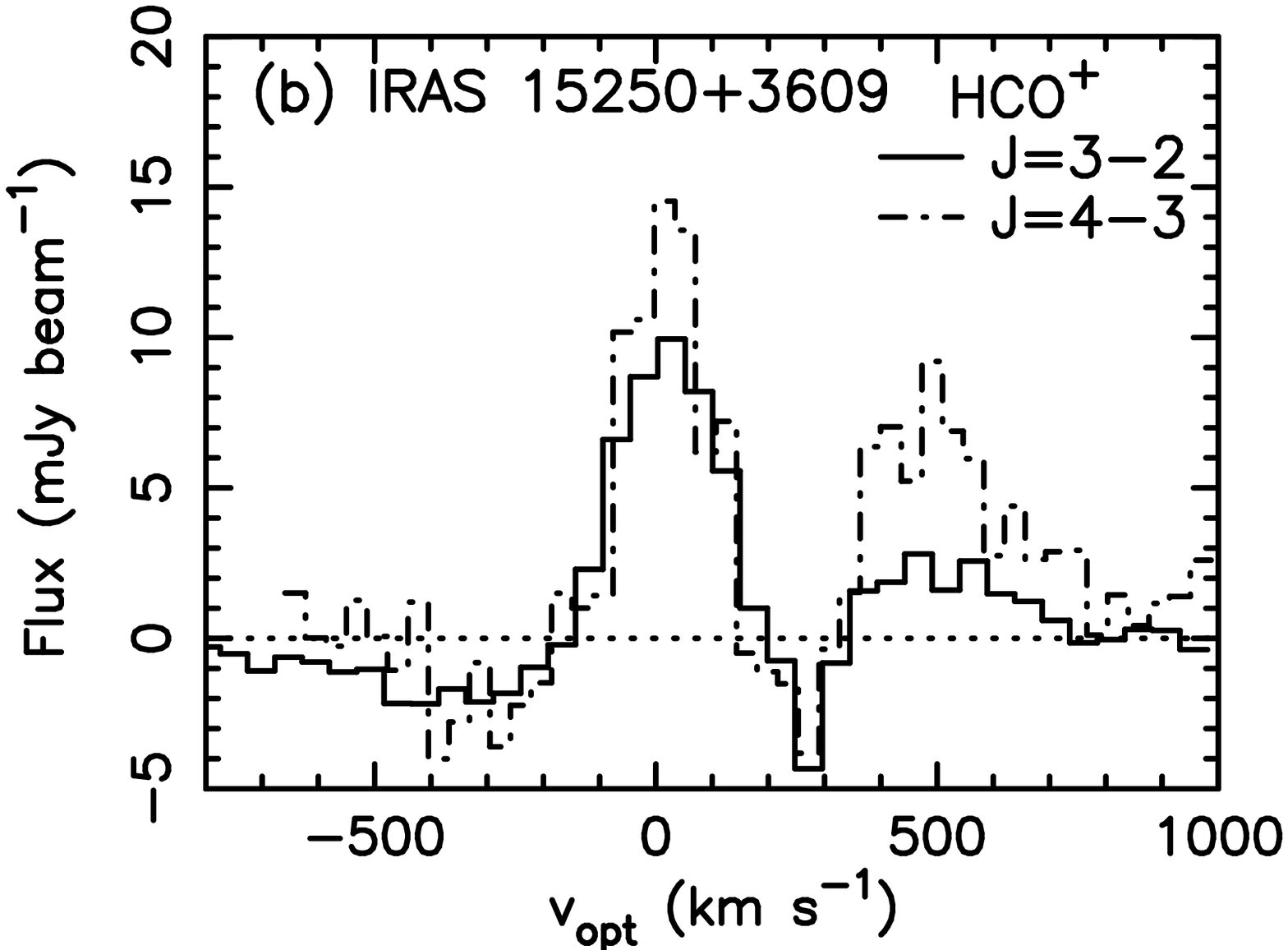}  
\includegraphics[angle=0,scale=.3]{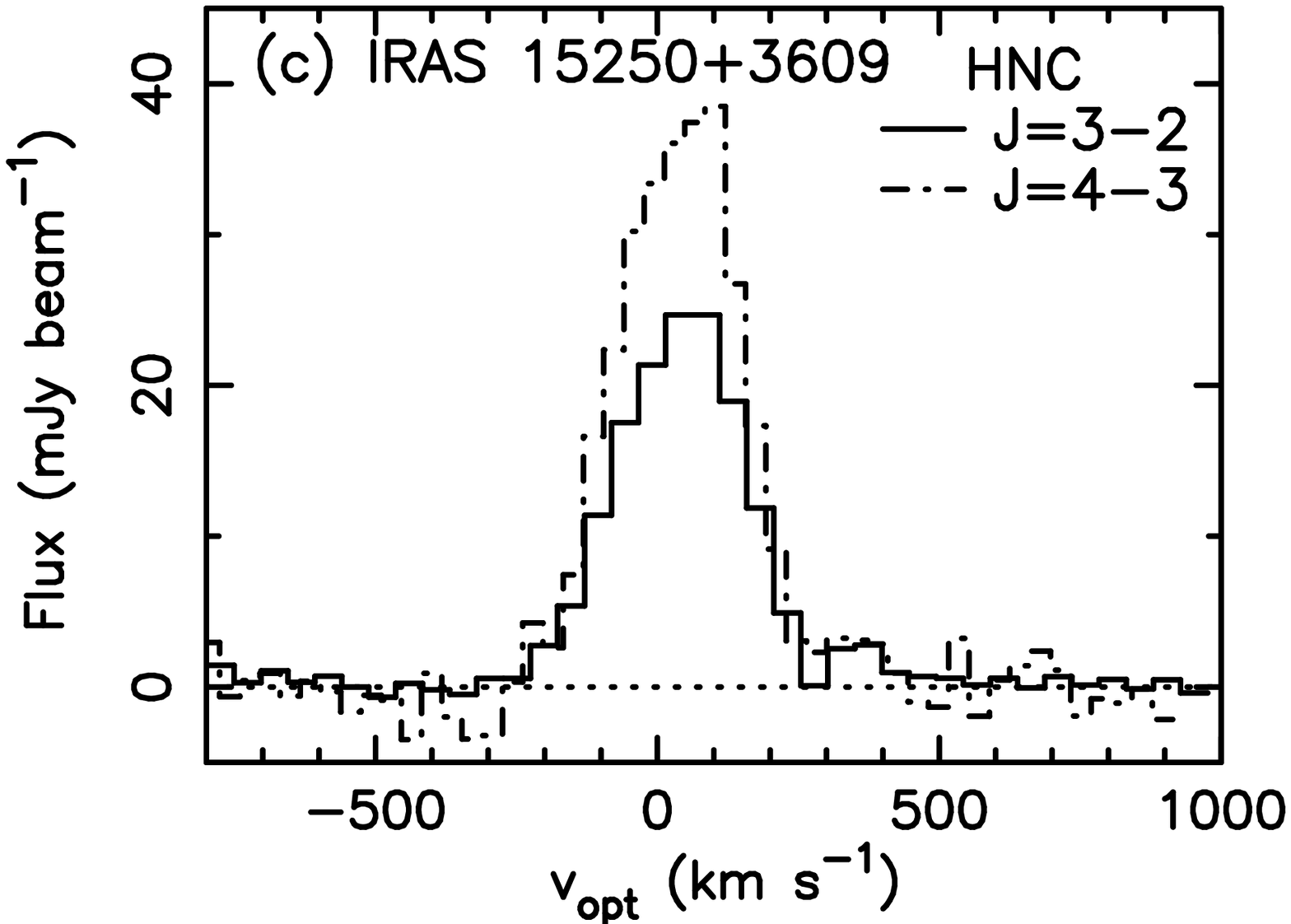} \\  
\caption{
Investigations of the emission and absorption profiles of HCN,
HCO$^{+}$, and HNC lines at J=3--2 and J=4--3 for IRAS 15250$+$3609.
HCN J=3--2 and HCO$^{+}$ J=3--2 data are taken from \citet{ima16c}.
The abscissa is velocity in (km s$^{-1}$), relative to the systemic
velocity (V$_{\rm sys}$ = 16560 km s$^{-1}$).
The ordinate is flux in (mJy beam$^{-1}$).
}
\end{center}
\end{figure}

\begin{figure}
\begin{center}
\includegraphics[angle=0,scale=.415]{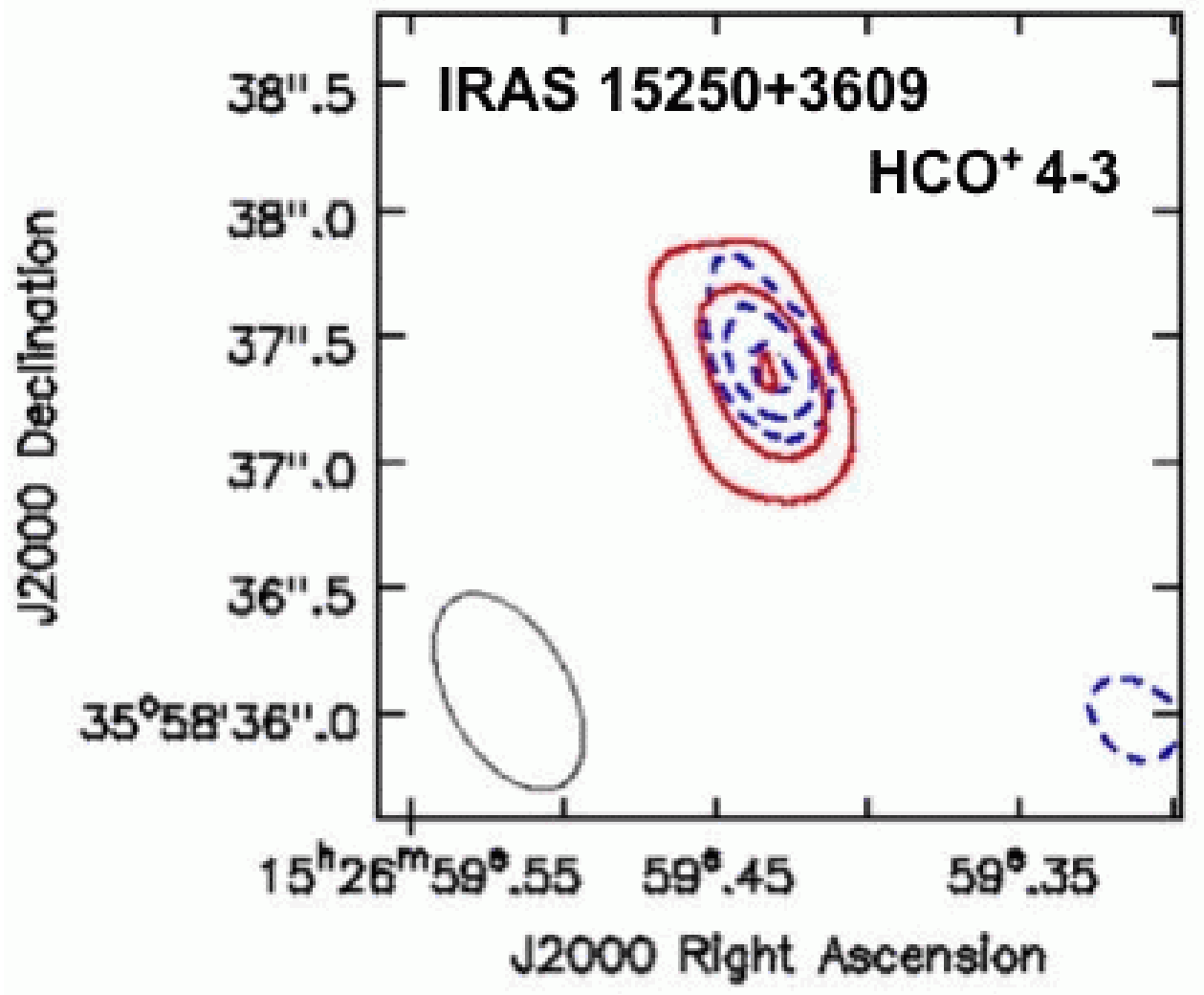}  
\includegraphics[angle=0,scale=.415]{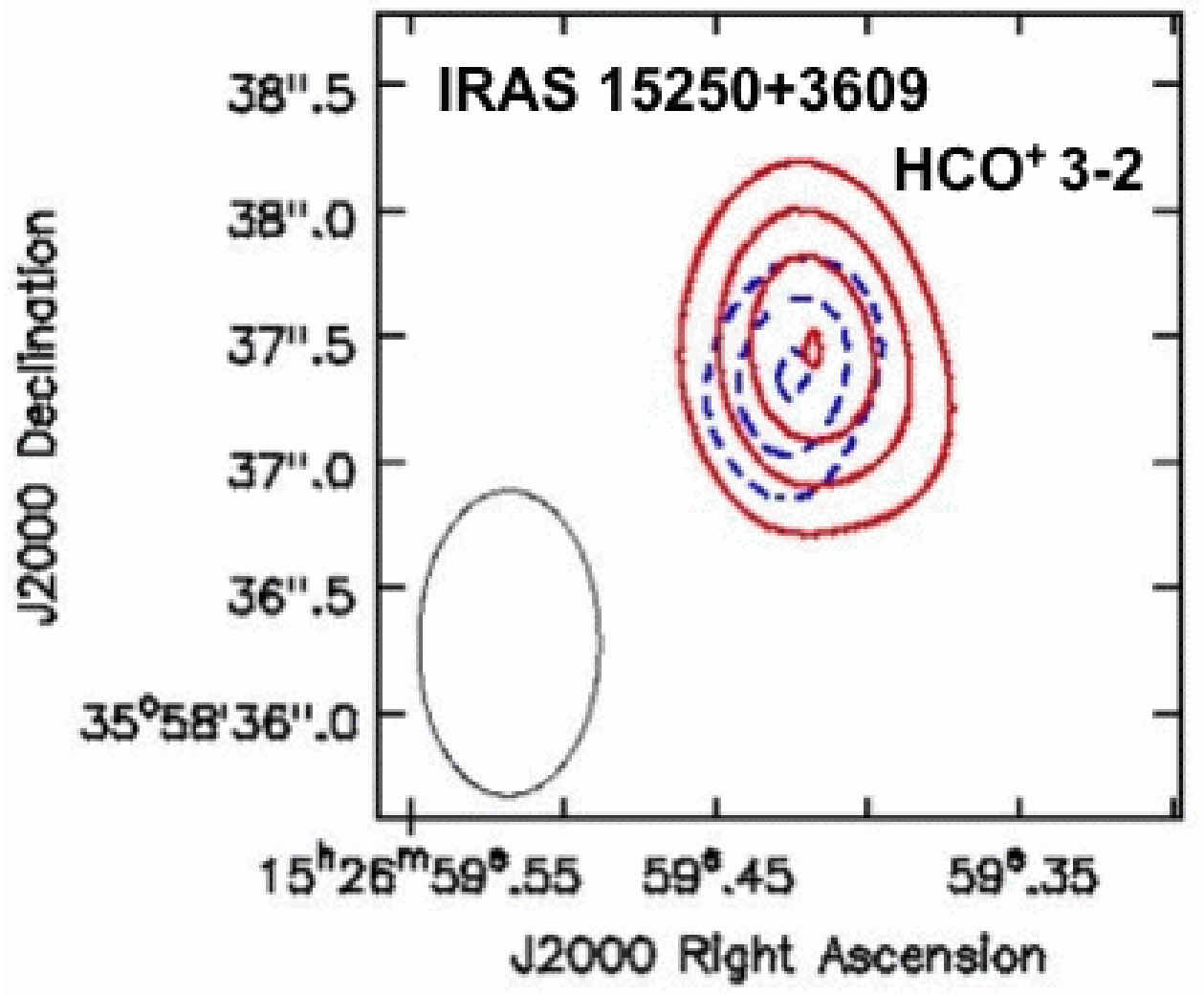} 
\includegraphics[angle=0,scale=.415]{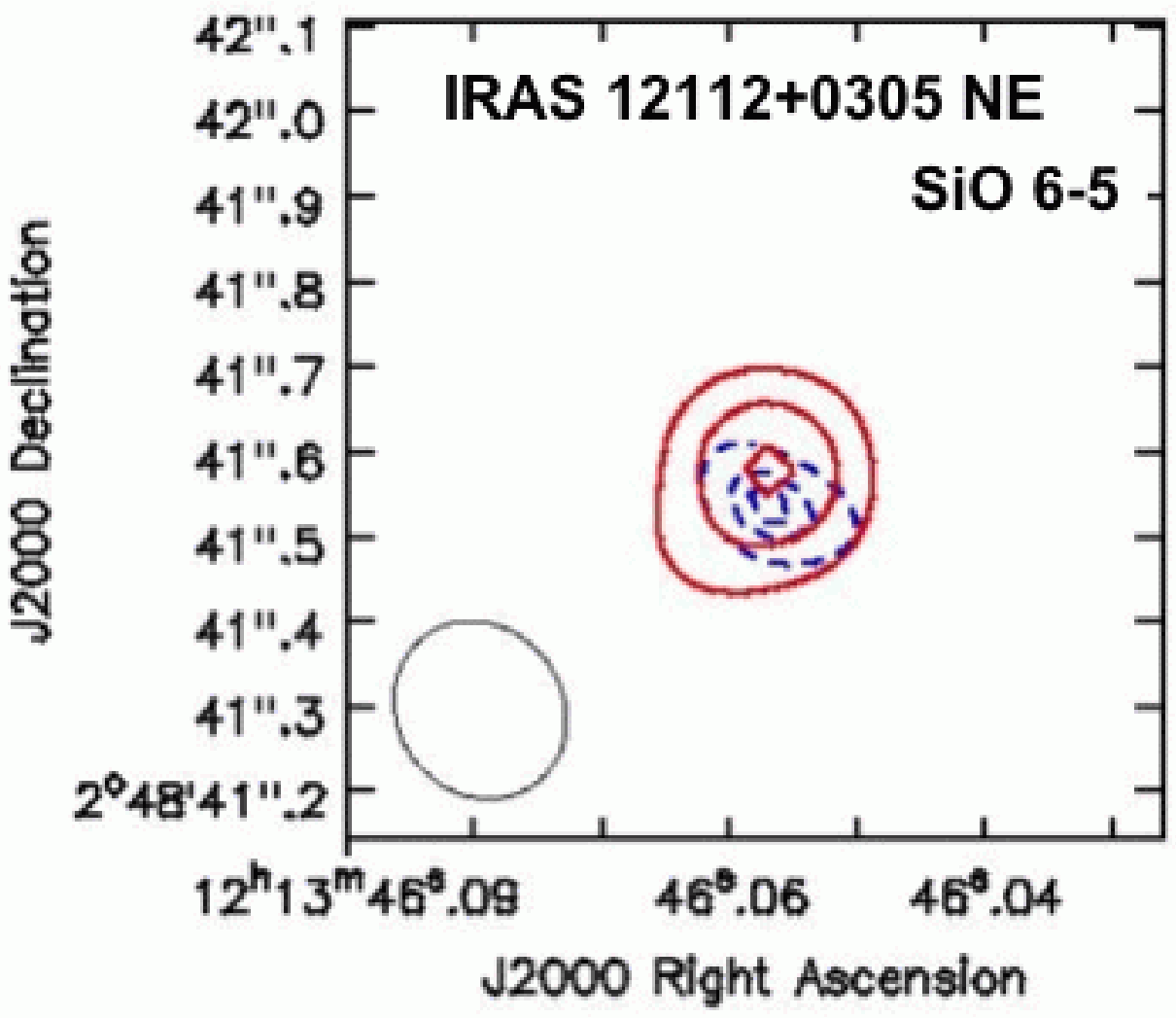}
\caption{
Spatial distribution of outflowing molecular gas indicated from the
blueshifted absorption and redshifted emission features. 
The abscissa and ordinate are R.A. (J2000) and decl. (J2000),
respectively. 
Redshifted emission and blueshifted absorption are shown as red solid
lines and blue dashed lines, respectively.
HCO$^{+}$ J=4--3 and J=3--2 of IRAS 15250$+$3609, and SiO J=6--5 of IRAS
12112$+$0305 NE are displayed, because both the redshifted emission 
and blueshifted absorption components are detected with $>$3$\sigma$. 
For HCO$^{+}$ J=4--3 of IRAS 15250$+$3609, contours are 4$\sigma$, 
7$\sigma$, and 10.5$\sigma$ for the redshifted emission component (1$\sigma$
= 0.18 Jy beam$^{-1}$ km s$^{-1}$), and $-$5$\sigma$, $-$4$\sigma$,
$-$3$\sigma$ for the blueshifted absorption component (1$\sigma$ = 0.091
Jy beam$^{-1}$ km s$^{-1}$).  
For HCO$^{+}$ J=3--2 of IRAS 15250$+$3609, contours are 3$\sigma$,
5$\sigma$, 7$\sigma$, and 9$\sigma$ for the redshifted emission component
(1$\sigma$ = 0.084 Jy beam$^{-1}$ km s$^{-1}$), and $-$4.8$\sigma$,
$-$4$\sigma$, and $-$3$\sigma$ for the blueshifted absorption component
(1$\sigma$ = 0.095 Jy beam$^{-1}$ km s$^{-1}$). 
For SiO J=6--5 of IRAS 12112$+$0305 NE, contours are 3$\sigma$,
5$\sigma$, and 7$\sigma$ for the redshifted emission component (1$\sigma$ =
0.025 Jy beam$^{-1}$ km s$^{-1}$), and $-$4.4$\sigma$, $-$4$\sigma$,
and $-$3$\sigma$ for the blueshifted absorption component (1$\sigma$ = 0.024
Jy beam$^{-1}$ km s$^{-1}$). 
Beam sizes are shown as open circles at the lower-left part.
}
\end{center}
\end{figure}

\begin{figure}
\begin{center}
\includegraphics[angle=0,scale=.65]{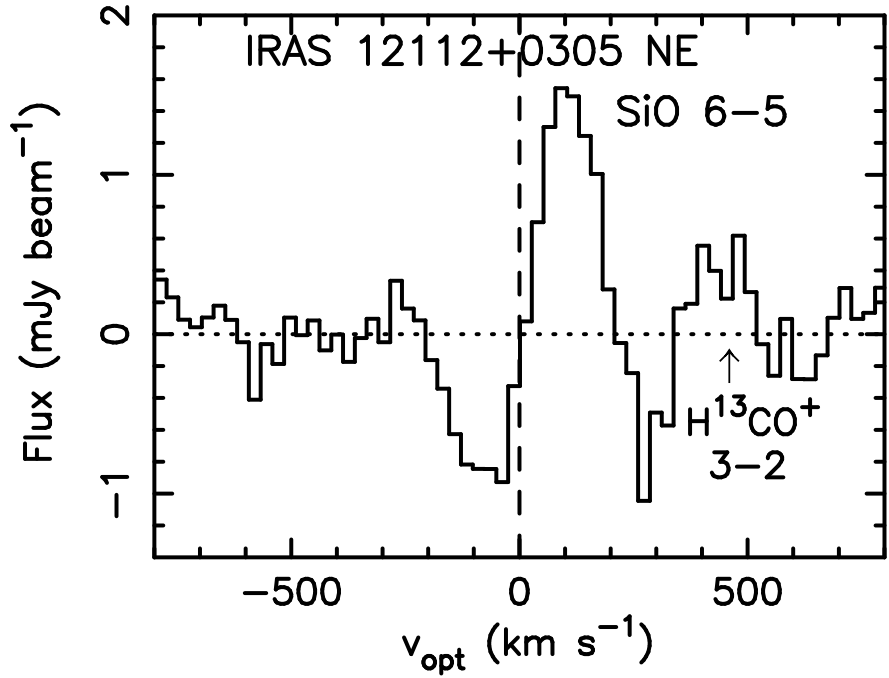}  
\caption{
Investigations of the P Cygni profile of SiO J=6--5 emission for IRAS
12112$+$0305 NE. 
The abscissa is velocity in (km s$^{-1}$), relative to the systemic
velocity (V$_{\rm sys}$ = 21900 km s$^{-1}$).
The ordinate is flux in (mJy beam$^{-1}$).
The horizontal dotted and vertical dashed lines are plotted at the zero
flux level and systemic velocity, respectively.
}
\end{center}
\end{figure}

\begin{figure}
\begin{center}
\includegraphics[angle=0,scale=.58]{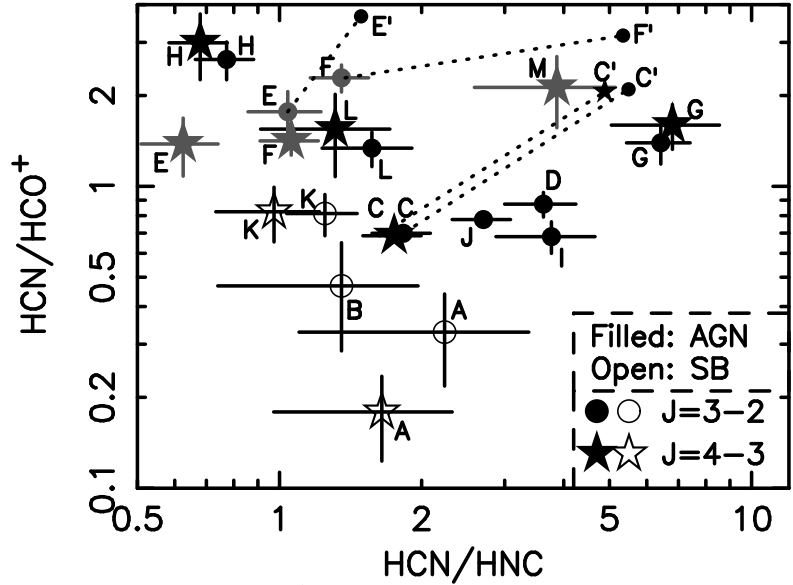}  \\
\vspace*{-0.3cm}
\caption{HCN-to-HNC (abscissa) and HCN-to-HCO$^{+}$ flux ratios 
(ordinate) at J=4--3 and J=3--2, shown in logarithmic scale.
Filled and open symbols mean AGN-important ULIRGs and
starburst-dominated galaxies, respectively. 
A: NGC 1614. 
B: IRAS 12112$+$0305 SW. 
C: IRAS 20551$-$4250.
D: IRAS 08572$+$3915. 
E: IRAS 12112$+$0305 NE. 
F: IRAS 22491$-$1808. 
G: Superantennae. 
H: IRAS 15250$+$3609. 
I: PKS 1345$+$12. 
J: IRAS 06035$-$7102. 
K: IRAS 13509$+$0442. 
L: IRAS 12127$-$1412.
M: IRAS 20414$-$1651. 
The (sub)millimeter-detected extremely deeply buried AGN candidates
(objects E, F, and M) are shown as gray to distinguish them from 
other ULIRGs with stronger luminous buried AGN
signatures in the infrared/X-ray spectra.
For objects C, E and F, H$^{13}$CN, H$^{13}$CO$^{+}$, and
HN$^{13}$C J=3--2 emission lines were detected in our Cycle 4 deep data. 
Line-opacity-corrected intrinsic flux ratios at J=3--2 are plotted
(denoted as C', E', and F'). 
For object C, a line-opacity-corrected intrinsic flux ratio at J=4--3
is also plotted (denoted as C'), by adopting a line opacity correction
factor of $\sim$3 \citep{ima16b,ima17}.
Objects A, B, and K do not display any obvious AGN signature in the
optical, infrared, and (sub)millimeter spectra.
}
\end{center}
\end{figure}

\begin{figure}
\begin{center}
\includegraphics[angle=0,scale=.58]{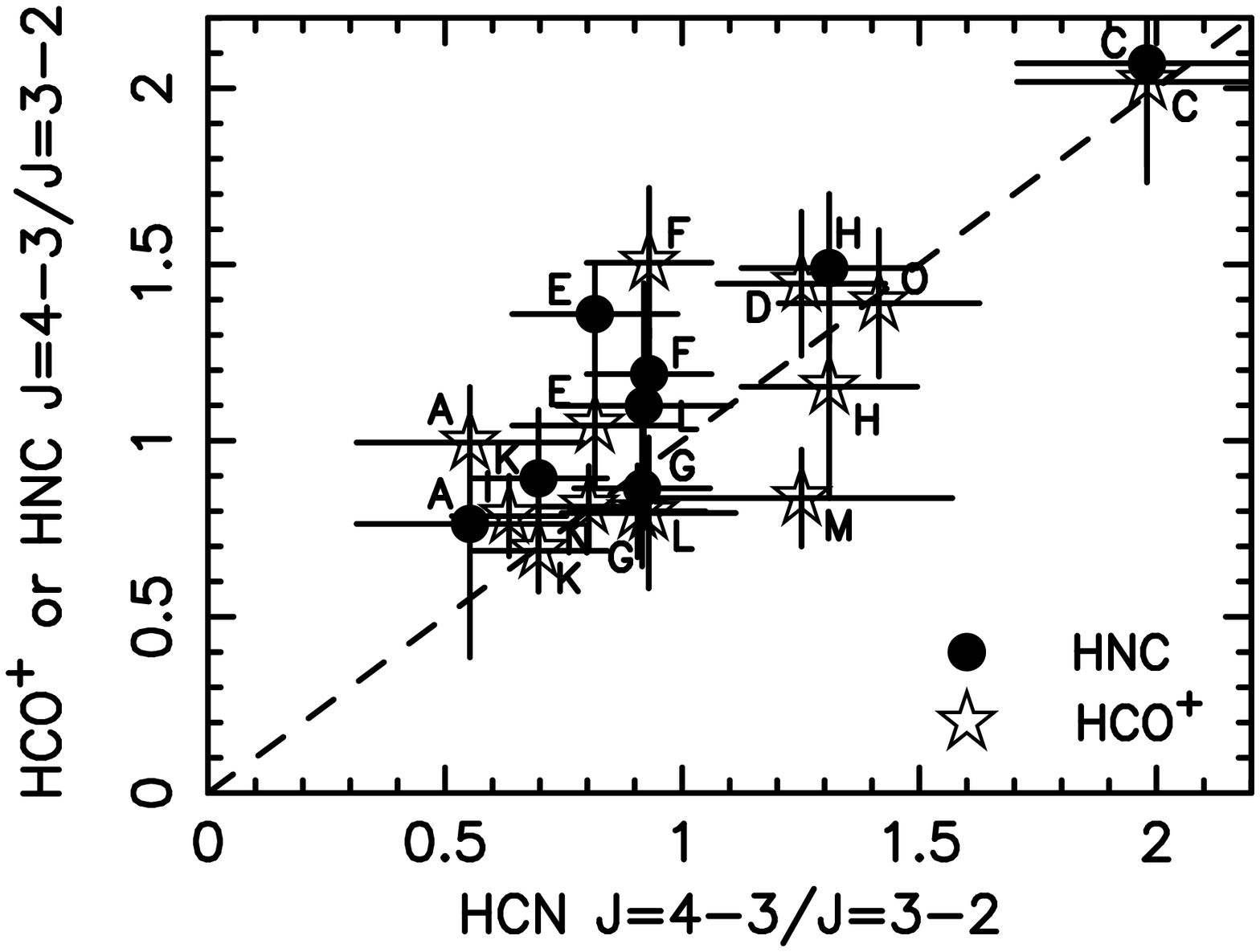}  
\vspace*{-0.3cm}
\caption{Ratio of J=4--3 to J=3--2 flux in (Jy km s$^{-1}$) for 
HCN (abscissa) and HCO$^{+}$ or HNC (ordinate).
The object symbols (A--M) are the same as in Figure 25.
We have added N: NGC 7469 \citep{izu15,ima16c} and O: NGC 4418
\citep{sak10}.   
The dashed line indicates the same J=4--3 to J=3--2 flux ratio between
the abscissa (HCN) and ordinate (HCO$^{+}$ and HNC).
Objects A and K (located at the lower-left part) are galaxies with no
clear AGN signatures in the optical, infrared, and (sub)millimeter
spectra. 
IRAS 12112$+$0305 SW has only upper limits for the J=4--3 to J=3--2 flux
ratios of HCN and HNC (Table 11).
This source is not plotted, because constraints are too loose to be
meaningful.
The open stars for objects G and L are overlapped. 
}
\end{center}
\end{figure}

\begin{figure}
\begin{center}
\includegraphics[angle=0,scale=.41]{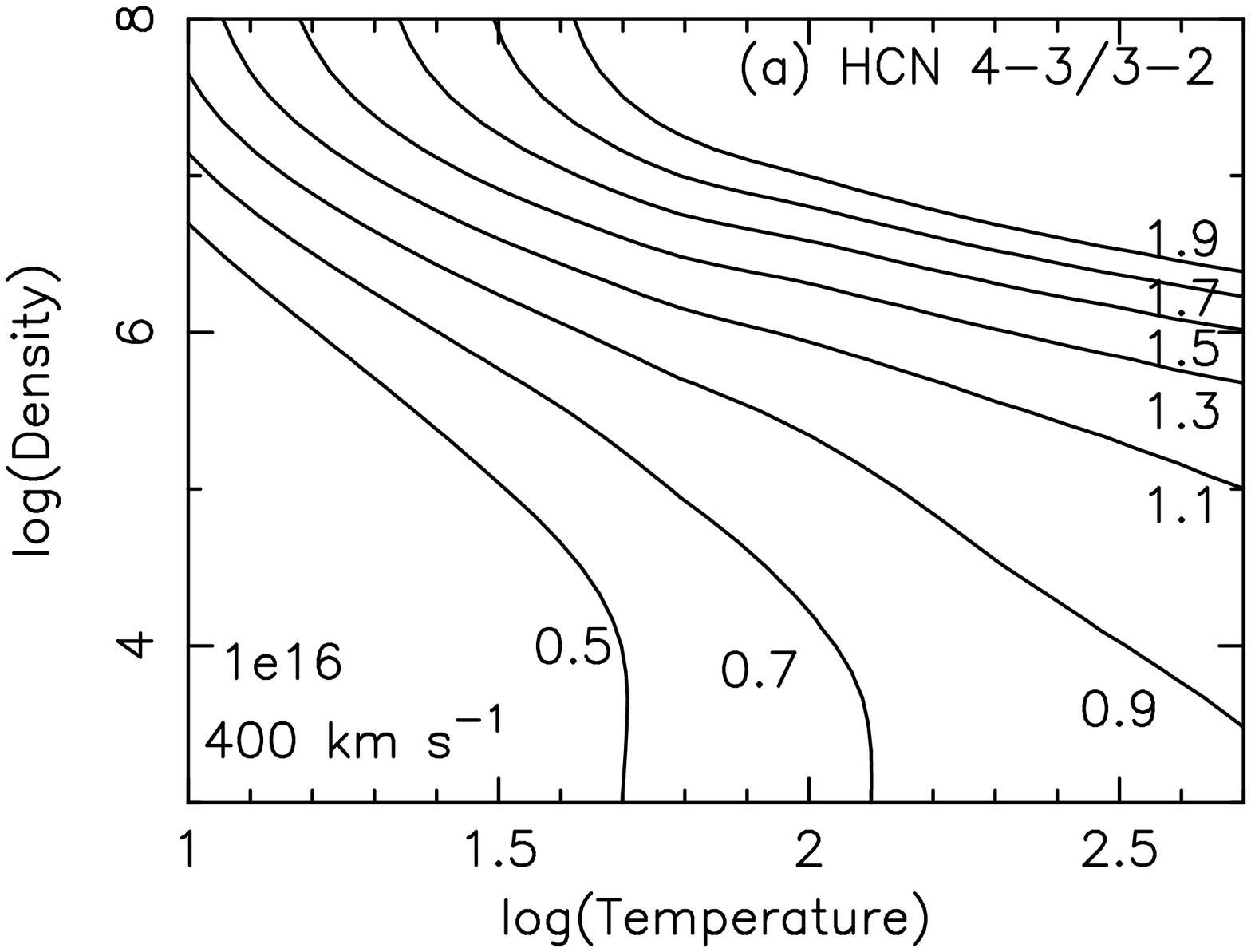}  
\includegraphics[angle=0,scale=.41]{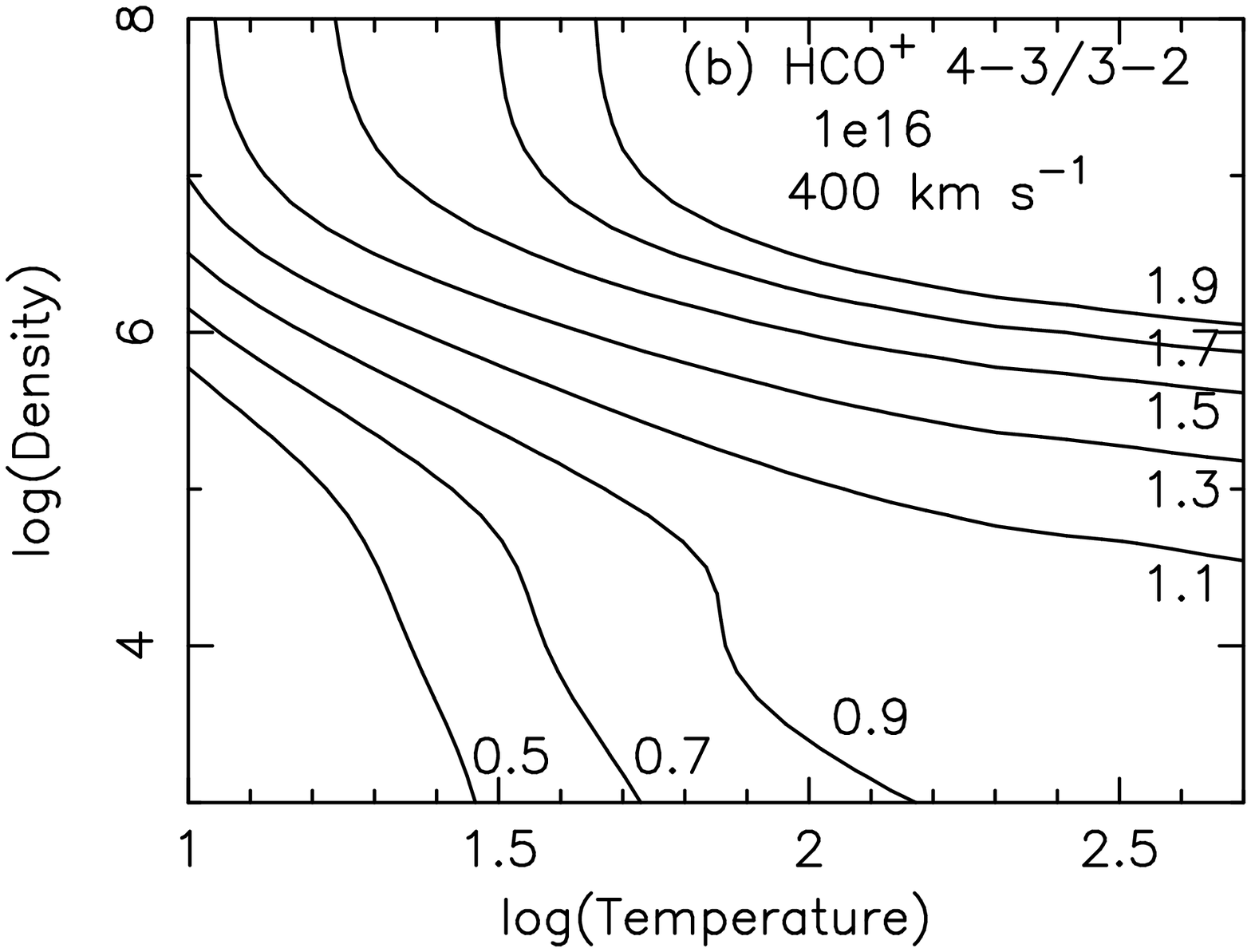} \\ 
\includegraphics[angle=0,scale=.41]{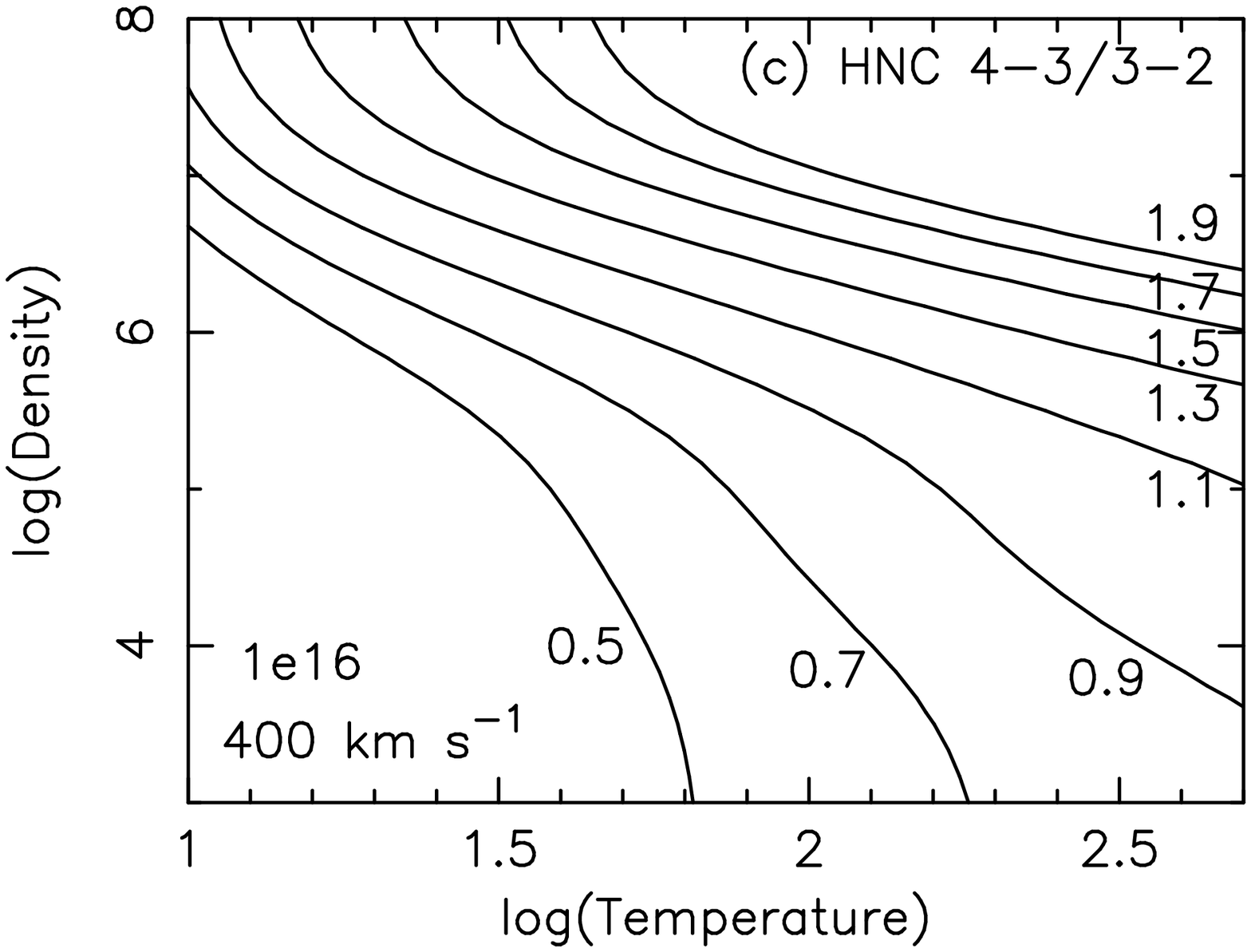} 
\includegraphics[angle=0,scale=.41]{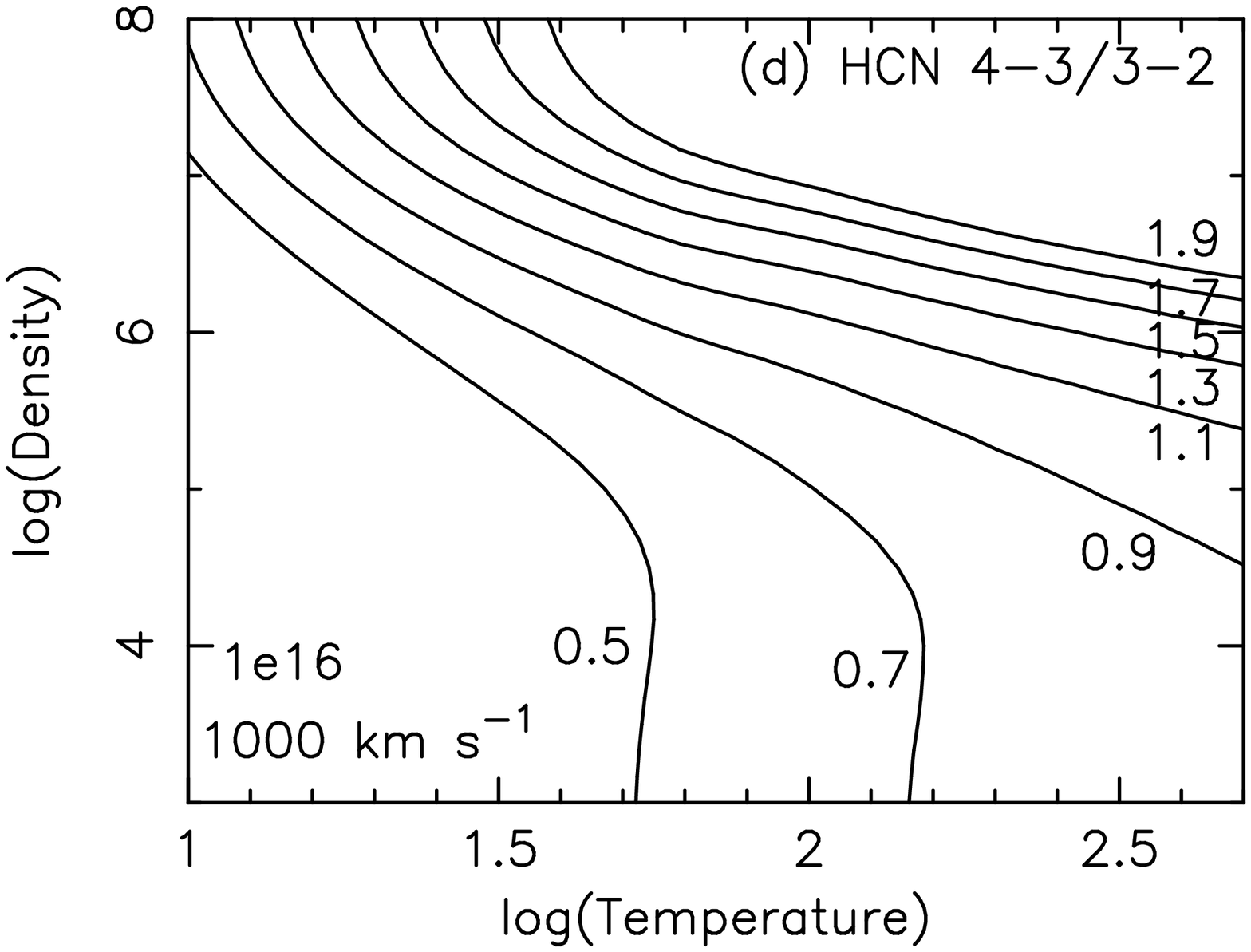} \\  
\includegraphics[angle=0,scale=.41]{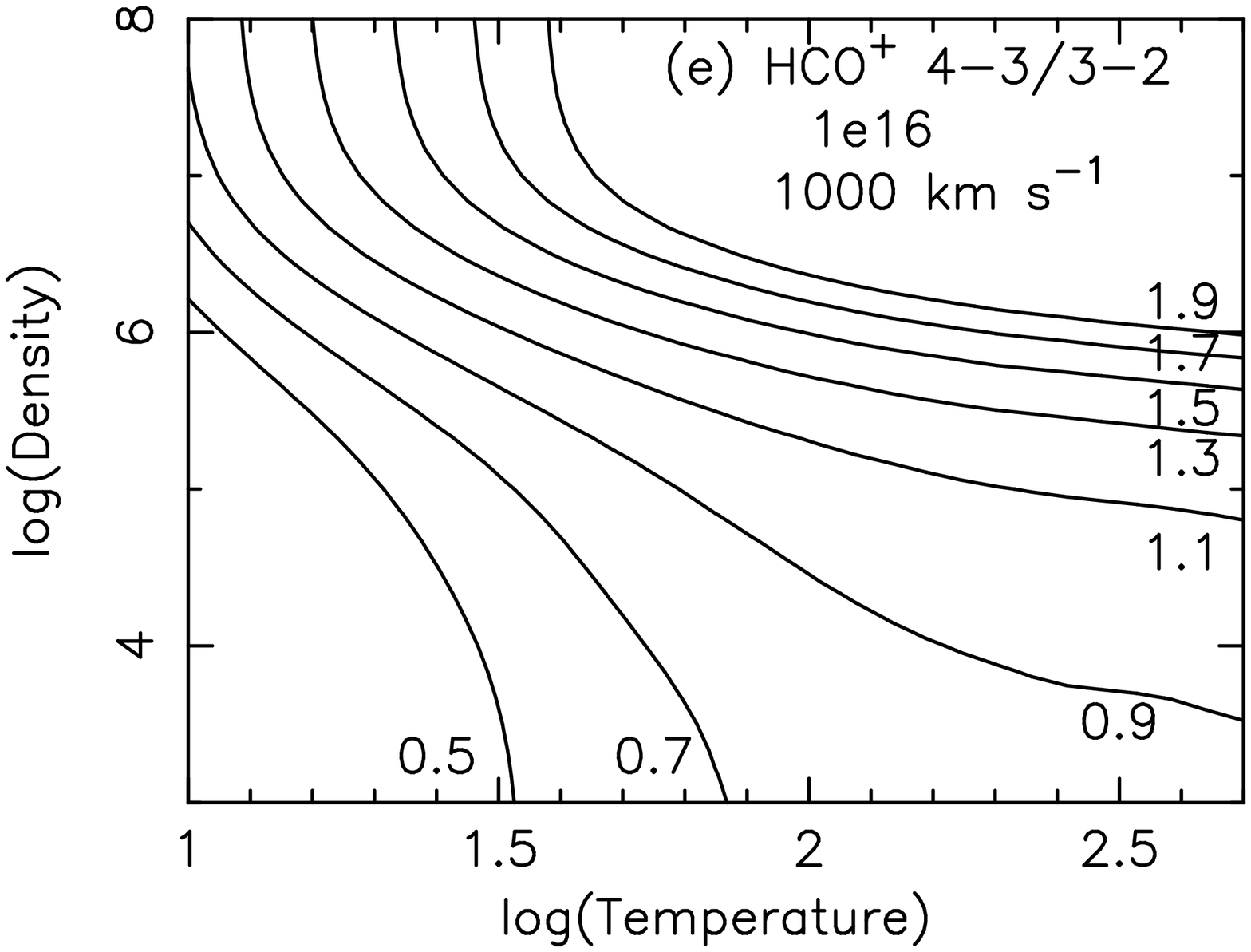} 
\includegraphics[angle=0,scale=.41]{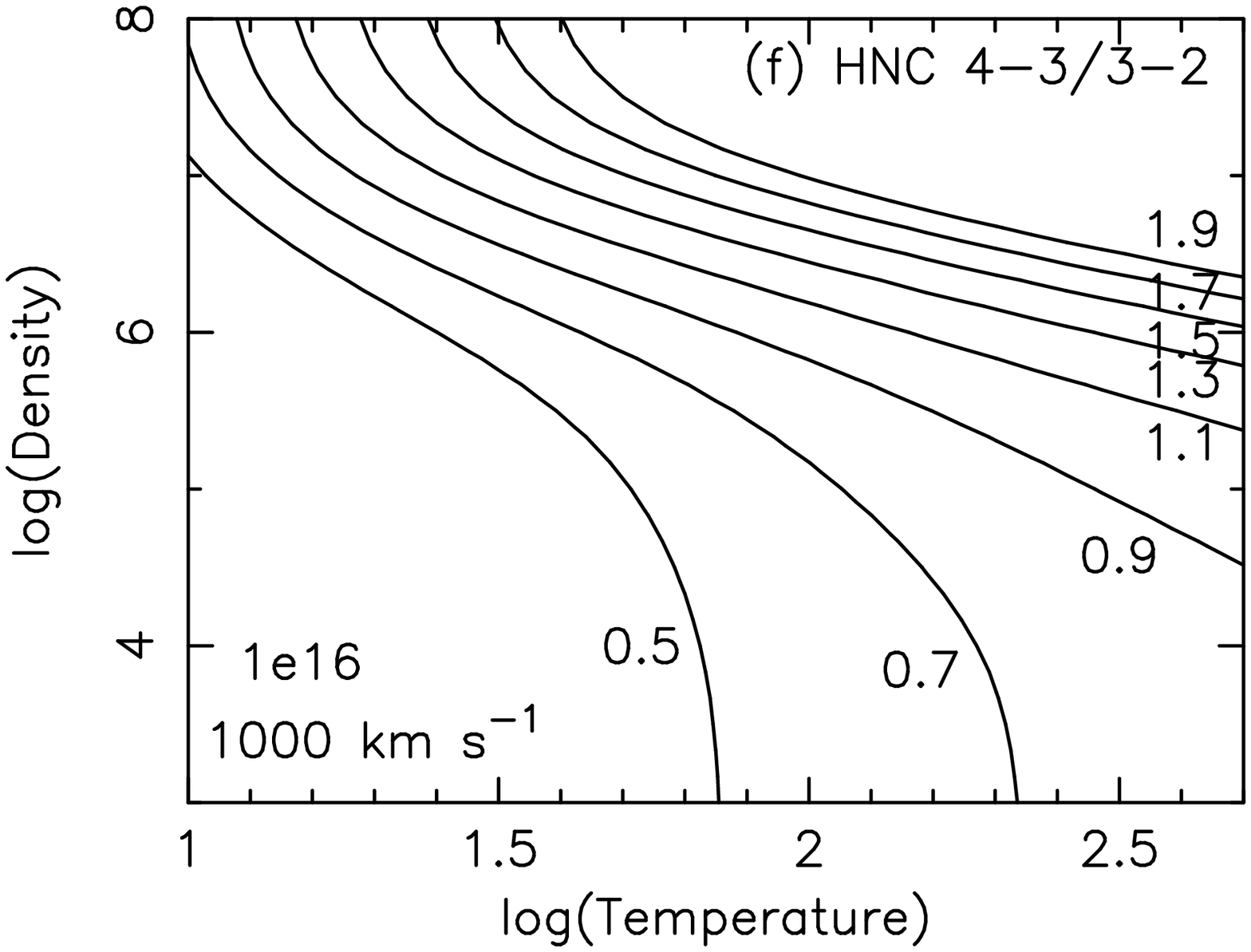} \\ 
\caption{RADEX calculations of the ratio of J=4--3 to J=3--2 flux in 
(Jy km s$^{-1}$), as a function of H$_{2}$ kinetic temperature and
volume number density, for HCN, HCO$^{+}$, and HNC.
The molecular column density with 1 $\times$ 10$^{16}$ cm$^{-2}$ is
assumed for HCN, HCO$^{+}$, and HNC.
Calculations are made for a molecular line width of 400 km s$^{-1}$
(a--c) and 1000 km s$^{-1}$ (d--f) to represent all galaxies (except
Superantennae) and Superantennae, respectively.
}
\end{center}
\end{figure}

\clearpage

\begin{figure}
\begin{center}
\includegraphics[angle=0,scale=.41]{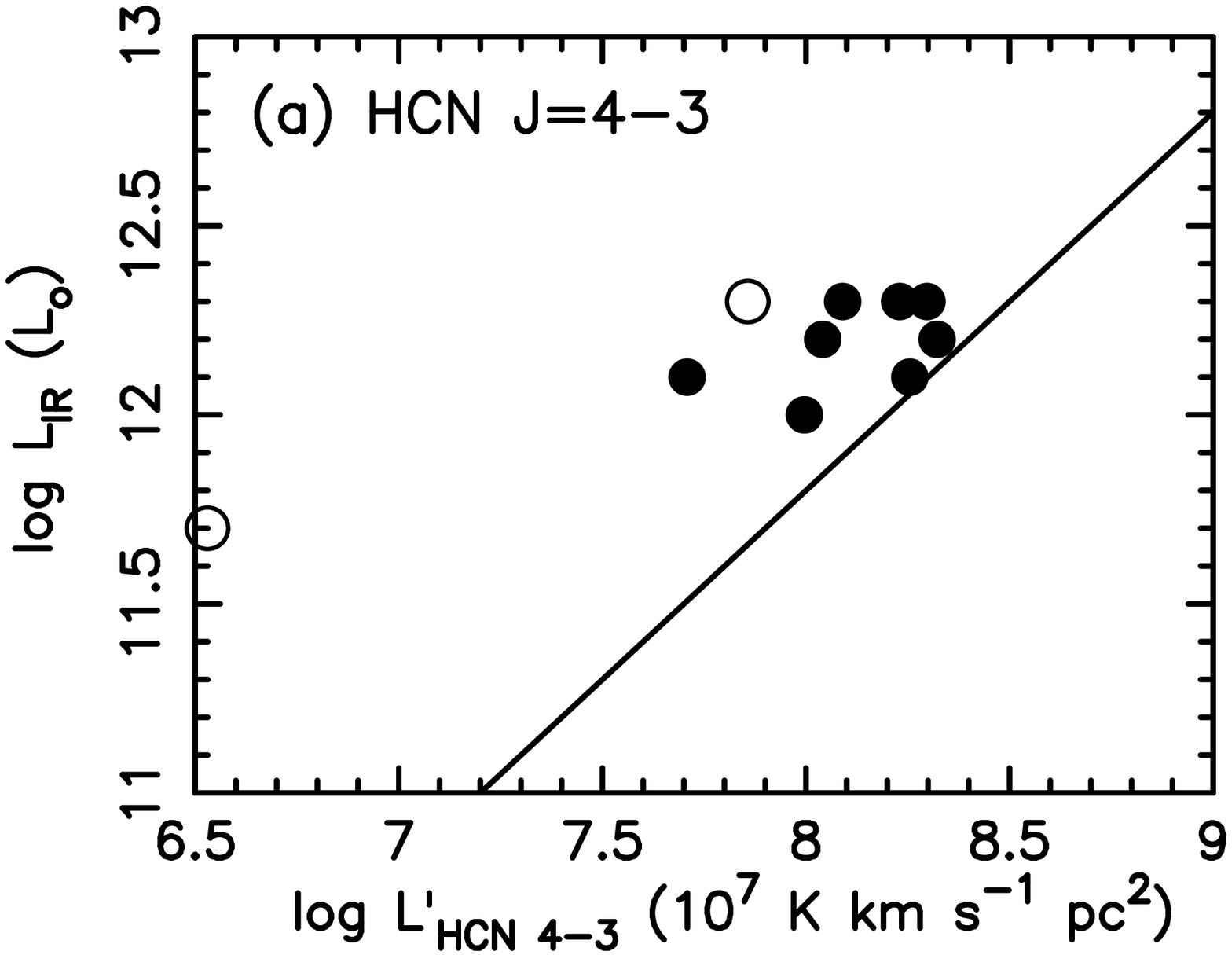}  
\includegraphics[angle=0,scale=.41]{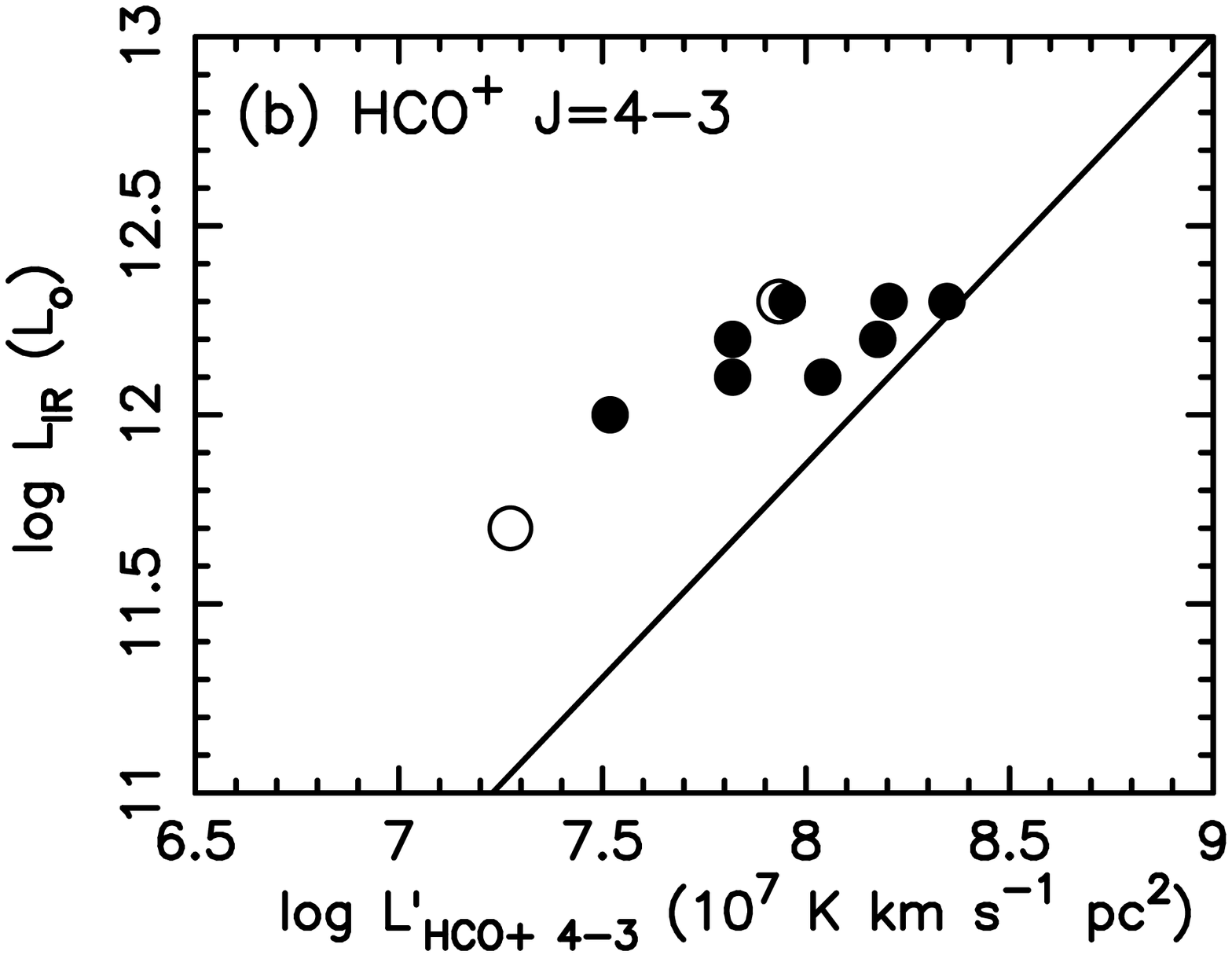} \\
\vspace*{-0.5cm}  
\includegraphics[angle=0,scale=.41]{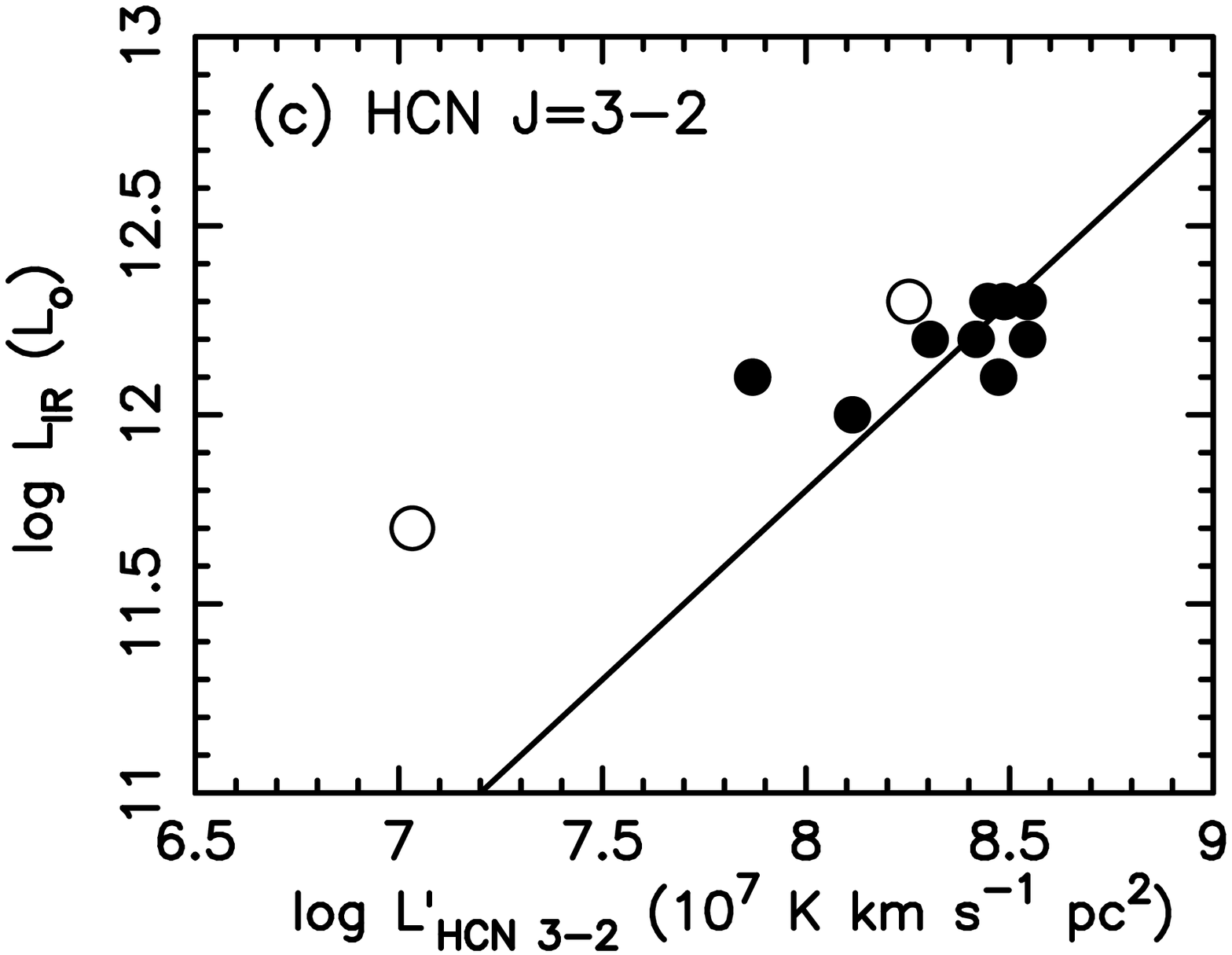}  
\includegraphics[angle=0,scale=.41]{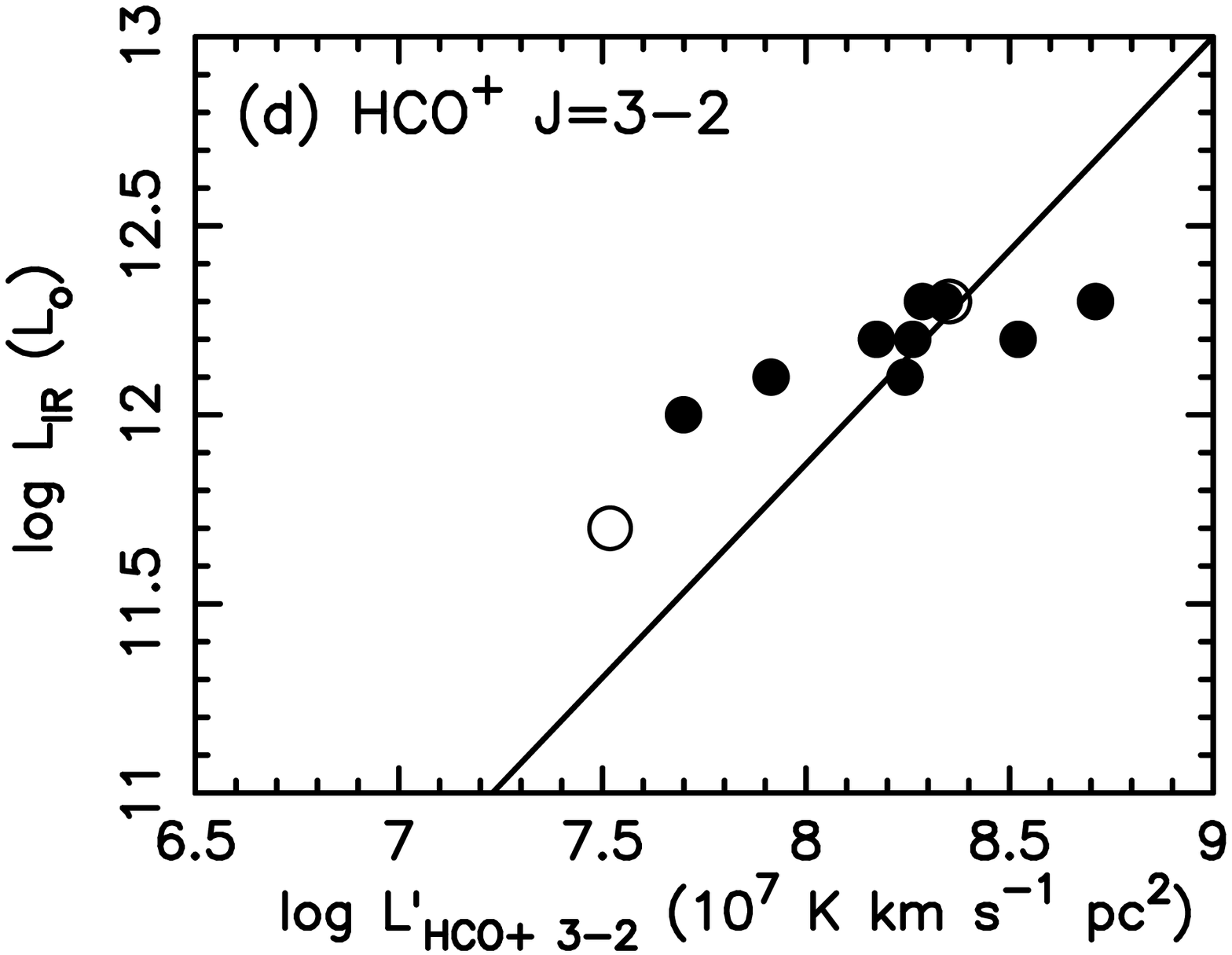}  
\caption{
Comparison of HCN and HCO$^{+}$ emission line luminosities at J=4--3 and
J=3--2 as measured with our ALMA interferometric observations
(abscissa), with infrared luminosity (ordinate). 
(a): HCN J=4--3. (b): HCO$^{+}$ J=4--3. (c): HCN J=3--2. (d): HCO$^{+}$
J=3--2. 
The solid lines in (a) and (b) are the best-fit lines for HCN
J=4--3 (log$L$$_{\rm IR}$ = 1.00log$L$$'$$_{\rm HCN(4-3)}$ + 3.80) and
HCO$^{+}$ J=4--3 (log$L$$_{\rm IR}$ = 1.13log$L$$'$$_{\rm HCO+(4-3)}$ + 2.83) 
for various types of galaxies \citep{tan17}. 
For optically thick and thermally excited molecular gas, the molecular
line luminosity in units of (K km s$^{-1}$ pc$^{2}$) should be the same
for J=4--3 and J=3--2.  
For HCN J=3--2 and HCO$^{+}$ J=3--2, the same best-fit lines as HCN
J=4--3 and HCO$^{+}$ J=4--3, respectively, are plotted as the solid
lines. 
Molecular line luminosities are estimated from the ALMA spectra within
the beam size in all sources, except NGC 1614 for which 
spatially integrated spectra are used, due to the obvious spatial
extent of molecular line emission in this source.
The starburst-dominated galaxies NGC 1614 and IRAS 13509$+$0442 are
shown as open circles, while the remaining ULIRGs with AGN signatures
are displayed as filled circles.
For IRAS 12112$+$0305, the HCO$^{+}$ J=4--3, HCN J=3--2, and HCO$^{+}$
J=3--2 emission lines are detected at both the NE and SW nuclei. 
Luminosities at both nuclei are added.
For the HCN J=4--3 emission, it is undetected in IRAS 12112$+$0305 SW
and the luminosity in IRAS 12112$+$0305 NE is plotted. 
At J=3--2, eleven sources are plotted. At J=4--3, ten sources are
plotted, because there are no J=4--3 data for IRAS 06035$-$7102. 
}
\end{center}
\end{figure}



\section{Appendix A}

Figure 29 displays the continuum emission of ULIRGs, which are dominated
by compact nuclear components. 
The observed properties of serendipitously detected molecular emission
lines are summarized in Table 16. 
Figures 30 and 31 show integrated intensity (moment 0) maps and Gaussian
fits in the spectra within the beam sizes of these molecular lines,
respectively.  
Intensity-weighted mean velocity (moment 1) and intensity-weighted
velocity dispersion (moment 2) maps of the CS J=5--4 emission line are 
displayed in Figures 32 and 33, respectively, for IRAS 12112$+$0305 NE
and IRAS 22491$-$1808.  

\begin{turnpage}
\begin{deluxetable}{lc|lcl|cccc}
\tabletypesize{\scriptsize}
\tablecaption{Flux of Other Faint Emission Lines \label{tbl-16}} 
\tablewidth{0pt}
\tablehead{
\colhead{Object} & \colhead{Line} & 
\multicolumn{3}{c}{Integrated intensity (moment 0) map} & 
\multicolumn{4}{c}{Gaussian line fit} \\  
\colhead{} & \colhead{} & \colhead{Peak} &
\colhead{rms} & \colhead{Beam} & \colhead{Velocity}
& \colhead{Peak} & \colhead{FWHM} & \colhead{Flux} \\ 
\colhead{} & \colhead{} & 
\multicolumn{2}{c}{Jy beam$^{-1}$ km s$^{-1}$} & 
\colhead{[$''$ $\times$ $''$] ($^{\circ}$)} &
\colhead{[km s$^{-1}$]} & \colhead{[mJy]} & \colhead{[km s$^{-1}$]} &
\colhead{[Jy km s$^{-1}$]} \\  
\colhead{(1)} & \colhead{(2)} & \colhead{(3)} & \colhead{(4)} & 
\colhead{(5)} & \colhead{(6)} & \colhead{(7)} & \colhead{(8)} &
\colhead{(9)} 
}
\startdata 
08572$+$3915 & CS J=5--4 (244.936) & 0.39 (9.6$\sigma$) & 0.041 &
0.40$\times$0.19 (16$^{\circ}$) & 17485$\pm$10 & 1.3$\pm$0.1
& 350$\pm$22 & 0.44$\pm$0.04 \\
 & SiO J=6--5 (260.518) & $<$0.11 ($<$3$\sigma$) & 0.036 &
0.37$\times$0.18 (16$^{\circ}$) & 17223$\pm$228 &
0.32$\pm$0.16 & 235 (fix) & 0.077$\pm$0.037 \\ 
 & SO 6(6)--5(5) (258.256) & $<$0.12 ($<$3$\sigma$) & 0.040 &
0.37$\times$0.18 (17$^{\circ}$) & 17474$\pm$18 & 0.44$\pm$0.06 &
269$\pm$45 & 0.12$\pm$0.03 \\ 
 & H$^{13}$CN J=3--2 (259.012) \tablenotemark{A} & $<$0.17 ($<$3$\sigma$) &
0.056 & 0.92$\times$0.51 (8$^{\circ}$) & --- & --- & --- & --- \\     
Superantennae & CH$_{3}$OH (364.859) & 0.67 (3.6$\sigma$) & 0.19 &
0.60$\times$0.49 (16$^{\circ}$) & 18418$\pm$60 & 1.5$\pm$0.4 &
418$\pm$165 & 0.62$\pm$0.30 \\       
 & H$^{13}$CN J=3--2 (259.012) \tablenotemark{A} & $<$0.21 ($<$3$\sigma$) &
0.068& 1.0$\times$0.62 (59$^{\circ}$) & --- & --- & --- & --- \\     
12112$+$0305 NE & H$^{13}$CN J=3--2 (259.012) \tablenotemark{A} &
0.28 (3.2$\sigma$) & 0.086 & 0.84$\times$0.57 (65$^{\circ}$) & 21759$\pm$48 &
1.0$\pm$0.3 & 322$\pm$98 & 0.32$\pm$0.14 \\      
 & HCN v$_{2}$=1f J=4--3 (356.256) & 1.1 (3.0$\sigma$) &
0.36 &  1.4$\times$0.55 (71$^{\circ}$) & --- & --- & --- & --- \\     
 & CH$_{3}$OH (364.859) & 4.3 (13$\sigma$) & 
0.33 & 1.0$\times$0.66 (82$^{\circ}$) & 21846$\pm$8 & 12$\pm$1
& 394$\pm$23 & 4.6$\pm$0.3 \\     
 & CS J=7--6 (342.883) & 1.4 (5.5$\sigma$) & 0.25
 & 1.4$\times$0.56 (71$^{\circ}$) & 21688, 21976  (fix) &
1.4$\pm$0.8, 5.1$\pm$1.1 (fix) & 315, 165 (fix) & 1.3$\pm$0.3 \\     
 & CS J=5--4 (244.936) & 1.4 (29$\sigma$) & 0.048 & 0.23$\times$0.20
(37$^{\circ}$) & 21660$\pm$4, 21972$\pm$2 & 3.3$\pm$0.1, 3.6$\pm$0.1 &
292$\pm$10, 136$\pm$5 & 1.4$\pm$0.1 \\   
 & HC$_{3}$N J=27--26 (245.606) & 0.53 (14$\sigma$) & 0.038 &
0.23$\times$0.20 (37$^{\circ}$) & 21765$\pm$16, 21982$\pm$10 &
1.7$\pm$0.1, 0.88$\pm$0.19 & 294$\pm$47, 124$\pm$29 & 0.59$\pm$0.17 \\
& SiO J=6--5 (260.518) & 0.18 (7.3$\sigma$) & 0.025 &
0.22$\times$0.19 (38$^{\circ}$) & --- & --- & --- & --- \\
 & SO 6(6)--5(5) (258.256) & 0.20 (4.2$\sigma$) & 0.048 &
0.22$\times$0.20 (39$^{\circ}$) & --- & --- & --- & --- \\
12112$+$0305 SW & CS J=5--4 (244.936) & 0.13 (5.0$\sigma$) & 0.026 &
0.23$\times$0.20 (37$^{\circ}$) & 21984$\pm$20 & 0.52$\pm$0.06 &
265$\pm$46 & 0.14$\pm$0.03 \\
 & H$^{13}$CN J=3--2 (259.012) \tablenotemark{A} &
$<$0.22 ($<$3$\sigma$) & 0.073 & 0.84$\times$0.57 (65$^{\circ}$) &
--- & --- & --- & --- \\ 
22491$-$1808 & H$^{13}$CN J=3--2 (259.012) \tablenotemark{A} &
0.42 (8.1$\sigma$) & 0.052 & 0.41$\times$0.37 (79$^{\circ}$) &
23274$\pm$24 & 2.2$\pm$0.2 & 336$\pm$46 & 0.72$\pm$0.13 \\      
& HC$_{3}$N J=30--29 (272.885) & 0.72 (8.8$\sigma$) & 0.081 & 0.40$\times$0.35
(79$^{\circ}$) & 23309$\pm$21 & 2.8$\pm$0.3 & 342$\pm$53 & 0.93$\pm$0.18 \\      
& CS J=5--4 (244.936) & 2.4 (46$\sigma$) & 0.052 & 0.19$\times$0.17
(59$^{\circ}$) & 23278$\pm$4 & 6.3$\pm$0.1 & 445$\pm$8 & 2.7$\pm$0.1 \\
& HC$_{3}$N J=27--26 (245.606) & 1.2 (23$\sigma$) & 0.050 & 0.19$\times$0.17
(59$^{\circ}$) & 23298$\pm$5 & 3.2$\pm$0.1 & 404$\pm$14 & 1.3$\pm$0.1 \\
& SiO J=6--5 (260.518) & 0.69 (19$\sigma$) & 0.037 & 0.16$\times$0.15
($-$80$^{\circ}$) & 23234$\pm$7 & 2.9$\pm$0.1 & 260$\pm$14 &
0.74$\pm$0.05 \\
& SO 6(6)--5(5) (258.256) & 0.23 (6.7$\sigma$) & 0.035 &
0.17$\times$0.15 (74$^{\circ}$) & 
--- & --- & --- & --- \\
NGC 1614 & H$^{13}$CN J=3--2 (259.012) \tablenotemark{A} &
$<$0.25 ($<$3$\sigma$) & 0.080 & 1.1$\times$0.40 ($-$67$^{\circ}$) & 
--- & --- & --- & --- \\ 
12127$-$1412 & H$^{13}$CN J=3--2 (259.012) \tablenotemark{B} & $<$0.14
($<$3$\sigma$) & 0.045 & 0.91$\times$0.76 (70$^{\circ}$) & & & & \\
15250$+$3609 & H$^{13}$CN J=3--2 (259.012) \tablenotemark{B} & 0.43 (4.7$\sigma$) &
0.091 & 1.3$\times$0.76 ($-$10$^{\circ}$) & 16615$\pm$24 &
2.3$\pm$0.6 & 182$\pm$49 & 0.43$\pm$0.16\\        
 & HNC v$_{2}$=1f J=4--3 (365.147) & 0.68 (5.3$\sigma$)
\tablenotemark{C} & 0.13 \tablenotemark{C} & 1.1$\times$0.55
($-$24$^{\circ}$) & 16508$\pm$35 & 4.0$\pm$0.9 & 
159$\pm$70 & 0.64$\pm$0.32 \\
 & CH$_{3}$OH (364.859) & 3.2 (12$\sigma$) & 0.26 & 1.1$\times$0.55
($-$24$^{\circ}$) & 16602$\pm$12 & 11$\pm$1 & 344$\pm$34 & 3.7$\pm$0.4 \\ 
 & HC$_{3}$N J=30--29 (272.885) & 0.81 (9.4$\sigma$) \tablenotemark{D} &
0.086 & 1.2$\times$0.72 ($-$11$^{\circ}$) & 16580$\pm$118 & 3.3$\pm$0.4 &
432$\pm$156 & 1.4$\pm$0.5 \\ 
PKS 1345$+$12 & CS J=7--6 (342.883) & 0.21 (3.4$\sigma$) & 0.062  &
0.66$\times$0.59 (53$^{\circ}$) & --- & --- & --- & --- \\        
& H$^{13}$CN J=3--2 (259.012) \tablenotemark{B} & $<$0.13 ($<$3$\sigma$) &
0.043 & 1.0$\times$0.85 ($-$49$^{\circ}$) & --- & --- & --- & --- \\ 
06035$-$7102 & H$^{13}$CN J=3--2 (259.012) \tablenotemark{B} &
$<$0.24 ($<$3$\sigma$) & 0.078 & 1.1$\times$0.76 (51$^{\circ}$) & --- &
--- & --- & --- \\ 
13509$+$0442 & CS J=7--6 (342.883) & 0.16 (3.3$\sigma$) & 0.050 &
0.68$\times$0.54 (47$^{\circ}$) & --- & --- & --- & --- \\        
 & CH$_{3}$CN (349.346) & 0.84 (9.5$\sigma$) & 0.089 &
0.69$\times$0.61 ($-$40$^{\circ}$) & --- & --- & --- & --- \\        
 & H$^{13}$CN J=3--2 (259.012) \tablenotemark{B} & $<$0.12 ($<$3$\sigma$)
& 0.040 & 0.95$\times$0.83 (57$^{\circ}$) & --- & --- & --- & --- \\
20414$-$1651  & CS J=7-6 (342.883) & 0.81 (6.7$\sigma$) & 0.12 &
0.38$\times$0.34 ($-$67$^{\circ}$) & 25844$\pm$20, 26278$\pm$18 &
2.1$\pm$0.2, 1.8$\pm$0.3 & 319$\pm$56, 189$\pm$53 & 1.0$\pm$0.2 \\ 

\enddata

\tablenotetext{A}{Based on ALMA Cycle 2 data.}

\tablenotetext{B}{Based on ALMA Cycle 3 data.}

\tablenotetext{C}{Data at $\nu_{\rm obs}$ = 346.00--346.19 GHz in 
Figure 8d are summed.}

\tablenotetext{D}{Not all of the HC$_{3}$N J=30--29 emission line is 
covered in the spectral window, which was used to create the moment 0
map. Thus, the estimated flux is smaller than the Gaussian fit in Figure
31.} 

\tablecomments{ 
Col.(1): Object.
Col.(2): Line. Rest-frame frequency in [GHz] is shown in parentheses.
Col.(3): Integrated intensity in [Jy beam$^{-1}$ km s$^{-1}$] at the 
emission peak. 
Detection significance relative to the rms noise (1$\sigma$) in the 
moment 0 map is shown in parentheses. 
Possible systematic uncertainty is not included. 
Col.(4): The rms noise (1$\sigma$) level in the moment 0 map in 
[Jy beam$^{-1}$ km s$^{-1}$], derived from the standard deviation 
of sky signals in each moment 0 map. 
Col.(5): Beam size in [arcsec $\times$ arcsec] and position angle in
[degree]. Position angle is 0$^{\circ}$ along the north-south direction, 
and increases counterclockwise. 
Cols.(6)--(9): Gaussian fits of emission lines in the spectra at the 
continuum peak position, within the beam size. 
Col.(6): Optical LSR velocity (v$_{\rm opt}$) of emission peak in 
[km s$^{-1}$]. 
Col.(7): Peak flux in [mJy]. 
Col.(8): Observed FWHM in [km s$^{-1}$].
Col.(9): Flux in [Jy km s$^{-1}$]. The observed FWHM in 
[km s$^{-1}$] in column 8 is divided by ($1+z$) to obtain the
intrinsic FWHM in [km s$^{-1}$]. 
}

\end{deluxetable}
\end{turnpage}

\begin{figure}
\begin{center}
\includegraphics[angle=0,scale=.406]{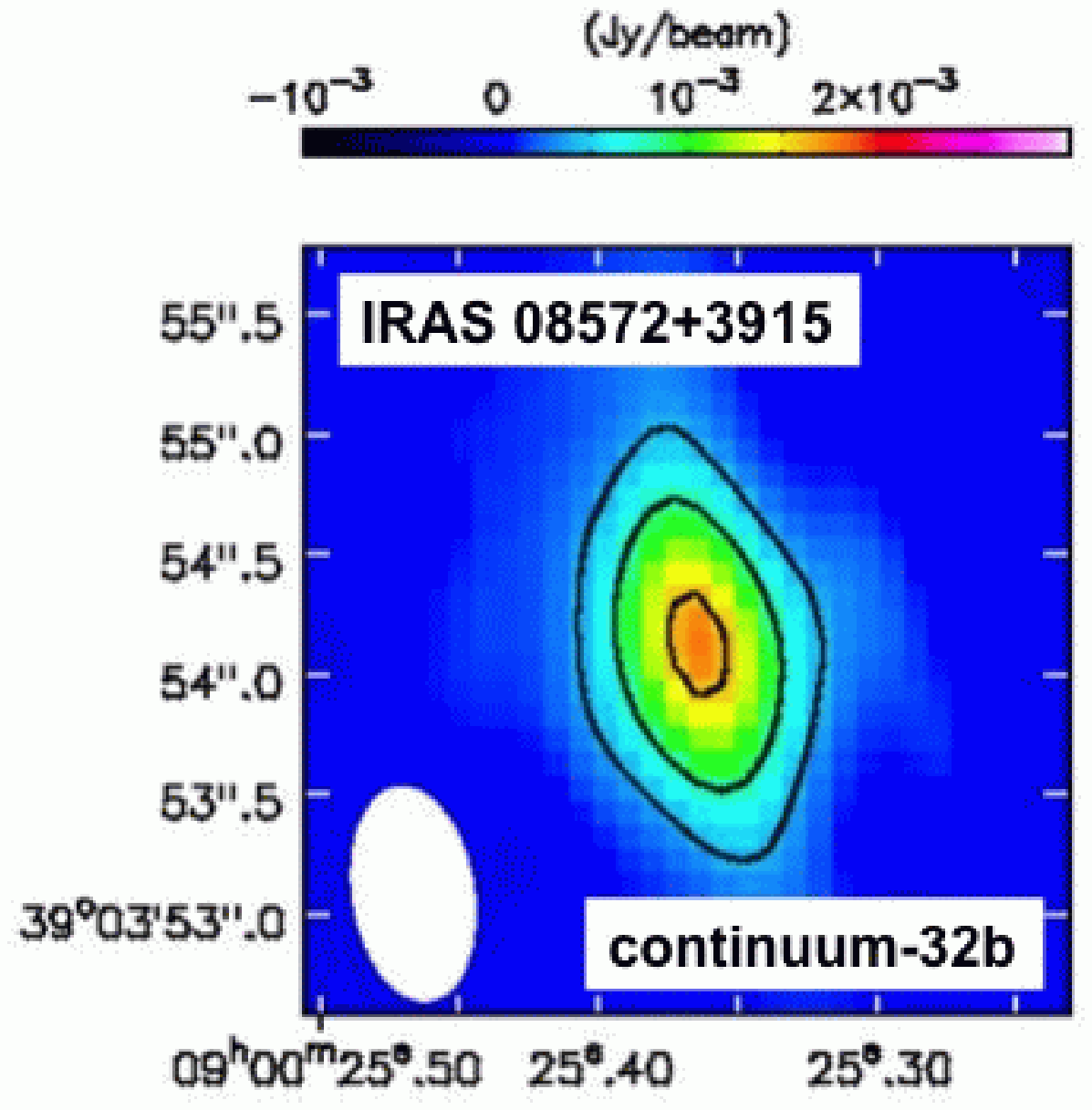} 
\includegraphics[angle=0,scale=.406]{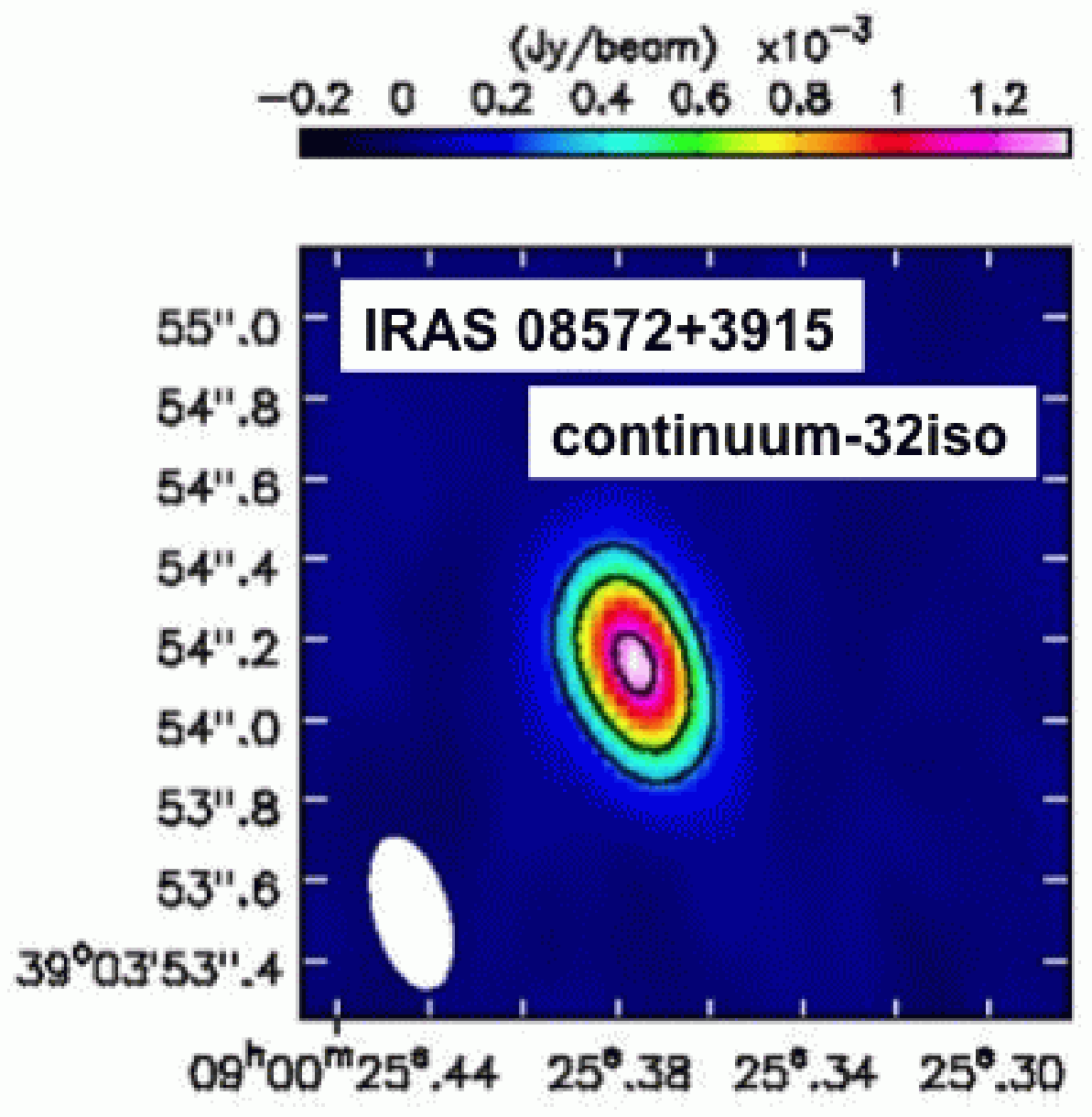} 
\includegraphics[angle=0,scale=.406]{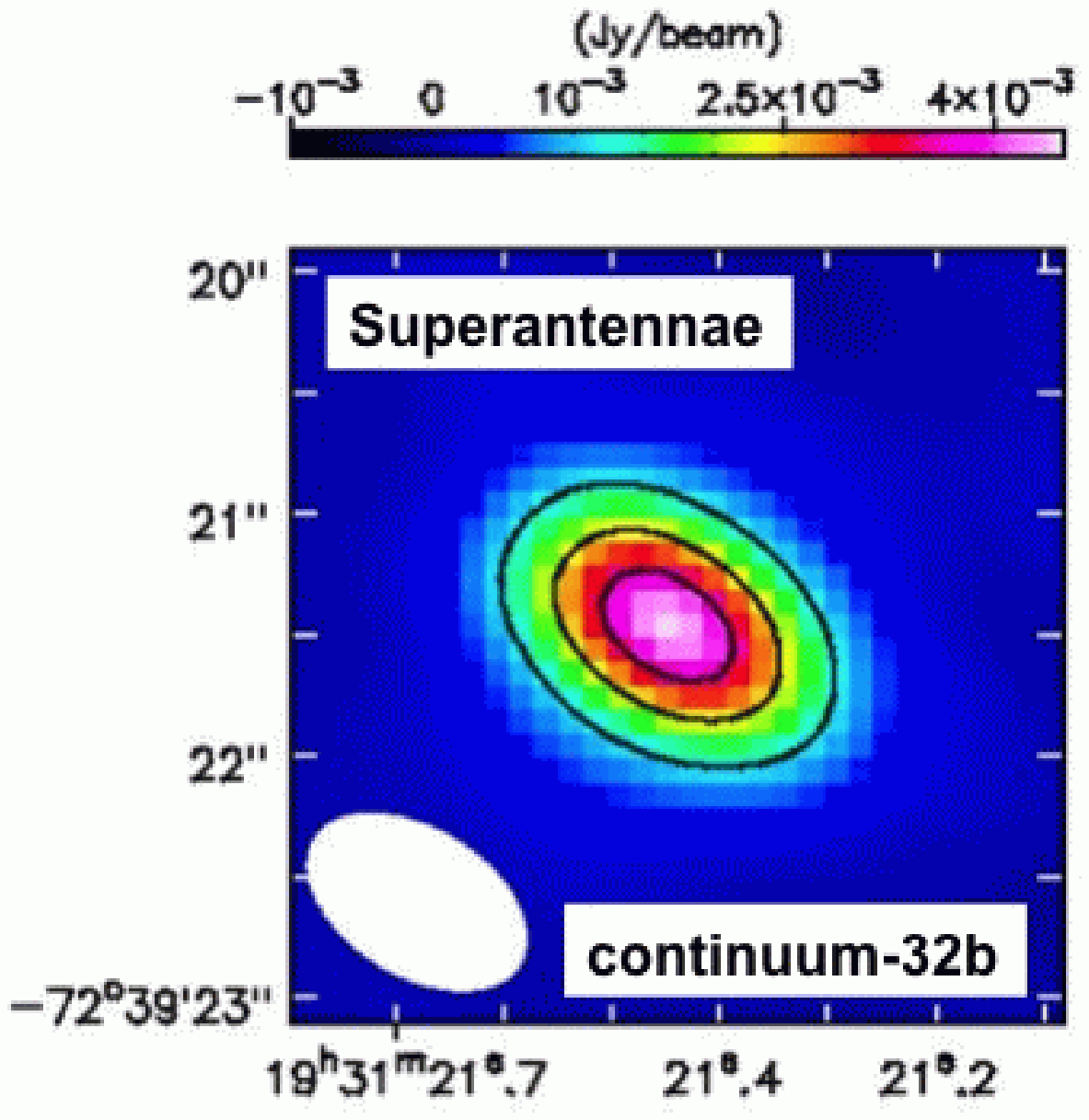} \\
\vspace{-1.3cm}
\includegraphics[angle=0,scale=.406]{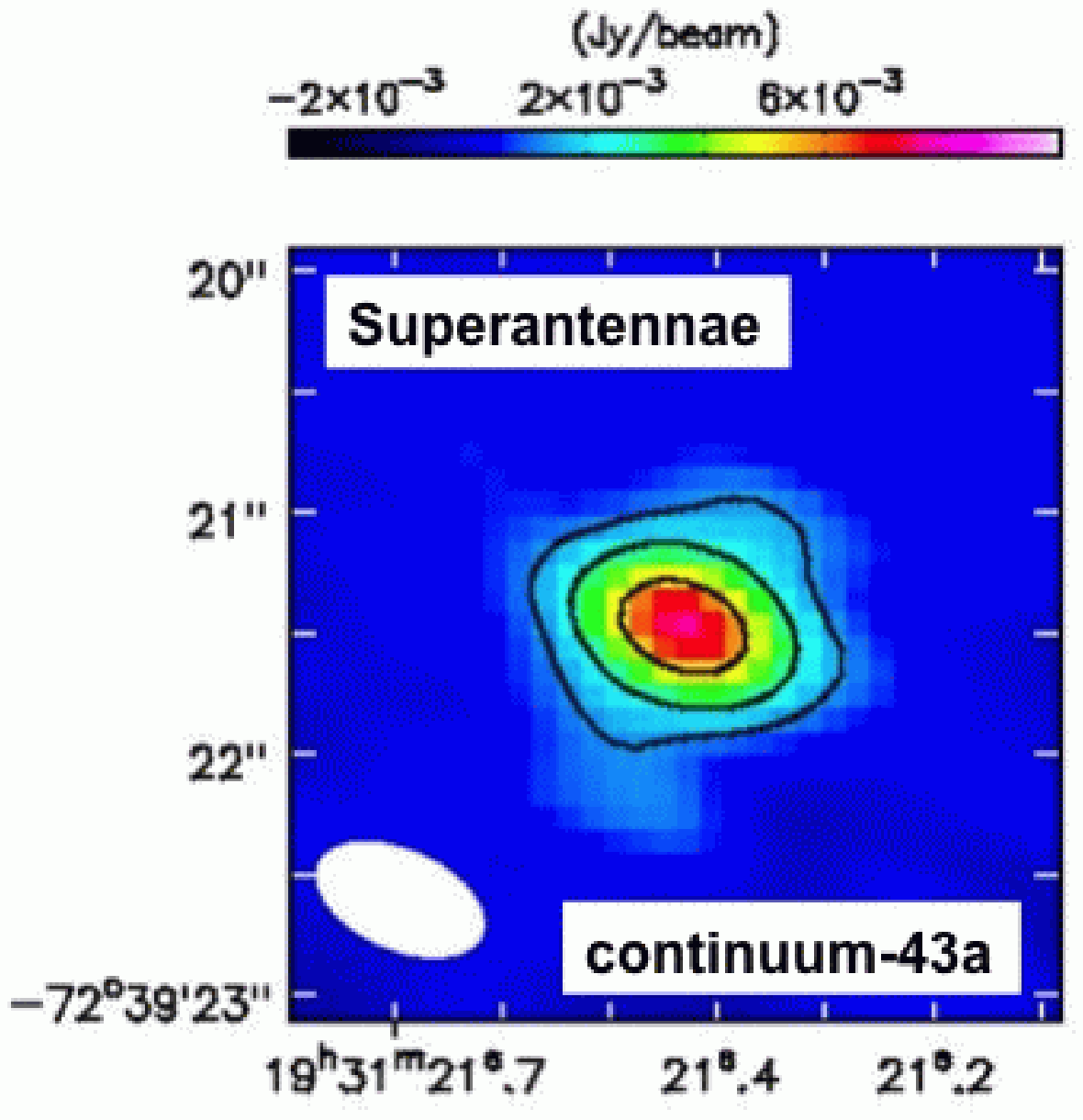} 
\includegraphics[angle=0,scale=.406]{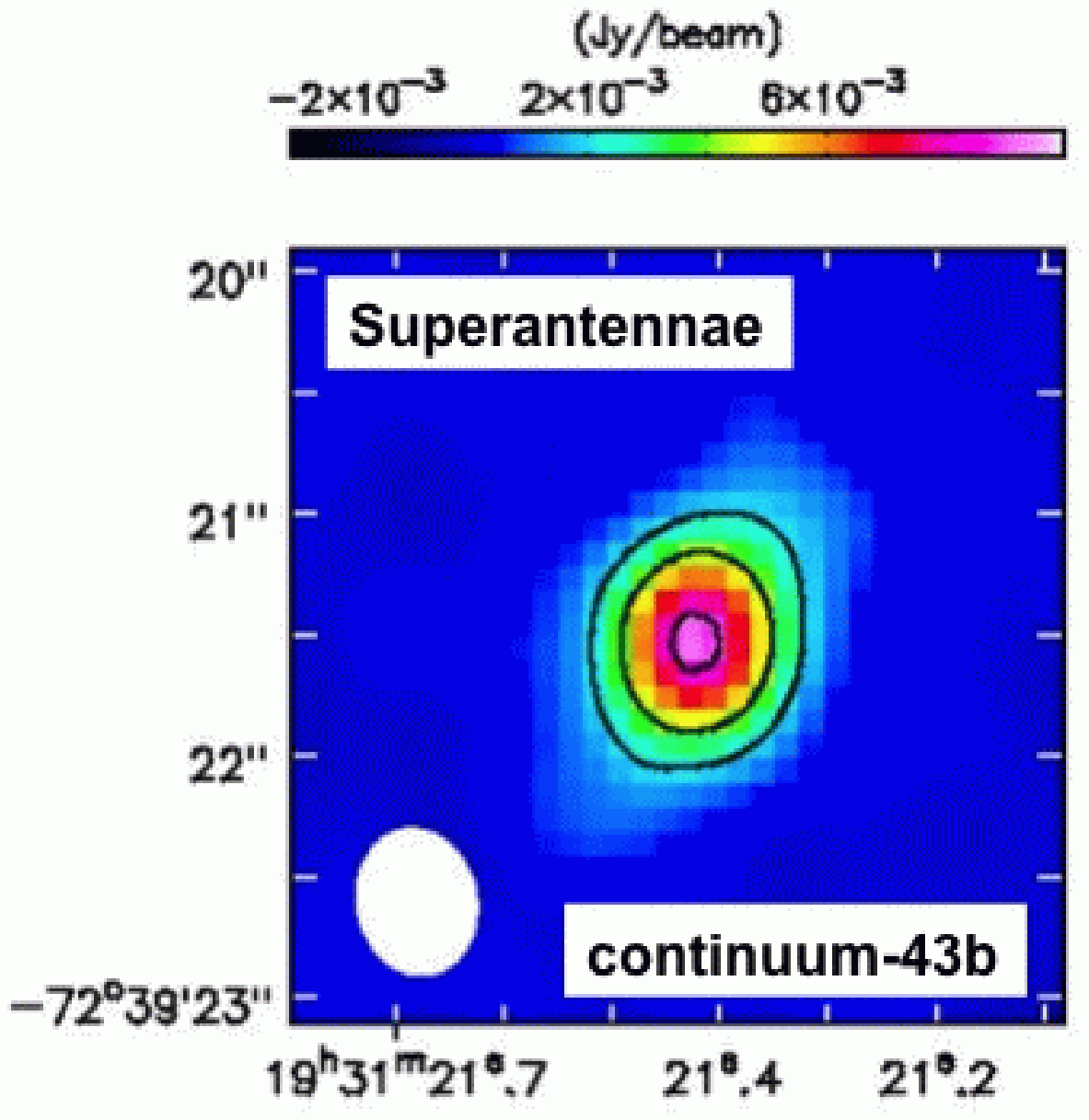} 
\includegraphics[angle=0,scale=.406]{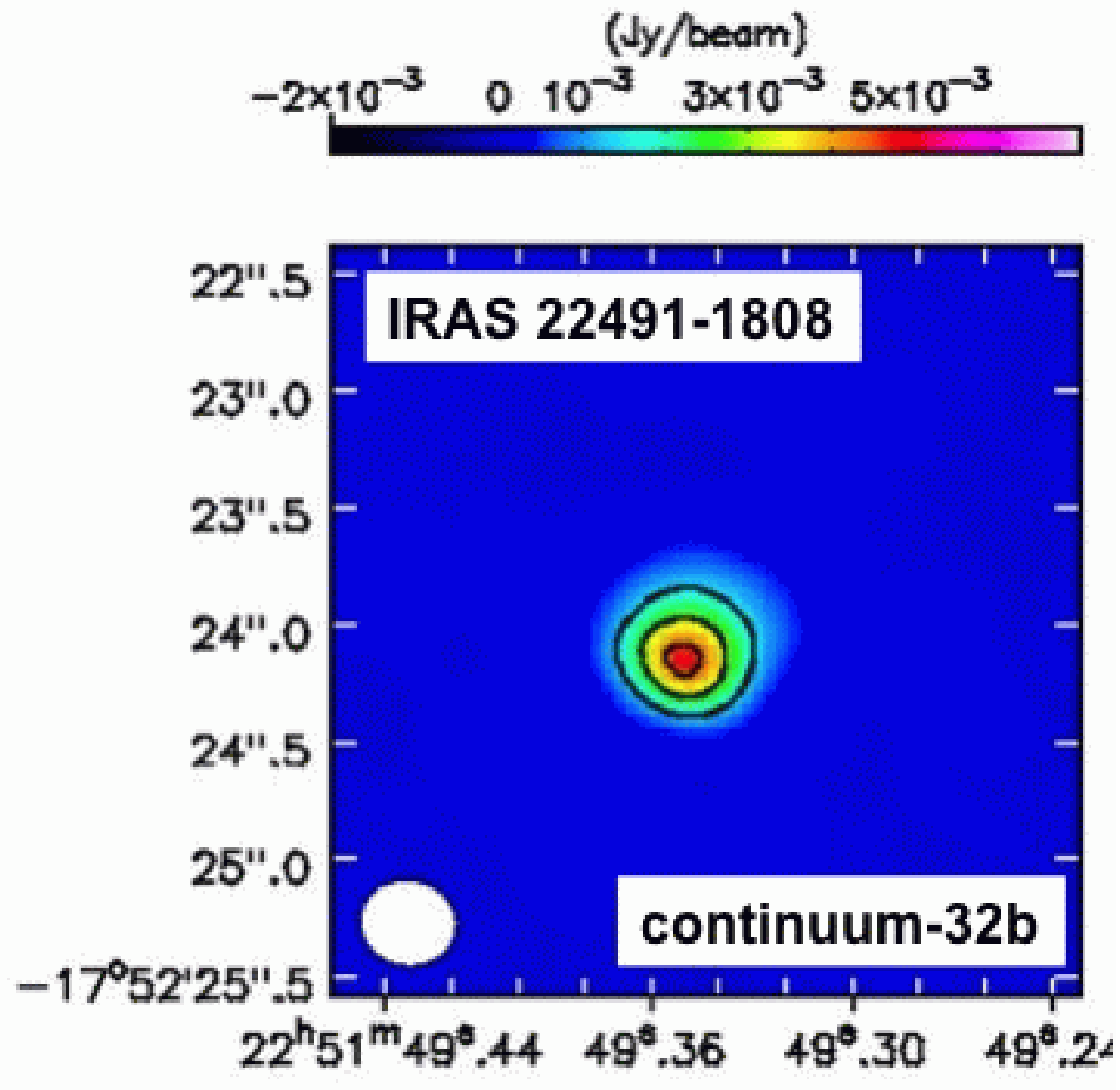} \\
\vspace{-1.3cm}
\includegraphics[angle=0,scale=.406]{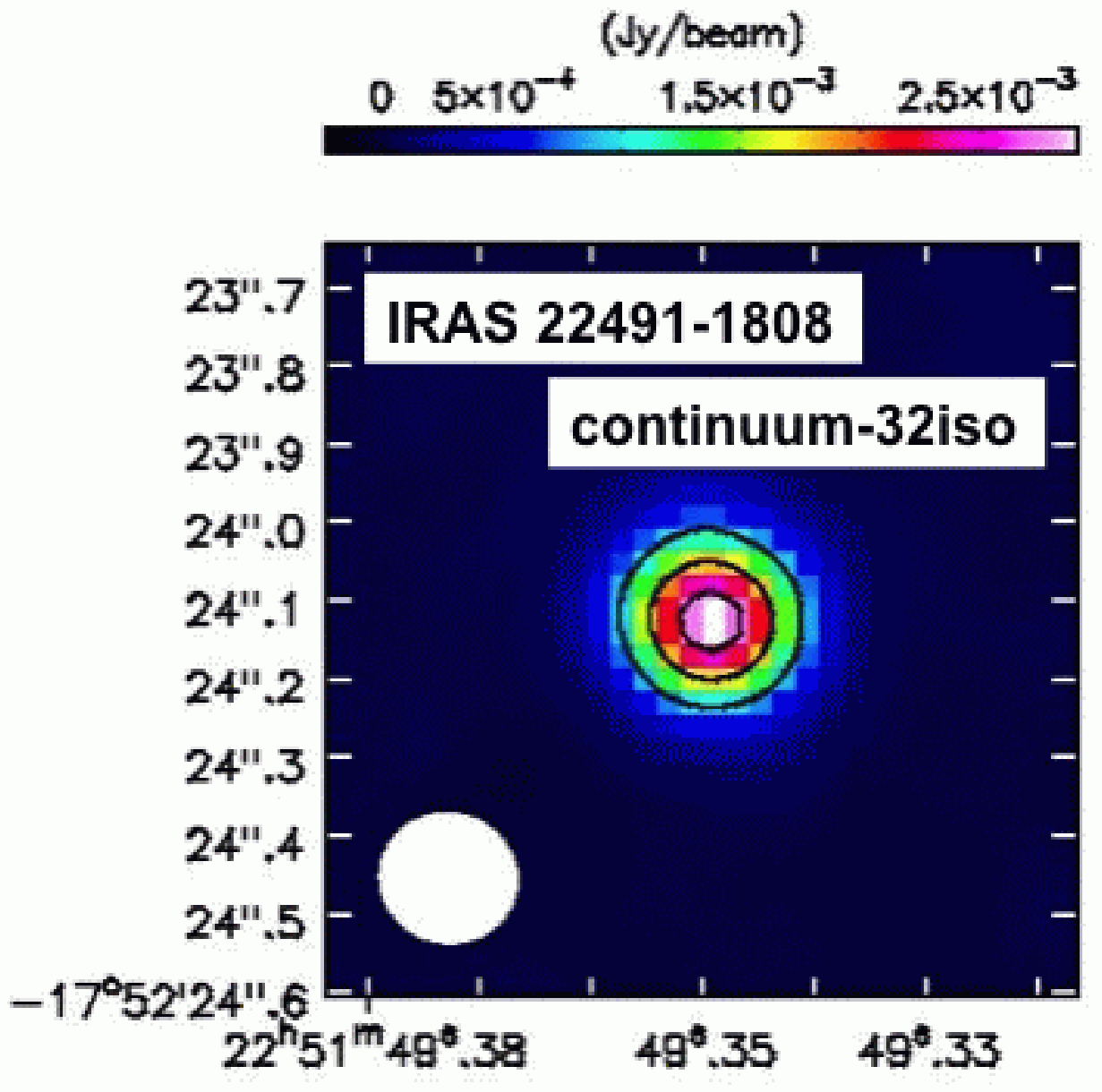} 
\includegraphics[angle=0,scale=.406]{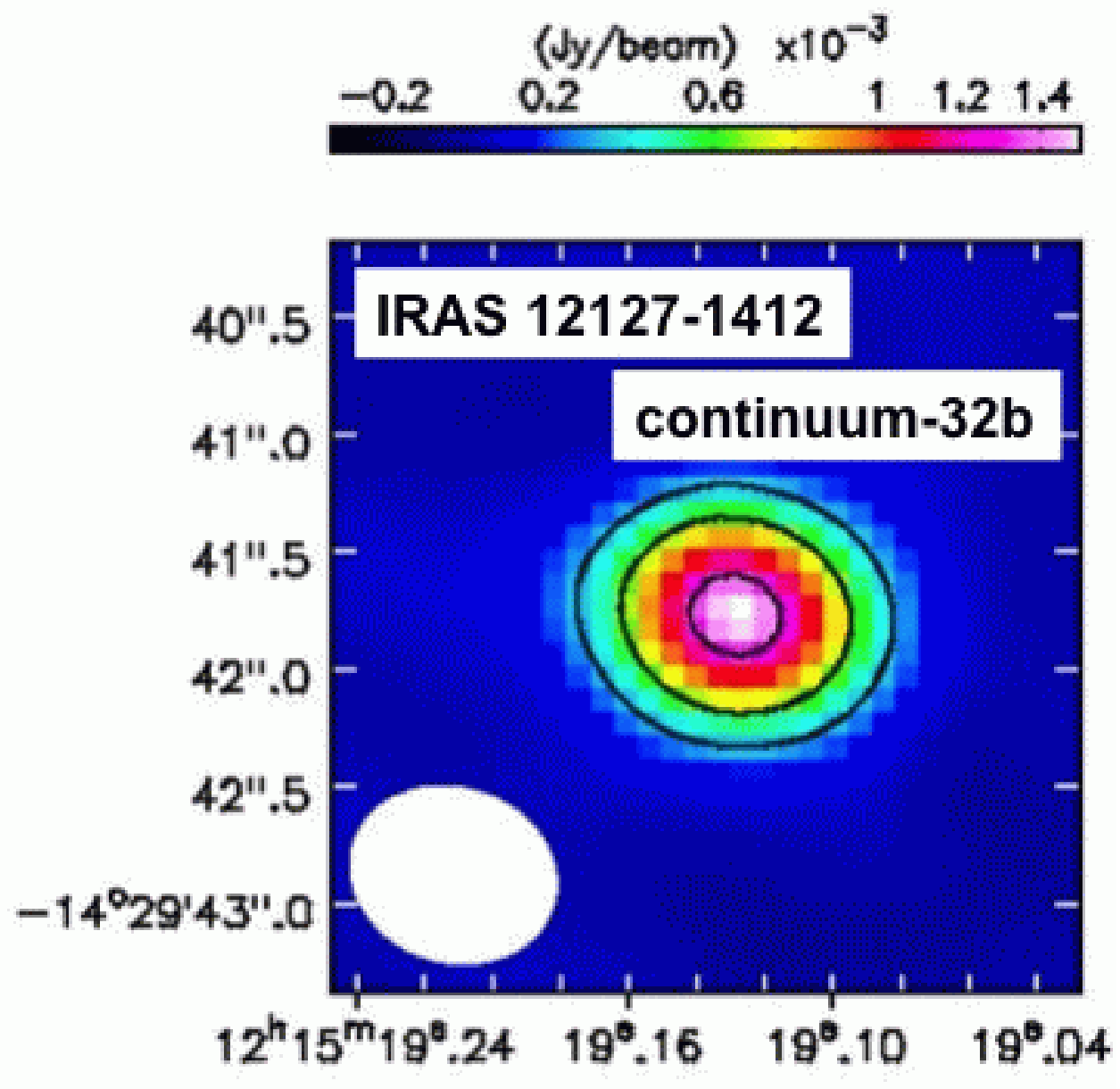} 
\includegraphics[angle=0,scale=.406]{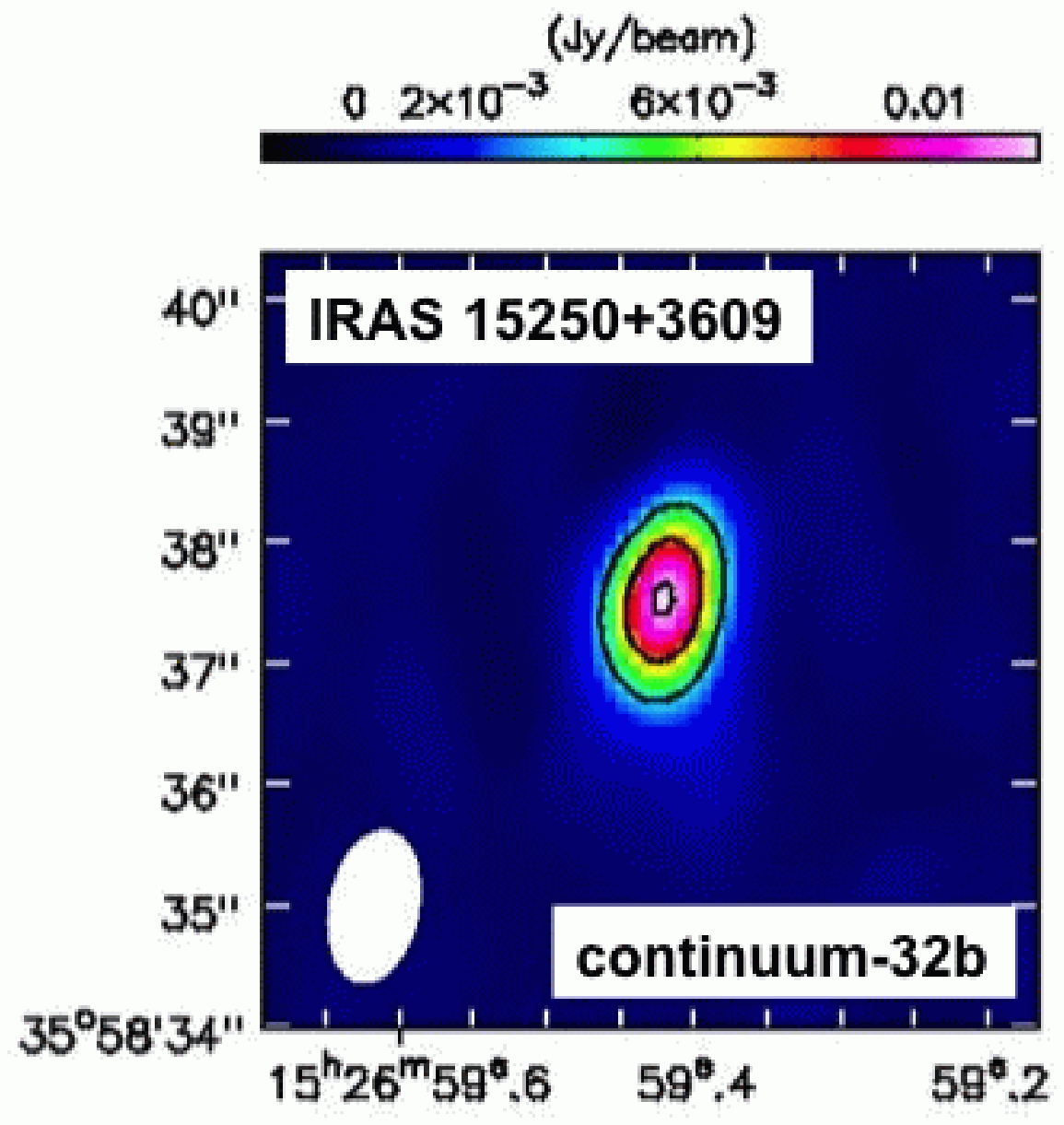} \\
\vspace{-1.3cm}
\includegraphics[angle=0,scale=.406]{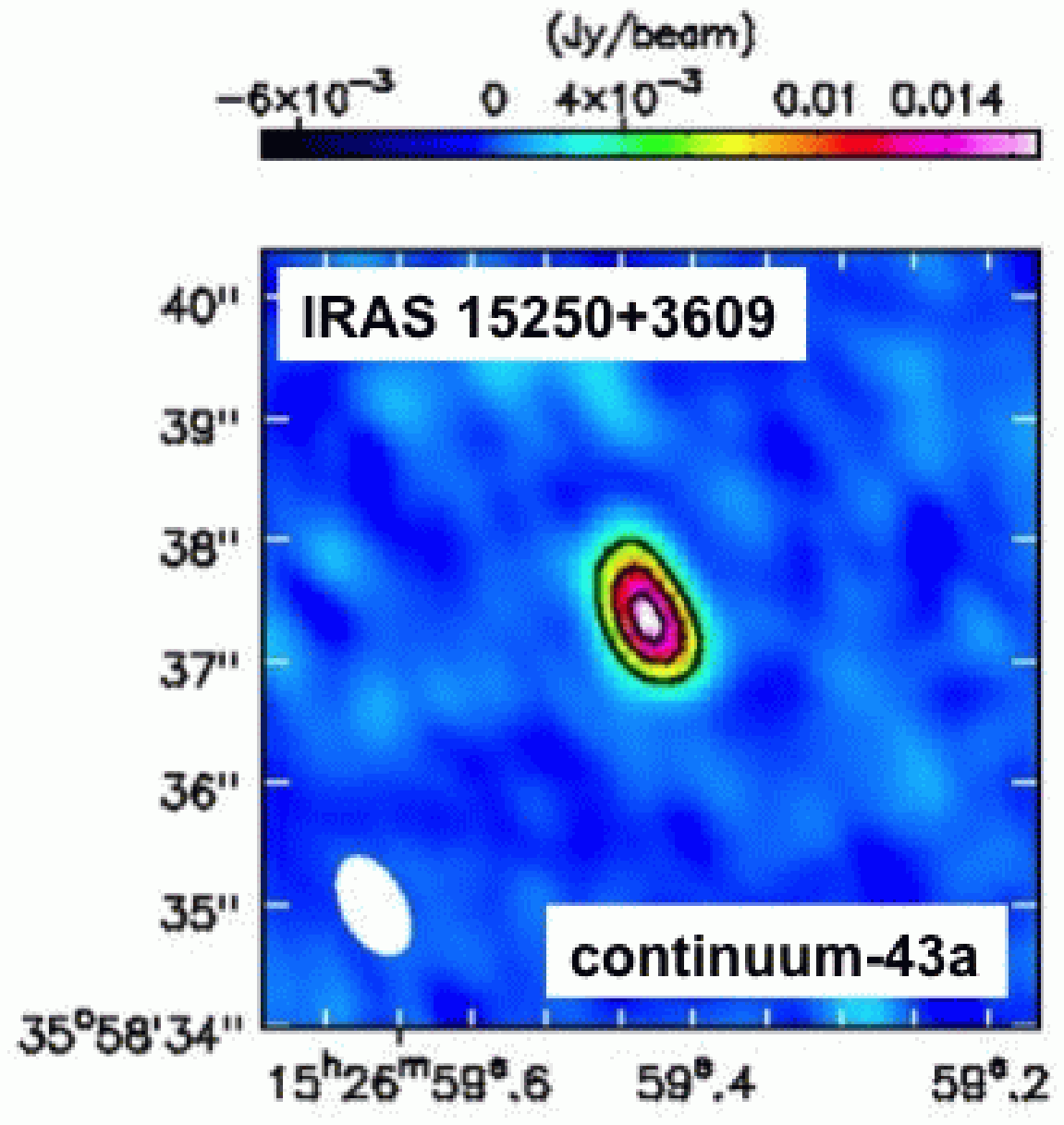}
\includegraphics[angle=0,scale=.406]{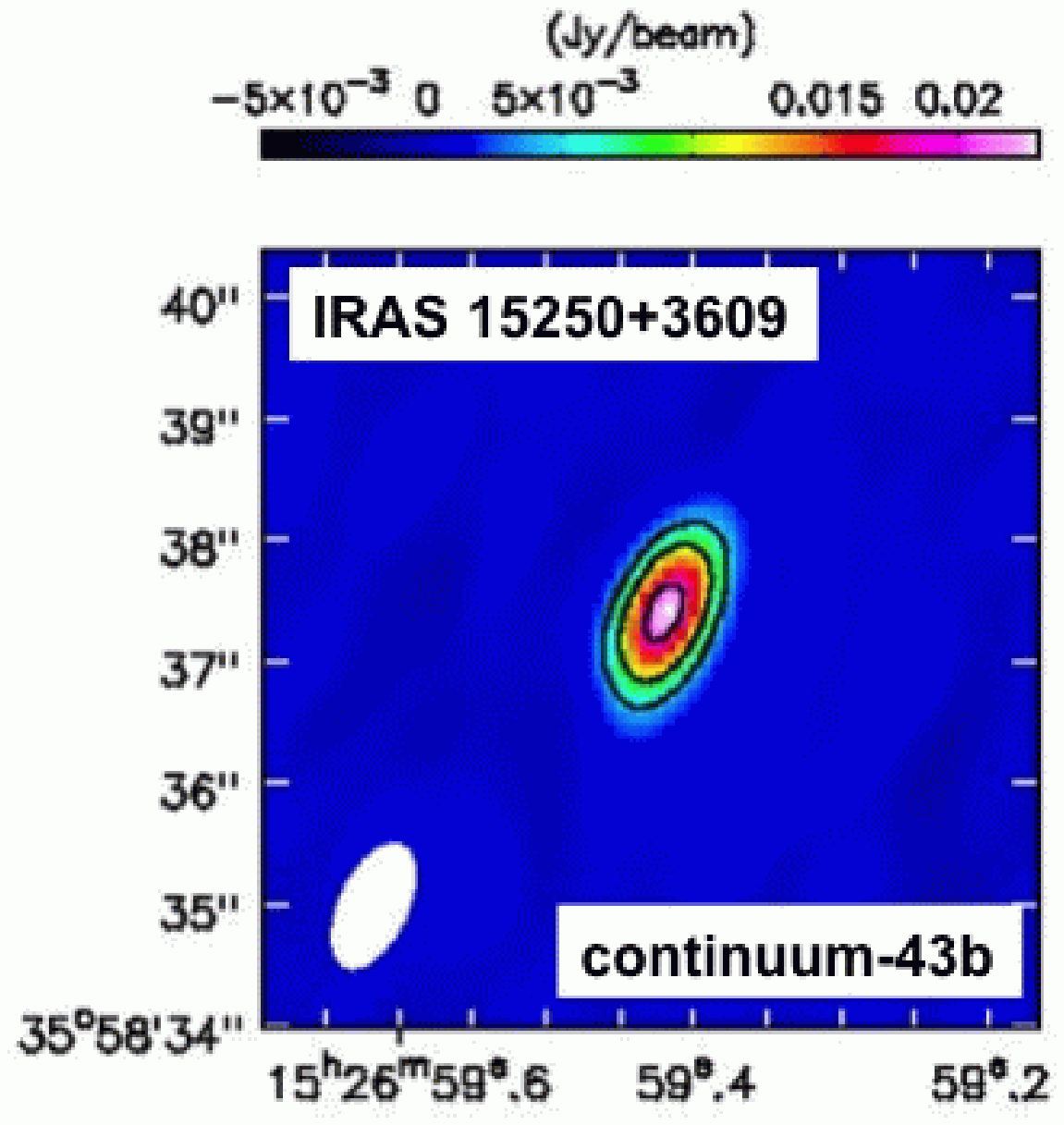}
\includegraphics[angle=0,scale=.406]{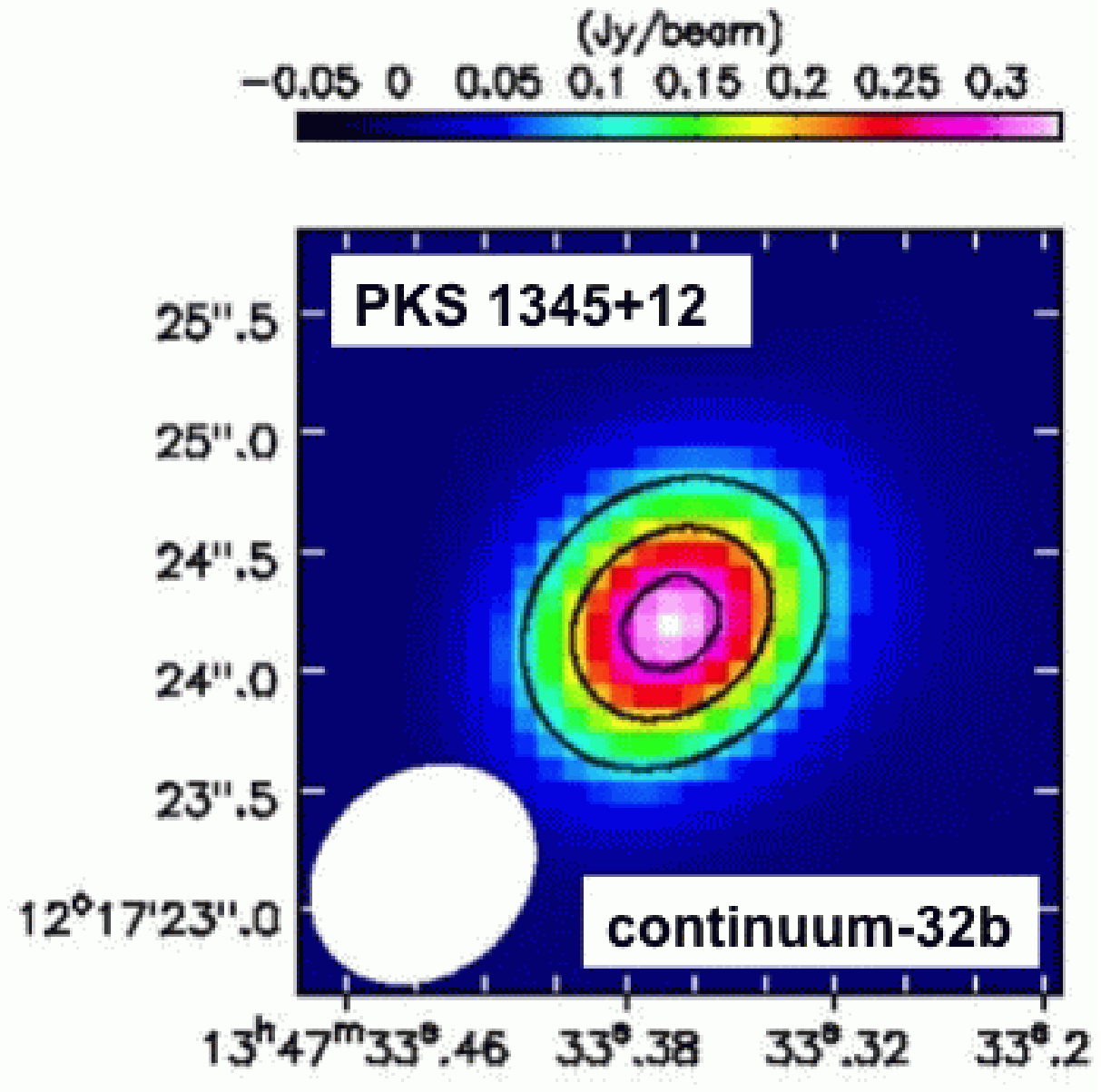} \\
\end{center}
\end{figure}

\clearpage

\begin{figure}
\begin{center}
\includegraphics[angle=0,scale=.406]{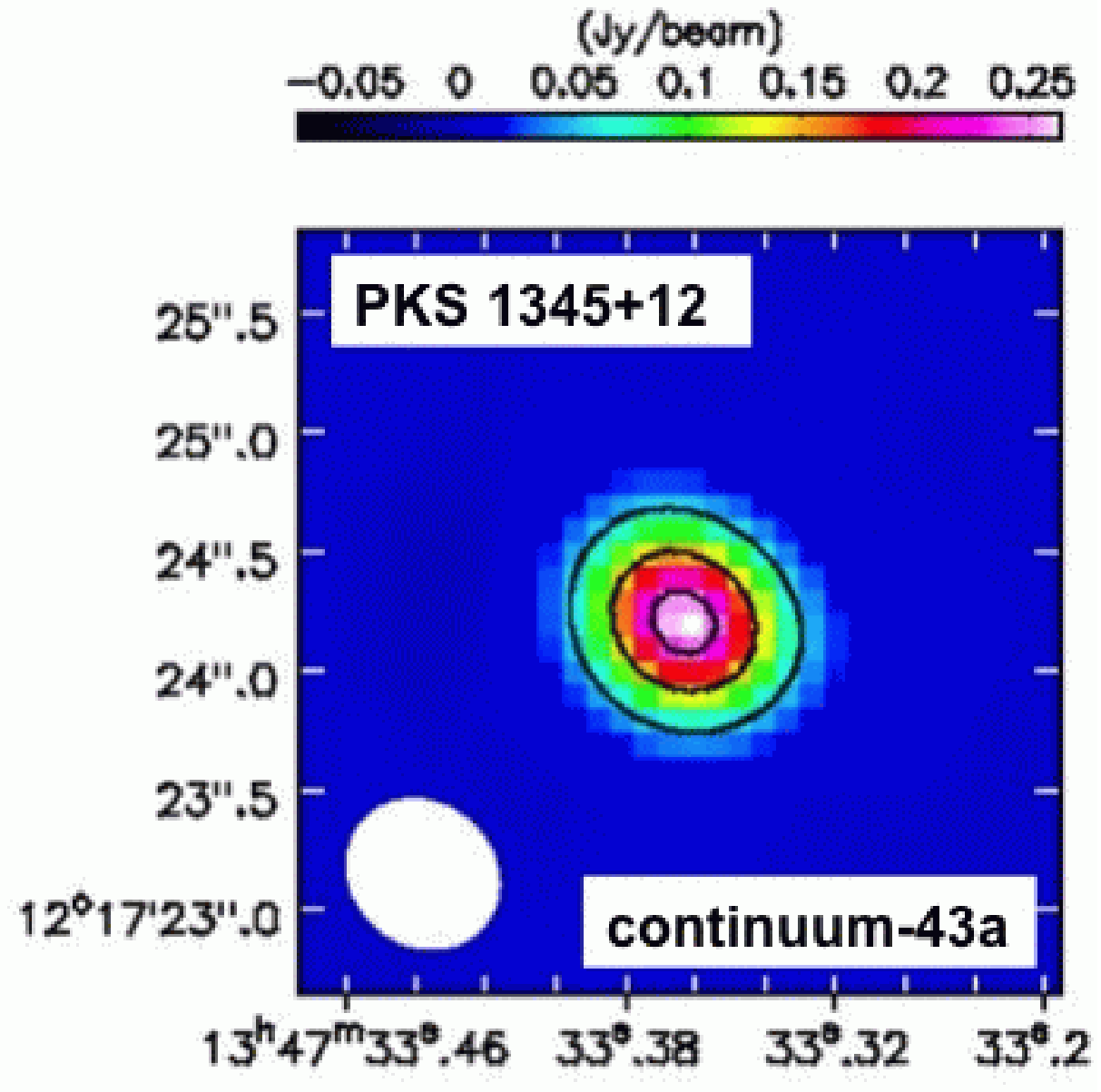} 
\includegraphics[angle=0,scale=.406]{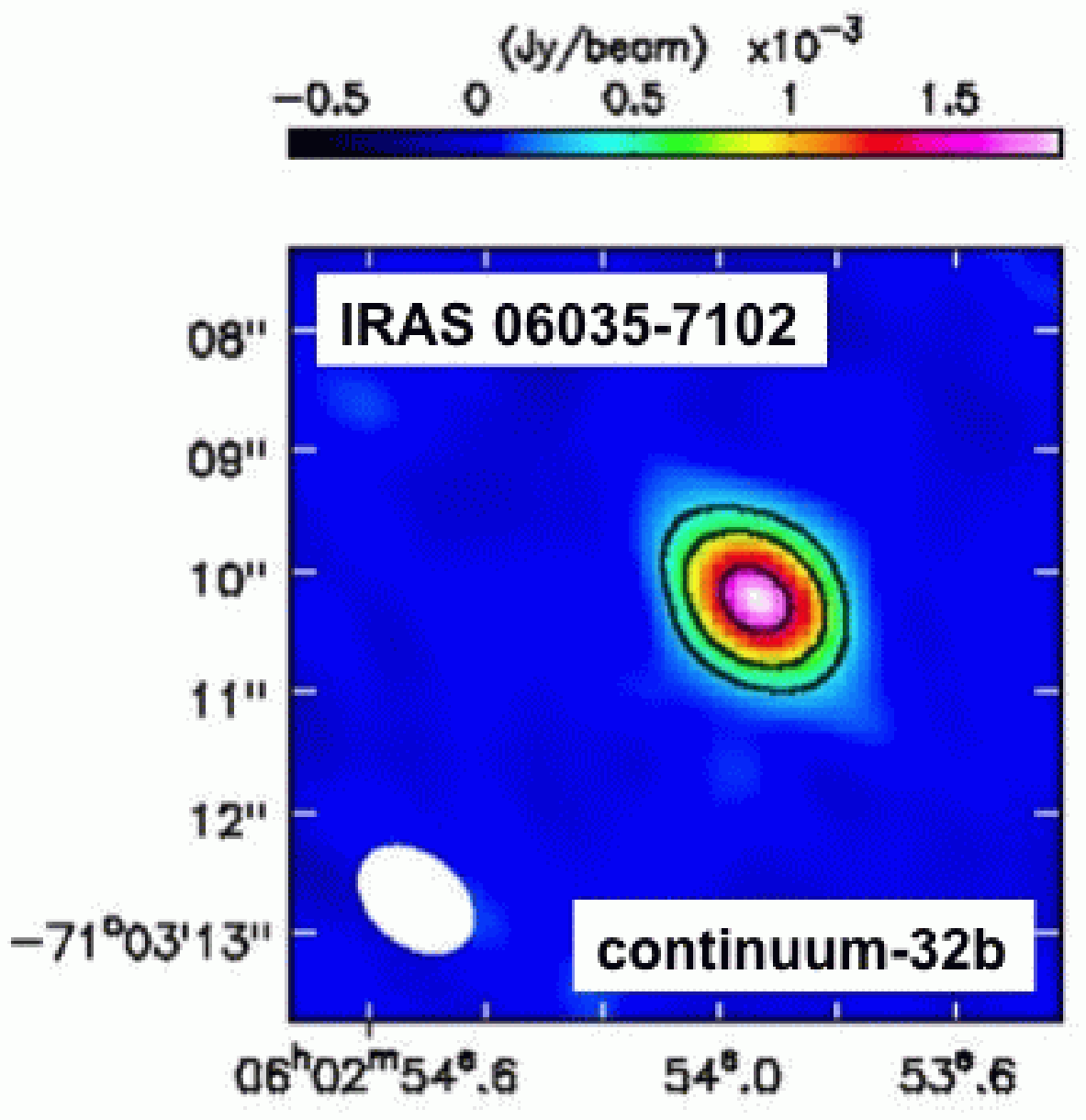} \\
\vspace{-1.3cm}
\includegraphics[angle=0,scale=.387]{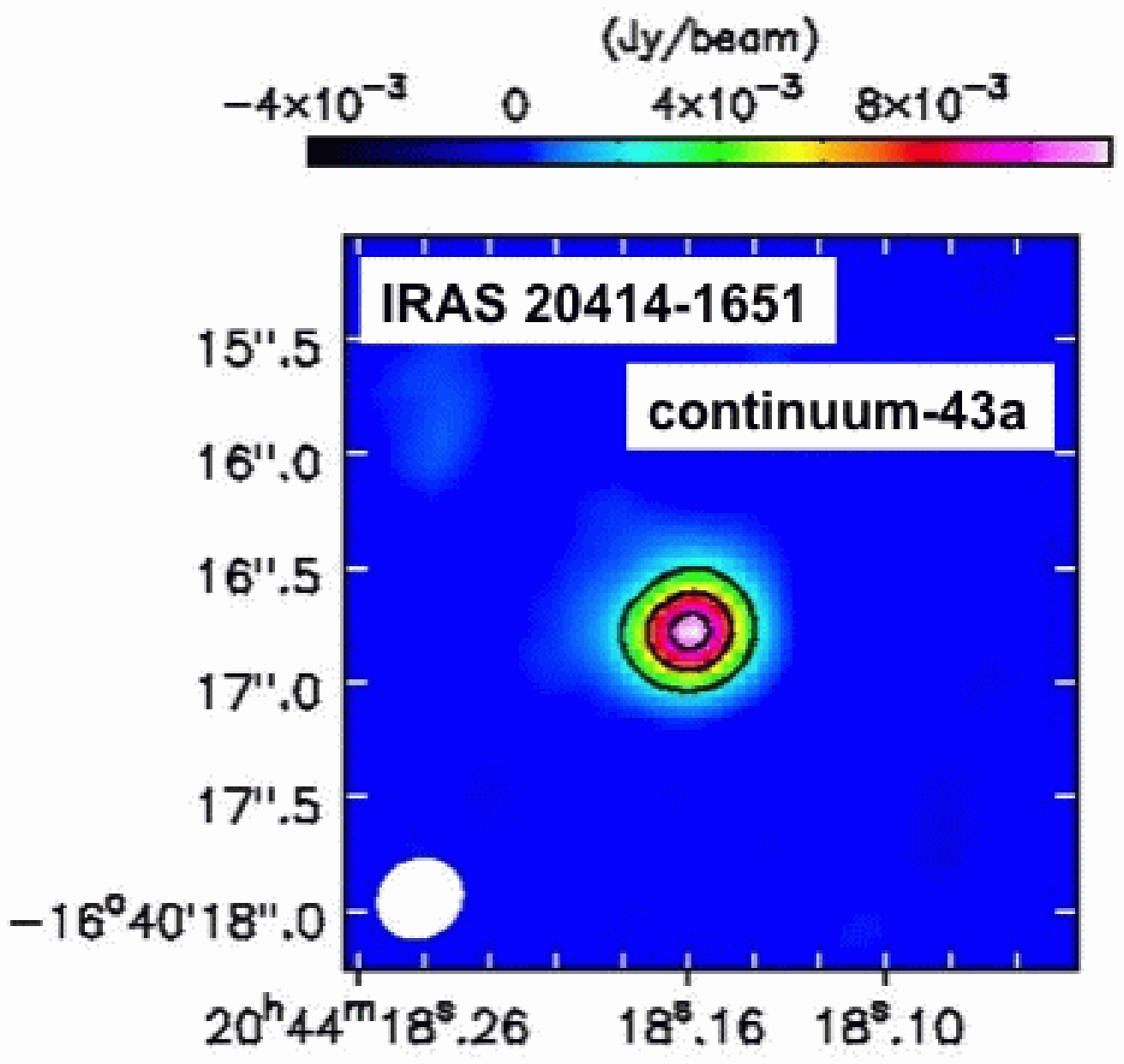}
\includegraphics[angle=0,scale=.387]{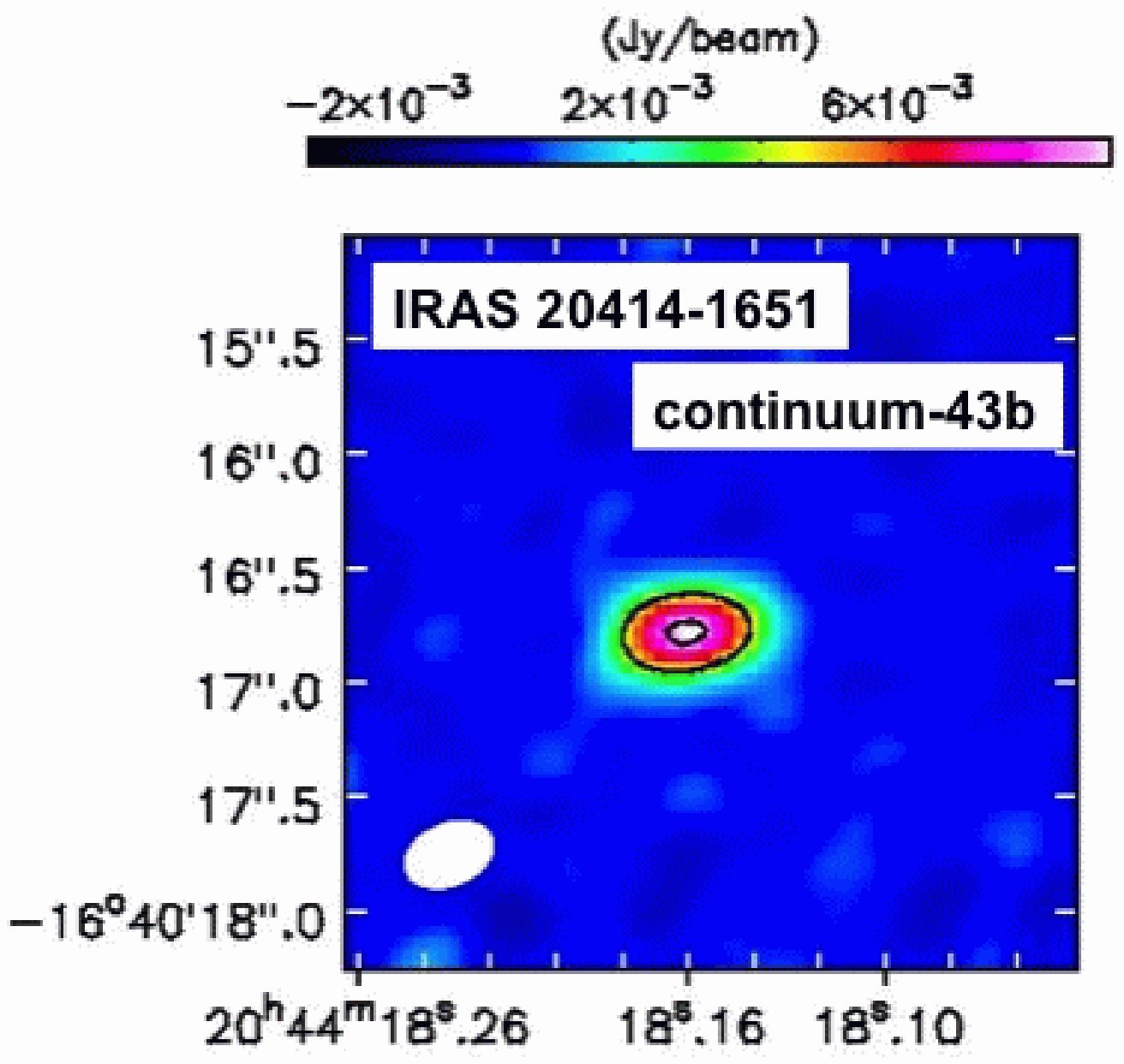} 
\end{center}
\caption{
Continuum emission maps of ULIRGs dominated by compact single nuclear
components.  
The abscissa and ordinate are R.A. (J2000) and decl. (J2000),
respectively. 
We denote continuum data taken with ``HNC J=3--2'',
``J=3--2 of H$^{13}$CN, H$^{13}$CO$^{+}$, and HN$^{13}$C'', 
``J=4--3 of HCN and HCO$^{+}$'', and ``HNC J=4--3'', 
as 32b, 32iso, 43a, and 43b, respectively.
The contours are 
5$\sigma$, 10$\sigma$, 20$\sigma$ for continuum-32b of IRAS 08572$+$3915, 
10$\sigma$, 20$\sigma$, 40$\sigma$ for continuum-32iso of IRAS 08572$+$3915, 
10$\sigma$, 20$\sigma$, 30$\sigma$ for continuum-32b of Superantennae,
5$\sigma$, 10$\sigma$, 20$\sigma$ for continuum-43a of Superantennae,
10$\sigma$, 20$\sigma$, 40$\sigma$ for continuum-43b of Superantennae,
20$\sigma$, 40$\sigma$, 60$\sigma$ for continuum-32b of IRAS 22491$-$1808, 
30$\sigma$, 55$\sigma$, 80$\sigma$ for continuum-32iso of IRAS 22491$-$1808, 
8$\sigma$, 16$\sigma$, 32$\sigma$ for continuum-32b of IRAS 12127$-$1412, 
15$\sigma$, 30$\sigma$, 45$\sigma$ for continuum-32b of IRAS 15250$+$3609,
10$\sigma$, 20$\sigma$, 30$\sigma$ for continuum-43a of IRAS 15250$+$3609,
10$\sigma$, 20$\sigma$, 40$\sigma$ for continuum-43b of IRAS 15250$+$3609,
20$\sigma$, 40$\sigma$, 60$\sigma$ for continuum-32b of PKS 1345$+$12,
20$\sigma$, 50$\sigma$, 80$\sigma$ for continuum-43a of PKS 1345$+$12,
5$\sigma$, 10$\sigma$, 20$\sigma$ for continuum-32b of IRAS 06035$-$7102,
20$\sigma$, 45$\sigma$, 70$\sigma$ for continuum-43a of IRAS
20414$-$1651, and 
20$\sigma$, 40$\sigma$ for continuum-43b of IRAS 20414$-$1651.
Beam sizes are shown as filled circles in the lower-left region.
}
\end{figure}

\begin{figure}
\begin{center}
\includegraphics[angle=0,scale=.82]{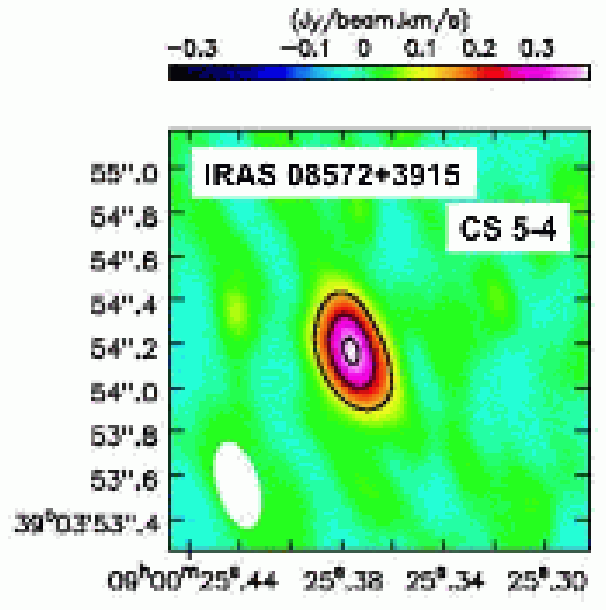} 
\includegraphics[angle=0,scale=.82]{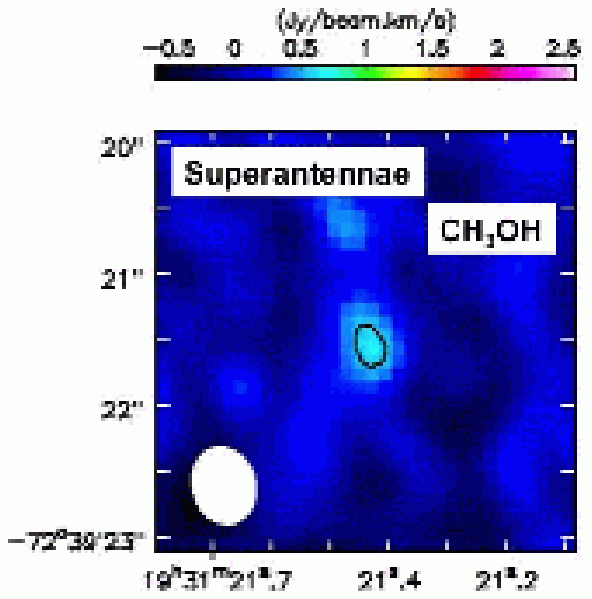} 
\includegraphics[angle=0,scale=.41]{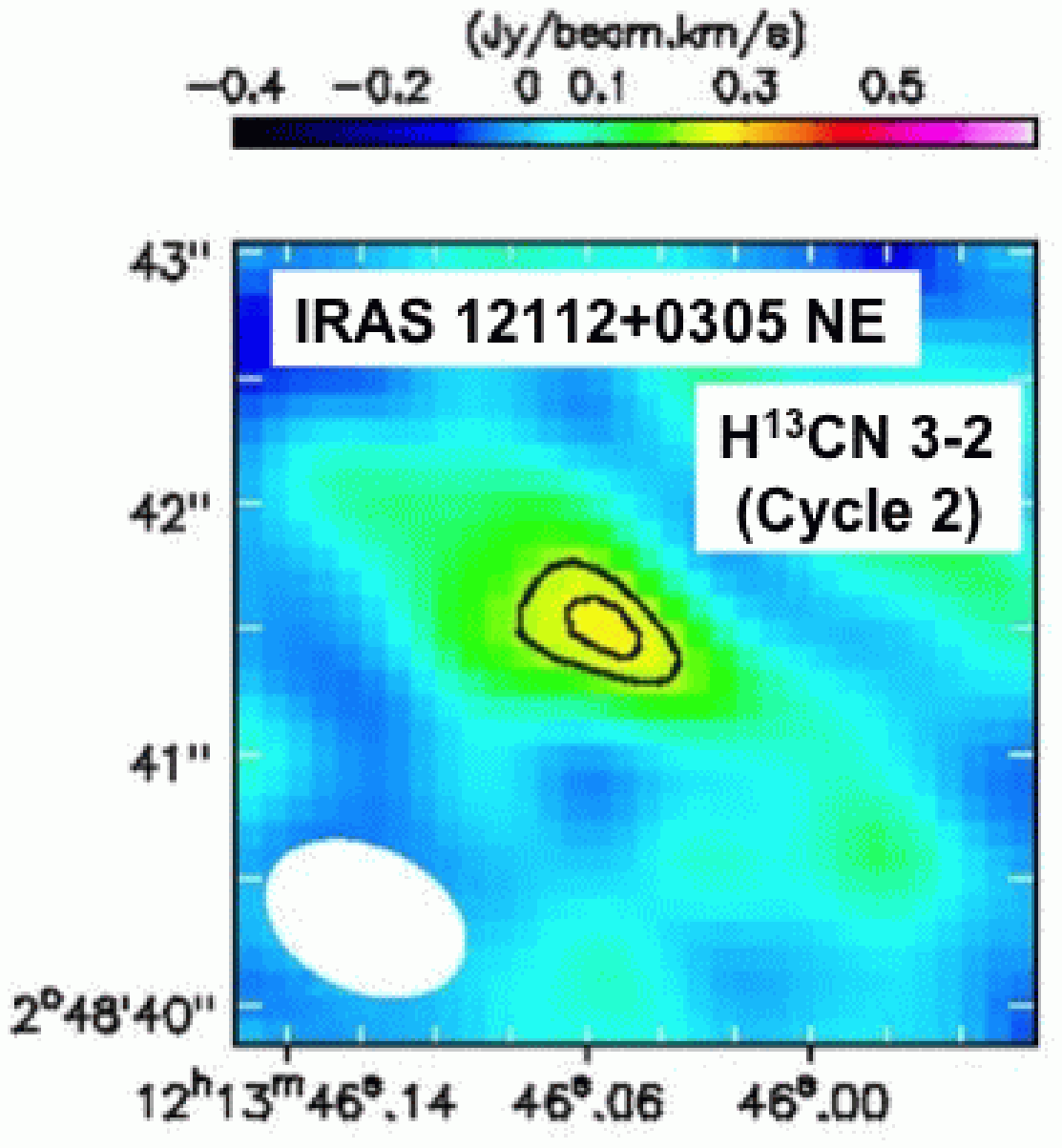} \\
\vspace{-1.3cm}
\includegraphics[angle=0,scale=.82]{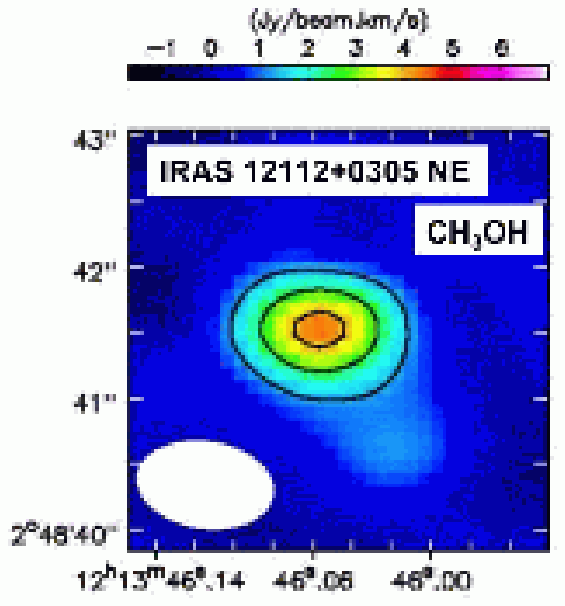} 
\includegraphics[angle=0,scale=.82]{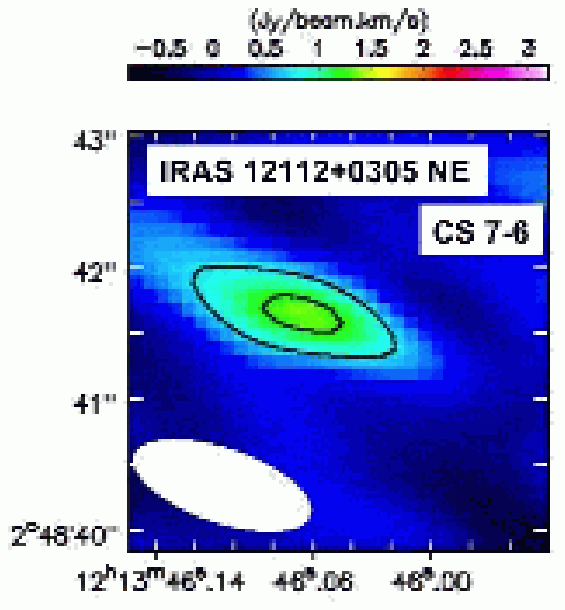} 
\includegraphics[angle=0,scale=.82]{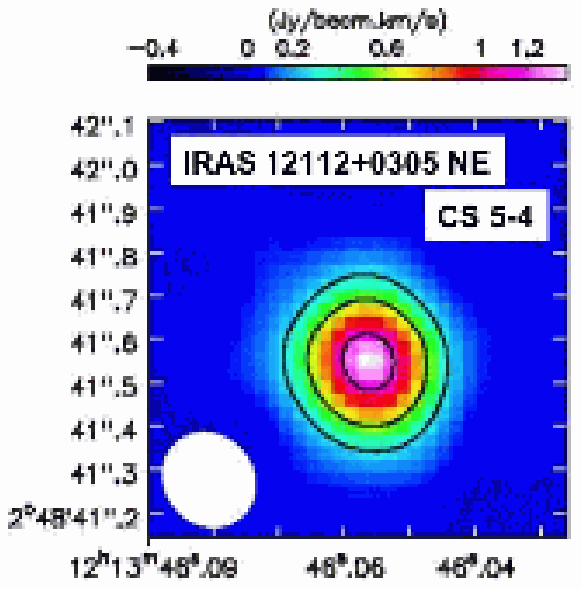} \\
\vspace{-1.3cm}
\includegraphics[angle=0,scale=.41]{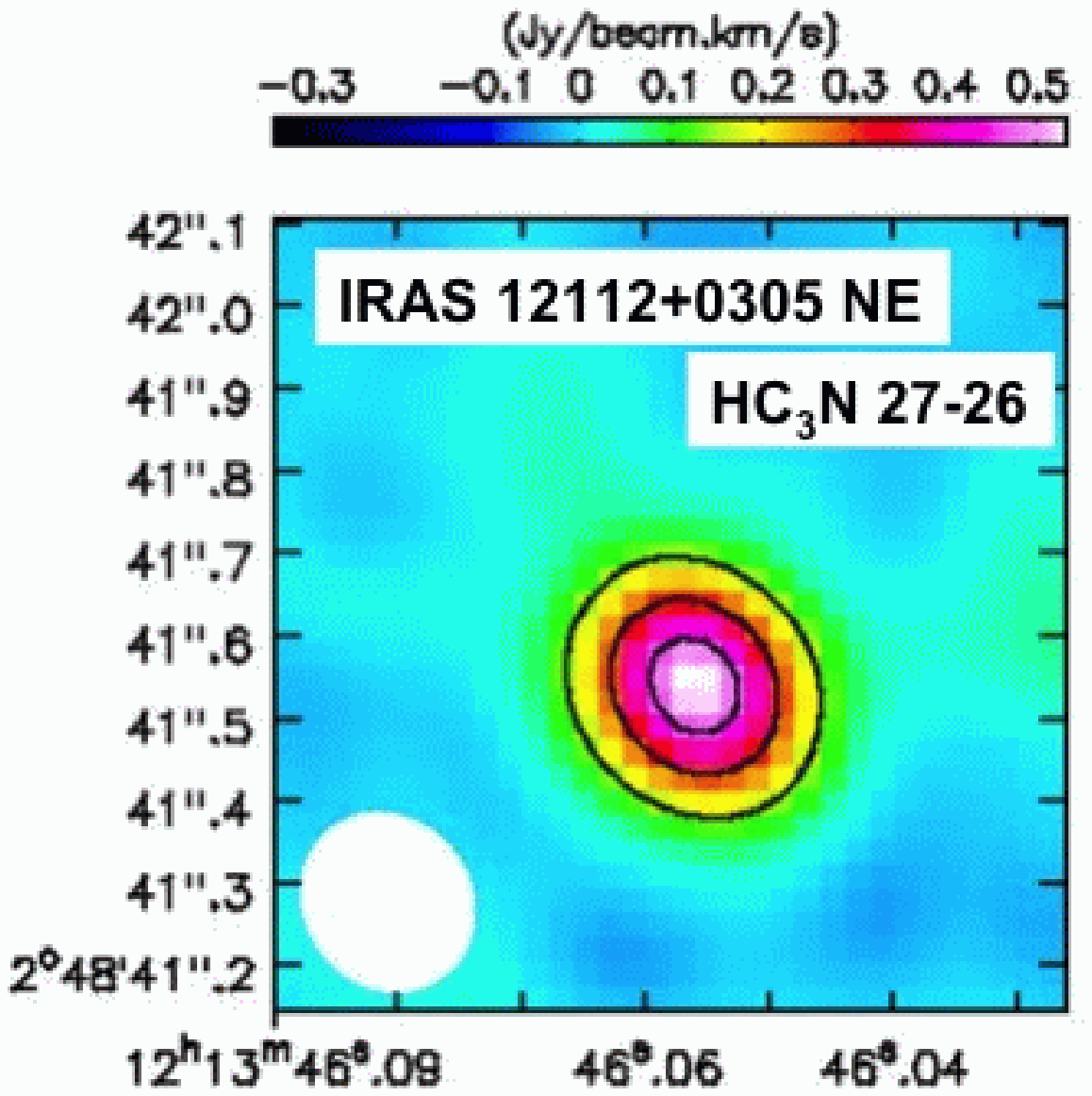} 
\includegraphics[angle=0,scale=.41]{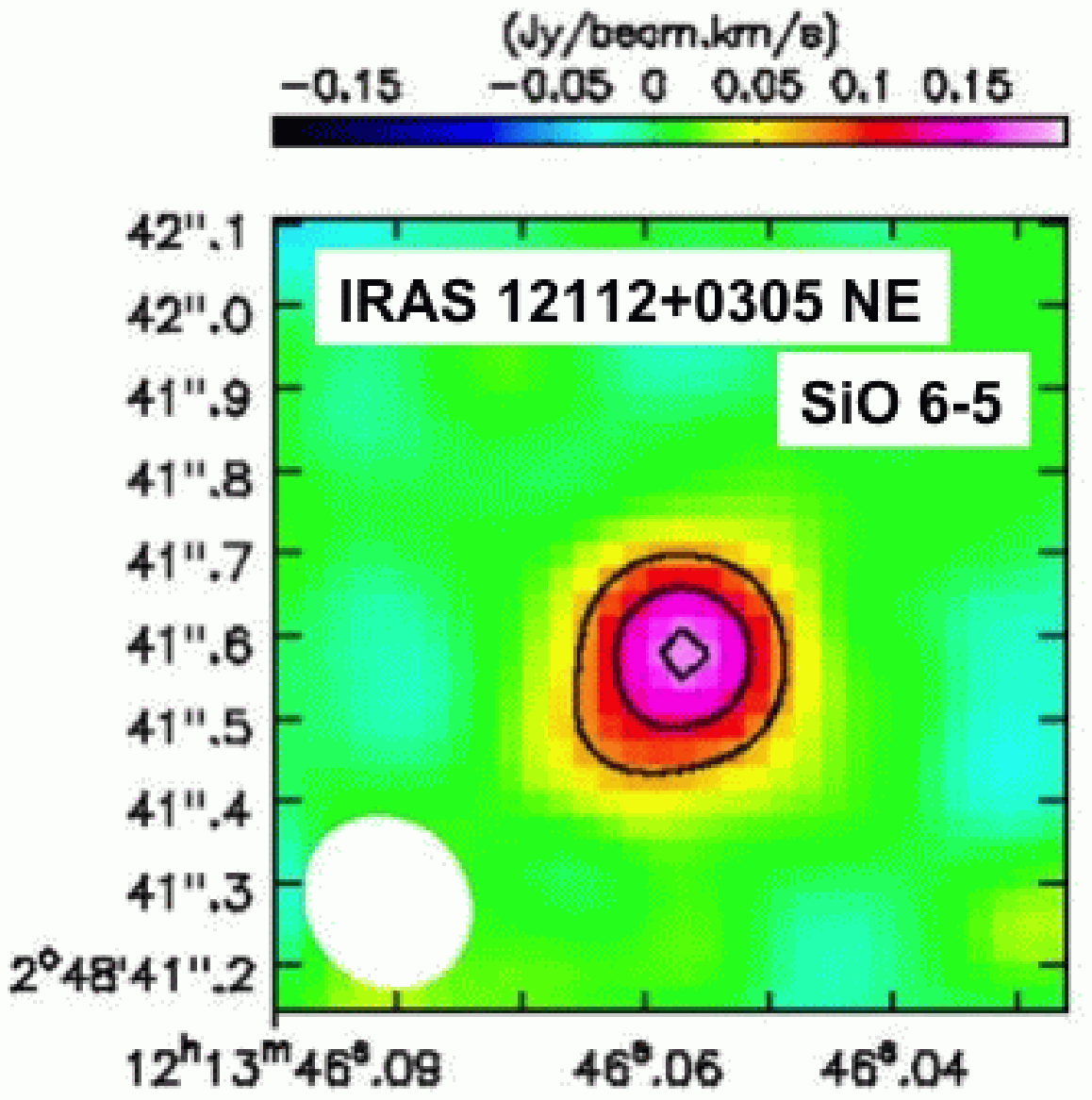} 
\includegraphics[angle=0,scale=.41]{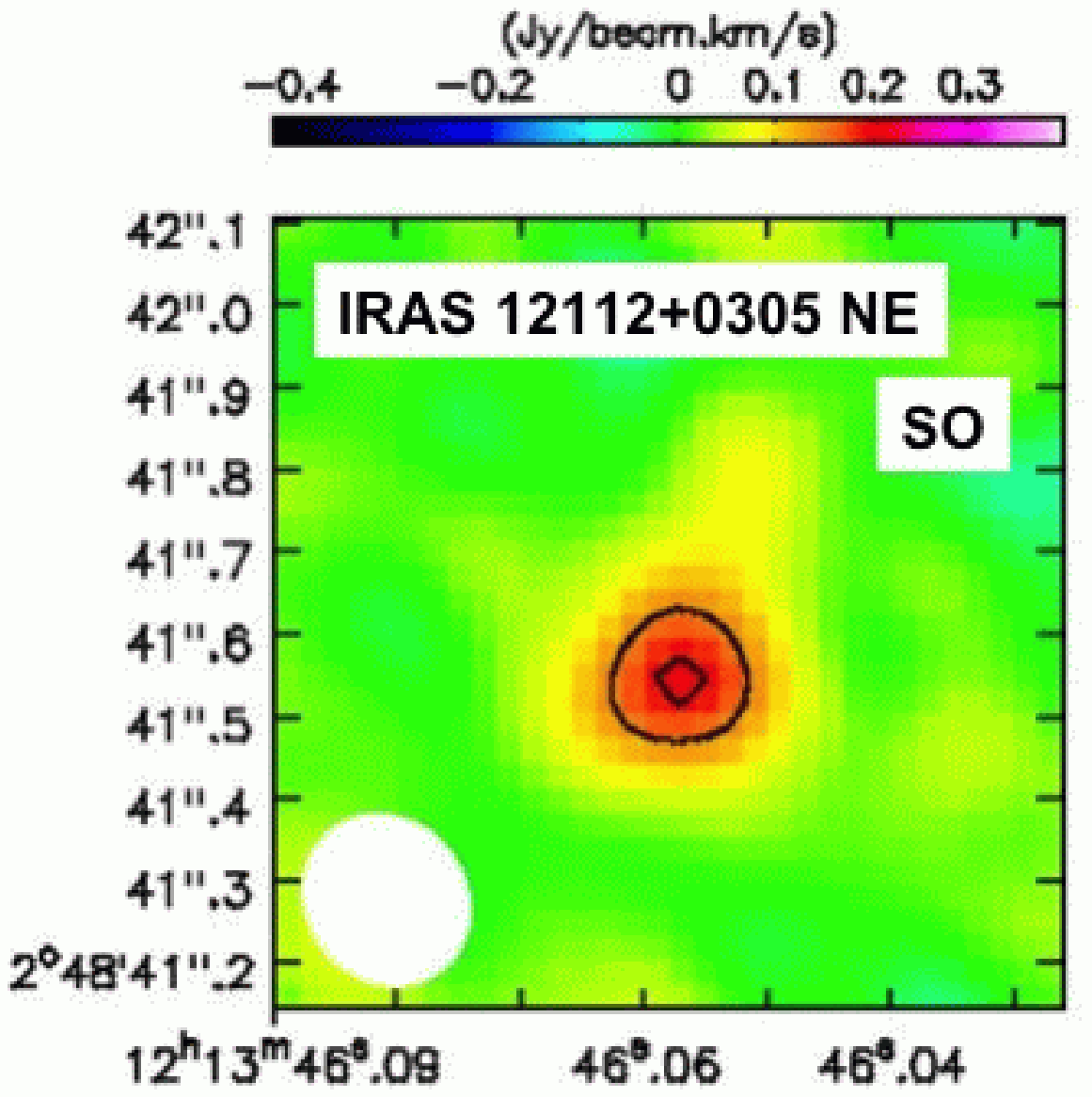} \\
\vspace{-1.3cm}
\includegraphics[angle=0,scale=.41]{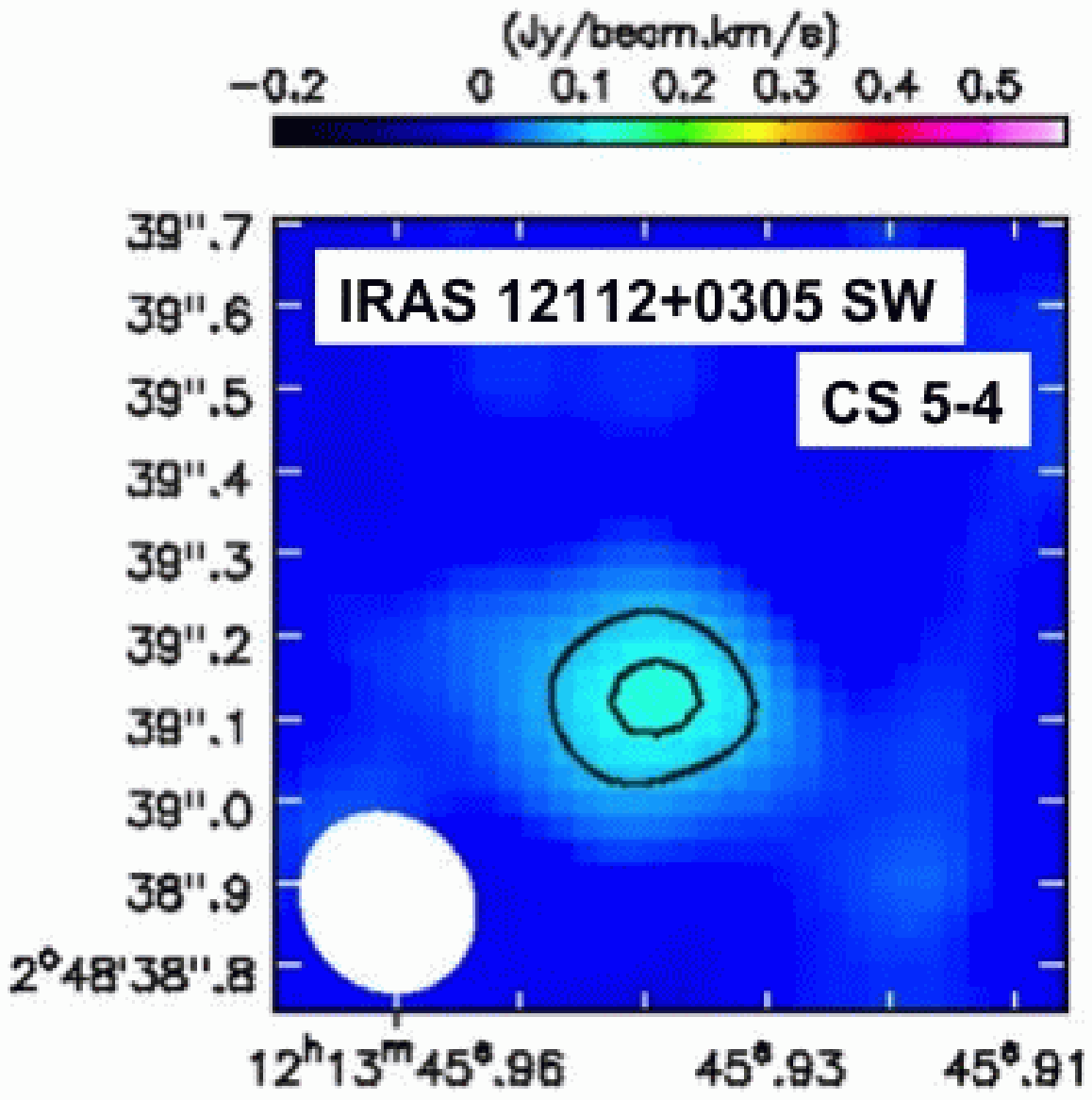} 
\includegraphics[angle=0,scale=.41]{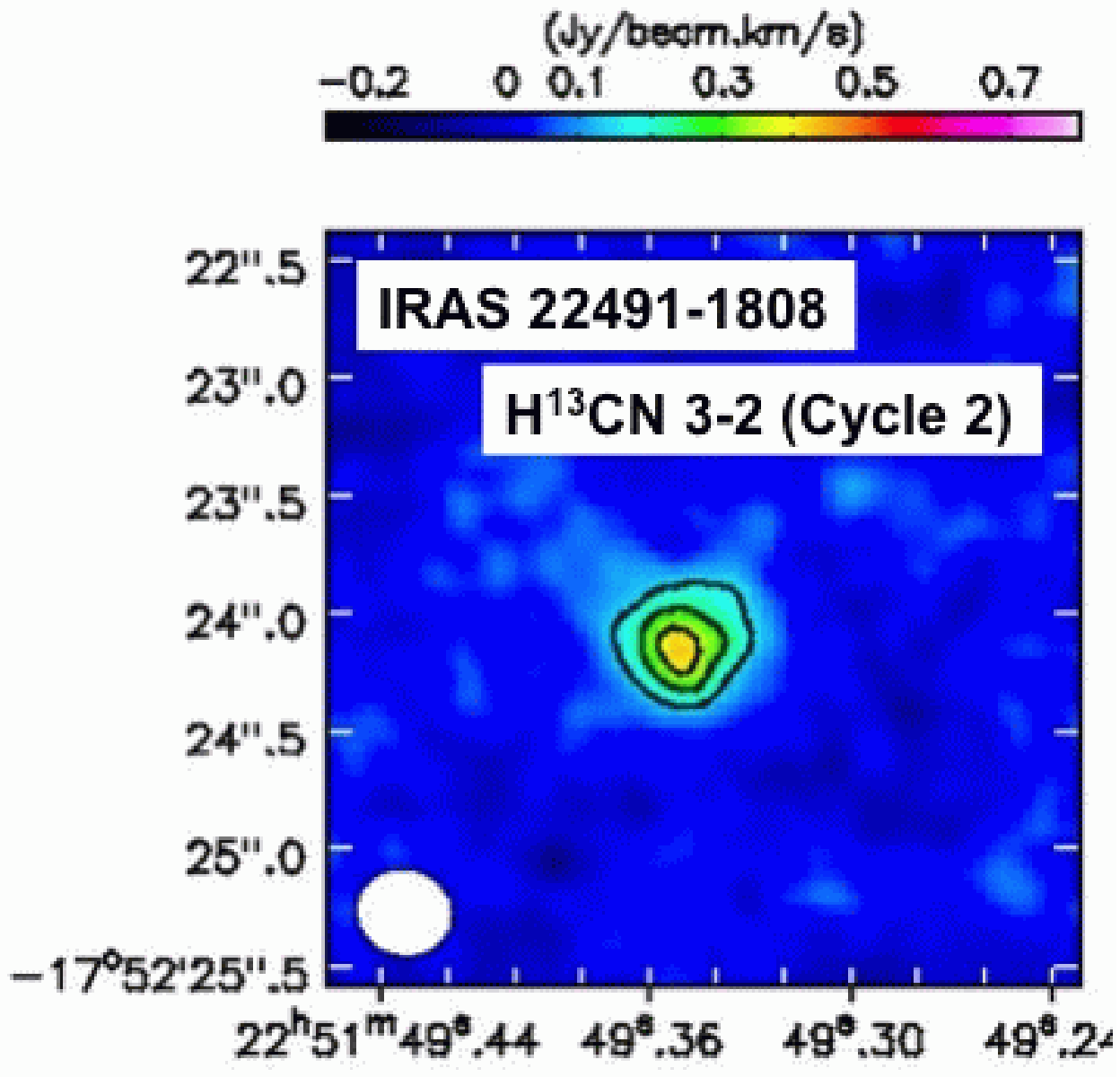} 
\includegraphics[angle=0,scale=.41]{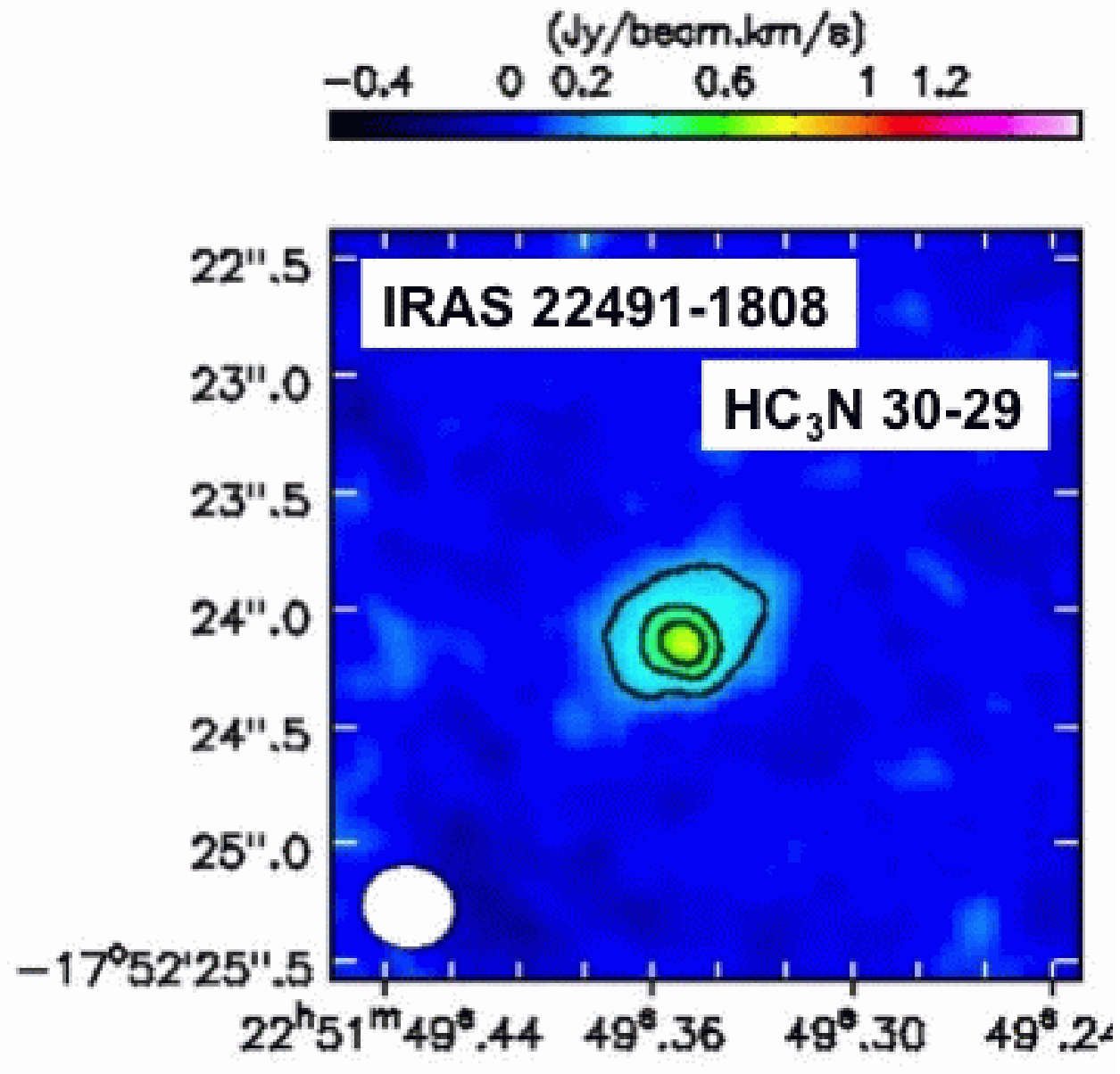} \\
\end{center}
\end{figure}

\clearpage

\begin{figure}
\begin{center}
\includegraphics[angle=0,scale=.41]{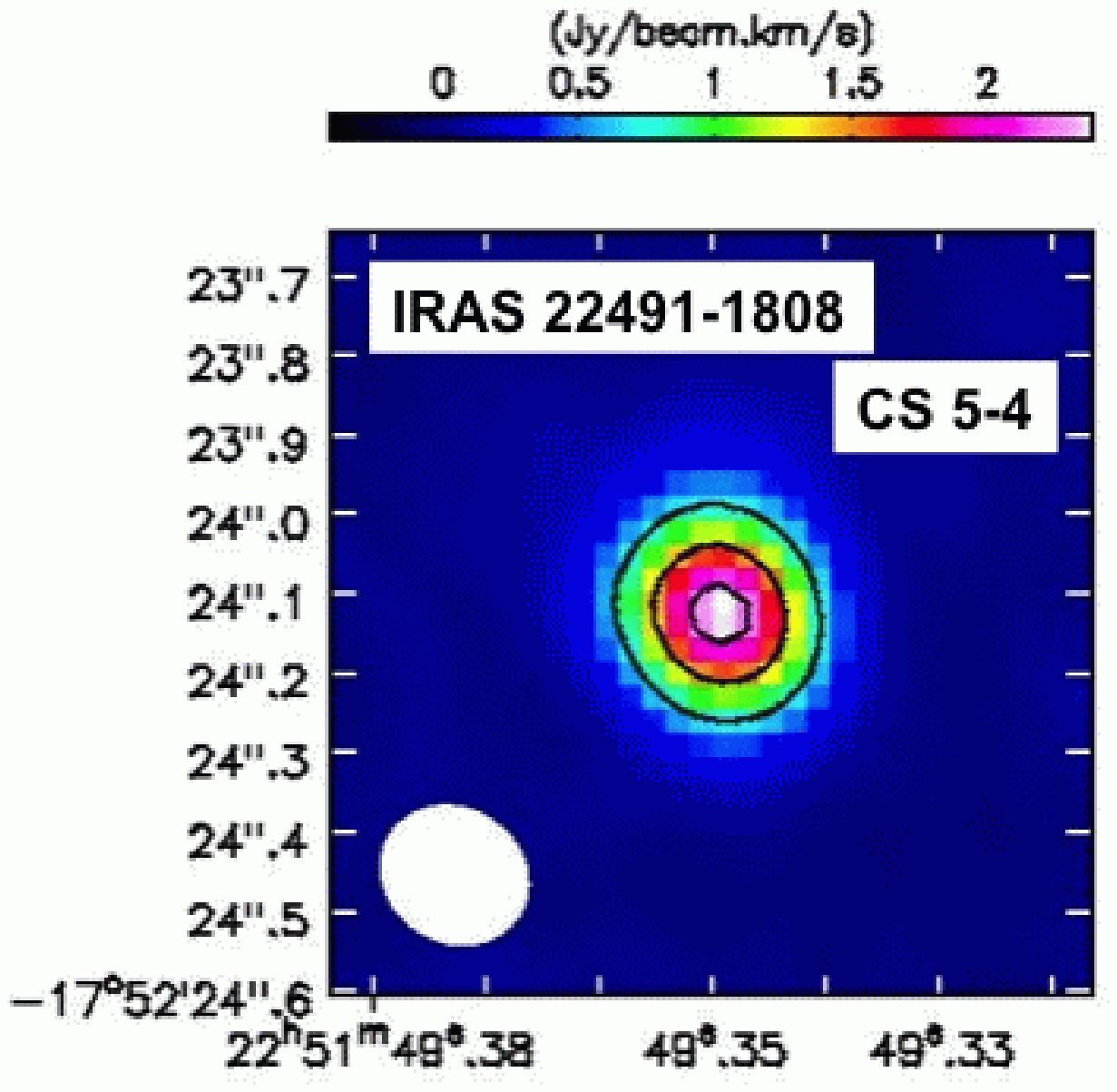} 
\includegraphics[angle=0,scale=.41]{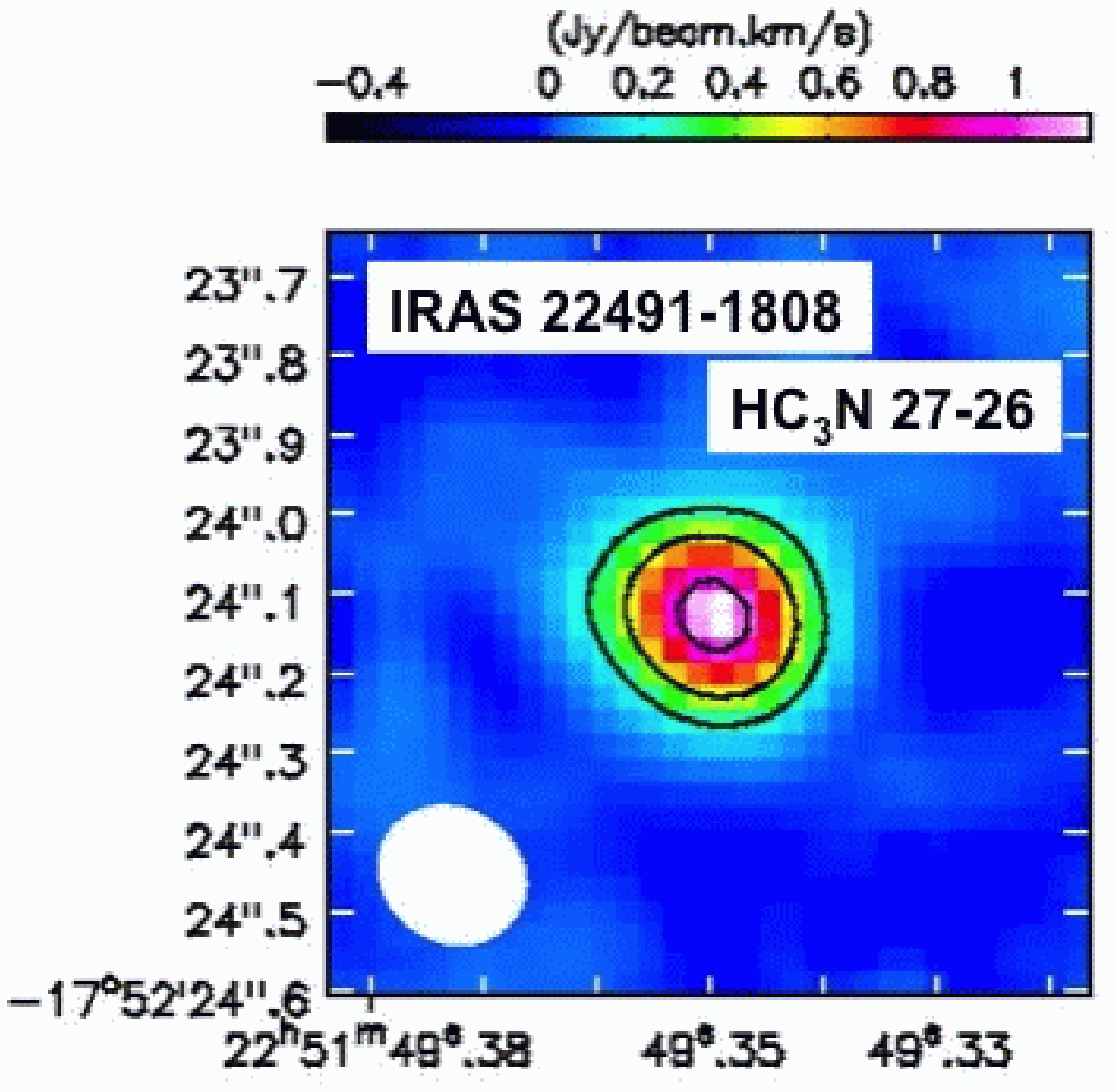} 
\includegraphics[angle=0,scale=.41]{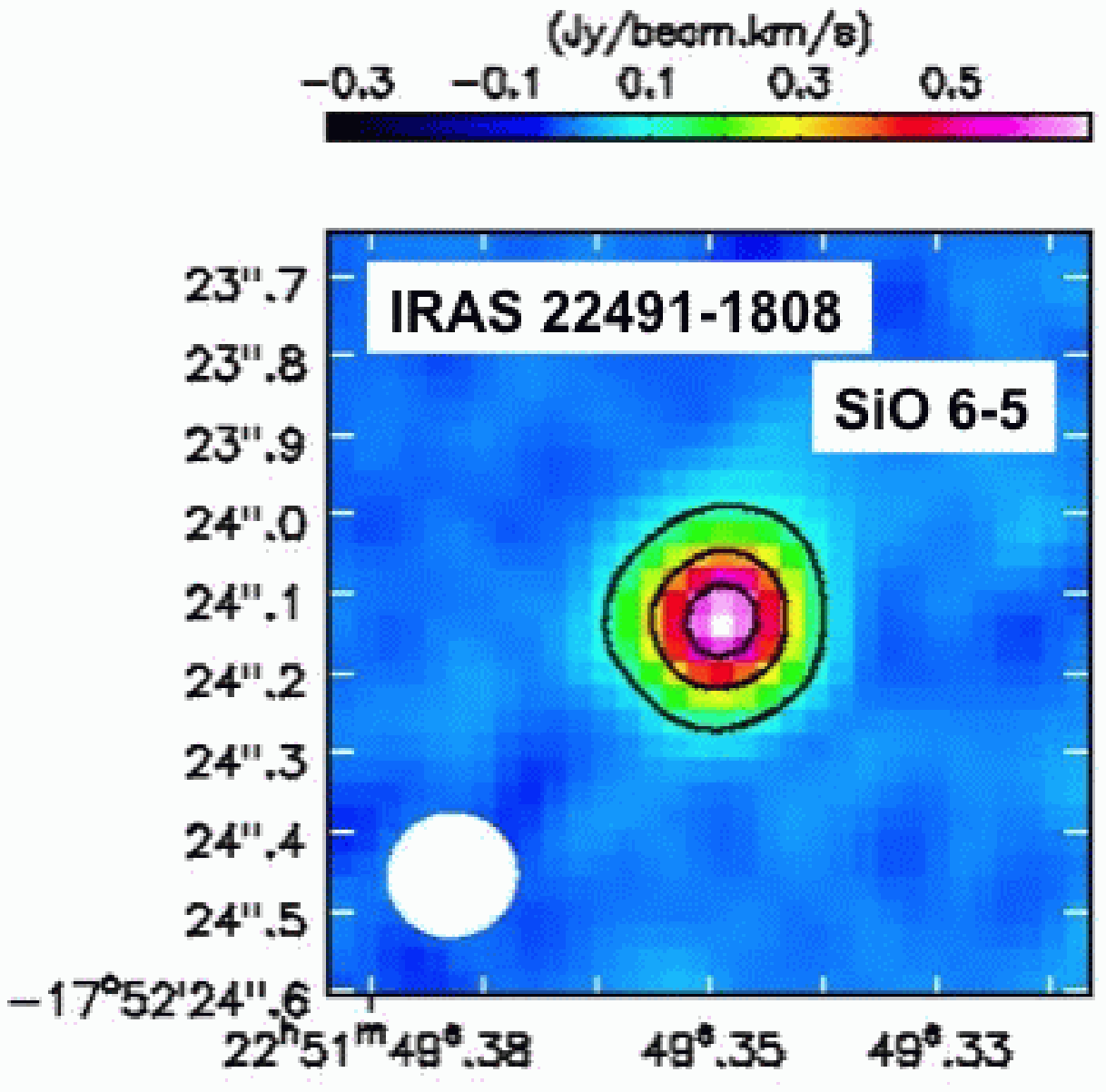} \\
\vspace{-1.3cm}
\includegraphics[angle=0,scale=.41]{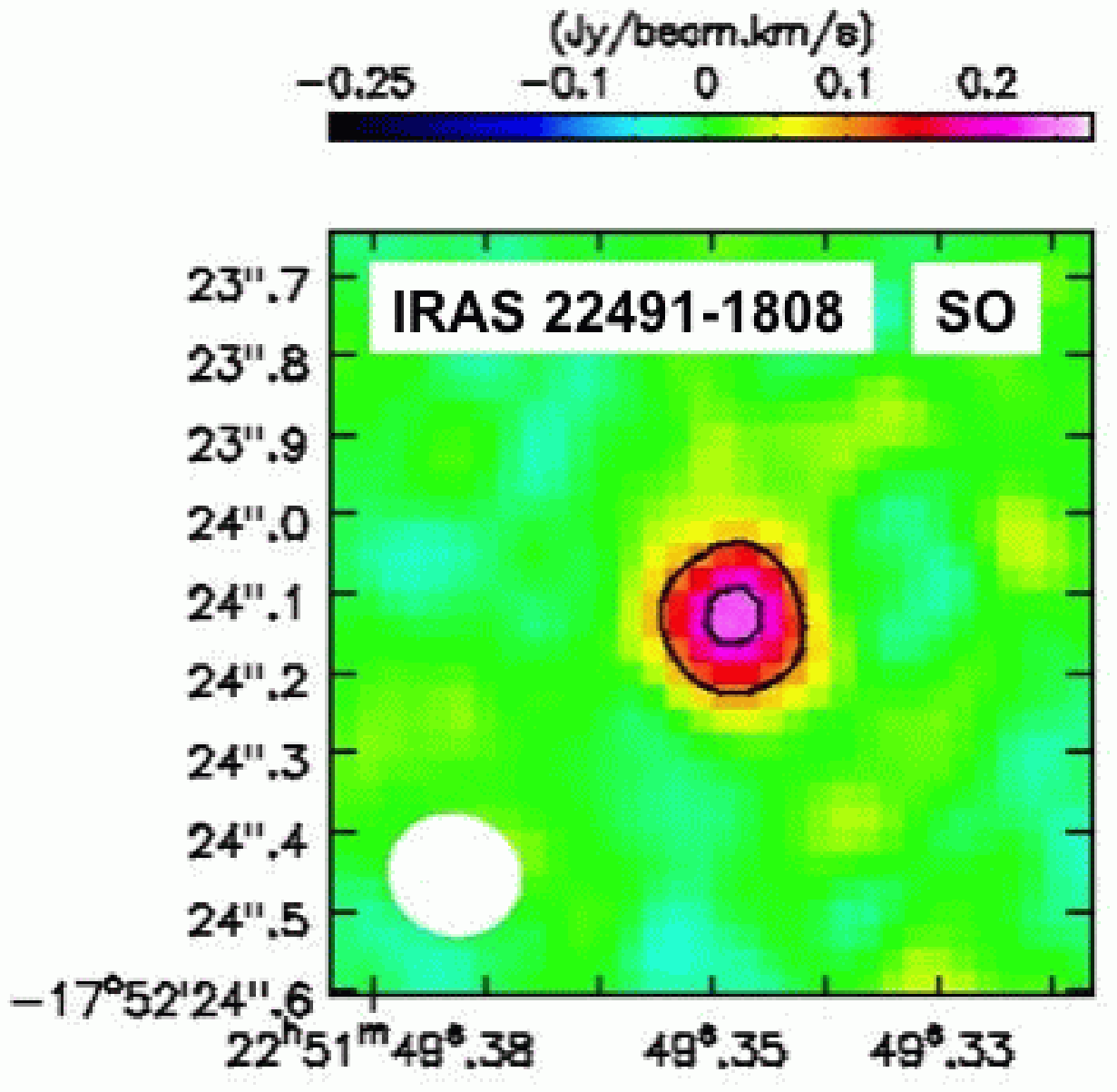} 
\includegraphics[angle=0,scale=.41]{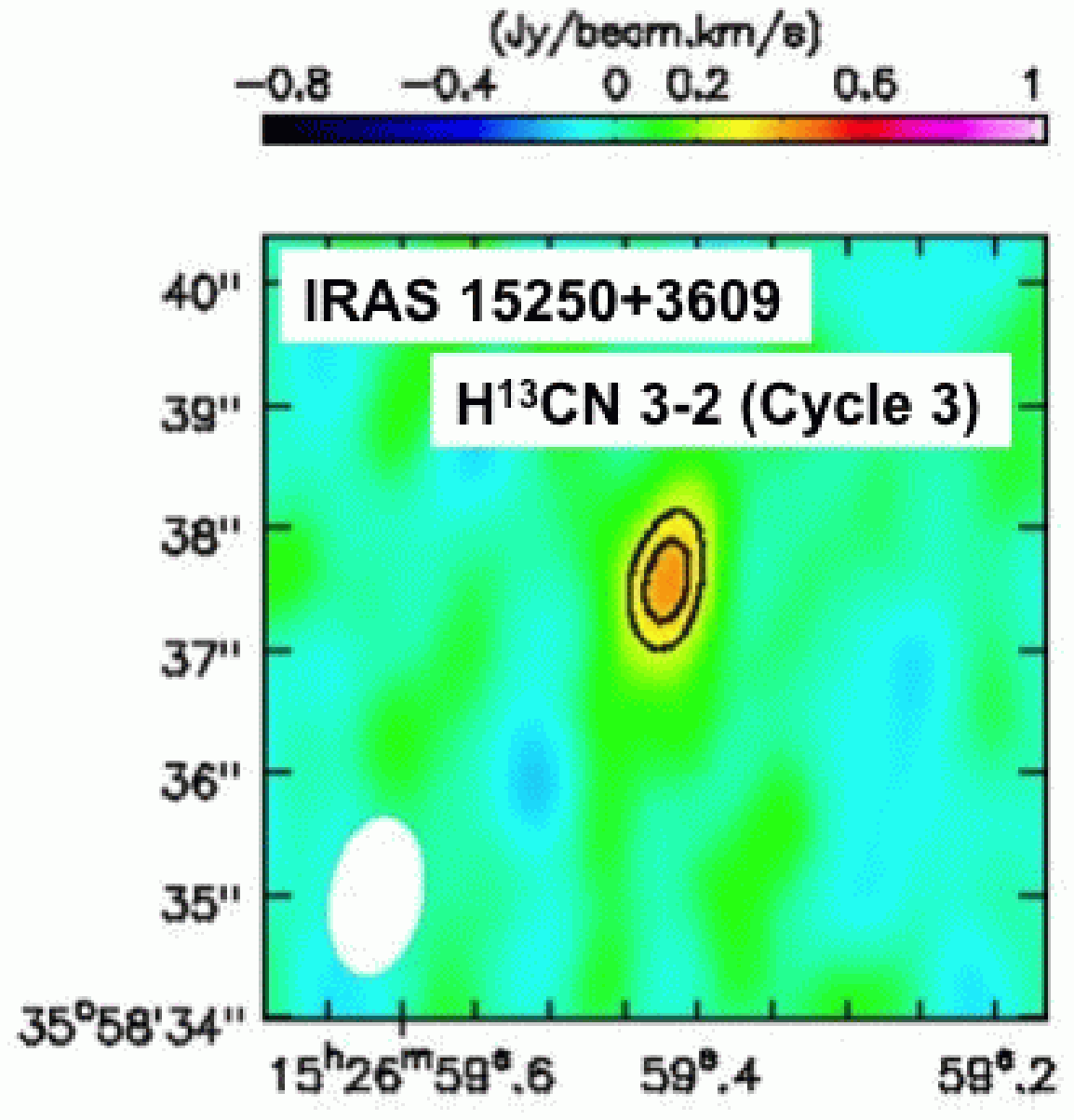} 
\includegraphics[angle=0,scale=.41]{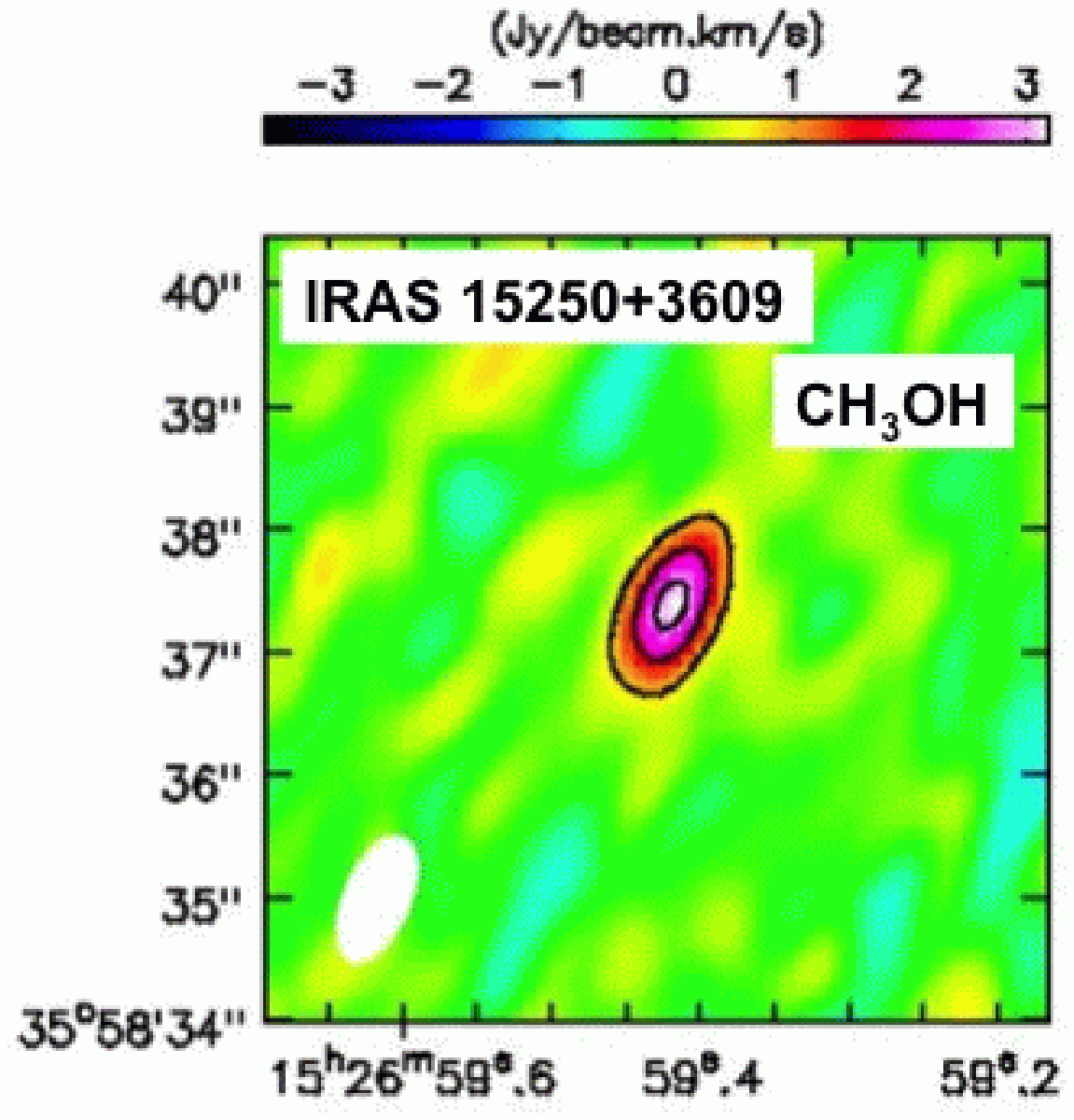} \\ 
\vspace{-1.3cm}
\includegraphics[angle=0,scale=.41]{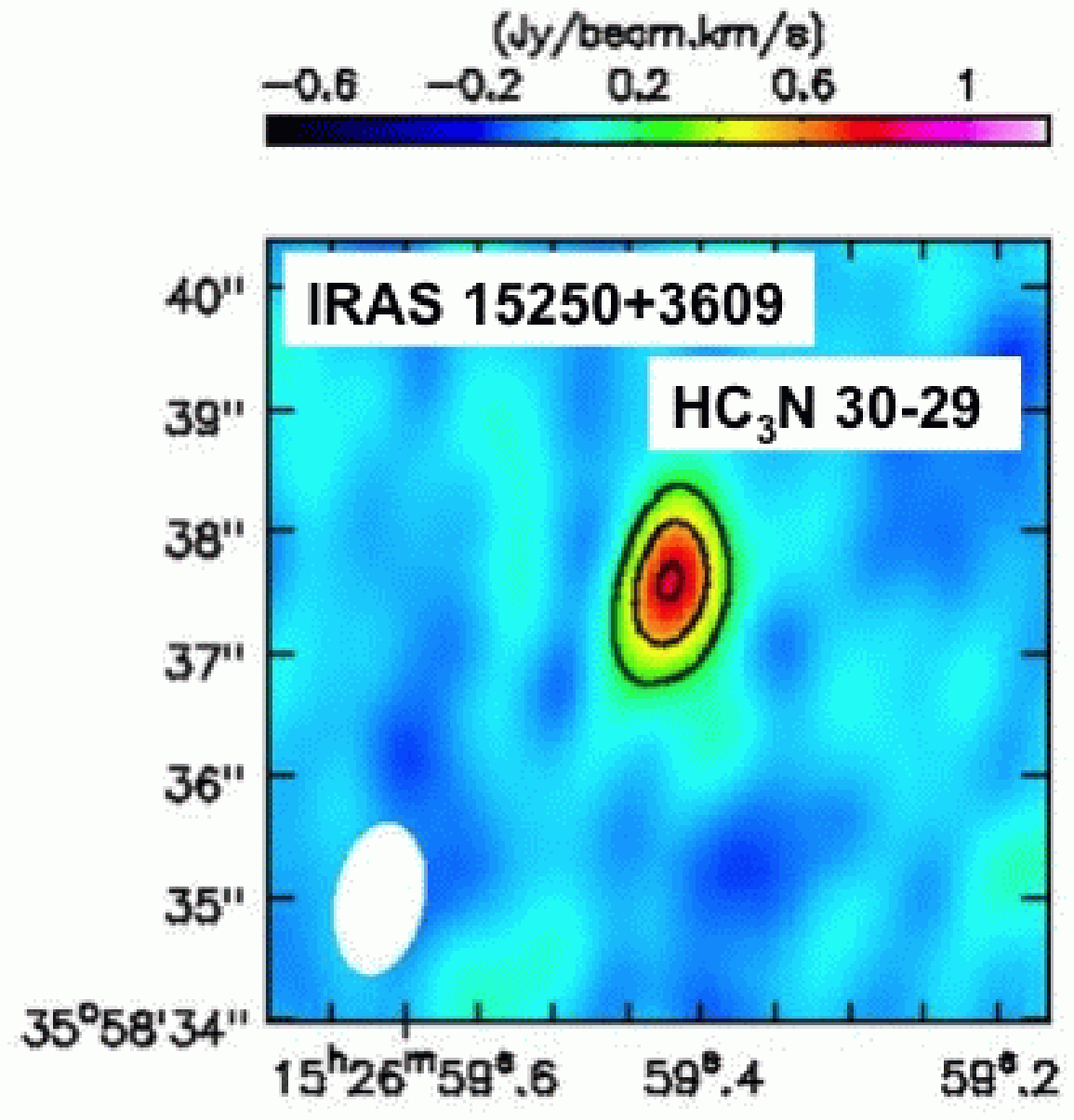} 
\includegraphics[angle=0,scale=.41]{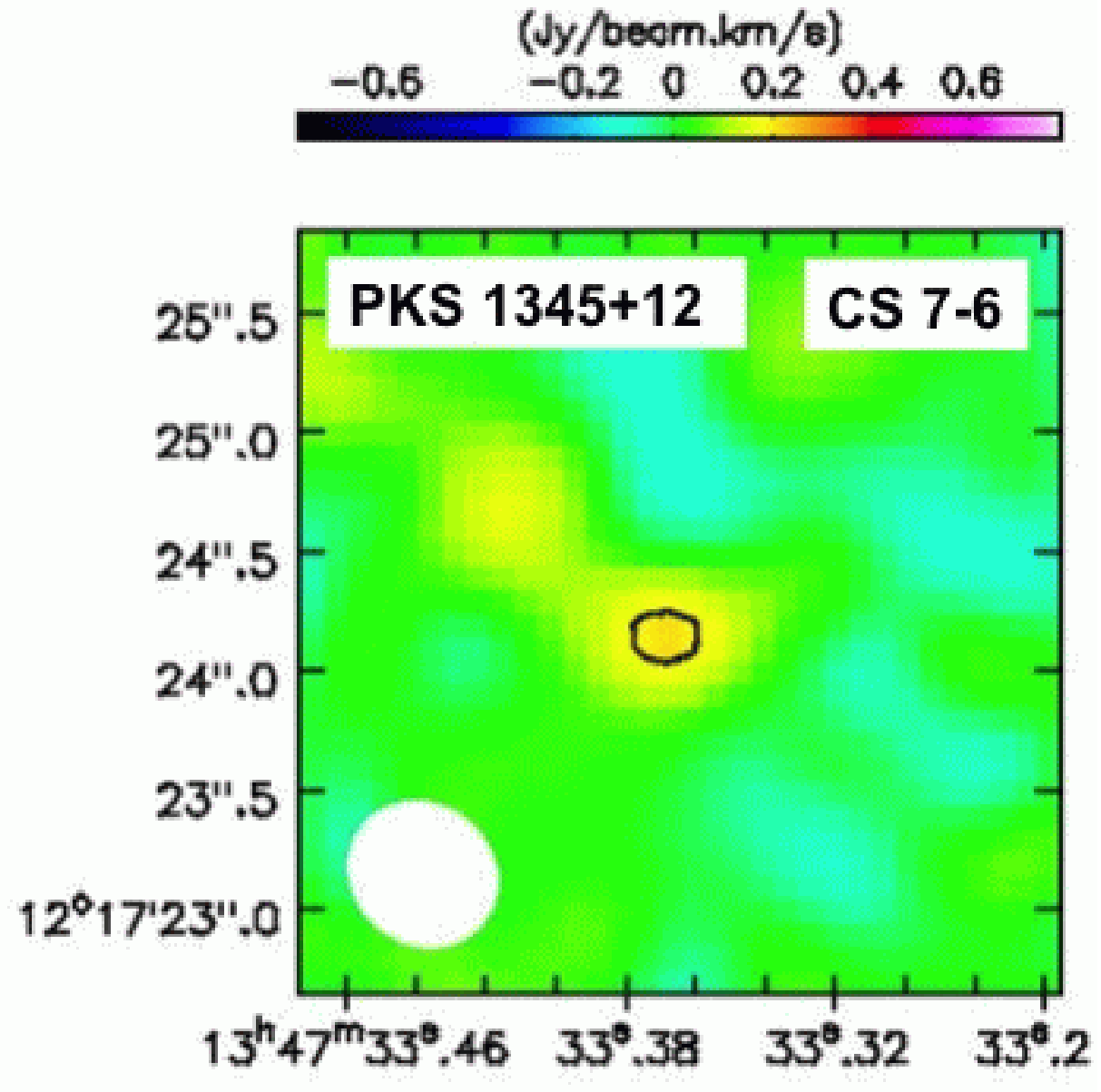} 
\includegraphics[angle=0,scale=.41]{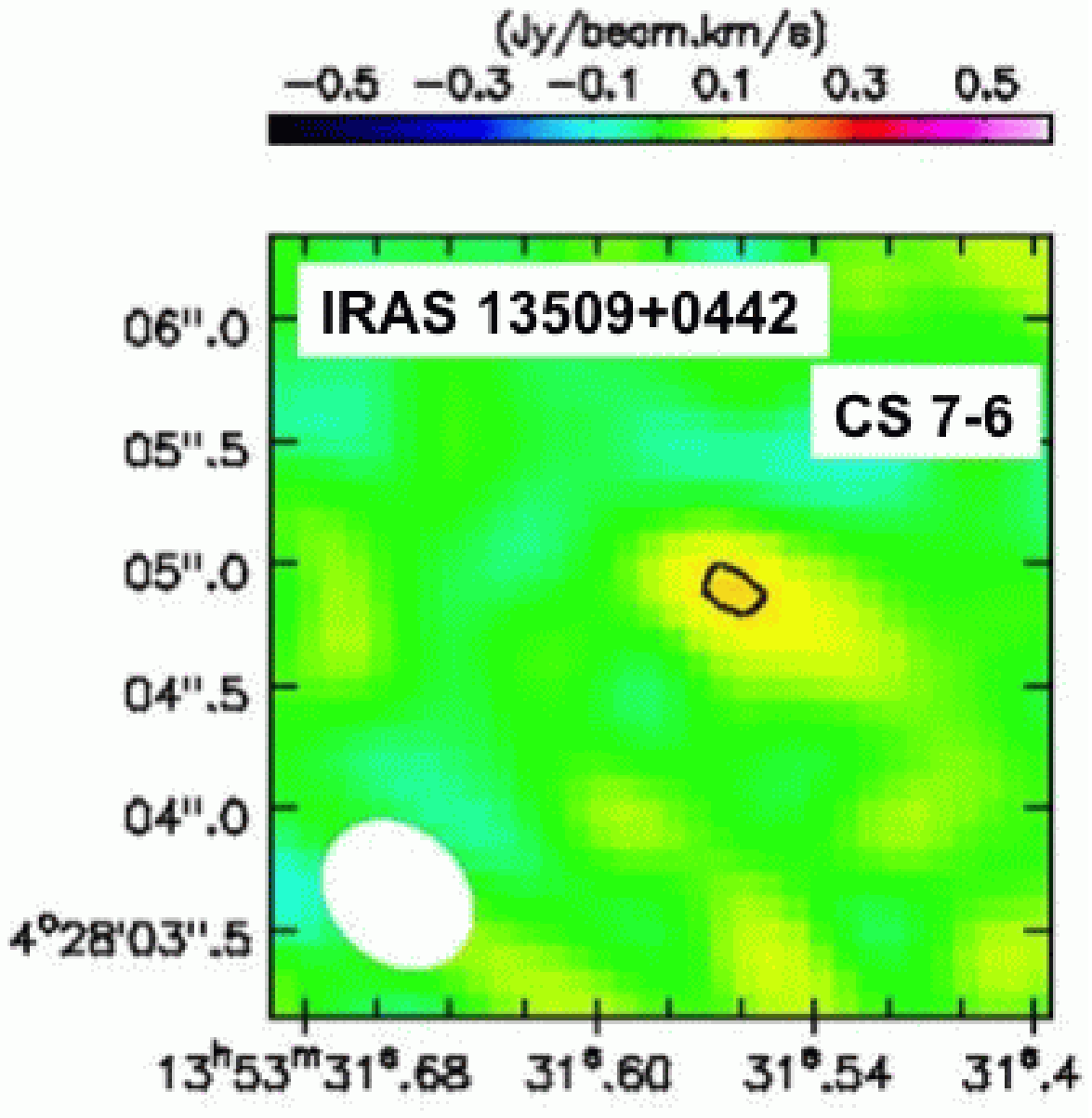} \\
\vspace{-1.3cm}
\includegraphics[angle=0,scale=.41]{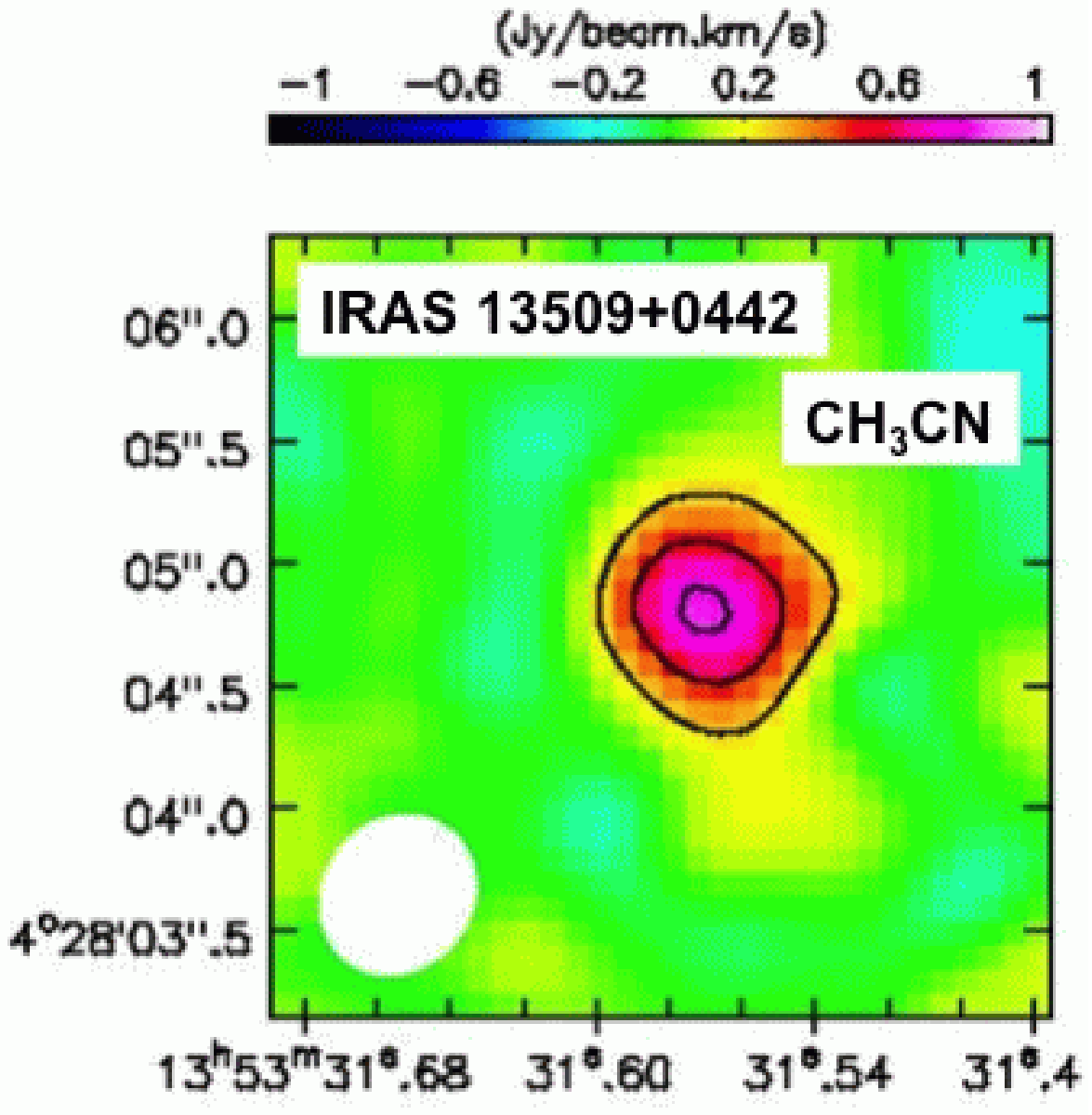} 
\includegraphics[angle=0,scale=.41]{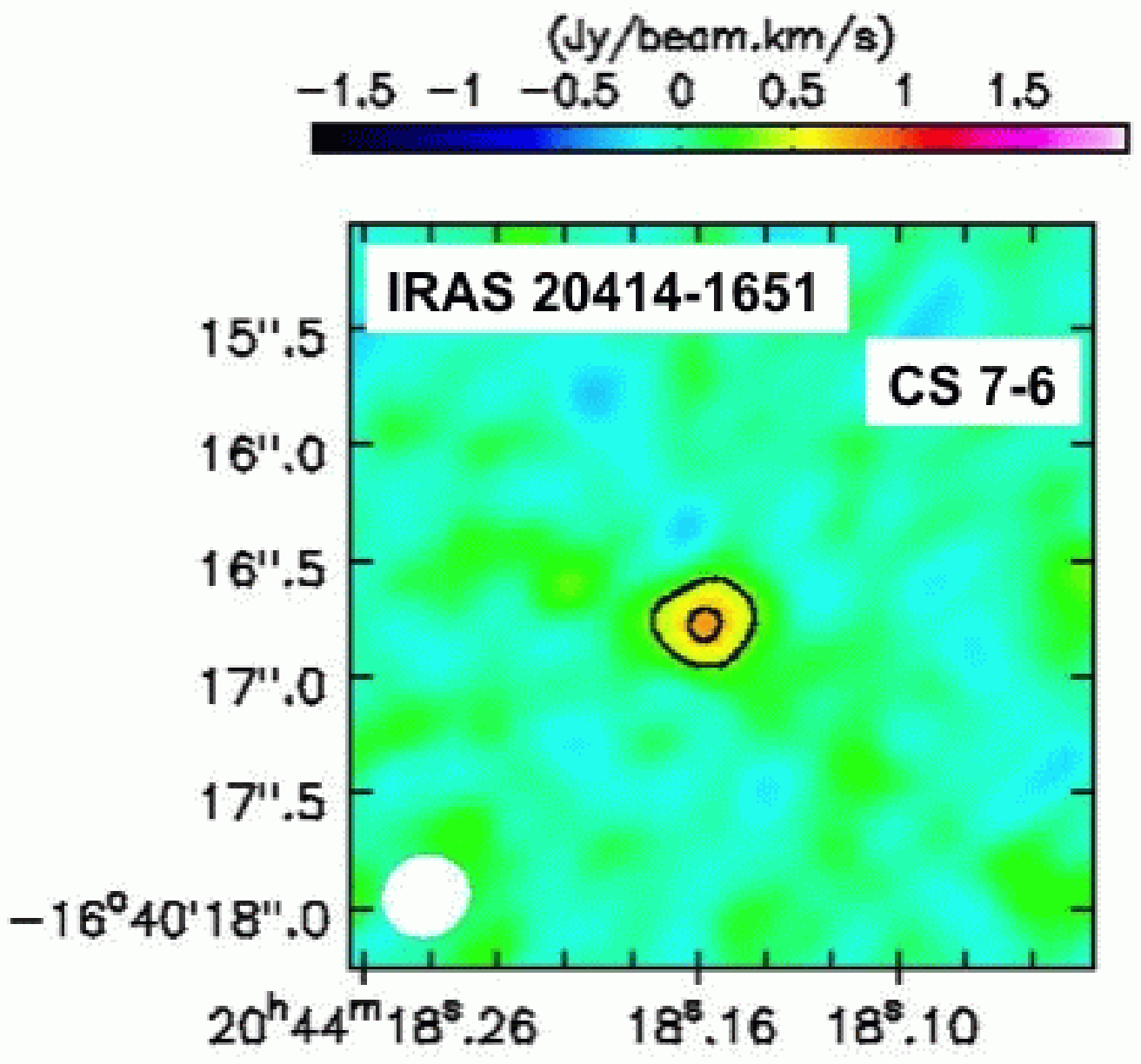} \\
\end{center}
\vspace{-1.0cm}
\end{figure}

\clearpage 

\begin{figure}
\caption{
Integrated intensity (moment 0) maps of serendipitously detected
molecular emission lines. 
The abscissa and ordinate are R.A. (J2000) and decl. (J2000),
respectively. 
The contours represent 
3$\sigma$, 6$\sigma$, 9$\sigma$ for IRAS 08572$+$3915 CS J=5--4, 
3$\sigma$ for Superantennae CH$_{3}$OH,
2.5$\sigma$, 3$\sigma$ for IRAS 12112$+$0305 NE H$^{13}$CN J=3--2 
(Cycle 2 data),
4$\sigma$, 8$\sigma$, 12$\sigma$ for IRAS 12112$+$0305 NE CH$_{3}$OH, 
3$\sigma$, 5$\sigma$ for IRAS 12112$+$0305 NE CS J=7--6,
6$\sigma$, 12$\sigma$, 24$\sigma$ for IRAS 12112$+$0305 NE CS J=5--4, 
4$\sigma$, 8$\sigma$, 12$\sigma$ for IRAS 12112$+$0305 NE HC$_{3}$N J=27--26,
3$\sigma$, 5$\sigma$, 7$\sigma$ for IRAS 12112$+$0305 NE SiO J=6--5,
3$\sigma$, 4$\sigma$ for IRAS 12112$+$0305 NE SO, 
3$\sigma$, 4.5$\sigma$ for IRAS 12112$+$0305 SW CS J=5--4, 
3$\sigma$, 5$\sigma$, 7$\sigma$ for IRAS 22491$-$1808 H$^{13}$CN J=3--2 (Cycle 2 data), 
3$\sigma$, 5$\sigma$, 7$\sigma$ for IRAS 22491$-$1808 HC$_{3}$N J=30--29,
15$\sigma$, 30$\sigma$, 45$\sigma$ for IRAS 22491$-$1808 CS J=5--4, 
5$\sigma$, 10$\sigma$, 20$\sigma$ for IRAS 22491$-$1808 HC$_{3}$N J=27--26, 
3$\sigma$, 9$\sigma$, 15$\sigma$ for IRAS 22491$-$1808 SiO J=6--5,
3$\sigma$, 6$\sigma$ for IRAS 22491$-$1808 SO, 
3$\sigma$, 4$\sigma$ for IRAS 15250$+$3609 H$^{13}$CN J=3--2 (Cycle 3 data), 
3$\sigma$, 7$\sigma$, 11$\sigma$ for IRAS 15250$+$3609 CH$_{3}$OH, 
3$\sigma$, 6$\sigma$, 9$\sigma$ for IRAS 15250$+$3609 HC$_{3}$N J=30--29, 
3$\sigma$ for PKS 1345$+$12 CS J=7--6, 
3$\sigma$ for IRAS 13509$+$0442 CS J=7--6,
3$\sigma$, 6$\sigma$, 9$\sigma$ for IRAS 13509$+$0442 CH$_{3}$CN,
and 3$\sigma$, 6$\sigma$ for IRAS 20414$-$1651 CS J=7--6.
Beam sizes are shown as filled circles in the lower-left region.
}
\end{figure}

\begin{figure}
\begin{center}
\includegraphics[angle=0,scale=.273]{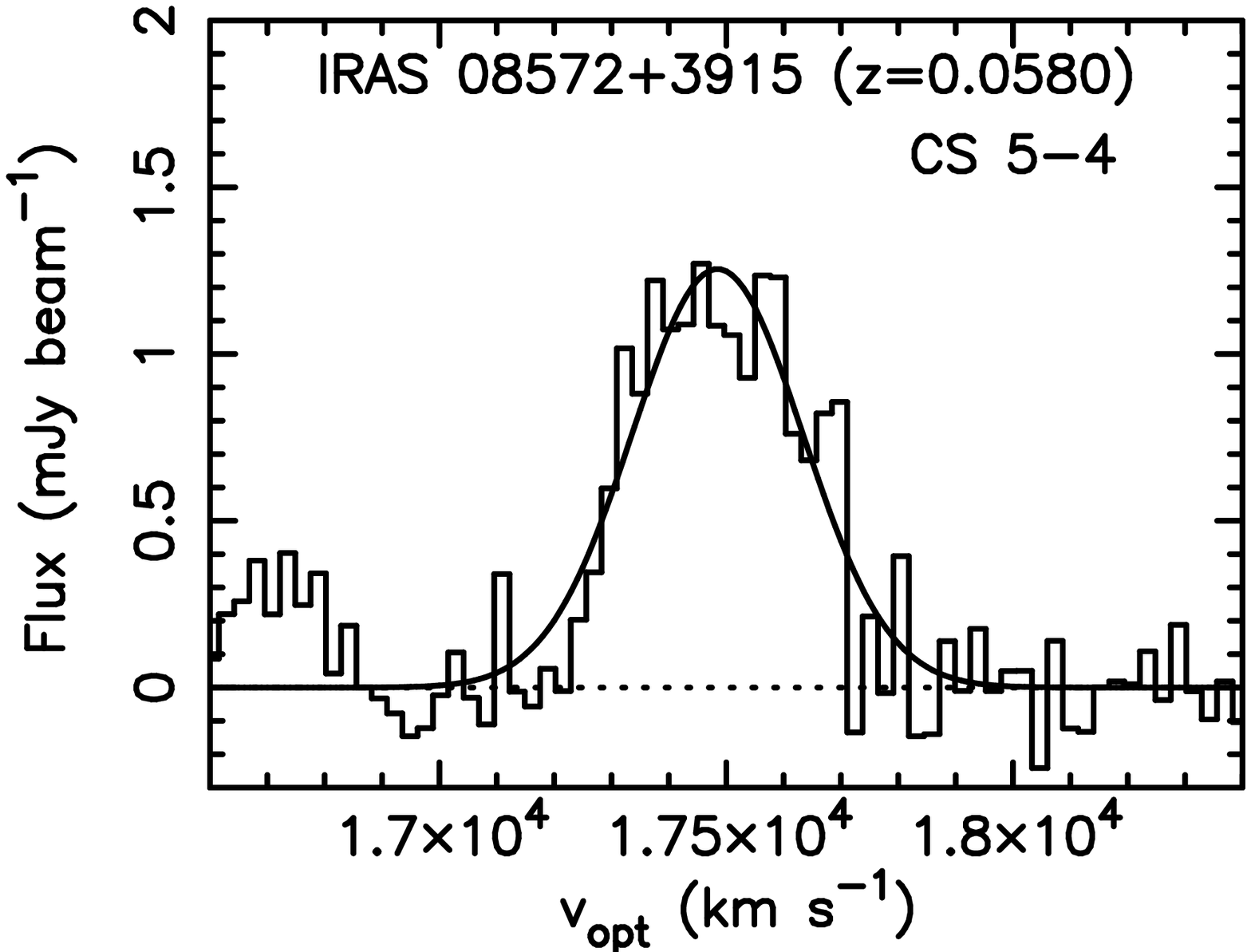}
\includegraphics[angle=0,scale=.273]{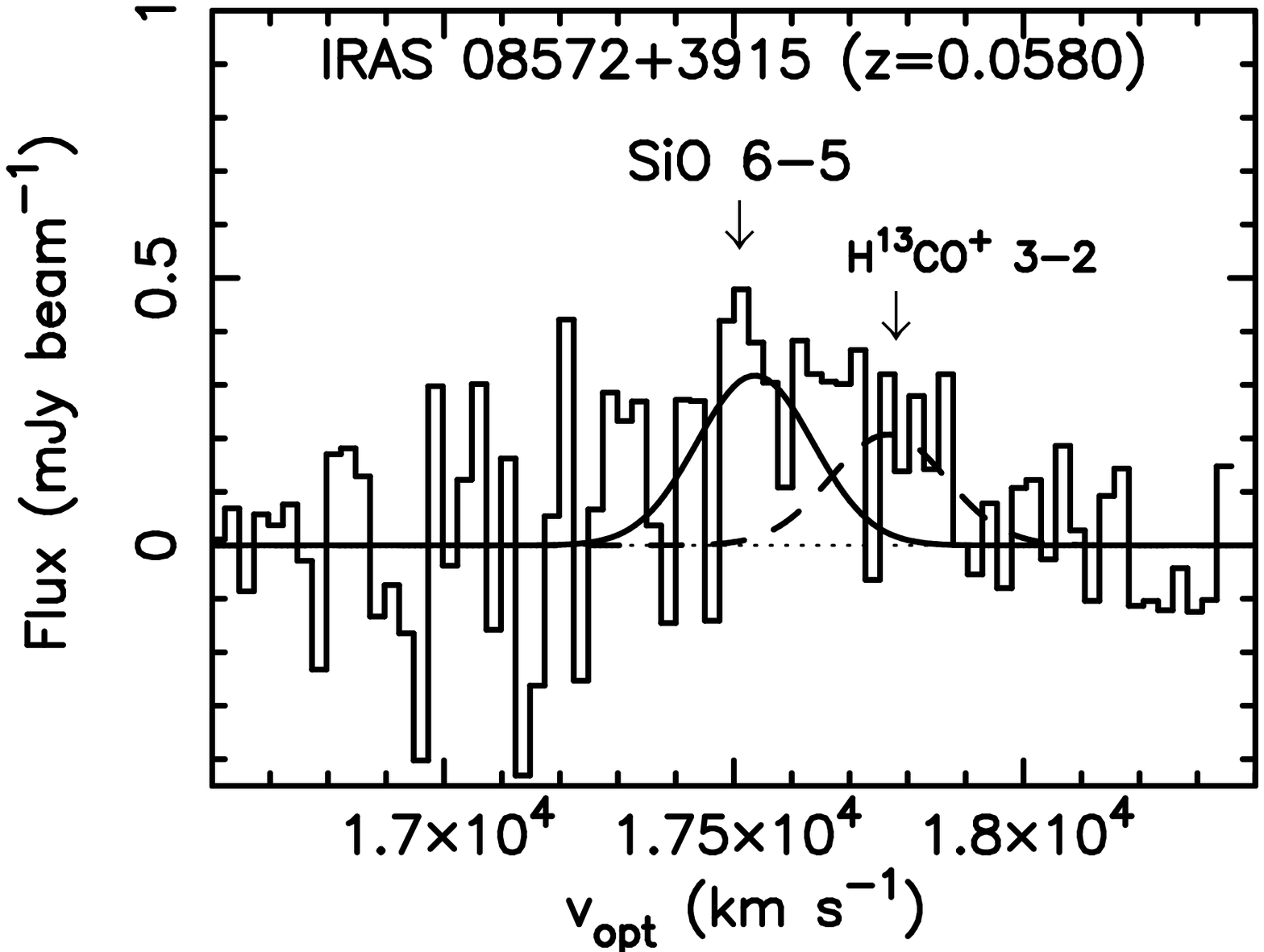} 
\includegraphics[angle=0,scale=.273]{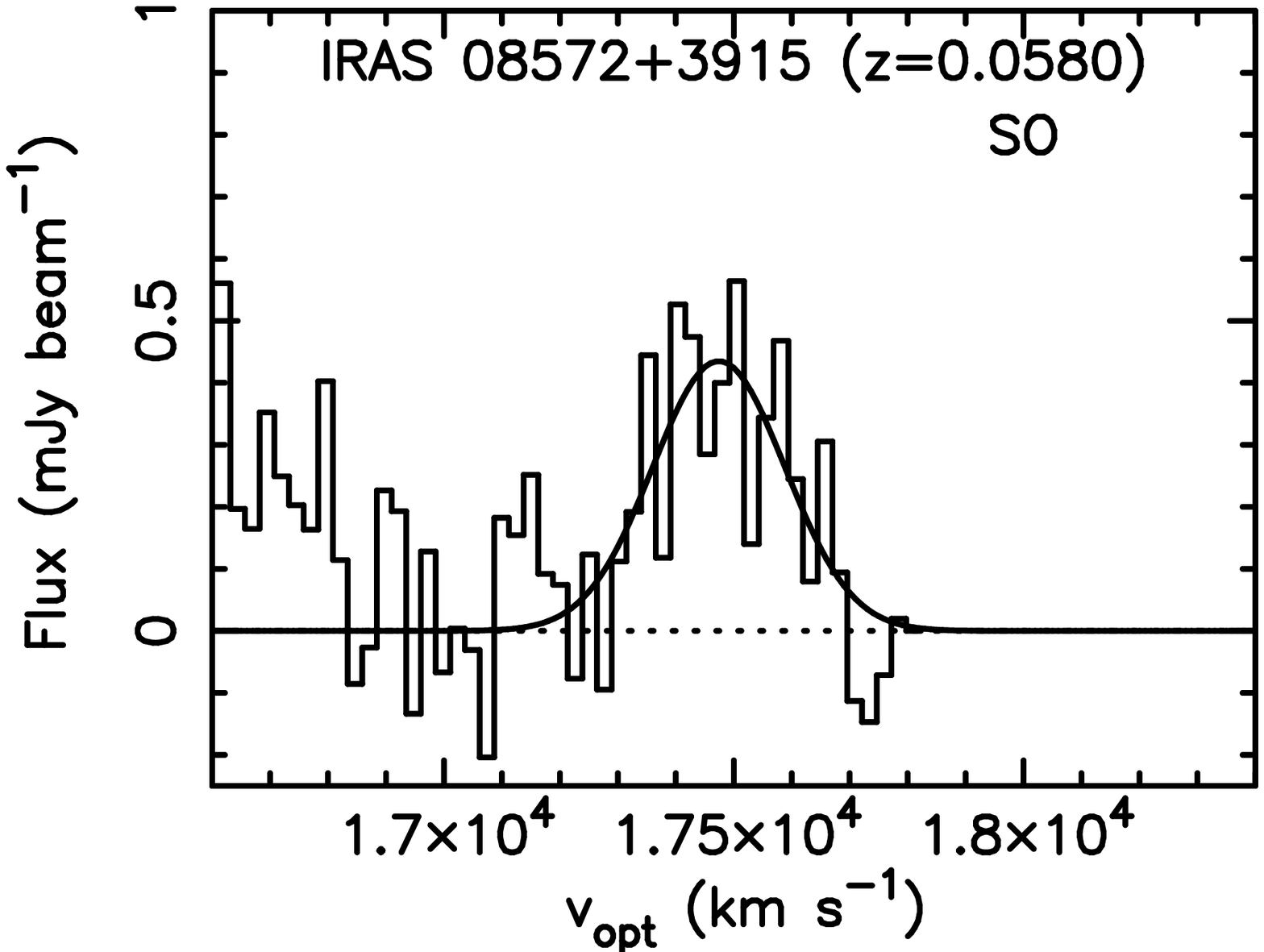} \\
\includegraphics[angle=0,scale=.273]{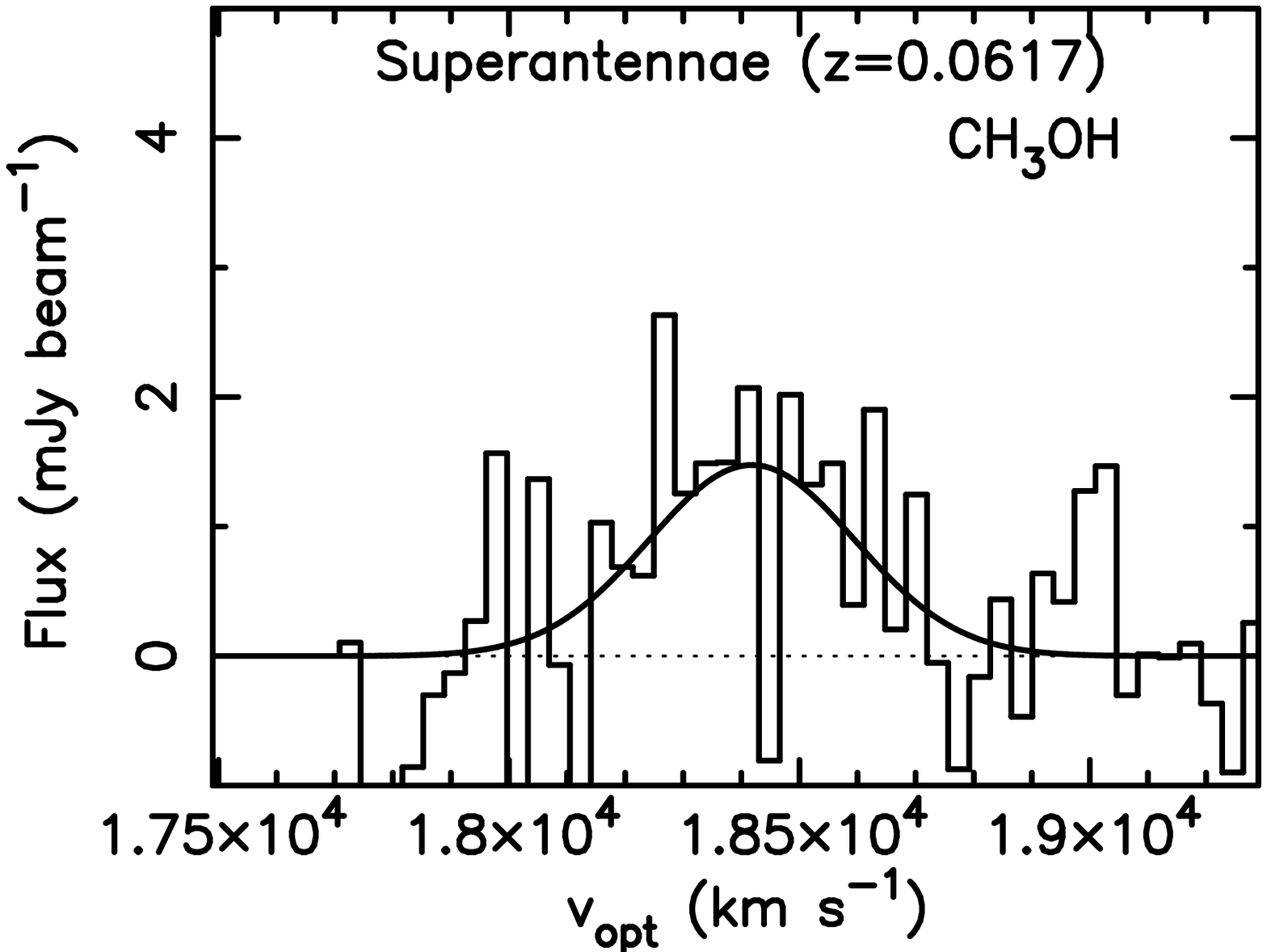} 
\includegraphics[angle=0,scale=.273]{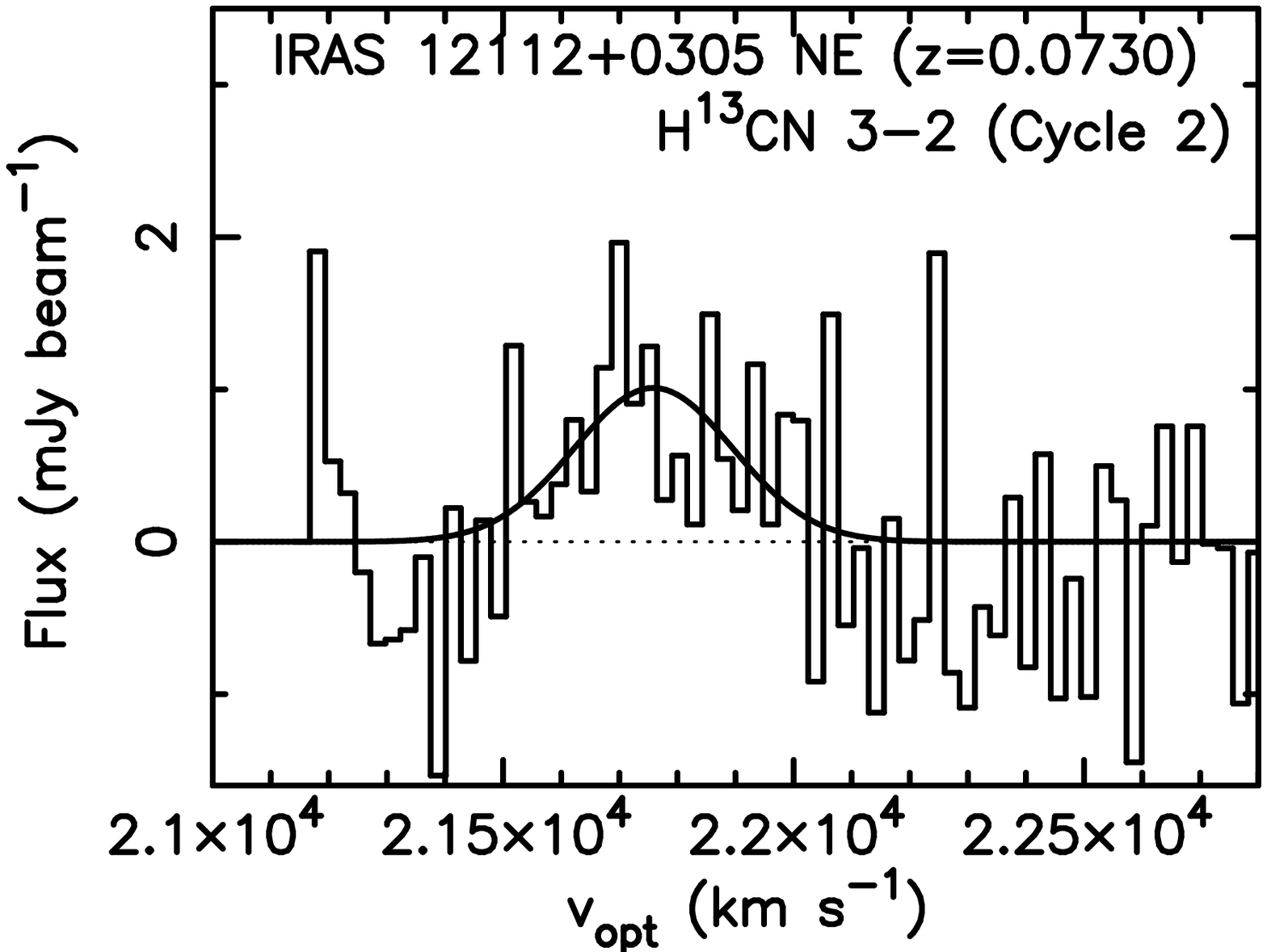} 
\includegraphics[angle=0,scale=.273]{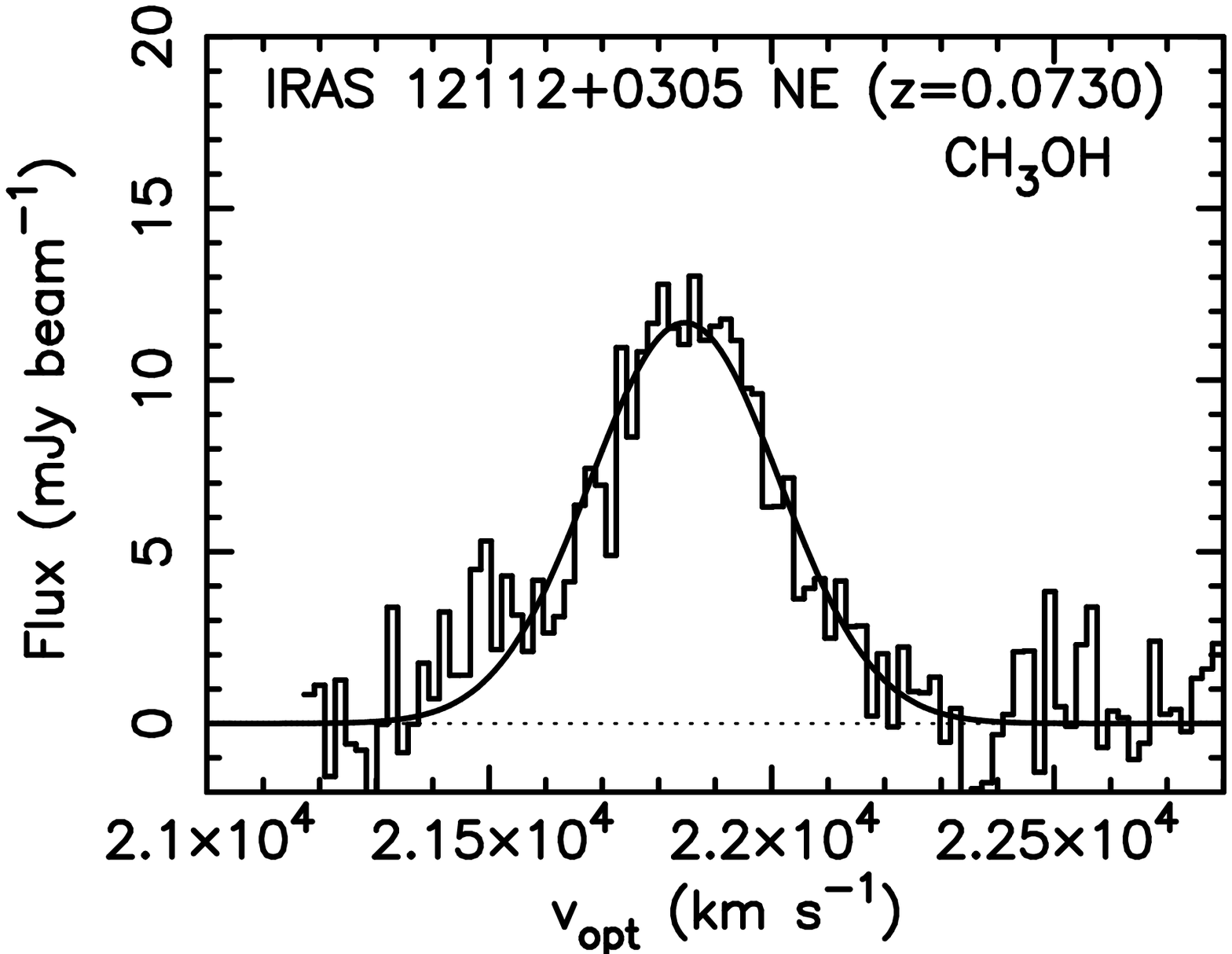} \\
\includegraphics[angle=0,scale=.273]{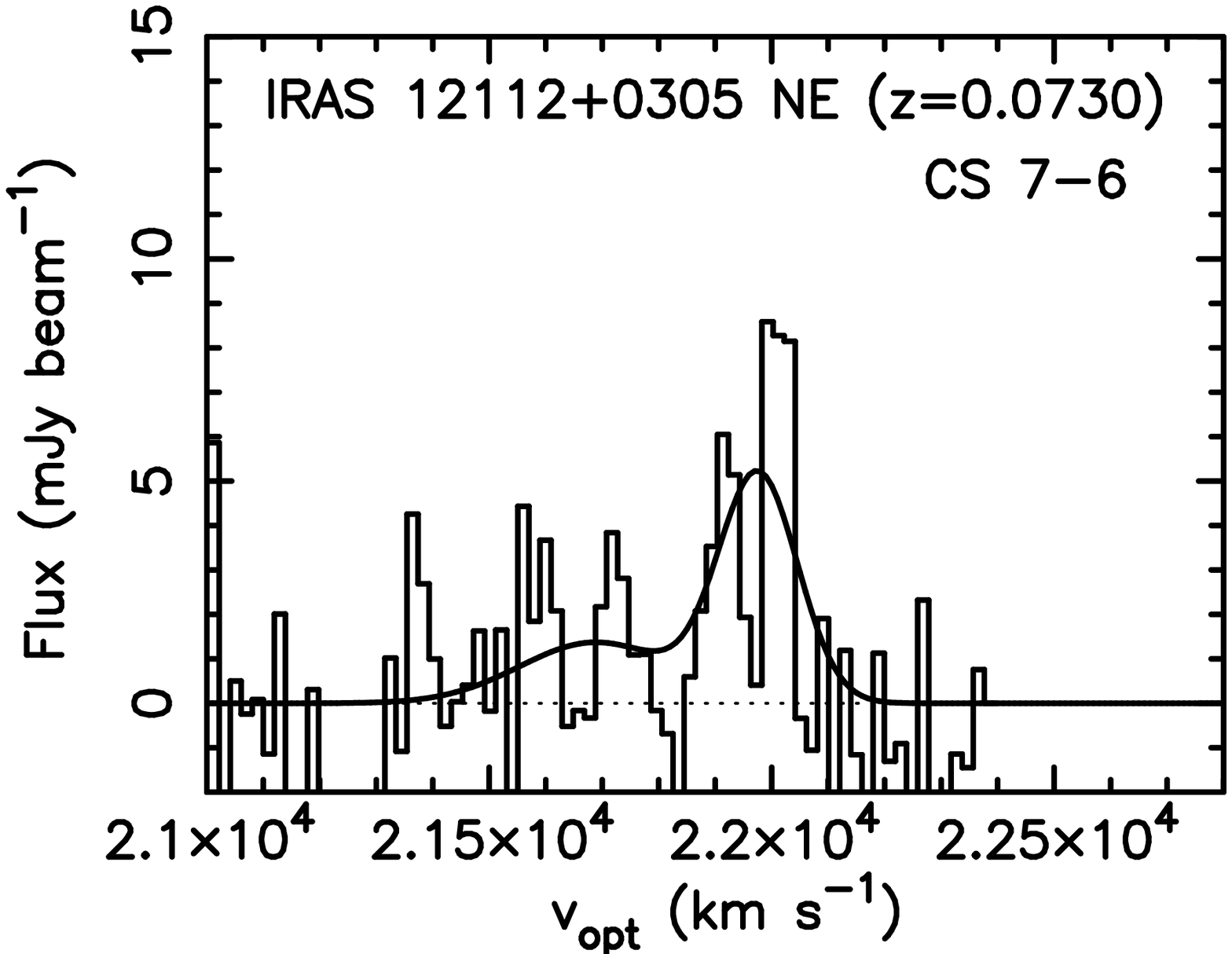} 
\includegraphics[angle=0,scale=.273]{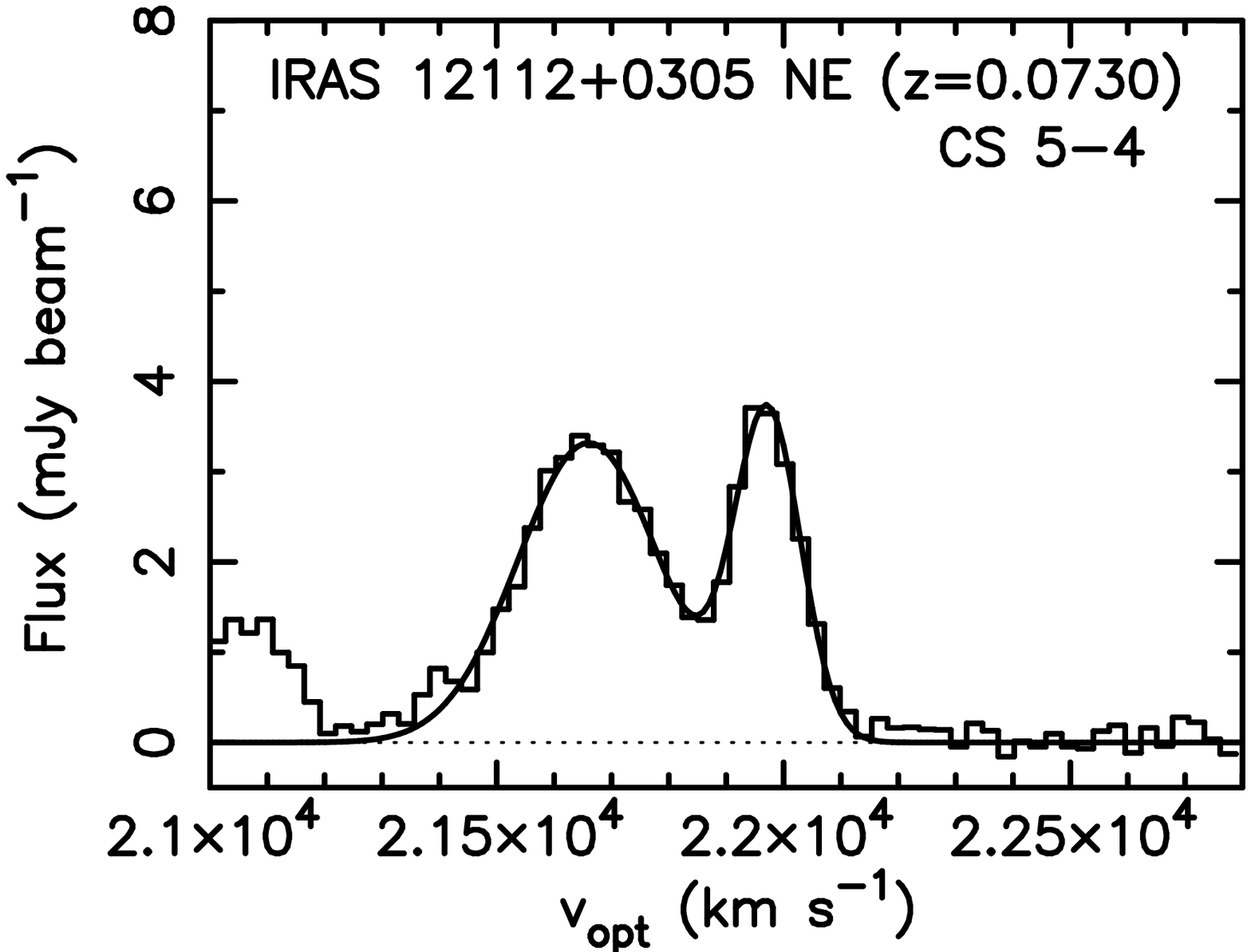} 
\includegraphics[angle=0,scale=.273]{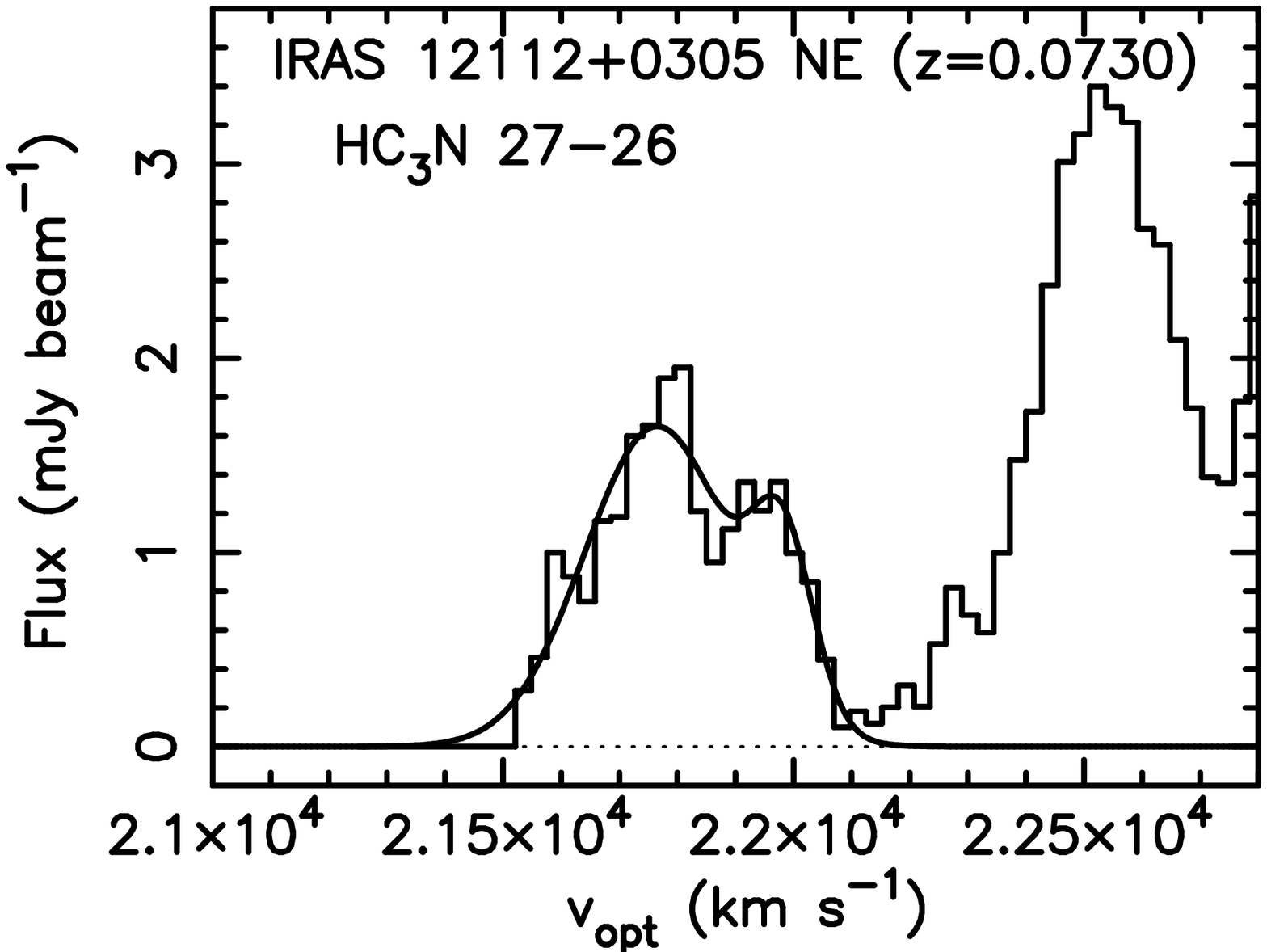} \\
\includegraphics[angle=0,scale=.273]{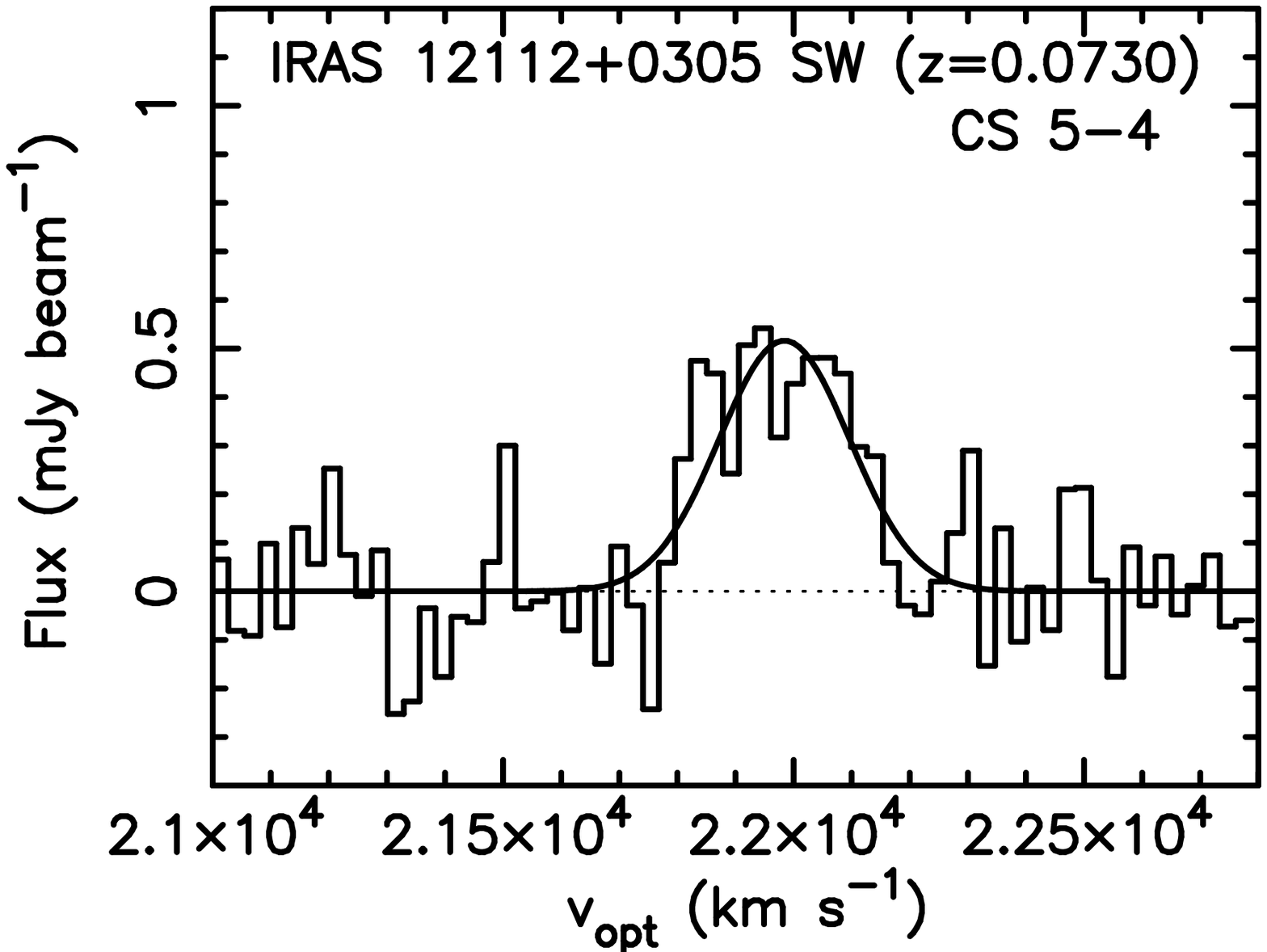} 
\includegraphics[angle=0,scale=.273]{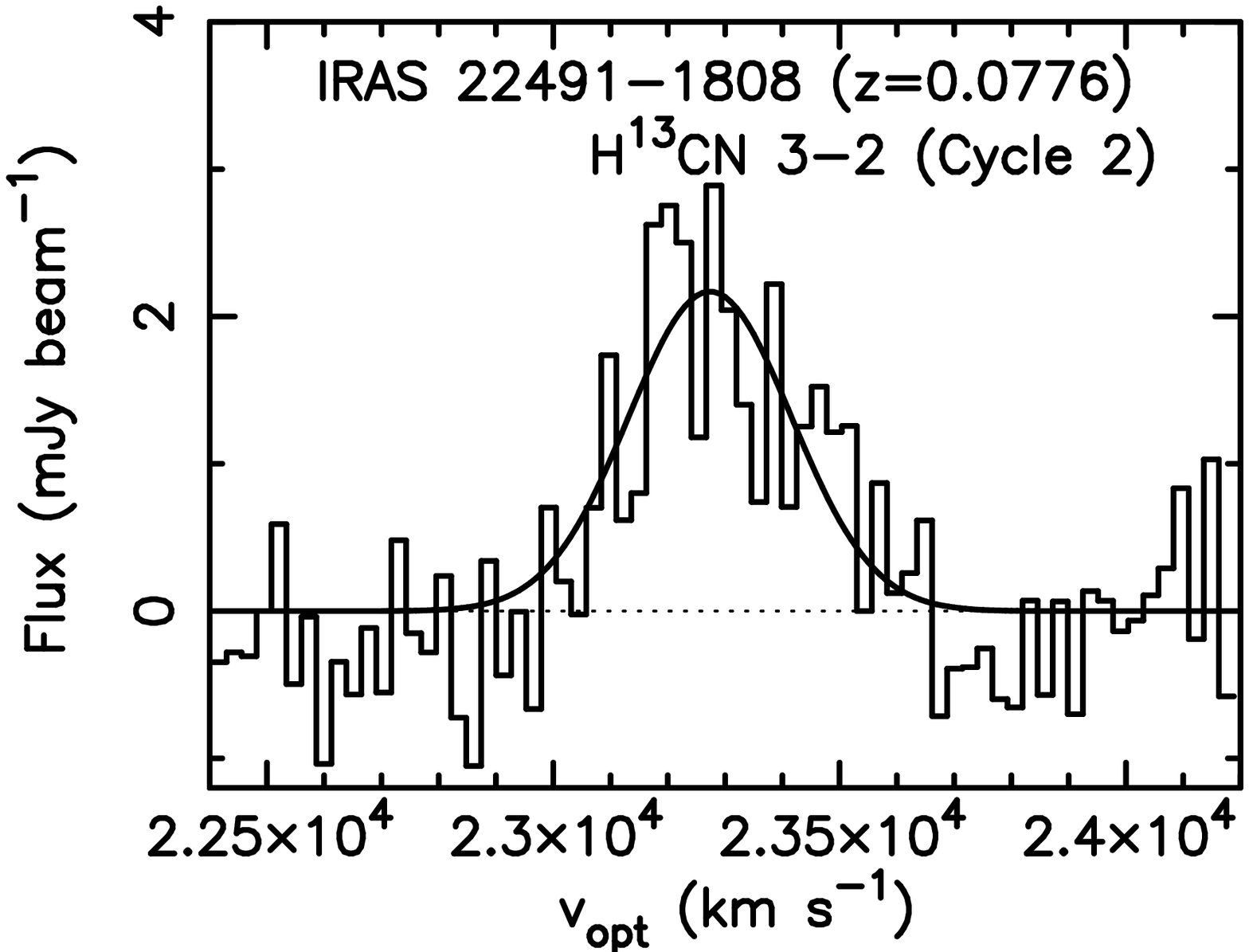} 
\includegraphics[angle=0,scale=.273]{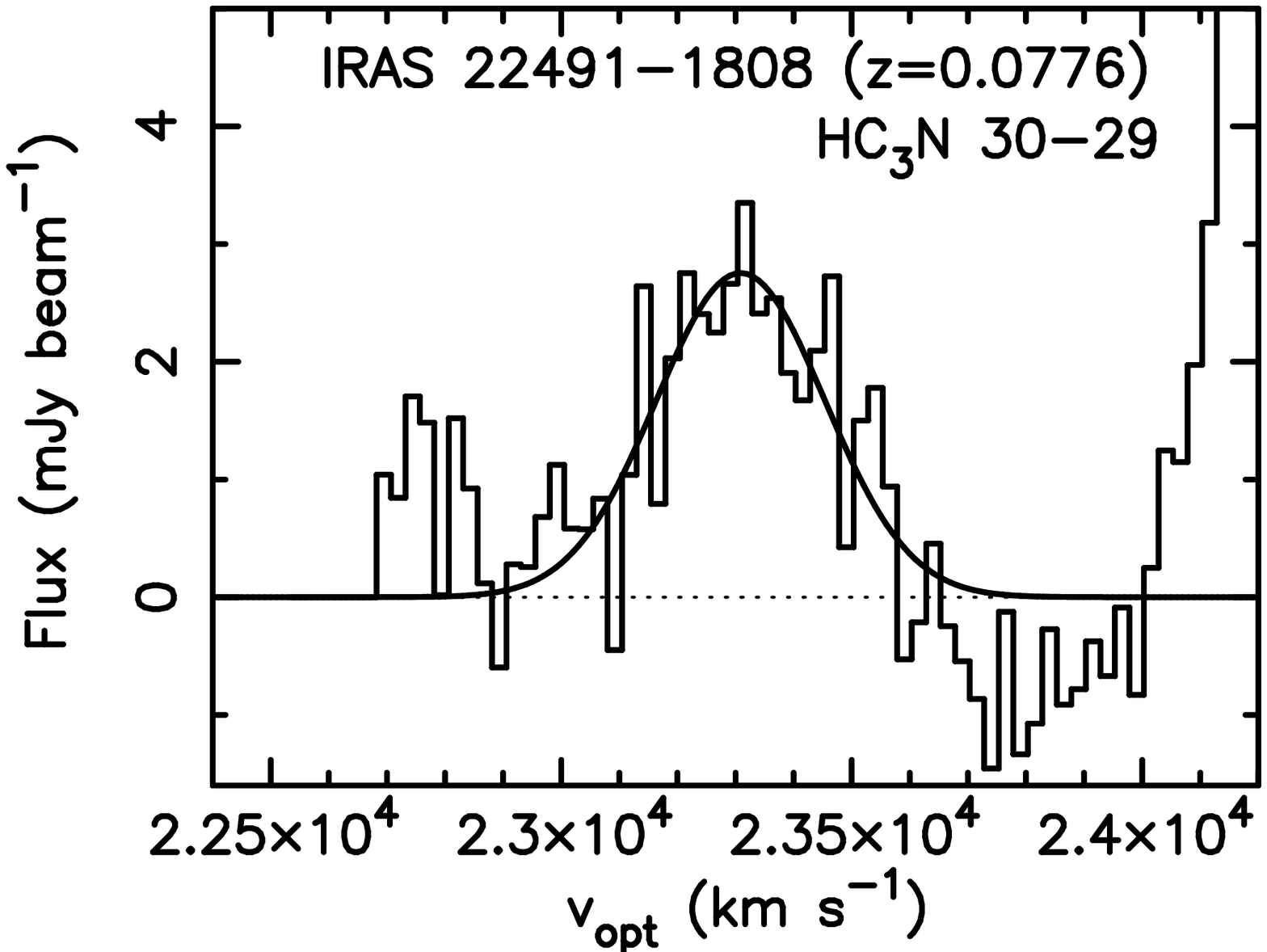} \\
\end{center}
\end{figure}

\clearpage

\begin{figure}
\begin{center}
\includegraphics[angle=0,scale=.273]{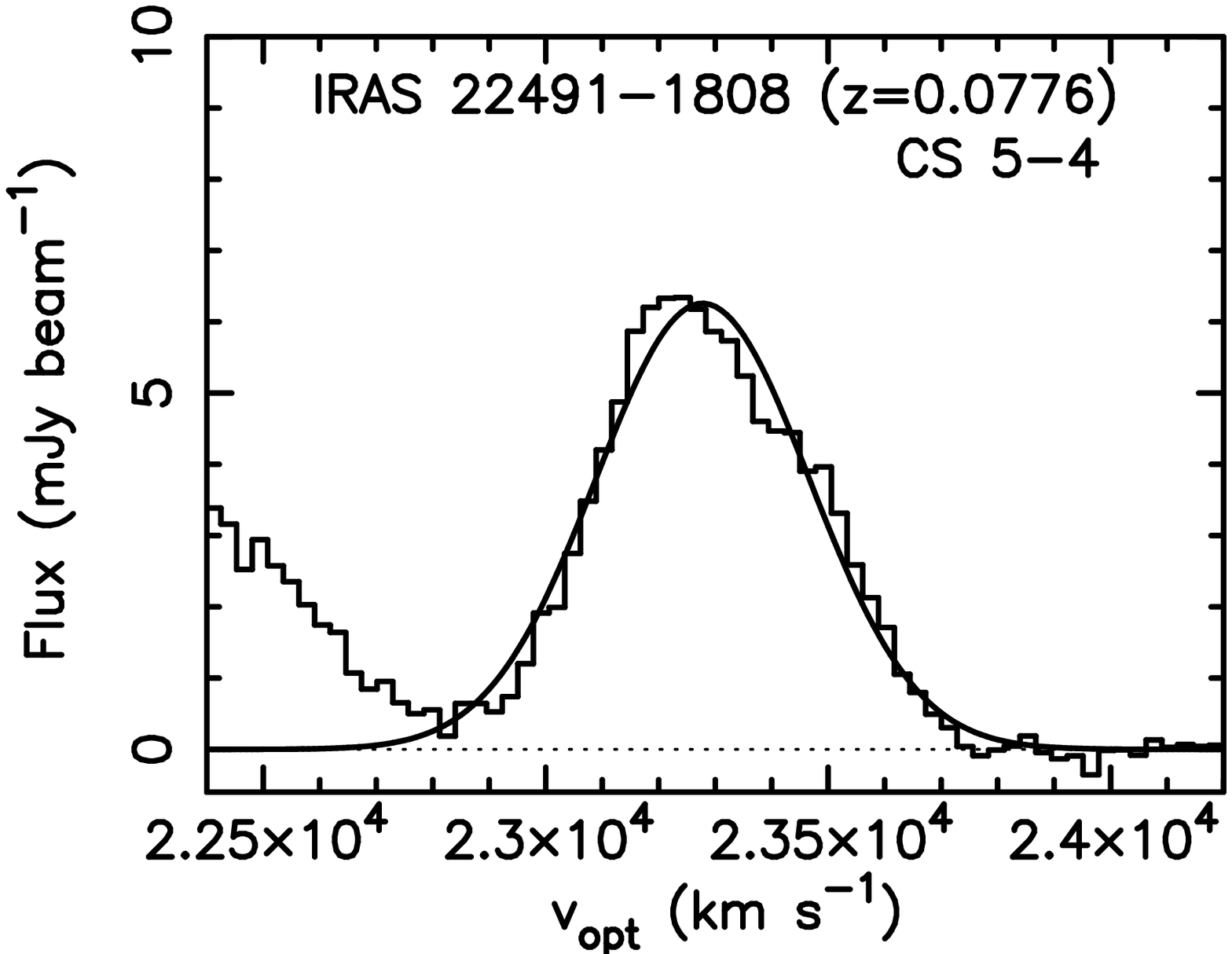}  
\includegraphics[angle=0,scale=.273]{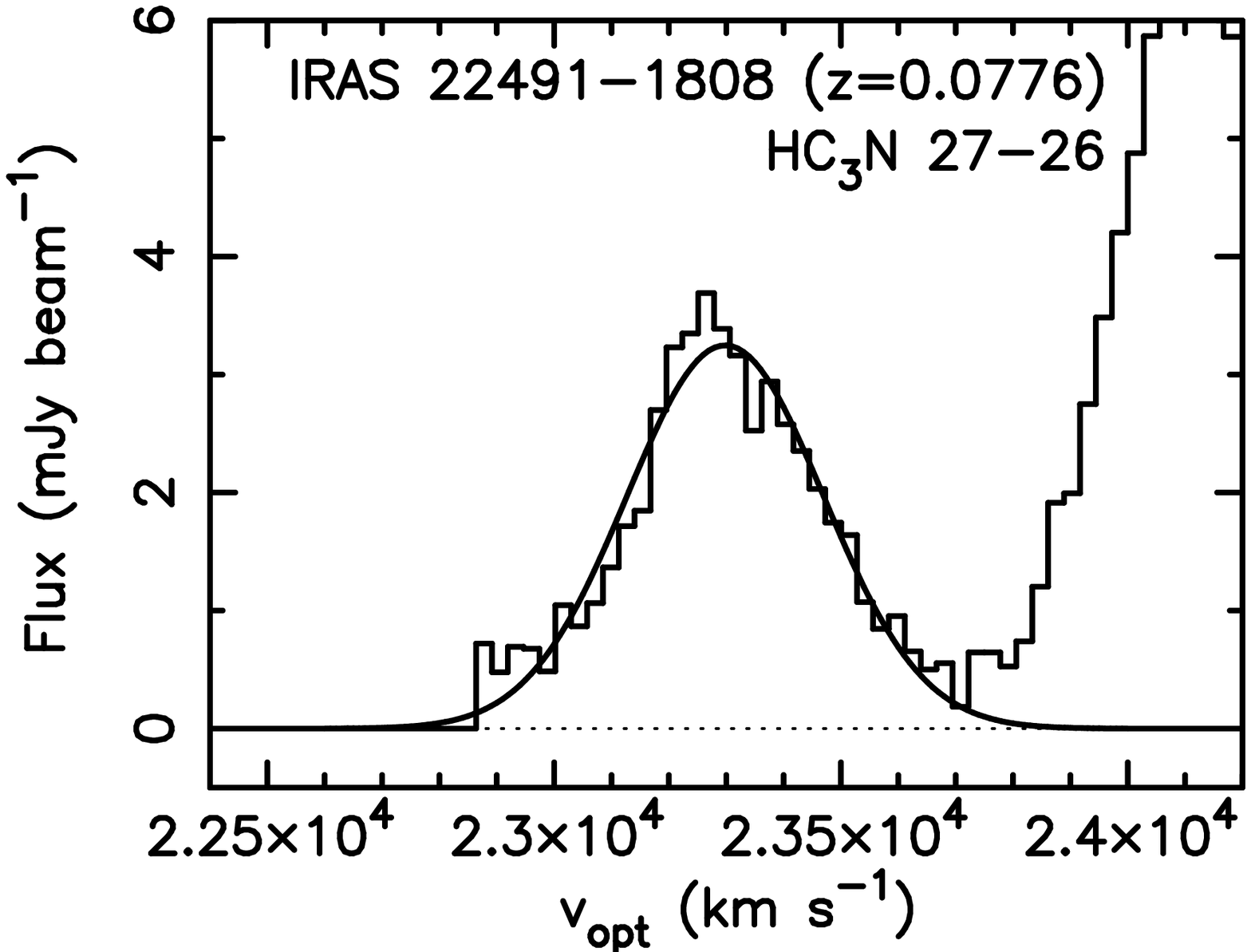} 
\includegraphics[angle=0,scale=.273]{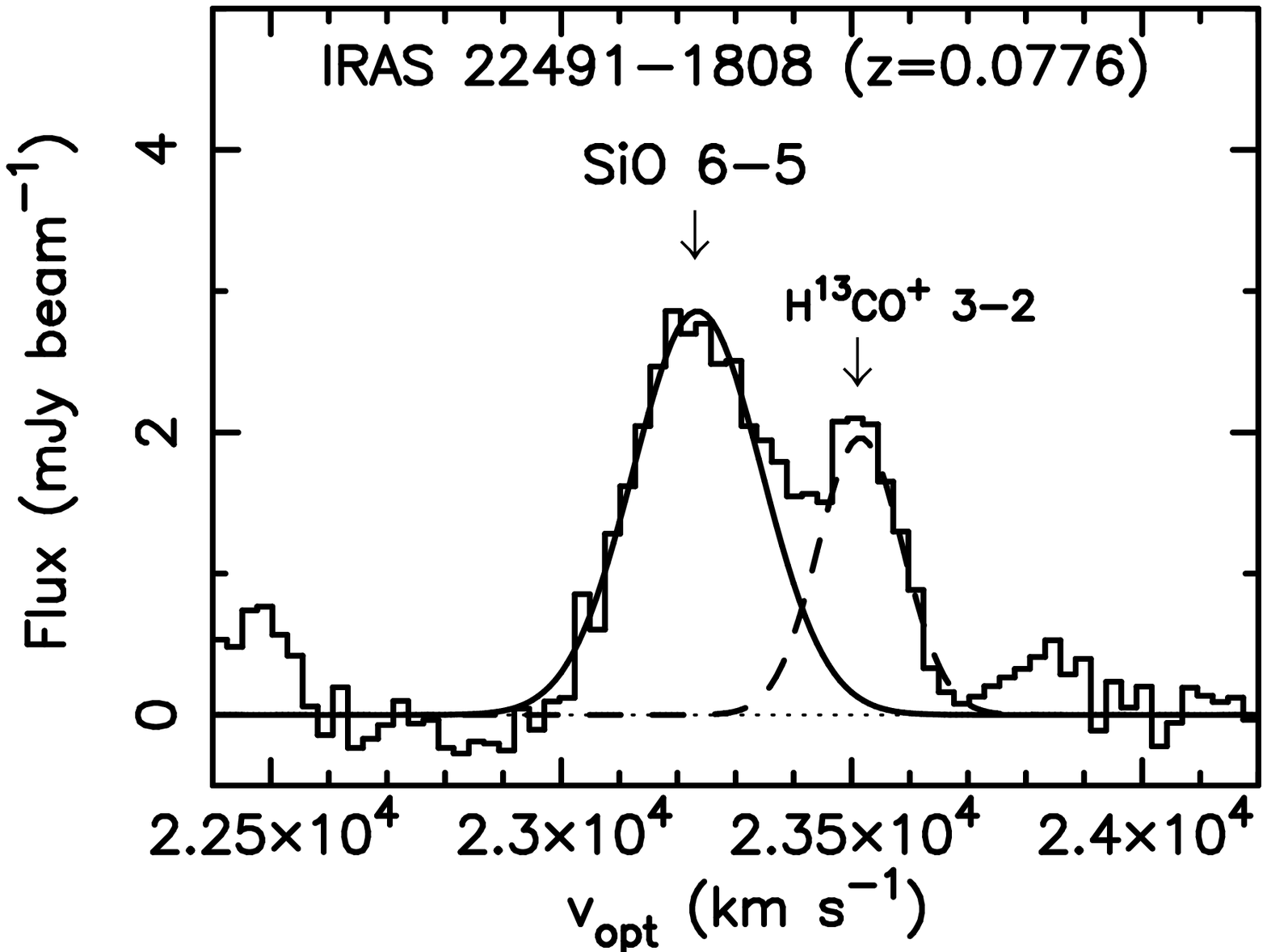} \\
\includegraphics[angle=0,scale=.273]{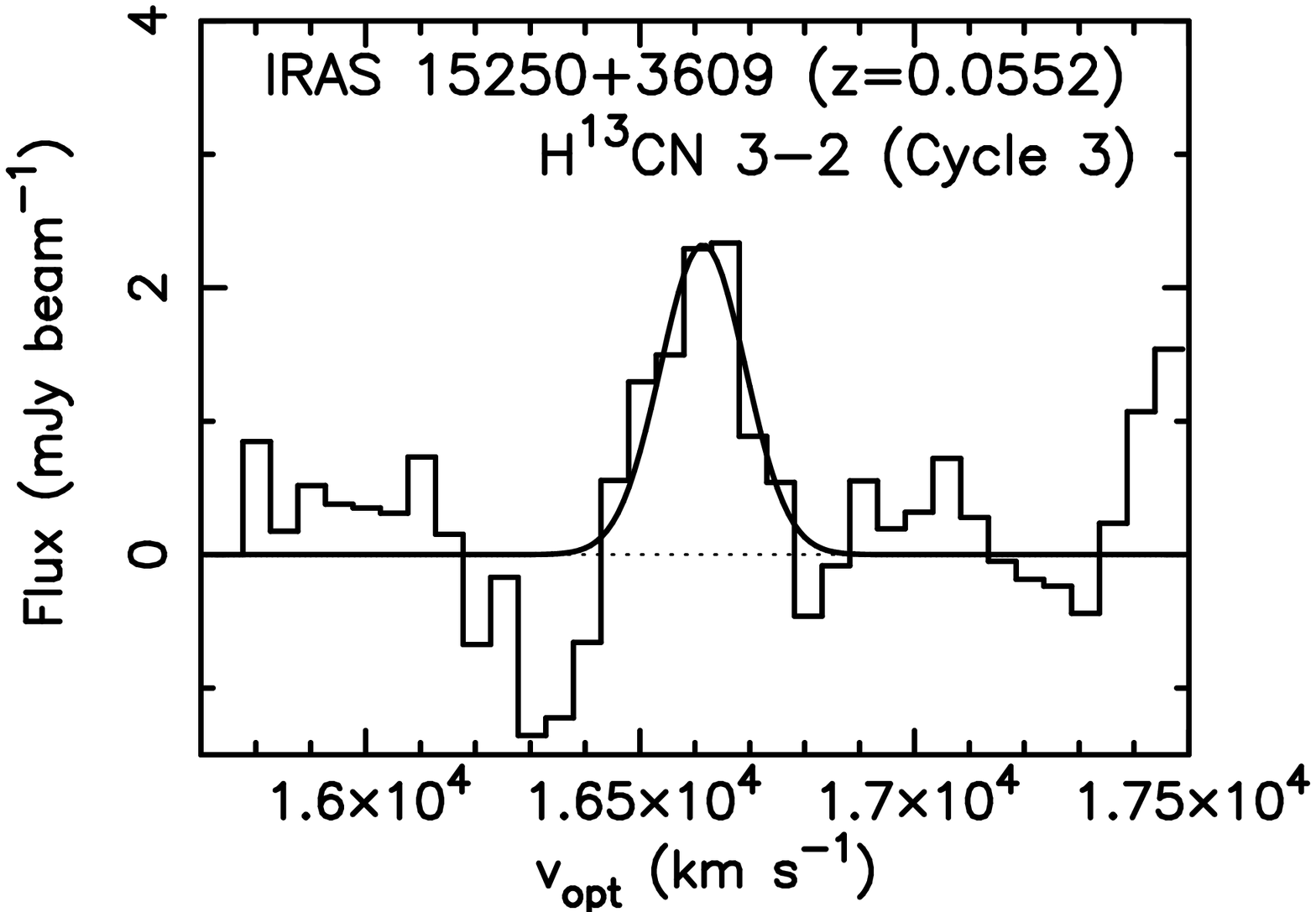} 
\includegraphics[angle=0,scale=.273]{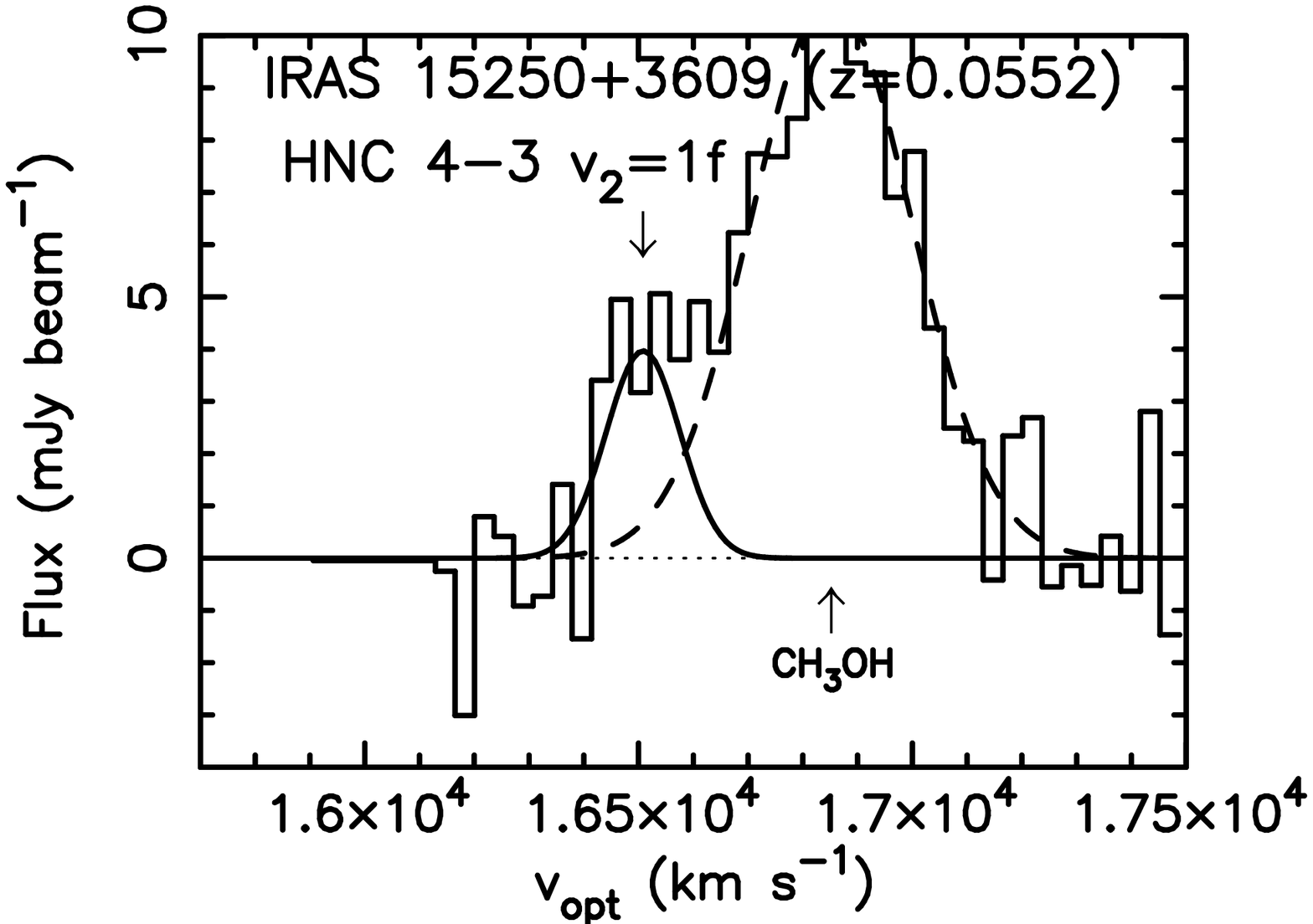}
\includegraphics[angle=0,scale=.273]{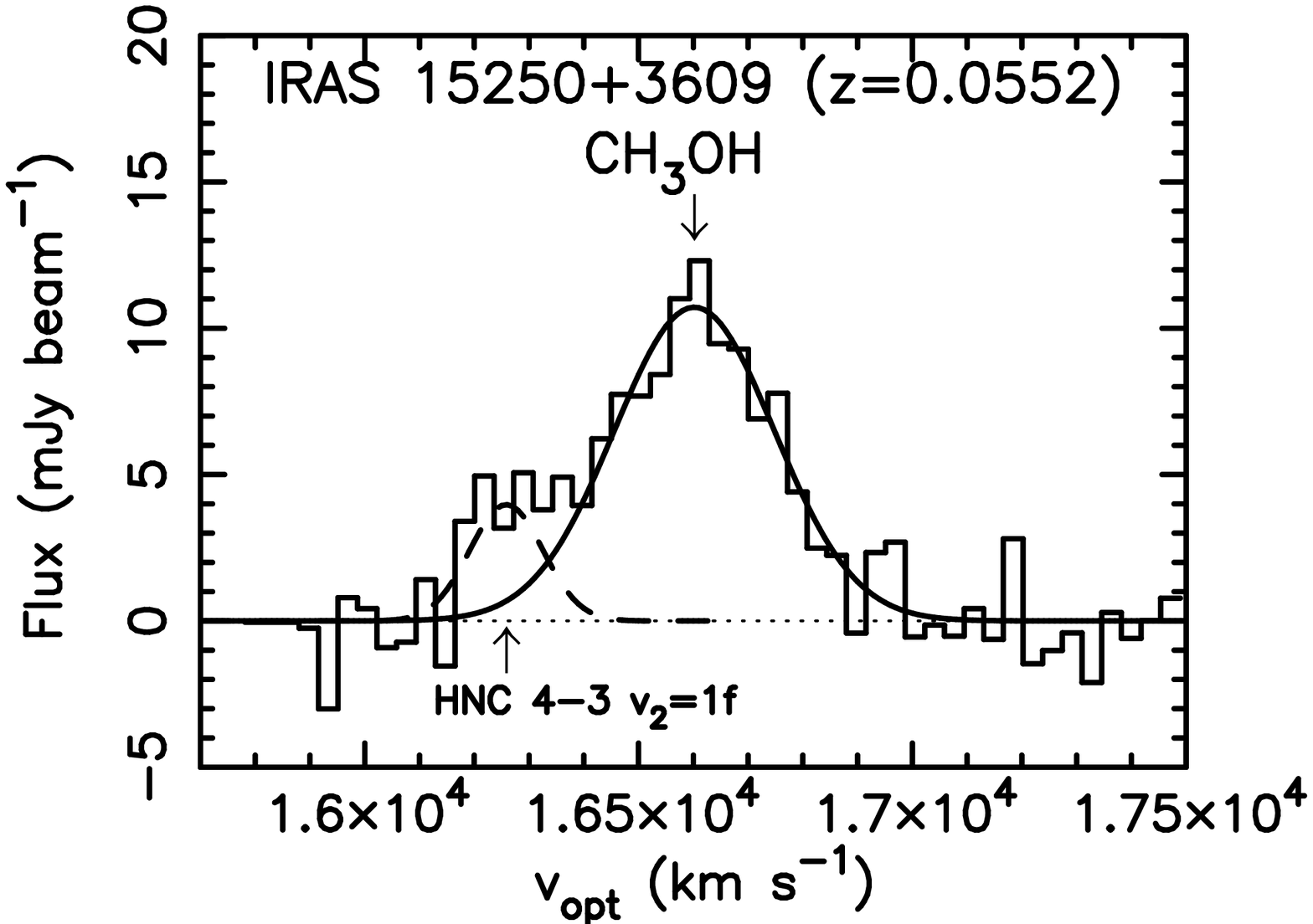} \\
\includegraphics[angle=0,scale=.273]{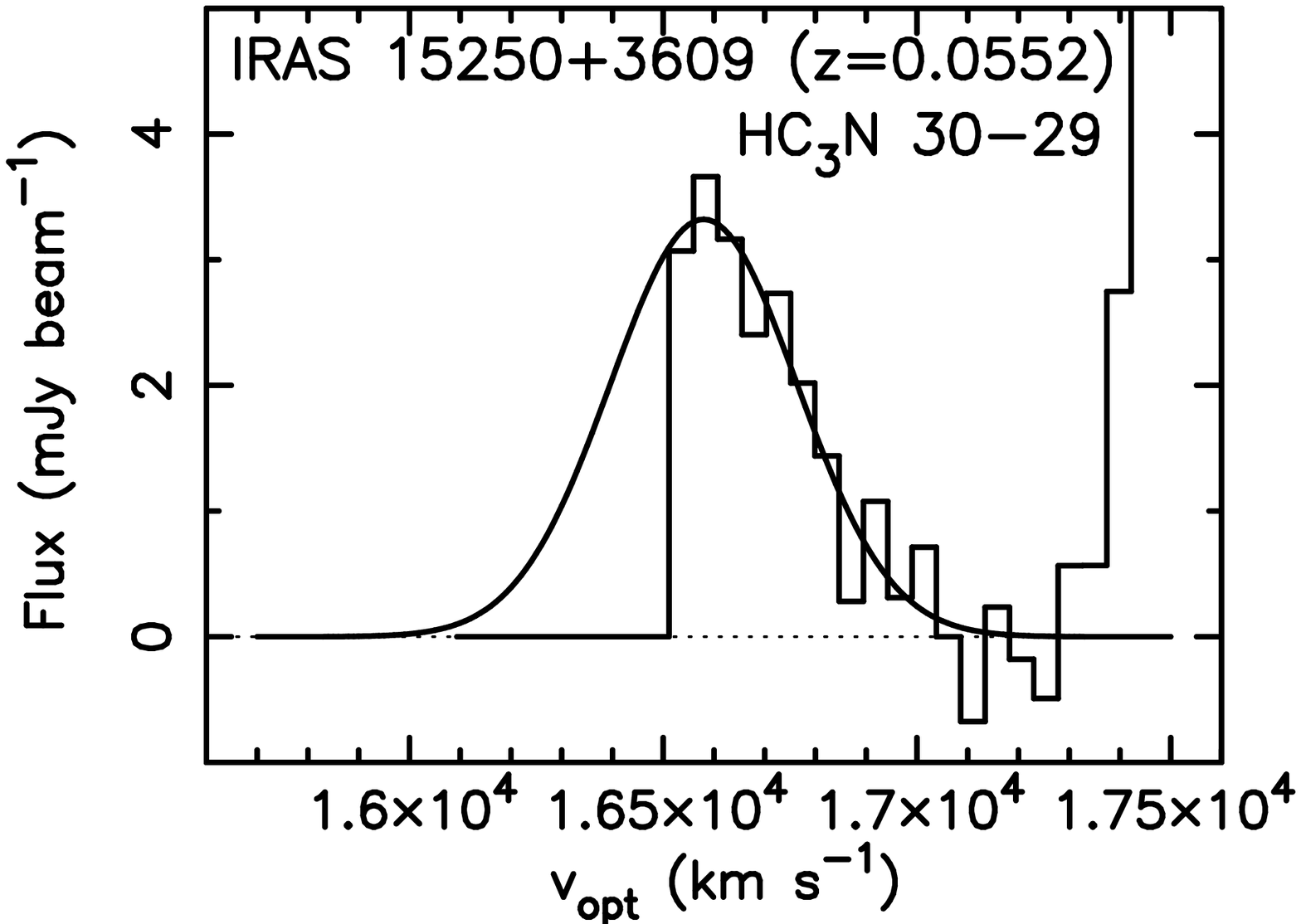} 
\includegraphics[angle=0,scale=.273]{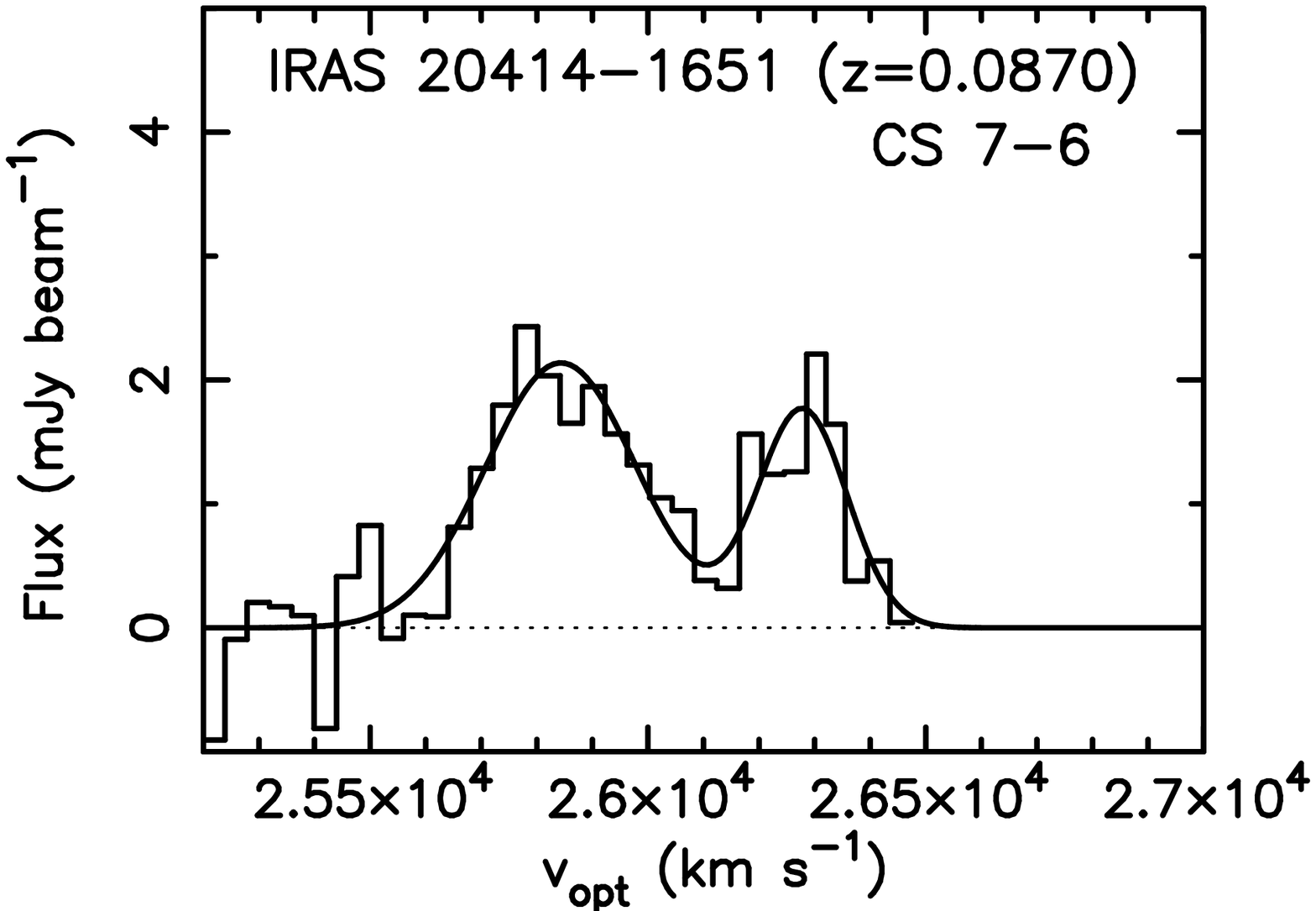} \\
\end{center}
\caption{
Gaussian fits for serendipitously detected emission lines.
The abscissa is optical LSR velocity in (km s$^{-1}$) and the ordinate
is flux in (mJy beam$^{-1}$). 
The presented H$^{13}$CN J=3--2 emission lines for IRAS 12112$+$0305 NE,
IRAS 22491$-$1808, and IRAS 15250$+$3609 are from Cycle 2 or 3 data 
which are shallower than our Cycle 4 data presented in Figure 18 for the 
former two sources. 
For the SiO J=6--5 lines of IRAS 08572$+$3915 and IRAS 22491$-$1808,
SiO J=6--5 and nearby H$^{13}$CO$^{+}$ J=3--2 emission lines were
simultaneously fit with two Gaussian components. 
The Gaussian fits of the H$^{13}$CO$^{+}$ J=3--2 emission lines are
shown as dashed curved lines. 
The HNC v$_{2}$=1f J=4--3 and CH$_{3}$OH lines in IRAS 15250$+$3609 
were also simultaneously fit.
For the fits of the HNC v$_{2}$=1f J=4--3 and CH$_{3}$OH lines, 
CH$_{3}$OH and HNC v$_{2}$=1f J=4--3 lines are displayed as the dashed
curved lines, respectively.
}
\end{figure}

\begin{figure}
\begin{center}
\includegraphics[angle=0,scale=.39]{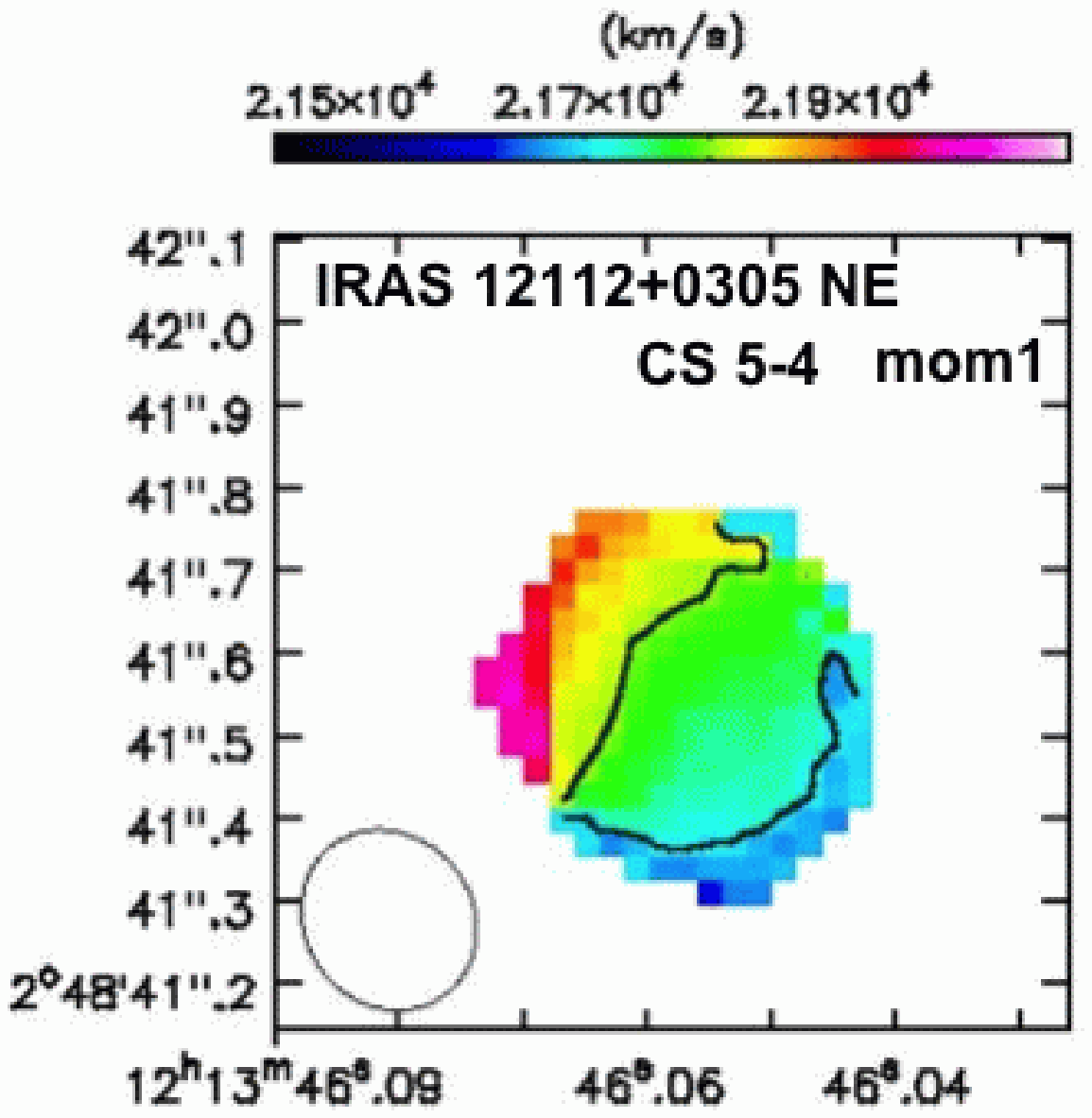} 
\includegraphics[angle=0,scale=.39]{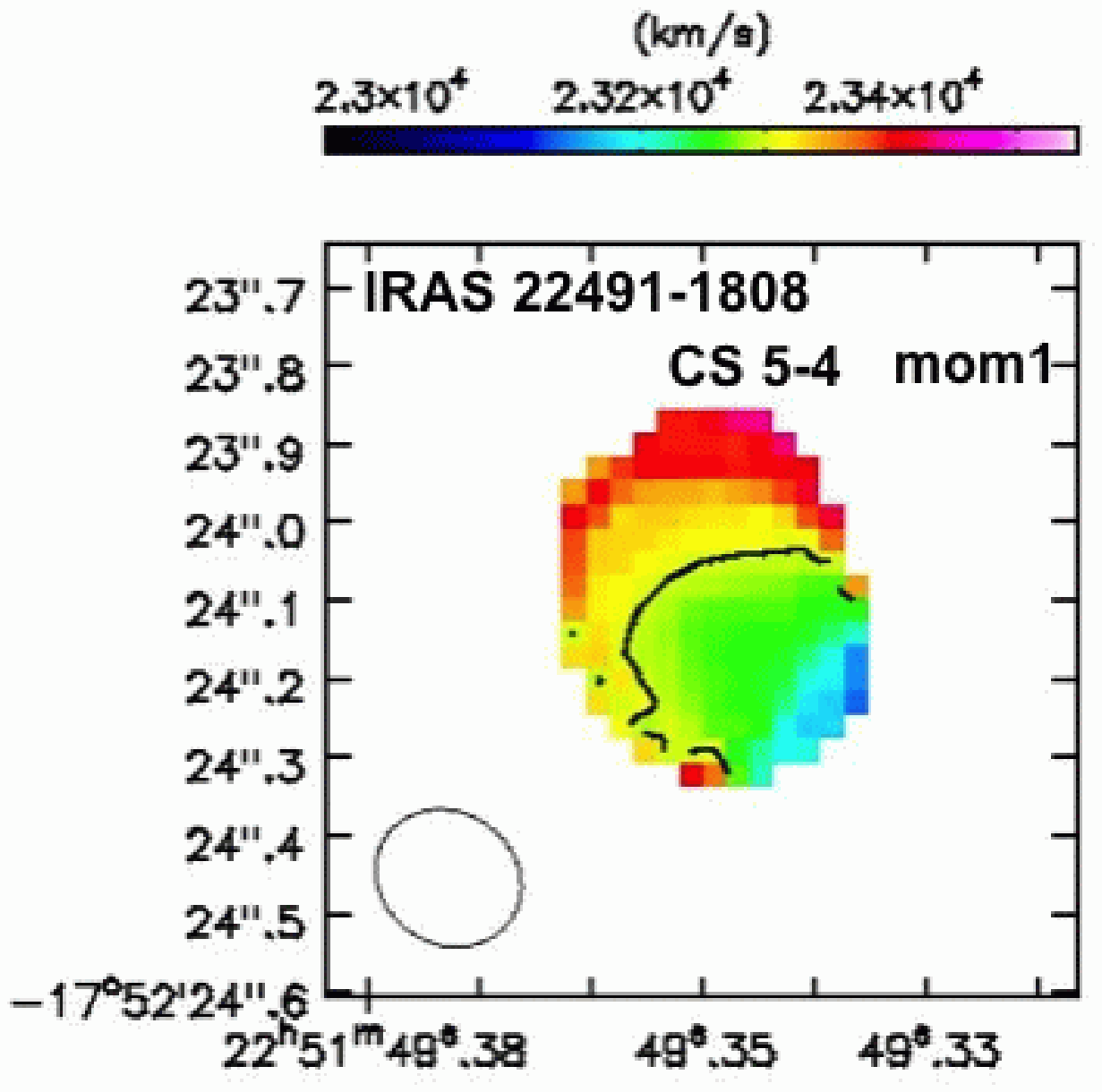} 
\caption{
Intensity-weighted mean velocity (moment 1) maps of the CS J=5--4
emission line for IRAS 12112$+$0305 NE and IRAS 22491$-$1808.
The abscissa and ordinate are R.A. (J2000) and decl. (J2000),
respectively. 
The contours represent 
21700, 21800 km s$^{-1}$ for IRAS 12112$+$0305 NE and 
23300 km s$^{-1}$ for IRAS 22491$-$1808.
Beam sizes are shown as open circles at the lower-left part.
An appropriate cut-off was chosen for these moment 1 maps.
}
\end{center}
\end{figure}

\begin{figure}
\begin{center}
\vspace*{-1.5cm}
\includegraphics[angle=0,scale=.39]{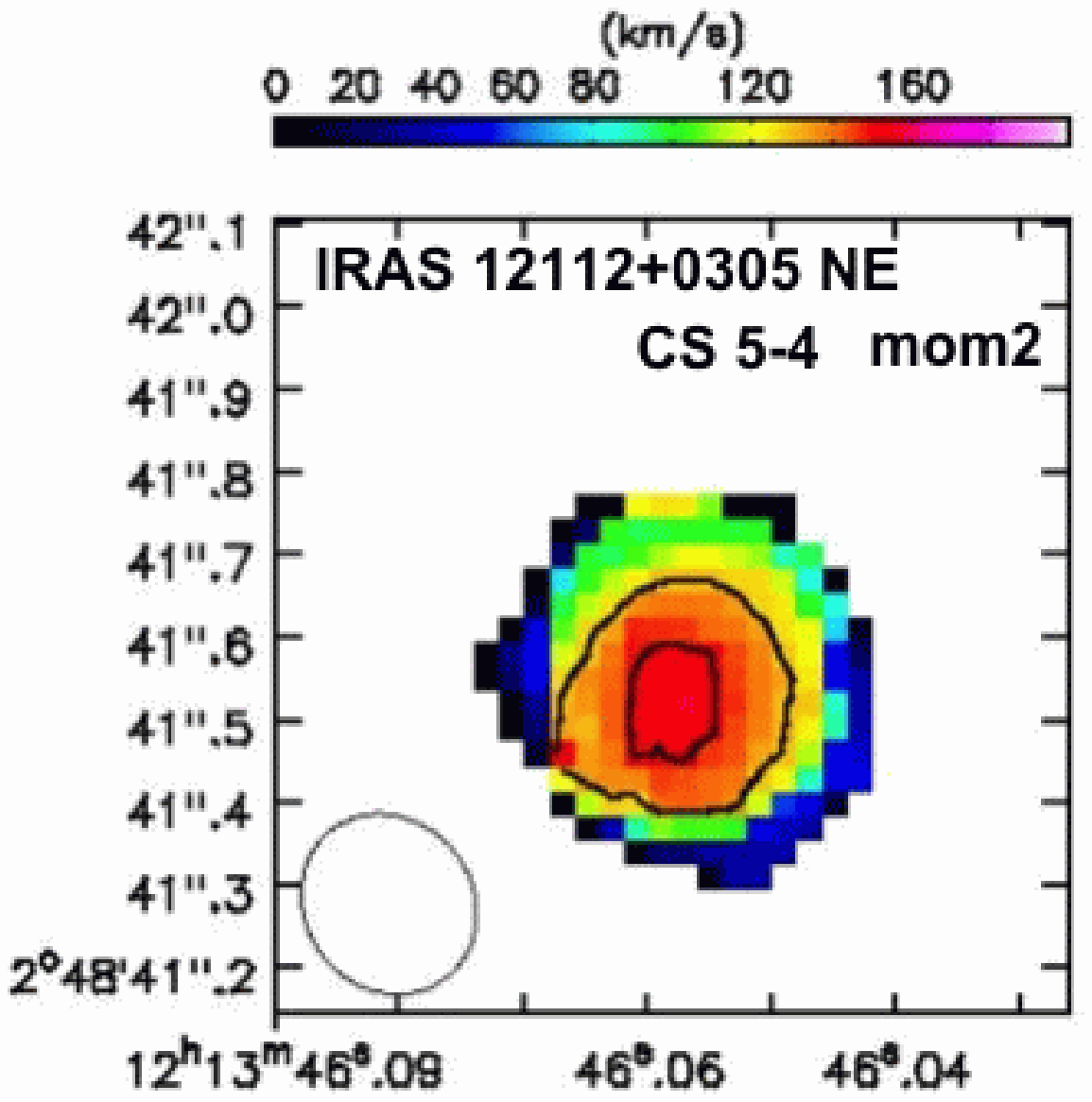} 
\includegraphics[angle=0,scale=.39]{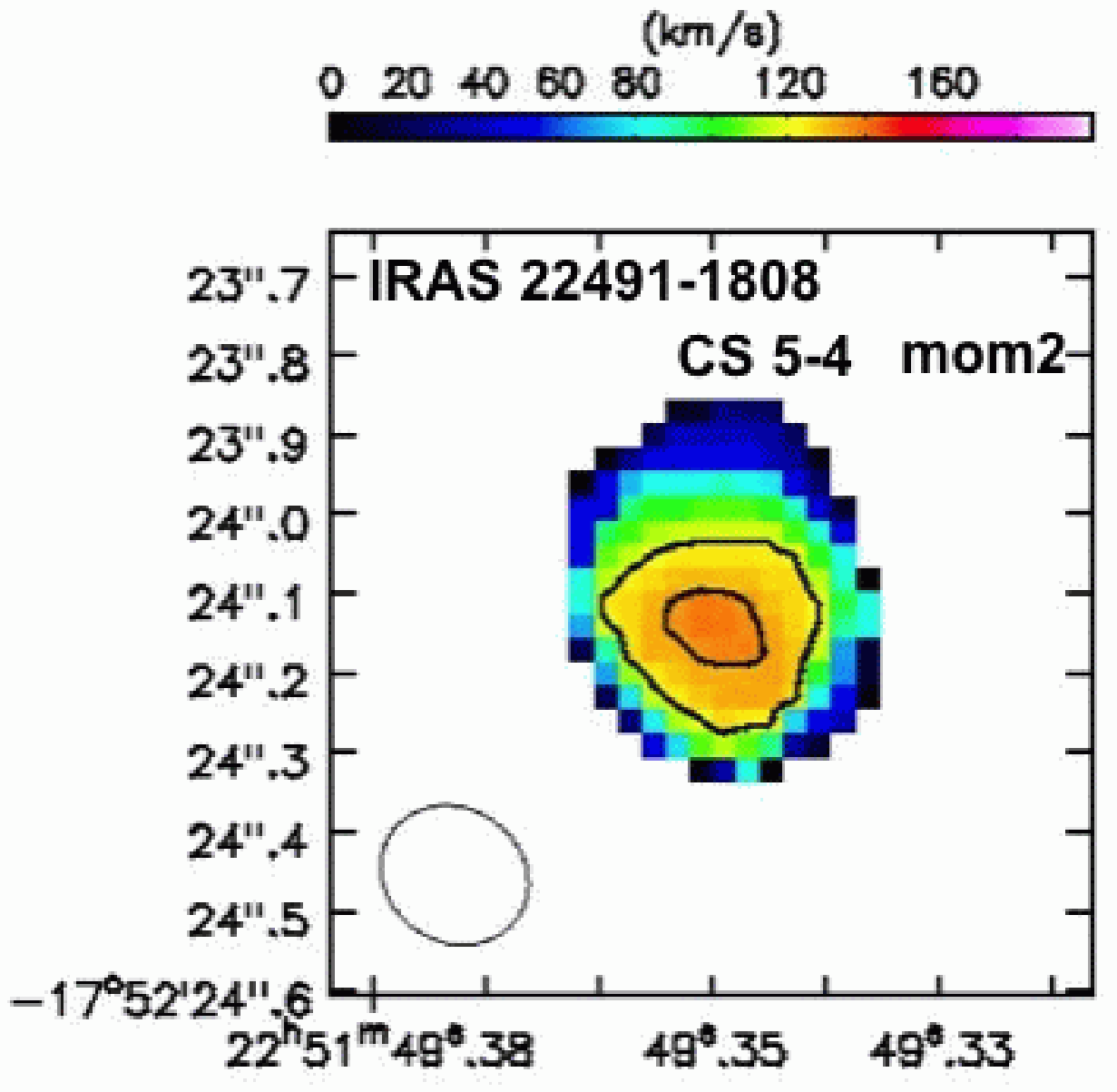} 
\caption{
Intensity-weighted velocity dispersion (moment 2) maps of the CS J=5--4 
emission line for IRAS 12112$+$0305 NE and IRAS 22491$-$1808. 
The abscissa and ordinate are R.A. (J2000) and decl. (J2000),
respectively. 
The contours represent 130, 150 km s$^{-1}$ for IRAS 12112$+$0305 NE and
120, 134 km s$^{-1}$ for IRAS 22491$-$1808.
Beam sizes are shown as open circles at the lower-left part.
An appropriate cut-off was chosen for these moment 2 maps.
}
\end{center}
\end{figure}

\end{document}